\documentclass[aps,prd,nofootinbib,preprintnumbers,superscriptaddress,floatfix]{revtex4-1}

\usepackage{epsfig}
\usepackage{wrapfig}
\usepackage{graphicx}
\usepackage{bm}
\usepackage{amssymb}
\usepackage{amsmath}
\usepackage{amsfonts}
\usepackage[dvipsnames]{xcolor}
\usepackage{dcolumn}
\usepackage{booktabs}
\usepackage[makeroom]{cancel}
\usepackage{float}
\usepackage{xcolor}
\usepackage{tensor}
\usepackage{caption}
\usepackage{subcaption}
\usepackage{array,longtable}
\usepackage{ltcaption}
\makeatletter
\expandafter\let\csname longtable*\endcsname\@undefined
\expandafter\let\csname endlongtable*\endcsname\@undefined
\makeatother

\usepackage[utf8]{inputenc}
\usepackage{soul}
\usepackage{amsmath}
\usepackage{multirow}
\usepackage{makecell}
\usepackage{amssymb}
\usepackage{amsthm}
\usepackage{amscd, amsfonts, mathrsfs}
\usepackage{cases}
\usepackage{verbatim} 
\usepackage{epsf}
\usepackage[bookmarksnumbered,pdfpagelabels=true,plainpages=false,colorlinks=true,
            linkcolor=black,citecolor=black,urlcolor=Blue]{hyperref}
\usepackage{cleveref}

\usepackage{ulem}

\usepackage{enumitem}

\usepackage{upgreek}

\usepackage{orcidlink}
\theoremstyle{plain}
\newtheorem{theorem}{Theorem}[section]

\newtheorem{lemma}{Lemma}[section]
\newtheorem{proposition}{Proposition}[section]

\newtheorem{assumption}{Assumption}[section]

\theoremstyle{definition}
\newtheorem{remark}{Remark}[section]
\newtheorem{definition}{Definition}[section]

\allowdisplaybreaks

\usepackage{appendix}
\newcommand{\cA}{\mathcal A}

\newcommand{\norm}[1]{\left\Vert#1\right\Vert}

\newcommand{\be}{\begin{equation}}
\newcommand{\ee}{\end{equation}}
\newcommand{\bea}{\begin{eqnarray}}
\newcommand{\eea}{\end{eqnarray}}

\newcommand{\bml}{\begin{subequations}}
\newcommand{\eml}{\end{subequations}}

\newcommand{\bbm}{\begin{bmatrix}}
\newcommand{\ebm}{\end{bmatrix}}
\newcommand{\bvm}{\begin{vmatrix}}
\newcommand{\evm}{\end{vmatrix}}

\definecolor{fashionfuchsia}{RGB}{193,84,193}

\newcommand{\GeV}{\rm{GeV}}

\usepackage{acronym}
\acrodef{EX}[EX]{\emph{Example}}
\acrodef{DOF}[DOF]{degree of freedom}
\acrodef{QCD}[QCD]{quantum chromodynamics}
\acrodef{QGP}[QGP]{quark-gluon plasma}
\acrodef{PBC}[PBC]{periodic boundary conditions}
\acrodef{MO}[MO]{mixed--order}
\acrodef{FO}[FO]{first--order}

\usepackage{centernot}

\newcounter{draftpart}

\renewcommand{\thedraftpart}{\Roman{draftpart}}

\newcommand{\draftpart}[1]{%
  \par\bigskip
  \refstepcounter{draftpart}%
  \addtocontents{toc}{\protect\addvspace{1em}}%
  \addcontentsline{toc}{section}{%
    \protect\textbf{PART \thedraftpart: \MakeUppercase{#1}}%
  }%
  \begin{center}
    \large\bfseries
    PART \thedraftpart: \MakeUppercase{#1}
  \end{center}
  \vspace{1em}
}

\begin{document}

\title{Mathematical foundations and numerical simulations of BDNK hydrodynamics: from heavy-ion collisions to astrophysics}

\begin{abstract}
    We present new developments in the mathematical foundations and numerical simulations of the causal and stable first-order relativistic viscous hydrodynamic theory by Bemfica, Disconzi, Noronha, and Kovtun (BDNK) in general spacetimes. We develop two independent flux-conservative algorithms for arbitrary 3+1-dimensional spacetime metrics. Both are hyperbolic, causal, and well-posed, pass convergence tests, and reproduce known results. The first uses a new first-order reduction with mathematically preserved constraints, whose numerical violations can be handled with cleaning fields (the resulting coupled system is proven hyperbolic and well-posed). The second uses a new mixed-order formulation, whose solutions are proven strictly equivalent to the original BDNK system, without cleaning fields. Our proofs rely on a new general theorem providing sufficient conditions for transforming a mixed-order system with arbitrary-order equations into a strictly hyperbolic system, strengthening a classical result by Choquet Bruhat for weak hyperbolicity. We introduce a new practical prescription for selecting admissible hydrodynamic frames for a broad class of equations of state, including nonconformal ones relevant to heavy-ion collisions and astrophysics. Our simulations capture viscous shock broadening and suppression of Kelvin--Helmholtz vortices, reproduce Bjorken and Gubser flows, and study departures from stationary ideal-fluid configurations in Michel accretion and rotating tori on fixed Schwarzschild and Kerr backgrounds. For heavy-ion collisions, we perform boost invariant simulations with a lattice-QCD-based equation of state and fluctuating TRENTo initial conditions, comparing BDNK with a minimal Israel--Stewart theory. The descriptions agree at small shear viscosity but can differ substantially at larger viscosities, particularly for early initialization with rapid longitudinal expansion.  
\end{abstract}

\date{\today}

\author{Teerthal Patel}
\email{teerthal.patel@vanderbilt.edu}
\affiliation{Department of Mathematics, Vanderbilt University, USA}

\author{Henry V. Hirvonen}
\email{henry.v.hirvonen@vanderbilt.edu}
\affiliation{Department of Mathematics, Vanderbilt University, USA}

\author{F\'abio S. Bemfica}
\email{fabio.bemfica@ufrn.br}
\affiliation{Escola de Ci\^encias e Tecnologia, Universidade Federal do Rio Grande do Norte, Brazil}

\author{Marcelo M. Disconzi}
\email{marcelo.disconzi@vanderbilt.edu}
\affiliation{Department of Mathematics, Vanderbilt University, USA}

\author{Jorge Noronha}
\email{jn0508@illinois.edu}
\affiliation{Illinois Center for Advanced Studies of the Universe \& Department of Physics,
University of Illinois Urbana-Champaign, USA}

\author{Jean-Fran\c{c}ois Paquet}
\email{jean-francois.paquet@vanderbilt.edu}
\affiliation{Department of Physics, Vanderbilt University, USA}

\maketitle

\tableofcontents

\section{Introduction}
\label{S:Introduction}

Relativistic fluid dynamics is a comprehensive framework employed in many areas of physics -- see the monographs \cite{Rezzolla-Zanotti-Book-2013,Romatschke:2017ejr,Weinberg:1972kfs,Landau-Lifshitz-Book-Fluids-1987} for a plethora of examples and applications. Natural to the study of fluids is the investigation of effects due to viscosity and dissipation. Models of relativistic viscous fluids go back to the early works of Eckart \cite{Eckart:1940te}, Landau and Lifshitz \cite[Chapter XV]{Landau-Lifshitz-Book-Fluids-1987}, and Lichnerowicz \cite[Chapter V]{Lichnerowicz-Book-1955} in the '40s and '50s. Such viscous effects, however, are often neglected in many applications, in particular in astrophysics. 
One area where the effect of dissipation has been found to be considerable is for the  \ac{QGP} produced in ultrarelativistic heavy-ion collisions~\cite{Adcox:2004mh,Back:2004je,Arsene:2004fa,STAR:2005gfr,Gyulassy:2004zy}.
The need for considering astrophysical models with viscosity became more prominent in view of numerical and theoretical evidence that viscosity may affect the gravitational waves emitted during binary neutron star coalescence \cite{Alford:2017rxf,Most:2021zvc,Most:2022yhe,Hammond:2022uua}, which may be within the sensitivity range of the next generation of gravitational wave detectors 
\cite{MUSES:2023hyz}. Such evidence has been only strengthened since then \cite{Chabanov:2023blf,Shibata:2017xht,Shibata:2017jyf,Alford:2020lla,Alford:2019kdw,Alford:2019qtm,Alford:2020pld,Ripley:2023qxo,Camelio:2022ljs,Camelio:2022fds,Huth:2021bsp,Dexheimer:2020zzs,MUSES:2023hyz,Huang:2022mqp,Ripley:2023lsq,Yang:2023ogo,HegadeKR:2023glb,Gavassino:2023xkt,Yang:2025yoo,HegadeKR:2026iou}. 

However, formulating sound theories of relativistic viscous fluids turned out to be a daunting enterprise (see \cite{Rocha:2023ilf} for a review). A large class of relativistic viscous fluid theories, including the aforementioned ones by Eckart, Landau-Lifshitz, and Lichnerowicz, turned out to be acausal, unstable \cite{Hiscock:1985zz}, and ill-posed \cite{Pichon-1965,Bemfica:2020zjp}. Causality is a fundamental postulate in relativity stating that information cannot propagate faster than the speed of light; stability is the statement that small deviations from thermodynamic equilibrium cannot grow over time; and well-posedness means that for each set of initial data, the equations of motion admit a solution that is unique and behaves well in a sense that can be made precise (ill-posedness, on the other hand, is the absence of well-posedness). We elaborate on these concepts below in Section~\ref{S:LWP_and_all_that}.

An approach that has fared better was proposed by Israel and Stewart in the '70s \cite{Stewart-1977,Israel-Stewart-1979,Stewart-Stewart-1976,Israel:1976tn,Israel:1979wp} building on a phenomenological approach that takes into account second-order deviations from equilibrium to the entropy current and a relativistic version of Grad's moments method \cite{Grad-1949} (see also earlier non-relativistic work by M\"uller \cite{Mueller-1967}). Modern incarnations of the Israel-Stewart theory frequently used in heavy-ion collision simulations are the DNMR theory developed by 
Denicol, Niemi, Molnar, and Rischke \cite{Denicol:2012cn} and the re-summed BRSSS theory \cite{Baier:2007ix} developed by
Baier, Romatschke, Son, Starinets, and Stephanov. Although these theories differ on specifics, they are often collectively referred to as Israel-Stewart theory(ies) or (transient) \emph{second-order} (relativistic viscous) theories because they all share the same basic core assumptions, as explained below.
Before discussing advancements brought about by second-order theories, let us highlight their key innovative aspect. 

A theory of relativistic fluids provides evolution laws for the fluid's energy density $\varepsilon$ and flow velocity $u^\mu$. These are encoded in an energy-momentum tensor $T^{\mu\nu}$ that establishes basic constitutive relations among $\varepsilon$ and $u^\mu$, and $T^{\mu\nu}$.
The equations of motion are the usual conservation of energy and momentum
\begin{align}
    \label{Basic_EoM_fluid}
    \nabla_\mu T^{\mu\nu} = 0.
\end{align}
A different choice of thermodynamic scalars, such as for example the fluid's temperature $T$ or its pressure $P$, can be made in place of $\varepsilon$. We can also include a conserved current $J^\mu$, but we avoid it in this introduction for the sake of simplicity (see Section~\ref{BDNK_theory_section}). Finally, $T^{\mu\nu}$ also involves the spacetime metric $g_{\mu\nu}$, which can be a dynamical variable when coupling to Einstein's equations is considered but here in this section we take it to be fixed for simplicity (although spacetime does not need to be flat).

The fields $\varepsilon,u^\mu$ (or their alternates $T,P$, etc.) are called \emph{hydrodynamic fields.} 
In the works by Eckart, Landau-Lifshitz, Lichnerowicz, and similar approaches, viscous and dissipative effects are modeled by constitutive relations involving derivatives of the hydrodynamic fields in $T^{\mu\nu}$. This is an approach inspired by the well-known non-relativistic Navier-Stokes equations, wherein the shear viscosity, for example, is modeled by the term $\nabla^2 u^i$ in the equations of motion (e.g. for incompressible fluids). Fluid theories where the only variables governing the evolution are the hydrodynamic fields, and with constitutive relations defined in terms of first-order derivatives of those fields,  are called \emph{first-order hydrodynamic theories}, and it was long assumed that all such theories were acausal and unstable. 

Second-order theories, such as in transient relativistic hydrodynamics, differ drastically from first-order theories in the way they parametrize the fluid's configuration space. In addition to the hydrodynamic fields, they introduce new fields $\Pi$, $\pi^{\mu\nu}$, and $q^\mu$ modeling, respectively, the fluid's bulk scalar, shear stress tensor, and energy diffusion (assume, for example, the Eckart frame), so that now $T^{\mu\nu} = T^{\mu\nu}(T,u^\mu, \Pi, \pi^{\mu\nu}, q^\mu)$. In addition to Equation~\eqref{Basic_EoM_fluid}, new equations of motion are then introduced to account for the new degrees of freedom $\Pi,\pi^{\mu\nu}, q^\mu$. Such equations of motion can be postulated based on physical principles or derived from a more microscopic approach (e.g., kinetic theory). The form of the equations of motion for $\Pi$, $\pi^{\mu\nu}$, and $q^\mu$ distinguishes  different second-order theories and, even though various types of equations are possible, the equations of motion are often nonlinear relaxation equations with relaxation times as additional second-order transport coefficients. These equations of motion can moreover involve dozens of other second-order transport coefficients~\cite{Denicol:2012cn,Molnar:2013lta,Denicol:2014vaa,Bhattacharyya:2007vjd,Baier:2007ix,Rocha:2023ilf}.

The causality and stability properties of second-order hydrodynamic theories have been studied extensively in the linear regime around local equilibrium as well as non-linearly~\cite{Hiscock:1983zz,Hiscock:1987zz,Olson:1990rzl,Denicol:2008ha,Pu:2009fj,Floerchinger:2017cii,Bemfica:2019cop,Bemfica:2020xym,Gavassino:2023odx}. It had long been known that certain constraints on the second-order transport coefficients, such as the relaxation times, had to be enforced to maintain causality and stability in the linear regime~\cite{Hiscock:1983zz,Hiscock:1987zz}, but recent works have obtained additional conditions for non-linear causality~\cite{Bemfica:2019cop,Bemfica:2020xym,Cordeiro:2023ljz,Cordeiro:2026jle}. Because these conditions also involve the dissipative stresses themselves\footnote{This follows from the fact that in such theories the principal part depends both on the hydrodynamic fields and the dissipative stresses, see for example the detailed explanation in \cite{Cordeiro:2026jle}.}, they cannot be enforced solely with constraints on the equation of state (EOS) or the transport coefficients. This implies that in such second-order theories, conditions obtained in the linear case near equilibrium \cite{Hiscock:1983zz,Hiscock:1987zz,Olson:1990rzl,Denicol:2008ha,Pu:2009fj} are not sufficient to ensure causality in the nonlinear regime.

For practical applications, in relativistic heavy-ion collisions for example, it can be challenging to ensure causal evolution in simulations of second-order hydrodynamic theories. At a minimum, one must verify that the initial conditions of the dynamical fields satisfy non-linear causality constraints \cite{Bemfica:2020xym,Cordeiro:2026jle} in addition to imposing that the transport coefficients and the EOS  satisfy the linear stability conditions. However, even under these conditions, acausal propagation can occur later during the hydrodynamic evolution~\cite{Gavassino:2025bsn}, and one in principle has to verify numerically that causality was maintained throughout the fluid's spacetime evolution (as first investigated in~\cite{Plumberg:2021bme,Chiu:2021muk,ExTrEMe:2023nhy}) to claim causal evolution in most second-order hydrodynamic formalisms.
Importantly, recent works~\cite{Bemfica:2020zjp,Gavassino:2021owo} have established a strong connection between causality and stability in Lorentz covariant theories, and the presence of spurious instabilities is a crucial consideration for physical models \cite{Gavassino:2025bsn}. 
In fact, because numerical implementations of the equations solve a discretized version of the PDEs, unstable modes may inadvertently be suppressed at finite spacetime resolution, or by various implementation choices~\cite{Gavassino:2025bsn}.

More recently, Bemfica, Disconzi and Noronha~\cite{Bemfica:2017wps,Bemfica:2019knx,Bemfica:2020zjp} and Kovtun~\cite{Kovtun:2019hdm} (BDNK) showed that a second-order formalism is not necessary to systematically obtain a causal and covariantly stable theory of relativistic viscous hydrodynamics. The BDNK formalism that they put forward is first-order in derivatives, does not have the additional 
independent dissipative degrees of freedom of second-order hydrodynamics, and
has been shown to have well-defined conditions under which non-linear causality and linear stability can be guaranteed. The key ingredient is the freedom in the choice of \emph{hydrodynamic frame}, which defines the hydrodynamic variables---temperature, fluid velocity, and, when applicable, chemical potential---away from equilibrium. For example, the flow velocity does not have to be defined as the flow of energy or conserved charge, as is done respectively in the Landau and Eckart frames. The BDNK formalism provides conditions that hydrodynamic frames have to satisfy in a first-order theory to remain causal and stable: the corresponding \emph{frame parameters} must be chosen to satisfy certain inequalities given the fluid's equation of state and transport coefficients. This defines an infinite set of admissible hydrodynamic frames that lead to causal and stable evolution still within the range of validity of first-order hydrodynamics.  

The BDNK formalism has already been used and studied broadly in recent years. For example, it was compared in detail with Israel-Stewart hydrodynamics for a Bjorken-symmetric fluid in Ref.\ \cite{Das:2020fnr}. Subsequently, extensive investigations by Pandya, Most and Pretorius \cite{Pandya-21,Pandya_2022,Pandya_2022_IG}  studied different numerical approaches and their performance for various fluid geometries, including shock tubes, Kelvin--Helmholtz instability and Bjorken expansion.
BDNK's connections with microscopic theories were studied in~\cite{Hoult:2021gnb,Bantilan:2022ech,Rocha:2022ind,Rocha:2023hts}.
BDNK hydrodynamics was also used to study the viscous hydrodynamic regime of neutron stars~\cite{Redondo-Yuste:2024vdb,Shum:2025jnl,caballero2025neutronstarradialperturbations,Keeble:2026zms} or other compact objects~\cite{redondoyuste2025superradiantamplificationrotatingviscous}, leveraging the approximate spherical symmetry of the corresponding systems.
A comparison between kinetic theory, Israel-Stewart, BDNK and Density Frame hydrodynamics  \cite{Armas:2020mpr} was performed in \cite{Bhambure:2024axa} for a Cartesian 1+1D system, while a different study 
in 1+1D was done in~\cite{bea2025relativisticnavierstokesdescriptionquarkgluon}. Other properties of BDNK that have been explored include its non-relativistic limit~\cite{HegadeKR:2023glb}, its relation to Israel-Stewart hydrodynamics in the Landau frame~\cite{Bhattacharyya:2023srn} and other hydrodynamic frames \cite{Noronha:2021syv}, chiral hydrodynamics~\cite{Abboud:2023hos,Speranza:2021bxf}, shock waves~\cite{Friestuhler-BDNK-shock,Clarisse:2025lli}, thermal fluctuations~\cite{Mullins:2023ott,Jain:2023obu,Gavassino:2024ufs,Gavassino:2024vyu}, its extension to dissipative magnetohydrodynamics~\cite{Armas:2022wvb,Hoult:2024qph,Lier:2026owd,Panda:2026lbp}, cosmology \cite{Bemfica:2022dnk}, as well as other analytical studies~\cite{Fantini:2025gnm,Schianchi:2026ayx,Armas:2026kfp}.
In relativistic heavy-ion physics, mature numerical frameworks~\cite{Song:2007ux,Song:2010aq,Karpenko:2013wva,DelZanna:2013eua,Shen:2014vra,Pang:2018zzo,Du:2019obx,Nijs:2020roc,Noronha-Hostler:2013gga,Noronha-Hostler:2014dqa,Pala:2025qoa,Schenke:2010nt,Schenke:2010rr,Paquet:2015lta} based on Israel-Stewart-type hydrodynamics have been extensively developed and validated, and are routinely employed to interpret experimental observables. 
While significant progress has been made in numerical simulations of BDNK in the works cited above, the development of comparably mature and general-purpose numerical frameworks remains an important goal that is advanced greatly by this work.

A crucial aspect from a numerical perspective is that,
in contrast to Israel-Stewart hydrodynamics, which is a system of first-order partial differential equations, BDNK is a set of second-order differential equations. Israel-Stewart has more degrees of freedom than BDNK --- ten for a fluid without additional conserved charges in the Landau frame ---  but as a first-order differential equation, it only requires initial values for these fields. In contrast, BDNK has only four degrees of freedom --- the energy density and the flow velocity --- but as a second-order differential system, it requires both the initial value and the initial time derivative of the fields.
Nevertheless, BDNK can be reformulated as a system of first-order differential equations in conservation (flux-conservative) form $\partial_t Q+\partial_a F_a(Q)=S(Q)$, see \cite{Bemfica:2020zjp}. This \emph{first-order reformulation} can then be solved using the well-developed high-resolution shock-capturing methods that have long been used to solve relativistic hydrodynamics~\cite{Rezzolla-Zanotti-Book-2013}. Alternatively, in this work we put forward a \emph{mixed-order formulation} of BDNK which is first-order in time and second-order in spatial gradients, and can be expressed in a conservation-like form $\partial_t Q+\partial_a F_a(Q,\partial_b Q)=S(Q,\partial_b Q)$, which can be solved with other well-studied methods. These developments build on previous works, especially from Pandya, Most and Pretorius~\cite{Pandya-21,Pandya_2022,Pandya_2022_IG}, and results using the first-order formalism were previously presented in \cite{Clarisse:2025lli} for a 1+1D system.

In this work, we present numerical implementations of BDNK for both the first-order and mixed-order formulations, with the goal of establishing a general numerical framework applicable in $3+1$ dimensions, without symmetry assumptions, and allowing for arbitrary spacetime metrics. Such a framework is an important step toward future fully dynamical simulations of BDNK coupled to Einstein's equations.
An additional advantage of developing two independent formulations, with distinct numerical properties, is that their solutions can be quantitatively compared to assess numerical reliability. This is particularly valuable for systems where analytical or semi-analytical solutions are unavailable and independent validation is difficult.
Importantly, we also provide a prescription for choosing BDNK hydrodynamic frame parameters that satisfy the causality conditions for several equations of state and transport coefficients, with certain restrictions, and which are relevant for both nuclear physics and astrophysics. 

This work presents a large number of new results both in the context of the mathematics of PDEs relevant for relativistic fluid dynamics, and in the corresponding numerical simulations of the BDNK equations for systems relevant for heavy-ion collisions and astrophysics. We summarize below the different sections and results contained in this paper, and provide a practical guide for the interested reader.

\subsection{How to read this work}
\label{S:How_to_read}

Table~\ref{T:Reading_guide} gives a guide to the contents of this work, in the order in which they appear. The topic abbreviations are \textbf{BG}: Background; \textbf{BF}: Basic formalism; \textbf{MP}: Math proofs; \textbf{NS}: Numerical simulations; and \textbf{PD}: Physical discussion. A slash indicates more than one topic. All entries in Part~\ref{Part:LWP_and_convergence} are classified as BG, and all entries in Part~\ref{Part:Math} as MP.

The Dependencies column lists the main reading prerequisites. Prerequisites of enclosing sections, and those of the listed sections, are understood; a dash means that no additional prerequisite is listed. The notation and conventions of Section~\ref{S:Notation_conventions} are used throughout. Forward references are included when the required material appears later in the paper. %
\newpage
\begingroup
\small
\setlength{\tabcolsep}{4pt}
\renewcommand{\arraystretch}{1.18}
\begin{longtable}{@{}>{\raggedright\arraybackslash}p{0.105\textwidth}>{\raggedright\arraybackslash}p{0.46\textwidth}>{\raggedright\arraybackslash}p{0.11\textwidth}>{\raggedright\arraybackslash}p{0.27\textwidth}@{}}
\caption{Guide to the sections and their reading prerequisites.}
\label{T:Reading_guide}\\
\hline\hline
\textbf{Section} & \textbf{Summary} & \textbf{Topic} & \textbf{Dependencies} \\
\hline
\endfirsthead
\multicolumn{4}{l}{\textit{Table~\ref{T:Reading_guide}: continued}}\\
\hline\hline
\textbf{Section} & \textbf{Summary} & \textbf{Topic} & \textbf{Dependencies} \\
\hline
\endhead
\hline
\multicolumn{4}{r}{\textit{Continued on next page}}\\
\endfoot
\hline\hline
\endlastfoot
\multicolumn{4}{@{}l}{\rule{0pt}{2.5ex}\textbf{Part I: Theoretical and implementation requirements}}\\*
\ref{S:Overview_LWP} & Overview of well-posedness and numerical reliability. & BG & --- \\
\ref{S:LWP_and_all_that} & Existence, uniqueness, and continuous dependence. & BG & --- \\
\ref{S:Existence_via_examples} & Local and global existence illustrated by examples. & BG & --- \\
\ref{S:Why_local_solutions} & Why existence arguments give local solutions. & BG & \ref{S:Existence_via_examples} \\
\ref{S:Function_spaces} & Initial data, regularity, and function spaces. & BG & \ref{S:Existence_via_examples}, \ref{S:Why_local_solutions} \\
\ref{S:Uniqueness} & Meaning and importance of uniqueness. & BG & \ref{S:Existence_via_examples} \\
\ref{S:Continuous_dependence} & Continuous dependence and predictability. & BG & \ref{S:Function_spaces}, \ref{S:Uniqueness} \\
\ref{Causality_section} & Causality and domains of dependence. & BG & \ref{S:Uniqueness} \\
\ref{Stability_section} & Linear, nonlinear, and mode stability. & BG & \ref{S:Function_spaces}, \ref{S:Continuous_dependence} \\
\ref{S:Importance_convergence} & Why numerical convergence tests are needed. & BG & --- \\
\ref{Sec:CC} & Divergence cleaning and auxiliary fields. & BG & --- \\
\ref{Reduction_toy_model_section} & First-order reduction and curl cleaning in a toy model. & BG & --- \\
\ref{S:LWP_example} & Toy-model well-posedness with derivative loss. & BG & \ref{Reduction_toy_model_section}, \ref{S:Function_spaces} \\
\ref{S:LWP_no_derivative_lost_example} & Direct toy-model proof without derivative loss. & BG & \ref{Reduction_toy_model_section}, \ref{S:Function_spaces} \\
\ref{S:Bounds_cleaning_speeds} & Cleaning speeds in different metrics, damping, and timestep restrictions. & BG & \ref{Causality_section}, \ref{Reduction_toy_model_section} \\
\multicolumn{4}{@{}l}{\rule{0pt}{2.5ex}\textbf{Part II: The basic formalism}}\\*
\ref{BDNK_theory_section} & BDNK equations, constitutive relations, and assumptions. & BF & --- \\
\ref{S:Auxiliary_definitions} & Variables and tensors used in the reformulations. & BF & --- \\
\ref{Sec:First_Order} & First-order flux-conservative BDNK system. & BF & \ref{S:Auxiliary_definitions} \\
\ref{Inverse_M_section} & Explicit inverse of the matrix $M$. & BF & \ref{S:Auxiliary_definitions} \\
\ref{Mixed_order_section} & Motivation and structure of the mixed-order system. & BF & \ref{S:Auxiliary_definitions} \\
\ref{Toy_model_example_mixed_order_section} & Mixed-order reduction illustrated by the wave equation. & BG/BF & \ref{S:LWP_and_all_that} \\
\ref{Mixed_order_per_se_section} & Mixed-order BDNK equations and fluxes. & BF & \ref{Inverse_M_section} \\
\ref{Sec:Cleaning_formulation} & Div-curl cleaning for the first-order BDNK system. & BF & \ref{Reduction_toy_model_section}, \ref{Sec:First_Order} \\
\ref{S:Equivalence_formulations} & Compatibility needed to recover the BDNK equations. & BF & \ref{BDNK_theory_section} \\
\ref{FO:Equivalence} & Constraint propagation and first-order equivalence. & BF/MP & \ref{Sec:First_Order}, \ref{Sec:Cleaning_formulation} \\
\ref{MO:Equivalence} & Mixed-order equivalence. & BF/MP & \ref{Mixed_order_per_se_section} \\
\multicolumn{4}{@{}l}{\rule{0pt}{2.5ex}\textbf{Part III: Mathematical results}}\\*
\ref{Sec:LWP} & Local well-posedness of the first-order cleaning system. & MP & \ref{Sec:First_Order}, \ref{Mixed_order_per_se_section}, \ref{Sec:Cleaning_formulation}, \ref{Sec:SH} \\
\ref{Sec:Proof_MO} & Local well-posedness of the mixed-order system. & MP & \ref{Mixed_order_per_se_section}, \ref{Sec:SH} \\
\ref{Sec:SH} & Strictly hyperbolic reduction and local well-posedness. & MP & \ref{S:Function_spaces} \\
\multicolumn{4}{@{}l}{\rule{0pt}{2.5ex}\textbf{Part IV: Numerical algorithms, implementation, and physical results}}\\*
\ref{S:Numerical_setup} & Ingredients needed to set up BDNK simulations. & BF/NS & \ref{BDNK_theory_section} \\
\ref{SS:EOS} & Transport and frame parameterization. & BF/PD & --- \\
\ref{Transport coefficients} & Transport and frame parameters subject to causality. & BF/PD & --- \\
\ref{SSS:conformal} & Conformal equation of state and parameter choices. & BF/PD & \ref{Transport coefficients} \\
\ref{SSS:QCD-EOS} & QCD equation of state and parameter choices. & BF/PD & \ref{SSS:conformal} \\
\ref{SSS:ideal-gas} & Ideal-gas equation of state and parameter choices. & BF/PD & \ref{Transport coefficients}, \ref{SSS:conformal} \\
\ref{Sec:Initial Conditions} & Initial data for both numerical formulations. & BF/NS & \ref{Sec:First_Order}, \ref{Mixed_order_per_se_section} \\
\ref{Sec:Initial Conditions, time derivatives} & Two prescriptions for initial time derivatives. & BF/NS & --- \\
\ref{S:Initial_data_summaries} & Initialization procedures for both formulations. & NS & \ref{Sec:Initial Conditions, time derivatives}, \ref{Sec:Cleaning_formulation} \\
\ref{Sec:Initial Conditions, discussion} & Physical interpretation and limitations of initial data. & PD & \ref{Sec:Initial Conditions, time derivatives} \\
\ref{Sec:Num-methods} & Finite-volume methods and time evolution algorithms. & BF/NS & \ref{Sec:First_Order}, \ref{Mixed_order_per_se_section}, \ref{Sec:Cleaning_formulation} \\
\ref{S:Flat_spacetime_tests} & Numerical tests on flat spacetime backgrounds. & NS/PD & \ref{Sec:Initial Conditions}, \ref{Sec:Num-methods} \\
\ref{S:Cartesian_tests} & Tests in Cartesian coordinates. & NS/PD & \ref{SSS:conformal} \\
\ref{S:Smooth_shock} & Smooth shock tests and viscosity dependence. & NS/PD & --- \\
\ref{S:Kelvin_Helmholtz} & Kelvin--Helmholtz evolution and viscosity dependence. & NS/PD & --- \\
\ref{S:Milne_coordinates} & Milne coordinates and expanding-fluid tests. & BF/NS/PD & --- \\
\ref{Subsubsec:Bjorken} & Bjorken-flow benchmarks and convergence tests. & NS/PD & \ref{SSS:conformal} \\
\ref{Subsubsec:Gubser} & Gubser-flow benchmarks and convergence tests. & NS/PD & \ref{SSS:conformal} \\
\ref{S:Curved_spacetime_tests} & Black-hole backgrounds and ideal-fluid benchmarks. & NS/PD & \ref{SSS:ideal-gas}, \ref{Sec:Initial Conditions}, \ref{Sec:Num-methods} \\
\ref{Sec:Michel-Accretion} & Michel accretion and viscous departures from it. & NS/PD & --- \\
\ref{Sec:Kerr-Fat-Disk} & Kerr-torus setup and viscous departures from stationarity. & NS/PD & --- \\
\ref{S:Israel_Stewart_comparisons} & Israel--Stewart equations and comparison setup. & BF/NS/PD & \ref{S:Milne_coordinates}, \ref{Sec:Initial Conditions}, \ref{Sec:Num-methods} \\
\ref{S:Transverse_Gaussian} & BDNK/Israel--Stewart comparisons for Gaussian initial conditions. & NS/PD & \ref{SSS:conformal}, \ref{SSS:QCD-EOS} \\
\ref{S:Trento_initial_conditions} & BDNK/Israel--Stewart comparisons for TRENTo initial conditions, and a 3D BDNK test. & NS/PD & \ref{SSS:QCD-EOS} \\
\multicolumn{4}{@{}l}{\rule{0pt}{2.5ex}\textbf{Part V: Conclusions and outlook}}\\*
Part~\ref{Part:Conclusion} & Conclusions and outlook. & PD & \ref{S:Flat_spacetime_tests}, \ref{S:Curved_spacetime_tests}, \ref{S:Israel_Stewart_comparisons} \\
\multicolumn{4}{@{}l}{\rule{0pt}{2.5ex}\textbf{Appendices}}\\*
\ref{Appendix:Rescaled EOMs} & Rescaled evolution equations for numerical implementation. & BF/NS & \ref{Sec:First_Order}, \ref{Mixed_order_per_se_section} \\
\ref{S:Rescaled_local_well_posedness} & Local well-posedness of the rescaled first-order system. & MP & \ref{Sec:LWP} \\
\ref{app:more-gubser-plots} & Gubser-flow derivative tests and cleaning-parameter dependence. & NS & \ref{Subsubsec:Gubser}, \ref{Sec:Cleaning_formulation} \\
\ref{S:Convergence_tables} & Error measures, convergence rates, and tables. & NS & \ref{S:Importance_convergence}, \ref{Sec:Num-methods},  \ref{Subsubsec:Gubser} \\
\ref{appendix:gubser-boundary-handling} & Gubser-flow boundary and initial-data treatment. & NS & \ref{Subsubsec:Gubser} \\
\ref{Appendix:kerr-atm-tor-smoothing} & Torus atmosphere and further numerical details. & NS & --- \\
\ref{app:atmosphere} & Construction of the atmosphere outside the torus. & NS & \ref{Sec:Kerr-Fat-Disk} \\
\ref{app:smoothing} & Smoothing at the torus--atmosphere interface. & NS & \ref{app:atmosphere} \\
\ref{Appendix:KT} & Kurganov--Tadmor fluxes for both formulations. & BF/NS & \ref{Sec:Num-methods} \\
\ref{S:Reconstruction} & Reconstruction of values at cell interfaces. & BF/NS & \ref{Sec:Num-methods} \\
\ref{Appendix:MINMOD} & Piecewise-linear reconstruction with the minmod limiter. & BF/NS & --- \\
\ref{Appendix:WENO-Z} & Fifth-order WENO-Z reconstruction. & BF/NS & --- \\
\end{longtable}
\endgroup

\newpage
\subsection{Notation and conventions} 
\label{S:Notation_conventions}
Throughout this work, we employ natural units in which
$\hbar = c = k_B = 1$ and all dimensionful quantities can be expressed in appropriate powers of energy.
We adopt additional conventions and normalizations for the different spacetimes in this work.
For flat Minkowski spacetime in Cartesian~(Section~\ref{S:Cartesian_tests}) and Milne coordinates~(Sections~\ref{S:Milne_coordinates} and \ref{S:Israel_Stewart_comparisons}),
length and time scales are measured in $\GeV^{-1}$, with $[L]=[t]=[\tau]=\GeV^{-1}$ and energy density units are $[\varepsilon]=[L]^{-4}=\GeV^4$.
In these sections, we use the shorthand $\log\varepsilon$ to denote $\log(\varepsilon/\GeV^4)$.
For the Schwarzschild and Kerr metrics spacetimes considered in section~\ref{S:Curved_spacetime_tests}, we further set $G=c=1$ and adopt the normalization for black hole mass~$M=1$, with lengths and times measured in units of $M$.
Consequently, in the corresponding discussions, $[L]=[t]=M$ and $\log\varepsilon$ denotes $\log(\varepsilon/M^{-4})$.

\newpage
\draftpart{Theoretical and implementation requirements}
\label{Part:LWP_and_convergence}

\section{Overview of local well-posedness, causality, stability, and numerical convergence}
\label{S:Overview_LWP}

The importance of well-posedness and convergent numerical schemes has already been emphasized in the Introduction. Here, we discuss these notions in more detail. Readers familiar with them can skip Sections~\ref{S:LWP_and_all_that} and \ref{S:Importance_convergence}.

\subsection{Well-posedness}
\label{S:LWP_and_all_that}

The importance of having well-posed equations for physical applications and their numerical implementation is well known. From a purely theoretical point of view, Courant and Hilbert state that a mathematical problem which is to correspond to physical reality ``should satisfy existence, uniqueness, and continuous dependence on the data, the latter requirement being essential for meaningful problems in mathematical physics" \cite[Chapter III, Section~6.2, p. 227]{Courant-Hilbert-Book-1989-2}. Wald emphasizes that without well-posedness, ``the theory would lose essentially all predictive power" \cite[Chapter 10]{wald1984general}, while Hawking and Ellis rely on well-posedness when they describe the Cauchy development as the region of spacetime ``which can be predicted from knowledge of data" \cite[Chapter 1]{Hawking-Ellis-Book-1973}. For numerical applications, Baumgarte and Shapiro summarize the importance of
well-posedness when they state that ``From a numerical perspective,
then, well-posedness is a necessary but not sufficient condition" \cite[Section~11.1]{Baumgarte:2010ndz}, whereas Rezzolla and Zanotti \cite[Section~4.1.1]{Rezzolla-Zanotti-Book-2013} point out that well-posedness is ``clearly a pre-requisite for any system
whose solutions should be computed numerically."

We begin with some terminological clarifications. First, what we have been referring to as well-posedness is always called in the mathematics literature \emph{local} well-posedness. In the physics literature, one often drops ``local" and refers simply to well-posedness. We will mostly adopt the physicists' convention, albeit we will occasionally say local well-posedness to emphasize its local aspect, whose meaning is explained below. Second, by well-posedness we always mean (local) well-posedness of the Cauchy problem (also known as the initial-value problem), meaning that we are always studying a PDE subject to initial conditions.

Well-posedness of a PDE refers to three properties: existence, uniqueness, and continuous dependence on the initial data, which we now discuss in detail.

\subsubsection{Existence: basic ideas via examples}
\label{S:Existence_via_examples}

Existence refers to the following very basic property: given a PDE and Cauchy data (also called initial data, sometimes abbreviated simply as data), does there exist a solution? I.e., does there exist a field (or collection of fields in the case of systems of PDEs) that satisfies the PDE and agrees with the given initial data at\footnote{When writing $t=0$, it is assumed that we are working in a pre-selected coordinate system. This does not mean that the problem of existence is coordinate dependent; it is possible to formulate it in a coordinate-independent fashion, but we will not need it here. See \cite[Chapter 10]{wald1984general}.} $t=0$?

It is hard to downplay the importance of existence. If the equations we employ do not have solutions, how can we use them to describe a physical phenomenon?\footnote{It should be noted that there are many non-trivial examples of PDEs that have no solutions, see \cite[Chapter 6]{Hormander-Book-1976}.}

Let us illustrate some key ideas about existence with the Cauchy problem for the well-known Riccati ordinary differential equation (ODE):
\begin{subequations}
\label{E:Riccati_CP}
    \begin{align}
        \dot{y}- y^2 &=0,
        \label{E:Riccati_ODE}\\
        y(0) & = y_0,
        \label{E:Riccati_IV}
    \end{align}
\end{subequations}
whose solution is
\begin{align}
    \label{E:Riccati_solution}
    y(t) &= \frac{y_0}{1-y_0 t}.
\end{align}
Indeed, we can easily verify that Equation~\eqref{E:Riccati_solution} satisfies Equation~\eqref{E:Riccati_ODE} and takes the correct initial data in Equation~\eqref{E:Riccati_IV} at $t=0$.

Observe that Equation~\eqref{E:Riccati_solution} blows up at time $t=\frac{1}{y_0}$, $y_0\neq0$ (we also say that it forms a singularity or becomes singular at $t=\frac{1}{y_0}$). Therefore, Equation~\eqref{E:Riccati_solution} is not defined for all times. Rather, for $y_0\neq 0$, it exists only for times on a proper interval\footnote{One could say that the solution in fact exists for all times except $t=\frac{1}{y_0}$. However, for several mathematical reasons it is natural to define the time of existence of a solution to be an interval, and clearly this interval must contain $t=0$ which is when the initial conditions are prescribed (or $t=t_0$ if the initial conditions are given at a different time). This also makes sense physically, as it is not clear what it would mean to say that we have a continuous-in-time evolution but somehow exclude a particular time value.} of the real line, namely, $(-\infty,\frac{1}{y_0})$ for $y_0>0$ and $(\frac{1}{y_0}, \infty)$ for $y_0<0$. (Here we consider both positive and negative $t$. Often we are interested only in positive time values, a situation we discuss below.) The solution in Equation~\eqref{E:Riccati_solution} is thus what we call a \emph{local-in-time} solution or simply a \emph{local solution}, meaning that its existence is asserted on some time interval containing $t=0$, without asserting existence for all times. A solution that is defined for all times is called a \emph{global-in-time} or simply \emph{global solution}. 

This is the meaning of local in local well-posedness: it means that, when it comes to existence, the equations admit local-in-time solutions, i.e., when we say ``solutions exist" we really mean ``local-in-time solutions exist" or simply ``local solutions exist." We similarly say ``global solutions exist" to mean that ``global-in-time solutions exist." We talk about global well-posedness when the existence part of well-posedness refers to global solutions. In this work, we will be concerned only with local well-posedness\footnote{The difference between global and local well-posedness is a good reason to avoid using simply the term well-posedness, without the local or global qualification. However, since it is common in the physics literature to abbreviate local well-posedness by well-posedness, we also use this convention when its meaning is clear.}, thus only with local solutions, but in the ensuing discussion we make some remarks on global solutions that are helpful to give perspective.

It is crucial to note that when we say that the equations ``admit local solutions," that ``local solutions exist," or that the Cauchy problem is ``locally solvable," we are not saying that global-in-time solutions do not exist. A local existence result asserts existence on some interval containing the initial time, but the solution might extend to all times. In the case of Equation~\eqref{E:Riccati_CP} with $y_0\neq 0$, the explicit solution in Equation~\eqref{E:Riccati_solution} shows that this extension is impossible in at least one time direction. Often, however, we can show that local solutions exist without knowing whether they blow up.

This can be seen in the following example. Consider
\begin{subequations}
\label{E:Local_global_example_CP}
\begin{align}
    (1+t^2) \dot{y} + 2ty &=0,
    \label{E:Local_global_example_ODE}
    \\
    y(0) & = y_0.
\end{align}
\end{subequations}
Because the ODE is linear and homogeneous, if $y_1(t)$ is a solution satisfying $y_1(0)=1$ then $y(t) = y_0 y_1(t)$ is a solution satisfying $y(0)=y_0$. Thus, it suffices to consider the case $y_0=1$. Seeking a power-series solution,
\begin{align}
    \label{E:Power_series_ansatz}
    y_{\mathrm{series}}(t) = \sum_{n=0}^\infty a_n t^n,
\end{align}
we find, after plugging Equation~\eqref{E:Power_series_ansatz} into Equation~\eqref{E:Local_global_example_ODE} and using $y_0=1$ to derive the coefficients $a_n$,
\begin{align}
    \label{E:Power_series_solution}
    y_{\mathrm{series}}(t) = \sum_{n=0}^\infty (-1)^n t^{2n}.
\end{align}
The right-hand side of Equation~\eqref{E:Power_series_solution} converges only for $|t|<1$. Thus,  
\begin{align*}
    y_{\mathrm{local}}(t)=y_0 y_{\mathrm{series}}(t)
\end{align*}
is a local solution to Equation~\eqref{E:Local_global_example_CP}, defined for $t \in (-1,1)$. However, one can see that
\begin{align}
\label{E:Example_global_series}
    y_{\mathrm{global}}(t)=\frac{y_0}{1+t^2}
\end{align}
also solves Equation~\eqref{E:Local_global_example_CP}, is defined for all $t \in \mathbb{R}$, and
\begin{align*}
    y_{\mathrm{global}}(t) = y_{\mathrm{local}}(t) \quad \text{for} \quad |t|<1.
\end{align*}
Therefore, the problem in Equation~\eqref{E:Local_global_example_CP} admits global-in-time solutions. The fact that we initially were able to establish only the existence of local solutions was an artifact of the method we used, namely, a power series approach.

The above global versus local example is not an outlier. In the theory of PDEs, methods that establish local existence often do not, by themselves, determine whether the solutions extend globally in time. 

Unfortunately, for complex, especially nonlinear, systems of PDEs, there is no hope of writing explicit solutions like Equations~\eqref{E:Riccati_solution} or \eqref{E:Example_global_series} (except in some very special cases, e.g., if one focuses on some special classes of solutions, like those with a high degree of symmetry, and even in such cases it is not guaranteed that explicit solutions can be found), so that we cannot in general decide by mere inspection whether local solutions turn out to also be global. 

The above examples point to the following important distinction. The problems of local and global existence of solutions are closely related but require different information. Since every global solution is in particular a local solution (since if it is defined for every $t \in \mathbb{R}$, it is in particular defined on a proper subset of the real line), a global existence result includes local existence. A common approach is to first establish local existence and then investigate whether the solutions can be continued for all times. But, as mentioned, a local existence theorem alone typically does not settle the question of global existence. In general, one has to develop new, more refined tools to determine whether local solutions blow up or turn out to be global. Another important difference is that while methods to establish local existence can often treat general classes of PDEs (e.g., symmetric hyperbolic or strongly hyperbolic systems), global existence techniques are often tailored to specific PDEs. This can again be illustrated with ODEs. One can show that the Cauchy problem
\begin{subequations}
\label{E:Illustration_general_CP_via_ODE}
\begin{align}
    \dot{y} &= F(t,y),
    \label{E:Illustration_general_ODE}
    \\
    y(0) &= y_0,
    \label{E:Illustration_general_IC}    
\end{align}
\end{subequations}
admits local solutions under very general assumptions on the function $F$ (which could be nonlinear), e.g., if $F$ is continuously differentiable; a sketch of the argument is given in Section~\ref{S:Why_local_solutions} and details can be found in \cite[Chapter II, Theorem 1.1]{Hartman-Book-2002}. But whether or not the corresponding solutions are global will depend on the specific structure of $F$, as can be seen from the above two examples: the existence theorem applies equally to both, but in the Riccati example there is blow-up for $y_0\neq 0$, whereas in the second, linear example, there is not.

If local existence is an absolute necessity for physical theories (again, how could we work with equations that have no solutions?), the question of global existence also has important physical implications. Often singularities that prevent solutions from existing globally have physical meaning, e.g., shocks in fluid dynamics or curvature singularities in general relativity. Even if the mechanism underlying a failure of global existence might not have a clear physical interpretation, knowing that solutions fail to be global in time can also shed light on a system's physical behavior. For example, it might indicate that the system is driven to states that are incompatible with assumptions made when deriving the model\footnote{It is crucial to emphasize that assumptions made to derive a model might not be preserved by the evolution. For example, we can derive a PDE model under assumptions that gradients are small and consider only initial data with small gradients. But there is no guarantee that solutions will maintain small gradients over time. We have the freedom to choose the PDE model we want to work with and the initial data we provide, but how solutions behave will be determined by the PDE and the data. For example, we can choose to work with the model in Equations~\eqref{E:Riccati_CP} or \eqref{E:Local_global_example_CP}, and can choose $y_0$, but solutions are necessarily given by Equations~\eqref{E:Riccati_solution} and \eqref{E:Example_global_series}, respectively. In the case of Equation~\eqref{E:Riccati_CP}, in particular, no matter how small $y_0>0$ is, $y(t)$ will not remain small as time increases.}, signaling, for example, that the dynamics reached the limit of validity of the theory.

Another feature worth emphasizing is that global existence is typically regime dependent. For example, solutions to Einstein's equations can develop singularities; the singularity theorems give conditions under which spacetime is geodesically incomplete, meaning that some freely falling observers or light rays cannot be continued for arbitrarily large values of their proper time or affine parameter \cite[Chapter 8]{Hawking-Ellis-Book-1973}. On the other hand, for asymptotically flat vacuum initial data sufficiently close to Minkowski data in appropriate weighted norms, Christodoulou and Klainerman show that the resulting spacetime is globally regular and geodesically complete in their celebrated work on the stability of Minkowski space \cite{Christodoulou-Klainerman-Book-1993}.
To see dependence on the initial data in the model in Equation~\eqref{E:Riccati_CP}, observe that solutions always blow up in one time direction for $y_0\neq 0$, but for $y_0=0$ we have the trivial solution $y(t) \equiv 0$, which exists globally. The example in Equation~\eqref{E:Riccati_CP} becomes more instructive if we consider only forward time evolution, i.e., restrict to non-negative time, as is often the case in applications. Then, Equation~\eqref{E:Riccati_solution} gives that solutions exist globally for initial data $y_0 \leq 0$, but blow up for $y_0 > 0$. In this forward-time setting, ``global" means defined for every $t\geq 0$. 
We observe that global existence for the BDNK equations has been obtained for small data in \cite{Sroczinski-2024,Freistuhler-Sroczinski-2025}.

\subsubsection{Existence: why local solutions}
\label{S:Why_local_solutions}
It is instructive to look more closely at why standard local existence theorems generally fail to provide information about global existence. Once more, this can be illustrated at the ODE level. Consider the problem in Equation~\eqref{E:Illustration_general_CP_via_ODE}. We can rewrite Equation~\eqref{E:Illustration_general_ODE} in integral form as
\begin{align}
\label{E:Illustration_general_ODE_integral_form}
y(t) &= y_0 + \int_0^t F(\tau,y(\tau)) \,\mathrm{d}\tau.
\end{align}
Equation~\eqref{E:Illustration_general_ODE_integral_form}  suggests a way of seeking solutions. Let us define
\begin{align}
\label{E:Illustration_general_ODE_integral_form_1}
    y_1(t) &:= y_0 + \int_0^t F(\tau,y_0) \,\mathrm{d}\tau.
\end{align}
While Equation~\eqref{E:Illustration_general_ODE_integral_form} is an equivalent formulation of the Cauchy problem and does not by itself establish the existence of a solution $y$, the right-hand side of Equation~\eqref{E:Illustration_general_ODE_integral_form_1} is known (since $y_0$ is assumed to be given), so that $y_1$ is a known function. Moreover, differentiating Equation~\eqref{E:Illustration_general_ODE_integral_form_1} gives
\begin{align}
\label{E:Illustration_general_ODE_1}
    \dot{y}_1 &= F(t,y_0).
\end{align}
But Equation~\eqref{E:Illustration_general_ODE_integral_form_1} also says that if time is very small, so is the integral and therefore
\begin{align}
    \label{E:Illustratoin_general_ODE_approx_solution_1}
    y_1(t) \approx y_0
\end{align}
so that using Equation~\eqref{E:Illustratoin_general_ODE_approx_solution_1} on the right-hand side of Equation~\eqref{E:Illustration_general_ODE_1} gives
\begin{align*}
    \dot{y}_1 &\approx F(t,y_1),
\end{align*}
i.e., $y_1$ is an approximate solution to Equation~\eqref{E:Illustration_general_CP_via_ODE} for small times. This suggests obtaining an actual solution to the problem via a limiting procedure, where we obtain better and better approximate solutions in iterative steps. Starting with the constant function $y_0(t)=y_0$, define inductively
\begin{align}
\label{E:Illustration_general_ODE_integral_form_n}
    y_{n+1}(t) &:= y_0 + \int_0^t F(\tau,y_n(\tau)) \,\mathrm{d}\tau, \quad n \geq 0,
\end{align}
and note that similarly to the $n=0$ case above, the right-hand side (and thus $y_{n+1}$) is known at each step, as long as the iterates remain in the domain of $F$. Note also that the term outside the integral on the right-hand side is always $y_0$ (and not $y_n$) because we want all the approximate solutions to have the same initial data $y_{n+1}(0)=y_0$, which is the initial data for the original problem. Differentiating Equation~\eqref{E:Illustration_general_ODE_integral_form_n} gives
\begin{align}
    \label{E:Illustration_general_ODE_n}
    \dot{y}_{n+1}  &= F(t,y_n).
\end{align}

If we can show that the sequence $\{ y_n \}$ converges, then putting\footnote{Here we are talking about limits of functions. The precise sense in which they converge, as well as the justification of the passage to the limit in Equation~\eqref{E:Illustration_general_ODE_n_passage_to_the_limit} and inside $F$, is given below.}
\begin{align*}
    y_\infty := \lim_{n\rightarrow \infty} y_n
\end{align*}
we can pass to the limit in Equation~\eqref{E:Illustration_general_ODE_n} to get
\begin{align}
\label{E:Illustration_general_ODE_n_passage_to_the_limit}
    \dot{y}_\infty = \lim_{n\rightarrow\infty} \dot{y}_{n+1} =    
    \lim_{n\rightarrow\infty} F(t,y_n)
    =
    F(t,\lim_{n\rightarrow\infty} y_n)
    = F(t,y_\infty),
\end{align}
satisfying Equation~\eqref{E:Illustration_general_ODE}, and
\begin{align*}
    y_\infty(0) = \lim_{n\rightarrow\infty} y_n(0) = y_0,
\end{align*}
satisfying Equation~\eqref{E:Illustration_general_IC}.

Thus, the key problem boils down to showing that the sequence of approximate solutions $\{y_n\}$ converges. To show this, it suffices to get difference estimates for
\begin{align*}
    y_{n+1} - y_n.
\end{align*}
An estimate for $ y_{n+1} - y_n$ can be iterated and turned into an estimate for $ y_{n+j} - y_n$, which is what is needed to analyze whether the sequence is a Cauchy sequence and convergent, see below.

First, we need to be precise about how we quantify the size of $y_{n+1} - y_n$. For continuous functions $f$ defined on a closed, finite-length interval, $[a,b]$, we define the norm
\begin{align*}
    \norm{f}_{C^0([a,b])} := \max_{t \in [a,b]} |f(t)|.
\end{align*}
This is a well-defined quantity since, by the extreme value theorem, a continuous function on a closed, finite-length interval always has a maximum and a minimum.  Thus, we will be estimating
\begin{align*}
    \norm{ y_{n+1} - y_n}_{C^0([0,T])},
\end{align*}
for some $T>0$ that will be chosen below\footnote{Here, we are restricting attention only to non-negative time.}.

We consider initial data $y_0$ for which 
\begin{align}
\label{E:Illustration_Picard_initial_data}
    |y_0| \leq B,
\end{align}
for some $B >0$ (in practice, given $y_0$ we choose $B$ such that this inequality holds).

For the function $F$, we assume that it is a continuously differentiable function, i.e., $F$ is continuous, and its first-order derivatives exist and are continuous.
Fix some $T_0>0$. It then follows that there exist constants $K>0$ and $M>0$ such that for all $z,z_1,z_2$ in $[-2B,2B]$ and all $t \in [0,T_0]$, we have
\begin{subequations}
\label{E:Illustration_Picard_assumptions}
   \begin{align}
   \label{E:Illustration_Picard_assumptions_boundedness}
       |F(t,z)| &\leq K,
       \\
       \label{E:Illustration_Picard_assumptions_Lipschitz}
    |F(t,z_1) - F(t,z_2)| & \leq M |z_1 - z_2|.
\end{align}
\end{subequations}
Indeed, Equation~\eqref{E:Illustration_Picard_assumptions_boundedness} follows from the continuity of $F$ and the extreme value theorem for functions of two variables applied to $[0,T_0]\times [-2B,2B]$. Equation~\eqref{E:Illustration_Picard_assumptions_Lipschitz} follows from the mean value theorem
\begin{align*}
    F(t,z_1)-F(t,z_2) = \frac{\partial F}{\partial z}(t,z_c) (z_1-z_2),
\end{align*}
where $z_c$ lies between $z_1$ and $z_2$. The continuity of $\partial_z F$ and the extreme value theorem give a constant $M$ such that $|\frac{\partial F}{\partial z}(t,z) |\leq M$ for all $(t,z)\in [0,T_0]\times [-2B,2B]$.

Choose $T_0^\prime=\min\{T_0,\frac{B}{K}\}$. Suppose inductively that $|y_n(t)| \leq 2B$ on $[0,T_0^\prime]$ and note that this condition holds for $n=0$. Then Equations~\eqref{E:Illustration_general_ODE_integral_form_n}, \eqref{E:Illustration_Picard_initial_data},  and \eqref{E:Illustration_Picard_assumptions_boundedness} give, for $0 \leq t \leq T_0^\prime$,
\begin{align*}
   |y_{n+1}(t)| &\leq |y_0| + \int_0^t | F(\tau,y_n(\tau))| \, \mathrm{d} \tau
   \\
   & \leq B + t K.
\end{align*}
By the choice of $T_0^\prime$, we obtain\footnote{Observe that restricting to a smaller interval does not change the value of the constants $K$ and $M$.}
\begin{align*}
    |y_{n+1}(t)| \leq B + B = 2B,
\end{align*}
and therefore $|y_n| \leq 2B$ for all $n$. 
Observe that our assumptions on $F$ and a simple induction show that $y_{n+1}$ is a continuously differentiable function, so in particular continuous, that satisfies Equation~\eqref{E:Illustration_general_ODE_n}. Thus, we also have
\begin{align}
    \label{E:Illustration_Picard_sequence_remains_bounded}
    \norm{y_n}_{C^0([0,T_0^\prime]) }\leq 2B
\end{align}
for all $n$.

By the foregoing, for $t \in [0,T_0^\prime]$ we can apply Equation~\eqref{E:Illustration_Picard_assumptions_Lipschitz} to the sequence $y_n$. Thus, for $n\geq 1$, Equation~\eqref{E:Illustration_general_ODE_integral_form_n} gives
\begin{align*}
\begin{split}
    |y_{n+1}(t)-y_n(t)| &= 
    | y_0 + \int_0^t F(\tau,y_n(\tau)) \,\mathrm{d}\tau - ( y_0 + \int_0^t F(\tau,y_{n-1}(\tau)) \,\mathrm{d}\tau) |
    \\
    & \leq \int_0^t 
    | F(\tau,y_n(\tau)) - F(\tau,y_{n-1}(\tau))| \,\mathrm{d}\tau
    \\&
    \leq \int_0^t 
    M | y_n(\tau) - y_{n-1}(\tau)| \,\mathrm{d}\tau.
\end{split}
\end{align*}
Fix a number $0 < \lambda < 1$. Restricting $t$ further to $[0,T]$ where $T\leq T_0^\prime$ is chosen such that $TM \leq \lambda$, we thus obtain
\begin{align*}
\begin{split}
    |y_{n+1}(t)-y_n(t)| 
    & \leq \int_0^t 
    M | y_n(\tau) - y_{n-1}(\tau)| \,\mathrm{d}\tau
    \\
    & \leq \int_0^t 
    M ( \max_{r \in [0,T]}| y_n(r) - y_{n-1}(r)| )\,\mathrm{d}\tau    
    \\
    &=  \norm{y_n - y_{n-1}}_{C^0([0,T])} M t
    \\
    & \leq \lambda \norm{y_n - y_{n-1}}_{C^0([0,T])}.
\end{split}
\end{align*}
Since this holds for all $t \in [0,T]$,
\begin{align}
\label{E:Illustration_Picard_sequence_remains_is_a_contraction}
    \norm{y_{n+1} - y_n}_{C^0([0,T])}
    \leq \lambda \norm{y_n - y_{n-1}}_{C^0([0,T])}.
\end{align}
Iterating Equation~\eqref{E:Illustration_Picard_sequence_remains_is_a_contraction} yields
\begin{align}
\label{E:Illustration_Picard_contraction}
    \norm{y_{n+1} - y_n }_{C^0([0,T])} \leq \lambda^n \norm{y_1 - y_{0} }_{C^0([0,T])}. 
\end{align}

For any two elements $y_n$ and $y_{n+j}$, $j > 0$, Equation~\eqref{E:Illustration_Picard_contraction} gives
\begin{align*}
    \begin{split}
        \norm{y_{n+j}-y_n}_{C^0([0,T])} & \leq 
        \norm{y_{n+j}-y_{n+j-1}}_{C^0([0,T])}
        +
        \norm{y_{n+j-1}-y_{n+j-2}}_{C^0([0,T])}
        + \cdots
        + \norm{y_{n+1}-y_{n}}_{C^0([0,T])}
        \\
        & 
        \leq \lambda^{n+j-1}\norm{y_{1}-y_{0}}_{C^0([0,T])}
        +
        \lambda^{n+j-2}\norm{y_{1}-y_{0}}_{C^0([0,T])}
        + \cdots +
        \lambda^{n}\norm{y_{1}-y_{0}}_{C^0([0,T])}
        \\
        &=
        \lambda^n \norm{y_{1}-y_{0}}_{C^0([0,T])} \sum_{k=0}^{j-1} \lambda^k
        \\
        &\leq 
        \lambda^n \norm{y_{1}-y_{0}}_{C^0([0,T])} \sum_{k=0}^{\infty} \lambda^k   
        \\
        & \leq 
                \lambda^n \norm{y_{1}-y_{0}}_{C^0([0,T])}  \frac{1}{1-\lambda}.
    \end{split}
\end{align*}
The bound $\frac{\lambda^n}{1-\lambda}\norm{y_1-y_0}_{C^0([0,T])}$ is independent of $j$. Thus, since $0 < \lambda < 1$, we can make the right-hand side, and thus
$\norm{y_{n+j}-y_n}_{C^0([0,T])}$, as small as we want by taking $n$ sufficiently large. Therefore, the sequence $y_n$ is a Cauchy sequence in the space 
\begin{align*}
   C^0([0,T]) := \{ f \colon [0,T] \rightarrow \mathbb{R} \, | \, f \text{ is continuous} \}, 
\end{align*}
and hence it converges\footnote{Here, we are using the crucial fact that the space $C^0([0,T])$ is complete, which means that every Cauchy sequence in this space converges to an element of $C^0([0,T])$, i.e., in the same space. To illustrate this, consider the sequence $x_n = \frac{1}{n}$. It satisfies $|x_{n+j} - x_n| = \frac{j}{n(n+j)} \leq \frac{1}{n}$, which can be made as small as we want by taking $n$ sufficiently large, thus it is a Cauchy sequence. Whether it converges, however, depends on the space in which we consider the sequence. If we consider $x_n$ as elements of $[0,1]$, then the sequence converges to $0$ which belongs to $[0,1]$. But if we consider $x_n$ as elements of $(0,1]$, then it converges to $0$ which is not an element of the space $(0,1]$ -- strictly speaking we say that $x_n$ does not converge in $(0,1]$. Thus, knowing that we are working in a complete space, one in which every Cauchy sequence converges within that space, is key for the argument. This also illustrates how the choice of function spaces is crucial for the study of existence of solutions, a topic we will discuss in more detail in Section~\ref{S:Function_spaces}.} to a limit: there exists a unique $y_\infty \in C^0([0,T])$ such that
\begin{align*}
    \lim_{n\rightarrow\infty} 
    \norm{y_n - y_\infty}_{C^0([0,T])} = 0.
\end{align*}

From Equations~\eqref{E:Illustration_general_ODE_n} and \eqref{E:Illustration_Picard_assumptions_Lipschitz}, for $n\geq 0$ and $j\geq 1$, we have
\begin{align*}
    \norm{\dot{y}_{n+j+1} - \dot{y}_{n+1}}_{C^0([0,T])} \leq M \norm{y_{n+j} - y_n}_{C^0([0,T])},
\end{align*}
and therefore $\dot{y}_n$ is also a Cauchy sequence in $C^0([0,T])$ and thus converges to a limit. This means that the sequence $\{ y_n \}$ in fact converges in the norm
\begin{align*}
    \norm{f}_{C^1([0,T])} := \max_{t \in [0,T]} |f(t)| + \max_{t \in [0,T]} |\dot{f}(t)|. 
\end{align*}
$C^1([0,T])$ is also a complete space, which means that $y_\infty$ is continuously differentiable and the limit of $\dot{y}_n$ is precisely $\dot{y}_\infty$. Hence, Equation~\eqref{E:Illustration_general_ODE_n_passage_to_the_limit} is justified and we note that  $y_\infty(0) = y_0$. We have thus established the desired existence result.

We can now identify why this proof guarantees only a local solution. We fixed a finite interval $[0,T_0]$ to obtain the bounds in Equations~\eqref{E:Illustration_Picard_assumptions}, and then restricted the time interval to keep the iterates bounded as in Equation~\eqref{E:Illustration_Picard_sequence_remains_bounded} and to obtain the contraction estimate in Equation~\eqref{E:Illustration_Picard_sequence_remains_is_a_contraction}. In particular, the proof works with
\begin{align*}
    T=\min\left\{T_0,\frac{B}{K},\frac{\lambda}{M}\right\}.
\end{align*}
The constants $K$ and $M$ depend on $F$, $B$, and $T_0$, so this formula gives no guarantee of existence for arbitrarily large times.

One might point out, correctly, that these restrictions arise from the method we employed. Nevertheless, similar restrictions occur in standard local existence proofs for the nonlinear hyperbolic equations relevant to relativistic fluids. One constructs approximate solutions\footnote{Common methods include approximating the nonlinear problem by linear problems, see, e.g., \cite[Chapter 9]{Ringstrom-Book-2009}; discretizing the evolution in time, see, e.g., \cite[Section~6]{Disconzi-Ifrim-Tataru-2022}; or obtaining solutions as limits of solutions in a more restrictive class, as in Sections~\ref{Sec:LWP}, \ref{Sec:Proof_MO}, and \ref{Sec:SH} of this work. In our ODE example, once $y_n$ is known, Equation~\eqref{E:Illustration_general_ODE_n} is a linear equation for $y_{n+1}$.} and proves estimates that ensure their convergence. These estimates typically hold on a sufficiently short time interval. Additional information about the particular equations is then needed to decide whether the solution extends globally.

The tradeoff is the typical one in mathematics of generality in the hypotheses versus strength of the conclusions. For the illustrative problem in Equation~\eqref{E:Illustration_general_CP_via_ODE}, we made very mild assumptions on the function $F$, which makes the result applicable to a large class of equations. The price to pay is that the conclusion does not provide much information about solutions; we conclude that there exists a solution defined for some time interval $[0,T]$, where $T$ depends on the initial-data bound and on $F$. This alone does not determine whether the solution is global. Indeed, both the system with blowing-up solutions in Equation~\eqref{E:Riccati_CP} and the system with global solutions in Equation~\eqref{E:Local_global_example_CP} fall under the scope of Equation~\eqref{E:Illustration_general_CP_via_ODE}. Imposing more assumptions on $F$ can lead to stronger conclusions but inevitably makes the applicability of the corresponding result limited to a smaller class of equations. Just to take an extreme case: if we assume that $F$ is a linear function of $y$ with smooth coefficients defined for every $t \in \mathbb{R}$, then solutions will be global in time, but such a result can only be applied to a very limited class of equations. The situation with PDEs is similar. Typically, we want theorems that guarantee local existence under very general, and hopefully easy to verify, conditions. First-order symmetric hyperbolic systems are an example: together with suitable regularity of the coefficients and initial data, symmetry and positive-definiteness of the matrix multiplying $\partial_t$ give a general local well-posedness theory, but do not by themselves imply global existence; see \cite[Chapter 9]{Ringstrom-Book-2009}.

A natural question at this point is whether we can obtain global solutions by applying a local existence result successively. Consider again the solution $y_\infty$ to Equation~\eqref{E:Illustration_general_CP_via_ODE}, defined on $[0,T]$. We can take $y_{0,1}:=y_\infty(T)$ as new initial data at $t=T$. Applying the same argument starting at this time gives a solution $y_{\infty,1}$ on $[T,T+T_1]$ for some $T_1>0$. Define
\begin{align*}
    \hat{y}_1(t) :=
    \begin{cases}
        y_\infty(t), & 0 \leq t \leq T,\\
        y_{\infty,1}(t), & T < t \leq T+T_1.
    \end{cases}
\end{align*}
This function is continuous because $y_{\infty,1}(T)=y_\infty(T)$. Its one-sided derivatives at $t=T$ also agree, since both equal $F(T,y_\infty(T))$ by Equation~\eqref{E:Illustration_general_ODE}. Thus, $\hat{y}_1$ is continuously differentiable and solves Equation~\eqref{E:Illustration_general_CP_via_ODE} on the larger interval $[0,T+T_1]$. Continuing in this way, we obtain a solution on $[0,T+T_1+\cdots+T_n]$, where each $T_j>0$. Why does this not necessarily give a global solution?

The simple reason why the above argument does not in general produce a global solution is that we do not have good control over the size of the existence times $T_n$. Remember that we need to choose $T$ small enough in the existence proof so that certain bounds hold. Applying the existence result successively as indicated above can produce times $T_n$ that become smaller for each new solution. For example, suppose the argument for producing $y_{\infty,n}$ gives $T_n \leq 2^{-n}$. Then
\begin{align*}
    [0,T+T_1+T_2+\cdots +T_n] \subset [0,T+1)
\end{align*}
for all $n$ and this argument would not give a solution defined beyond $T+1$. 

We picked $T_n \leq 2^{-n}$ only as an illustration, but the important thing is that we know that something like this can and will in certain cases happen. Indeed, if we apply the above argument to Equation~\eqref{E:Riccati_CP} with $y_0>0$, we already know that at best we can obtain a solution defined on $[0,\frac{1}{y_0})$. To better understand what is behind this, 
recall that the permitted time step depends on bounds for $F$ and its derivative over a region containing the initial data. These bounds can grow as the solution grows, causing the guaranteed time steps to shrink. For the Riccati example, on $|y|\leq 2B$ we can take $K=4B^2$ and $M=4B$, so that the restrictions $TK\leq B$ and $TM\leq\lambda$ permit a time step proportional to $\frac{1}{B}$. As the positive solution grows toward its blow-up time, restarting the argument requires larger values of $B$ and gives smaller time steps.

The underlying message is that continuing local solutions for all times requires quantitative control that the local existence argument alone does not provide. The same issue arises for PDEs: general local existence theorems accommodate large classes of equations, while global existence typically requires further estimates that use the particular structure of the equations.

\subsubsection{Existence: choice of initial data and function spaces}
\label{S:Function_spaces}
When asking whether the Cauchy problem for a PDE can be solved, we are interested in knowing whether it can be solved for general data. General, here, means that the initial data should satisfy a pre-specified set of criteria but be otherwise arbitrary. 
In the example in Equation~\eqref{E:Illustration_general_CP_via_ODE} that we discussed in Sections~\ref{S:Existence_via_examples}
and 
\ref{S:Why_local_solutions}, we showed how local solutions exist for any given $y_0$. If there are natural restrictions on $y$, such restrictions are then imposed on the initial data, but we otherwise consider arbitrary data satisfying them. For example, if $y$ in Equation~\eqref{E:Illustration_general_CP_via_ODE} is a quantity that must be non-negative, we take only initial conditions $y_0 \geq 0$, and seek to establish existence of local solutions for any such data. If we have, for example, a result that establishes existence of solutions only for $y_0 \in [0,1]$, whereas all $y_0 \geq 0$ are allowed in the model, that would address only part of the question. Moreover, if non-negativity is required physically, we must also check that the evolution preserves it; imposing it on the initial data does not by itself guarantee this.

An important qualitative difference between initial data for PDEs and those for the ODE example in Equation~\eqref{E:Illustration_general_CP_via_ODE} is that initial data for PDEs are functions. Let us illustrate the discussion with the initial-value problem for the Schr\"odinger equation:
\begin{subequations}
\label{E:Schrodinger_example}
    \begin{align}
        \label{E:Schrodinger_example_equation}
        i \hbar \partial_t \Psi & = -\frac{\hbar^2}{2m} \Delta \Psi + V \Psi,\\
        \Psi(0,x) &= \Psi_0(x).
    \end{align}
\end{subequations}
What properties should we assume about $\Psi_0$ in order to ensure that local solutions to Equation~\eqref{E:Schrodinger_example} exist?

The first observation is that we must specify what kinds of initial data and solutions we are considering. Even when a linear PDE admits solutions for very general data, such as distributions or generalized functions, this remains a specified class of data and a specified meaning of solution. We cannot simply ask for a solution for ``any function" without making these choices. 

In the example in Equation~\eqref{E:Schrodinger_example}, the wave function of a normalized state is square-integrable, and thus we consider data $\Psi_0$ such that the $L^2(\mathbb{R}^3)$ norm
\begin{align*}
    \norm{\Psi_0}_{L^2(\mathbb{R}^3)}^2 := 
    \int_{\mathbb{R}^3} |\Psi_0(x)|^2 \, \mathrm{d}
x\end{align*}
is finite, i.e., 
\begin{align*}
    \norm{\Psi_0}_{L^2(\mathbb{R}^3)} < \infty.
\end{align*}
Thus, a natural function space to consider in the study of Equation~\eqref{E:Schrodinger_example} is the \emph{$L^2(\mathbb{R}^3)$ space}\footnote{Here we consider complex-valued functions because of the Schr\"odinger equation. The physical fields in the fluid equations considered in this work are real valued. In the definition of $L^2$ and of the Sobolev spaces below, functions are understood to be measurable, and functions that agree almost everywhere are identified.}
\begin{align*}
    L^2(\mathbb{R}^3) := \{ \Psi \colon \mathbb{R}^3 \rightarrow \mathbb{C} \, | \, \norm{\Psi}_{L^2(\mathbb{R}^3)} < \infty \}.
\end{align*}
In the Schr\"odinger problem in Equation~\eqref{E:Schrodinger_example}, the solutions we are interested in are those whose $L^2(\mathbb{R}^3)$ norm is equal to one. We take $m>0$ and a real, time-independent potential $V=V(x)$ that is smooth, with $V$ and all its derivatives bounded. These are convenient sufficient assumptions for our example. Under these assumptions, Equation~\eqref{E:Schrodinger_example_equation} preserves the $L^2(\mathbb{R}^3)$ norm. Since the equation is linear, any solution with nonzero initial data can then be normalized by dividing it by its initial $L^2$ norm. From the point of view of PDE theory, it is better to work with the $L^2(\mathbb{R}^3)$ space instead of working only with functions with norm one because $L^2(\mathbb{R}^3)$ is a Hilbert space, with the inner product
\begin{align*}
    \langle \Psi , \Phi \rangle := 
    \int_{\mathbb{R}^3} \Psi(x) \bar{\Phi}(x) \, \mathrm{d}x,
\end{align*}
so that tools from Hilbert space theory are available to study existence. 

On the notational side, we observe that the $L^2(\mathbb{R}^3)$ norm is taken over space, while solutions to Equation~\eqref{E:Schrodinger_example_equation} are functions of space and time. To emphasize this, we sometimes write
$ \norm{\Psi(t)}_{L^2(\mathbb{R}^3)}$
or 
$ \norm{\Psi(t,\cdot)}_{L^2(\mathbb{R}^3)}$, meaning
\begin{align*}
    \norm{\Psi(t)}_{L^2(\mathbb{R}^3)}^2 \equiv 
    \norm{\Psi(t,\cdot)}_{L^2(\mathbb{R}^3)}^2=
    \int_{\mathbb{R}^3} |\Psi(t,x)|^2 \, \mathrm{d}x.    
\end{align*}
Another common notation is to write simply
\begin{align*}
    \norm{\Psi}_{L^2(\Sigma_t)},
\end{align*}
where
\begin{align*}
    \Sigma_t := \{ (\tau,x) \in \mathbb{R} \times \mathbb{R}^3 \, | \, \tau = t \} \approx \mathbb{R}^3.
\end{align*}
Similar conventions apply to other function spaces defined over spatial variables applied to functions that also depend on time; see below.

When seeking solutions to Equation~\eqref{E:Schrodinger_example}, we need to take into consideration the obvious fact that derivatives of $\Psi$ are involved (it would not be a differential equation otherwise). This suggests that $L^2(\mathbb{R}^3)$ is not yet the correct space for the problem if we seek solutions with classical derivatives, since being square-integrable does not guarantee differentiability\footnote{There are two technical points worth mentioning here. First, for the theory of function spaces to work properly, the integral defining the $L^2(\mathbb{R}^3)$ norm, as well as other integral norms introduced below, has to be the Lebesgue integral, which generalizes the Riemann integral one learns in standard calculus \cite[Chapter V]{Halmos-Book-1974-1}. With the Lebesgue integral, the function $f \colon [0,1] \rightarrow \mathbb{R}$ given by $f(x) = 0$ if $x \in \mathbb{Q}$ and $f(x)=1$ if $x \notin \mathbb{Q}$ is square-integrable with $\norm{f}_{L^2([0,1])} =1$, but $f$ is not differentiable at any point on $[0,1]$. Hence, square-integrability is not enough if we want our functions to be differentiable. Second, there is a sense in which we can talk about solutions to Equation~\eqref{E:Schrodinger_example} for $\Psi$ belonging to $L^2(\mathbb{R}^3)$ without any differentiability requirements. But such solutions are abstract in the sense that they arise as $\Psi(t) = e^{-\frac{itH}{\hbar}} \Psi_0$ for suitable self-adjoint Hamiltonians $H$, where $e^{-\frac{itH}{\hbar}}$ is a one-parameter group of unitary transformations \cite[Theorem 5.1]{Teschl-Book-2014}. Such solutions, however, do not in general solve Equation~\eqref{E:Schrodinger_example_equation} in the classical sense, i.e., Equation~\eqref{E:Schrodinger_example_equation} will not in general make sense as a pointwise equality, i.e., something we can evaluate at a point on the left- and right-hand sides and state that both sides are equal, not even in an ``almost everywhere" measure-theoretical sense. Such abstract formulations, therefore, will not be considered here; moreover, typically they do not generalize to the types of nonlinear hyperbolic equations we are interested in here.}. A class of spaces that has proven useful in PDE theory, while retaining square-integrability, is that of \emph{Sobolev spaces}. For a non-negative integer $s$, we define
\begin{align*}
    H^s(\mathbb{R}^3) := 
    \{ \Psi \colon \mathbb{R}^3 \rightarrow \mathbb{C} \, | \, \norm{\Psi}_{H^s(\mathbb{R}^3)} < \infty \},
\end{align*}
where the Sobolev norm $\norm{\cdot}_{H^s(\mathbb{R}^3)}$ is defined as
\begin{align*}
    \norm{\Psi}_{H^s(\mathbb{R}^3)}^2 &:= \sum_{k=0}^s \norm{ \partial^k \Psi}_{L^2(\mathbb{R}^3)}^2
    \\
    &= \sum_{k=0}^s \int_{\mathbb{R}^3} |\partial^k \Psi(x)|^2 \, \mathrm{d}x,
\end{align*}
where $\partial^k \Psi$ abbreviates all derivatives of $\Psi$ of order $k$ with respect to standard coordinates in $\mathbb{R}^3$ and $|\partial^k \Psi|$ its Euclidean norm, with each distinct mixed derivative counted once. For example,
\begin{align*}
    |\partial^2 \Psi (x)|^2 = 
    |\partial_{x^1}^2 \Psi(x)|^2 + |\partial_{x^1 x^2}^2 \Psi(x)|^2 + \cdots + |\partial_{x^3}^2 \Psi(x)|^2.
\end{align*}
Thus, we see that the Sobolev norm generalizes the $L^2$ norm by requiring that not only $\Psi$ be square-integrable, but also that all its derivatives, assumed to exist\footnote{
\label{FT:Sobolev_derivatives}
Another technical point is that in the definition of the Sobolev norm the derivatives are derivatives in the sense of distributions; these agree with ordinary derivatives for functions that are differentiable in the usual sense, see \cite[Section~5.2]{evans2010pde}.}, up to order $s$ are square-integrable. The Sobolev spaces are also Hilbert spaces, with inner product
\begin{align*}
    \langle \Psi , \Phi \rangle_s := 
    \sum_{k=0}^s \int_{\mathbb{R}^3} \partial^k \Psi(x) \cdot \overline{\partial^k \Phi(x)} \, \mathrm{d}x.
\end{align*}

For the potentials $V$ specified above and any integer $s\geq 4$, initial data $\Psi_0\in H^s(\mathbb{R}^3)$ give a solution to Equation~\eqref{E:Schrodinger_example}\footnote{See \cite[Sections~7.2--7.4]{Teschl-Book-2014} and \cite[Chapter III, Theorem 1.3; Chapter II, Proposition 6.2]{Engel-Nagel-Book-2000}.}. Such solutions in fact exist globally in time, but we will consider their restriction to a finite interval $[0,T]$ to illustrate the local existence statements relevant to this work.\footnote{The choice $s\geq 4$ is a convenient sufficient condition but is not a threshold for existence of less regular solutions.}

As in the case of the $L^2$ norm, the Sobolev norm is defined relative to spatial variables at a fixed time, and we write 
$ \norm{\Psi(t)}_{H^s(\mathbb{R}^3)}$,
$ \norm{\Psi(t,\cdot)}_{H^s(\mathbb{R}^3)}$, or $ \norm{\Psi}_{H^s(\Sigma_t)}$
to emphasize the time dependence of $\Psi$. The Sobolev norm measures differentiability properties of functions relative to spatial coordinates in $\mathbb{R}^3$. Since Equation~\eqref{E:Schrodinger_example_equation} involves differentiation in time, we need information about the time regularity of solutions. For a non-negative integer $k$, we introduce the space of functions that are $k$ times continuously differentiable in $t$ with values in a Sobolev space, where continuity and differentiation are understood in the $H^s$ norm:
\begin{align*}
    C^k([0,T],H^s(\mathbb{R}^3)) := \{ \Psi \colon [0,T] \rightarrow
H^s(\mathbb{R}^3) \, | \, & \Psi = \Psi(t,x) \text{ is } k \text{ times continuously differentiable in } t
\\
& 
\text{and } \norm{\Psi}_{C^k([0,T],H^s(\mathbb{R}^3))} < \infty \} ,   
\end{align*}
with norm
\begin{align*}
    \norm{\Psi}_{C^k([0,T],H^s(\mathbb{R}^3))} := \sum_{j=0}^k \,\sup_{0\leq t\leq T} \norm{\partial_t^j \Psi(t,\cdot)}_{H^s(\mathbb{R}^3)}.
\end{align*}
One can show that the aforementioned solutions to the Schr\"odinger equation have regularity
\begin{align}
\label{E:Schrodinger_solution_regularity}
    \Psi \in C^0([0,T],H^s(\mathbb{R}^3)) \cap C^1([0,T],H^{s-2}(\mathbb{R}^3)) .
\end{align}
Thus, such solutions are continuous in time with spatial derivatives up to order $s$ being square-integrable in $\mathbb{R}^3$ and they are continuously differentiable in time with the corresponding spatial derivatives of $\partial_t\Psi$ up to order $s-2$ being square-integrable in $\mathbb{R}^3$. The difference between $s$ and $s-2$ is natural because Equation~\eqref{E:Schrodinger_example_equation} says that each time derivative corresponds to two spatial derivatives. The precise relation between the time and spatial regularity of solutions depends on the particular equation being studied.

The above discussion with the Schr\"odinger equation illustrates the following basic philosophy to study PDEs: while we cannot expect local existence for the Cauchy problem for \emph{completely} arbitrary data (e.g., ``any" function), we would like to prove local existence for data that is arbitrary within a predetermined set of initial conditions. \emph{A natural way of selecting such a predetermined set of functions is to specify a function space.} Not only is this convenient but also important for the applications of many PDE tools. Function spaces form the arena in which PDE investigations take place.

Function spaces other than Sobolev spaces are also possible. For example, we can consider spaces of bounded continuously differentiable functions
\begin{align*}
        C_b^k(\mathbb{R}^3) := \{ \Psi \colon \mathbb{R}^3 \rightarrow \mathbb{C} \, | \, &\Psi \text{ is } k \text{ times continuously differentiable}\\
        &\text{and } \norm{\Psi}_{C_b^k(\mathbb{R}^3)}< \infty
        \}
\end{align*}
and 
\begin{align*}
        C_b^k([0,T]\times\mathbb{R}^3) := \{ \Psi \colon [0,T] \times  \mathbb{R}^3 \rightarrow \mathbb{C} \, | \, & \Psi \text{ is } k \text{ times continuously differentiable}\\
        &\text{and } \norm{\Psi}_{C_b^k([0,T]\times \mathbb{R}^3)} < \infty
        \}
\end{align*}
where the norms are given by
\begin{align*}
    \norm{\Psi}_{C_b^k(\mathbb{R}^3)} := \sum_{j=0}^k \,\sup_{x \in \mathbb{R}^3} |\partial^j \Psi(x)|
\end{align*}
and
\begin{align*}
    \norm{\Psi}_{C_b^k([0,T]\times \mathbb{R}^3)} := \sum_{j=0}^k \,\sup_{(t,x) \in [0,T]\times \mathbb{R}^3} |\partial^j \Psi(t,x)|.
\end{align*}
The subscript $b$ indicates that the function and its derivatives through order $k$ are bounded. Here, $k$-times continuous differentiability refers to differentiability relative to spatial coordinates $(x^1,x^2, x^3)$ on $\mathbb{R}^3$ in the first case and relative to spacetime coordinates 
$(t,x^1,x^2, x^3)$ on $[0,T]\times\mathbb{R}^3$ in the second case. Similarly, $\partial^j$ denotes all derivatives of order $j$ with respect to $(x^1,x^2, x^3)$ in the first case and all derivatives of order $j$ with respect to 
$(t,x^1,x^2, x^3)$ in the second case.
We have already encountered the corresponding spaces and norms on a compact interval in Section~\ref{S:Why_local_solutions}.

There are many other function spaces used in PDEs, both in $\mathbb{R}^3$ and $[0,T] \times \mathbb{R}^3$. For example, spaces of infinitely differentiable functions, $C^\infty(\mathbb{R}^3)$ or $C^\infty([0,T]\times\mathbb{R}^3)$, spaces of analytic functions,  $C^\omega(\mathbb{R}^3)$ or $C^\omega([0,T]\times\mathbb{R}^3)$, Gevrey spaces (which are ``intermediate" spaces between $C^\omega$ and $C^\infty$ \cite[Chapter I]{Rodino-Book-1993}), H\"older spaces (which, for an exponent strictly between zero and one, give intermediate regularity between $C^{k-1}$ and $C^k$ \cite[Section~5.1]{evans2010pde}), spaces of distributions, and several others.

A natural and important question is: \emph{which function space do we choose to study a given PDE?} There is no single or ``correct" answer to this question. The relevance of a function space depends on several aspects of the problem, including but not limited to the type of PDE, the level of generality one is aiming at, and the types of questions about the model one is interested in investigating. In our example of the Schr\"odinger equation, we motivated the use of Sobolev spaces by pointing out the physical requirement that the wave function be square integrable, complemented by the natural requirement that derivatives be well defined. But even this does not tell us \emph{which} Sobolev space $H^s$  to work with. As we saw, the choice $s\geq 4$ gives classical solutions under our assumptions on $V$. But suppose that we are interested in computing observables that involve several derivatives of the wave functions. In that case, we might have to work in $H^s$ with $s$ values significantly larger than four.

As another example, let us consider the Cauchy problem for Burgers' equation\footnote{In order to avoid overcharging the notation, we will continue to use $\Psi$ in our discussion of Section~\ref{S:LWP_and_all_that} even if it does not denote a solution to Schr\"odinger's equation. The distinction will always be clear from the context.} 
\begin{subequations}
\label{E:Burgers_example}
    \begin{align}
        \label{E:Burgers_example_equation}
        \partial_t \Psi + \Psi \partial_x \Psi &= 0,\\
        \Psi(0,x) &= \Psi_0(x).
    \end{align}
\end{subequations}
Here $\Psi$ and $\Psi_0$ are real valued. The problem in Equation~\eqref{E:Burgers_example} is locally solvable for initial data in $C_b^k(\mathbb{R})$, $k\geq 1$, and also in $H^s(\mathbb{R})$ for integer $s\geq 2$. The first statement follows by applying the method of characteristics to Equation~\eqref{E:Burgers_example_equation}, see \cite[Section~3.2]{evans2010pde}, and the second is a special case of the Sobolev existence theory for symmetric hyperbolic systems, see \cite[Chapter 16, Section~1]{taylor2011pde3}.
Which choice we make depends on the task at hand. If we want a quick route to local existence by following characteristics, $C_b^k$ spaces are convenient, since their definition requires only basic calculus. But suppose we are studying Equation~\eqref{E:Burgers_example} as a warm-up to investigate a similar equation in higher dimensions. In that case, Sobolev spaces are more useful for comparison with general symmetric hyperbolic systems, whose standard local existence theory is based on estimates in Sobolev norms \cite[Chapter 9]{Ringstrom-Book-2009}.
Sobolev spaces have been shown to be particularly useful in the study of hyperbolic PDEs, which are the types of PDEs most relevant in relativity theory.  

In sum, \emph{local existence is always local existence in a function space.} While there is no general rule telling which function space should be employed for a given PDE, Sobolev spaces have been shown to be well adapted to the study of hyperbolic PDEs, with a comprehensive existence theory built on them.

We saw in our discussion about local existence of solutions to Equation~\eqref{E:Illustration_general_CP_via_ODE} in Section~\ref{S:Why_local_solutions} that typical methods to establish existence of solutions guarantee existence only on a finite time interval whose size depends on the size of the initial data. Something similar happens in PDEs, where the size of the initial data is measured in the norm of the function spaces employed. For example, applying standard PDE theory to Equation~\eqref{E:Burgers_example} yields a solution on a time interval $[0,T]$ with $T$ depending on $M$, $M$ being a bound on the initial data, e.g., $\norm{\Psi_0}_{C_b^1(\mathbb{R})} \leq M$ or $\norm{\Psi_0}_{H^s(\mathbb{R})}\leq M$ in the example in Equation~\eqref{E:Burgers_example}. As in the ODE discussion, this in general does not preclude the existence of global solutions, but further estimates are needed to establish it, and in many cases we know that global solutions are ruled out\footnote{We recall, as emphasized in Section~\ref{S:Existence_via_examples}, that global results are typically regime dependent.}. For example, every nonzero, compactly supported $C^1$ initial datum for Equation~\eqref{E:Burgers_example} has negative derivative somewhere. Following characteristics shows that $\partial_x\Psi$ becomes unbounded in finite positive time, while $\Psi$ remains bounded; see \cite[Section~3.4.1]{evans2010pde}. Thus, it is the classical solution that breaks down. This does not rule out continuing the solution in a weaker sense that allows shocks. 

We now turn our attention to a slightly more technical aspect of local existence, one that is relevant for Theorem \ref{Theorem_1_precise}. Let us continue to illustrate the main point with the Schr\"odinger equation. As we saw in 
Equation~\eqref{E:Schrodinger_solution_regularity}, initial data $\Psi_0 \in H^s(\mathbb{R}^3)$ yields a solution that, for each fixed time, belongs to $H^s(\mathbb{R}^3)$, i.e.,
\begin{align*}
    \Psi(t,\cdot) \in H^s(\mathbb{R}^3).
\end{align*}
Thus, the solution has the same spatial regularity as the initial data.

Suppose instead that we use a method that produces local solutions with control of one fewer spatial derivative at fixed $t>0$ than is assumed for the initial data, i.e., 
\begin{align*}
    \Psi_0 \in H^s(\mathbb{R}^3)
    \Longrightarrow
    \Psi(t,\cdot) \in H^{s-1}(\mathbb{R}^3).
\end{align*}
In such a case, there is a loss of information in the conclusion of the existence result: we start with initial data at $t=0$ that has all derivatives up to order $s$ square integrable, but the result guarantees only that derivatives up to order $s-1$ remain square-integrable for $t>0$. This need not mean that the solution actually loses a derivative; it may reflect a limitation of the proof. We say that we have established \emph{local existence with a derivative or regularity loss}.

When existence without regularity loss is available, a result with regularity loss gives less information about the solution. For other equations, a loss may be unavoidable. Whether such loss of information is relevant depends on the particular application one has in mind. For example, if all observables we are interested in computing from solutions involve $\Psi$ but not its derivatives, the loss of information about derivatives of order equal to $s$ does not preclude the computation of such observables. In this regard, it is useful to know that typical methods for local existence generally require a lower bound on the value of $s$ (e.g., the sufficient choices $s \geq 4$ and $s \geq 2$ made above), but leave $s$ otherwise free. If our application involves only derivatives of solutions up to a fixed order $\ell$, then we can pick\footnote{Something similar is already true without derivative loss. If we want to compute up to $\ell$ derivatives of the solution we need to choose $s \geq \ell$. 
Here we are illustrating regularity loss by one derivative, i.e., data in $H^s(\mathbb{R}^3)$ produce solutions in  
$H^{s-1}(\mathbb{R}^3)$, but derivative losses can be larger, i.e., data in $H^s(\mathbb{R}^3)$ can produce solutions in  
$H^{s-k}(\mathbb{R}^3)$ for some fixed $k<s$. Such losses can occur in existence results proved using Nash--Moser iteration, a method designed to handle derivative loss in intermediate estimates; a loss in those estimates does not by itself imply that the final solution is less regular than the data \cite{Secchi-2016}. In such cases we would then choose $s$ such that $s-k \geq \ell$.
We should point out here that we are using the term ``derivative" as implied in the definition of Sobolev spaces, i.e., in the sense of distributions (see Footnote \ref{FT:Sobolev_derivatives}). But if we insist that those derivatives up to order $\ell$ are classical derivatives as defined in calculus, we can rely on a standard result known as the Sobolev embedding theorem which implies that spatial derivatives up to order $\ell$ of the solution are in fact classical derivatives if $s-1>\ell+\frac{3}{2}$ in this three-dimensional example \cite[Section~5.6]{evans2010pde}.} $s$ such that $s-1 \geq \ell$. Thus, while results with derivative loss should be avoided when possible, they do not prevent us from obtaining solutions that can be used for many physical applications.

\subsubsection{Uniqueness}
\label{S:Uniqueness}
Uniqueness is the statement that there exists at most one local solution to the Cauchy problem within the class of solutions under consideration. Because we are talking about local solutions, a more appropriate term would be local uniqueness, but we will continue to refer simply to uniqueness if no confusion can arise. In the example in Equation~\eqref{E:Schrodinger_example}, this means the following. Imagine that we fix\footnote{We observe that we are talking about uniqueness of the Cauchy problem. Without imposing initial data, there is no reason to expect uniqueness to hold. This can be seen in the very simple case of linear ODEs where, without prescribing initial data, general solutions are not unique as they carry arbitrary constants of integration.} $\Psi_0$ and we have two local solutions $\Psi_1$ and $\Psi_2$ to Equation~\eqref{E:Schrodinger_example_equation} defined on the time intervals $[0,T_1]$ and $[0,T_2]$, respectively, with the regularity stated in Equation~\eqref{E:Schrodinger_solution_regularity}, and taking the same initial data, i.e., $\Psi_1(0,x)=\Psi_0(x)=\Psi_2(0,x)$. Uniqueness is then the statement that both solutions agree on their common interval of existence, i.e., $\Psi_1(t,x) = \Psi_2(t,x)$ for all $t \in [0,T_1] \cap [0,T_2]$ and all $x \in \mathbb{R}^3$.

The importance of uniqueness for physical applications should be clear. If for given initial data $\Psi_0$ we get two different solutions $\Psi_1$ and $\Psi_2$, which one do we use for applications? This naturally raises questions about the consistency of the model as a predictive tool\footnote{That uniqueness can fail is a real possibility. For example, already at the ODE level, Equation~\eqref{E:Illustration_general_CP_via_ODE} can admit non-unique solutions if we assume that $F$ is only continuous. With $F(t,y)=3|y|^{\frac{2}{3}}$ and $y_0=0$, both $y(t)\equiv0$ and $y(t)=t^3$ are solutions. See also \cite[Section~II.2]{Hartman-Book-2002} and the heat-equation example below.}. On some occasions, we might be able to disregard a solution as non-physical. But this approach is hard to implement in practice. For complex nonlinear systems of PDEs, we do not have direct access to solutions provided by local existence theorems, so that it is not clear how one would go about inspecting whether or not they are physical. Moreover, we might in this fashion miss some genuinely important physics that could be mistaken as ``non-physical"\footnote{A remarkable example concerns the Schwarzschild solution, which is now at the center of black-hole theory. In 1939, Einstein argued, on the basis of a stationary model with particles moving on circular orbits, that the ``Schwarzschild singularity" could not occur in a physical system \cite[Section~5]{Einstein-1939}. This argument did not address dynamical gravitational collapse.}.

Even if we could, in simple situations, inspect non-unique solutions after the fact and discard those deemed non-physical, such a process would be very ad hoc. Instead, we should start with a class of initial data considered physical (e.g., square-integrable functions in quantum mechanics, with possibly additional assumptions as discussed in Section~\ref{S:Function_spaces}) and specify the corresponding class of solutions in which uniqueness is to be established.

Non-uniqueness can, in fact, happen even for simple equations and very ``nice" initial data. For example, the one-dimensional heat equation $\partial_t\Psi-\partial_x^2\Psi=0$ on $\mathbb{R}$ admits nonzero smooth solutions with zero initial data if no restriction is imposed on their growth at spatial infinity \cite[Section~1]{Tychonoff-1935}. The zero solution has the same initial data, so uniqueness fails in this class. By contrast, for integer $s\geq 2$ and initial data in $H^s(\mathbb{R})$, uniqueness holds among solutions belonging to $C^0([0,T],H^s(\mathbb{R}))\cap C^1([0,T],H^{s-2}(\mathbb{R}))$ \cite[Chapter II, Section~2.13 and Proposition 6.2]{Engel-Nagel-Book-2000}. The zero initial datum belongs to every $H^s(\mathbb{R})$, but the nonzero solutions in Tychonoff's example do not belong to this class of solutions. This illustrates that, as in the case of local existence, uniqueness is always \emph{uniqueness in a function space.}

\subsubsection{Continuous dependence on the initial data}
\label{S:Continuous_dependence}

Continuous dependence (of the Cauchy problem) on the initial data, also referred to as continuous dependence on the data or simply continuous dependence, is roughly the statement that ``small changes in the initial conditions produce small changes in the corresponding solutions." In order to explain what this means in more quantitative terms, let us once more rely on Equation~\eqref{E:Schrodinger_example} for illustration purposes. Suppose that we have a sequence of initial data $\Psi_{0,n} \in H^s(\mathbb{R}^3)$ converging to a limit $\Psi_{0,\infty} \in H^s(\mathbb{R}^3)$, for an integer $s\geq 4$, i.e.,
\begin{align}
    \label{E:Continuous_dependence_Hs_Schrodiner_data}
    \lim_{n\rightarrow\infty} \norm{\Psi_{0,n} - \Psi_{0,\infty}}_{H^s(\mathbb{R}^3)} = 0.
\end{align}
Let $\Psi_n \in C^0([0,T],H^s(\mathbb{R}^3))$
and $\Psi_\infty \in C^0([0,T],H^s(\mathbb{R}^3))$ be the corresponding solutions. Continuous dependence on the initial data is then the statement that
\begin{align}
    \label{E:Continuous_dependence_Hs_Schrodiner_solutions}
    \lim_{n\rightarrow\infty} \norm{\Psi_{n} - \Psi_{\infty}}_{C^0([0,T],H^s(\mathbb{R}^3))} = 0.
\end{align}
In simple terms, continuous dependence on the initial data says that \emph{if the initial data converge, so do the solutions.} Here the convergence of solutions in the $H^s$ norm is uniform for $t\in[0,T]$.

We can say ``the" solutions above because uniqueness holds in this setting. In this Schr\"odinger equation example, we can take all the solutions to be defined on the same time interval $[0,T]$ because such solutions are global in time. For nonlinear problems, local existence theory generally produces solutions $\Psi_n$ defined on a time interval $[0,T_n]$ that depends on $n$. This is because, as we saw in Section~\ref{S:Function_spaces}, the time of existence supplied by local-in-time theorems typically depends on the size of the initial data (measured in the relevant norm of the problem), and $\norm{\Psi_{0,n}}_{H^s(\mathbb{R}^3)}$ may change with $n$. However, as we also saw in Section~\ref{S:Function_spaces}, many such theorems provide a positive time of existence depending only on a bound on $\norm{\Psi_{0,n}}_{H^s(\mathbb{R}^3)}$, i.e., on $M$ where
$\norm{\Psi_{0,n}}_{H^s(\mathbb{R}^3)} \leq M$,
rather than on $\norm{\Psi_{0,n}}_{H^s(\mathbb{R}^3)}$ itself. By Equation~\eqref{E:Continuous_dependence_Hs_Schrodiner_data}, there is an $M$ such that $\norm{\Psi_{0,n}}_{H^s(\mathbb{R}^3)} \leq M$ for all $n$ and $\norm{\Psi_{0,\infty}}_{H^s(\mathbb{R}^3)} \leq M$. When the local existence theorem supplies such a uniform time, there exists a common interval $[0,T]$ on which all solutions are defined and hence we can talk about continuous dependence on the initial data also in such nonlinear settings. Any additional conditions required by the theorem, such as a positive lower bound on the density, must also be controlled uniformly.

\begin{figure}
    \centering
    \includegraphics[scale=0.5]{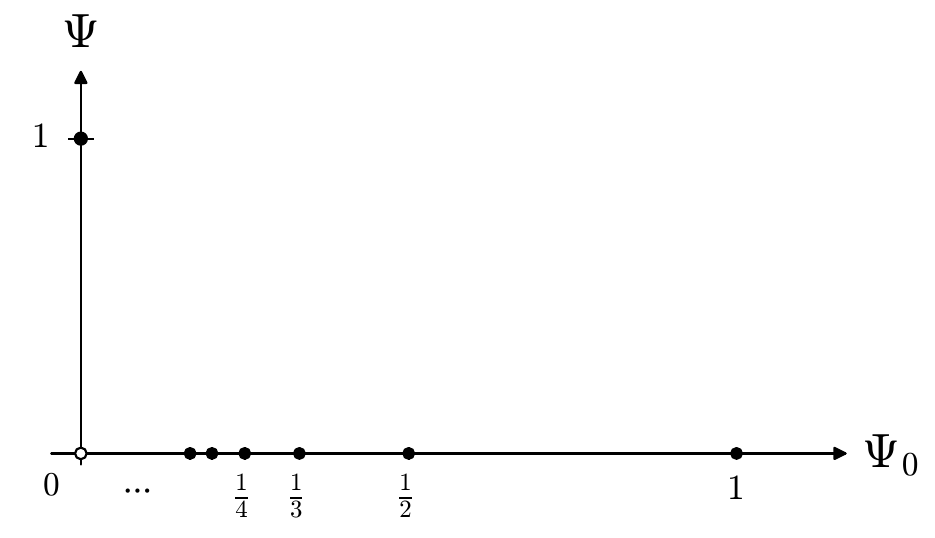}
    \caption{Illustration of discontinuous dependence on the initial data.}
    \label{fig:continuous_dependence_illustration}
\end{figure}

Continuous dependence on the data says that solutions respond continuously to changes in the initial data, with sufficiently small changes in the data producing small changes in the solutions on a fixed common time interval. This does not exclude, for instance, chaotic behavior or sensitivity to initial conditions over long times. Even exponential or double-exponential growth of an initial error is compatible with continuous dependence on a fixed time interval; a longer interval may require a smaller initial error to obtain the same accuracy. What continuous dependence excludes is a discontinuity in the response to changes in the initial data. Suppose we have a problem where $\Psi_0 \equiv 0$ produces the solution $\Psi \equiv 0$, and consider a sequence $\Psi_{0,n} \rightarrow 0$, with corresponding solutions $\Psi_n$. Continuous dependence on the data requires $\Psi_n \rightarrow 0$, with both convergences understood in the relevant norms. If $\Psi_n \centernot{\longrightarrow} 0$, this would indicate a ``jump" in the behavior of solutions as the data approach zero (see Figure~\ref{fig:continuous_dependence_illustration}).

Continuous dependence is important for physical applications because of unavoidable errors in measurements and numerical simulations. In particular, errors in prescribing the initial data lead us to consider solutions with slightly different initial conditions. We want the numerical solutions to stay close to the mathematical solution and the approximation to improve as discretization errors are reduced. Think roughly of the $\Psi_n$ in the last paragraph as exact solutions obtained with errors in the initial data and of the limit $n\rightarrow\infty$ as the limit in which these errors vanish. Numerical solutions also incur discretization and round-off errors during the evolution, and their control requires properties of the numerical method. We stress that, despite its importance, continuous dependence on the data is far from a sufficient condition to guarantee the reliability of numerical simulations, and several other ingredients are required, see Section~\ref{S:Importance_convergence}.

Furthermore, continuous dependence is also important to support common modeling approximations. Imagine that we are studying a rotationally invariant system with nearly spherically symmetric initial data $\Psi_0$ and corresponding solution $\Psi$. It is reasonable to approximate it using spherically symmetric data $\Psi_{0,\mathrm{sym}}$ and the corresponding solution $\Psi_{\mathrm{sym}}$ of the same equations. But for this we need to know that the closer the initial data are to spherical symmetry in the relevant norm, the closer the solutions, which is precisely the kind of comparison supplied by continuous dependence on the data.

Similarly, we often consider taking special limits, e.g., the conformal limit or the limit of zero shear viscosity. When we do so, we need the solutions to converge to the desired limit. For example, if $\Psi_\eta$ denotes solutions with coefficient of shear viscosity $\eta$ and $\Psi_{\eta\equiv 0}$ the solution with $\eta\equiv 0$, a natural physical requirement would be $\lim_{\eta\rightarrow 0} \Psi_\eta = \Psi_{\eta\equiv 0}$. Here, however, changing $\eta$ changes the equations. This limit requires control of the dependence on the coefficients of the equations as well as on the initial data; it does not follow from continuous dependence on the initial data for each fixed $\eta$. In particular, the zero-viscosity limit can be singular and may require estimates uniform in $\eta$ and a separate convergence argument.

One important aspect of continuous dependence to be emphasized is the following. In applications, we have freedom to choose the initial data, but not the solutions themselves. Solutions are determined by the PDEs and data, and will behave in accordance with the physical laws encoded in the equations of motion. In the zero-data example above, we can construct data $\Psi_{0,n}$ with $\Psi_{0,n} \rightarrow 0$ as $n \rightarrow \infty$, but whether the corresponding solutions $\Psi_n$ satisfy $\Psi_n \rightarrow 0$ is something that needs to be proven, not assumed. Knowing that the fixed system satisfies continuous dependence on the data gives precisely this conclusion.

We next turn to another technical aspect that will be relevant for Theorems \ref{Theorem_1_precise}, \ref{T:Precise_mixed_order_local_well_posedness}, and \ref{Theorem_SH_LWP}. Continuous dependence as stated in Equations~\eqref{E:Continuous_dependence_Hs_Schrodiner_data} and \eqref{E:Continuous_dependence_Hs_Schrodiner_solutions}
says that if the data converge in $H^s$, so do solutions -- or, more precisely, the solutions and their spatial derivatives up to order $s$ converge in $L^2$, uniformly for $t\in[0,T]$, since $H^s$ is defined relative to the spatial variables only. But we could have a weaker statement
\begin{align}
\label{E:Continuous_dependence_below_top_order_illustration}
    \lim_{n\rightarrow\infty} \norm{\Psi_{0,n} - \Psi_{0,\infty}}_{H^s(\mathbb{R}^3)} = 0
    \Longrightarrow
        \lim_{n\rightarrow\infty} \norm{\Psi_{n} - \Psi_{\infty}}_{C^0([0,T],H^{s-1}(\mathbb{R}^3))} = 0,
\end{align}
i.e., solutions are proven to converge in $H^{s-1}$, while convergence in $H^s$ has not been established. As in the discussion of Section~\ref{S:Function_spaces} about derivative loss, Equation~\eqref{E:Continuous_dependence_below_top_order_illustration} involves a loss of information: we start with data whose derivatives up to order $s$ converge in $L^2$, but conclude for solutions that only derivatives up to order $s-1$ converge in $L^2$, uniformly for $t\in[0,T]$. The same philosophy discussed in Section~\ref{S:Function_spaces} applies here; whether this loss of information is relevant for physical applications depends on the particular problem. If we are interested only in observables that depend continuously on the solution and its spatial derivatives up to order $\ell$, we can take $s$ large enough to control those observables, provided the data have the required regularity and the existence and continuous-dependence results apply at that regularity.\footnote{For example, in three space dimensions, $s-1>\ell+\frac{3}{2}$ ensures that convergence in $C^0([0,T],H^{s-1}(\mathbb{R}^3))$ implies uniform convergence of spatial derivatives up to order $\ell$, by Sobolev embedding \cite[Section~5.6]{evans2010pde}. This is sufficient, for example, for polynomial expressions in these derivatives. Observables involving time derivatives require additional control of those derivatives, typically obtained from the equations.} The perturbations of the data must still be small in $H^s$; small errors in a weaker norm need not be small in this stronger norm. See \cite[Sections~5--6]{Ifrim-Tataru-2023} for the distinction between convergence below and at top order, and \cite[Section~2.3]{Rendall-2002} and \cite[Sections~3.1 and 7.1]{Sarbach-Tiglio-2012} for related discussions of the choice of regularity and its physical and numerical significance.

\subsubsection{Causality}
\label{Causality_section}

\begin{figure}
    \centering
  \includegraphics[scale=0.7]{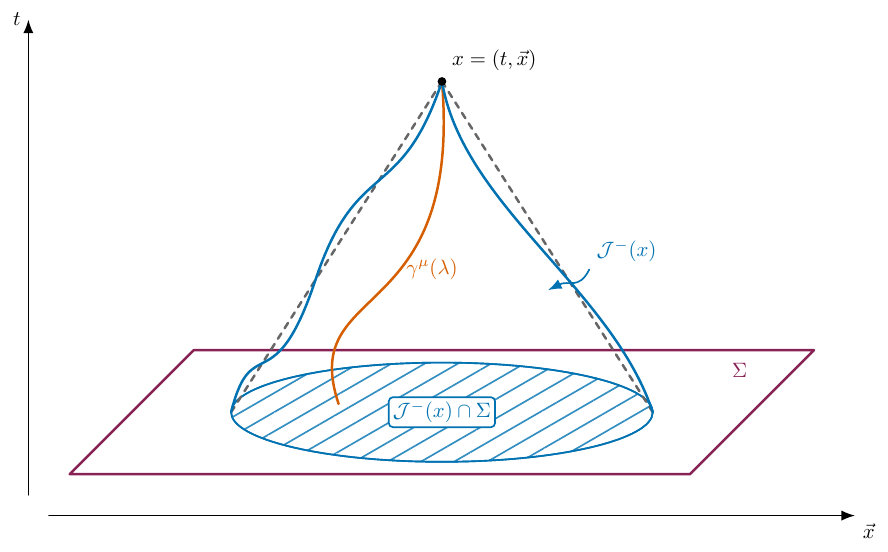}
    \caption{Sketch of the domain of influence. The Cauchy surface $\Sigma$ usually corresponds to the initial data, with $\mathcal{J}^-(x)\cap\Sigma$ being those points on $\Sigma$ that lie in the causal past of point $x$. The timelike curve $\gamma$ parameterized by some affine parameter $\lambda$ represents an instance of causal information transport between $\Sigma$ and $x$. A key facet of this definition is that any curves outside of $\mathcal{J}_-(x)$ do not influence the solution at $x$.}
    \label{fig:causality}
\end{figure}

Causality is a fundamental principle of relativity \cite{Hawking-Ellis-Book-1973} and hence must be respected by any theory that aspires to be relativistic. Consider a globally hyperbolic spacetime\footnote{We recall that spacetimes that arise as a solution to the initial-value problem for Einstein's equations (with or without matter coupling) as well as Minkowski space are globally hyperbolic. Global hyperbolicity precludes the existence of closed timelike curves, although it must be stressed that this by itself is not enough to ensure causality, since there are many examples, such as Eckart's theory, that are defined in Minkowski space but violate causality.} and let $\Psi$ be a field defined on spacetime. We say that $\Psi$ propagates causally if for any spacetime point $x$, $\Psi(x)$ depends only on the causal past of $x$, noting that the causal past is well-defined in view of the global hyperbolicity assumption. In other words, any physical  quantity can only be influenced by its causally connected past. Thus, it is completely oblivious to information from other points in spacetime that lie outside the lightcone. We highlight the definition of causality in Figure~\ref{fig:causality}. The key for this definition is to make precise what it means to ``depend only on the causal past of $x$," which is formalized in Definition \ref{def:causality}. 

\begin{definition}[Causality]\label{def:causality}
Let $(\mathcal{M}, g)$ be a globally hyperbolic spacetime. We are concerned with the Cauchy problem for a system of differential equations
\be
\label{E:PDE_causality_def}
\mathcal{D}\Psi = \boldsymbol{0},
\ee
where $\mathcal{D}$ is a differential operator, possibly depending on $\Psi$ itself (that is the case for nonlinear equations), where $\Psi$ is the  unknown. 
Let $\Psi_1$ be a solution to Equation~\eqref{E:PDE_causality_def} with initial data given along a Cauchy surface $\Sigma\subset\mathcal{M}$. Let $x \in \mathcal{M}$ be in the future\footnote{The future of $\Sigma$ is well defined in view of global hyperbolicity.} of $\Sigma$. We say that $\Psi_1(x)$ depends only on the causal past of $x$, denoted $\mathcal{J}^-(x)$, if the following happens. Let
$\Psi_2$ be a solution to Equation~\eqref{E:PDE_causality_def} with initial data given along the same Cauchy surface $\Sigma$. Assume that the initial data for $\Psi_1(x)$ and $\Psi_2(x)$ agree on $\mathcal{J}^-(x) \cap \Sigma$. Then, $\Psi_1(y) = \Psi_2(y)$ for every $y \in \mathcal{J}^-(x)$ and $y$ to the future of $\Sigma$. We say that Equation~\eqref{E:PDE_causality_def} is causal if solutions to Equation~\eqref{E:PDE_causality_def} depend only on the causal past of $x$ for every $x \in \mathcal{M}$ in the future of $\Sigma$.
\end{definition}

The physical interpretation of Definition \ref{def:causality} is simple. Since no causal curve can connect $\Sigma \backslash (\mathcal{J}^-(x) \cap \Sigma)$ to $x$ (the complement of $\mathcal{J}^-(x) \cap \Sigma$ inside $\Sigma$), the value of a solution at $x$ cannot depend on the data in the region  $\Sigma \backslash (\mathcal{J}^-(x) \cap \Sigma)$. Thus, two solutions whose data agree on $\mathcal{J}^-(x) \cap \Sigma$ (but might otherwise be different in $\Sigma \backslash (\mathcal{J}^-(x) \cap \Sigma)$) cannot be distinguished within $\mathcal{J}^-(x)$, which is the region composed of all causal curves that start on $\Sigma$ and reach $x$.

We observe that causality is a \emph{conditional statement:} given solutions to Equation~\eqref{E:PDE_causality_def}, it states a property of such solutions, namely, that the solution fields depend only on the causal past of $x$. It makes no claim about whether such solutions in fact exist. Thus, causality can be studied independently of the problem of existence of solutions.

In order to determine whether a PDE system is causal, the key ingredient is to show that the characteristic surfaces $\{ \varphi(x) = 0 \}$ of the PDE emanating from $x$ lie inside the lightcone with vertex at $x$ (for every $x$ in $\mathcal{M}$). This, in turn, boils down to showing that the roots  of the characteristic determinant are always real and spacelike \cite[Appendix~B.2]{Disconzi-2024}. 

\subsubsection{Stability}
\label{Stability_section}

In the context of relativistic viscous hydrodynamics, stability means \emph{mode stability} of solutions to the equations of motion linearized about global equilibrium states. Thus, one linearizes the equations about a state with vanishing dissipation, seeks plane-wave solutions to the resulting linear equations, and carries out a mode analysis to exclude exponentially growing solutions.

The basic idea is that the linearized equations work as a proxy for the behavior of the system near equilibrium. The expectation is that small disturbances away from equilibrium will decay back to equilibrium or at least stay near equilibrium states due to the effect of dissipation. While this picture is physically appealing, it is important to stress that it makes several implicit and often unproven assumptions about the behavior of solutions to the full, nonlinear equations of motion describing relativistic viscous fluids. Essentially, it assumes that, for initial data very close to equilibrium, the behavior of nonlinear solutions is well approximated by the behavior of the corresponding linearized equations. While this can in principle be true for specific models, it is not difficult to see that it is not an expected property of nonlinear PDEs.

Consider the following problem
\begin{subequations}{\label{E:Linear_stability_example}}
\begin{align}
\label{E:Linear_stability_example_eq}
    \partial_t \Psi + \partial_x \Psi 
    & = \frac{\Psi^2}{1+ \Psi^2},
    \\
\label{E:Linear_stability_example_IC}    
    \Psi(0,x) &= \Psi_0(x),
\end{align}
\end{subequations}
on the one-dimensional torus, $x \in \mathbb{T} := \mathbb{R}/(2\pi \mathbb{Z})$, and for convenience take the measure of integration to be $\frac{dx}{2\pi}$ so $\mathbb{T}$ has volume one. Observe that Equation~\eqref{E:Linear_stability_example_eq} is a causal hyperbolic PDE with characteristic speed equal to the speed of light. Note also that $\Psi\equiv 0$ is a solution for zero data.

Consider constant initial data
\begin{align}
\label{E:Linear_stability_example_constant_data} 
    \Psi_0(x) = \alpha, \quad 0 < \alpha < \frac{1}{2}.
\end{align}
It is not difficult to see that the problem in Equation~\eqref{E:Linear_stability_example} admits uniqueness of sufficiently regular solutions. Thus, to solve the problem with the constant data in Equation~\eqref{E:Linear_stability_example_constant_data}, it suffices to seek $x$-independent solutions $\Psi(t,x) = \Phi(t)$. The problem in Equation~\eqref{E:Linear_stability_example} then becomes the ODE problem
\begin{align*}
    \dot{\Phi}  
    & = \frac{\Phi^2}{1+ \Phi^2},
    \\   
    \Phi(0) &= \alpha.
\end{align*}
Because $\dot{\Phi}  
     = \frac{\Phi^2}{1+ \Phi^2}>0$ and $\Phi(0)>0$, we have $\Phi(t)>0$ and can thus multiply the ODE by $1+\Phi^{-2}$, rewriting the ODE as
\begin{align}
    \frac{d}{dt}\left(\Phi(t) - \frac{1}{\Phi(t)} \right) = 1,
\end{align}
which gives
\begin{align*}
  \Phi(t) - \frac{1}{\Phi(t)} = t + \alpha - \frac{1}{\alpha}.  
\end{align*}
It follows that no matter how small $\alpha$ is, at time
\begin{align*}
    t_*= \frac{1}{\alpha} - \alpha -\frac{3}{2} >0 
\end{align*}
we obtain
\begin{align*}
    \Phi(t_*) = \frac{1}{2}.
\end{align*}
In terms of norms, taking $\alpha$ small, we obtain
\begin{align*}
    \norm{\Phi(0)}_{L^2(\mathbb{T})} = \alpha \ll \frac{1}{2}, \quad \text{but} \quad
    \norm{\Phi(t_*)}_{L^2(\mathbb{T})} = \frac{1}{2}.
\end{align*}
The conclusion is that the zero solution of Equation~\eqref{E:Linear_stability_example_eq} is \emph{nonlinearly unstable:} no matter how small we take $\alpha$, eventually the solution becomes effectively $O(1)$, far larger than the initial condition.

Consider now the linearization of Equation~\eqref{E:Linear_stability_example_eq} about zero with the same initial data:
\begin{subequations}{\label{E:Linear_stability_example_linearization}}
\begin{align}
\label{E:Linear_stability_example_linearization_eq}
    \partial_t \delta\Psi + \partial_x \delta\Psi 
    & = 0,
    \\
\label{E:Linear_stability_example_linearization_IC}    
    \delta\Psi(0,x) &= \alpha,
\end{align}
\end{subequations}

To analyze the modes of Equation~\eqref{E:Linear_stability_example_linearization_eq}, we seek solutions of the form
\begin{align}
\label{E:Linear_stability_example_mode}
    \delta\Psi(t,x) = A e^{i(\omega t + kx)},
\end{align}
where $A\in\mathbb{C}\backslash\{0\}$, $\omega\in\mathbb{C}$, and $k\in\mathbb{Z}$. The restriction $k\in\mathbb{Z}$ ensures that the modes are $2\pi$-periodic in $x$. Substitution of Equation~\eqref{E:Linear_stability_example_mode} into Equation~\eqref{E:Linear_stability_example_linearization_eq} gives
\begin{align*}
    i(\omega+k)A e^{i(\omega t+kx)} = 0.
\end{align*}
Since $A\neq 0$, the dispersion relation is
\begin{align*}
    \omega=-k\in\mathbb{R}.
\end{align*}
With the convention in Equation~\eqref{E:Linear_stability_example_mode}, the time dependence of the amplitude is
\begin{align*}
    |e^{i\omega t}|=e^{-\operatorname{Im}(\omega)t}=1.
\end{align*}
Thus, every periodic Fourier mode has constant amplitude: none grows or decays exponentially. This proves mode stability of the zero solution of Equation~\eqref{E:Linear_stability_example_linearization_eq}.

The mode calculation also gives stability in $L^2(\mathbb{T})$. For arbitrary initial data $\delta\Psi_0\in L^2(\mathbb{T})$, let $c_k$ denote its Fourier coefficients, so that
\begin{align*}
    \delta\Psi_0(x)&=\sum_{k\in\mathbb{Z}}c_k e^{ikx}.
\end{align*}
The solution and its norm are then given by
\begin{align*}
    \delta\Psi(t,x)&=\sum_{k\in\mathbb{Z}}c_k e^{-ikt}e^{ikx},
    \\
    \norm{\delta\Psi(t,\cdot)}_{L^2(\mathbb{T})}^2
    &=\sum_{k\in\mathbb{Z}}|c_k e^{-ikt}|^2
    =\sum_{k\in\mathbb{Z}}|c_k|^2
    =\norm{\delta\Psi_0}_{L^2(\mathbb{T})}^2,
\end{align*}
where the series converge in $L^2(\mathbb{T})$ and the norm identity follows from Parseval's identity. This is neutral stability, with no decay of the $L^2$ norm.

In particular, the constant initial data in Equation~\eqref{E:Linear_stability_example_linearization_IC} excite only the mode $k=0$, for which $\omega=0$. Hence,
\begin{align*}
    \delta\Psi(t,x)&=\alpha,
    \\
    \norm{\delta\Psi(t,\cdot)}_{L^2(\mathbb{T})}&=\alpha
\end{align*}
for every $t\geq 0$. At time $t_*$, the nonlinear solution has $L^2$ norm $\frac{1}{2}$, while the linearized solution still has $L^2$ norm $\alpha$. Thus, linear mode stability does not imply nonlinear stability, even for arbitrarily small initial data.

(A simpler way of concluding stability of the linearized problem is to note that solutions can be written as $\Psi(t,x)=\Psi_0(x-t)$, so that $\norm{\Psi(t,\cdot)}_{L^2(\mathbb{T})} = \norm{\Psi_0}_{L^2(\mathbb{T})}$.)

The message is that linear mode analysis should be viewed as a simple test for the behavior of the original nonlinear equations, but we should be careful about generalizing its conclusions to the nonlinear problem that we are ultimately interested in.

\subsection{Numerical convergence}
\label{S:Importance_convergence}
The importance of verifying numerical results through convergence tests is also well known. Alcubierre emphasizes that ``We should simply never believe a numerical calculation for which no convergence studies have been made" \cite[Section~9.11]{Alcubierre-Book-2008}. In the computational fluid dynamics literature, Roache states that ``Systematic grid-convergence studies are the most common, most straightforward and arguably constitute the most reliable technique for the quantification of numerical uncertainty" \cite[pages 156ff]{Roache-1997}. Furthermore, Baumgarte and Shapiro point out that, even when neither an analytical solution nor a constraint is available, ``we can always verify that a numerical code converges self-consistently," while cautioning that this provides ``a necessary, but not sufficient, test of a numerical code" \cite[Appendix~B.4]{Baumgarte-Shapiro-Book-2021}.

\section{Overview of div--curl cleaning tools}
\label{Sec:CC}

In this Section, we review the main ideas behind the div-curl cleaning method introduced in the 
pioneering work of Dedner et al. \cite{DEDNER2002645} on generalized Lagrange multiplier (GLM) methods (readers familiar with the div-curl cleaning method can skip this Section). 
These techniques address numerical errors that grow exponentially in magnetohydrodynamics (MHD) due to the divergence constraint $\nabla\cdot\mathbf{B}=0$ for magnetic fields. While mathematically this constraint is preserved under exact evolution, numerical rounding errors in simulations inevitably violate it, potentially leading to exponentially growing artifacts. The GLM approach introduces auxiliary fields, called \emph{cleaning fields,} governed by hyperbolic equations to control these errors through relaxation mechanisms, albeit at the cost of additional computational complexity. 

Consider the magnetic field equation $\partial_t\mathbf{B}-\nabla\times \mathbf{E}=0$, which lacks any dynamical enforcement of $\nabla\cdot\mathbf{B}=0$. To maintain this constraint numerically, we introduce a Lagrange multiplier $\phi$, called a divergence-cleaning field, with the modified system
\begin{subequations}
\begin{align}
\partial_t\mathbf{B}-\nabla\times\mathbf{E}+\nabla \phi&=0
\label{CC_1a}
\\
\partial_t\phi+c_h^2\nabla\cdot\mathbf{B}&=-\frac{\phi}{\tau}.
\label{CC_1b}
\end{align}  
\end{subequations}
where $c_h$ represents the propagation speed of $\phi$ and $\tau$ is the damping timescale, both of which are constants whose values have to be chosen depending on the application (see section~\ref{S:Bounds_cleaning_speeds}
).
The limit $c_h\rightarrow\infty$ formally enforces $\nabla\cdot \mathbf{B}=0$. 

Observe that Equation~\eqref{CC_1b} implies that $\nabla \cdot \mathbf{B} \approx 0$ if $\phi$ and its derivatives are small compared to $c_h$ and $\tau$. Thus, one way of numerically enforcing the divergence-free constraint of the magnetic field is to ensure that $\phi$ and its time derivative do not grow over time; we also choose the initial condition for $\phi$ to be zero in numerical simulations.

Differentiating Equation~\eqref{CC_1b} with respect to $t$ and using Equation~\eqref{CC_1a} decouples $\phi$, yielding the damped wave equation
\begin{align}
\label{phi_wave_eq_B_example}
\partial_t^2\phi-c_h^2\nabla^2\phi=-\frac{\partial_t\phi}{\tau},
\end{align}    
which is hyperbolic for all $c_h>0$ with characteristics $\xi_0=\pm c_h\sqrt{\xi_i\xi_i}$.

For $\left. \phi \right|_{t=0} = \left. \partial_t \phi \right|_{t=0} = 0$ (the latter condition being compatible with Equation~\eqref{CC_1b} and zero divergence of $\mathbf{B}$ at time zero), and under boundary conditions for which integration by parts produces no boundary contribution, the energy functional
\begin{align*}
E(t)&=\frac{1}{2}\int \left[(\partial_t\phi)^2+c_h^2|\nabla\phi|^2\right]d^3x
\end{align*}
satisfies 
\begin{align*}
\partial_tE(t)&=-\frac{1}{\tau}\int(\partial_t\phi)^2d^3x\leq 0.
\end{align*}

Therefore, we see that $\phi$ and its time derivative cannot grow, hence guaranteeing that $\nabla\cdot \mathbf{B} \approx 0$. (We note that we can differentiate Equation~\eqref{phi_wave_eq_B_example} to show that derivatives of $\phi$ cannot grow either, obtaining stronger control over the boundedness of $\phi$ and its derivatives.)

\begin{remark}
While $c_h>1$ preserves mathematical well-posedness, it permits acausal propagation for cleaning fields -- a permissible trade-off since these fields are non-physical. In practice, large $c_h$ values are often necessary to outpace numerical error growth.
\end{remark}

\begin{remark}
It is implicit in the above introduction of a cleaning field that the cleaning speed $c_h$ and the damping timescale $\tau$ are positive constants. Our discussion in this Section assumes that all cleaning speeds and damping timescales are positive constants, in particular for $c_\varphi,c_\phi,\tau_1,\tau_2$ introduced below. We use $\bm{\varphi}$ for the vector curl-cleaning field and $\phi$ for the scalar divergence-cleaning field.
\end{remark}

\subsection{First-Order Reduction and Curl Cleaning}
\label{Reduction_toy_model_section}

We will now work out an example more akin to the BDNK reduction of Section~\ref{Sec:First_Order}, which involves
curl constraints (e.g., $\mathbf{X}=\nabla\psi$ implies $\nabla\times\mathbf{X}=0$). Numerical errors violate these constraints, necessitating a technique similar to the one discussed above but for the curl instead of the divergence of a vector. 

Consider the damped wave equation
\begin{align}
\label{Example_damped_wave}
\partial^2_t\psi-\nabla^2\psi&=-\psi.
\end{align}

Introducing variables 
\begin{subequations}
\label{Example_curl}
\begin{align}
Y&=\partial_t\psi
\label{Example_curl_Y}
\\
\mathbf{X}&=\nabla\psi.
\label{Example_curl_X}
\end{align}
\end{subequations}
yields
\begin{subequations}
\label{CC_2}
\begin{align}
\partial_t Y-\nabla\cdot \mathbf{X}&=-\psi
\label{CC_2a}
\\
\partial_t\mathbf{X}-\nabla Y&=0
\label{CC_2b}
\\
\partial_t\psi&=Y.
\label{CC_2c}
\end{align}
\end{subequations}

The system in Equations~\eqref{CC_2} admits more solutions than 
Equation~\eqref{Example_damped_wave} in the sense that every solution to Equation~\eqref{Example_damped_wave} yields a solution to Equations~\eqref{CC_2} upon defining $Y$ and $\mathbf{X}$ by Equations~\eqref{Example_curl}, but not every solution to Equation~\eqref{CC_2} gives a solution to Equation~\eqref{Example_damped_wave} when we consider $Y,\mathbf{X},\psi$ as independent variables, so that a solution $\mathbf{X}$ to Equations~\eqref{CC_2} need not be of the form in Equation~\eqref{Example_curl_X}. However, for properly chosen initial data we can recover solutions to Equation~\eqref{Example_damped_wave} from solutions to Equations~\eqref{CC_2} as follows. Combining Equations~\eqref{CC_2b} and \eqref{CC_2c} gives
\begin{align*}
\partial_t(\mathbf{X}-\nabla\psi)=0,
\end{align*} 
with general solution $\mathbf{X}(t,\mathbf{x})=\nabla\psi(t,\mathbf{x})+f(\mathbf{x})$. If the initial data is such that $\mathbf{X}(0,\mathbf{x})=\nabla\psi(0,\mathbf{x})$, then $f(\mathbf{x}) \equiv 0$, and we obtain from Equation~\eqref{CC_2a} that $\psi$ is also a solution to Equation~\eqref{Example_damped_wave}, since Equation~\eqref{Example_curl_Y} is satisfied because $\psi$ and $Y$ are solutions to Equation~\eqref{CC_2c}.

In other words, $\mathbf{X} = \nabla \psi$ is a \emph{constraint that needs to be propagated} for solutions to Equations~\eqref{CC_2} to yield solutions to Equation~\eqref{Example_damped_wave}, meaning that we impose $\mathbf{X} = \nabla \psi$ to hold at time zero and then need to prove that $\mathbf{X} = \nabla \psi$ remains true for $t>0$ for solutions to Equations~\eqref{CC_2}. 

Since $\nabla \times \nabla \psi = 0$, the constraint $\mathbf{X}=\nabla\psi$ implies the curl constraint $\nabla\times\mathbf{X}=0$. The converse is not true in general: $\nabla\times\mathbf{X}=0$ implies locally that $\mathbf{X}=\nabla\widetilde{\psi}$ for some scalar $\widetilde{\psi}$, but it does not imply that $\widetilde{\psi}=\psi$ up to a constant. Thus, $\nabla\times\mathbf{X}=0$ is a necessary compatibility condition for $\mathbf{X}=\nabla\psi$, rather than an equivalent statement of the full constraint. For solutions to Equations~\eqref{CC_2}, the full constraint is propagated by the identity $\partial_t(\mathbf{X}-\nabla\psi)=0$ derived above.

In numerical simulations, we will in fact have $\mathbf{X}=\nabla\psi+\mathbf{e}$, where $\mathbf{e}$ is an error due to numerical imprecision. If the error remains very small, then $\mathbf{X}\approx\nabla\psi$ and $\nabla\times\mathbf{X}\approx0$. But without a control mechanism, $\mathbf{e}$ can grow over time so that in the end $|\mathbf{X}-\nabla\psi|\gg1$ and numerical solutions to Equation~\eqref{CC_2} will not give numerical solutions to Equation~\eqref{Example_damped_wave}.

This difficulty is addressed with a similar method as the above example for $\nabla \cdot \mathbf{B} = 0$, but now involving the curl. Following \cite{DUMBSER2020109088}, we introduce a vector 
curl-cleaning field $\bm{\varphi}$:
\begin{subequations}
\begin{align}
\partial_t\mathbf{X}-\nabla Y+\nabla\times\bm{\varphi}&=0
\label{CC_3a}
\\
\partial_t\bm{\varphi}-c_\varphi^2\nabla\times \mathbf{X}&=-\frac{\bm{\varphi}}{\tau_1}.
\label{CC_3b}
\end{align}
\end{subequations}
where $c_\varphi$ and $\tau_1$ are the propagation speed and damping timescale, taken to be constants (again, their specific values depend on applications, see Section~\ref{S:Bounds_cleaning_speeds}), 
 and Equations~\eqref{CC_2a} and \eqref{CC_2c} are also part of the system.

As before, we want to prevent $\bm{\varphi}$
from growing over time. Equations~\eqref{CC_3a} and \eqref{CC_3b} imply
\begin{align}
\label{Example_complicated_wave_cleaning}
\partial_t^2\bm{\varphi}-c_\varphi^2\nabla^2\bm{\varphi}+c_\varphi^2\nabla(\nabla\cdot\bm{\varphi})&=-\frac{\partial_t\bm{\varphi}}{\tau_1},
\end{align}
which does not seem immediately amenable to a simple growth analysis. However, we can analyze $\bm{\varphi}$ by analyzing its curl and divergence separately. Taking the curl and divergence of Equation~\eqref{Example_complicated_wave_cleaning} yields 

\begin{subequations}
\label{Example_complicated_wave_div_curl}
\begin{align}
\label{Example_complicated_wave_curl}
\partial_t^2(\nabla\times\bm{\varphi})-c_\varphi^2\nabla^2(\nabla\times\bm{\varphi})&=-\frac{\partial_t(\nabla\times\bm{\varphi})}{\tau_1}\\
\label{Example_complicated_wave_div}
\partial_t^2(\nabla\cdot\bm{\varphi})&=-\frac{\partial_t(\nabla\cdot\bm{\varphi})}{\tau_1}.
\end{align}
\end{subequations}
Equation~\eqref{Example_complicated_wave_curl}
is a damped wave equation for $\nabla \times \bm{\varphi}$. Thus, $\nabla \times \bm{\varphi}$ will remain zero if it and its time derivative are initially zero (the latter condition being compatible with Equation~\eqref{CC_3b} and $\nabla\times\mathbf{X}=0$ at time zero).
Equation~\eqref{Example_complicated_wave_div} admits solutions of the form
\begin{align*}
\nabla\cdot\bm{\varphi}&=f(\mathbf{x})+\tau_1 h(\mathbf{x})e^{-\frac{t}{\tau_1}}.
\end{align*}
However, the solutions to Equation~\eqref{Example_complicated_wave_div} that arise from the first-order Equation~\eqref{CC_3b} satisfy an additional compatibility condition. Indeed, setting $D=\nabla\cdot\bm{\varphi}$ and taking the divergence of Equation~\eqref{CC_3b} gives
\begin{align*}
\partial_tD=-\frac{D}{\tau_1},
\end{align*}
because $\nabla\cdot(\nabla\times\mathbf{X})=0$. Therefore,
\begin{align*}
D(t,\mathbf{x})=e^{-\frac{t}{\tau_1}}D(0,\mathbf{x}).
\end{align*}
Equivalently, the compatibility condition $\partial_tD(0,\mathbf{x})=-\frac{D(0,\mathbf{x})}{\tau_1}$ forces $f(\mathbf{x})=0$ in the general solution to Equation~\eqref{Example_complicated_wave_div}. In particular, for initial conditions such that $\bm{\varphi}(0,\mathbf{x})=0$ and $\partial_t\bm{\varphi}(0,\mathbf{x})=0$, we obtain $\nabla\cdot\bm{\varphi}=0$.

Thus, the time-independent term $f(\mathbf{x})$ is absent for solutions to the continuum system in Equations~\eqref{CC_3a}--\eqref{CC_3b}. Nevertheless, the discrete divergence and curl operators need not satisfy the discrete analogue of $\nabla\cdot(\nabla\times\mathbf{X})=0$ exactly, and numerical divergence errors in $\bm{\varphi}$ can therefore be generated. To propagate and damp these errors, following \cite{DUMBSER2020109088}, we introduce a further scalar cleaning field $\phi$, which is a divergence-cleaning field for $\nabla\cdot\bm{\varphi}$.

In sum, following \citep{DUMBSER2020109088}, the system to be numerically simulated that controls violations of the constraint $\mathbf{X} = \nabla \psi$ is 
\begin{subequations}
\label{CC_4}
\begin{align}
\partial_t\mathbf{X}-\nabla Y+\nabla\times\bm{\varphi}&=0
\label{CC_4a}
\\
\partial_t\bm{\varphi}-c_\varphi^2\nabla\times \mathbf{X}+\nabla\phi&=-\frac{\bm{\varphi}}{\tau_1}
\label{CC_4b}
\\
\partial_t\phi+c_\phi^2\nabla\cdot \bm{\varphi}&=-\frac{\phi}{\tau_2}.
\label{CC_4c}
\end{align}
\end{subequations}
in addition to Equations~\eqref{CC_2a} and \eqref{CC_2c}.
It then follows that, under boundary conditions for which integration by parts produces no boundary contribution, the energy
\begin{align*}
E(t)&=\frac{1}{2}\int\left[|\partial_t\bm{\varphi}|^2+c_\varphi^2|\nabla\times\bm{\varphi}|^2+c_\phi^2|\nabla\cdot\bm{\varphi}|^2+\frac{1}{c_\phi^2}(\partial_t\phi)^2+|\nabla\phi|^2\right]d^3x
\end{align*}
satisfies
\begin{align*}
\partial_tE(t)&=-\int\left[\frac{|\partial_t\bm{\varphi}|^2}{\tau_1}+\frac{(\partial_t\phi)^2}{c_\phi^2\tau_2}+\frac{|\nabla\phi|^2}{\tau_2}+\frac{c_\phi^2(\nabla\cdot\bm{\varphi})^2}{\tau_1}\right]d^3x\leq 0,
\end{align*}
Thus, $E(t)\leq E(0)$, and each norm appearing in $E(t)$ remains bounded in terms of the initial energy. As in the previous example with the magnetic field, stronger control of the cleaning fields can be obtained by applying spatial derivatives to Equations~\eqref{CC_4} and repeating the same energy estimate.

To explain precisely how the curl-cleaning field controls the original constraint, define
\begin{align*}
\mathbf{C}&=\mathbf{X}-\nabla\psi
\\
\mathbf{Q}&=\nabla\times\mathbf{X}=\nabla\times\mathbf{C}
\\
\mathbf{W}&=\nabla\times\bm{\varphi}.
\end{align*}
Equations~\eqref{CC_4a} and \eqref{CC_2c}, together with the curls of Equations~\eqref{CC_4a} and \eqref{CC_4b}, give
\begin{align*}
\partial_t\mathbf{C}&=-\mathbf{W}
\\
\partial_t\mathbf{Q}+\nabla\times\mathbf{W}&=0
\\
\partial_t\mathbf{W}-c_\varphi^2\nabla\times\mathbf{Q}&=-\frac{\mathbf{W}}{\tau_1}.
\end{align*}
The scalar divergence-cleaning field $\phi$ does not appear in the equations for $\mathbf{Q}$ and $\mathbf{W}$ because $\nabla\times\nabla\phi=0$. Thus, the curl of $\bm{\varphi}$ is the part of the cleaning field that acts directly on the error $\mathbf{C}$, whereas $\phi$ controls the numerical divergence errors in $\bm{\varphi}$ discussed above. Moreover, the energy
\begin{align*}
E_{\mathrm{curl}}(t)&=\frac{1}{2}\int\left[c_\varphi^2|\mathbf{Q}|^2+|\mathbf{W}|^2\right]d^3x
\end{align*}
satisfies
\begin{align*}
\partial_tE_{\mathrm{curl}}(t)&=-\frac{1}{\tau_1}\int|\mathbf{W}|^2d^3x\leq0,
\end{align*}
again under boundary conditions for which integration by parts produces no boundary contribution. Thus, $E_{\mathrm{curl}}(t)\leq E_{\mathrm{curl}}(0)$, so the $L^2$ norms of $\mathbf{Q}$ and $\mathbf{W}$ remain bounded in terms of the initial curl energy.

Although the first equation for $\mathbf{C}$ gives
\begin{align*}
\mathbf{C}(t)=\mathbf{C}(0)-\int_0^t\mathbf{W}(s)\,ds,
\end{align*}
decay of $\mathbf{W}$ by itself does not imply that this integral is small; an integrable-in-time estimate for $\mathbf{W}$ would be needed for that conclusion. Instead, taking the divergence of $\partial_t\mathbf{C}=-\mathbf{W}$ gives
\begin{align*}
\partial_t(\nabla\cdot\mathbf{C})=0.
\end{align*}
Therefore, compatible initial data $\mathbf{C}(0)=0$ gives $\nabla\cdot\mathbf{C}(t)=0$. On a periodic spatial domain, the spatial mean of $\mathbf{C}$ also remains zero, since the spatial mean of a curl vanishes. For every integer $s\geq1$, one then has
\begin{align*}
\|\mathbf{C}\|_{H^s}&\leq C_s\left(\|\nabla\times\mathbf{C}\|_{H^{s-1}}+\|\nabla\cdot\mathbf{C}\|_{H^{s-1}}\right),
\end{align*}
and consequently
\begin{align*}
\|\mathbf{X}-\nabla\psi\|_{H^s}&\leq C_s\|\nabla\times\mathbf{X}\|_{H^{s-1}}.
\end{align*}
On a domain with boundary, the corresponding conclusion requires boundary conditions that give the analogous estimate and exclude nonzero vector fields satisfying both zero curl and zero divergence. Thus, controlling $\nabla\times\mathbf{X}$ controls the full constraint error under these additional conditions. Without them, $\nabla\times\mathbf{X}\approx0$ alone does not imply $\mathbf{X}\approx\nabla\psi$.

For exact compatible initial data $\mathbf{X}(0)=\nabla\psi(0)$ and zero initial cleaning fields, the identically zero cleaning fields and $\mathbf{C}=0$ solve the continuum system. The uniqueness established in the next Section therefore implies that the full constraint is preserved exactly for such data.

\subsection{Well-Posedness Analysis}
\label{S:LWP_example}
Above we discussed the role of the cleaning fields in taming errors that can arise from numerical constraint violations. This assumes that one can solve the system of Equations~\eqref{CC_2a}, \eqref{CC_2c}, and \eqref{CC_4}, and that its solutions are unique. In this Section we establish existence and uniqueness using a decoupling of the cleaning fields, which is the method employed to prove Theorem \ref{Theorem_1}. The resulting existence statement loses one Sobolev derivative in the variables $Y,\mathbf{X},\psi$, as stated explicitly below. The present argument does not give a same-order estimate for the complete first-order system (see, however, Section~\ref{S:LWP_no_derivative_lost_example}). 

Differentiating Equations~\eqref{CC_4b} and \eqref{CC_4c} with respect to $t$ and using Equation~\eqref{CC_4a} produces
\begin{subequations}
\label{CC_5}
\begin{align}
\partial_t^2\bm{\varphi}-c_\varphi^2\nabla^2\bm{\varphi}+(c_\varphi^2-c_\phi^2)\nabla(\nabla\cdot\bm{\varphi})&=-\frac{\partial_t\bm{\varphi}}{\tau_1}+\frac{\nabla\phi}{\tau_2}
\label{CC_5a}
\\
\partial_t^2\phi-c_\phi^2\nabla^2\phi&=-\frac{\partial_t\phi}{\tau_2}+\frac{c_\phi^2\nabla\cdot \bm{\varphi}}{\tau_1}.
\label{CC_5b}
\end{align}
\end{subequations}

Originally, $\bm{\varphi}$ and $\phi$ satisfy first-order equations so that we prescribe their values at time zero. For the second-order Equations~\eqref{CC_5}, we need to also prescribe their time derivatives. These are obtained from Equations~\eqref{CC_4b} and \eqref{CC_4c}. 

For $c_\varphi=c_\phi$, set $c:=c_\varphi=c_\phi$ and
\begin{align*}
V
&=
(\varphi^1,\varphi^2,\varphi^3,\phi),
\\
V_{(0)}
&=
\left.V\right|_{t=0},
\\
V_{(1)}
&=
\left.\partial_tV\right|_{t=0}.
\end{align*}
For every integer $s\geq0$, define
\begin{align*}
\mathcal{E}_s(t)
&:=
\frac{1}{2}
\left(
\left\|
\partial_t\bm{\varphi}(t)
\right\|_{H^s}^2
+
\left\|
\partial_t\phi(t)
\right\|_{H^s}^2
\right.
\\
&\qquad\left.
+
c^2
\left\|
\nabla\bm{\varphi}(t)
\right\|_{H^s}^2
+
c^2
\left\|
\nabla\phi(t)
\right\|_{H^s}^2
+
\left\|
\bm{\varphi}(t)
\right\|_{H^s}^2
+
\left\|
\phi(t)
\right\|_{H^s}^2
\right).
\end{align*}
Taking the $H^s$ inner products of Equations~\eqref{CC_5a} and \eqref{CC_5b} with $\partial_t\bm{\varphi}$ and $\partial_t\phi$, respectively, and integrating the Laplacian terms by parts give
\begin{align*}
\frac{\mathrm{d}}{\mathrm{d}t}\mathcal{E}_s(t)
&=
-\frac{1}{\tau_1}
\left\|
\partial_t\bm{\varphi}
\right\|_{H^s}^2
-\frac{1}{\tau_2}
\left\|
\partial_t\phi
\right\|_{H^s}^2
\\
&\quad
+\frac{1}{\tau_2}
\left\langle
\partial_t\bm{\varphi},
\nabla\phi
\right\rangle_{H^s}
+\frac{c^2}{\tau_1}
\left\langle
\partial_t\phi,
\nabla\cdot\bm{\varphi}
\right\rangle_{H^s}
\\
&\quad
+
\left\langle
\bm{\varphi},
\partial_t\bm{\varphi}
\right\rangle_{H^s}
+
\left\langle
\phi,
\partial_t\phi
\right\rangle_{H^s}.
\end{align*}
The Cauchy--Schwarz and Young inequalities therefore imply
\begin{align*}
\frac{\mathrm{d}}{\mathrm{d}t}\mathcal{E}_s(t)
&\leq
C\mathcal{E}_s(t),
\end{align*}
where $C$ depends only on $c$, $\tau_1$, and $\tau_2$. Since
\begin{align*}
\mathcal{E}_s(t)
&\approx
\left\|
\partial_tV(t)
\right\|_{H^s}^2
+
\left\|
V(t)
\right\|_{H^{s+1}}^2
\end{align*}
on $\mathbb{T}^3$, where $\approx$ means equivalence\footnote{I.e., $A\approx B$ means that there exist constants $C_1,C_2 > 0$, independent of $A$ and $B$ but possibly depending on other fixed parameters in the problem, such that $A\leq C B$ and $B\leq C A$}, Gronwall's inequality gives
\begin{align}
\label{E:CC_Equal_speed_energy_estimate}
\sup_{0\leq t\leq T}
\left(
\left\|
V(t)
\right\|_{H^{s+1}}^2
+
\left\|
\partial_tV(t)
\right\|_{H^s}^2
\right)
&\leq
C_T
\left(
\left\|
V_{(0)}
\right\|_{H^{s+1}}^2
+
\left\|
V_{(1)}
\right\|_{H^s}^2
\right).
\end{align}
For each $N\in\mathbb{N}$, project $V_{(0)}$ and $V_{(1)}$ onto the spatial Fourier modes $k\in\mathbb{Z}^3$ with $|k|\leq N$ and solve the resulting finite-dimensional system. Applying Equation~\eqref{E:CC_Equal_speed_energy_estimate} to the difference of two such solutions shows that the sequence converges in
\begin{align*}
C\big([0,T];H^{s+1}(\mathbb{T}^3)\big)
\cap
C^1\big([0,T];H^s(\mathbb{T}^3)\big).
\end{align*}

For $c_\varphi\neq c_\phi$, Equation~\eqref{CC_5a} is not strictly hyperbolic because the characteristics $\xi_0=\pm c_\varphi\sqrt{\xi_i\xi_i}$ have multiplicity 2. This repeated root does not prevent the matrix $h_2^{-1}A$ introduced below from being diagonalizable. In order to obtain local well-posedness in Sobolev spaces for the case $c_\varphi\neq c_\phi$, we will show that Equation~\eqref{CC_5a} can be reduced to a system of strictly hyperbolic PDEs of fourth order. In doing so we will employ Theorem
\ref{SH:Theorem1}, which will also be used to establish local well-posedness of the mixed-order formulation of Section~\ref{Mixed_order_section}. 

To construct $V$ in this remaining case, we assume $c_\varphi\neq c_\phi$ until the application of Theorem \ref{Theorem_SH_LWP} to the full system in Equations~\eqref{CC_5} has been completed.

Let us rewrite Equation~\eqref{CC_5a} as
\begin{align*}
h(\partial^2)^i_j \varphi^j&=B^i(\partial\bm{\varphi}, \partial\phi)=-\frac{\partial_t \varphi^i}{\tau_1}+\frac{\partial_i \phi}{\tau_2},
\end{align*}
where $h(\partial^2)^i_j = (\partial_t^2 - c_\varphi^2 \partial_k \partial_k)\delta_{ij} + (c_\varphi^2 - c_\phi^2)\partial_i \partial_j$. The Leray indices of Equation~\eqref{CC_5a} are $m_i = 2$ for all variables and $n_i = 0$ for all equations (we recall the definition of Leray indices in Section~\ref{Sec:SH}, noting that we employ the standard definition used in the literature, as in e.g. \cite[Appendix~IV]{Choquet-Bruhat-Book-2009}
or \cite[Appendix~B]{Disconzi-2024}). The characteristic determinant of the system in Equation~\eqref{CC_5a} is then
\begin{align*}
\mathcal{P}(\xi_0)&=\det[h(\xi)]=(\xi_0^2-c_\varphi^2\xi_k\xi_k)^2(\xi_0^2-c_\phi^2\xi_k\xi_k),
\end{align*}
which is homogeneous in $\xi_\mu$ and, for every $\bm{\xi}\neq0$, has the real characteristic roots $\xi_0 = \pm c_\varphi \sqrt{\xi_k \xi_k}$ with multiplicity 2 and $\xi_0 = \pm c_\phi \sqrt{\xi_k \xi_k}$ with multiplicity 1, in accordance with Assumption \ref{SH:Assumption 1}. 

Following the assumptions and notation of Theorem \ref{SH:Theorem1} introduced in Section~\ref{Sec:SH}, the highest-order time derivative appearing in the principal part of the system is $r = 2$, the degree of $\mathcal{P}(\xi_0)$ in $\xi_0$ is $Nr = 3(2) = 6$, $h_2 = I_3$, while the matrix $A$ has components 
\begin{align*}
A^i_j = -c_\varphi^2 \xi_k \xi_k \delta_{ij} + (c_\varphi^2 - c_\phi^2)\xi_i \xi_j.  
\end{align*}

The eigenvalues of $h_2^{-1}A$ are obtained from the equation
\begin{align*}
\det[h_2^{-1}A - \lambda I_3] = -(\lambda + c_\varphi^2 \xi_k \xi_k)^2 (\lambda + c_\phi^2 \xi_k \xi_k)=0   
\end{align*}
and are $\lambda_1 = -c_\varphi^2 \xi_k \xi_k$ with multiplicity 2 and eigenspace spanned by the basis $\left\{ w^j_a \right\}_{a=1}^2$, where $w^i_1$ and $w^i_2$ are any two linearly independent vectors orthogonal to $\xi_i$, i.e., $w^i_a \xi_i = 0$, while $\lambda_2 = -c_\phi^2 \xi_k \xi_k$ has multiplicity 1 with eigenspace spanned by the eigenvector $\xi_i$. 

The displayed characteristic polynomial has degree $6=Nr$, $h_2=I_3$, and all its roots are real, so Assumption \ref{SH:Assumption 1} holds. Since $c_\varphi,c_\phi>0$ and $c_\varphi\neq c_\phi$, the characteristic polynomial has exactly four distinct roots, while $h_2^{-1}A=A$ has $m=2$ distinct eigenvalue functions. Therefore, $4=m\,r$, which verifies Assumption \ref{SH:Assumption 2}. Moreover, in Equation~\eqref{E:xi_0_polynomial_h_r_inv_A}, $B_0=A$ and $B_1=0$. The matrix $A$ is real and symmetric, so $B_0$ and $B_1$ are simultaneously diagonalizable, as required by Assumption \ref{SH:Assumption_polynomials}. Finally,
\begin{align*}
m\,r=4\geq \max_I m_I-\min_J n_J+1=3,
\end{align*}
which verifies Assumption \ref{SH:Assumption3}. Theorem \ref{SH:Theorem1} thus shows that Equation~\eqref{CC_5a} implies the decoupled strictly hyperbolic equations of order 4 given by
\begin{align}
\label{Strict_hyperboli_4th_order_example}
\mathcal{G}^{\mu\nu\alpha\beta} \partial_\mu \partial_\nu \partial_\alpha \partial_\beta \bm{\varphi} = \mathbf{B}(\partial^3 \bm{\varphi}, \partial^3 \phi),    
\end{align}
where 
\begin{align*}
\mathcal{G}^{\mu\nu\alpha\beta} \partial_\mu \partial_\nu \partial_\alpha \partial_\beta
&=(\delta^\mu_0 \delta^\nu_0-c_\varphi^2\delta^\mu_k\delta^\nu_k)(\delta^\alpha_0 \delta^\beta_0-c_\phi^2\delta^\alpha_\ell\delta^\beta_\ell)\partial_\mu\partial_\nu\partial_\alpha\partial_\beta
\end{align*}
is a strictly hyperbolic operator when $c_\varphi \neq c_\phi$, both assumed strictly positive. 

For completeness, Theorem \ref{Theorem_SH_LWP} can be applied directly to the full system in Equations~\eqref{CC_5}. Set
\begin{align*}
V=(\varphi^1,\varphi^2,\varphi^3,\phi)
\end{align*}
and take $m_I=2$ and $n_I=0$ for $I=1,\dots,4$. The principal symbol is the block-diagonal matrix with blocks $h(\xi)$ and $\xi_0^2-c_\phi^2\xi_k\xi_k$, and hence its determinant is
\begin{align*}
(\xi_0^2-c_\varphi^2\xi_k\xi_k)^2
(\xi_0^2-c_\phi^2\xi_k\xi_k)^2.
\end{align*}
The coefficient of $\xi_0^2$ is $I_4$. In Equation~\eqref{E:xi_0_polynomial_h_r_inv_A}, the two matrices are
\begin{align*}
B_0&=\operatorname{diag}\big(A,-c_\phi^2\xi_k\xi_k\big),
\\
B_1&=0.
\end{align*}

Thus, $B_0$ and $B_1$ are simultaneously diagonalizable, the two distinct eigenvalue functions are $-c_\varphi^2\xi_k\xi_k$ and $-c_\phi^2\xi_k\xi_k$, and the characteristic polynomial has the four distinct real roots $\pm c_\varphi\sqrt{\xi_k\xi_k}$ and $\pm c_\phi\sqrt{\xi_k\xi_k}$. Therefore, Assumptions \ref{SH:Assumption 1}, \ref{SH:Assumption 2}, and \ref{SH:Assumption_polynomials} hold with $r=2$ and $m=2$, and Assumption \ref{SH:Assumption3} follows again from $m\,r=4\geq3$. The coefficients of Equations~\eqref{CC_5} are constant, and their right-hand sides are linear functions of $V$ and its first derivatives. They are therefore real analytic on the entire jet space $\mathcal{J}$ appearing in Theorem \ref{Theorem_SH_LWP}, so we may take $\mathcal{U}=\mathcal{J}$. On $\Sigma=\mathbb{T}^3$, the initial jet is continuous by Sobolev embedding and its range is compact; there are no compatibility conditions because $n_I=0$ for every $I$; and, for $N\in\mathbb{N}$, truncating the Fourier series of each initial datum to the spatial modes $k\in\mathbb{Z}^3$ with $|k|\leq N$ gives real-analytic approximations that converge in the Sobolev spaces in Equation~\eqref{E:SH_Initial_data_regularity}.

Since $p=m\,r=4$ and $m_*=2$, Theorem \ref{Theorem_SH_LWP} gives a unique solution of Equations~\eqref{CC_5} for
\begin{align*}
V_{(0)}
&\in
H^{\ell+4}(\mathbb{T}^3),
\\
V_{(1)}
&\in
H^{\ell+3}(\mathbb{T}^3),
\qquad
\ell\geq3,
\end{align*}
such that
\begin{align*}
\partial_t^jV
&\in
L^\infty\big([0,T];H^{\ell+4-j}(\mathbb{T}^3)\big),
\qquad
0\leq j\leq4.
\end{align*}
Moreover,
\begin{align*}
\partial_t^jV
&\in
C\big([0,T];H^s(\mathbb{T}^3)\big)
\end{align*}
for every $s<\ell+4-j$ and $0\leq j\leq4$.

For the application to the original first-order system, suppose that
\begin{align*}
Y_{(0)},
\quad
\mathbf{X}_{(0)},
\quad
\psi_{(0)},
\quad
\bm{\varphi}_{(0)},
\quad
\phi_{(0)}
&\in
H^{\ell+4}(\mathbb{T}^3).
\end{align*}
The initial value $V_{(1)}$ is determined by Equations~\eqref{CC_4b} and \eqref{CC_4c}. Since these equations contain at most one spatial derivative of $\mathbf{X}_{(0)}$, $\bm{\varphi}_{(0)}$, and $\phi_{(0)}$, they give
\begin{align*}
V_{(1)}
&\in
H^{\ell+3}(\mathbb{T}^3),
\end{align*}
as required above.

Having solved for $\bm{\varphi}$ and $\phi$, we can now use $\bm{\varphi}$ in
Equations~\eqref{CC_2a}, \eqref{CC_2c}, and \eqref{CC_4a}, treating these equations as a system for $Y,\mathbf{X}, \psi$  with $\bm{\varphi}$ as a source term. This can be seen directly. Let $U=(Y,\mathbf{X},\psi)^{\mathrm{T}}$, and let $\mathbf{e}_j$ be the $j$th standard basis vector of $\mathbb{R}^3$. Equations~\eqref{CC_2a}, \eqref{CC_2c}, and \eqref{CC_4a} are
\begin{align*}
\partial_tU+A^j\partial_jU
&=
\begin{pmatrix}
-\psi\\
-\nabla\times\bm{\varphi}\\
Y
\end{pmatrix},
\qquad
A^j=
\begin{pmatrix}
0&-\mathbf{e}_j^{\mathrm{T}}&0\\
-\mathbf{e}_j&0_{3\times3}&0\\
0&0&0
\end{pmatrix}.
\end{align*}
The coefficient of $\partial_tU$ is $I_5$, and each matrix $A^j$ is constant and symmetric. Hence this is a constant-coefficient symmetric-hyperbolic system. The scalar field $\phi$ does not occur in these three equations, while $-\nabla\times\bm{\varphi}$ is a prescribed source. For every integer $s\geq0$, the $H^s$ energy identity, in which the terms $-\psi$ and $Y$ cancel, gives
\begin{align*}
\|U(t)\|_{H^s}
&\leq
\|U(0)\|_{H^s}
+\int_0^t\|\nabla\times\bm{\varphi}(r)\|_{H^s}\,dr.
\end{align*}

For $N\in\mathbb{N}$, project the initial data and the prescribed source onto the spatial Fourier modes $k\in\mathbb{Z}^3$ with $|k|\leq N$. Solving the resulting finite-dimensional systems and using the preceding estimate to pass to the limit as $N\rightarrow\infty$ give existence and uniqueness in $C([0,T];H^s(\mathbb{T}^3))$ whenever $U(0)\in H^s(\mathbb{T}^3)$ and
\begin{align*}
\nabla\times\bm{\varphi}
&\in
L^1\big([0,T];H^s(\mathbb{T}^3)\big).
\end{align*}
The solution of Equations~\eqref{CC_5} obtained above satisfies
\begin{align*}
\bm{\varphi}
&\in
L^\infty\big([0,T];H^{\ell+4}(\mathbb{T}^3)\big),
\end{align*}
and therefore
\begin{align*}
\nabla\times\bm{\varphi}
&\in
L^1\big([0,T];H^{\ell+3}(\mathbb{T}^3)\big).
\end{align*}
Taking $s=\ell+3$ gives
\begin{align*}
U
&\in
C\big([0,T];H^{\ell+3}(\mathbb{T}^3)\big).
\end{align*}
Thus, although $U(0)\in H^{\ell+4}(\mathbb{T}^3)$, the present proof constructs $U$ only in $H^{\ell+3}(\mathbb{T}^3)$. This is the one-derivative loss in the existence argument. An $H^{\ell+4}$ estimate for $U$ would require $\nabla\times\bm{\varphi}\in L^1([0,T];H^{\ell+4}(\mathbb{T}^3))$, which is not supplied by the regularity obtained above.

It remains to show that Equations~ 
\eqref{CC_4b} and \eqref{CC_4c} are satisfied, so that we end with a solution to Equations~
\eqref{CC_2a}, \eqref{CC_2c}, and \eqref{CC_4},
as desired. To do so, note that at this point we have fields $Y,\mathbf{X},\psi, \bm{\varphi},\phi$ satisfying Equations~\eqref{CC_2a}, \eqref{CC_2c}, \eqref{CC_4a}, and \eqref{CC_5}, and we only need to show that they also satisfy Equations~\eqref{CC_4b} and \eqref{CC_4c}.

Define
\begin{align*}
\mathcal{S}_{\bm{\varphi}} &= \partial_t \bm{\varphi} - c_\varphi^2 \nabla \times \mathbf{X} + \nabla \phi + \frac{\bm{\varphi}}{\tau_1}
\\
\mathcal{S}_\phi & = \partial_t \phi + c_\phi^2 \nabla \cdot \bm{\varphi} + \frac{\phi}{\tau_2}.
\end{align*}

More explicitly, let $\mathbf{X}_{(0)}$, $\bm{\varphi}_{(0)}$, and $\phi_{(0)}$ denote the initial values of $\mathbf{X}$, $\bm{\varphi}$, and $\phi$. As stated after Equations~\eqref{CC_5}, the initial time derivatives for that second-order system are defined by Equations~\eqref{CC_4b} and \eqref{CC_4c}, namely,
\begin{align*}
\left.\partial_t\bm{\varphi}\right|_{t=0}
&=c_\varphi^2\nabla\times\mathbf{X}_{(0)}
-\nabla\phi_{(0)}-\frac{\bm{\varphi}_{(0)}}{\tau_1},
\\
\left.\partial_t\phi\right|_{t=0}
&=-c_\phi^2\nabla\cdot\bm{\varphi}_{(0)}
-\frac{\phi_{(0)}}{\tau_2}.
\end{align*}
Substitution into the definitions of $\mathcal{S}_{\bm{\varphi}}$ and $\mathcal{S}_\phi$ gives
\begin{align*}
\left.\mathcal{S}_{\bm{\varphi}}\right|_{t=0}
&=0,
\\
\left.\mathcal{S}_\phi\right|_{t=0}
&=0.
\end{align*}

Equation~\eqref{CC_4a} and the identity $\nabla\times(\nabla\times\bm{\varphi})=\nabla(\nabla\cdot\bm{\varphi})-\nabla^2\bm{\varphi}$ give
\begin{align*}
\nabla\times\partial_t\mathbf{X}
&=\nabla\times\big(\nabla Y-\nabla\times\bm{\varphi}\big)
\\
&=\nabla^2\bm{\varphi}-\nabla(\nabla\cdot\bm{\varphi}).
\end{align*}
Consequently, Equation~\eqref{CC_5a} implies
\begin{align*}
\partial_t\mathcal{S}_{\bm{\varphi}}
&=\partial_t^2\bm{\varphi}
-c_\varphi^2\nabla\times\partial_t\mathbf{X}
+\nabla\partial_t\phi
+\frac{\partial_t\bm{\varphi}}{\tau_1}
\\
&=\partial_t^2\bm{\varphi}
-c_\varphi^2\nabla^2\bm{\varphi}
+(c_\varphi^2-c_\phi^2)\nabla(\nabla\cdot\bm{\varphi})
+c_\phi^2\nabla(\nabla\cdot\bm{\varphi})
+\nabla\partial_t\phi
+\frac{\partial_t\bm{\varphi}}{\tau_1}
\\
&=\nabla\left(\partial_t\phi
+c_\phi^2\nabla\cdot\bm{\varphi}
+\frac{\phi}{\tau_2}\right)
\\
&=\nabla\mathcal{S}_\phi.
\end{align*}
Moreover, $\nabla\cdot(\nabla\times\mathbf{X})=0$ and Equation~\eqref{CC_5b} give
\begin{align*}
\partial_t\mathcal{S}_\phi-c_\phi^2\nabla\cdot\mathcal{S}_{\bm{\varphi}}
&=\partial_t^2\phi+c_\phi^2\nabla\cdot\partial_t\bm{\varphi}
+\frac{\partial_t\phi}{\tau_2}
-c_\phi^2\left(\nabla\cdot\partial_t\bm{\varphi}
+\nabla^2\phi+\frac{\nabla\cdot\bm{\varphi}}{\tau_1}\right)
\\
&=\partial_t^2\phi-c_\phi^2\nabla^2\phi
+\frac{\partial_t\phi}{\tau_2}
-\frac{c_\phi^2\nabla\cdot\bm{\varphi}}{\tau_1}
\\
&=0.
\end{align*}
Thus,
\begin{subequations}
\label{Cleaning_fields_satisfy_original_eq_propagation}
\begin{align}
\partial_t \mathcal{S}_{\bm{\varphi}} - \nabla \mathcal{S}_\phi &= 0
\label{Cleaning_fields_satisfy_original_eq_propagation_a}
\\
\partial_t \mathcal{S}_\phi - c_\phi^2\nabla \cdot \mathcal{S}_{\bm{\varphi}} & = 0.
\label{Cleaning_fields_satisfy_original_eq_propagation_b}
\end{align}
\end{subequations}
On $\mathbb{T}^3$, define
\begin{align*}
E_{\mathcal{S}}(t)
&=\frac{1}{2}\int_{\mathbb{T}^3}
\left[c_\phi^2|\mathcal{S}_{\bm{\varphi}}|^2
+|\mathcal{S}_\phi|^2\right]d^3x.
\end{align*}Equations~\eqref{Cleaning_fields_satisfy_original_eq_propagation_a} and \eqref{Cleaning_fields_satisfy_original_eq_propagation_b}, followed by integration by parts on $\mathbb{T}^3$, give
\begin{align*}
\partial_tE_{\mathcal{S}}(t)
&=\int_{\mathbb{T}^3}
\left(c_\phi^2\mathcal{S}_{\bm{\varphi}}\cdot\partial_t\mathcal{S}_{\bm{\varphi}}
+\mathcal{S}_\phi\partial_t\mathcal{S}_\phi\right)d^3x
\\
&=c_\phi^2\int_{\mathbb{T}^3}
\left(\mathcal{S}_{\bm{\varphi}}\cdot\nabla\mathcal{S}_\phi
+\mathcal{S}_\phi\nabla\cdot\mathcal{S}_{\bm{\varphi}}\right)d^3x
\\
&=0.
\end{align*}
The initial identities above imply $E_{\mathcal{S}}(0)=0$. Therefore, $E_{\mathcal{S}}(t)=0$ for every $t\in[0,T]$. Since $c_\phi>0$, this gives $\mathcal{S}_{\bm{\varphi}}=0$ and $\mathcal{S}_\phi=0$, which are precisely Equations~\eqref{CC_4b} and \eqref{CC_4c}.

Let us finish this Section with some remarks:
\begin{itemize}
    \item The above method of decoupling the cleaning fields is used for proving local well-posedness of Equations~\eqref{CC_2a}, \eqref{CC_2c}, and \eqref{CC_4}. This decoupling, i.e., Equations~\eqref{CC_5}, however, is not used in the numerical simulations. We would simulate directly Equations~\eqref{CC_2a}, \eqref{CC_2c}, and \eqref{CC_4}.
    \item In simulations, we initialize $\bm{\varphi}$ and $\phi$ as zero. For exact solutions with compatible initial data, $\nabla\times\mathbf{X}=0$ and the zero cleaning fields are preserved. However, the whole point is that the constraint $\nabla\times\mathbf{X}=0$ will not in general be exactly satisfied in simulations, so that $\bm{\varphi}$ and $\phi$ will in general not remain zero either. Rather, they will take values that control violations of the constraint.
    \item Although we initialize $\bm{\varphi}$, $\phi$, and $\nabla\times\mathbf{X}$ as zero in simulations, the proof allows arbitrary initial values satisfying the Sobolev assumptions stated above. For initial fields in $H^{\ell+4}(\mathbb{T}^3)$, the argument gives the cleaning fields in $L^\infty([0,T];H^{\ell+4}(\mathbb{T}^3))$ and gives $Y,\mathbf{X},\psi$ in $C([0,T];H^{\ell+3}(\mathbb{T}^3))$. The restriction to compatible initial data and zero cleaning fields is imposed only when recovering solutions of the original second-order equation.
\end{itemize}

\subsection{Direct well-posedness analysis}
\label{S:LWP_no_derivative_lost_example}

The proof in the preceding Section first solves the second-order system in Equations~\eqref{CC_5} for the cleaning fields and then reconstructs $Y,\mathbf{X},\psi$. The conclusion is a local well-posedness with a derivative loss, as stated. We followed this route in Section~\ref{S:LWP_example} because it mimics the argument used to prove Theorem \ref{Theorem_1}, which also incurs a derivative loss, see a precise statement in Theorem \ref{Theorem_1_precise}. We have not tried to optimize regularity statements in this work and thus leave the conclusion of Theorem  \ref{Theorem_1} as it is. The derivative loss, however, seems to be tied to our method rather than to fundamental structural properties of Equations~\eqref{Complete_set_Eq}. We illustrate this point here by providing a local well-posedness result without derivative loss for the toy-model  system in Equations~\eqref{CC_2a}, \eqref{CC_2c}, and \eqref{CC_4},

\begin{proposition}
Assume that $c_\varphi,c_\phi,\tau_1,\tau_2$ are positive constants, and let $s\geq1$ be an integer. If
\begin{align*}
Y_{(0)},\mathbf{X}_{(0)},\psi_{(0)},\bm{\varphi}_{(0)},\phi_{(0)}
&\in
H^s(\mathbb{T}^3),
\end{align*}
where the Sobolev space for a vector field is understood componentwise, then, for every $T>0$, the system in Equations~\eqref{CC_2a}, \eqref{CC_2c}, and \eqref{CC_4} has a unique solution satisfying
\begin{align*}
Y,\mathbf{X},\psi,\bm{\varphi},\phi
&\in
C\left([0,T];H^s(\mathbb{T}^3)\right)
\cap
C^1\left([0,T];H^{s-1}(\mathbb{T}^3)\right).
\end{align*}
Thus, the result is global for this linear system and, in particular, gives local existence and uniqueness at the same Sobolev order as the initial fields.
\end{proposition}

\begin{proof}
For scalar or vector fields $f$ and $g$, write
\begin{align*}
\left\langle f,g\right\rangle_{H^s}
&:=
\sum_{|\alpha|\leq s}
\int_{\mathbb{T}^3}
\partial^\alpha f\cdot\partial^\alpha g\,d^3x,
\end{align*}
with the usual scalar product when $f$ and $g$ are scalar. Define
\begin{align}
\label{E:CC_full_first_order_energy}
\mathcal{E}_s(t)
&:=
\frac{1}{2}
\left(
\left\|Y(t)\right\|_{H^s}^2
+\left\|\mathbf{X}(t)\right\|_{H^s}^2
+\left\|\psi(t)\right\|_{H^s}^2
+\frac{1}{c_\varphi^2}
\left\|\bm{\varphi}(t)\right\|_{H^s}^2
+\frac{1}{c_\varphi^2c_\phi^2}
\left\|\phi(t)\right\|_{H^s}^2
\right).
\end{align}
Because $c_\varphi,c_\phi>0$, this energy is equivalent to the squared product $H^s$ norm of the five fields.

All coefficients are constant, so every spatial derivative $\partial^\alpha$ commutes with the equations. For a smooth solution, summing over $|\alpha|\leq s$ gives
\begin{align*}
\frac{d}{dt}\mathcal{E}_s
&=
\left\langle Y,\nabla\cdot\mathbf{X}-\psi\right\rangle_{H^s}
+\left\langle \mathbf{X},\nabla Y-\nabla\times\bm{\varphi}\right\rangle_{H^s}
+\left\langle \psi,Y\right\rangle_{H^s}
\\
&\quad
+\frac{1}{c_\varphi^2}
\left\langle
\bm{\varphi},
c_\varphi^2\nabla\times\mathbf{X}
-\nabla\phi
-\frac{\bm{\varphi}}{\tau_1}
\right\rangle_{H^s}
\\
&\quad
+\frac{1}{c_\varphi^2c_\phi^2}
\left\langle
\phi,
-c_\phi^2\nabla\cdot\bm{\varphi}
-\frac{\phi}{\tau_2}
\right\rangle_{H^s}.
\end{align*}
Integration by parts on $\mathbb{T}^3$ gives
\begin{align*}
\left\langle Y,\nabla\cdot\mathbf{X}\right\rangle_{H^s}
&=
-\left\langle \nabla Y,\mathbf{X}\right\rangle_{H^s},
\\
\left\langle \mathbf{X},\nabla\times\bm{\varphi}\right\rangle_{H^s}
&=
\left\langle \nabla\times\mathbf{X},\bm{\varphi}\right\rangle_{H^s},
\\
\left\langle \bm{\varphi},\nabla\phi\right\rangle_{H^s}
&=
-\left\langle \nabla\cdot\bm{\varphi},\phi\right\rangle_{H^s}.
\end{align*}
Consequently, the $Y$--$\mathbf{X}$, $Y$--$\psi$, $\mathbf{X}$--$\bm{\varphi}$, and $\bm{\varphi}$--$\phi$ terms cancel in pairs, and
\begin{align}
\label{E:CC_full_first_order_dissipation}
\frac{d}{dt}\mathcal{E}_s(t)
&=
-\frac{1}{c_\varphi^2\tau_1}
\left\|\bm{\varphi}(t)\right\|_{H^s}^2
-\frac{1}{c_\varphi^2c_\phi^2\tau_2}
\left\|\phi(t)\right\|_{H^s}^2
\leq0.
\end{align}
In particular, Equation~\eqref{E:CC_full_first_order_energy} controls every field at the same Sobolev order.

We now construct a solution. Let $P_N$ denote the Fourier projection onto the modes $k\in\mathbb{Z}^3$ satisfying $|k|\leq N$, and apply $P_N$ to every initial field. The corresponding finite Fourier system is a finite-dimensional linear ordinary differential equation, so it has a smooth solution for all times. Since $P_N$ commutes with the constant-coefficient spatial operators, the calculation above applies to this solution and gives Equation~\eqref{E:CC_full_first_order_dissipation}.

Let $N,M\in\mathbb{N}$, and let $\mathcal{E}^{N,M}_s$ denote the energy in Equation~\eqref{E:CC_full_first_order_energy} evaluated on the componentwise difference of the $N$th and $M$th solutions. That difference satisfies the same homogeneous system. Hence
\begin{align*}
\sup_{0\leq t\leq T}
\mathcal{E}^{N,M}_s(t)
&\leq
\mathcal{E}^{N,M}_s(0)
\longrightarrow0
\end{align*}
as $N,M\longrightarrow\infty$. The equivalence of Equation~\eqref{E:CC_full_first_order_energy} with the product $H^s$ norm shows that the approximate solutions converge in $C\left([0,T];H^s(\mathbb{T}^3)\right)$, componentwise. Each right-hand side in Equations~\eqref{CC_2a}, \eqref{CC_2c}, and \eqref{CC_4} contains at most one spatial derivative. The right-hand sides therefore converge in $C\left([0,T];H^{s-1}(\mathbb{T}^3)\right)$, and the integrated equations pass to the limit. The limit belongs to $C^1\left([0,T];H^{s-1}(\mathbb{T}^3)\right)$ and solves the full first-order system with the prescribed initial fields.

For uniqueness, let two solutions in the stated class have the same initial fields, and take their difference. Applying $P_N$ to this difference gives a finite Fourier solution of the homogeneous system with zero initial data. Equation~\eqref{E:CC_full_first_order_dissipation} implies that its energy remains zero, so its five components vanish. Letting $N\longrightarrow\infty$ shows that the original difference vanishes. This proves uniqueness.

Taking $s=\ell+4$ shows that initial fields in $H^{\ell+4}(\mathbb{T}^3)$ produce all five solution fields in $C\left([0,T];H^{\ell+4}(\mathbb{T}^3)\right)$; no component is reconstructed by taking a spatial derivative of another component. This is local well-posedness without continuous dependence relative to the top norm, as  used here. 
\end{proof}

\subsection{Bounds on Cleaning Field Speeds}
\label{S:Bounds_cleaning_speeds}

From the fact that the curl-cleaning and divergence-cleaning fields are not physical quantities, they are allowed to have superluminal propagation speeds as well as to obey non-covariant evolution equations. Therefore, it is usual to keep Equations~\eqref{CC_4} unchanged regardless of the underlying metric. The local well-posedness proof, as well as the energy estimates, remain the same, as desired. 

As a matter of fact, it is intuitive that the cleaning field propagation modes must exceed any propagation mode from the physical fields so that the cleaning fields can correct curl/divergence errors faster than the evolution of the physical fields can create new ones. Consequently, by choosing $c_h$, we expect that no physical mode is greater than the cleaning modes, i.e., $|\xi_0^c| = c_h \sqrt{\xi_i \xi_i}>|\xi_0^{\text{phys}}(\bm{\xi})|$ for any $\bm{\xi}$. 

Now, suppose that the physical modes $\xi_\mu$ obey $\xi_\mu \xi^\mu \geq 0$, so that the propagation modes $\xi_0^{\text{phys}} = \xi_0(\bm{\xi})$ are causal. Because $g^{00} < 0$ and, therefore, $\xi_\mu \xi^\mu = g^{00} \xi_0^2 + 2 g^{0i} \xi_0 \xi_i + g^{ij} \xi_i \xi_j$ is a polynomial in $\xi_0$ with downward concavity, this means that $\xi_0^{\text{phys}}$ must lie between the roots of $\xi_\mu \xi^\mu = 0$, i.e.,
\begin{align*}
\frac{g^{0i} \xi_i - \sqrt{(g^{0i} \xi_i)^2 + |g^{00}| g^{ij} \xi_i \xi_j}}{|g^{00}|} \leq \xi_0^{\text{phys}} \leq \frac{g^{0i} \xi_i + \sqrt{(g^{0i} \xi_i)^2 + |g^{00}| g^{ij} \xi_i \xi_j}}{|g^{00}|}.
\end{align*}

Because we require that $|\xi_0^c| > |\xi_0^{\text{phys}}|$, we obtain
\begin{align*}
c_h \sqrt{\xi_i \xi_i} \geq \frac{g^{0i} \xi_i + \sqrt{(g^{0i} \xi_i)^2 + |g^{00}| g^{ij} \xi_i \xi_j}}{|g^{00}|}
\end{align*}
for all $\bm{\xi}$. In the Minkowski metric, this is automatically satisfied for any $c_h \geq 1$. However, in a curved metric, this may not hold, and $c_h$ may need to be larger than 1 to some extent. 

Let us write the metric in terms of the lapse function $N$ and shift vector $N^i$, where $g_{00} = -N^2 + N^i N_i$, $g_{0i} = N_i$, $q_{ij} = g_{ij}$, $g^{00} = -1/N^2$, $g^{0i} = N^i/N^2$, $q^{ij} q_{jk} = \delta^i_k$, $N_i = q_{ij} N^j$, and $g^{ij} = q^{ij} - N^i N^j / N^2$. Then, the above inequality may be rewritten as
\begin{align*}
c_h \sqrt{\xi_i \xi_i} \geq N^i \xi_i + N \sqrt{q^{ij} \xi_i \xi_j}.
\end{align*}

Since this must hold for any $\bm{\xi}$, we may take $\xi_i$ as the eigenvector of $q^{ij}$ associated with its largest eigenvalue, which we denote as $1/\lambda_{\mathrm{min}}(g_{ij})$, where $\lambda_{\mathrm{min}}(g_{ij})$ is the smallest eigenvalue of $g_{ij} = q_{ij}$. Then, using the inequality $|N^i \xi_i| \leq \sqrt{q^{ij}N_i N_j} \sqrt{q^{ij} \xi_i \xi_j}=N\sqrt{g^{0i}g_{0i}} \sqrt{\xi_i \xi_i}/\sqrt{\lambda_{\mathrm{min}}(g_{ij})}$, we obtain
\begin{align*}
N^i \xi_i + N \sqrt{q^{ij} \xi_i \xi_j} \leq \frac{N\sqrt{g^{0i}g_{0i}} + N}{\sqrt{\lambda_{\mathrm{min}}(g_{ij})}} \sqrt{\xi_i \xi_i}.
\end{align*}

Hence, a sufficient bound on $c_h$ is
\begin{align*}
c_h \geq \frac{N}{\sqrt{\lambda_{\mathrm{min}}(g_{ij})
}}
\left (\sqrt{g^{0i}g_{0i}} + 1\right ),
\end{align*}
where we used that $N=\frac{1}{\sqrt{-g^{00}}}$.

Therefore, both non-diagonal metrics and cases where $N > \lambda_{\mathrm{min}}(g_{ij})$ must be taken into account. In particular, the condition $N > \lambda_{\mathrm{min}}(g_{ij})$ frequently arises in highly curved or dynamical spacetimes. Examples include:
\begin{itemize}
    \item Kerr (rotating) black holes, where frame-dragging modifies the metric structure~\cite{kerr1963gravitational},
    \item Cosmological metrics (e.g., FLRW) in which spatial expansion suppresses $g_{ij}$ while $N = 1$ for comoving time~\cite{weinberg1972gravitation}, and
    \item Anisotropic spacetimes like Taub-NUT~\cite{taub1951empty} or Gödel~\cite{godel1949example}, where exact solutions exhibit nontrivial relationships between $g_{00}$ and $g_{ij}$, with $|g_{00}|$ exceeding the smallest eigenvalue of $g_{ij}$.
\end{itemize}
The results in \cite{alic2012z4} corroborate the above bound on $c_h$ for extreme curvature (near singularities or anisotropic regions).

The practical considerations involved in choosing the cleaning propagation speed $c_h$ can be illustrated by works involving the $\nabla \cdot \mathbf{B} = 0$ constraint (the first example we discussed at the beginning of Section~\ref{Sec:CC}).
The timestep must satisfy the CFL condition, $\Delta t\leq\Delta x/c_h$ \cite{courant1928partial,courant1967partial}, requiring smaller steps for larger $c_h$.
The prescription~$\tau^{-1}\sim c_h/\Delta x$, relating the damping and cleaning propagation has been proposed in~\cite{DEDNER2002645}.

 \newpage
\draftpart{The basic formalism}
\label{Part:Basic_formalism}

\section{The BDNK theory}
\label{BDNK_theory_section}

We consider in $3+1$ dimensions the BDNK energy-momentum tensor~\cite{Bemfica:2020zjp}
\be
\label{BDNKTmunu}
T_{\mu\nu} = \left(\varepsilon +\cA \right) u_\mu u_\nu + (P+\Pi)\Delta_{\mu\nu}-2\eta \sigma_{\mu\nu}+u_\mu \mathcal{Q}_\nu + u_\nu \mathcal{Q}_\mu
\ee
with the ideal (baryon) current 
\begin{align}\label{idealcurrent}
J^\mu=nu^\mu,
\end{align} 
where $g_{\mu\nu}$ is the spacetime metric, $\Delta_{\mu\nu} = g_{\mu\nu}+u_\mu u_\nu$, $P=P(\varepsilon,n)$ is the pressure, $\varepsilon$ is the energy density, $n$ is the number density, and $\sigma^{\mu\nu}=\Delta^{\mu\nu\alpha\beta}\nabla_\alpha u_\beta$ is the shear tensor. The tensor $\Delta^{\mu\nu\alpha\beta} = \left(\Delta^{\alpha\mu} \Delta^{\beta\nu} + \Delta^{\alpha\nu} \Delta^{\beta\mu}\right)/2 - \Delta^{\mu\nu}\Delta^{\alpha\beta}/3$ projects a two-tensor onto its symmetric and trace-free part. The  out-of-equilibrium contributions $\mathcal{A}$, $\Pi$, and $\mathcal{Q}_\mu$, which vanish in the absence of dissipation, are chosen so that the second law of thermodynamics is not violated up to the order of validity of the theory. More specifically, they  are described by the correction to the energy
\be
\cA = \tau_\varepsilon\left[ u^\mu \nabla_\mu \varepsilon + (\varepsilon+P)\nabla_\mu u^\mu\right] = -\tau_\varepsilon \left(u_\alpha \nabla_\beta T^{\alpha\beta}_0\right),
\label{defA}
\ee
the correction to the pressure
\be
\Pi =  \frac{\tau_P}{\tau_\varepsilon} \mathcal{A} - \zeta \nabla_\mu u^\mu = - \zeta \nabla_\mu u^\mu - \tau_P \left(u_\alpha \nabla_\beta T^{\alpha\beta}_0\right),
\label{definePi}
\ee
and the heat flux
\begin{align}
\label{defQ}
\begin{split}
\mathcal{Q}_\nu &= \sigma T\frac{\varepsilon+P}{n}\Delta^\alpha_\nu\nabla_\alpha\left (\frac{\mu}{T}\right )
+\tau_Q \left[(\varepsilon+P)u^\lambda \nabla_\lambda u_\nu + \Delta_{\nu}^\lambda \nabla_\lambda P\right]\\
&=\tau_Q (\varepsilon+P)u^\lambda \nabla_\lambda u_\nu + \beta_\varepsilon\Delta_{\nu}^\lambda \nabla_\lambda \varepsilon+ \beta_n\Delta_{\nu}^\lambda \nabla_\lambda n.
\end{split}
\varepsilon
\end{align}

In Equations~\eqref{defA},\eqref{definePi}, and \eqref{defQ}, $T^{\alpha\beta}_0=\varepsilon  u^\alpha u^\beta + P\Delta^{\alpha\beta}$ abbreviates the expression for the energy-momentum of an ideal fluid, %
 $\mu$ is the chemical potential, $T$ is the temperature, 
$\eta=\eta(\varepsilon,n)$ is the shear viscosity, $\zeta=\zeta(\varepsilon,n)$ is the bulk viscosity, and $\sigma=\sigma(\varepsilon,n)$ defines the thermal conductivity coefficient $\kappa=(\varepsilon+P)^2 \sigma/nT$. 
The quantities $\beta_{\varepsilon}$ and $\beta_{n}$ are defined by
\begin{align}
\label{def:beta_epsilon}
\beta_{\varepsilon}&=\tau_Q P_{\varepsilon}+\frac{\rho\sigma T}{n}\left (\frac{\partial(\mu/T)}{\partial\varepsilon}\right )_n\\
\label{def:beta_n}
\beta_{n}&=\tau_Q P_{n}+\frac{\rho\sigma T}{n}\left (\frac{\partial(\mu/T)}{\partial n}\right )_{\varepsilon},
\end{align}
 where $\rho=\varepsilon+P$, $P'_\varepsilon=\partial P/\partial\varepsilon$, $P'_n=\partial P/\partial n$.

 The transport coefficients $\{\tau_\varepsilon,\tau_P,\tau_Q,\sigma\}$ are assumed to be functions of $\varepsilon$ and $n$ that parametrize our definition of temperature and flow velocity out of equilibrium (i.e., they define the hydrodynamic frame used). Allowed values for these parameters, coming from causality and linear stability, can be found in \cite{Bemfica:2020zjp} (see also \cite{,Kovtun:2019hdm,Bemfica:2019knx,Bemfica:2020zjp}). A simple description of values of transport coefficients imposed by causality and stability can be given in the conformal regime, wherein $\varepsilon \sim T^4$, $P = \varepsilon/3$, all the BDNK parameters scale as $1/T$ and $\tau_P = \tau_\varepsilon/3$ (note that $\zeta=\sigma = 0$ and, to match with \citep{Bemfica:2017wps}, $\lambda=\rho\tau_Q$ and $\chi=\rho\tau_\varepsilon$). 

The equations of motion are given by the usual divergence-free and normalization conditions

\begin{subequations}{\label{BDNK_EoM}}
    \begin{align}
        \nabla_\mu T^{\mu\nu} &= 0, 
        \label{div_T_mumnu_DNK}
        \\
        \nabla_\mu J^\mu & = 0,
        \label{div_J_mu_BDNK}
        \\
        g_{\mu\nu} u^\mu u^\nu & =-1.    \label{normalization_u_BDNK}
    \end{align}
\end{subequations}

\medskip
We introduce the following:
\begin{definition}
The speed of sound is defined by the equation 
\begin{align}
\label{soundspeed}
c_s^2 = \left (\frac{dP}{d\varepsilon}\right )_{\bar{s}}=P'_\varepsilon+\frac{n}{\rho}P'_n,
\end{align}
where $\bar{s}=S/N$ is the specific entropy. 
Moreover, we follow the notation of \cite{Bemfica:2020zjp} and write
\begin{align}
\label{kappa_s_def}
\kappa_s=\frac{T\rho^2}{n}\left (\frac{\partial(\mu/T)}{\partial\varepsilon}\right )_{\bar{s}}
=\frac{T\rho^2}{n}\left (\frac{\partial(\mu/T)}{\partial\varepsilon}\right )_{n}+T\rho \left (\frac{\partial(\mu/T)}{\partial n}\right )_{\varepsilon}.
\end{align}
\end{definition}

We remind the reader that $c_s$ is not the true speed of propagation of sound waves in a viscous fluid in the nonlinear regime. Rather, it is the sound speed in an ideal fluid, but it receives corrections in the viscous setting. However, $c_s^2$ as defined in Equation~\eqref{soundspeed} figures prominently in many expressions involving causality and well-posedness of the BDNK equations, hence the advantage of introducing it here.

The reader is referred to \cite[Theorem I]{Bemfica:2020zjp} for precise causality conditions for the BDNK theory (which will be satisfied in our numerical simulations). Here, we will need conditions for well-posedness of the formulations we will introduce for numerical simulations. To state these conditions, it is convenient to introduce the following.
 
\begin{definition}
Let us define the characteristic velocities squared and the discriminant as
\begin{subequations}
\label{Characteristics}
\begin{align}
c_1^2&=\frac{\eta}{\rho\tau_Q},\\
c_\pm^2&=\frac{\tau_\varepsilon \left(\rho  c_s^2 \tau _Q+\zeta +\frac{4 \eta}{3}+\sigma\kappa _s\right)+\rho  \tau _P \tau _Q\pm\sqrt{\Delta}}{2\rho \tau_Q\tau_\varepsilon},\\
\Delta&=\left [\tau_\varepsilon \left(\rho  c_s^2 \tau _Q+\zeta +\frac{4 \eta}{3}+\sigma\kappa _s\right)+\rho  \tau _P \tau _Q\right ]^2+4\rho\tau_Q\tau_\varepsilon
\left [\beta_\varepsilon \left (\zeta +\frac{4 \eta}{3}\right )-\tau _P\left(\rho c_s^2 \tau_Q+\sigma  \kappa _s\right)\right ]. 
\end{align}
\end{subequations}
\end{definition}

\begin{assumption}
\label{Assumption_LWP}
We assume that $\tau_\varepsilon,\rho,\tau_Q,\tau_P>0$, $\Delta>0$, $c_1,c_+,c_-\in(0,1)$, and $c_a\ne c_b$ for $a\ne b$ ($a,b=1,+,-$). 
\end{assumption} 

\begin{assumption}
\label{Assumption_physics}
    We assume that $\eta,\zeta,\sigma \geq 0$ and that $n,T > 0$ (see, however, Remark \ref{Remark_assumptions_LWP_physics}).
\end{assumption}

\begin{remark}
\label{Remark_assumptions_LWP_physics}
Assumption \ref{Assumption_LWP} is key for our proof of well-posedness. The assumptions on $\eta$, $\zeta$, and $\sigma$ in Assumption \ref{Assumption_physics} are made on physical grounds since these coefficients are never negative, thus there is no need to consider this possibility in our analysis ($\eta \geq 0$ follows from Assumption \ref{Assumption_LWP} but it is convenient to state it explicitly). The positivity of $n$ and $T$ guarantees that the fractions with them in the denominators, e.g. Equation~\eqref{kappa_s_def}, are well defined. Strictly speaking, we could have $n$ or $T$ vanishing as long as those fractions remain well-defined, which would be the case if other quantities vanish at the same or faster rate, but it is simpler to take $n,T>0$; $n<0$ would also be allowed. That said, we remark that \emph{our proof goes through with very minor modifications in the case $n=0$ if we set the chemical potential to zero at the very beginning.}
\end{remark}

\begin{remark}
    We have written Assumption \ref{Assumption_LWP} in a form that is convenient for our purposes here, making it look different than the assumptions for well-posedness in  \cite[Theorem II]{Bemfica:2020zjp}. One can show, however, that the conditions are equivalent.
\end{remark}

\subsection{Auxiliary definitions}
\label{S:Auxiliary_definitions}
In applications such as a conformal fluid where $\varepsilon\sim T^4$, small temperatures correspond to much smaller energy density. This can lead to numerical complications when the temperature is very small. To avoid it, we introduce 
the new variable $\theta$ such that $\varepsilon=\varepsilon_0 e^{\theta}$, where $\varepsilon_0$ is a constant.

Then, we may rewrite the energy-momentum tensor in Equation~\eqref{BDNKTmunu} as follows:
\begin{align}
\label{defineBDNKnovo}
T^{\mu\nu} &= T_0^{\mu\nu} + \frac{1}{\varepsilon}H^{\mu\nu\alpha\beta}\nabla_\alpha (\varepsilon u_\beta)+2\beta_n u^{(\mu} \Omega^{j\nu)}\partial_j n\nonumber\\
&= T_0^{\mu\nu} + H^{\mu\nu\alpha\beta}\left (\nabla_\alpha  u_\beta+u_\beta\nabla_\alpha\theta\right )+2\beta_n u^{(\mu} \Omega^{j\nu)}\partial_j n,
\end{align}
where we defined the three co-vectors 
\begin{align}
\label{Omega}
\Omega^i_\alpha=\delta^i_\alpha-\frac{\delta^0_\alpha u^i}{u^0}=\Delta^i_\alpha-\frac{\Delta^0_\alpha u^i}{u^0}\quad \left( \Omega^{i\mu}=g^{i\mu}-\frac{g^{0\mu}u^i}{u^0}=\Delta^{i\mu}-\frac{\Delta^{0\mu}u^i}{u^0}\right )
\end{align} 
and used $u^{(\mu} \Omega^{j\nu)}=(u^{\mu} \Omega^{j\nu}+u^{\nu} \Omega^{j\mu})/2$. Moreover, we have explicitly used the conservation of $J^\mu$ to obtain 
\begin{align*}
\partial_t n=-\frac{u^j\partial_j n}{u^0}-\frac{n\Delta^{\alpha\beta}\nabla_\alpha (\varepsilon u_\beta)}{\varepsilon u^0}
\end{align*}
so that $\partial_t n$ was replaced in the expression for $T^{\mu\nu}$. The tensor $H^{\mu\nu\alpha\beta}$ is defined by
\begin{align}
H^{\mu\nu\alpha\beta} &= \tau_\varepsilon u^\mu u^\nu \left[(\varepsilon+P) \Delta^{\alpha\beta}-\varepsilon u^\alpha u^\beta\right]+\Delta^{\mu\nu}\left[(\tau_P(\varepsilon+P)-\zeta) \Delta^{\alpha\beta}-\varepsilon \tau_P u^\alpha u^\beta\right]- 2\eta \Delta^{\mu\nu\alpha\beta} \nonumber \\ 
&+ \tau_Q(\varepsilon+P)u^\mu u^\alpha \Delta^{\beta\nu}+\tau_Q(\varepsilon+P) u^\nu u^\alpha \Delta^{\mu\beta}-\varepsilon\beta_\varepsilon u^\mu u^\beta \Delta^{\alpha\nu}-\varepsilon\beta_\varepsilon u^\nu u^\beta \Delta^{\mu\alpha}\nonumber\\
&-\frac{n\beta_n}{u^0}\left (u^\mu \Delta^{\nu0}+u^\nu \Delta^{\mu0}\right )\Delta^{\alpha\beta}
\label{defineBDNKnew}
\end{align} 
Since $u^0$ is not a dynamical variable, it is obtained from the normalization $g_{\mu\nu}u^\mu u^\nu=-1$ for an arbitrary metric $g$ and reads
\begin{align}
\label{u0}
u^0=\frac{g_{0i}u^i+\sqrt{(g_{0i}g_{0j}-g_{00}g_{ij})u^i u^j-g_{00}}}{-g_{00}},
\end{align}
where we considered only the positive solution for $u^0$ (future-pointing vector). From the conservation equation $\nabla_\nu T^{\mu\nu}=0$, we obtain that 
\begin{align*}
H^{\mu 00 \beta}\partial_t^2(\varepsilon u_\beta)=r^\mu,
\end{align*} 
where $r^\mu$ contains all lower-order terms in time derivatives. Let us define the matrix
\begin{align}
\label{M}
M^{\mu\beta}\equiv H^{\mu 00 \beta}.
\end{align}
Existence of a solution requires that $\det[M]\ne 0$. We will show in Section~\ref{Sec:LWP} that this is always the case under assumption \ref{Assumption_LWP}. 

For the rest of the paper, we are going to rely on the use of some important quantities that we now define separately for easy access.  
\begin{definition}
Let $C_i^\mu$ be the three linearly independent (LI) 3-vectors orthogonal to $u_\mu$ defined by
\begin{align}
\label{C}
C_i^\mu=\delta^\mu_i-\frac{\delta^\mu_0 u_i}{u_0}=\Delta^\mu_i-\frac{\Delta^\mu_0 u_i}{u_0},\quad \left (C_{i\mu}=g_{\alpha\mu}C_i^\alpha=g_{\mu i}-\frac{g_{\mu 0} u_i}{u_0}\right ).
\end{align}
\end{definition}
For any vector $v^\alpha$ orthogonal to $u^\alpha$ ($u_\alpha v^\alpha=0$), one can write $v^\alpha=v^i C_i^\alpha$. In particular, $\nabla_\alpha u^\beta=C_i^\beta\nabla_\alpha u^i$ and $\Delta^{\mu\nu}=\Delta^{ij}C^\mu_i C^\nu_j$. Unless otherwise specified, repeated Greek indices sum from 0 to 3, while repeated $i,j,k$, and $\ell$ sum from 1 to 3.
\begin{definition}
Let $P_{A\mu}$ and $\tilde{P}^{A\mu}$ be two $4\times 4$ matrices defined by
\begin{align}
\label{Matrices_P}
P_{A\mu}=\begin{cases} u_\mu\text{ if } A=\theta\\ C_{i\mu} \text{ if } A=i\end{cases}\quad\text{and}\quad 
\tilde{P}^{A\mu}=\begin{cases} -u^\mu\text{ if } A=\theta\\ \Delta^{i\mu}\text{ if } A=i\end{cases}.
\end{align}
\end{definition}
In particular, their product obeys $\tilde{P}^{A\mu}P_{B\mu}=\delta^A_B$ and $\tilde{P}^{A\mu}P_{A\nu}=\delta^\mu_\nu$. Unless specified, repeated indices $A$ and $B$ sum over the values $\theta,1,2$, and 3.
\begin{definition}\label{def:U^A}
For brevity, let us write 
\[U^{A}=\begin{cases}\theta,\; A=\theta\\ u^i,\; A=i \end{cases}.\]
\end{definition}
The above definitions allow us to rewrite Equation~\eqref{defineBDNKnovo} as
\begin{align}
\label{defineBDNKnovo2}
T^{\mu\nu} = T_0^{\mu\nu} + H^{\mu\nu\alpha\beta}P_{B\beta}\nabla_\alpha U^B+2\beta_n u^{(\mu} \Omega^{j\nu)}\partial_j n,
\end{align}
which will be used to define the subsequent formulations of BDNK.

\section{First-order flux conservative formulation of BDNK}
\label{Sec:First_Order}

In this Section, we introduce a first-order formulation of the BDNK equations that is the basis for our first numerical algorithm. First-order reductions are very standard, thus we will not motivate it here. However, readers looking for a simple example that motivates the procedure can consult Section~\ref{Reduction_toy_model_section}.

The idea is to reduce the second-order equations coming from the conservation equations for $T^{\mu\nu}$ and $J^\mu$ to a set of first-order equations, written in a flux-conservative formulation. To do this, we define the new quantities $X_{i}^A$, $Y^A$, and $Z_i$, with $A=\theta,1,2,3$, such that
\bml
\label{New_variables}
\begin{align}
Y^A&\equiv\partial_t U^A,\label{Y}\\
X_{i}^A&\equiv \partial_i U^A, \label{X}\\
Z_i&\equiv \partial_i n.\label{Z}
\end{align}
\eml   
We may now expand Equation~\eqref{defineBDNKnovo2} as 
\begin{align}
\label{New_Tmunu}
T^{\mu\nu} &= T_0^{\mu\nu} + H^{\mu\nu 0\beta} P_{A\beta}Y^A+H^{\mu\nu j \beta} P_{A\beta}X_j^A
+2\beta_n u^{(\mu} \Omega^{j\nu)} Z_j 
+ H^{\mu\nu\alpha\beta} C_{j\beta} \Gamma^j_{\alpha \lambda}u^\lambda
\end{align}

The price to pay for this purely first-order formulation is that we now have a set of 24 variables $\{Y^A,X_{i}^A,Z_i,U^A,n\}$. The dynamical equations for $Y_A$ come from $\nabla_\nu T^{\mu\nu}=0$, while the equation for $n$ comes from $\nabla_\mu J^\mu=0$, and those for $U^A$ come from Equation~\eqref{Y}. We need to provide supplemental dynamical equations for the remaining 12 variables $X_{i}^A$ and the 3 variables $Z_i$. We use the fact that $\partial_\mu \partial_\nu U^A-\partial_\nu\partial_\mu U^A=0$ and $\partial_\mu \partial_\nu n-\partial_\nu\partial_\mu n=0$, which corresponds to 6 constraints for each of the variables $U^A$ and $n$ (see below for a discussion of the constraints), totaling 30 equations. We may split these identities to obtain
\bml
\label{supplemental_EQ}
\begin{align}
\partial_t X_{i}^A-\partial_i Y^A=0,\label{supplemental_EQ_a}\\
\epsilon_{ijk}\partial_j X_{k}^A=0,\label{supplemental_EQ_b}\\
\partial_t Z_i-\partial_i S_n=0,\label{supplemental_EQ_c}\\
\epsilon_{ijk}\partial_j Z_k=0,\label{supplemental_EQ_d}
\end{align}
\eml
where
\begin{align}
S_n&=-\frac{u^j Z_j}{u^0}-\frac{n X_{j}^j}{u^0}+\frac{n u_j Y^j}{u^0u_0}
-\frac{n C^\alpha_j\Gamma^j_{\alpha \lambda}u^\lambda}{u^0}\label{Sn},
\end{align}
which comes from $\partial_t n=S_n$, and $\epsilon_{ijk}$ is the totally antisymmetric Levi-Civita symbol with $\epsilon_{123}=+1$.
Equations~\eqref{supplemental_EQ_a} and \eqref{supplemental_EQ_c} serve as the missing 15 dynamical equations for $X_{i}^A$ and $Z_i$. On the other hand, the other 15 Equations~\eqref{supplemental_EQ_b} and \eqref{supplemental_EQ_d} are constraints. Mathematically, they need no extra treatment from the point of view of the evolution of the first-order system, although we need to show that these constraints are preserved in order for solutions to the first-order system to reproduce solutions to the original, second-order, BDNK Equations~\eqref{BDNK_EoM}. The propagation of the constraints is established in Section~\ref{FO:Equivalence}.

However, in simulations, numerical errors may jeopardize the correct evolution of these constraints, introducing small violations that can grow over time. Therefore, we must add extra fields, called \emph{cleaning fields,} that regulate these constraints and keep their violations small. Readers not familiar with the cleaning field procedure are referred to Section~\ref{Sec:CC} and references therein. 

For our computations, we need to show that the matrix 
\begin{align*}
    M^{\mu\beta}=H^{\mu00\beta}
\end{align*}
is invertible and compute its inverse. This is done in Section~\ref{Inverse_M_section}. From the invertible nature of the matrix $M^{\mu\beta}$, we may promote $T^{0\mu}$ as dynamical variables instead of $Y^A$. Using the inverse $M^{-1}_{\lambda \mu}$ together with $\tilde{P}^{A\mu}$ defined in Equation~\eqref{Matrices_P}, we can express $Y^A$ as a function of $T^{0\mu}$ as well as the other variables to obtain the following relations: 
\begin{align}
\label{Relations}
Y^A  &=\tilde{P}^{A\lambda}M^{-1}_{\lambda\mu}\left [T^{0\mu}- T_0^{0\mu} - H^{\mu0 k \beta}P_{B\beta}X_k^B-2\beta_n u^{(\mu} \Omega^{k0)} Z_k - H^{\mu 0 \alpha\beta} C_{k\beta} \Gamma^k_{\alpha \lambda}u^\lambda\right ],\\
\label{Relations-Tij-FO}
T^{ij} &= T_0^{ij} + H^{ij 0\beta} P_{A\beta} Y^A
+H^{ij k \beta}P_{A\beta}X_k^A+2\beta_n u^{(i} \Omega^{kj)} Z_k + H^{ij\alpha\beta} C_{k\beta} \Gamma^k_{\alpha \lambda}u^\lambda.
\end{align}

The first-order evolution equations for the set of variables $\mathfrak{B}=\{T^{0\mu},X_{i}^A,Z_i, U^A,n\}$ read
\bml
\label{Set_Eq}
\begin{align}
&\partial_t T^{0\mu}+\partial_i T^{i\mu}=-\Gamma^\nu_{\nu 0}T^{0\mu}-\Gamma^\mu_{0\nu}T^{0\nu}
-\Gamma^\nu_{\nu i}T^{i\mu}-\Gamma^\mu_{i\nu}T^{i\nu},\label{Set_Eq_a}\\
&\partial_t X_i^A-\partial_i Y^A=0,\label{Set_Eq_b}\\
&\partial_t Z_i-\partial_i S_n=0,\label{Set_Eq_d}\\
&\partial_t U^A =Y^A,\label{Set_Eq_c}\\
&\partial_t n=S_n.\label{Set_Eq_e}
\end{align}
\eml
As mentioned, we need to provide extra fields to guarantee that the constraints in Equations~\eqref{supplemental_EQ_b} and \eqref{supplemental_EQ_d} are correctly evolved in the numerical simulation. Below we just give the final form of the equations containing already the constraint cleaning fields. See Section~\ref{Sec:Cleaning_formulation} for the details. Also, see Section~\ref{FO:Equivalence} for the proof of equivalence between the second-order and first-order systems of PDEs. With the extra cleaning fields, the full set of variables is $\{T^{0\mu},X_i^A,U^A,n,Z_i,\varphi_i^A,\phi^A,\Psi_i,\psi\}$ with evolution equations given by
\bml
\label{Complete_set_Eq}
\begin{align}
&\partial_t T^{0\mu}+\partial_i T^{i\mu}=-\Gamma^\nu_{\nu 0}T^{0\mu}-\Gamma^\mu_{0\nu}T^{0\nu}
-\Gamma^\nu_{\nu i}T^{i\mu}-\Gamma^\mu_{i\nu}T^{i\nu},\label{Complete_set_Eqa}\\
&\partial_t X_{i}^A+\epsilon_{ijk}\partial_j \varphi_{k}^A-\partial_i Y^A=0,\label{Complete_set_Eqb}\\
&\partial_t Z_{i}+\epsilon_{ijk}\partial_j \Psi_{k}-\partial_i S_n=0,\label{Complete_set_Eqc}\\
&\partial_t U^A =Y^A=Y^A(T^{0\mu}),\label{Complete_set_Eq_d}\\
&\partial_t n=S_n,\label{Complete_set_Eq_e}\\
&\partial_t \varphi_{i}^A -\kappa_{1}^A\epsilon_{ijk} \partial_j X_{k}^A +\partial_i \phi^A =-\beta_{1}^A \varphi_{i}^A, \quad (\text{no sum over } A) \label{Complete_set_Eq_f}\\
&\partial_t \phi^A+\kappa_{2}^A\partial_i \varphi_{i}^A=-\beta_{2}^A \phi^A, \quad (\text{no sum over } A) \label{Complete_set_Eq_g}\\
&\partial_t \Psi_i -\lambda_{1} \epsilon_{ijk} \partial_j Z_k +\partial_i \psi =-\sigma_{1} \Psi_i, \label{Complete_set_Eq_h}\\
&\partial_t \psi +\lambda_{2}\partial_i \Psi_i=-\sigma_{2} \psi, \label{Complete_set_Eq_i}
\end{align}
\eml
with $T^{ij}$ depending on $Y^A=Y^A(T^{0\nu})$ as can be seen in Equation~\eqref{Relations}. As discussed in Section~\ref{FO:Equivalence}, the initial data for the extra cleaning fields must be $\varphi_{i}^A=\phi^A=\Psi_i=\psi=0$. The coefficients $\kappa_1^A,\kappa_2^A,\lambda_1,\lambda_2,\beta_1^A,\beta_2^A,\sigma_1,\sigma_2$ are all constants. The choice of these constants may depend on the application. See the discussion in Section~\ref{Sec:CC} for useful bounding conditions for them.
\begin{assumption}
\label{Assumption_velocities}
We assume that $\kappa_1^A,\kappa_2^A,\lambda_1,\lambda_2,\beta_1^A,\beta_2^A,\sigma_1,\sigma_2$ are all strictly positive constants.
\end{assumption}
\begin{theorem}
\label{Theorem_1}
Under assumptions \ref{Assumption_LWP}, \ref{Assumption_physics}, and \ref{Assumption_velocities}, Equations~\eqref{Complete_set_Eq} are locally well-posed in Sobolev spaces.
\end{theorem}
A more precise statement of Theorem \ref{Theorem_1} and its proof can be found in Section~\ref{Sec:LWP}.

In order to verify the flux-conservative form of Equations~\eqref{Complete_set_Eq}, we observe that in matrix form they can be written as
\begin{align}
\label{Matrix}
\partial_t \mathcal{Q}+\partial_i F_i=\mathcal{S},
\end{align}
where 
\begin{align}
\label{FO:Matrices}
\mathcal{Q}=\bbm T^{0\mu}\\ X_{1}^A\\ X_{2}^A \\ X_{3}^A \\ Z_1 \\ Z_2 \\ Z_3 \\ U^A \\ n \\  \varphi_{1}^A \\ \varphi_{2}^A \\ \varphi_{3}^A \\ \phi^A \\  \Psi_1\\ \Psi_2\\ \Psi_3\\ \psi 
\ebm, \quad 
F_i=\bbm T^{i\mu} \\ 
\epsilon_{1ik}\varphi_{k}^A-\delta_{1i}Y^A \\ 
\epsilon_{2ik}\varphi_{k}^A-\delta_{2i}Y^A \\ 
\epsilon_{3ik}\varphi_{k}^A-\delta_{3i}Y^A \\ 
\epsilon_{1ik}\Psi_k-\delta_{1i}S_n\\
\epsilon_{2ik}\Psi_k-\delta_{2i}S_n\\
\epsilon_{3ik}\Psi_k-\delta_{3i}S_n\\
0_{4\times 1}\\
0\\
-\kappa_{1}^A \epsilon_{1ik}X_{k}^A+\delta_{1i}\phi^A\\ 
-\kappa_{1}^A \epsilon_{2ik}X_{k}^A+\delta_{2i}\phi^A\\
-\kappa_{1}^A \epsilon_{3ik}X_{k}^A+\delta_{3i}\phi^A\\
\kappa_{2}^A \varphi_{i}^A \\ 
-\lambda_{1} \epsilon_{1ik}  Z_k +\delta_{1i}\psi\\
-\lambda_{1} \epsilon_{2ik}  Z_k +\delta_{2i}\psi\\
-\lambda_{1} \epsilon_{3ik}  Z_k +\delta_{3i}\psi\\
\lambda_{2}\Psi_i
 \ebm,\quad\text{and}\quad 
\mathcal{S}=\bbm S^\mu \\ 0_{4\times 1} \\ 0_{4\times 1} \\ 0_{4\times 1} \\  0 \\ 0 \\ 0  \\  Y^A \\ S_n \\ -\beta_{1}^A \varphi_{1}^A \\ -\beta_{1}^A \varphi_{2}^A \\ -\beta_{1}^A \varphi_{3}^A \\ 
-\beta_{2}^A \phi^A \\-\sigma_{1} \Psi_1\\-\sigma_{1} \Psi_2\\-\sigma_{1} \Psi_3\\-\sigma_{2} \psi \ebm
\end{align} 
are real $44\times 1$ column vectors, $S_n$ is given in Equation~\eqref{Sn}, and 
\[S^\mu\equiv -\Gamma^\nu_{\nu 0}T^{0\mu}-\Gamma^\mu_{0\nu}T^{0\nu}
-\Gamma^\nu_{\nu i}T^{i\mu}-\Gamma^\mu_{i\nu}T^{i\nu}.\]

\subsection{Inverse of $M$}
\label{Inverse_M_section}
The matrix $M^{\mu\beta}=H^{\mu00\beta}$ may be written as 
\begin{align*}
M^{\mu\beta} &=-\varepsilon\left [\tau_\varepsilon (u^0)^2+\beta_\varepsilon \Delta^{00} \right ]u^\mu  u^\beta
+ \frac{(\varepsilon+P)(\tau_\varepsilon+\tau_Q)(u^0)^2-n\beta_n \Delta^{0 0}}{ u^0} u^\mu\Delta^{0\beta} 
-\varepsilon u^0 (\tau_P+\beta_\varepsilon)  \Delta^{\mu 0} u^\beta\\
&+\left [\tau_P(\varepsilon+P)-\zeta-\frac{\eta}{3}-n\beta_n \right ] \Delta^{0\mu}\Delta^{0\beta}
+ \left [\tau_Q(\varepsilon+P) (u^0)^2-\eta \Delta^{00} \right ] \Delta^{\mu\beta}
\end{align*}
where we used that
\begin{align*}
2\Delta^{\mu 0 0 \beta}&=\Delta^{0\mu}\Delta^{0\beta}+\Delta^{00}\Delta^{\mu\beta}-\frac{2}{3}\Delta^{0\mu}\Delta^{0\beta}\\
&=\frac{1}{3}\Delta^{0\mu}\Delta^{0\beta}+\Delta^{00}\Delta^{\mu\beta}.
\end{align*}

Under Assumption \ref{Assumption_LWP}, the inverse $M^{-1}_{\lambda\mu}$ is 
\be
\label{Inverse_M}
M^{-1}_{\lambda\mu}=A u_\lambda u_\mu+B u_\lambda \Delta^0_\mu+C\Delta^0_\lambda u_\mu+D\Delta^0_\lambda \Delta^0_\mu+E\Delta_{\lambda\mu},
\ee
where
\bml\label{Inverse_M_coefficients}
\bea
A&=&\frac{\Delta^{00} \left(3\rho \tau_P-3 \zeta -4\eta-3n\beta_n \right)+3(u^0)^2 \rho\tau_Q}{\varepsilon F},\\
B&=&\frac{3\left [(u^0)^2 \rho\left(\tau_Q+\tau_{\varepsilon }\right)-\Delta^{00}n \beta _n\right ]}{\varepsilon u^0 F},\\
C&=& \frac{- 3 u^0 (\beta_\varepsilon+\tau_P)}{F},\\
D&=&\frac{ \left\{(u^0)^2  \left[3 \rho\tau_Q\left (\tau_P + \beta _{\varepsilon }\right )+\tau_{\varepsilon } \left(3 \zeta+3 \rho \beta_s +\eta\right ) \right]+\Delta^{00}\left [\beta_{\varepsilon } \left(3\zeta +\eta \right)- 3\rho\tau_P \beta_s\right ]\right\}}{(\Delta^{00} \eta -(u^0)^2 \rho\tau_Q)F},\\
E&=& \frac{1}{\rho\tau_Q(u^0)^2-\eta \Delta^{00}}\\
F&=&(\Delta^{00})^2\left [\beta_{\varepsilon } \left(3 \zeta +4 \eta \right)-3 \rho\tau_P\beta_s\right ]+\Delta^{00}(u^0)^2 \left[\tau_{\varepsilon } \left(3 \zeta+3 \rho\beta _{s} +4 \eta \right)+3 \rho\tau _P \tau_Q\right ]-3 (u^0)^4 \rho\tau_Q \tau_{\varepsilon },
\eea
\eml
and we used the definitions $\beta_s=\beta_\varepsilon+n \beta_n/(\varepsilon+P)$.

\section{Mixed-order flux-conservative formulation of BDNK}
\label{Mixed_order_section}

In this section, we present a formulation of the equations that is first-order in time derivatives but second-order in spatial derivatives. This is achieved by introducing auxiliary fields to reduce the temporal order while preserving the spatial order, significantly reducing, as compared to the first-order formulation of Section~\ref{Sec:First_Order}, the number of dynamical variables to 9 -- \emph{without the need for curl-cleaning constraints.} Crucially, despite the presence of second-order spatial derivatives in the equations, these
do not require explicit computation in numerical simulations. 
This is because the system is written in a flux-conservative form, where the fluxes $F_i$ naturally incorporate first-order spatial derivatives.

This approach aligns with established methods for handling mixed-order systems in central schemes. For instance, the Kurganov-Tadmor (KT) second-order central scheme \cite{Tadmor2000} and the Kurganov-Levy (KL) third-order central scheme \cite{KurganovLevy2000} inherently accommodate such fluxes, treating the derivative terms through numerical dissipation and Riemann solvers without ever requiring direct computation of second-order spatial derivatives. Thus, the formulation retains the efficiency and robustness of standard high-resolution central schemes. 

Because mixed-order reductions such as the one we present here are not as common as first-order reductions such as the one employed in Section~\ref{Sec:First_Order}, in Section~\ref{Toy_model_example_mixed_order_section} we first present a simple example illustrating the procedure. We also use the example of Section~\ref{Toy_model_example_mixed_order_section} to illustrate some points concerning well-posedness of the mixed-order reduction we finally introduce in Section~\ref{Mixed_order_per_se_section}. 

\subsection{A toy-model example}
\label{Toy_model_example_mixed_order_section}
Here we illustrate some points of our mixed-order reduction using a simple example. Readers uninterested in these kinds of illustration can skip this Section.

Let us consider the Cauchy problem for the wave equation,
\begin{subequations}{\label{Wave_illustration_mixed_order}}
    \begin{align}
        -\partial_t^2 \psi + \partial_x^2 \psi & = 0,
\label{Wave_illustration_mixed_order_eq} 
\\
        \left. \psi \right|_{t=0} & = \psi_0,
 \label{Wave_illustration_mixed_order_psi_0}       \\
        \left. \partial_t \psi \right|_{t=0} & = \psi_1.
\label{Wave_illustration_mixed_order_psi_1} 
    \end{align}
\end{subequations}

A standard first-order reduction consists in introducing
\begin{align*}
     u & = \partial_t \psi,\\
     v & = \partial_x \psi,
\end{align*}
so that Equation~\eqref{Wave_illustration_mixed_order_eq} gives
\begin{align*}
    \begin{bmatrix}
     1 & 0 \\
     0 & 1
    \end{bmatrix}
    \partial_t
    \begin{bmatrix}
        u\\
        v
    \end{bmatrix}
    -
    \begin{bmatrix}
     0 & 1 \\
     1 & 0
    \end{bmatrix}
    \partial_x
    \begin{bmatrix}
        u\\
        v
    \end{bmatrix}
    =     
    \begin{bmatrix}
        0\\
        0
    \end{bmatrix},
\end{align*}
with the initial conditions from Equations~\eqref{Wave_illustration_mixed_order_psi_0}-\eqref{Wave_illustration_mixed_order_psi_1} becoming
\begin{align*}
    \left. u \right|_{t=0} & = \psi_1 \\
    \left. v \right|_{t=0} & = \partial_x\psi_0. \\    
\end{align*}

The first-order reduction of Section~\ref{Sec:First_Order} is akin to the above standard reduction of the wave equation. Our mixed-order reduction, on the contrary, considers a reduction of order only in the time derivatives. To see how it works with Equation~\eqref{Wave_illustration_mixed_order}, we again consider the new variable 
\begin{align*}
    u = \partial_t \psi,
\end{align*}
but keep $\psi$ as a variable without introducing $v = \partial_x \psi$. In this case, Equation~\eqref{Wave_illustration_mixed_order_eq} gives
\begin{subequations}{\label{Illustration_mixed_order}}
\begin{align}
    \partial_t u - \partial_x^2 \psi & = 0,
    \label{Illustration_mixed_order_u}
    \\
    \partial_t \psi - u & = 0,
    \label{Illustration_mixed_order_psi}
\end{align}
\end{subequations}
with the initial conditions from
Equations~\eqref{Wave_illustration_mixed_order_psi_0}-\eqref{Wave_illustration_mixed_order_psi_1} becoming
\begin{align*}
    \left. \psi \right|_{t=0} & = \psi_0. \\ 
    \left. u \right|_{t=0} & = \psi_1.
\end{align*}

In order to understand the mathematical structure of Equations~\eqref{Illustration_mixed_order}, we begin by identifying its principal part. Since Equation~\eqref{Illustration_mixed_order} is a mixed-order system, to correctly identify its principal part we need to find Leray indices for each unknown and each equation in Equations~\eqref{Illustration_mixed_order} (we recall the definition of Leray indices in Section~\ref{Sec:SH}, noting that we employ the standard definition used in the literature, as in e.g. \cite[Appendix~IV]{Choquet-Bruhat-Book-2009}
or \cite[Appendix~B]{Disconzi-2024}). We find
\begin{align*}
    m_u & = 1, 
    n_{\eqref{Illustration_mixed_order_u}} = 0,
    \\
    m_\psi & = 2, 
    n_{\eqref{Illustration_mixed_order_psi}} = 1. 
\end{align*}

It follows that the principal part of Equation~\eqref{Illustration_mixed_order} is the following mixed-order matrix comprised of second-, first-, and zeroth-order operators
\begin{align*}
    \begin{bmatrix}
        \partial_t &  -\partial_x^2 \\
        -1 & \partial_t
    \end{bmatrix},
\end{align*}
from which we obtain the characteristics
\begin{align*}
    \det
    \begin{bmatrix}
        \xi_0 &  -\xi_1^2 \\
        -1 & \xi_0
    \end{bmatrix}
    = \xi_0^2 - \xi_1^2 = 0,
\end{align*}
so that the characteristics of Equations~\eqref{Illustration_mixed_order} are the same as the original wave equation, i.e., the light-cones.

Since Equation~\eqref{Illustration_mixed_order_u} involves one time derivative and two spatial derivatives, we might wonder if the system in Equations~\eqref{Illustration_mixed_order} has a structure akin to the heat equation
\begin{align}
    \label{Illustration_mixed_order_heat}
    \partial_t \varphi - \partial_x^2 \varphi = 0.
\end{align}
The heat equation is of parabolic type. Thus, the pairing of one time derivative with two spatial derivatives in Equation~\eqref{Illustration_mixed_order_u}  raises the question of whether Equation~\eqref{Illustration_mixed_order} has a \emph{parabolic structure.} A reduction that introduces parabolicity would not work for our purposes due to well-known features of parabolic equations: their Cauchy problem is ill-posed and not causal. Moreover, for parabolic equations, the Cauchy-Kowalevskaya theorem, which gives existence and uniqueness for analytic data, does not apply.  
 While the ability to work within the analytic class is not a necessary condition for well-posedness,  for ``standard" hyperbolic PDEs (e.g., those not presenting degeneracies, see \cite[Section~5]{Disconzi-2024}), applicability of the Cauchy-Kowalevskaya and the corresponding existence and uniqueness in the analytic class are viewed as an important test case for well-posedness. In fact, well-posedness can be obtained by combining the Cauchy-Kowalevskaya with energy estimates via the so-called Schauder method \cite[Section~5.3]{John-Book-1982}.

That Equation~\eqref{Illustration_mixed_order} is a well-posed hyperbolic system can be anticipated from its characteristics, the light-cones. But it is instructive to understand this from a more pedestrian point of view. The \emph{intuitive feature} of Equations~\eqref{Illustration_mixed_order} that ensures hyperbolicity and well-posedness despite its mixed-order structure is that time derivatives are never paired with two spatial derivatives \emph{of the same variable in a given equation.} 
Contrast Equation~\eqref{Illustration_mixed_order_u}, where $\partial_t u$ and $\partial_x^2 \psi$ are combined, with the heat Equation~\eqref{Illustration_mixed_order_heat}, where $\partial_t \varphi$ is paired with $\partial_x^2 \varphi$.
This otherwise simple feature of Equation~\eqref{Illustration_mixed_order} allows us to ensure its hyperbolic character. As we will see in Section~\ref{Mixed_order_per_se_section}, that is also a feature of our mixed-order formulation: we will derive a mixed-order system for an unknown vector $\psi = (\psi^1,\dots, \psi^N)$ involving first-order time derivatives and second-order spatial derivatives so that we never have
\begin{align*}
    \partial_t \psi^I 
    + \cdots+ a^{ij}(\psi) \partial_{ij}^2 \psi^J 
    + \cdots = 0,
\end{align*}
with $I=J$.

That said, we highlight that simply pairing one time derivative with two spatial derivatives of a different variable is not enough to ensure hyperbolicity. For example,
\begin{align*}
    \partial_t u - \partial_{x}^2v &=0,
    \\
    \partial_t v - \partial_x^2 u&= 0,
\end{align*}
contains a backward heat equation for $u-v$. Thus, an analysis of the reduced system and its characteristics is still needed to verify hyperbolicity.

\begin{remark}
    At first sight, one might think it is natural that Equation~\eqref{Illustration_mixed_order} turns out to be a well-posed hyperbolic system whose characteristics are the light-cones, since it came from reducing the wave equation. It is important to remember, however, that the whole difficulty with reductions is to ensure that they preserve features like hyperbolicity, as discussed in the introduction. The BDNK Equations~\eqref{BDNK_EoM} are already known to be hyperbolic, well-posed, and causal, as shown in \cite{Bemfica:2020zjp}. Our task here is to produce a reduction of the BDNK that preserves these features while also being suitable for numerical simulations.
\end{remark}

\subsection{The mixed-order formulation}
\label{Mixed_order_per_se_section}

Let us consider the energy-momentum tensor written in the form in Equation~\eqref{defineBDNKnovo} and define the auxiliary variables
\begin{align}
\label{Y_Mixed}
Y^A=\partial_t U^A.
\end{align} 
By means of the three vectors $C^\mu_i$, we may write the energy-momentum tensor as
\begin{align}
\label{Tmunu_Mixed}
T^{\mu\nu} &= T_0^{\mu\nu} + H^{\mu\nu 0\beta}P_{A\beta} Y^A
+H^{\mu\nu j \beta}P_{A\beta} \partial_j U^A +2\beta_n u^{(\mu} \Omega^{j\nu)} \partial_j n + H^{\mu\nu\alpha\beta} C_{j\beta} \Gamma^j_{\alpha \lambda}u^\lambda.
\end{align}
Following the same definitions for the matrix $M^{\mu\beta}=H^{\mu00\beta}$, the equations of motion in the variables $\{Y^A,U^A,J^0\}$ are simply
\bml
\label{MO:Eqs}
\begin{align}
\partial_t T^{0\mu}+\partial_i T^{i\mu}&=-\Gamma^\nu_{\nu 0}T^{0\mu}-\Gamma^\mu_{0\nu}T^{0\nu}
-\Gamma^\nu_{\nu i}T^{i\mu}-\Gamma^\mu_{i\nu}T^{i\nu},\\
\partial_t U^A &=Y^A,\\
\partial_t J^0+\partial_i J^i & =-\Gamma^\nu_{\nu\mu}n u^\mu, 
\end{align}
\eml
where, from the invertible nature of the matrix $M^{\mu\beta}$, 
\bml
\label{MO:Transformations}
\begin{align}
n&=J^0/u^0,\\
J^i&=n u^i=\frac{J^0}{u^0} u^i,\\
Y^A  &=\tilde{P}^{A\lambda}M^{-1}_{\lambda\mu}\left [T^{0\mu}- T_0^{0\mu} - H^{\mu0 j \beta}P_{B\beta}\partial_j U^B-2\beta_n u^{(\mu} \Omega^{j0)} \partial_jn - H^{\mu 0 \alpha\beta} C_{j\beta} \Gamma^j_{\alpha \sigma}u^\sigma\right ],\label{FO:YA}\\
T^{ij} &= T_0^{ij} + H^{ij 0\beta} P_{A\beta} Y^A
+H^{ij k \beta}P_{A\beta}\partial_k U^A+2\beta_n u^{(i} \Omega^{kj)} \partial_k n + H^{ij\alpha\beta} C_{k\beta} \Gamma^k_{\alpha \lambda}u^\lambda.
\end{align}
\eml
In matrix form:
\begin{align*}
\partial_t \mathcal{Q}+\partial_i F_i=\mathcal{S},
\end{align*}
where 
\begin{align}\mathcal{Q}=\bbm T^{0\mu}\\ U^A\\ J^0\ebm, \quad 
F_i=\bbm T^{i\mu} \\ 
0_{4\times 1}\\
\frac{J^0}{u^0} u^i \ebm,\quad\text{and}\quad 
\mathcal{S}=\bbm S^\mu \\  Y^A \\ -n \Gamma^\nu_{\nu\lambda}u^\lambda \ebm\label{MO:eoms-Matrix}\end{align}
are real $9\times 1$ column vectors and 
\[S^\mu\equiv -\Gamma^\nu_{\nu 0}T^{0\mu}-\Gamma^\mu_{0\nu}T^{0\nu}
-\Gamma^\nu_{\nu i}T^{i\mu}-\Gamma^\mu_{i\nu}T^{i\nu}.\] 
\begin{theorem}
\label{Theorem_2}
Suppose that Assumptions \ref{Assumption_LWP} and \ref{Assumption_physics} hold. Then Equations~\eqref{MO:Eqs} are locally well-posed in Sobolev spaces for initial data and coefficients satisfying the conditions stated precisely in Theorem \ref{T:Precise_mixed_order_local_well_posedness}.
\end{theorem}

The proof is given in Section~\ref{Sec:Proof_MO}, and the equivalence of the mixed-order formulation and the original BDNK equations is proved in Section~\ref{MO:Equivalence}.

\section{Implementation of div-curl cleaning for the first-order formulation of Section~\ref{Sec:First_Order}}
\label{Sec:Cleaning_formulation}

The reduction to first-order carried out in Section~\ref{Sec:First_Order} introduces constraints that need to be propagated by the flow. Such propagation is shown in Section~\ref{FO:Equivalence}. In a numerical simulation, however, such constraints might not be exactly preserved due to rounding errors. These constraint violations can increase over time and spoil the numerical simulation. To avoid this, we implement the method of div-curl constraint cleaning explained in Section~\ref{Sec:CC}.

Similarly to the example worked out in Section~\ref{Reduction_toy_model_section}, the new variables $X^A_i$~\eqref{X} and $Z_i$~\eqref{Z}
require special treatment because the equivalence between first-order and second-order equations demands that their curl must vanish, i.e., the constraint Equations~\eqref{supplemental_EQ_b} and \eqref{supplemental_EQ_d} 
must be satisfied. We may directly apply Equations~\eqref{CC_4} to each of these variables, introducing five new curl-cleaning three-vector fields $\varphi_{i}^A$ and $\Psi_{i}$, along with their corresponding five divergence-cleaning fields $\phi_A$ and $\psi$. 

Then, Equations~\eqref{supplemental_EQ} can be rewritten as
\bml
\label{supplemental_EQ_II}
\begin{align}
\partial_t X_{i}^A+\textcolor{blue}{\epsilon_{ijk}\partial_j \varphi_{k}^A}-\partial_i Y^A&=0,\label{supplemental_II_EQ_a}\\
\textcolor{blue}{\partial_t \varphi_{i}^A} -\kappa_{1}^A\epsilon_{ijk} \partial_j X_{k}^A \textcolor{red}{+\partial_i \phi^A} &=\textcolor{blue}{-\beta_{1}^A \varphi_{i}^A}, \quad (\text{no sum over} A)\label{supplemental_II_EQ_b}\\
\textcolor{red}{\partial_t \phi^A} \textcolor{blue}{+\kappa_{2}^A\partial_i \varphi_{i}^A}&=\textcolor{red}{-\beta_{2}^A \phi^A},\quad (\text{no sum over} A)\\
\partial_t Z_{i}+\textcolor{blue}{\epsilon_{ijk}\partial_j \Psi_{k}}-\partial_i S_n&=0,\label{supplemental_II_EQ_c}\\
\textcolor{blue}{\partial_t \Psi_i} -\lambda_{1} \epsilon_{ijk} \partial_j Z_k \textcolor{red}{+\partial_i \psi} &=\textcolor{blue}{-\sigma_{1} \Psi_i}, \label{supplemental_II_EQ_d}\\
\textcolor{red}{\partial_t \psi} \textcolor{blue}{+\lambda_{2}\partial_i \Psi_i}&=\textcolor{red}{-\sigma_{2} \psi},
\end{align}
\eml
where $\kappa_1^A$, $\kappa_2^A$, $\lambda_1$, and $\lambda_2 > 0$ represent the squared propagation speeds of each cleaning field, while $\beta_1^A$, $\beta_2^A$, $\sigma_1$, and $\sigma_2 > 0$ control the damping strength of each cleaning field. The latter parameters correspond to the inverse of the damping timescale. Following \citet{DUMBSER2020109088}, we have highlighted in blue the curl-cleaning terms and in red the divergence-cleaning terms, while the original equations remain in black. Therefore, Equations~\eqref{Complete_set_Eq} correspond to Equations~\eqref{Set_Eq} augmented with Equations~\eqref{supplemental_EQ_II}.

\section{Equivalence of the first- and mixed-order formulations and the original second-order BDNK equations}
\label{S:Equivalence_formulations}
In this Section we established that the new formulations introduced in Sections~\ref{Sec:First_Order} and \ref{Mixed_order_section}
lead to solutions to the original second-order BDNK Equations~\eqref{BDNK_EoM}, provided the data for the new formulations is compatible with initial data for the original equations, i.e., the data for the new formulation is derived from the original data for the second-order system.

\subsection{Equivalence of the first-order formulation of Section~\ref{Sec:First_Order} and the original second-order BDNK equations}
\label{FO:Equivalence}

Although the first-order system \eqref{Complete_set_Eq} for the variables $\{T^{0\mu},X_i^A,Z_i,U^A,n,\varphi_i^A,\phi^A,\Psi_i,\psi\}$ was derived from the equations
\begin{subequations}
\label{Second_order_C}
\begin{align}
\nabla_\nu T^{\mu\nu}
&=\nabla_\nu\left(T_{(0)}^{\mu\nu}
+H^{\mu\nu\alpha\beta}P_{A\beta}\nabla_\alpha U^A
+2\beta_n u^{(\mu}\Omega^{j\nu)}\partial_j n\right)=0,
\\
\nabla_\mu(nu^\mu)&=0,
\end{align}
\end{subequations}
the converse requires compatibility conditions on the initial data. Here, \eqref{Second_order_C} is \eqref{BDNK_EoM} written using Equation~\eqref{defineBDNKnovo2}; in particular, $\nabla_\alpha U^\theta=\partial_\alpha\theta$ and $\nabla_\alpha U^i=\nabla_\alpha u^i$. For the constraint-propagation argument, we use Assumption \ref{Assumption_velocities} and the periodic solution class of Theorem \ref{Theorem_1_precise}. Since \eqref{Complete_set_Eq} evolves a larger set of variables, its solutions need not satisfy the original equations. The normalization \eqref{normalization_u_BDNK}, however, holds automatically because $u^0$ is defined by Equation~\eqref{u0}.

The identity $Y^A=\partial_tU^A$ is imposed by Equation~\eqref{Complete_set_Eq_d}. We must also show that the variables $X_i^A$ and $Z_i$ retain the meanings prescribed in Equations~\eqref{X} and \eqref{Z}. Subtracting the spatial derivative of Equation~\eqref{Complete_set_Eq_d} from Equation~\eqref{Complete_set_Eqb}, and the spatial derivative of Equation~\eqref{Complete_set_Eq_e} from Equation~\eqref{Complete_set_Eqc}, gives
\begin{subequations}
\label{Equivalence_EQ}
\begin{align}
\partial_t\left(X_i^A-\partial_iU^A\right)
+\epsilon_{ijk}\partial_j\varphi_k^A&=0,
\\
\partial_t\left(Z_i-\partial_i n\right)
+\epsilon_{ijk}\partial_j\Psi_k&=0.
\end{align}
\end{subequations}

We impose the compatible initial data
\begin{subequations}
\label{E:FO_Compatible_initial_data}
\begin{align}
X_i^A(0,\mathbf{x})&=\partial_iU^A(0,\mathbf{x}),
\\
Z_i(0,\mathbf{x})&=\partial_i n(0,\mathbf{x}),
\\
\varphi_i^A(0,\mathbf{x})
&=\phi^A(0,\mathbf{x})=\Psi_i(0,\mathbf{x})=\psi(0,\mathbf{x})=0.
\end{align}
\end{subequations}
To prove that these conditions propagate, write the constraint errors as in Equations~\eqref{E:FO_CD_definitions},
\begin{align*}
C_i^A&=X_i^A-\partial_iU^A,
\\
D_i&=Z_i-\partial_i n.
\end{align*}
Commutation of spatial derivatives gives
\begin{align*}
\epsilon_{ijk}\partial_jX_k^A
&=\epsilon_{ijk}\partial_jC_k^A,
\\
\epsilon_{ijk}\partial_jZ_k
&=\epsilon_{ijk}\partial_jD_k.
\end{align*}
Consequently, Equations~\eqref{Equivalence_EQ} and \eqref{Complete_set_Eq_f}--\eqref{Complete_set_Eq_i} give the homogeneous system \eqref{E:FO_CD_cleaning_system} for $C_i^A,D_i$ and the cleaning fields. The data \eqref{E:FO_Compatible_initial_data} make every component of this system zero initially. By Equation~\eqref{E:FO_CD_cleaning_energy_identity}, the energy defined in Equation~\eqref{E:FO_CD_cleaning_energy} satisfies
\begin{align*}
0\leq E_0(t)\leq E_0(0)=0.
\end{align*}
All weights in this energy are strictly positive by Assumption \ref{Assumption_velocities}. It follows that
\begin{align*}
C_i^A&=0,
\\
D_i&=0,
\\
\varphi_i^A&=\phi^A=\Psi_i=\psi=0
\end{align*}
throughout the existence interval. Thus, $X_i^A=\partial_iU^A$ and $Z_i=\partial_i n$, and \eqref{Complete_set_Eq} reduces to Equations~\eqref{Set_Eq}. Substitution into Equation~\eqref{New_Tmunu} recovers the constitutive expression in Equations~\eqref{Second_order_C}, while Equations~\eqref{Sn} and \eqref{Set_Eq_e} recover $\nabla_\mu(nu^\mu)=0$. Hence Equations~\eqref{Set_Eq_a} and \eqref{Set_Eq_e} give the original BDNK equations~\eqref{Second_order_C}.

Conversely, let $T>0$, let $(U^A,n)$ be a $C^2([0,T]\times\mathbb{T}^3)$ solution of Equations~\eqref{Second_order_C}, and suppose that the matrix \eqref{M} is invertible on $[0,T]\times\mathbb{T}^3$. Define $Y^A,X_i^A,Z_i$ by Equations~\eqref{New_variables}, take $T^{0\mu}$ from the constitutive expression in Equations~\eqref{Second_order_C}, and set all cleaning fields to zero. Commutation of derivatives gives Equations~\eqref{supplemental_EQ}, and current conservation gives $\partial_t n=S_n$. The constitutive identity \eqref{New_Tmunu} gives Equation~\eqref{Relations}, while energy-momentum conservation gives Equation~\eqref{Set_Eq_a}. These identities verify every equation in Equations~\eqref{Complete_set_Eq} and give the compatible data \eqref{E:FO_Compatible_initial_data}.

\begin{remark}
Numerical errors can violate the identities $X_i^A=\partial_iU^A$ and $Z_i=\partial_i n$ and generate nonzero cleaning fields. Equation~\eqref{Equivalence_EQ} shows that the errors $C_i^A$ and $D_i$ can then depend on time. They cannot, in general, be represented by fixed functions of the spatial coordinates alone.

The estimate \eqref{E:FO_CD_cleaning_energy_identity} bounds the continuum constraint and cleaning energy by its initial value, but it does not imply decay of every constraint error. Indeed, any time-independent curl-free fields $C_i^A,D_i$, together with zero cleaning fields, solve Equations~\eqref{E:FO_CD_cleaning_system}. Thus, damping the cleaning fields does not by itself enforce the full gradient identities. Theorem \ref{Theorem_1_precise} establishes local well-posedness of the augmented continuum system without requiring the constraint errors or cleaning fields to vanish initially. Stability and convergence estimates for a numerical discretization are separate from that theorem.
\end{remark}

\subsection{Equivalence of the mixed-order formulation of Section~\ref{Mixed_order_section} and the original second-order BDNK equations}
\label{MO:Equivalence}

Here, we show that the mixed-order system \eqref{MO:Eqs}, with the algebraic relations \eqref{MO:Transformations}, is equivalent to the original equations~\eqref{BDNK_EoM}. We work in the coordinate and state domain used in Section~\ref{Mixed_order_per_se_section}, where $u^0$ and $u_0$ are nonzero and the matrix $M$ is invertible. The normalization of $u^\mu$ holds by Equation~\eqref{u0}.

Let $(T^{0\mu},U^A,J^0)$ solve Equations~\eqref{MO:Eqs}, and define $n=J^0/u^0$ and $J^i=nu^i$. The equation for $J^0$ gives
\begin{align*}
\nabla_\mu(nu^\mu)
&=\partial_tJ^0+\partial_iJ^i
+\Gamma^\nu_{\nu\lambda}nu^\lambda=0.
\end{align*}
Consequently,
\begin{align}
\label{E:MO_Equivalence_density_time}
\partial_tn
=-\frac{u^j}{u^0}\partial_jn
-\frac{n}{u^0}\nabla_\alpha u^\alpha.
\end{align}
Using Equation~\eqref{Omega}, we obtain
\begin{align}
\label{E:MO_Equivalence_density_projection}
\begin{split}
\Delta^{\nu\alpha}\partial_\alpha n
&=\Delta^{\nu0}\partial_tn+\Delta^{\nu j}\partial_jn\\
&=\Omega^{j\nu}\partial_jn
-\frac{n}{u^0}\Delta^{\nu0}\nabla_\alpha u^\alpha.
\end{split}
\end{align}

The normalization and metric compatibility imply
\begin{align*}
u_\beta\nabla_\alpha u^\beta&=0,\\
\nabla_\alpha u^0&=-\frac{u_i}{u_0}\nabla_\alpha u^i,\\
\nabla_\alpha u_\beta&=C_{i\beta}\nabla_\alpha u^i.
\end{align*}
It follows from Equation~\eqref{Matrices_P} that
\begin{align}
\label{E:MO_Equivalence_derivatives}
\nabla_\alpha u_\beta+u_\beta\partial_\alpha\theta
=P_{A\beta}\partial_\alpha U^A
+C_{i\beta}\Gamma^i_{\alpha\sigma}u^\sigma.
\end{align}

To recover the temporal components of the constitutive tensor, we use
\begin{align*}
M^{\mu\beta}P_{A\beta}
\tilde{P}^{A\lambda}M^{-1}_{\lambda\nu}
=M^{\mu\lambda}M^{-1}_{\lambda\nu}
=\delta^\mu_\nu.
\end{align*}
Multiplication of Equation~\eqref{FO:YA} by $M^{\mu\beta}P_{A\beta}$ therefore gives
\begin{align}
\label{E:MO_Equivalence_temporal_tensor}
\begin{split}
T^{0\mu}
={}&T_0^{0\mu}
+H^{\mu00\beta}P_{A\beta}Y^A
+H^{\mu0j\beta}P_{B\beta}\partial_jU^B\\
&+2\beta_nu^{(\mu}\Omega^{j0)}\partial_jn
+H^{\mu0\alpha\beta}C_{j\beta}
\Gamma^j_{\alpha\sigma}u^\sigma.
\end{split}
\end{align}
Since $\partial_tU^A=Y^A$, equations~\eqref{E:MO_Equivalence_derivatives} and \eqref{E:MO_Equivalence_temporal_tensor}, together with the formula for $T^{ij}$ in Equations~\eqref{MO:Transformations} and $T^{i0}=T^{0i}$, yield
\begin{align}
\label{E:MO_Equivalence_constitutive}
\begin{split}
T^{\mu\nu}
={}&T_0^{\mu\nu}
+H^{\mu\nu\alpha\beta}
\big(\nabla_\alpha u_\beta+u_\beta\partial_\alpha\theta\big)\\
&+2\beta_nu^{(\mu}\Omega^{j\nu)}\partial_jn.
\end{split}
\end{align}
Thus, Equation~\eqref{E:MO_Equivalence_constitutive} is Equation~\eqref{defineBDNKnovo}. The substitution of $\partial_tn$ used in deriving Equation~\eqref{defineBDNKnovo} is valid by Equation~\eqref{E:MO_Equivalence_density_time}. More explicitly, since
\begin{align*}
\Delta^{\alpha\beta}
\big(\nabla_\alpha u_\beta+u_\beta\partial_\alpha\theta\big)
=\nabla_\alpha u^\alpha,
\end{align*}
the terms involving $\beta_n$ in Equations~\eqref{defineBDNKnew} and \eqref{E:MO_Equivalence_constitutive} combine, by Equation~\eqref{E:MO_Equivalence_density_projection}, into
\begin{align*}
&2\beta_nu^{(\mu}\Omega^{j\nu)}\partial_jn
-\frac{2n\beta_n}{u^0}
u^{(\mu}\Delta^{\nu)0}\nabla_\alpha u^\alpha\\
&\qquad=2\beta_nu^{(\mu}\Delta^{\nu)\alpha}\partial_\alpha n.
\end{align*}
Hence $T^{\mu\nu}$ is the original constitutive tensor \eqref{BDNKTmunu}. The equation for $T^{0\mu}$ in Equations~\eqref{MO:Eqs} now gives
\begin{align*}
\nabla_\nu T^{\mu\nu}
&=\partial_tT^{0\mu}+\partial_iT^{i\mu}
+\Gamma^\nu_{\nu\lambda}T^{\lambda\mu}
+\Gamma^\mu_{\nu\lambda}T^{\nu\lambda}=0.
\end{align*}
This proves that the mixed-order solution satisfies Equations~\eqref{BDNK_EoM}.

Conversely, let $(\theta,u^i,n)$ solve Equations~\eqref{BDNK_EoM}, with $u^0$ given by Equation~\eqref{u0}. Take $U^A$ from Definition~\ref{def:U^A}, set $J^0=nu^0$, and take $T^{0\mu}$ from Equation~\eqref{BDNKTmunu}. Particle conservation gives Equation~\eqref{E:MO_Equivalence_density_time}, so the constitutive tensor has the form \eqref{E:MO_Equivalence_constitutive}. Equation~\eqref{E:MO_Equivalence_derivatives} then gives Equation~\eqref{Tmunu_Mixed} with $Y^A=\partial_tU^A$. Its temporal components are \eqref{E:MO_Equivalence_temporal_tensor}; multiplication by $\tilde{P}^{A\lambda}M^{-1}_{\lambda\mu}$ recovers Equation~\eqref{FO:YA}. Its spatial components give the formula for $T^{ij}$ in Equations~\eqref{MO:Transformations}. Finally, the coordinate expressions of $\nabla_\nu T^{\mu\nu}=0$ and $\nabla_\mu(nu^\mu)=0$ give Equations~\eqref{MO:Eqs}.

For the initial-value problems, the mixed-order data are obtained from the original data by setting $J^0|_{t=0}=(nu^0)|_{t=0}$ and evaluating $T^{0\mu}|_{t=0}$ from the constitutive tensor. Conversely, Equation~\eqref{FO:YA} determines $\partial_tU^A|_{t=0}$ from the mixed-order data, and Equation~\eqref{E:MO_Equivalence_density_time} determines $\partial_tn|_{t=0}$. No independent spatial-derivative variables are introduced, so no additional spatial constraints need to be propagated.

\newpage
\draftpart{Mathematical results}
\label{Part:Math}

\section{Proof of Theorem \ref{Theorem_1}}
\label{Sec:LWP}

In this Section, we prove Theorem \ref{Theorem_1}, which we state precisely below.

For the proof of the next theorem, we recall that we work in coordinates for which $\partial_t$ is timelike and the hypersurfaces $t=\mathrm{constant}$ are spacelike. Consequently, the coordinate vector fields $\partial_i$ are spacelike and tangent to these hypersurfaces, and $g^{00}=g^{-1}(\mathrm{d}t,\mathrm{d}t)<0$.

\begin{theorem}
\label{Theorem_1_precise}

Let $\Sigma=\mathbb{T}^3$,  let $\ell\geq4$ be an integer, and let $T_g>0$. Let $g_{\mu\nu}$ be a fixed Lorentzian metric whose components are real-analytic functions of $(t,x)$ on an open neighborhood of $[0,T_g]\times\Sigma$. Assume that the functions
\begin{align*}
P,
\quad
T,
\quad
\mu,
\quad
\eta,
\quad
\zeta,
\quad
\sigma,
\quad
\tau_\varepsilon,
\quad
\tau_P,
\quad
\tau_Q
\end{align*}
are real-analytic functions of $(\varepsilon,n)$ on an open set $\mathcal{O}\subset\mathbb{R}^2$. Suppose that Assumption \ref{Assumption_velocities} holds.

For
\begin{align*}
V
&=
\big(T^{0\mu},U^A,J^0\big),
\\
\mathcal{Z}_{\mathrm{FO}}[V]
&=
\big(T^{0\mu},U^A,J^0,\partial_\alpha U^A,
\partial_\alpha J^0\big),
\end{align*}
where $\mu,\alpha=0,1,2,3$ and $A=\theta,1,2,3$, suppose that there exists an open set $\mathcal{U}_{\mathrm{FO}}$ in the finite-dimensional range of $\mathcal{Z}_{\mathrm{FO}}$ with the following properties. For every $z\in\mathcal{U}_{\mathrm{FO}}$ and every $(t,x)$ in an open neighborhood of $[0,T_g]\times\Sigma$, the quantities defined in Equations~\eqref{u0}, \eqref{Tmunu_Mixed}, and \eqref{MO:Transformations} are well-defined, $(\varepsilon,n)\in\mathcal{O}$, and Assumptions \ref{Assumption_LWP} and \ref{Assumption_physics} hold. Assume also that every coefficient and every right-hand side in Equations~\eqref{MO:LWP-Eqs}, after the substitutions in Equations~\eqref{MO:Transformations}, is a real-analytic function of $((t,x),z)$ on this set.

Prescribe real initial data for Equations~\eqref{Complete_set_Eq} satisfying
\begin{subequations}{\label{E:FO_CD_complete_initial_data}}
\begin{align}
T^{0\mu}_{(0)},X_{i,(0)}^A,Z_{i,(0)}
&\in
H^{\ell+7}(\Sigma),
\\
U^A_{(0)},n_{(0)}
&\in
H^{\ell+8}(\Sigma),
\\
C_{i,(0)}^A
&:=
X_{i,(0)}^A-\partial_iU^A_{(0)},
\\
D_{i,(0)}
&:=
Z_{i,(0)}-\partial_i n_{(0)},
\\
\mathcal{W}_{(0)}
&:=
\big(
C_{(0)},\bm{\varphi}_{(0)},\phi_{(0)},
D_{(0)},\bm{\Psi}_{(0)},\psi_{(0)}
\big)
\in
\qquad H^{\ell+7}(\Sigma).
\end{align}
\end{subequations}

The preceding regularity assumptions imply that $C_{(0)},D_{(0)}\in H^{\ell+7}(\Sigma)$. Thus, the last equation in Equations~\eqref{E:FO_CD_complete_initial_data} imposes only the additional conditions
\begin{align*}
\bm{\varphi}_{(0)},\phi_{(0)},\bm{\Psi}_{(0)},\psi_{(0)}
&\in
H^{\ell+7}(\Sigma).
\end{align*}

Set
\begin{align*}
u^0_{(0)}
&=
u^0(U_{(0)}),
\\
J^0_{(0)}
&=
n_{(0)}u^0_{(0)},
\\
J^i_{(0)}
&=
n_{(0)}u^i_{(0)}.
\end{align*}
Define the first time traces, which are not independent initial data, by
\begin{subequations}{\label{E:FO_CD_theorem_initial_traces}}
\begin{align}
U^A_{(1)}
&=
Y_{\mathrm{MO},(0)}^A
+\mathcal{L}^A[C_{(0)},D_{(0)}],
\\
J^0_{(1)}
&=
-\partial_iJ^i_{(0)}
-\Gamma^\nu_{\nu\mu}n_{(0)}u^\mu_{(0)}
-u^i_{(0)}D_{i,(0)}
-n_{(0)}C_{i,(0)}^i,
\end{align}
\end{subequations}
where $Y_{\mathrm{MO},(0)}^A$ is obtained by evaluating the third equation in Equations~\eqref{MO:Transformations} on the initial data. Suppose that there exists a compact set $\mathcal{K}\Subset\mathcal{U}_{\mathrm{FO}}$ such that
\begin{align}
\label{E:FO_CD_theorem_initial_containment}
\overline{
\mathcal{Z}_{C,D,0}(\Sigma)
}
&\subset
\mathcal{K},
\end{align}
where $\mathcal{Z}_{C,D,0}$ is the function defined in Equation~\eqref{E:FO_CD_modified_initial_jet} using the traces in Equations~\eqref{E:FO_CD_theorem_initial_traces}.

Then there exist $T\in(0,T_g]$ and a solution
\begin{align*}
\big(
T^{0\mu},X_i^A,Z_i,U^A,n,
\varphi_i^A,\phi^A,\Psi_i,\psi
\big)
\end{align*}
of Equations~\eqref{Complete_set_Eq} with the prescribed initial data. If
\begin{align*}
J^0
&=
nu^0(U),
\\
C_i^A
&=
X_i^A-\partial_iU^A,
\\
D_i
&=
Z_i-\partial_i n,
\\
\mathcal{W}
&=
\big(C,\bm{\varphi},\phi,D,\bm{\Psi},\psi\big),
\end{align*}
then the solution satisfies

\begin{subequations}{\label{E:FO_CD_theorem_solution_regularity}}
\begin{align}
\partial_t^jT^{0\mu}
&\in
L^\infty\big([0,T];H^{\ell+6-j}(\Sigma)\big),
&&
0\leq j\leq7,
\\
\partial_t^jU^A,
\ \partial_t^jn,
\ \partial_t^jJ^0
&\in
L^\infty\big([0,T];H^{\ell+7-j}(\Sigma)\big),
&&
0\leq j\leq7,
\\
\partial_t^jX_i^A,
\ \partial_t^jZ_i
&\in
L^\infty\big([0,T];H^{\ell+6-j}(\Sigma)\big),
&&
0\leq j\leq7,
\\
\partial_t^j\mathcal{W}
&\in
C\big([0,T];H^{\ell+7-j}(\Sigma)\big),
&&
0\leq j\leq7.
\end{align}
\end{subequations}
For $0\leq j\leq7$, the solution also satisfies

\begin{subequations}{\label{E:FO_CD_theorem_time_continuity}}
\begin{align}
\partial_t^jT^{0\mu}
&\in
C\big([0,T];H^s(\Sigma)\big)
&&
\text{for every }s<\ell+6-j,
\\
\partial_t^jU^A,
\ \partial_t^jn,
\ \partial_t^jJ^0
&\in
C\big([0,T];H^s(\Sigma)\big)
&&
\text{for every }s<\ell+7-j,
\\
\partial_t^jX_i^A,
\ \partial_t^jZ_i
&\in
C\big([0,T];H^s(\Sigma)\big)
&&
\text{for every }s<\ell+6-j.
\end{align}
\end{subequations}
Moreover, $T$ may be chosen so that there exists a compact set $\mathcal{K}_1\Subset\mathcal{U}_{\mathrm{FO}}$ for which
\begin{align*}
\mathcal{K}
&\subset
\operatorname{int}\mathcal{K}_1,
\\
\overline{
\mathcal{Z}_{\mathrm{FO}}[V]
\big([0,T]\times\Sigma\big)
}
&\subset
\mathcal{K}_1.
\end{align*}

The solution is unique among solutions satisfying Equations~\eqref{E:FO_CD_theorem_solution_regularity} and \eqref{E:FO_CD_theorem_time_continuity} whose physical jets $\mathcal{Z}_{\mathrm{FO}}[V]$ take values in $\mathcal{U}_{\mathrm{FO}}$.

Let
\begin{align*}
\big(
T^{0\mu}_{(0),q},
X_{i,(0),q}^A,
Z_{i,(0),q},
U^A_{(0),q},
n_{(0),q},
\mathcal{W}_{(0),q}
\big),
\qquad
q\in\mathbb{N},
\end{align*}
be a sequence of initial data satisfying the hypotheses of this theorem, whose initial physical jets take values in $\mathcal{K}_1$, and suppose that
\begin{align*}
T^{0\mu}_{(0),q}
&\rightarrow
T^{0\mu}_{(0)}
\quad\text{in }H^{\ell+7}(\Sigma),
\\
X_{i,(0),q}^A
&\rightarrow
X_{i,(0)}^A
\quad\text{in }H^{\ell+7}(\Sigma),
\\
Z_{i,(0),q}
&\rightarrow
Z_{i,(0)}
\quad\text{in }H^{\ell+7}(\Sigma),
\\
U^A_{(0),q}
&\rightarrow
U^A_{(0)}
\quad\text{in }H^{\ell+8}(\Sigma),
\\
n_{(0),q}
&\rightarrow
n_{(0)}
\quad\text{in }H^{\ell+8}(\Sigma),
\\
\mathcal{W}_{(0),q}
&\rightarrow
\mathcal{W}_{(0)}
\quad\text{in }H^{\ell+7}(\Sigma).
\end{align*}
After decreasing $T$ if necessary, the corresponding solutions are defined on $[0,T]\times\Sigma$ for every sufficiently large $q$ and, for $0\leq j\leq6$, satisfy
\begin{align*}
\partial_t^jT_q^{0\mu}
&\rightarrow
\partial_t^jT^{0\mu}
\quad\text{in }C\big([0,T];H^s(\Sigma)\big)
&&
\text{for every }s<\ell+6-j,
\\
\partial_t^jU_q^A
&\rightarrow
\partial_t^jU^A
\quad\text{in }C\big([0,T];H^s(\Sigma)\big)
&&
\text{for every }s<\ell+7-j,
\\
\partial_t^jn_q
&\rightarrow
\partial_t^jn
\quad\text{in }C\big([0,T];H^s(\Sigma)\big)
&&
\text{for every }s<\ell+7-j,
\\
\partial_t^jJ_q^0
&\rightarrow
\partial_t^jJ^0
\quad\text{in }C\big([0,T];H^s(\Sigma)\big)
&&
\text{for every }s<\ell+7-j,
\\
\partial_t^jX_{i,q}^A
&\rightarrow
\partial_t^jX_i^A
\quad\text{in }C\big([0,T];H^s(\Sigma)\big)
&&
\text{for every }s<\ell+6-j,
\\
\partial_t^jZ_{i,q}
&\rightarrow
\partial_t^jZ_i
\quad\text{in }C\big([0,T];H^s(\Sigma)\big)
&&
\text{for every }s<\ell+6-j.
\end{align*}
In addition, for $0\leq j\leq7$,
\begin{align*}
\partial_t^j\mathcal{W}_q
&\rightarrow
\partial_t^j\mathcal{W}
\quad\text{in }C\big([0,T];H^{\ell+7-j}(\Sigma)\big).
\end{align*}

\end{theorem}

\begin{remark}
The assumption in Theorem \ref{Theorem_1_precise} that the components of the fixed metric are real-analytic functions of $(t,x)$ is a hypothesis for the prescribed-background problem considered in that theorem. It does not imply that a corresponding result for the coupled Einstein--BDNK system would be restricted to real-analytic Cauchy data. In the coupled system, the components of $g_{\mu\nu}$ would be included among the unknowns, rather than prescribed as functions of the spacetime coordinates. As explained in Remark \ref{R:Analyticity_hypothesis}, after Einstein's equations are written in wave coordinates, the coefficient functions and right-hand sides of the reduced Einstein--matter system are real-analytic functions of the relevant finite collection of unknowns and their derivatives, provided that the equation of state and the transport coefficients are real analytic on the set under consideration. Thus, for the coupled system, the real-analytic dependence required by Theorem \ref{Theorem_SH_LWP} would concern the dependence of the nonlinearities on the jet variables, not the regularity of $g_{\mu\nu}$ as a function of the spacetime coordinates. The Cauchy data themselves would remain Sobolev data; the analytic data in Equation~\eqref{E:SH_Analytic_data_approximation} are compatible approximations converging to those data in the stated Sobolev spaces. This observation concerns only the real-analytic dependence required by Theorem \ref{Theorem_SH_LWP}; all of its remaining hypotheses would have to be verified separately for the coupled system.
\end{remark}

\begin{proof}[Proof of Theorem \ref{Theorem_1_precise}]

\medskip
\noindent \underline{Step 1} (\textbf{The equivalent forced mixed-order system}).\nopagebreak[4]

For the proof, we rewrite the first-order system in Equations~\eqref{Complete_set_Eq} in mixed-order variables while retaining the constraint errors and cleaning fields allowed by Theorem \ref{Theorem_1_precise}; none of these fields is assumed to vanish. Recall that
\begin{subequations}{\label{E:FO_CD_definitions}}
\begin{align}
C_i^A
&:=X_i^A-\partial_iU^A,
\\
D_i
&:=Z_i-\partial_i n.
\end{align}
\end{subequations}
Thus,
\begin{subequations}{\label{E:FO_CD_reconstruction}}
\begin{align}
\label{E:FO_CD_reconstruction_X}
X_i^A
&=\partial_iU^A+C_i^A,
\\
\label{E:FO_CD_reconstruction_Z}
Z_i
&=\partial_i n+D_i.
\end{align}
\end{subequations}
For $C=(C_i^A)$ and $D=(D_i)$, set
\begin{subequations}{\label{E:FO_CD_corrections}}
\begin{align}
\label{E:FO_CD_Q_definition}
\mathcal{Q}^{\mu}[C,D]
&:=
H^{\mu0k\beta}P_{A\beta}C_k^A
+2\beta_n u^{(\mu}\Omega^{k0)}D_k,
\\
\label{E:FO_CD_L_definition}
\mathcal{L}^A[C,D]
&:=
-\widetilde{P}^{A\lambda}M^{-1}_{\lambda\mu}
\mathcal{Q}^{\mu}[C,D],
\\
\label{E:FO_CD_R_definition}
\mathcal{R}^{ij}[C,D]
&:=
H^{ij0\beta}P_{A\beta}\mathcal{L}^A[C,D]
+H^{ijk\beta}P_{A\beta}C_k^A
+2\beta_n u^{(i}\Omega^{kj)}D_k.
\end{align}
\end{subequations}
We also define
\begin{align}
\label{E:FO_CD_R_time_definition}
\mathcal{R}^{i0}[C,D]
&:=0.
\end{align}
The definition in Equation~\eqref{E:FO_CD_R_time_definition} is required because $T^{i0}=T^{0i}$ is an independent variable and is not reconstructed by Equation~\eqref{Relations-Tij-FO}. Every coefficient in Equation~\eqref{E:FO_CD_corrections} is evaluated at $(U^A,n)$.

Let $Y_{\mathrm{MO}}^A$ and $T_{\mathrm{MO}}^{ij}$ denote the right-hand sides of the third and fourth equations in Equations~\eqref{MO:Transformations}, respectively. Substituting Equation~\eqref{E:FO_CD_reconstruction} into Equations~\eqref{Relations} and \eqref{Relations-Tij-FO} gives
\begin{subequations}{\label{E:FO_CD_constitutive_identities}}
\begin{align}
\label{E:FO_CD_Y_identity}
Y^A
&=Y_{\mathrm{MO}}^A+\mathcal{L}^A[C,D],
\\
\label{E:FO_CD_T_identity}
T^{ij}
&=T_{\mathrm{MO}}^{ij}+\mathcal{R}^{ij}[C,D].
\end{align}
\end{subequations}
When a spacetime index is used below, we set $T_{\mathrm{MO}}^{i0}:=T^{0i}$.

Set $J^0=nu^0$ and $J^i=nu^i$. Equations~\eqref{Complete_set_Eqa}, \eqref{Complete_set_Eq_d}, and \eqref{Complete_set_Eq_e} are equivalent to
\begin{subequations}{\label{E:FO_CD_modified_mixed_order_system}}
\begin{align}
\label{E:FO_CD_modified_T_equation}
\partial_tT^{0\mu}
+\partial_i\big(T_{\mathrm{MO}}^{i\mu}+\mathcal{R}^{i\mu}[C,D]\big)
&=
-\Gamma^\nu_{\nu0}T^{0\mu}
-\Gamma^\mu_{0\nu}T^{0\nu}
\nonumber\\
&\quad
-\Gamma^\nu_{\nu i}\big(T_{\mathrm{MO}}^{i\mu}+\mathcal{R}^{i\mu}[C,D]\big)
\nonumber\\
&\quad
-\Gamma^\mu_{i\nu}\big(T_{\mathrm{MO}}^{i\nu}+\mathcal{R}^{i\nu}[C,D]\big),
\\
\label{E:FO_CD_modified_U_equation}
\partial_tU^A
&=Y_{\mathrm{MO}}^A+\mathcal{L}^A[C,D],
\\
\label{E:FO_CD_modified_J_equation}
\partial_tJ^0+\partial_iJ^i
&=-\Gamma^\nu_{\nu\mu}nu^\mu-u^iD_i-nC_i^i.
\end{align}
\end{subequations}

Although Equation~\eqref{E:FO_CD_modified_mixed_order_system} is written in divergence form, it is a mixed-order system. Indeed, the third and fourth equations in Equations~\eqref{MO:Transformations}, together with $n=J^0/u^0$, show that $T_{\mathrm{MO}}^{i\ell}$ depends on the first spatial derivatives of $U^A$ and $J^0$. Hence, $\partial_iT_{\mathrm{MO}}^{i\ell}$ contains second spatial derivatives of $U^A$ and $J^0$. The corresponding terms are displayed in the principal matrix in Equation~\eqref{E:FO_CD_principal_matrix}.

To verify Equation~\eqref{E:FO_CD_modified_J_equation}, denote by $S_{n,\mathrm{MO}}$ the expression on the right-hand side of Equation~\eqref{Sn} after the replacements $X_i^A=\partial_iU^A$ and $Z_i=\partial_i n$, with the same $Y^A=\partial_tU^A$. Equations~\eqref{Sn} and \eqref{E:FO_CD_reconstruction} give
\begin{align}
\label{E:FO_CD_Sn_difference}
S_n-S_{n,\mathrm{MO}}
&=-\frac{u^iD_i+nC_i^i}{u^0}.
\end{align}
The derivation of Equation~\eqref{Sn}, with $X_i^A=\partial_iU^A$ and $Z_i=\partial_i n$, gives
\begin{align*}
\partial_tJ^0+\partial_iJ^i
&=-\Gamma^\nu_{\nu\mu}nu^\mu
\end{align*}
when $\partial_tn=S_{n,\mathrm{MO}}$. Multiplication of Equation~\eqref{E:FO_CD_Sn_difference} by $u^0$ gives the last two terms in Equation~\eqref{E:FO_CD_modified_J_equation}. This calculation uses Equation~\eqref{Sn} without imposing any condition on the coordinate derivatives of the metric.

Equations~\eqref{Complete_set_Eqb}--\eqref{Complete_set_Eq_i} and \eqref{E:FO_CD_definitions} give
\begin{subequations}{\label{E:FO_CD_cleaning_system}}
\begin{align}
\partial_t\bm{C}^A+\nabla\times\bm{\varphi}^A
&=0,
\\
\partial_t\bm{\varphi}^A
-\kappa_1^A\nabla\times\bm{C}^A
+\nabla\phi^A
&=-\beta_1^A\bm{\varphi}^A,
\\
\partial_t\phi^A+\kappa_2^A\nabla\cdot\bm{\varphi}^A
&=-\beta_2^A\phi^A,
\\
\partial_t\bm{D}+\nabla\times\bm{\Psi}
&=0,
\\
\partial_t\bm{\Psi}
-\lambda_1\nabla\times\bm{D}
+\nabla\psi
&=-\sigma_1\bm{\Psi},
\\
\partial_t\psi+\lambda_2\nabla\cdot\bm{\Psi}
&=-\sigma_2\psi.
\end{align}
\end{subequations}
No field in Equation~\eqref{E:FO_CD_cleaning_system} is required to vanish. Conversely, suppose that Equations~\eqref{E:FO_CD_modified_mixed_order_system} and \eqref{E:FO_CD_cleaning_system} hold and define $X_i^A,Z_i$ by Equations~\eqref{E:FO_CD_reconstruction_X} and \eqref{E:FO_CD_reconstruction_Z}. The first and fourth equations in Equations~\eqref{E:FO_CD_cleaning_system} give Equations~\eqref{Complete_set_Eqb} and \eqref{Complete_set_Eqc}. Since the curl of a spatial gradient vanishes, the remaining four equations in Equations~\eqref{E:FO_CD_cleaning_system} give Equations~\eqref{Complete_set_Eq_f}--\eqref{Complete_set_Eq_i}. Equations~\eqref{E:FO_CD_constitutive_identities} and \eqref{E:FO_CD_modified_mixed_order_system} give Equations~\eqref{Complete_set_Eqa}, \eqref{Complete_set_Eq_d}, and \eqref{Complete_set_Eq_e}. Therefore, the first-order system in Equations~\eqref{Complete_set_Eq} and the coupled system consisting of Equations~\eqref{E:FO_CD_modified_mixed_order_system} and \eqref{E:FO_CD_cleaning_system} are equivalent wherever the changes of variables used in Equations~\eqref{Relations} and \eqref{MO:Transformations} are defined and invertible.

More precisely, no equation is replaced by only one of its differential consequences. Define
\begin{align*}
\mathcal{E}_U^A
&:=
\partial_tU^A-Y^A,
\\
\mathcal{E}_{X,i}^A
&:=
\partial_tX_i^A
+\epsilon_{ijk}\partial_j\varphi_k^A
-\partial_iY^A,
\\
\mathcal{E}_{C,i}^A
&:=
\partial_tC_i^A
+\epsilon_{ijk}\partial_j\varphi_k^A.
\end{align*}
The equations $\mathcal{E}_U^A=0$ and $\mathcal{E}_{X,i}^A=0$ are Equations~\eqref{Complete_set_Eq_d} and \eqref{Complete_set_Eqb}, respectively, whereas $\mathcal{E}_{C,i}^A=0$ is the equation for $\bm{C}^A$ in Equations~\eqref{E:FO_CD_cleaning_system}.
The definition of $C_i^A$ in Equation~\eqref{E:FO_CD_definitions} gives
\begin{align*}
\mathcal{E}_{C,i}^A
&=
\mathcal{E}_{X,i}^A-\partial_i\mathcal{E}_U^A,
\\
\mathcal{E}_{X,i}^A
&=
\mathcal{E}_{C,i}^A+\partial_i\mathcal{E}_U^A.
\end{align*}
By Equation~\eqref{E:FO_CD_Y_identity}, Equation~\eqref{E:FO_CD_modified_U_equation} is equivalent to $\mathcal{E}_U^A=0$. Consequently, the pair of equations
\begin{align*}
\mathcal{E}_U^A
&=0
\\
\mathcal{E}_{X,i}^A
&=0
\end{align*}
is equivalent to
\begin{align*}
\mathcal{E}_U^A
&=0
\\
\mathcal{E}_{C,i}^A
&=0.
\end{align*}

Similarly, define
\begin{align*}
\mathcal{E}_n
&:=
\partial_tn-S_n,
\\
\mathcal{E}_{Z,i}
&:=
\partial_tZ_i
+\epsilon_{ijk}\partial_j\Psi_k
-\partial_iS_n,
\\
\mathcal{E}_{D,i}
&:=
\partial_tD_i
+\epsilon_{ijk}\partial_j\Psi_k,
\\
\mathcal{E}_J
&:=
\partial_tJ^0+
\partial_iJ^i+
\Gamma^\nu_{\nu\mu}nu^\mu+
u^iD_i+
nC_i^i.
\end{align*}
The equations $\mathcal{E}_n=0$ and $\mathcal{E}_{Z,i}=0$ are Equations~\eqref{Complete_set_Eq_e} and \eqref{Complete_set_Eqc}, respectively, whereas $\mathcal{E}_{D,i}=0$ is the equation for $\bm{D}$ in Equations~\eqref{E:FO_CD_cleaning_system} and $\mathcal{E}_J=0$ is Equation~\eqref{E:FO_CD_modified_J_equation}.
Substitution of $X_i^A=\partial_iU^A$, $Z_i=\partial_i n$, and $Y^A=\partial_tU^A$ into Equation~\eqref{Sn}, together with $J^0=nu^0$, $J^i=nu^i$, and Equation~\eqref{E:FO_CD_Sn_difference}, gives
\begin{align*}
\mathcal{E}_J
&=
u^0\big(\partial_tn-S_{n,\mathrm{MO}}\big)
+u^iD_i+nC_i^i
\\
&=
u^0\mathcal{E}_n.
\end{align*}
Since $u^0\neq0$, Equation~\eqref{E:FO_CD_modified_J_equation} is equivalent to $\mathcal{E}_n=0$. The definition of $D_i$ in Equation~\eqref{E:FO_CD_definitions} gives
\begin{align*}
\mathcal{E}_{D,i}
&=
\mathcal{E}_{Z,i}-\partial_i\mathcal{E}_n,
\\
\mathcal{E}_{Z,i}
&=
\mathcal{E}_{D,i}+\partial_i\mathcal{E}_n.
\end{align*}
It follows that the pair of equations
\begin{align*}
\mathcal{E}_n
&=0
\\
\mathcal{E}_{Z,i}
&=0
\end{align*}
is equivalent to
\begin{align*}
\mathcal{E}_J
&=0
\\
\mathcal{E}_{D,i}
&=0.
\end{align*}
Moreover, Equations~\eqref{E:FO_CD_reconstruction_X} and \eqref{E:FO_CD_reconstruction_Z} give
\begin{align*}
\epsilon_{ijk}\partial_jX_k^A
&=
\epsilon_{ijk}\partial_jC_k^A,
\\
\epsilon_{ijk}\partial_jZ_k
&=
\epsilon_{ijk}\partial_jD_k.
\end{align*}
Thus, Equations~\eqref{Complete_set_Eq_f}--\eqref{Complete_set_Eq_i} agree with the equations for $\bm{\varphi}^A,\phi^A,\bm{\Psi},\psi$ in Equations~\eqref{E:FO_CD_cleaning_system} after the reconstruction in Equations~\eqref{E:FO_CD_reconstruction}. Finally, Equations~\eqref{E:FO_CD_constitutive_identities} transform Equations~\eqref{Complete_set_Eqa} and \eqref{Complete_set_Eq_d} into Equations~\eqref{E:FO_CD_modified_T_equation} and \eqref{E:FO_CD_modified_U_equation}, and conversely. Therefore, the equivalence uses algebraically invertible substitutions in the jet variables and the invertible differential transformations displayed above. No equation in Equation~\eqref{Complete_set_Eq} is replaced only by a differential consequence, and the transformation neither introduces an additional constraint nor permits a solution that does not reconstruct to a solution of Equations~\eqref{Complete_set_Eq}.

\medskip
\noindent \underline{Step 2} (\textbf{The cleaning system and its Sobolev estimates}).\nopagebreak[4]

For every integer $r\geq0$, define
\begin{align}
\label{E:FO_CD_cleaning_energy}
E_r(t)
&:=\frac{1}{2}\sum_{|\alpha|\leq r}
\int_{\mathbb{T}^3}
\left[
\sum_{A\in\{\theta,1,2,3\}}
\left(
\kappa_1^A|\partial^{\alpha}\bm{C}^A|^2
+|\partial^{\alpha}\bm{\varphi}^A|^2
+\frac{1}{\kappa_2^A}|\partial^{\alpha}\phi^A|^2
\right)
\right.
\nonumber\\
&\hspace{42mm}\left.
+\lambda_1|\partial^{\alpha}\bm{D}|^2
+|\partial^{\alpha}\bm{\Psi}|^2
+\frac{1}{\lambda_2}|\partial^{\alpha}\psi|^2
\right]\,\mathrm{d}^3x.
\end{align}
Applying $\partial^{\alpha}$ to Equations~\eqref{E:FO_CD_cleaning_system}, taking the $L^2(\mathbb{T}^3)$ inner products with the fields occurring in Equation~\eqref{E:FO_CD_cleaning_energy}, and integrating by parts gives
\begin{align}
\label{E:FO_CD_cleaning_energy_identity}
\frac{\mathrm{d}}{\mathrm{d}t}E_r(t)
&=-\sum_{|\alpha|\leq r}
\int_{\mathbb{T}^3}
\left[
\sum_{A\in\{\theta,1,2,3\}}
\left(
\beta_1^A|\partial^{\alpha}\bm{\varphi}^A|^2
+\frac{\beta_2^A}{\kappa_2^A}|\partial^{\alpha}\phi^A|^2
\right)
\right.
\nonumber\\
&\hspace{42mm}\left.
+\sigma_1|\partial^{\alpha}\bm{\Psi}|^2
+\frac{\sigma_2}{\lambda_2}|\partial^{\alpha}\psi|^2
\right]\,\mathrm{d}^3x
\leq0.
\end{align}
Indeed, the two curl terms cancel because
\begin{align*}
\int_{\mathbb{T}^3}\bm{a}\cdot(\nabla\times\bm{b})\,\mathrm{d}^3x
&=\int_{\mathbb{T}^3}\bm{b}\cdot(\nabla\times\bm{a})\,\mathrm{d}^3x,
\end{align*}
and the two gradient-divergence terms cancel because
\begin{align*}
\int_{\mathbb{T}^3}\bm{a}\cdot\nabla f\,\mathrm{d}^3x
&=-\int_{\mathbb{T}^3}f\nabla\cdot\bm{a}\,\mathrm{d}^3x.
\end{align*}
Assumption \ref{Assumption_velocities} makes all coefficients in Equation~\eqref{E:FO_CD_cleaning_energy} positive. Hence Equation~\eqref{E:FO_CD_cleaning_energy_identity} controls the $H^r$ norm of every field in Equations~\eqref{E:FO_CD_cleaning_system} by its initial $H^r$ norm.

It remains to count the derivatives of $C,D$ in Equations~\eqref{E:FO_CD_modified_mixed_order_system}. The functions $\mathcal{L}^A$ and $\mathcal{R}^{i\mu}$ in Equations~\eqref{E:FO_CD_corrections} contain no derivative of $C,D$. Moreover,
\begin{align}
\label{E:FO_CD_R_chain_rule}
\partial_i\mathcal{R}^{i\mu}
&=
\frac{\partial\mathcal{R}^{i\mu}}{\partial x^i}
+\frac{\partial\mathcal{R}^{i\mu}}{\partial U^A}\partial_iU^A
+\frac{\partial\mathcal{R}^{i\mu}}{\partial n}\partial_i n
\nonumber\\
&\quad
+\frac{\partial\mathcal{R}^{i\mu}}{\partial C_k^A}\partial_iC_k^A
+\frac{\partial\mathcal{R}^{i\mu}}{\partial D_k}\partial_iD_k,
\end{align}
where $\frac{\partial\mathcal{R}^{i\mu}}{\partial x^i}$ is taken with $(U,n,C,D)$ fixed. Thus, Equation~\eqref{E:FO_CD_modified_T_equation} contains at most one spatial derivative of $C,D$, whereas Equations~\eqref{E:FO_CD_modified_U_equation} and \eqref{E:FO_CD_modified_J_equation} contain no derivative of $C,D$.  With the Leray indices specified below Equations~\eqref{MO:LWP-Eqs}, the terms containing $C,D$ do not enter the principal matrix. The principal matrix, its characteristic polynomial, and its eigenspaces are calculated directly in Step 4.

\medskip
\noindent \underline{Step 3} (\textbf{The real-analytic approximations}).\nopagebreak[4]

We now construct the analytic approximations used below.  We work on $\mathbb{T}^3$ under the coefficient and compact-containment hypotheses stated in Theorem \ref{Theorem_1_precise}.

Let the initial data for Equations~\eqref{Complete_set_Eq} satisfy
\begin{align*}
T^{0\mu}_{(0)},X_{i,(0)}^A,Z_{i,(0)}
&\in H^{\ell+7}(\mathbb{T}^3),
\\
U^A_{(0)},n_{(0)}
&\in H^{\ell+8}(\mathbb{T}^3).
\end{align*}
Define the initial constraint errors before any approximation by
\begin{subequations}{\label{E:FO_CD_initial_constraint_errors}}
\begin{align}
C_{i,(0)}^A
&:=
X_{i,(0)}^A-\partial_iU^A_{(0)},
\\
D_{i,(0)}
&:=
Z_{i,(0)}-\partial_i n_{(0)}.
\end{align}
\end{subequations}
Assume, in addition, that
\begin{align}
\label{E:FO_CD_full_initial_regularity}
\mathcal{W}_{(0)}
:=
\big(
C_{(0)},\bm{\varphi}_{(0)},\phi_{(0)},
D_{(0)},\bm{\Psi}_{(0)},\psi_{(0)}
\big)
\in
\qquad H^{\ell+7}(\mathbb{T}^3).
\end{align}

The regularity assumed for $U^A_{(0)},n_{(0)},X_{i,(0)}^A,Z_{i,(0)}$ already implies that the fields in Equations~\eqref{E:FO_CD_initial_constraint_errors} belong to $H^{\ell+7}(\mathbb{T}^3)$. Thus, Equation~\eqref{E:FO_CD_full_initial_regularity} imposes no additional derivative on these combinations.

Set
\begin{align*}
u^0_{(0)}
&=
u^0(U_{(0)}),
\\
J^0_{(0)}
&=
n_{(0)}u^0_{(0)}.
\end{align*}
Define the first physical traces by
\begin{subequations}{\label{E:FO_CD_modified_initial_first_traces}}
\begin{align}
U^A_{(1)}
&=
Y_{\mathrm{MO},(0)}^A
+\mathcal{L}^A[C_{(0)},D_{(0)}],
\label{E:FO_CD_modified_initial_U_trace}
\\
J^0_{(1)}
&=
-\partial_iJ^i_{(0)}
-\Gamma^\nu_{\nu\mu}n_{(0)}u^\mu_{(0)}
-u^i_{(0)}D_{i,(0)}
-n_{(0)}C_{i,(0)}^i,
\label{E:FO_CD_modified_initial_J_trace}
\end{align}
\end{subequations}
where $Y_{\mathrm{MO},(0)}^A$ is obtained from the third equation in Equations~\eqref{MO:Transformations} and $J^i_{(0)}=n_{(0)}u^i_{(0)}$. Suppose that the closure of the range of
\begin{align}
\label{E:FO_CD_modified_initial_jet}
\mathcal{Z}_{C,D,0}
:=
\big(
T^{0\mu}_{(0)},
U^A_{(0)},
J^0_{(0)},
U^A_{(1)},
\partial_iU^A_{(0)},
J^0_{(1)},
\partial_iJ^0_{(0)}
\big)
\end{align}
is contained in a compact set $\mathcal{K}\Subset\mathcal{U}_{\mathrm{FO}}$.

Choose a real-valued even function $a\in C_c^\infty(\mathbb{R}^3)$ that is equal to one on a neighborhood of the origin. For $k\in\mathbb{Z}^3$, define
\begin{align}
\label{E:FO_CD_analytic_smoothing}
\widehat{\mathcal{S}_{\nu}f}(k)
&=
a\left(\frac{k}{\nu}\right)\widehat{f}(k).
\end{align}
The function $\mathcal{S}_{\nu}f$ has only finitely many nonzero Fourier coefficients and is therefore real analytic on $\mathbb{T}^3$. Since $a$ is real-valued and even, $\mathcal{S}_{\nu}f$ is real-valued whenever $f$ is real-valued. Dominated convergence gives
\begin{align}
\label{E:FO_CD_smoothing_convergence}
\mathcal{S}_{\nu}f
&\rightarrow
f
\quad\text{in }H^s(\mathbb{T}^3)
\end{align}
for every $f\in H^s(\mathbb{T}^3)$.

Define the physical initial approximations by
\begin{align*}
T^{0\mu}_{(0),\nu}
&=
\mathcal{S}_{\nu}T^{0\mu}_{(0)},
\\
U^A_{(0),\nu}
&=
\mathcal{S}_{\nu}U^A_{(0)},
\\
J^0_{(0),\nu}
&=
\mathcal{S}_{\nu}J^0_{(0)},
\end{align*}
and define all six initial fields in Equation~\eqref{E:FO_CD_full_initial_regularity} by
\begin{align}
\label{E:FO_CD_cleaning_initial_approximations}
\mathcal{W}_{(0),\nu}
&=
\mathcal{S}_{\nu}\mathcal{W}_{(0)}.
\end{align}
Here Equation~\eqref{E:FO_CD_analytic_smoothing} is applied componentwise. Set
\begin{subequations}{\label{E:FO_CD_reconstructed_analytic_data}}
\begin{align}
u^0_{(0),\nu}
&=
u^0(U_{(0),\nu}),
\\
n_{(0),\nu}
&=
\frac{J^0_{(0),\nu}}{u^0_{(0),\nu}},
\\
X_{i,(0),\nu}^A
&=
\partial_iU^A_{(0),\nu}
+C_{i,(0),\nu}^A,
\label{E:FO_CD_reconstructed_X_data}
\\
Z_{i,(0),\nu}
&=
\partial_i n_{(0),\nu}
+D_{i,(0),\nu}.
\label{E:FO_CD_reconstructed_Z_data}
\end{align}
\end{subequations}
For all sufficiently large $\nu$, the ranges of $U_{(0),\nu}$ lie in a compact subset on which $u^0$ and $\frac{1}{u^0}$ are analytic. It follows that every field in Equation~\eqref{E:FO_CD_reconstructed_analytic_data} is real analytic. Equations~\eqref{E:FO_CD_reconstructed_X_data} and \eqref{E:FO_CD_reconstructed_Z_data} give the exact identities
\begin{align*}
X_{i,(0),\nu}^A-\partial_iU^A_{(0),\nu}
&=
C_{i,(0),\nu}^A,
\\
Z_{i,(0),\nu}-\partial_i n_{(0),\nu}
&=
D_{i,(0),\nu};
\end{align*}
hence no compatibility condition setting $C_{(0),\nu}$ or $D_{(0),\nu}$ equal to zero has been imposed.

Equations~\eqref{E:FO_CD_smoothing_convergence} and \eqref{E:FO_CD_full_initial_regularity}, followed by the Sobolev product and analytic-composition estimates, give
\begin{subequations}{\label{E:FO_CD_full_data_convergence}}
\begin{align}
T^{0\mu}_{(0),\nu}
&\rightarrow
T^{0\mu}_{(0)}
\quad\text{in }H^{\ell+7},
\\
U^A_{(0),\nu}
&\rightarrow
U^A_{(0)}
\quad\text{in }H^{\ell+8},
\\
J^0_{(0),\nu}
&\rightarrow
J^0_{(0)}
\quad\text{in }H^{\ell+8},
\\
n_{(0),\nu}
&\rightarrow
n_{(0)}
\quad\text{in }H^{\ell+8},
\\
X_{i,(0),\nu}^A
&\rightarrow
X_{i,(0)}^A
\quad\text{in }H^{\ell+7},
\\
Z_{i,(0),\nu}
&\rightarrow
Z_{i,(0)}
\quad\text{in }H^{\ell+7},
\\
\mathcal{W}_{(0),\nu}
&\rightarrow
\mathcal{W}_{(0)}
\quad\text{in }
\qquad H^{\ell+7}.
\end{align}
\end{subequations}

Let $\mathcal{W}_{\nu}$ be the solution of Equation~\eqref{E:FO_CD_cleaning_system} with initial data in Equation~\eqref{E:FO_CD_cleaning_initial_approximations}. Since the system has constant coefficients and $\mathcal{W}_{(0),\nu}$ has finite Fourier support, $\mathcal{W}_{\nu}$ is real analytic in $(t,x)$. Equations~\eqref{E:FO_CD_cleaning_energy_identity} and \eqref{E:FO_CD_shifted_time_derivative_formula} give
\begin{align}
\label{E:FO_CD_cleaning_approximation_bound}
\sup_{\nu}
\sup_{0\leq t\leq T}
\sum_{b=0}^{7}
\left\|
\partial_t^b\mathcal{W}_{\nu}(t)
\right\|_{
\;H^{\ell+7-b}
}^2
&<\infty
\end{align}
for every $T>0$. Applying the same identities to
$\mathcal{W}_{\nu}-\mathcal{W}_{\mu}$ gives
\begin{align}
\label{E:FO_CD_cleaning_approximation_difference}
\sup_{0\leq t\leq T}
\sum_{b=0}^{7}
\left\|
\partial_t^b
\big(
\mathcal{W}_{\nu}-\mathcal{W}_{\mu}
\big)(t)
\right\|_{
\;H^{\ell+7-b}
}^2
&\leq
K
\left\|
\big(
\mathcal{S}_{\nu}-\mathcal{S}_{\mu}
\big)\mathcal{W}_{(0)}
\right\|_{
\;H^{\ell+7}
}^2.
\end{align}
The right-hand side tends to zero as $\nu,\mu\rightarrow\infty$ by Equation~\eqref{E:FO_CD_smoothing_convergence}.

For every sufficiently large $\nu$, define $U^A_{(1),\nu}$ and $J^0_{(1),\nu}$ by Equations~\eqref{E:FO_CD_modified_initial_first_traces}, with every field replaced by its approximating field. The convergence in Equations~\eqref{E:FO_CD_full_data_convergence}, the fact that $H^{\ell+7}(\mathbb{T}^3)$ is an algebra, and the analytic-composition estimates imply
\begin{subequations}{\label{E:FO_CD_first_trace_convergence}}
\begin{align}
U^A_{(1),\nu}
&\rightarrow
U^A_{(1)}
\quad\text{in }H^{\ell+7},
\\
J^0_{(1),\nu}
&\rightarrow
J^0_{(1)}
\quad\text{in }H^{\ell+7}.
\end{align}
\end{subequations}
Consequently, the approximating versions $\mathcal{Z}_{C,D,0,\nu}$ of Equation~\eqref{E:FO_CD_modified_initial_jet} converge uniformly to $\mathcal{Z}_{C,D,0}$. There is therefore a compact set $\mathcal{K}_0\Subset\mathcal{U}_{\mathrm{FO}}$ that contains the ranges of $\mathcal{Z}_{C,D,0,\nu}$ for every sufficiently large $\nu$.

\medskip
\noindent \underline{Step 4} (\textbf{The principal symbol and the hypotheses of Theorem \ref{Theorem_SH_LWP}}).\nopagebreak[4]

We calculate the principal symbol of Equations~\eqref{E:FO_CD_modified_mixed_order_system} directly. The Leray indices specified below Equations~\eqref{MO:LWP-Eqs} are
\begin{align*}
m_T
&=1,
\\
m_U=m_J
&=2,
\\
n_T
&=0,
\\
n_U=n_J
&=1.
\end{align*}
The terms containing $W=(C,D)$ occur only on the right-hand side and have the derivative orders established in Equation~\eqref{E:FO_CD_R_chain_rule}. With the rows and columns ordered as $(T^{00},T^{0k},U^B,J^0)$, the principal differential-operator matrix is therefore denoted by $h_{\mathrm{MO}}$. It is given by
\begin{align}
\label{E:FO_CD_principal_matrix}
h_{\mathrm{MO}}(\partial V,\partial)
&=
\bbm
\partial_t & \partial_k & 0_{1\times 4} & 0
\\
H^{\ell j 0\beta} M^{-1}_{\beta 0} \partial_j
& \delta^\ell_k\partial_t +H^{\ell j 0\beta} M^{-1}_{\beta k} \partial_j
& \mathfrak{H}^{\ell j k \beta}P_{B\beta}\partial_j\partial_k
& \mathfrak{O}^{\ell j k \beta}\partial_j\partial_k
\\
-\tilde{P}^{A\lambda}M^{-1}_{\lambda 0 }
& -\tilde{P}^{A\lambda}M^{-1}_{\lambda k}
& \delta^A_B\partial_t +\tilde{P}^{A\lambda}M^{-1}_{\lambda\mu}\tilde{H}^{\mu0 j \beta}P_{B\beta}\partial_j
& \tilde{P}^{A\lambda}M^{-1}_{\lambda\mu}\frac{2\beta_n}{u^0} u^{(\mu} \Omega^{j0)} \partial_j
\\
0
& 0_{1\times 3}
& n\left(\Delta^{j\beta}+\frac{u_i\Delta^{i\beta}u^j}{u^0u_0}\right)P_{B\beta}\partial_j
& \partial_t +\frac{u^j}{u^0}\partial_j
\ebm,
\end{align}
where
\begin{align*}
\mathfrak{H}^{\ell j k \beta}
&=
\tilde{H}^{\ell j k \beta}
-H^{\ell j 0\lambda} M^{-1}_{\lambda\mu} \tilde{H}^{\mu0 k \beta},
\\
\mathfrak{O}^{\ell j k \beta}
&=
\frac{2\beta_n}{u^0}
\left(
u^{(\ell} \Omega^{kj)}
-H^{\ell j 0\beta} M^{-1}_{\beta\mu} u^{(\mu} \Omega^{k0)}
\right).
\end{align*}
The matrix in Equation~\eqref{E:FO_CD_principal_matrix} contains first-order pure time derivatives and no pure time derivative of higher order. Thus, the integer $r$ in Definition \ref{SH:Definition 1} is one.

The change from $n$ to $J^0$ also differentiates $u^0$. At fixed metric, normalization of $u$ and Equation~\eqref{Matrices_P} give
\begin{align}
\label{E:MO_State_change_differential}
\mathrm{d}n=\frac{1}{u^0}\mathrm{d}J^0-\frac{n}{u^0}\Delta^{0\beta}P_{B\beta}\mathrm{d}U^B.
\end{align}
Since $u_\mu\Delta^{\mu\beta}=0$, the tensor already used in Equation~\eqref{E:FO_CD_principal_matrix} satisfies
\begin{align}
\label{E:MO_Tilde_H_state_change}
\tilde{H}^{\mu\nu j\beta}
=H^{\mu\nu j\beta}-\frac{2n\beta_n}{u^0}u^{(\mu}\Omega^{j\nu)}\Delta^{0\beta}.
\end{align}
These two identities account for the $U$-column in the change of variables.

We next calculate its determinant. After substituting $\partial_\mu$ by $\xi_\mu$, write
\begin{align*}
h_{\mathrm{MO}}(\partial V,\xi)
&=
\widetilde{h}_{\mathrm{MO}}(\partial V,\xi)\mathcal{C},
\end{align*}
where
\begin{align*}
\widetilde{h}_{\mathrm{MO}}(\partial V,\xi)
&=
\bbm
H^{\mu\sigma 0\beta}P_{B\beta}\xi_\sigma
& H^{\mu\sigma j \beta}P_{B\beta}\xi_\sigma \xi_j
& 2\beta_n u^{(\mu} \Omega^{j\alpha)}\xi_\alpha \xi_j
\\
-\delta^A_B
& \delta^A_B\xi_0
& 0_{4\times 1}
\\
0_{1\times 4}
& n \Delta^{\alpha\beta}P_{B\beta}\xi_\alpha
& u^\alpha\xi_\alpha
\ebm,
\\
\mathcal{C}
&=
\bbm
\tilde{P}^{A\lambda}M^{-1}_{\lambda\nu}
& -\tilde{P}^{A\lambda}M^{-1}_{\lambda\sigma} \tilde{H}^{\sigma 0 j \beta}P_{B\beta}\xi_j
& -\frac{2\beta_n}{u^0} \tilde{P}^{A\lambda}M^{-1}_{\lambda\sigma} u^{(\sigma} \Omega^{j0)} \xi_j
\\
0_{4\times 4}
& I_4
& 0_{4\times 1}
\\
0_{1\times 4}
& -\frac{n}{u^0}\Delta^{0\beta}P_{B\beta}
& \frac{1}{u^0}
\ebm.
\end{align*}
The transformations used in Equations~\eqref{Relations} and \eqref{MO:Transformations} are invertible on $\mathcal{U}_{\mathrm{FO}}$, and
\begin{align*}
\det[\mathcal{C}]
&=
\frac{1}{u^0}
\det[\tilde{P}^{A\lambda}]
\det[M^{-1}_{\lambda\nu}]
\neq0.
\end{align*}

Set
\begin{align*}
a
&=
u^\alpha\xi_\alpha,
\\
v^\mu
&=
\Delta^{\mu\alpha}\xi_\alpha,
\\
v_\beta
&=
\Delta^\alpha_\beta\xi_\alpha,
\\
q
&=
v^\alpha v_\alpha,
\\
L^\mu_\beta
&=
\xi_\alpha\xi_\nu\tensor{H}{^\mu^\nu^\alpha_\beta},
\\
Q^\mu
&=
2\beta_nu^{(\mu}\Omega^{k\alpha)}\xi_k\xi_\alpha.
\end{align*}
Adding $H^{\mu\sigma0\beta}P_{B\beta}\xi_\sigma$ times the second block row of $\widetilde{h}_{\mathrm{MO}}$ to its first block row and taking the Schur complement of its final scalar entry gives, when $a\neq0$,
\begin{align}
\label{E:FO_CD_determinant_reduction}
\det[h_{\mathrm{MO}}(\partial V,\xi)]
&=
\frac{a}{u^0}
\det[(M^\mu_\nu)^{-1}]
\det[K^\mu_\beta],
\end{align}
where
\begin{align*}
K^\mu_\beta
&=
L^\mu_\beta-\frac{n}{a}Q^\mu v_\beta.
\end{align*}
Substitution of Equations~\eqref{defineBDNKnew} and \eqref{Omega} gives
\begin{align}
\label{E:FO_CD_K_rank_two_form}
K^\mu_\beta
&=
\mathbb{A}\delta^\mu_\beta
+W_1^\mu u_\beta
+W_2^\mu v_\beta,
\end{align}
where
\begin{align*}
\mathbb{A}
&=
\rho\tau_Qa^2-\eta q,
\\
W_1^\mu
&=
\left[
\rho\tau_Qa^2
-\varepsilon\beta_\varepsilon q
-\varepsilon\tau_\varepsilon a^2
-\eta q
\right]u^\mu
-\varepsilon(\tau_P+\beta_\varepsilon)av^\mu,
\\
W_2^\mu
&=
\left[
\rho(\tau_\varepsilon+\tau_Q)a
-\frac{n\beta_nq}{a}
\right]u^\mu
\\
&\quad
+\left[
\rho\tau_P-\zeta-\frac{\eta}{3}-n\beta_n
\right]v^\mu.
\end{align*}
For $\mathbb{A}\neq0$, the determinant formula for the rank-two perturbation in Equation~\eqref{E:FO_CD_K_rank_two_form} gives
\begin{align*}
\det[K^\mu_\beta]
&=
\mathbb{A}^2
\left[
\mathbb{A}^2
+\mathbb{A}
\big(W_1^\mu u_\mu+W_2^\mu v_\mu\big)
\right.
\\
&\qquad\left.
+\big(W_1^\mu u_\mu\big)
\big(W_2^\nu v_\nu\big)
-\big(W_2^\mu u_\mu\big)
\big(W_1^\nu v_\nu\big)
\right].
\end{align*}
Both sides of this identity are polynomials in the entries of Equation~\eqref{E:FO_CD_K_rank_two_form}; hence the identity also holds when $\mathbb{A}=0$.
To evaluate the expression in brackets, set
\begin{align*}
z
&=
a^2,
\\
d_0
&=
\zeta+\frac{4\eta}{3},
\\
b_0
&=
\rho\beta_\varepsilon+n\beta_n.
\end{align*}
Equations~\eqref{def:beta_epsilon}, \eqref{def:beta_n}, \eqref{soundspeed}, and \eqref{kappa_s_def} give
\begin{align*}
b_0
&=
\rho c_s^2\tau_Q+\sigma\kappa_s.
\end{align*}
Using $u^\mu u_\mu=-1$ and $u^\mu v_\mu=0$, the four contractions in the determinant are
\begin{align*}
W_1^\mu u_\mu
&=
\big(-\rho\tau_Q+\varepsilon\tau_\varepsilon\big)z
+\big(\varepsilon\beta_\varepsilon+\eta\big)q,
\\
W_1^\mu v_\mu
&=
-\varepsilon(\tau_P+\beta_\varepsilon)aq,
\\
W_2^\mu u_\mu
&=
-\rho(\tau_\varepsilon+\tau_Q)a
+\frac{n\beta_nq}{a},
\\
W_2^\mu v_\mu
&=
\left(
\rho\tau_P-\zeta-\frac{\eta}{3}-n\beta_n
\right)q.
\end{align*}
Substitution of these four identities and $\mathbb{A}=\rho\tau_Qz-\eta q$ gives
\begin{align*}
&\mathbb{A}^2
+\mathbb{A}
\big(W_1^\mu u_\mu+W_2^\mu v_\mu\big)
+\big(W_1^\mu u_\mu\big)
\big(W_2^\nu v_\nu\big)
-\big(W_2^\mu u_\mu\big)
\big(W_1^\nu v_\nu\big)
\\
&\qquad
=
\varepsilon\rho\tau_Q\tau_\varepsilon z^2
-\varepsilon
\left[
\tau_\varepsilon(b_0+d_0)+\rho\tau_P\tau_Q
\right]zq
+\varepsilon
\left[
\tau_Pb_0-\beta_\varepsilon d_0
\right]q^2.
\end{align*}
The definitions in Equations~\eqref{Characteristics} imply
\begin{align*}
c_+^2+c_-^2
&=
\frac{
\tau_\varepsilon(b_0+d_0)+\rho\tau_P\tau_Q
}{
\rho\tau_Q\tau_\varepsilon
},
\\
c_+^2c_-^2
&=
\frac{
\tau_Pb_0-\beta_\varepsilon d_0
}{
\rho\tau_Q\tau_\varepsilon
}.
\end{align*}
Since $c_1^2=\eta/(\rho\tau_Q)$, Equations~\eqref{E:FO_CD_determinant_reduction}--\eqref{E:FO_CD_K_rank_two_form} therefore give
\begin{align}
\label{E:FO_CD_characteristic_factorization}
\det[h_{\mathrm{MO}}(\partial V,\xi)]
&=
\frac{1}{u^0}
\det[(M^\mu_\nu)^{-1}]
(u^\alpha\xi_\alpha)
\varepsilon\tau_\varepsilon(\rho\tau_Q)^3
\nonumber\\
&\quad\times
\left[
(u^\mu\xi_\mu)^2
-c_1^2\Delta^{\mu\nu}\xi_\mu\xi_\nu
\right]^2
\nonumber\\
&\quad\times
\left[
(u^\mu\xi_\mu)^2
-c_+^2\Delta^{\mu\nu}\xi_\mu\xi_\nu
\right]
\nonumber\\
&\quad\times
\left[
(u^\mu\xi_\mu)^2
-c_-^2\Delta^{\mu\nu}\xi_\mu\xi_\nu
\right].
\end{align}
Both sides of Equation~\eqref{E:FO_CD_characteristic_factorization} are polynomials in $\xi_\mu$ and agree when $a\neq0$; hence they agree for every $\xi_\mu$.

The coefficient $h_1$ of $\xi_0$ in Equation~\eqref{E:FO_CD_principal_matrix} is
\begin{align*}
h_1
&=
I_9.
\end{align*}
Thus, $\det[h_1]=1$. It remains to determine the roots and the eigenspaces. For $b\in\{1,+,-\}$, set
\begin{align*}
G_b^{\mu\nu}
&=
u^\mu u^\nu-c_b^2\Delta^{\mu\nu},
\\
A_b
&=
G_b^{00},
\\
B_b
&=
G_b^{0i}\xi_i,
\\
C_b
&=
G_b^{ij}\xi_i\xi_j.
\end{align*}

Fix a spacetime point. Since $g_{\mu\nu}u^\mu u^\nu=-1$, there exists a $g$-orthonormal basis of the tangent space at this point whose timelike basis vector is $u$. Relative to this basis and its dual basis,
\begin{align*}
u^{\widehat{\mu}}
&=
(1,0,0,0),
\\
G_b^{\widehat{\mu}\widehat{\nu}}
&=
\operatorname{diag}\big(1,-c_b^2,-c_b^2,-c_b^2\big).
\end{align*}
Thus, $G_b^{\widehat{\mu}\widehat{\nu}}$ has one positive and three negative eigenvalues. This basis is used only to determine the inertia of $G_b^{\mu\nu}$ at the chosen point; no coordinate system in a neighborhood of that point is being changed. By Sylvester's law of inertia, an invertible change of basis preserves the numbers of positive and negative eigenvalues of the matrix representing a symmetric bilinear form. Therefore, $G_b^{\mu\nu}$ has one positive and three negative eigenvalues in the original coordinate basis. From this point onward, we again use the original coordinates. In these coordinates,
\begin{align*}
A_b
&=
(1-c_b^2)(u^0)^2-c_b^2g^{00}
>0
\end{align*}
by Assumption \ref{Assumption_LWP}. Sylvester's law of inertia applied to the Schur complement
\begin{align*}
S_b^{ij}
&=
G_b^{ij}-\frac{G_b^{i0}G_b^{0j}}{A_b}
\end{align*}
shows that $S_b^{ij}$ is negative definite. Hence, for every $\bm{\xi}\neq0$,
\begin{align*}
\mathcal{Z}_b
&:=
B_b^2-A_bC_b
=
-A_bS_b^{ij}\xi_i\xi_j
>0.
\end{align*}
The two roots of $G_b^{\mu\nu}\xi_\mu\xi_\nu$ are therefore $\xi_0=-\Lambda_b^\pm$, where
\begin{align*}
\Lambda_b^\pm
&=
\frac{B_b\pm\sqrt{\mathcal{Z}_b}}{A_b}.
\end{align*}
They are real and distinct. If a root associated with $c_b$ were also a root associated with $c_{b'}$, where $b\neq b'$, subtraction would give
\begin{align*}
\big(c_b^2-c_{b'}^2\big)
\Delta^{\mu\nu}\xi_\mu\xi_\nu
&=0.
\end{align*}
At any root of a quadratic factor in Equation~\eqref{E:FO_CD_characteristic_factorization}, $\Delta^{\mu\nu}\xi_\mu\xi_\nu>0$ when $\bm{\xi}\neq0$; otherwise that factor would also give $u^\mu\xi_\mu=0$, and
\begin{align*}
\xi_\mu
&=
-\big(u^\alpha\xi_\alpha\big)u_\mu
+\Delta_\mu^\alpha\xi_\alpha
\end{align*}
would give $\xi_\mu=0$. Since the three numbers $c_1,c_+,c_-$ are distinct by Assumption \ref{Assumption_LWP}, roots belonging to different quadratic factors cannot coincide. The remaining root, $u^\mu\xi_\mu=0$, cannot coincide with a root of a quadratic factor for the same reason. Thus, Equation~\eqref{E:FO_CD_characteristic_factorization} has seven distinct real roots: the two roots associated with $c_1$ have algebraic multiplicity two, and the other five roots have algebraic multiplicity one. Its degree is nine, which is $Nr$ for $N=9$ and $r=1$. This verifies Assumption \ref{SH:Assumption 1}.

We finally verify diagonalizability. Write
\begin{align*}
\widetilde{h}_{\mathrm{MO}}
&=
\widetilde{h}_1\xi_0+\widetilde{A}.
\end{align*}
Since $\mathcal{C}$ is invertible and independent of $\xi_0$,
\begin{align*}
h_1^{-1}A
&=
\mathcal{C}^{-1}\widetilde{h}_1^{-1}\widetilde{A}\mathcal{C}.
\end{align*}
Fix one of the roots $\xi_0=-\Lambda_1^\pm$ and set
\begin{align*}
a
&=
u^\mu\xi_\mu,
\\
v^\mu
&=
\Delta^{\mu\nu}\xi_\nu.
\end{align*}
The root relation is
\begin{align}
\label{E:FO_CD_c1_root_relation}
a^2
&=
c_1^2v^\mu v_\mu.
\end{align}
For $\bm{\xi}\neq0$, both $a$ and $v^\mu$ are nonzero. The covectors $w_\nu$ satisfying
\begin{align*}
u^\nu w_\nu
&=0,
\\
v^\nu w_\nu
&=0
\end{align*}
form a two-dimensional vector space. Choose a basis $w_{1,\nu}^\pm,w_{2,\nu}^\pm$ and define
\begin{align*}
y_s^A
&=
w_{s,\nu}^\pm\tilde{P}^{A\nu},
\\
\widetilde{V}_{1,s}^\pm
&=
\bbm
-\Lambda_1^\pm y_s^A
\\
y_s^A
\\
0
\ebm,
\qquad
s=1,2.
\end{align*}
The second block row of $\widetilde{h}_{\mathrm{MO}}(\partial V,\xi)\widetilde{V}_{1,s}^\pm$ vanishes because $\xi_0=-\Lambda_1^\pm$, and the third block row vanishes because $v^\nu w_{s,\nu}^\pm=0$. Equation~\eqref{E:FO_CD_c1_root_relation} gives $\mathbb{A}=0$. Since $u^\nu w_{s,\nu}^\pm=v^\nu w_{s,\nu}^\pm=0$, Equation~\eqref{E:FO_CD_K_rank_two_form} gives
\begin{align*}
K^\mu_\beta w_s^{\pm,\beta}
&=0.
\end{align*}
The definition of $K^\mu_\beta$ and $v^\nu w_{s,\nu}^\pm=0$ then give
\begin{align*}
\xi_\alpha\xi_\nu H^{\mu\nu\alpha\beta}w_{s,\beta}^\pm
&=0,
\end{align*}
which is precisely the vanishing of the first block row. Thus, $\widetilde{V}_{1,1}^\pm$ and $\widetilde{V}_{1,2}^\pm$ are linearly independent eigenvectors for $\Lambda_1^\pm$. Each of the five remaining eigenvalues has algebraic multiplicity one by Equation~\eqref{E:FO_CD_characteristic_factorization} and therefore has a one-dimensional eigenspace. The seven eigenspaces consequently have total dimension
\begin{align*}
2+2+1+1+1+1+1
&=9.
\end{align*}
Hence $h_1^{-1}A$ is diagonalizable for every $\bm{\xi}\neq0$, which verifies Assumption \ref{SH:Assumption_polynomials}. It has $m=7$ distinct eigenvalue functions, and the characteristic polynomial has $7=mr$ distinct roots. This verifies Assumption \ref{SH:Assumption 2}. Moreover,
\begin{align*}
mr
&=7
\geq
\max_I m_I-\min_I n_I+1
=3,
\end{align*}
which verifies Assumption \ref{SH:Assumption3}. Thus,
\begin{align*}
p
&=mr
=7,
\\
m_*
&=
\min_I m_I
=1.
\end{align*}
The hypotheses of Theorem \ref{Theorem_1_precise} impose Assumption \ref{Assumption_LWP} at every state in $\mathcal{U}_{\mathrm{FO}}$, so the same values $r=1$, $m=7$, and $p=7$ apply throughout that set.

\medskip
\noindent \underline{Step 5} (\textbf{The application of Theorem \ref{Theorem_SH_LWP}}).\nopagebreak[4]

We now apply Theorem \ref{Theorem_SH_LWP} directly to Equations~\eqref{E:FO_CD_modified_mixed_order_system}, with
\begin{align*}
V
&=
\big(T^{0\mu},U^A,J^0\big)
\end{align*}
as the unknown vector and
\begin{align*}
W
&=
\big(C,D\big)
\end{align*}
as the prescribed function. We use the integer $\ell-1$ in place of the integer denoted by $\ell$ in Theorem \ref{Theorem_SH_LWP}. This is permitted because $\ell\geq4$ and the spatial dimension is $d=3$.

We first verify the regularity required of $W$. The Fourier coefficients of Equations~\eqref{E:FO_CD_cleaning_system} satisfy a finite-dimensional constant-coefficient system of ordinary differential equations for each spatial frequency. Thus, for the initial value $\mathcal{W}_{(0)}$ in Equation~\eqref{E:FO_CD_full_initial_regularity}, this system has a unique solution
\begin{align*}
\mathcal{W}
&=
\big(C,\bm{\varphi},\phi,D,\bm{\Psi},\psi\big)
\in
C\big([0,T_g];H^{\ell+7}(\mathbb{T}^3)\big).
\end{align*}
Since Equation~\eqref{E:FO_CD_cleaning_system} is first order with constant coefficients, for every integer $a\geq0$ there exists a constant-coefficient spatial differential operator $\mathcal{A}_a$ of order at most $a$ such that
\begin{align}
\label{E:FO_CD_shifted_time_derivative_formula}
\partial_t^a\mathcal{W}
&=
\mathcal{A}_a\mathcal{W}.
\end{align}
Indeed, Equation~\eqref{E:FO_CD_cleaning_system} gives Equation~\eqref{E:FO_CD_shifted_time_derivative_formula} for $a=1$, and differentiation with respect to $t$, followed by another use of Equations~\eqref{E:FO_CD_cleaning_system}, proves the assertion by induction. Equations~\eqref{E:FO_CD_shifted_time_derivative_formula} and \eqref{E:FO_CD_cleaning_energy_identity}, with $r=\ell+7$, give
\begin{align}
\label{E:FO_CD_shifted_cleaning_regularity}
\sup_{0\leq t\leq T_g}
\sum_{a=0}^{7}
\left\|
\partial_t^a\mathcal{W}(t)
\right\|_{H^{\ell+7-a}}^2
&\leq
K
\left\|
\mathcal{W}_{(0)}
\right\|_{H^{\ell+7}}^2,
\end{align}
and, for every $0\leq a\leq7$, the corresponding function on the left-hand side belongs to
\begin{align*}
C\big([0,T_g];H^{\ell+7-a}(\mathbb{T}^3)\big).
\end{align*}

For the prescribed function $W=(C,D)$, the $T^{0\mu}$-rows of Equations~\eqref{E:FO_CD_modified_mixed_order_system} contain at most one derivative of $W$, by Equation~\eqref{E:FO_CD_R_chain_rule}, whereas the $U^A$- and $J^0$-rows contain no derivative of $W$, by Equations~\eqref{E:FO_CD_modified_U_equation} and \eqref{E:FO_CD_modified_J_equation}. Hence the integers $\chi_I$ in Theorem \ref{Theorem_SH_LWP} are
\begin{align*}
\chi_T
&=1,
\\
\chi_U
&=0,
\\
\chi_J
&=0.
\end{align*}
The Leray indices specified below Equations~\eqref{MO:LWP-Eqs} give
\begin{align*}
m_T
&=1,
\\
m_U=m_J
&=2,
\\
n_T
&=0,
\\
n_U=n_J
&=1.
\end{align*}
 Step 4 gives $p=7$ and $m_*=1$. Consequently, the indices in Equation~\eqref{E:SH_Forcing_indices}, with the integer in Theorem \ref{Theorem_SH_LWP} replaced by $\ell-1$, are
\begin{subequations}{\label{E:FO_CD_shifted_forcing_indices}}
\begin{align}
\chi_*
&=
\max\{n_T+\chi_T,n_U+\chi_U,n_J+\chi_J\}
=1,
\\
d_W
&=
p-m_*+\chi_*
=7,
\\
S_W
&=
(\ell-1)+p-m_*+\chi_*+1
=\ell+7.
\end{align}
\end{subequations}
Thus, Equation~\eqref{E:FO_CD_shifted_cleaning_regularity} is exactly the regularity required in Equation~\eqref{E:SH_Forcing_regularity}.

For completeness, the derivative count in the reduced equations is as follows. The application of the $(I,J)$-entry of $\widehat{H}$ to the $J$th right-hand side produces at most
\begin{align*}
p-m_I+n_J+\chi_J
\end{align*}
spacetime derivatives of $W$. If $a$ of these derivatives are time derivatives, then estimating the resulting term in the space $H^{S_I-p+1}$ requires
\begin{align}
\label{E:FO_CD_shifted_forcing_count}
S_I-p+1+p-m_I+n_J+\chi_J-a
&=
(\ell-1)+p-m_*+n_J+\chi_J+1-a
\nonumber
\\
&\leq
(\ell-1)+p-m_*+\chi_*+1-a
\nonumber
\\
&=
\ell+7-a
\end{align}
spatial Sobolev derivatives of $W$. Every such norm is bounded by Equation~\eqref{E:FO_CD_shifted_cleaning_regularity}. This proves directly, without an additional derivative assumption on $\mathcal{W}_{(0)}$, that the forcing terms satisfy the derivative requirement in Equation~\eqref{E:SH_Forcing_reduced_derivative_count}.

The functions $\mathcal{L}^A$ and $\mathcal{R}^{i\mu}$ in Equations~\eqref{E:FO_CD_corrections} are linear in $W$ and have coefficients that are real-analytic functions of the physical jet on $\mathcal{U}_{\mathrm{FO}}$. Equation~\eqref{E:FO_CD_R_chain_rule} shows that the right-hand sides in Equations~\eqref{E:FO_CD_modified_mixed_order_system} are real-analytic functions of the physical jet, $W$, and its first derivatives.  The same equation shows that the terms containing $W$ do not change the principal matrix in Equation~\eqref{E:FO_CD_principal_matrix}. Step 4 verifies directly that this matrix satisfies Assumptions \ref{SH:Assumption 1}, \ref{SH:Assumption 2}, \ref{SH:Assumption3}, and \ref{SH:Assumption_polynomials} with the values required by Theorem \ref{Theorem_SH_LWP}. The explicit dependence on $(t,x)$ through the fixed metric is permitted by the remark following Theorem \ref{Theorem_SH_LWP}.

The value of $\chi_{\max}$ in Theorem \ref{Theorem_SH_LWP} is one. By Equation~\eqref{E:FO_CD_shifted_cleaning_regularity} and Sobolev embedding, $\mathcal{Y}[W]$, which consists of $W$ and its first spacetime derivatives, is continuous on $[0,T_g]\times\mathbb{T}^3$. Its range is therefore contained in a compact subset $\mathcal{L}$ of the finite-dimensional space $\mathcal{J}_{\mathcal{W}}$ defined in Theorem \ref{Theorem_SH_LWP}. Since the right-hand sides are linear in these variables, they are real analytic for every value in $\mathcal{J}_{\mathcal{W}}$, and the open set $\mathcal{V}$ in Theorem \ref{Theorem_SH_LWP} can be taken to be $\mathcal{J}_{\mathcal{W}}$.

The analytic approximations required in Equation~\eqref{E:SH_Analytic_forcing_approximation} are supplied by the solutions $\mathcal{W}_\nu$ constructed above. Indeed, applying Equations~\eqref{E:FO_CD_cleaning_energy_identity} and \eqref{E:FO_CD_shifted_time_derivative_formula} to $\mathcal{W}_\nu-\mathcal{W}$ gives
\begin{align}
\label{E:FO_CD_shifted_cleaning_approximation}
\sup_{0\leq t\leq T_g}
\sum_{a=0}^{7}
\left\|
\partial_t^a
\big(
\mathcal{W}_\nu-\mathcal{W}
\big)(t)
\right\|_{H^{\ell+7-a}}^2
&\leq
K
\left\|
\big(
\mathcal{S}_\nu-I
\big)
\mathcal{W}_{(0)}
\right\|_{H^{\ell+7}}^2
\rightarrow0.
\end{align}
Since $\mathcal{W}_{(0),\nu}$ has finite Fourier support, $\mathcal{W}_\nu$ is real analytic in $(t,x)$. Equation~\eqref{E:FO_CD_shifted_cleaning_approximation} and Sobolev embedding also give the compact containment required in Equation~\eqref{E:SH_Analytic_forcing_containment}.

It remains to identify the physical Sobolev indices and to verify the initial-data hypotheses. The definitions of $S_I$ in Theorem \ref{Theorem_SH_LWP} give
\begin{subequations}{\label{E:FO_CD_shifted_physical_indices}}
\begin{align}
S_T
&=
(\ell-1)+p+m_T-m_*
=\ell+6,
\\
S_U=S_J
&=
(\ell-1)+p+m_U-m_*
=\ell+7.
\end{align}
\end{subequations}
The data in Equations~\eqref{E:FO_CD_complete_initial_data} are one derivative more regular than the spaces required by Equations~\eqref{E:FO_CD_shifted_physical_indices}. Moreover, Equations~\eqref{E:FO_CD_full_data_convergence} and \eqref{E:FO_CD_first_trace_convergence} imply
\begin{subequations}{\label{E:FO_CD_shifted_initial_approximation}}
\begin{align}
T^{0\mu}_{(0),\nu}
&\rightarrow
T^{0\mu}_{(0)}
\quad\text{in }H^{\ell+6},
\\
U^A_{(0),\nu},\ J^0_{(0),\nu}
&\rightarrow
U^A_{(0)},\ J^0_{(0)}
\quad\text{in }H^{\ell+7},
\\
U^A_{(1),\nu},\ J^0_{(1),\nu}
&\rightarrow
U^A_{(1)},\ J^0_{(1)}
\quad\text{in }H^{\ell+6}.
\end{align}
\end{subequations}
The $T^{0\mu}$-rows have $n_T=0$ and impose no compatibility condition. Since $n_U=n_J=1$, the two remaining compatibility conditions in Equation~\eqref{E:SH_Compatibility_conditions} are precisely the identities used to define $U^A_{(1),\nu}$ and $J^0_{(1),\nu}$ from Equations~\eqref{E:FO_CD_modified_initial_first_traces}. The compact containment of the approximating initial jets was proved after Equations~\eqref{E:FO_CD_first_trace_convergence}. Thus, all the initial-data and approximation hypotheses of Theorem \ref{Theorem_SH_LWP} hold.

Applying Theorem \ref{Theorem_SH_LWP}, with $T_0=T_g$, gives $T\in(0,T_g]$ and a solution $V$ of Equations~\eqref{E:FO_CD_modified_mixed_order_system} such that
\begin{subequations}{\label{E:FO_CD_shifted_physical_output}}
\begin{align}
\partial_t^jT^{0\mu}
&\in
L^\infty\big([0,T];H^{\ell+6-j}(\mathbb{T}^3)\big),
&&
0\leq j\leq7,
\\
\partial_t^jU^A,\ \partial_t^jJ^0
&\in
L^\infty\big([0,T];H^{\ell+7-j}(\mathbb{T}^3)\big),
&&
0\leq j\leq7.
\end{align}
\end{subequations}
For the same values of $j$, these fields are continuous in time with values in every Sobolev space whose exponent is strictly smaller than the exponent displayed in Equations~\eqref{E:FO_CD_shifted_physical_output}. The physical jet takes values in a compact set $\mathcal{K}_1\Subset\mathcal{U}_{\mathrm{FO}}$ containing $\mathcal{K}$ in its interior.

\medskip
\noindent \underline{Step 6} (\textbf{Reconstruction and time regularity}).\nopagebreak[4]

Define
\begin{align*}
n
&=
\frac{J^0}{u^0(U)},
\\
X_i^A
&=
\partial_iU^A+C_i^A,
\\
Z_i
&=
\partial_i n+D_i.
\end{align*}
The analytic-composition and Sobolev product estimates applied to $n=J^0/u^0(U)$ and Equations~\eqref{E:FO_CD_shifted_physical_output} give
\begin{align*}
\partial_t^jn
&\in
L^\infty\big([0,T];H^{\ell+7-j}(\mathbb{T}^3)\big),
&&
0\leq j\leq7.
\end{align*}
More explicitly, $\partial_t^jn$ is a finite sum of products of derivatives of $J^0$, derivatives of $U$, and derivatives of the analytic function $1/u^0$. In a term containing $\partial_t^jJ^0$ or $\partial_t^jU$, that factor belongs to $H^{\ell+7-j}$ and every other factor belongs to a Sobolev algebra embedded into $L^\infty$. In every other term all factors have strictly more Sobolev regularity than $H^{\ell+7-j}$. This proves the displayed assertion for $n$.

Equations~\eqref{E:FO_CD_reconstruction}, \eqref{E:FO_CD_shifted_cleaning_regularity}, and \eqref{E:FO_CD_shifted_physical_output} give
\begin{align*}
\partial_t^jX_i^A
&=
\partial_i\partial_t^jU^A+\partial_t^jC_i^A,
\\
\partial_t^jZ_i
&=
\partial_i\partial_t^jn+\partial_t^jD_i,
\end{align*}
and hence
\begin{align*}
\partial_t^jX_i^A,\ \partial_t^jZ_i
&\in
L^\infty\big([0,T];H^{\ell+6-j}(\mathbb{T}^3)\big),
&&
0\leq j\leq7.
\end{align*}
 The time-continuity conclusion of Theorem \ref{Theorem_SH_LWP} gives continuity of $\partial_t^jT^{0\mu}$ in $H^s$ only for $s<\ell+6-j$, and continuity of $\partial_t^jU^A$ and $\partial_t^jJ^0$ only for $s<\ell+7-j$. The analytic-composition argument for $n=J^0/u^0(U)$ gives the same conclusion for $\partial_t^jn$ when $s<\ell+7-j$. Combining these strict inequalities with Equation~\eqref{E:FO_CD_shifted_cleaning_regularity} and the reconstruction formulas gives continuity of $\partial_t^jX_i^A$ and $\partial_t^jZ_i$ only for $s<\ell+6-j$. These are exactly the assertions in Equations~\eqref{E:FO_CD_theorem_time_continuity}; no continuity in time is asserted there at the top solution exponents displayed in Equations~\eqref{E:FO_CD_theorem_solution_regularity}. The definitions of $C_{(0)},D_{(0)}$, together with Equations~\eqref{E:FO_CD_shifted_initial_approximation}, show that the reconstructed fields assume the prescribed initial values.

Finally, the equivalence proved after Equation~\eqref{E:FO_CD_cleaning_system} shows that
\begin{align*}
\big(
T^{0\mu},X_i^A,Z_i,U^A,n,
\varphi_i^A,\phi^A,\Psi_i,\psi
\big)
\end{align*}
satisfies every equation in Equations~\eqref{Complete_set_Eq}. Equations~\eqref{E:FO_CD_shifted_cleaning_regularity} and \eqref{E:FO_CD_shifted_physical_output}, together with the reconstruction formulas, give exactly Equations~\eqref{E:FO_CD_theorem_solution_regularity}. This completes the existence and reconstruction portion of the proof of Theorem \ref{Theorem_1_precise}.

\medskip
\noindent \underline{Step 7} (\textbf{Uniqueness}).\nopagebreak[4]

 Let
\begin{align*}
\big(V,\mathcal{W}\big)
\quad\text{and}\quad
\big(\widetilde{V},\widetilde{\mathcal{W}}\big)
\end{align*}
be two solutions in the uniqueness class with the same initial data. By the equivalence following Equations~\eqref{E:FO_CD_cleaning_system}, both $\mathcal{W}$ and $\widetilde{\mathcal{W}}$ satisfy Equations~\eqref{E:FO_CD_cleaning_system}. Apply Equation~\eqref{E:FO_CD_cleaning_energy_identity} to $\mathcal{W}-\widetilde{\mathcal{W}}$. Its initial value is zero, and every coefficient in Equation~\eqref{E:FO_CD_cleaning_energy} is positive by Assumption \ref{Assumption_velocities}. Hence
\begin{align}
\label{E:FO_CD_cleaning_uniqueness}
\mathcal{W}
&=
\widetilde{\mathcal{W}}
\end{align}
on the common lifespan.

By Equation~\eqref{E:FO_CD_cleaning_uniqueness}, the prescribed functions $W=(C,D)$ and $\widetilde{W}=(\widetilde{C},\widetilde{D})$ coincide. The equivalence following Equation~\eqref{E:FO_CD_cleaning_system} shows that $V$ and $\widetilde{V}$ therefore solve the same forced mixed-order system in Equations~\eqref{E:FO_CD_modified_mixed_order_system}, with the same prescribed function $W$ and the same initial data. Both solutions belong to the uniqueness class of Theorem \ref{Theorem_SH_LWP}. Its uniqueness conclusion consequently gives
\begin{align*}
V
&=
\widetilde{V}
\end{align*}
on the common lifespan. In particular, the definition $n=J^0/u^0(U)$ gives $n=\widetilde{n}$. Equation~\eqref{E:FO_CD_cleaning_uniqueness} and the reconstruction formulas in Equations~\eqref{E:FO_CD_reconstruction} then give $X=\widetilde{X}$ and $Z=\widetilde{Z}$. Hence every original variable in Equations~\eqref{Complete_set_Eq} agrees with its tilded counterpart, which proves uniqueness.

\medskip
\noindent \underline{Step 8} (\textbf{Continuous dependence below the top Sobolev orders}).\nopagebreak[4]
Let the initial data converge as in the continuous-dependence assertion of the theorem. Applying Equations~\eqref{E:FO_CD_cleaning_energy_identity} and \eqref{E:FO_CD_shifted_time_derivative_formula} to $\mathcal{W}_q-\mathcal{W}$ gives
\begin{align*}
\sup_{0\leq t\leq T_g}
\sum_{a=0}^{7}
\left\|
\partial_t^a(\mathcal{W}_q-\mathcal{W})(t)
\right\|_{H^{\ell+7-a}}^2
&\rightarrow
0.
\end{align*}
The continuous-dependence conclusion of Theorem \ref{Theorem_SH_LWP}, applied with the indices in Equations~\eqref{E:FO_CD_shifted_physical_indices}, therefore gives the asserted convergence of $T_q^{0\mu}$, $U_q^A$, and $J_q^0$ on a common interval $[0,T]$. The analytic-composition estimates applied to $n_q=J_q^0/u^0(U_q)$ give the asserted convergence of $n_q$. Equations~\eqref{E:FO_CD_reconstruction_X} and \eqref{E:FO_CD_reconstruction_Z}, together with the displayed convergence of $\mathcal{W}_q$, then give the asserted convergence of $X_q$ and $Z_q$. This proves the continuous-dependence assertion.

\end{proof}

\section{Proof of theorem \ref{Theorem_2} }
\label{Sec:Proof_MO}

In this Section, we prove Theorem \ref{Theorem_2}, which we state precisely below.

For the proof of the next theorem, we recall that we work in coordinates for which $\partial_t$ is timelike and the hypersurfaces $t=\mathrm{constant}$ are spacelike. Consequently, the coordinate vector fields $\partial_i$ are spacelike and tangent to these hypersurfaces, and $g^{00}=g^{-1}(\mathrm{d}t,\mathrm{d}t)<0$.

\begin{theorem}
\label{T:Precise_mixed_order_local_well_posedness}
Let $\Sigma=\mathbb{T}^3$, let $\ell\geq3$ be an integer, and let $g_{\mu\nu}$ be a fixed Lorentzian metric whose components are real-analytic functions of $(t,x)$ on an open neighborhood of $\{0\}\times\Sigma$. Assume that the functions
\begin{align*}
P,
\quad
T,
\quad
\mu,
\quad
\eta,
\quad
\zeta,
\quad
\sigma,
\quad
\tau_\varepsilon,
\quad
\tau_P,
\quad
\tau_Q
\end{align*}
are real-analytic functions of $(\varepsilon,n)$ on an open set $\mathcal{O}\subset\mathbb{R}^2$.

Set
\begin{align*}
\Psi
&=
\big(T^{0\mu},U^A,J^0\big),
\\
\mathcal{Z}_{\mathrm{MO}}[\Psi]
&=
\big(T^{0\mu},U^A,J^0,\partial_\alpha U^A,
\partial_\alpha J^0\big),
\end{align*}
where $\mu,\alpha=0,1,2,3$ and $A=\theta,1,2,3$. Suppose that there exists an open set $\mathcal{U}_{\mathrm{MO}}$ in the finite-dimensional range of $\mathcal{Z}_{\mathrm{MO}}$ with the following properties. For every $z\in\mathcal{U}_{\mathrm{MO}}$ and every $(t,x)$ in the coordinate domain under consideration, the quantities defined in Equations~\eqref{u0}, \eqref{Tmunu_Mixed}, and \eqref{MO:Transformations} are well-defined, $(\varepsilon,n)\in\mathcal{O}$, and Assumptions \ref{Assumption_LWP} and \ref{Assumption_physics} hold. Assume also that every coefficient and every right-hand side in Equations~\eqref{MO:LWP-Eqs}, after the substitutions in Equations~\eqref{MO:Transformations}, is a real-analytic function of $((t,x),z)$ on this set.

Let the initial data satisfy
\begin{align*}
T^{0\mu}_{(0)}
&\in
H^{\ell+7}(\Sigma),
\\
U^A_{(0)}
&\in
H^{\ell+8}(\Sigma),
\\
J^0_{(0)}
&\in
H^{\ell+8}(\Sigma).
\end{align*}
Define $U^A_{(1)}$ by evaluating the right-hand side of Equation~\eqref{FO:YA} on $\{t=0\}$ using these data, and define
\begin{align}
\label{E:MO_Initial_J_trace}
J^0_{(1)}
&=
-\left.
\left[
\partial_i\left(\frac{J^0}{u^0}u^i\right)
+\Gamma^\nu_{\nu\mu}\frac{J^0}{u^0}u^\mu
\right]
\right|_{t=0}.
\end{align}
Thus, $U^A_{(1)}$ and $J^0_{(1)}$ are the first time derivatives prescribed by the second and third equations in Equations~\eqref{MO:Eqs}; they are not independent initial data. Let
\begin{align*}
\mathcal{Z}_{\mathrm{MO},0}
=
\big(
T^{0\mu}_{(0)},
U^A_{(0)},
J^0_{(0)},
U^A_{(1)},
\partial_iU^A_{(0)},
J^0_{(1)},
\partial_iJ^0_{(0)}
\big).
\end{align*}
Assume that there exists a compact set $\mathcal{K}\Subset\mathcal{U}_{\mathrm{MO}}$ such that
\begin{align*}
\overline{\mathcal{Z}_{\mathrm{MO},0}(\Sigma)}
\subset
\mathcal{K}.
\end{align*}

Then there exist $T>0$ and a unique solution $\Psi$ of Equations~\eqref{MO:Eqs}, with $T^{i\mu}$, $Y^A$, $J^i$, and $n$ given by Equations~\eqref{Tmunu_Mixed} and \eqref{MO:Transformations}, such that
\begin{align*}
\partial_t^jT^{0\mu}
&\in
L^\infty\big([0,T];H^{\ell+7-j}(\Sigma)\big),
&&
0\leq j\leq7,
\\
\partial_t^jU^A
&\in
L^\infty\big([0,T];H^{\ell+8-j}(\Sigma)\big),
&&
0\leq j\leq7,
\\
\partial_t^jJ^0
&\in
L^\infty\big([0,T];H^{\ell+8-j}(\Sigma)\big),
&&
0\leq j\leq7.
\end{align*}
For $0\leq j\leq7$, the solution also satisfies
\begin{align*}
\partial_t^jT^{0\mu}
&\in
C\big([0,T];H^s(\Sigma)\big)
&&
\text{for every }s<\ell+7-j,
\\
\partial_t^jU^A,
\ \partial_t^jJ^0
&\in
C\big([0,T];H^s(\Sigma)\big)
&&
\text{for every }s<\ell+8-j.
\end{align*}
Moreover, $T$ may be chosen so that there exists a compact set $\mathcal{K}_1\Subset\mathcal{U}_{\mathrm{MO}}$ for which
\begin{align*}
\mathcal{K}
\subset
\operatorname{int}\mathcal{K}_1,
\qquad
\overline{
\mathcal{Z}_{\mathrm{MO}}[\Psi]
\big([0,T]\times\Sigma\big)
}
\subset
\mathcal{K}_1.
\end{align*}

The solution is unique among solutions with the regularity stated above whose values of $\mathcal{Z}_{\mathrm{MO}}$ remain in $\mathcal{U}_{\mathrm{MO}}$. The same $T$ may be chosen for initial data in a sufficiently small neighborhood of
\begin{align*}
\big(T^{0\mu}_{(0)},U^A_{(0)},J^0_{(0)}\big)
\end{align*}
in
\begin{align*}
H^{\ell+7}(\Sigma;\mathbb{R}^4)
\times
H^{\ell+8}(\Sigma;\mathbb{R}^4)
\times
H^{\ell+8}(\Sigma)
\end{align*}
whose initial jets take values in $\mathcal{K}_1$.

Let
\begin{align*}
\big(
T^{0\mu}_{(0),q},
U^A_{(0),q},
J^0_{(0),q}
\big),
\qquad
q\in\mathbb{N},
\end{align*}
be a sequence of initial data satisfying the hypotheses of this theorem, whose initial jets take values in $\mathcal{K}_1$, and such that
\begin{align*}
T^{0\mu}_{(0),q}
&\rightarrow
T^{0\mu}_{(0)}
\quad\text{in }H^{\ell+7}(\Sigma),
\\
U^A_{(0),q}
&\rightarrow
U^A_{(0)}
\quad\text{in }H^{\ell+8}(\Sigma),
\\
J^0_{(0),q}
&\rightarrow
J^0_{(0)}
\quad\text{in }H^{\ell+8}(\Sigma).
\end{align*}
After decreasing $T$ if necessary, the corresponding solutions are defined on $[0,T]\times\Sigma$ for every sufficiently large $q$ and, for $0\leq j\leq6$, satisfy
\begin{align*}
\partial_t^jT^{0\mu}_q
&\rightarrow
\partial_t^jT^{0\mu}
\quad\text{in }C\big([0,T];H^s(\Sigma)\big)
&&
\text{for every }s<\ell+7-j,
\\
\partial_t^jU^A_q
&\rightarrow
\partial_t^jU^A
\quad\text{in }C\big([0,T];H^s(\Sigma)\big)
&&
\text{for every }s<\ell+8-j,
\\
\partial_t^jJ^0_q
&\rightarrow
\partial_t^jJ^0
\quad\text{in }C\big([0,T];H^s(\Sigma)\big)
&&
\text{for every }s<\ell+8-j.
\end{align*}
\end{theorem}

\begin{remark}
The assumption in Theorem \ref{T:Precise_mixed_order_local_well_posedness} that the components of the fixed metric are real-analytic functions of $(t,x)$ has the same prescribed-background interpretation. It does not imply that a corresponding result for the coupled Einstein--BDNK system would be restricted to real-analytic Cauchy data. When $g_{\mu\nu}$ is included among the unknowns, the real-analytic dependence required by Theorem \ref{Theorem_SH_LWP} concerns the dependence of the coefficient functions and right-hand sides on the jet variables. Remark \ref{R:Analyticity_hypothesis} explains this point for Einstein's equations in wave coordinates: on the open set of nondegenerate Lorentzian metrics, the inverse metric and the nonlinearities in the reduced Einstein--matter system are real-analytic functions of the relevant components of the unknowns and their derivatives, provided that the equation of state and the transport coefficients are real analytic on the set under consideration. This does not require the Sobolev Cauchy data, or the metric obtained from them, to be analytic functions of the spatial or spacetime coordinates. As in Theorem \ref{Theorem_SH_LWP}, compatible real-analytic approximations of the Sobolev data would still be required, and all the remaining hypotheses of that theorem would have to be verified separately for the coupled system.
\end{remark}

\begin{proof}
\noindent \underline{Step 1} (\textbf{The Leray indices and the initial data}).\nopagebreak[4]
Using Equations~\eqref{MO:Transformations}, we first write Equations~\eqref{MO:Eqs} in terms of the nine unknowns
\begin{align*}
\Psi
&=
\big(T^{00},T^{0k},U^B,J^0\big)
\end{align*}
as follows:

\bml
\label{MO:LWP-Eqs}
\begin{align}
&\partial_t T^{00}+\partial_j T^{0j}=B_{T^{00}}(\Psi,\partial^2 g),\label{LWP-MO_Eqs_a}\\
&\partial_t T^{0\ell}+H^{\ell j 0\beta} M^{-1}_{\beta\mu} \partial_j T^{0\mu}+\left (\tilde{H}^{\ell j k \beta} - H^{\ell j 0\lambda} M^{-1}_{\lambda\mu} \tilde{H}^{\mu0 k \beta}\right )P_{A\beta}\partial_j\partial_k U^A\\
&+\frac{2\beta_n}{u^0} \left (u^{(\ell} \Omega^{kj)}-H^{\ell j 0\beta} M^{-1}_{\beta\mu} u^{(\mu} \Omega^{k0)}\right )\partial_j\partial_k J^0=B_{T^{0m}}(T^{0\mu},\partial U^B,\partial J^0,\partial^2 g),\label{LWP-MO_Eqs_b}\\
&\partial_t U^A -\tilde{P}^{A\lambda}M^{-1}_{\lambda\mu}\left [T^{0\mu} - \tilde{H}^{\mu0 j \beta}P_{A\beta}\partial_j U^A-\frac{2\beta_n}{u^0} u^{(\mu} \Omega^{j0)} \partial_jJ^0 \right ]=B_{U^B}^A(U^B,J^0,\partial^2g),\label{LWP-MO_Eqs_c}\\
&\partial_t J^0+\frac{u^j}{u^0}\partial_j J^0+n\left (\Delta^{j\beta}+\frac{u_i\Delta^{i\beta}u^j}{u^0u_0}\right )P_{B\beta}\partial_j U^B  =B_{J^0}(U^B,J^0,\partial^2g),\label{LWP-MO_Eqs_d} 
\end{align} 
\eml
where $n=J^0/u^0$ and
\begin{align*}
\tilde{H}^{\mu\nu k\beta}=H^{\mu\nu k\beta}+\frac{2n\beta_n u^{(\mu}\Omega^{k \nu)}u_i\Delta^{i\beta}}{u^0u_0}.
\end{align*}
The Leray indices are $m_{T^{0\mu}}=1$, $m_{U^B}=m_{J^0}=2$, $n_I=0$ for Equations~\eqref{LWP-MO_Eqs_a} and \eqref{LWP-MO_Eqs_b}, and $n_I=1$ for Equations~\eqref{LWP-MO_Eqs_c} and \eqref{LWP-MO_Eqs_d}. 

For the four Equations~\eqref{LWP-MO_Eqs_a}--\eqref{LWP-MO_Eqs_b}, for which $n_I=0$, the principal differential operators have order at most one on $T^{0\mu}$ and at most two on $U^B$ and $J^0$. Their coefficients and right-hand sides depend on $T^{0\mu}$, on at most one derivative of $U^B$ and $J^0$, and on the fixed metric. These are exactly the orders $m_{T^{0\mu}}-n_I-1=0$ and $m_{U^B}-n_I-1=m_{J^0}-n_I-1=1$ allowed in Equation~\eqref{SH:1}. For the five Equations~\eqref{LWP-MO_Eqs_c}--\eqref{LWP-MO_Eqs_d}, for which $n_I=1$, the principal differential operators have order zero on $T^{0\mu}$ and order one on $U^B$ and $J^0$. The coefficients in these five rows depend on $U^B$ and $J^0$, but not on their derivatives or on $T^{0\mu}$, in agreement with
\begin{align*}
m_{T^{0\mu}}-n_I-1&=-1,
\\
m_{U^B}-n_I-1
&=
m_{J^0}-n_I-1
=0.
\end{align*}
Thus, Equation~\eqref{MO:LWP-Eqs} has the form in Equation~\eqref{SH:1} with the stated Leray indices.

The symbol calculation in Step 6 below gives $m=7$; since $r=1$, this gives $p=m\,r=7$. Moreover,
\begin{align*}
m_*
&=
\min_I m_I
=1.
\end{align*}
Consequently, the numbers $S_I=\ell+p+m_I-m_*$ in Theorem \ref{Theorem_SH_LWP} are
\begin{align*}
S_{T^{0\mu}}
&=
\ell+7,
\\
S_{U^B}
&=
S_{J^0}
=
\ell+8.
\end{align*}
The Sobolev assumptions in Theorem \ref{T:Precise_mixed_order_local_well_posedness} therefore give precisely the regularity in Equation~\eqref{E:SH_Initial_data_regularity} for $T^{0\mu}_{(0)}$, $U^B_{(0)}$, and $J^0_{(0)}$. The formulas in Equations~\eqref{FO:YA} and \eqref{E:MO_Initial_J_trace} contain at most one spatial derivative of $U^B_{(0)}$ and $J^0_{(0)}$. Since $\ell\geq3$, the Sobolev product and composition estimates on $\mathbb{T}^3$ give
\begin{align*}
U^B_{(1)},
\ J^0_{(1)}
\in
H^{\ell+7}(\mathbb{T}^3).
\end{align*}
This is the remaining part of Equation~\eqref{E:SH_Initial_data_regularity}. The compact-containment assumption on $\mathcal{Z}_{\mathrm{MO},0}$ in Theorem \ref{T:Precise_mixed_order_local_well_posedness} is exactly Equation~\eqref{E:SH_Initial_state_containment} with $\mathcal{U}=\mathcal{U}_{\mathrm{MO}}$.

This verifies the Leray indices, Equation~\eqref{E:SH_Initial_data_regularity}, and Equation~\eqref{E:SH_Initial_state_containment}.

\medskip
\noindent \underline{Step 2} (\textbf{The initial compatibility conditions}).\nopagebreak[4]
Only the five Equations~\eqref{LWP-MO_Eqs_c}--\eqref{LWP-MO_Eqs_d} have $n_I>0$, and they all have $n_I=1$. Thus, Equation~\eqref{E:SH_Compatibility_conditions} requires only the residuals of these five equations to vanish on $\{t=0\}$. For Equation~\eqref{LWP-MO_Eqs_c}, this condition is the equation that defines $\partial_tU^A|_{t=0}$. Restoring the lower-order terms included in $B_{U^B}^A$, this equation is precisely Equation~\eqref{FO:YA} evaluated on $\{t=0\}$, which is the definition of $U^A_{(1)}$ in Theorem \ref{T:Precise_mixed_order_local_well_posedness}. For Equation~\eqref{LWP-MO_Eqs_d}, the condition is
\begin{align*}
J^0_{(1)}
&=
-\left.
\left[
\partial_i\left(\frac{J^0}{u^0}u^i\right)
+\Gamma^\nu_{\nu\mu}\frac{J^0}{u^0}u^\mu
\right]
\right|_{t=0},
\end{align*}
which is Equation~\eqref{E:MO_Initial_J_trace}. The four Equations~\eqref{LWP-MO_Eqs_a}--\eqref{LWP-MO_Eqs_b} have $n_I=0$ and impose no condition in Equation~\eqref{E:SH_Compatibility_conditions}. Hence all the compatibility conditions required by Theorem \ref{Theorem_SH_LWP} hold.

This concludes the verification of Equation~\eqref{E:SH_Compatibility_conditions}.

\medskip
\noindent \underline{Step 3} (\textbf{Compatible real-analytic approximations}).\nopagebreak[4]
For a function $f$ on $\mathbb{T}^3$, define
\begin{align*}
\mathcal{S}_\nu f(x)
&=
\sum_{k\in\mathbb{Z}^3,\,|k|\leq\nu}
\widehat{f}(k)e^{\mathrm{i}k\cdot x}.
\end{align*}
Set
\begin{align*}
T^{0\mu}_{(0),\nu}
&=
\mathcal{S}_\nu T^{0\mu}_{(0)},
\\
U^A_{(0),\nu}
&=
\mathcal{S}_\nu U^A_{(0)},
\\
J^0_{(0),\nu}
&=
\mathcal{S}_\nu J^0_{(0)}.
\end{align*}
These functions are real analytic on $\mathbb{T}^3$, and
\begin{align*}
T^{0\mu}_{(0),\nu}
&\rightarrow
T^{0\mu}_{(0)}
\quad
\text{in }H^{\ell+7}(\mathbb{T}^3),
\\
U^A_{(0),\nu}
&\rightarrow
U^A_{(0)}
\quad
\text{in }H^{\ell+8}(\mathbb{T}^3),
\\
J^0_{(0),\nu}
&\rightarrow
J^0_{(0)}
\quad
\text{in }H^{\ell+8}(\mathbb{T}^3).
\end{align*}
Define $U^A_{(1),\nu}$ by evaluating Equation~\eqref{FO:YA} on $\{t=0\}$ using these three approximating fields, and define $J^0_{(1),\nu}$ by the right-hand side of Equation~\eqref{E:MO_Initial_J_trace} using the same approximating fields. By Step 2, these five traces satisfy Equation~\eqref{E:SH_Compatibility_conditions} exactly for every sufficiently large $\nu$.

Let
\begin{align*}
\mathcal{Y}_{\mathrm{MO},0}
&=
\big(
T^{0\mu}_{(0)},
U^A_{(0)},
J^0_{(0)},
\partial_iU^A_{(0)},
\partial_iJ^0_{(0)}
\big),
\end{align*}
and define $\mathcal{Y}_{\mathrm{MO},0,\nu}$ from the approximating fields in the same way. The range of $\mathcal{Y}_{\mathrm{MO},0}$ is contained in the projection of $\mathcal{K}$ onto these components. This projection is compact, and every denominator in Equations~\eqref{FO:YA} and \eqref{E:MO_Initial_J_trace} is bounded away from zero there. The three convergences above and Sobolev embedding imply that $\mathcal{Y}_{\mathrm{MO},0,\nu}$ converges uniformly to $\mathcal{Y}_{\mathrm{MO},0}$. Hence, for all sufficiently large $\nu$, the ranges of $\mathcal{Y}_{\mathrm{MO},0,\nu}$ lie in one compact set on which these denominators are bounded away from zero and all the derivatives of the coefficient functions needed in the Sobolev composition estimates are bounded. Those estimates give
\begin{align*}
U^A_{(1),\nu}
&\rightarrow
U^A_{(1)}
\quad
\text{in }H^{\ell+7}(\mathbb{T}^3),
\\
J^0_{(1),\nu}
&\rightarrow
J^0_{(1)}
\quad
\text{in }H^{\ell+7}(\mathbb{T}^3).
\end{align*}
It follows, in particular, that the complete approximating jets $\mathcal{Z}_{\mathrm{MO},0,\nu}$ converge uniformly to $\mathcal{Z}_{\mathrm{MO},0}$. Since $\overline{\mathcal{Z}_{\mathrm{MO},0}(\mathbb{T}^3)}\subset\mathcal{K}\Subset\mathcal{U}_{\mathrm{MO}}$, there exists a compact set $\mathcal{K}_0\Subset\mathcal{U}_{\mathrm{MO}}$ containing the ranges of $\mathcal{Z}_{\mathrm{MO},0,\nu}$ for all sufficiently large $\nu$.
The fixed metric and the coefficient functions are real analytic by the hypotheses of Theorem \ref{T:Precise_mixed_order_local_well_posedness}. Hence the compositions defining $U^A_{(1),\nu}$ and $J^0_{(1),\nu}$ are real analytic. The five convergences displayed in this step are precisely Equation~\eqref{E:SH_Analytic_data_approximation} for all the traces prescribed by the Leray indices.

This concludes the verification of the real-analytic approximation hypothesis in Theorem \ref{Theorem_SH_LWP}.

\medskip
\noindent \underline{Step 4} (\textbf{The coefficient functions and the principal symbol}).\nopagebreak[4]
Since $\min_I n_I=0$, the collection $\mathcal{Z}[\Psi]$ in Theorem \ref{Theorem_SH_LWP} consists of the fields $T^{0\mu}$ and of the fields $U^A,J^0$ together with their first spacetime derivatives. It is therefore exactly $\mathcal{Z}_{\mathrm{MO}}[\Psi]$ in Theorem \ref{T:Precise_mixed_order_local_well_posedness}. By the definition of $\mathcal{U}_{\mathrm{MO}}$ in that theorem, every coefficient and every right-hand side in Equation~\eqref{MO:LWP-Eqs} is real analytic in $((t,x),\mathcal{Z}_{\mathrm{MO}}[\Psi])$. We use here the coordinate-dependent version of Theorem \ref{Theorem_SH_LWP} stated in the second remark after its proof. Thus, the coefficient-analyticity hypothesis of Theorem \ref{Theorem_SH_LWP} holds on $\mathcal{U}_{\mathrm{MO}}$. Since the spatial manifold is $\mathbb{T}^3$, the additional hypotheses imposed in Theorem \ref{Theorem_SH_LWP} only when the spatial manifold is $\mathbb{R}^3$ do not arise.

With the rows and columns ordered as $(T^{00},T^{0k},U^B,J^0)$, the principal differential-operator matrix is

\begin{align}
\label{MO:Matrix_h}
h(\partial\Psi,\partial)=\bbm
\partial_t & \partial_k & 0_{1\times 4} & 0\\
H^{\ell j 0\beta} M^{-1}_{\beta 0} \partial_j  & \delta^\ell_k\partial_t +H^{\ell j 0\beta} M^{-1}_{\beta k} \partial_j & \mathfrak{H}^{\ell j k \beta}P_{B\beta}\partial_j\partial_k & \mathfrak{O}^{\ell j k \beta}\partial_j\partial_k \\
-\tilde{P}^{A\lambda}M^{-1}_{\lambda 0 } & -\tilde{P}^{A\lambda}M^{-1}_{\lambda k} & \delta^A_B\partial_t +\tilde{P}^{A\lambda}M^{-1}_{\lambda\mu}\tilde{H}^{\mu0 j \beta}P_{B\beta}\partial_j & \tilde{P}^{A\lambda}M^{-1}_{\lambda\mu}\frac{2\beta_n}{u^0} u^{(\mu} \Omega^{j0)} \partial_j \\
0 & 0_{1\times 3} & n\left (\Delta^{j\beta}+\frac{u_i\Delta^{i\beta}u^j}{u^0u_0}\right )P_{B\beta}\partial_j & \partial_t +\frac{u^j}{u^0}\partial_j 
\ebm,
\end{align}
where $A$ and $\ell$ distinguish between different rows, $B$ and $k$ distinguish columns, and we defined
\begin{align*}
\mathfrak{H}^{\ell j k \beta}&=\tilde{H}^{\ell j k \beta} - H^{\ell j 0\lambda} M^{-1}_{\lambda\mu} \tilde{H}^{\mu0 k \beta},\\
\mathfrak{O}^{\ell j k \beta}&=\frac{2\beta_n}{u^0} \left (u^{(\ell} \Omega^{kj)}-H^{\ell j 0\beta} M^{-1}_{\beta\mu} u^{(\mu} \Omega^{k0)}\right ).
\end{align*}

\begin{remark}
   Some of the calculations in Steps 5 and 6 are similar to calculations used for the first-order system in Section~\ref{Sec:LWP}. We give the complete calculations here because the principal symbol considered in this section is the $9\times 9$ mixed-order symbol, rather than the symbol coming from the  $24 \times 24$ system from Section~\ref{Sec:LWP}, and because the proof in this section is independent of Section~\ref{Sec:LWP}. 
\end{remark}

\medskip
\noindent \underline{Step 5} (\textbf{Verification of Assumption \ref{SH:Assumption 1}}).\nopagebreak[4]
The matrix in Equation~\eqref{MO:Matrix_h} contains first-order pure time derivatives and no pure time derivative of higher order. Thus, the integer in Definition \ref{SH:Definition 1} is $r=1$. Substituting $\partial_\mu$ by $\xi_\mu$ in Equation~\eqref{MO:Matrix_h} gives
\begin{align*}
h(\partial\Psi,\xi)=\bbm
\xi_0 & \xi_k & 0_{1\times 4} & 0\\
H^{\ell j 0\beta} M^{-1}_{\beta 0} \xi_j  & \delta^\ell_k\xi_0 +H^{\ell j 0\beta} M^{-1}_{\beta k} \xi_j & \mathfrak{H}^{\ell j k \beta}P_{B\beta}\xi_j\xi_k & \mathfrak{O}^{\ell j k \beta}\xi_j\xi_k \\
-\tilde{P}^{A\lambda}M^{-1}_{\lambda 0 } & -\tilde{P}^{A\lambda}M^{-1}_{\lambda k} & \xi_0\delta^A_B +\tilde{P}^{A\lambda}M^{-1}_{\lambda\mu}\tilde{H}^{\mu0 j \beta}P_{B\beta}\xi_j & \tilde{P}^{A\lambda}M^{-1}_{\lambda\mu}\frac{2\beta_n}{u^0} u^{(\mu} \Omega^{j0)} \xi_j \\
0 & 0_{1\times 3} & n\left (\Delta^{j\beta}+\frac{u_i\Delta^{i\beta}u^j}{u^0u_0}\right )P_{B\beta}\xi_j & \xi_0 +\frac{u^j}{u^0}\xi_j 
\ebm.
\end{align*}
For simplicity, let us rewrite the matrix $h(\partial\Psi,\xi)$ as the product
\[h(\partial\Psi,\xi)=\tilde{h}(\partial\Psi,\xi)\mathcal{C},\]
where
\begin{align*}
\tilde{h}(\partial\Psi,\xi)=\bbm
H^{\mu\sigma 0\beta}P_{B\beta}\xi_\sigma & H^{\mu\sigma j \beta}P_{B\beta}\xi_\sigma \xi_j  & 2\beta_n u^{(\mu} \Omega^{j\alpha)}\xi_\alpha \xi_j\\
-\delta^A_B & \delta^A_B\xi_0 & 0_{4\times 1} \\
0_{1\times 4} & n \Delta^{\alpha\beta}P_{B\beta}\xi_\alpha & u^\alpha\xi_\alpha 
\ebm
\end{align*}
and 
\begin{align*}
\mathcal{C}=\bbm
\tilde{P}^{A\lambda}M^{-1}_{\lambda\nu} & -\tilde{P}^{A\lambda}M^{-1}_{\lambda\sigma} \tilde{H}^{\sigma 0 j \beta}P_{B\beta}\xi_j & -\frac{2\beta_n}{u^0} \tilde{P}^{A\lambda}M^{-1}_{\lambda\sigma} u^{(\sigma} \Omega^{j0)} \xi_j  \\
0_{4\times 4} & I_4 & 0_{4\times 1}\\
0_{1\times 4} & -\frac{n}{u^0}\Delta^{0\beta}P_{B\beta} & \frac{1}{u^0}
\ebm
\end{align*} is an invertible matrix whose inverse is given by
\begin{align*}
\mathcal{C}^{-1}=\bbm
M^{\mu\beta}P_{B\beta}  & H^{\mu 0 j \beta}P_{B\beta} \xi_j & 2\beta_n u^{(\mu} \Omega^{j 0)} \xi_j \\
0_{4\times 4} & I_4 & 0_{4\times 1}\\
0_{1\times 4} & n\Delta^{0\beta}P_{B\beta} & u^0
\ebm,
\end{align*}
where, whenever $\mu$ and $A$ appear in the above matrices, they distinguish between different rows, while $\nu$ and $B$ distinguish different columns. 

The identities \eqref{E:MO_State_change_differential} and \eqref{E:MO_Tilde_H_state_change} give the nonzero $(3,2)$ blocks of $\mathcal{C}$ and $\mathcal{C}^{-1}$. In multiplying the two matrices, the additional term in $\tilde{H}$ cancels the product of the $(1,3)$ and $(3,2)$ blocks. This is why the $(1,2)$ block of $\mathcal{C}^{-1}$ contains $H$, whereas that of $\mathcal{C}$ contains $\tilde{H}$. The coefficient of $\xi_0$ in $h$ is $I_9$; consequently the coefficient of $\xi_0$ in $\tilde{h}$ is $\mathcal{C}^{-1}$. Thus the eigenvectors of the two symbols are related by $V=\mathcal{C}^{-1}\widetilde{V}$. In particular, if $\widetilde{V}=(\xi_0y,y,0)$, then its $J^0$ component in the original variables is $n\Delta^{0\beta}P_{B\beta}y^B$.

To understand how the first 4 columns and rows in the matrix $\tilde{h}(\partial\Psi,\xi)$ split into the first 1+3 rows and 1+3 columns, let us take as an example the product of the first 4 rows of matrix $\tilde{h}(\partial\Psi,\xi)$ by the first 4 columns of matrix $\mathcal{C}$. The product reads
\begin{align*}
&H^{\mu\sigma  0\beta}P_{B\beta}\xi_\sigma \tilde{P}^{B\lambda}M^{-1}_{\lambda\nu}=
H^{\mu\sigma  0\beta}\xi_\sigma M^{-1}_{\beta\nu}
=\bbm
H^{0\sigma  0\beta}\xi_\sigma M^{-1}_{\beta 0} & H^{0\sigma  0\beta}\xi_\sigma M^{-1}_{\beta k}\\
H^{\ell \sigma  0\beta}\xi_\sigma M^{-1}_{\beta 0} & H^{\ell\sigma  0\beta}\xi_\sigma M^{-1}_{\beta k}
\ebm\\
&=\bbm
\xi_0 & \xi_k\\
H^{\ell \sigma  0\beta}\xi_\sigma M^{-1}_{\beta 0} & H^{\ell\sigma  0\beta}\xi_\sigma M^{-1}_{\beta k}
\ebm,
\end{align*}
where we used the fact that $H^{0\sigma  0\beta}\xi_\sigma M^{-1}_{\beta\nu}=H^{\sigma 0 0\beta}\xi_\sigma M^{-1}_{\beta\nu}=M^{\sigma \beta}\xi_\sigma M^{-1}_{\beta\nu}=\xi_\nu$. Note that 
\begin{align*}
\det[\mathcal{C}]
&=
\frac{1}{u^0}
\det[\tilde{P}^{A\lambda}]
\det[M^{-1}_{\lambda\nu}]
\end{align*}
is not zero by Assumption \ref{Assumption_LWP} and does not depend on $\xi_\mu$. 

We calculate the determinant directly. Set
\begin{align*}
a
&=
u^\alpha\xi_\alpha,
\\
v^\mu
&=
\Delta^{\mu\alpha}\xi_\alpha,
\\
v_\beta
&=
\Delta^\alpha_\beta\xi_\alpha,
\\
v^2
&=
v^\alpha v_\alpha
=
\Delta^{\alpha\beta}\xi_\alpha\xi_\beta,
\end{align*}
and define
\begin{align*}
L^\mu_\beta
&=
\xi_\alpha\xi_\nu\tensor{H}{^\mu^\nu^\alpha_\beta},
\\
Q^\mu
&=
2\beta_n u^{(\mu}\Omega^{k\alpha)}\xi_k\xi_\alpha.
\end{align*}
Add $H^{\mu\sigma0\beta}P_{B\beta}\xi_\sigma$ times the second block row of $\tilde{h}(\partial\Psi,\xi)$ to its first block row. For $a\neq0$, expansion using the first block column and then the Schur complement of the scalar entry $a$ gives
\begin{align*}
\det[\tilde{h}(\partial\Psi,\xi)]
&=
\det\bbm
0_{4\times4}
&
L^\mu_\beta\tensor{P}{_B^\beta}
&
Q^\mu
\\
-I_4
&
\xi_0I_4
&
0_{4\times1}
\\
0_{1\times4}
&
nv_\beta\tensor{P}{_B^\beta}
&
a
\ebm
\\
&=
a\det[\tensor{P}{_B^\beta}]\det[K^\mu_\beta],
\end{align*}
where
\begin{align*}
K^\mu_\beta
&=
L^\mu_\beta-\frac{n}{a}Q^\mu v_\beta
\\
&=
\xi_\alpha\xi_\nu\tensor{H}{^\mu^\nu^\alpha_\beta}
-\frac{2n\beta_n}{a}u^{(\mu}\Omega^{k\alpha)}
\xi_k\xi_\alpha v_\beta.
\end{align*}
There is no sign in the second equality for the determinant: interchanging the two blocks of four rows gives the factor $(-1)^{16}=1$, and $\det[-I_4]=1$. Multiplication by $\det[\mathcal{C}]$, followed by the inverse relations between $P_{B\beta}$ and $\tilde{P}^{A\beta}$, gives
\begin{align}
\label{E:MO_Determinant_reduction}
\det[h(\partial\Psi,\xi)]
&=
\frac{a}{u^0}
\det[(M^\mu_\nu)^{-1}]
\det[K^\mu_\beta].
\end{align}

Substitution of Equations~\eqref{defineBDNKnew} and \eqref{Omega} into $K^\mu_\beta$, followed by the collection of the terms proportional to $\delta^\mu_\beta$, $u_\beta$, and $v_\beta$, gives
\begin{align}
\label{E:MO_K_rank_two_form}
K^\mu_\beta
&=
\mathbb{A}\delta^\mu_\beta
+W_1^\mu u_\beta
+W_2^\mu v_\beta,
\end{align}
where
\begin{align*}
\mathbb{A}
&=
\rho\tau_Qa^2-\eta v^2,
\\
W_1^\mu
&=
\left[
\rho\tau_Qa^2
-\varepsilon\beta_\varepsilon v^2
-\varepsilon\tau_\varepsilon a^2
-\eta v^2
\right]u^\mu
-\varepsilon(\tau_P+\beta_\varepsilon)av^\mu,
\\
W_2^\mu
&=
\left[
\rho(\tau_\varepsilon+\tau_Q)a
-\frac{n\beta_n v^2}{a}
\right]u^\mu
\\
&\quad
+\left[
\rho\tau_P-\zeta-\frac{\eta}{3}-n\beta_n
\right]v^\mu.
\end{align*}
The determinant of Equation~\eqref{E:MO_K_rank_two_form} is
\begin{align}
\label{E:MO_K_determinant_rank_two}
\det[K^\mu_\beta]
&=
\mathbb{A}^2
\left[
\mathbb{A}^2
+\mathbb{A}
\big(W_1^\mu u_\mu+W_2^\mu v_\mu\big)
\right.
\nonumber\\
&\qquad\left.
+\big(W_1^\mu u_\mu\big)
\big(W_2^\nu v_\nu\big)
-\big(W_2^\mu u_\mu\big)
\big(W_1^\nu v_\nu\big)
\right].
\end{align}
Indeed, Equation~\eqref{E:MO_K_determinant_rank_two} follows by applying the determinant formula for a rank-two perturbation of $\mathbb{A}I_4$ when $\mathbb{A}\neq0$, and then by polynomial continuation in $\mathbb{A}$.

We evaluate the expression in brackets. Set
\begin{align*}
z
&=
a^2,
\\
q
&=
v^2,
\\
d_0
&=
\zeta+\frac{4\eta}{3},
\\
b_0
&=
\rho\beta_\varepsilon+n\beta_n.
\end{align*}
Equations~\eqref{def:beta_epsilon}, \eqref{def:beta_n}, \eqref{soundspeed}, and \eqref{kappa_s_def} give
\begin{align*}
b_0
&=
\tau_Q\big(\rho P'_\varepsilon+nP'_n\big)
+\frac{\rho^2\sigma T}{n}
\left(\frac{\partial(\mu/T)}{\partial\varepsilon}\right)_n
+\rho\sigma T
\left(\frac{\partial(\mu/T)}{\partial n}\right)_\varepsilon
\\
&=
\rho c_s^2\tau_Q+\sigma\kappa_s.
\end{align*}
Since $u^\mu u_\mu=-1$ and $u^\mu v_\mu=0$, the four contractions in Equation~\eqref{E:MO_K_determinant_rank_two} are
\begin{align*}
W_1^\mu u_\mu
&=
\big(-\rho\tau_Q+\varepsilon\tau_\varepsilon\big)z
+\big(\varepsilon\beta_\varepsilon+\eta\big)q,
\\
W_1^\mu v_\mu
&=
-\varepsilon(\tau_P+\beta_\varepsilon)aq,
\\
W_2^\mu u_\mu
&=
-\rho(\tau_\varepsilon+\tau_Q)a
+\frac{n\beta_nq}{a},
\\
W_2^\mu v_\mu
&=
\left(
\rho\tau_P-\zeta-\frac{\eta}{3}-n\beta_n
\right)q.
\end{align*}
Using these identities and $\mathbb{A}=\rho\tau_Qz-\eta q$, the expression in brackets in Equation~\eqref{E:MO_K_determinant_rank_two} is
\begin{align*}
&\varepsilon\rho\tau_Q\tau_\varepsilon z^2
-\varepsilon
\left[
\tau_\varepsilon(b_0+d_0)+\rho\tau_P\tau_Q
\right]zq
\\
&\qquad
+\varepsilon
\left[
\tau_Pb_0-\beta_\varepsilon d_0
\right]q^2.
\end{align*}
The definition in Equations~\eqref{Characteristics} gives
\begin{align*}
c_+^2+c_-^2
&=
\frac{
\tau_\varepsilon(b_0+d_0)+\rho\tau_P\tau_Q
}{
\rho\tau_Q\tau_\varepsilon
},
\\
c_+^2c_-^2
&=
\frac{
\tau_Pb_0-\beta_\varepsilon d_0
}{
\rho\tau_Q\tau_\varepsilon
}.
\end{align*}
Consequently, the expression in brackets in Equation~\eqref{E:MO_K_determinant_rank_two} equals
\begin{align*}
\varepsilon\rho\tau_Q\tau_\varepsilon
\big(z-c_+^2q\big)
\big(z-c_-^2q\big).
\end{align*}
Since $c_1^2=\eta/(\rho\tau_Q)$, we also have
\begin{align*}
\mathbb{A}
&=
\rho\tau_Q\big(z-c_1^2q\big).
\end{align*}
It follows that
\begin{align}
\label{E:MO_K_determinant_factorization}
\det[K^\mu_\beta]
&=
\varepsilon\tau_\varepsilon(\rho\tau_Q)^3
\big(z-c_1^2q\big)^2
\big(z-c_+^2q\big)
\big(z-c_-^2q\big).
\end{align}
Combining Equations~\eqref{E:MO_Determinant_reduction} and \eqref{E:MO_K_determinant_factorization} yields
\begin{align}
\label{E:MO_Characteristic_factorization}
\det[h(\partial\Psi,\xi)]
&=
\frac{1}{u^0}
\det[(M^\mu_\nu)^{-1}]
(u^\alpha\xi_\alpha)
\varepsilon\tau_\varepsilon(\rho\tau_Q)^3
\nonumber\\
&\quad\times
\big[
(u^\mu\xi_\mu)^2
-c_1^2\Delta^{\mu\nu}\xi_\mu\xi_\nu
\big]^2
\nonumber\\
&\quad\times
\big[
(u^\mu\xi_\mu)^2
-c_+^2\Delta^{\mu\nu}\xi_\mu\xi_\nu
\big]
\nonumber\\
&\quad\times
\big[
(u^\mu\xi_\mu)^2
-c_-^2\Delta^{\mu\nu}\xi_\mu\xi_\nu
\big].
\end{align}
The block calculation was carried out for $a\neq0$. Both sides of Equation~\eqref{E:MO_Characteristic_factorization} are polynomials in $\xi_\mu$ and agree on the open set $a\neq0$. They therefore agree for every $\xi_\mu$. In particular, the intermediate matrix $K^\mu_\beta$, which contains $a^{-1}$, need not be defined when $a=0$.

Thus, the multiplicities in the characteristic polynomial are
\begin{align*}
s_1
&=
2,
\\
s_+
&=
1,
\\
s_-
&=
1.
\end{align*}

The right-hand side of Equation~\eqref{E:MO_Characteristic_factorization} is homogeneous in $\xi_\mu$. Since $s_1=2$ and $s_+=s_-=1$, its degree is
\begin{align*}
1+2s_1+2s_++2s_-
&=
9
=N\,r.
\end{align*}
The coefficient $h_1$ of $\xi_0$ is $h_1=\tilde{h}_1\mathcal{C}$, where
\begin{align*}
\tilde{h}_1=\frac{\partial \tilde{h}}{\partial \xi_0}
=\bbm
M^{\mu\beta}P_{B\beta} & H^{\mu 0 j \beta}P_{B\beta} \xi_j  & 2\beta_n u^{(\mu} \Omega^{j 0)} \xi_j\\
0_{4\times 4} & \delta^A_B & 0_{4\times 1} \\
0_{1\times 4} & n \Delta^{0\beta}P_{B\beta} & u^0 
\ebm.
\end{align*}

Directly from Equation~\eqref{MO:Matrix_h}, the coefficient of $\xi_0$ is the identity matrix:
\begin{align*}
h_1
&=
I_9,
\\
\det[h_1]
&=
1
\neq0.
\end{align*}

We now prove directly that the roots of the three quadratic factors are real. Fix $a=1,+,-$ and set
\begin{align*}
G_a^{\mu\nu}
&=
u^\mu u^\nu-c_a^2\Delta^{\mu\nu}.
\end{align*}
In a $g$-orthonormal frame in which $u^\mu=(1,0,0,0)$, the matrix of $G_a^{\mu\nu}$ is
\begin{align*}
\operatorname{diag}\big(1,-c_a^2,-c_a^2,-c_a^2\big).
\end{align*}
Consequently, $G_a^{\mu\nu}$ has one positive and three negative eigenvalues. In the coordinates used in the theorem, define
\begin{align*}
A_a
&=
G_a^{00},
\\
B_a
&=
G_a^{0i}\xi_i,
\\
C_a
&=
G_a^{ij}\xi_i\xi_j.
\end{align*}
By the convention for the Lorentzian coordinates used throughout, $dt$ is timelike and therefore $g^{00}=g^{-1}(dt,dt)<0$. Together with Assumption \ref{Assumption_LWP}, this gives
\begin{align*}
A_a
&=
(1-c_a^2)(u^0)^2-c_a^2g^{00}
>
0.
\end{align*}
Set
\begin{align*}
S_a^{ij}
&=
G_a^{ij}-\frac{G_a^{i0}G_a^{0j}}{A_a}.
\end{align*}
The identity
\begin{align*}
&\bbm
1 & 0_{1\times3}
\\
-\frac{G_a^{i0}}{A_a} & I_3
\ebm
\bbm
A_a & G_a^{0j}
\\
G_a^{i0} & G_a^{ij}
\ebm
\bbm
1 & -\frac{G_a^{0j}}{A_a}
\\
0_{3\times1} & I_3
\ebm
=
\bbm
A_a & 0_{1\times3}
\\
0_{3\times1} & S_a^{ij}
\ebm
\end{align*}
and Sylvester's law of inertia show that $S_a^{ij}$ is negative definite. Hence, for every $\bm{\xi}\neq0$, the discriminant of
\begin{align*}
G_a^{\mu\nu}\xi_\mu\xi_\nu
&=
A_a\xi_0^2+2B_a\xi_0+C_a
\end{align*}
satisfies
\begin{align*}
\mathcal{Z}_a
&:=
B_a^2-A_aC_a
=
-A_aS_a^{ij}\xi_i\xi_j
>
0.
\end{align*}
The two roots of this factor are therefore $\xi_0=-\Lambda_a^\pm$, where
\begin{align*}
\Lambda_a^\pm
&=
\frac{B_a\pm\sqrt{\mathcal{Z}_a}}{A_a}
=
\frac{-c_a^2\Delta^{0i}\xi_i+u^0u^i\xi_i\pm\sqrt{\mathcal{Z}_a}}
{(u^0)^2-c_a^2\Delta^{00}}.
\end{align*}
Thus, $-\Lambda_a^+$ and $-\Lambda_a^-$ are real and distinct. The remaining root is $\xi_0=-\Lambda_2$, where $\Lambda_2=u^i\xi_i/u^0$, and it is real. The two roots $-\Lambda_1^\pm$ each have algebraic multiplicity two, and the remaining five roots have algebraic multiplicity one. Thus, the characteristic polynomial is homogeneous of degree $N\,r=9$, $\det[h_1]\neq0$, and all nine roots counted with multiplicity are real. This verifies Assumption \ref{SH:Assumption 1}.

This concludes the verification of Assumption \ref{SH:Assumption 1}.

\medskip
\noindent \underline{Step 6} (\textbf{Verification of Assumptions \ref{SH:Assumption 2} and \ref{SH:Assumption_polynomials}}).
We first verify that the seven roots listed in Step 5 are distinct for every $\bm{\xi}\neq0$. For each fixed $a=1,+,-$, the identity $\mathcal{Z}_a=B_a^2-A_aC_a=-A_aS_a^{ij}\xi_i\xi_j>0$ proved in Step 5 shows that the two roots $-\Lambda_a^+$ and $-\Lambda_a^-$ are distinct. If a root of the quadratic factor with velocity $c_a$ were also a root of the factor with velocity $c_b$, where $a\neq b$, subtracting the two equations would give
\begin{align*}
\big(c_a^2-c_b^2\big)
\Delta^{\mu\nu}\xi_\mu\xi_\nu
&=
0.
\end{align*}
At a root of any of the quadratic factors,
\begin{align*}
\Delta^{\mu\nu}\xi_\mu\xi_\nu
&>
0
\end{align*}
when $\bm{\xi}\neq0$: otherwise that factor would also give $u^\mu\xi_\mu=0$, and the decomposition
\begin{align*}
\xi_\mu
&=
-\big(u^\alpha\xi_\alpha\big)u_\mu
+\Delta_\mu^\alpha\xi_\alpha
\end{align*}
would give $\xi_\mu=0$. Since $c_a\neq c_b$ by Assumption \ref{Assumption_LWP}, roots belonging to different quadratic factors cannot coincide. The root of $u^\mu\xi_\mu$ cannot be a root of a quadratic factor either, since the two equations would again imply $\Delta^{\mu\nu}\xi_\mu\xi_\nu=0$. Hence the characteristic polynomial has exactly seven distinct roots for every $\bm{\xi}\neq0$.

Let $A=h(\partial\Psi,\xi)-h_1 \xi_0$ be a $9\times 9$ matrix according to Definition \ref{SH:Definition 1}. We need to show that $h_1^{-1}A$ is diagonalizable. Because $\mathcal{C}$ is invertible, we may use $\tilde{h}(\partial\Psi,\xi)=\tilde{h}_1\xi_0+\tilde{A}$, where $\tilde{A}=A\mathcal{C}^{-1}$, to prove it. Note that $h_1^{-1}A=\mathcal{C}^{-1}\tilde{h}_1^{-1}\tilde{A}\mathcal{C}$ are similar and, therefore, have the same eigenvalues and their eigenvectors are connected by $\mathcal{C}$. The eigenvalue problem $\tilde{h}_1^{-1}\tilde{A}\tilde{V}=\Lambda\tilde{V}$ may be rewritten as $\tilde{h}(\partial\Psi,\xi)\tilde{V}=0$, where $\xi_\mu=(-\Lambda,\bm{\xi})$.  The factorization in Equation~\eqref{E:MO_Characteristic_factorization} and the roots calculated in Step 5 show directly that the eigenvalues are $\Lambda_1^{\pm}$ with algebraic multiplicity two, $\Lambda_+^\pm$ and $\Lambda_-^\pm$ with algebraic multiplicity one, and $\Lambda_2=u^i\xi_i/u^0$ with algebraic multiplicity one. The preceding distinctness argument shows that these are seven distinct eigenvalues for every $\bm{\xi}\neq0$; hence $m=7$. The last thing is to prove that $\tilde{h}_1^{-1}\tilde{A}$ has 9 linearly independent eigenvectors, i.e., that the sum of the dimensions of the null space of $\tilde{h}(\partial\Psi,\xi)|_{\Lambda}$ for each $\Lambda$ is 9.  

When $\bm{\xi}=0$, the matrix $A=h(\partial\Psi,0)$ is
\begin{align*}
A
&=
\bbm
0_{4\times4} & 0_{4\times4} & 0_{4\times1}
\\
-\tilde{P}^{A\lambda}M^{-1}_{\lambda\nu} & 0_{4\times4} & 0_{4\times1}
\\
0_{1\times4} & 0_{1\times4} & 0
\ebm,
\end{align*}
where $\tilde{P}^{A\lambda}M^{-1}_{\lambda\nu}$ is invertible. Therefore, $A\neq0$ and $A^2=0$, so $A$ is not diagonalizable. Assumption \ref{SH:Assumption_polynomials}, however, requires diagonalizability only for $\bm{\xi}\neq0$. Thus, no verification at $\bm{\xi}=0$ is required. 

We now determine the eigenspaces directly for $\bm{\xi}\neq0$. Fix one of the two eigenvalues $\Lambda_1^\pm$ and set
\begin{align*}
\xi_\mu
&=
(-\Lambda_1^\pm,\bm{\xi}),
\\
a
&=
u^\mu\xi_\mu,
\\
v^\mu
&=
\Delta^{\mu\nu}\xi_\nu.
\end{align*}
The equation for this characteristic root is
\begin{align}
\label{E:MO_c1_root_relation}
a^2
&=
c_1^2v^\mu v_\mu.
\end{align}
In particular, $a\neq0$ and $v^\mu\neq0$. Indeed, the restriction of $g_{\mu\nu}$ to the vectors orthogonal to the timelike vector $u^\mu$ is positive definite. Hence, if either $a$ or $v^\mu$ vanished, Equation~\eqref{E:MO_c1_root_relation} and $c_1>0$ would imply that both vanish. The identity
\begin{align*}
\xi_\mu
&=
-au_\mu+v_\mu
\end{align*}
would then give $\xi_\mu=0$, contrary to $\bm{\xi}\neq0$. Since $u^\mu$ is timelike, $v^\mu$ is nonzero and orthogonal to $u^\mu$, and the two conditions
\begin{align*}
u^\nu w_\nu
&=
0,
\\
v^\nu w_\nu
&=
0
\end{align*}
define a two-dimensional space of covectors $w_\nu$. Choose a basis $w_{1,\nu}^\pm,w_{2,\nu}^\pm$ of that space and set
\begin{align*}
y_a^A
&=
w_{a,\nu}^\pm\tilde{P}^{A\nu},
\qquad
a=1,2,
\\
\tilde{V}_{1,a}^\pm
&=
\bbm
-\Lambda_1^\pm y_a^A
\\
y_a^A
\\
0
\ebm.
\end{align*}
We verify each of the three block rows of
\begin{align*}
\tilde{h}(\partial\Psi,\xi)\tilde{V}_{1,a}^\pm
&=
0.
\end{align*}
The second block row vanishes because $\xi_0=-\Lambda_1^\pm$. Since $P_{B\beta}y_a^B=w_{a,\beta}^\pm$, the third block row is
\begin{align*}
nv^\beta P_{B\beta}y_a^B
&=
nv^\beta w_{a,\beta}^\pm
\\
&=
0.
\end{align*}
For the first block row, Equation~\eqref{E:MO_c1_root_relation} gives
\begin{align*}
\mathbb{A}
&=
\rho\tau_Q
\big(a^2-c_1^2v^\mu v_\mu\big)
\\
&=
0.
\end{align*}
Equation~\eqref{E:MO_K_rank_two_form} and $K^\mu_\beta=L^\mu_\beta-nQ^\mu v_\beta/a$ therefore give
\begin{align*}
\xi_\alpha\xi_\nu H^{\mu\nu\alpha\beta}w_{a,\beta}^\pm
&=
K^\mu_\beta w_a^{\pm,\beta}
+\frac{n}{a}Q^\mu v_\beta w_a^{\pm,\beta}
\\
&=
\mathbb{A}w_a^{\pm,\mu}
+W_1^\mu u_\beta w_a^{\pm,\beta}
+W_2^\mu v_\beta w_a^{\pm,\beta}
\\
&=
0.
\end{align*}
Thus, $\tilde{V}_{1,1}^\pm$ and $\tilde{V}_{1,2}^\pm$ belong to the eigenspace of $\Lambda_1^\pm$. They are linearly independent because $w_{1,\nu}^\pm,w_{2,\nu}^\pm$ are linearly independent and $\tilde{P}^{A\nu}$ is invertible. Since the algebraic multiplicity of $\Lambda_1^\pm$ is two, its eigenspace has dimension exactly two.

Each of $\Lambda_+^+$, $\Lambda_+^-$, $\Lambda_-^+$, $\Lambda_-^-$, and $\Lambda_2$ has algebraic multiplicity one by Equation~\eqref{E:MO_Characteristic_factorization}. Each is a root of the characteristic polynomial and hence has a nonzero eigenvector. Its eigenspace therefore has dimension one. Eigenvectors belonging to distinct eigenvalues are linearly independent. Consequently, the sum of the dimensions of the seven eigenspaces is
\begin{align*}
2+2+1+1+1+1+1
&=
9.
\end{align*}

The above analysis completes the 9 linearly independent eigenvectors and, consequently, the matrix $h_1^{-1}A$ is diagonalizable. More precisely, the eigenspaces for $\Lambda_1^+$ and $\Lambda_1^-$ have dimension two, the eigenspaces for $\Lambda_+^+$, $\Lambda_+^-$, $\Lambda_-^+$, and $\Lambda_-^-$ have dimension one, and the eigenspace for $\Lambda_2$ has dimension one. Their dimensions therefore sum to
\begin{align*}
2+2+1+1+1+1+1
&=
9.
\end{align*}
Since $r=1$, the matrix $B_0$ in Equation~\eqref{E:xi_0_polynomial_h_r_inv_A} is $h_1^{-1}A$. Its diagonalizability verifies Assumption \ref{SH:Assumption_polynomials}. The seven distinct eigenvalue functions give $m=7$, and Step 5 showed that the characteristic polynomial has seven distinct roots. Therefore, the number of distinct roots is $7=m\,r$, which verifies Assumption \ref{SH:Assumption 2}. Since Assumption \ref{Assumption_LWP} holds at every state in $\mathcal{U}_{\mathrm{MO}}$, the same integer $m=7$ applies throughout that set.

This concludes the verification of Assumptions \ref{SH:Assumption 2} and \ref{SH:Assumption_polynomials}.

\medskip
\noindent \underline{Step 7} (\textbf{Assumption \ref{SH:Assumption3} and the application of Theorem \ref{Theorem_SH_LWP}}).\nopagebreak[4]
For the Leray indices fixed in Step 1,
\begin{align*}
m\,r
&=
7
\geq
\max_I m_I-\min_I n_I+1
=
2-0+1
=3.
\end{align*}
Thus, Assumption \ref{SH:Assumption3} holds. All the hypotheses of Theorem \ref{Theorem_SH_LWP} have now been verified.

\begin{samepage}
According to Theorem \ref{SH:Theorem1}, the Equations~\eqref{MO:Eqs} may be rewritten as a strictly hyperbolic system of ($m=7$) seventh-order PDEs in the variables $\Psi$. In particular, the strictly hyperbolic equations will be, up to an overall nonzero factor,
\begin{align*}
\mathcal{G}(\partial^{\tilde{m}_I-\tilde{n}_I})\Psi^I=(u^{\mu_1}u^{\mu_2}-c_1^2\Delta^{\mu_1\mu_2})(u^{\mu_3}u^{\mu_4}-c_+^2 \Delta^{\mu_3\mu_4})(u^{\mu_5}u^{\mu_6}-c_-^2\Delta^{\mu_5\mu_6})u^{\mu_7}\partial_{\mu_1}\cdots\partial_{\mu_7}\Psi^I\\
=\tilde{B}^I
(\partial^{\tilde{m}_K-\tilde{n}_I-1}\Psi^K).
\end{align*}
\end{samepage}
According to the result in Section~\ref{Sec:SH}, we may set the new Leray's indices as $\tilde{n}_{T^{0\mu}}=0$ (the equations for $T^{0\mu}$), $\tilde{m}_{T^{0\mu}}=7$, $\tilde{n}_{U^B}=\tilde{n}_{J^0}=1$ (equations for $U^B$ and $J^0$), and $\tilde{m}_{U^B}=\tilde{m}_{J^0}=8$.  

Applying Theorem \ref{Theorem_SH_LWP} with
\begin{align*}
d&=3,
&
N&=9,
&
r&=1,
&
m&=p=7,
&
m_*&=1
\end{align*}
gives a solution on a time interval $[0,T]$ with exactly the Sobolev regularity, lower-order time continuity, compact containment, uniqueness, and uniform choice of $T$ stated in Theorem \ref{T:Precise_mixed_order_local_well_posedness}.

Finally, Equation~\eqref{MO:LWP-Eqs} was obtained from Equations~\eqref{MO:Eqs} by the substitutions in Equations~\eqref{MO:Transformations}, which are well-defined on $\mathcal{U}_{\mathrm{MO}}$. Hence the solution furnished by Theorem \ref{Theorem_SH_LWP} solves Equations~\eqref{MO:Eqs}, with $T^{i\mu}$, $Y^A$, $J^i$, and $n$ given by Equations~\eqref{Tmunu_Mixed} and \eqref{MO:Transformations}.

\medskip
\noindent \underline{Step 8} (\textbf{Continuous dependence below the top Sobolev orders}).\nopagebreak[4]
Let the initial data converge as in the continuous-dependence assertion of the theorem. The right-hand side of Equation~\eqref{FO:YA} evaluated on $\{t=0\}$ and Equation~\eqref{E:MO_Initial_J_trace}, together with the Sobolev product and analytic-composition estimates, give convergence of $U^A_{(1)}$ and $J^0_{(1)}$ in $H^{\ell+7}(\Sigma)$. Thus, the initial data for Theorem \ref{Theorem_SH_LWP} converge in the top Sobolev spaces with $S_T=\ell+7$ and $S_U=S_J=\ell+8$, as computed in Step 7. The continuous-dependence conclusion of that theorem gives exactly the convergence asserted here for $0\leq j\leq6$ and every Sobolev exponent strictly below the corresponding exponent in the top-order bounds. This proves the continuous-dependence assertion.

This concludes the proof of Theorem \ref{T:Precise_mixed_order_local_well_posedness}, and hence of Theorem \ref{Theorem_2}.
\end{proof}

\begin{remark}
The restriction to $\Sigma=\mathbb{T}^3$ in Theorem \ref{T:Precise_mixed_order_local_well_posedness} is convenient for the numerical applications considered below. We explain here how an analogous statement on $\mathbb{R}^3$ can be obtained and which additional hypotheses are needed.

Theorem \ref{Theorem_SH_LWP} imposes two additional hypotheses when $\Sigma=\mathbb{R}^3$: the open set $\mathcal{U}$ must contain the zero jet, and the right-hand side of every equation must vanish at that jet. These hypotheses do not hold for Equations~\eqref{MO:LWP-Eqs} in the unshifted variables
\begin{align*}
\Psi
&=
\big(T^{0\mu},U^A,J^0\big).
\end{align*}
Indeed, $J^0=0$ gives $n=0$ by Equations~\eqref{MO:Transformations}, whereas $n>0$ in Assumption \ref{Assumption_physics}. The simultaneous conditions $T^{0\mu}=0$, $U^A=0$, and $J^0=0$ also need not define a point at which the transformations in Equations~\eqref{MO:Transformations} are well-defined. This difficulty cannot be removed by choosing an open set that excludes the zero jet. If the unshifted data satisfied Equation~\eqref{E:SH_Initial_data_regularity}, then every component of the initial jet would belong to $H^s(\mathbb{R}^3)$ for some $s>\frac{3}{2}$. Since $H^s(\mathbb{R}^3)\subset C_0(\mathbb{R}^3)$ for such $s$, the initial jet would converge to zero as $|x|\rightarrow\infty$. The closure of its range would therefore contain the zero jet. Thus, Equation~\eqref{E:SH_Initial_state_containment} would force $0\in\mathcal{U}_{\mathrm{MO}}$, which is incompatible with the physical open set used in Theorem \ref{T:Precise_mixed_order_local_well_posedness}.

Suppose first that the components of the background metric are constant in the coordinates under consideration. Choose constants $\overline{U}^A$ and $\overline{n}>0$ satisfying Assumptions \ref{Assumption_LWP} and \ref{Assumption_physics}, and define
\begin{align*}
\overline{J}^0
&=
\overline{n}\,\overline{u}^0,
\\*
\overline{T}^{0\mu}
&=
\overline{T}_0^{0\mu},
\\*
\overline{\Psi}
&=
\big(\overline{T}^{0\mu},\overline{U}^A,\overline{J}^0\big).
\end{align*}
Here, all the barred quantities on the right-hand sides are evaluated at the constant state determined by $\overline{U}^A$ and $\overline{n}$. Since the Christoffel symbols and all derivatives of the barred variables vanish, Equations~\eqref{Tmunu_Mixed} and \eqref{MO:Transformations} give
\begin{align*}
\overline{Y}^A
&=0,
\\
\overline{T}^{\mu\nu}
&=
\overline{T}_0^{\mu\nu},
\\
\overline{J}^{\mu}
&=
\overline{n}\,\overline{u}^{\mu}.
\end{align*}
It follows directly from Equations~\eqref{MO:Eqs} that $\overline{\Psi}$ is a constant solution. Assume that its jet
\begin{align*}
\overline{z}
&=
\mathcal{Z}_{\mathrm{MO}}[\overline{\Psi}]
\end{align*}
belongs to $\mathcal{U}_{\mathrm{MO}}$, and set
\begin{align*}
V
&=
\Psi-\overline{\Psi},
\\
\widehat{\mathcal{U}}_{\mathrm{MO}}
&=
\big\{z\,|\,\overline{z}+z\in\mathcal{U}_{\mathrm{MO}}\big\}.
\end{align*}
Then $0\in\widehat{\mathcal{U}}_{\mathrm{MO}}$ and
\begin{align*}
\mathcal{Z}_{\mathrm{MO}}[\Psi]
&=
\overline{z}+\mathcal{Z}_{\mathrm{MO}}[V].
\end{align*}

To verify the second additional hypothesis in Theorem \ref{Theorem_SH_LWP}, write Equations~\eqref{MO:LWP-Eqs} in the form in Equation~\eqref{SH:1}, with principal differential-operator matrix $h$ and right-hand side $B$. The equation for $V$ has coefficient matrix and right-hand side
\begin{align*}
\widehat{h}(z,\partial)^I_J
&=
h(\overline{z}+z,\partial)^I_J,
\\
\widehat{B}^I(z)
&=
B^I(\overline{z}+z)
-h(\overline{z}+z,\partial)^I_J\overline{\Psi}^J.
\end{align*}
The second term in the definition of $\widehat{B}^I$ cannot in general be omitted because the matrix in Equation~\eqref{MO:Matrix_h} has entries of order zero. If
\begin{align*}
\mathcal{E}^I[\Psi]
&=
h(\mathcal{Z}_{\mathrm{MO}}[\Psi],\partial)^I_J\Psi^J
-B^I(\mathcal{Z}_{\mathrm{MO}}[\Psi]),
\end{align*}
then
\begin{align*}
\widehat{B}^I(0)
&=
-\mathcal{E}^I[\overline{\Psi}]
=0,
\end{align*}
where the last equality follows because $\overline{\Psi}$ solves Equations~\eqref{MO:Eqs}. Thus, both additional hypotheses in Theorem \ref{Theorem_SH_LWP} hold for the translated system.

The initial data for the translated system are required to satisfy
\begin{align*}
T^{0\mu}_{(0)}-\overline{T}^{0\mu}
&\in
H^{\ell+7}(\mathbb{R}^3),
\\
U^A_{(0)}-\overline{U}^A
&\in
H^{\ell+8}(\mathbb{R}^3),
\\
J^0_{(0)}-\overline{J}^0
&\in
H^{\ell+8}(\mathbb{R}^3).
\end{align*}
The first time traces are still determined by Equations~\eqref{FO:YA} and \eqref{E:MO_Initial_J_trace}; since the reference solution is constant, their translated values are $U^A_{(1)}$ and $J^0_{(1)}$. The compact-containment condition becomes
\begin{align*}
\mathcal{Z}_{\mathrm{MO},0}[V]
&=
\big(
T^{0\mu}_{(0)}-\overline{T}^{0\mu},
U^A_{(0)}-\overline{U}^A,
J^0_{(0)}-\overline{J}^0,
U^A_{(1)},
\partial_iU^A_{(0)},
J^0_{(1)},
\partial_iJ^0_{(0)}
\big),
\end{align*}
and
\begin{align*}
\overline{
\big(\overline{z}+\mathcal{Z}_{\mathrm{MO},0}[V]\big)(\mathbb{R}^3)
}
&\subset
\mathcal{K}
\Subset
\mathcal{U}_{\mathrm{MO}}.
\end{align*}

To verify Equation~\eqref{E:SH_Analytic_data_approximation}, choose a real-valued, even function $a\in C_c^\infty(\mathbb{R}^3)$ such that $a=1$ on a neighborhood of the origin, and define
\begin{align*}
\mathcal{S}_\nu f
&=
\mathcal{F}^{-1}
\big[
a(\frac{k}{\nu})\widehat{f}(k)
\big].
\end{align*}
Set
\begin{align*}
T^{0\mu}_{(0),\nu}
&=
\overline{T}^{0\mu}
+\mathcal{S}_\nu
\big(T^{0\mu}_{(0)}-\overline{T}^{0\mu}\big)
\\
U^A_{(0),\nu}
&=
\overline{U}^A
+\mathcal{S}_\nu
\big(U^A_{(0)}-\overline{U}^A\big)
\\
J^0_{(0),\nu}
&=
\overline{J}^0
+\mathcal{S}_\nu
\big(J^0_{(0)}-\overline{J}^0\big).
\end{align*}
Because $a$ is real-valued and even, the identity $\widehat{f}(-k)=\overline{\widehat{f}(k)}$ for real-valued $f$ implies that $\mathcal{S}_\nu f$ is real-valued. Its Fourier transform has compact support, so $\mathcal{S}_\nu f$ is real analytic. Dominated convergence gives
\begin{align*}
T^{0\mu}_{(0),\nu}-\overline{T}^{0\mu}
&\rightarrow
T^{0\mu}_{(0)}-\overline{T}^{0\mu}
\quad\text{in }H^{\ell+7}(\mathbb{R}^3)
\\
U^A_{(0),\nu}-\overline{U}^A
&\rightarrow
U^A_{(0)}-\overline{U}^A
\quad\text{in }H^{\ell+8}(\mathbb{R}^3)
\\
J^0_{(0),\nu}-\overline{J}^0
&\rightarrow
J^0_{(0)}-\overline{J}^0
\quad\text{in }H^{\ell+8}(\mathbb{R}^3).
\end{align*}
Consequently,
\begin{align*}
\big(&
T^{0\mu}_{(0),\nu}-\overline{T}^{0\mu},
U^A_{(0),\nu}-\overline{U}^A,
J^0_{(0),\nu}-\overline{J}^0,
\partial_iU^A_{(0),\nu},
\partial_iJ^0_{(0),\nu}
\big)
\end{align*}
converges uniformly on $\mathbb{R}^3$ to
\begin{align*}
\big(&
T^{0\mu}_{(0)}-\overline{T}^{0\mu},
U^A_{(0)}-\overline{U}^A,
J^0_{(0)}-\overline{J}^0,
\partial_iU^A_{(0)},
\partial_iJ^0_{(0)}
\big)
\end{align*}
by Sobolev embedding. The right-hand sides of Equations~\eqref{FO:YA} and \eqref{E:MO_Initial_J_trace} depend only on these quantities after the barred constants have been added to the first three entries. The closure of the range of the limiting unshifted tuple is contained in the projection of $\mathcal{K}$ onto these entries. This projection is compact, and every denominator in Equations~\eqref{FO:YA} and \eqref{E:MO_Initial_J_trace} is bounded away from zero there. The uniform convergence therefore places the ranges of the approximating unshifted tuples, for all sufficiently large $\nu$, in one compact set on which these denominators are bounded away from zero and all derivatives of the coefficient functions needed in the Sobolev composition estimates are bounded.

For every such $\nu$, define $U^A_{(1),\nu}$ by evaluating Equation~\eqref{FO:YA} on $\{t=0\}$ using $T^{0\mu}_{(0),\nu}$, $U^A_{(0),\nu}$, and $J^0_{(0),\nu}$, and define $J^0_{(1),\nu}$ by evaluating the right-hand side of Equation~\eqref{E:MO_Initial_J_trace} using the same fields. These five traces are real analytic and satisfy Equation~\eqref{E:SH_Compatibility_conditions}. Both formulas contain at most one spatial derivative of $U^A_{(0)}$ and $J^0_{(0)}$, and both right-hand sides vanish at the constant equilibrium. Since $\ell\geq3$, $H^{\ell+7}(\mathbb{R}^3)$ is an algebra. The Sobolev product estimates and the composition estimates applied to the difference between each coefficient and its value at the equilibrium therefore give
\begin{align*}
U^A_{(1),\nu}
&\rightarrow
U^A_{(1)}
\quad\text{in }H^{\ell+7}(\mathbb{R}^3)
\\
J^0_{(1),\nu}
&\rightarrow
J^0_{(1)}
\quad\text{in }H^{\ell+7}(\mathbb{R}^3).
\end{align*}
Define
\begin{align*}
\mathcal{Z}_{\mathrm{MO},0,\nu}[V]
&=
\big(
T^{0\mu}_{(0),\nu}-\overline{T}^{0\mu},
U^A_{(0),\nu}-\overline{U}^A,
J^0_{(0),\nu}-\overline{J}^0,
U^A_{(1),\nu},
\partial_iU^A_{(0),\nu},
J^0_{(1),\nu},
\partial_iJ^0_{(0),\nu}
\big).
\end{align*}
The displayed Sobolev convergences and Sobolev embedding imply
\begin{align*}
\mathcal{Z}_{\mathrm{MO},0,\nu}[V]
&\rightarrow
\mathcal{Z}_{\mathrm{MO},0}[V]
\quad\text{uniformly on }\mathbb{R}^3.
\end{align*}
Since
\begin{align*}
\overline{
\big(\overline{z}+\mathcal{Z}_{\mathrm{MO},0}[V]\big)(\mathbb{R}^3)
}
&\subset
\mathcal{K}
\Subset
\mathcal{U}_{\mathrm{MO}},
\end{align*}
there exists a compact set $\mathcal{K}_0\Subset\mathcal{U}_{\mathrm{MO}}$ containing the ranges of $\overline{z}+\mathcal{Z}_{\mathrm{MO},0,\nu}[V]$ for all sufficiently large $\nu$. The five Sobolev convergences above are precisely the convergences required in Equation~\eqref{E:SH_Analytic_data_approximation}. Theorem \ref{Theorem_SH_LWP} therefore gives the conclusion of Theorem \ref{T:Precise_mixed_order_local_well_posedness} on $\mathbb{R}^3$, with each zeroth-order unknown replaced in the Sobolev assertions by its difference from the constant equilibrium. In particular, the solution belongs to the affine spaces specified by

\begin{align*}
T^{0\mu}-\overline{T}^{0\mu}
&\in
L^\infty\big([0,T];H^{\ell+7}(\mathbb{R}^3)\big),
\\*
U^A-\overline{U}^A
&\in
L^\infty\big([0,T];H^{\ell+8}(\mathbb{R}^3)\big),
\\*
J^0-\overline{J}^0
&\in
L^\infty\big([0,T];H^{\ell+8}(\mathbb{R}^3)\big).
\end{align*}
For $1\leq j\leq7$, one has $\partial_t^jV=\partial_t^j\Psi$, and these time derivatives satisfy the assertions in Theorem \ref{T:Precise_mixed_order_local_well_posedness}. The lower-order time continuity, uniqueness, and uniform existence time for neighboring admissible data hold in the translated Sobolev spaces.

There is an additional issue when the fixed analytic background metric has nonconstant coordinate components. A constant physical state need not solve Equations~\eqref{MO:Eqs}: the last term in Equation~\eqref{Tmunu_Mixed} contains the Christoffel symbols, and the right-hand sides in Equations~\eqref{MO:Eqs} also contain Christoffel symbols. In this case the calculation above gives
\begin{align*}
\widehat{B}^I((t,x),0)
&=
-\mathcal{E}^I[\overline{\Psi}](t,x),
\end{align*}
which is generally nonzero. Real analyticity of the metric does not make this residual vanish. Moreover, on $\mathbb{R}^3$, real analyticity alone does not give the uniform bounds in $x$ for the metric, its inverse, its derivatives, and the hyperbolicity constants entering Equation~\eqref{E:SH_Uniform_energy_estimate}.

One way to recover a direct application of Theorem \ref{Theorem_SH_LWP} is to prescribe an exact reference solution $\overline{\Psi}(t,x)$ of Equations~\eqref{MO:Eqs} on a neighborhood of $\{0\}\times\mathbb{R}^3$ and apply the theorem to $V=\Psi-\overline{\Psi}$. In this case, with $\overline{z}(t,x)=\mathcal{Z}_{\mathrm{MO}}[\overline{\Psi}](t,x)$, the translated right-hand side is
\begin{align*}
\widehat{B}^I((t,x),z)
&=
B^I\big((t,x),\overline{z}(t,x)+z\big)
\\
&\quad
-h\big((t,x),\overline{z}(t,x)+z,\partial\big)^I_J
\overline{\Psi}^J(t,x),
\end{align*}
and $\widehat{B}^I((t,x),0)=0$ because the reference solution satisfies Equations~\eqref{MO:LWP-Eqs}. To use the coordinate-dependent version of Theorem \ref{Theorem_SH_LWP} described after its proof, one must additionally assume that the metric and the reference solution have the real-analytic dependence needed for the translated coefficients, that the closure of the range of $\overline{z}$ is contained in a compact subset of $\mathcal{U}_{\mathrm{MO}}$, and that the coefficient bounds and hyperbolicity constants entering Equation~\eqref{E:SH_Uniform_energy_estimate} are uniform in $x$. The initial perturbations must satisfy Equations~\eqref{E:SH_Initial_data_regularity}, \eqref{E:SH_Compatibility_conditions}, \eqref{E:SH_Initial_state_containment}, and \eqref{E:SH_Analytic_data_approximation} for the translated system. An arbitrary nonconstant analytic background metric does not by itself provide such a reference solution or these uniform bounds.

A different extension could be considered when the metric approaches a constant metric at spatial infinity. Translation by a constant equilibrium then produces the nonzero source $-\mathcal{E}^I[\overline{\Psi}](t,x)$ displayed above, which may decay as $|x|\rightarrow\infty$. Treating that residual as an inhomogeneous source would require a version of Theorem \ref{Theorem_SH_LWP} in which the condition $B^I(0)=0$ is replaced by Sobolev hypotheses on the residuals produced in Equation~\eqref{SH:4} that suffice for Equation~\eqref{E:SH_Uniform_energy_estimate}. Those hypotheses are not formulated or proved here. Therefore, Theorem \ref{T:Precise_mixed_order_local_well_posedness} is stated only on $\mathbb{T}^3$, while the $\mathbb{R}^3$ conclusions described in this remark are conditional on the additional reference-state hypotheses just specified.
\end{remark}

\section{A new theorem on strictly hyperbolic reduction}
\label{Sec:SH}

Consider a PDE system of equations in $d$ spatial dimensions for the unknowns $\Psi=(\Psi^1, \cdots,\Psi^N)$. 

Let $N_{\mathcal{W}}\geq0$ be an integer. When $N_{\mathcal{W}}\geq1$, let
\begin{align*}
\mathcal{W}
&=
\big(\mathcal{W}^1,\dots,\mathcal{W}^{N_{\mathcal{W}}}\big)
\end{align*}
be a prescribed function, and, for each $I=1,\dots,N$, let $\chi_I\geq0$ be an integer. The notation $\partial^{\chi_I}\mathcal{W}$ denotes the collection of all spacetime derivatives of the components of $\mathcal{W}$ of order at most $\chi_I$. When $N_{\mathcal{W}}=0$, every argument and every hypothesis involving $\mathcal{W}$ is omitted.

The system is
\begin{align}
\label{SH:1}
\nonumber
\\
h(\partial^{m_K-n_I-1}\Psi^K,\partial^{m_J-n_I})^I_J\Psi^J
=
B^I(\partial^{m_K-n_I-1}\Psi^K;\partial^{\chi_I}\mathcal{W}).
\end{align}

In Equation~\eqref{SH:1}, we employ the notation of Leray indices \cite[Appendix~IV]{Choquet-Bruhat-Book-2009}, \cite[Appendix~B]{Disconzi-2024}, which works as follows. We assigned to each unknown $\Psi^J$ a non-negative integer $m_J$ and to each equation, numbered from $1$ to $N$, an integer $n_I$. The indices $m_J$ and $n_I$ are called Leray indices.
$h(\partial^{m_K-n_I-1}\Psi^K,\partial^{m_J-n_I})^I_J$ is a differential operator of order  $m_J-n_I$ (that could be zero) whose coefficients depend on at most  $m_K-n_I-1$ derivatives of $\Psi^K$ for each $K=1,\dots,N$. The sum convention is adopted unless stated otherwise. Thus, there is a sum over $J$ in $h(\partial^{m_K-n_I-1}\Psi^K,\partial^{m_J-n_I})^I_J\Psi^J$. However, in accordance with what was just stated, when a repeated index appears in a Leray index in the argument of $h^I_J$, like $m_K$ in $\partial^{m_K-n_I-1}\Psi^K$, it does not indicate a sum but rather a dependence on at most $m_K-n_I-1$ derivatives of $\Psi^K$ for each $K=1,\dots,N$.  Similarly, for each $I=1,\dots,N$, the function $B^I(\partial^{m_K-n_I-1}\Psi^K;\partial^{\chi_I}\mathcal{W})$ depends on at most $m_K-n_I-1$ derivatives of $\Psi^K$, $K=1,\dots,N$, and on at most $\chi_I$ derivatives of the prescribed function $\mathcal{W}$. The prescribed function does not occur in the coefficients of $h$ and therefore does not change the principal part. If $m_J-n_I < 0$, the corresponding term is absent (consistent with the differential operator $h(\partial^{m_K-n_I-1}\Psi^K,\partial^{m_J-n_I})^I_J$ being allowed to be zero).

For a given system of PDEs, when it is possible to find indices $m_J,n_I$ as described above, we say that the system forms a Leray system. We observe that the Leray indices are defined up to an overall additive integer. In order to fix them, we adopt the convention that we make the choice with the smallest possible values of $n_I \geq 0$, $I=1,\dots,N$.

We are interested in the Cauchy problem for Equation~\eqref{SH:1} wherein we are given as initial conditions the values of $\Psi^I$ and of its derivatives up to order $m_J-1$ along $\{ t = 0\}$, i.e., 
\begin{align*}
    \left. \Psi^J \right|_{t=0},\dots, \left. \partial^{m_J -1} \Psi^J \right|_{t=0}, \, J=1,\dots,N.
\end{align*}

\begin{remark}
    If $n_I > 0$ for some $I$, then the data needs to satisfy some compatibility conditions. These conditions will be automatically satisfied for the problems we consider in this paper because the data for higher-order equations derived from the first- and second-order systems we are studying are obtained by using the (first- and second-order) equations to solve for higher-order time derivatives at time zero in terms of the originally given Cauchy data. See \cite[Appendix~B]{Disconzi-2024} for details and a more comprehensive discussion of the Cauchy problem for Leray systems.
\end{remark}

We observe that Equation~\eqref{SH:1} is a quasilinear mixed-order system of PDEs and that there is no restriction on the orders of the equations involved. The principal part of Equation~\eqref{SH:1} is defined by the matrix of differential operators $h(\partial^{m_K-n_I-1}\Psi^K,\partial^{m_J-n_I})^I_J$. One can verify that this agrees with the usual definition of principal part for a system of fixed order, i.e., when all the differential operators $h(\partial^{m_K-n_I-1}\Psi^K,\partial^{m_J-n_I})^I_J$ have the same order. In the mixed-order case, it can be shown that this definition of principal part retains the key features of the familiar definition of principal part of a system of fixed order, namely: characteristics can be defined from the principal part; it allows us to determine higher-order time derivatives inductively in terms of the data (for non-characteristic data); the Cauchy-Kovalevskaya theorem can be applied in the analytic case (under natural assumptions on 
$h(\partial^{m_K-n_I-1}\Psi^K,\partial^{m_J-n_I})^I_J$ and $B^I(\partial^{m_K-n_I-1}\Psi^K;\partial^{\chi_I}\mathcal{W})$); local well-posedness (in different classes of spaces)
\cite{Leray-Ohya-1964,Leray-Ohya-1967,Leray-Book-1953,Lichnerowicz-Book-1967,Choquet-Bruhat-1966,Mizohata-Book-1985,Mizohata-1965,Hormander-Book-1976} can be determined solely from properties of the principal part. 

Unfortunately, the mixed-order system of interest in this paper does not fall under the assumptions of the results mentioned in the previous paragraph. This is our main motivation to establish Theorems \ref{SH:Theorem1} and \ref{Theorem_SH_LWP}, although these are general theorems about the system in Equation~\eqref{SH:1} and thus of independent interest.

\begin{definition}
We denote by $r$ the highest-order time derivative that appears in the principal part of Equation~\eqref{SH:1}.
\end{definition}

We observe that the highest-order derivative appearing in the principal part of Equation~\eqref{SH:1} is $\max_K m_K- \inf_K n_K$. Thus
\begin{align}
\label{time_derivative_and_all_derivatives_inequality}
    r\le \max_K m_K- \inf_K n_K.
\end{align}
The inequality in Equation~\eqref{time_derivative_and_all_derivatives_inequality}  reflects the fact that the system may be of mixed order in time and spatial derivatives.

In what follows, we will have several quantities depending on derivatives of $\Psi$. To simplify the notation, we will indicate this dependence by $\partial \Psi$. This should not be interpreted as saying that those quantities depend only on first-order derivatives of $\Psi$.

\begin{definition}
\label{SH:Definition 1}
We denote by $h(\partial\Psi,\partial)$ the differential operator matrix with entries  $h(\partial^{m_K-n_I-1}\Psi^K,\partial^{m_J-n_I})^I_J$, i.e., the principal part of Equation~\eqref{SH:1}. The characteristic polynomial of $h$ is defined as
\begin{align*}
   \mathcal{P}(\xi)=\det[h(\partial\Psi,\xi)] 
\end{align*}
where the substitution $\partial_\mu\to \xi_\mu$ has been performed. We also define the expansion
\begin{align}
\label{SH:2}
h(\partial\Psi,\xi)=h_r(\partial\Psi) \xi_0^r+A(\partial\Psi,\xi_0^{\mathfrak{m}},\bm{\xi}^{\mathfrak{n}}),
\end{align}
where $\bm{\xi}=(\xi_1,\dots, \xi_d)$, $d$ being the spatial dimension; $h_r$ and $A$ are $N\times N$ real matrices defined by 
\begin{align*}
h_r(\partial\Psi,\bm{\xi})=\frac{1}{r!}\frac{\partial^r h(\partial\Psi,\xi)}{\partial\xi_0^r},
\end{align*}
and $A(\partial\Psi,\xi_0^{\mathfrak{m}},\bm{\xi}^{\mathfrak{n}})=h(\partial\Psi,\xi)-h_r(\partial\Psi,\bm{\xi}) \xi_0^r$, where the entry $A(\partial\Psi,\xi_0^{\mathfrak{m}},\bm{\xi}^{\mathfrak{n}})^I_J$ may have powers $0\le \mathfrak{m}\le r-1$ in $\xi_0$ such that $\mathfrak{m}+\mathfrak{n}= m_J-n_I$.
\end{definition}

We remark that $\mathcal{P}(\xi)$ is a homogeneous polynomial of order $\sum_I m_I-\sum_J n_J$ in $\xi$. Because we are interested in determining $\xi_0 = \xi_0(\bm{\xi})$ for each given $\bm{\xi} \neq 0$, we will often write $\mathcal{P}(\xi_0)$ for $\mathcal{P}(\xi)$. From now on, we will be always considering $\bm{\xi} \neq 0$ unless stated otherwise.

\begin{assumption}
\label{SH:Assumption 1} 
We assume that $\mathcal{P}(\xi_0)$ is a homogeneous polynomial of order $Nr$, that\newline $\det[h_r(\partial\Psi,\bm{\xi})]\neq0$ for $\bm{\xi}\neq0$, and that the roots $\xi_0=r_i(\bm{\xi})$, $i=1,\cdots,N r$ of $\mathcal{P}(\xi_0)$ are all real.
\end{assumption}

We observe that $\det[h_r(\partial\Psi,\bm{\xi})]\neq0$ implies that there is a $\partial_t^r$ in each equation of the system. Otherwise, $\frac{\partial^r h(\partial\Psi,\xi)}{\partial\xi_0^r}$ would produce a row of zeros. 

Under Assumption \ref{SH:Assumption 1}, the characteristic determinant is of order $N r$, which means that $\sum_Im_I-\sum_Jn_J=N r$. Furthermore, the characteristic polynomial may be written as
\begin{align*}
\mathcal{P}(\xi_0)&=W(\partial^s\Psi)^{\mu_1\cdots\mu_{Nr}}\xi_{\mu_1}\cdots\xi_{\mu_{Nr}}\\
&=F(\partial^{s}\Psi)\Pi_{i=1}^{Nr}(\xi_0-r_i(\bm{\xi})),
\end{align*}
where the roots $r_i(\bm{\xi})$, $F$, and $W^{\mu_1\cdots\mu_{Nr}}$ (with $W^{00\cdots 0}\ne 0$) are functions of $\Psi$ and its derivatives up to order $s=\max_K m_K-\inf_K n_K-1$. Note that, although the system of equations may be mixed in space and time derivatives, the characteristic determinant must be of degree $N r$ under Assumption \ref{SH:Assumption 1}. 

From now on, let us use 
\begin{align*}
h_r:= h_r(\partial\Psi,\bm{\xi})\quad\text{and}\quad A:=A(\partial\Psi,\xi_0^{\mathfrak{m}},\bm{\xi}^\mathfrak{n})
\end{align*}
for brevity. Because $h_r$ is assumed to be invertible, we can define the eigenvalue problem for the matrix $h_r^{-1}A$. 
\begin{definition}
\label{D:Distinct_eigenvalues}
Let $\lambda_k(\xi)$, $k=1,\cdots, m\le N$, be the $m$ distinct eigenvalues of the matrix $h_r^{-1}A$ considered as functions of $\xi$ (see Remark \ref{R:Distinct_eigenvalues_functions}) and write 
\begin{align}
\label{SH:Characteristic_Eigenval}
\det[h_r^{-1}A-\lambda I_N]=\Pi_{k=1}^{m}[\lambda_k(\xi)-\lambda]^{\alpha_k},
\end{align}
where $\alpha_k$ is the multiplicity with which the eigenvalue function $\lambda_k$ occurs. Define
\begin{align}
\label{SH:Polynomials}
p_k(\xi_0):=\xi_0^r+\lambda_k(\xi).
\end{align}
\end{definition}

\begin{remark}
    \label{R:Distinct_eigenvalues_functions}
    We emphasize that in Definition \ref{D:Distinct_eigenvalues}, the eigenvalues are taken as distinct if they are distinct functions of $\xi$, but this does not prevent distinct eigenvalue functions from coinciding at certain values of $\xi$, nor does the proof require them to be unequal at every $\xi$. For example, if $\lambda_1(\xi) = \xi_0 +1$ and $\lambda_2(\xi) = \xi_0 + 1$, then we take $\lambda(\xi) = \xi_0 +1$ to be an eigenvalue of multiplicity two since $\lambda_1$ and $\lambda_2$ are identically the same function of $\xi$. If on the other hand we have $\lambda_1(\xi) = \xi_0+1$ and $\lambda_2(\xi)=-\xi_0+1$, then $\lambda_1$ and $\lambda_2$ are taken to be distinct eigenvalues, even if  for certain values in their domain, namely, $\xi=(0,\bm{\xi})$, their values agree, $\lambda_1(0,\bm{\xi}) = 1 =\lambda_2(0,\bm{\xi})$.
\end{remark}

\begin{assumption}
\label{SH:Assumption 2} 
Assume that the number of distinct roots (which we already know to be real) of the characteristic polynomial is $m\,r$, where $m$ is the number of distinct eigenvalues in Equation~\eqref{SH:Characteristic_Eigenval}.  
\end{assumption}

\begin{assumption}
\label{SH:Assumption3}
We assume that $m\,r\ge \max_Im_I-\min_Jn_J+1$. 
\end{assumption}

Observe that we can write
\begin{align}
\label{E:xi_0_polynomial_h_r_inv_A}
    h_r^{-1} A = \sum_{j=0}^{r-1} B_j(\bm{\xi})\xi_0^j.
\end{align}

\begin{assumption}
\label{SH:Assumption_polynomials}
Assume that the matrices $B_j(\bm{\xi})$ in Equation~\eqref{E:xi_0_polynomial_h_r_inv_A} are simultaneously diagonalizable for every $\bm{\xi}\neq0$. Observe that when $r=1$, this assumption states that $h_r^{-1} A=B_0$ is diagonalizable.
\end{assumption}

\begin{lemma}
Under Assumptions
\ref{SH:Assumption 1} and \ref{SH:Assumption_polynomials}, the functions $p_k(\xi_0)$ defined in Equation~\eqref{SH:Polynomials} are polynomials in $\xi_0$ of degree $r$.
\end{lemma}
\begin{proof}
$h_r$ does not depend on $\xi_0$ and $A$ has entries with order up to $\xi_0^{r-1}$. Thus the result is immediate for $r=1$. 
Let $P=P(\bm{\xi})$ be the matrix that simultaneously diagonalizes the $B_j$ matrices according to Assumption \ref{SH:Assumption_polynomials}, so that
\begin{align*}
    P^{-1} B_j P = \operatorname{diag}(b_{j1}, \dots, b_{jN}), \quad j=0,\dots, r-1.
\end{align*}
We thus obtain the following polynomial eigenvalue functions of $P^{-1} h_r^{-1} A P$ and also of $h_r^{-1} A$ ,
\begin{align*}
    \Lambda_k(\xi) = \sum_{j=0}^{r-1} b_{jk}(\bm{\xi}) \xi_0^j,\quad k=1,\dots,N.
\end{align*}
In accordance with Definition
\ref{D:Distinct_eigenvalues}, denote the distinct functions among them $\lambda_1,\dots,\lambda_m$.
Permuting the columns of $P$ if necessary, we can assume that $\Lambda_k = \lambda_k$, $k=1,\dots,m$.
It thus follows that 
\begin{align*}
    \lambda_k(\xi) = \sum_{j=0}^{r-1} b_{jk}(\bm{\xi}) \xi_0^j,
\end{align*}
which implies the result.
\end{proof}

\begin{remark}
    Without Assumption \ref{SH:Assumption_polynomials}, Lemma XI.1 may fail for $r \geq 2$. To see it, consider the following example with $N=2$ and $r=2$,
    \begin{align*}
        h(\xi_0,\bm{\xi}) &= 
        \begin{bmatrix}
            \xi_0^2 - \bm{\xi}^2 & \xi_0 \xi_1 \\
            \xi_0 \xi_1 & \xi_0^2-4\bm{\xi}^2
        \end{bmatrix}
        \\
        & = I \xi_0^2 + 
        \begin{bmatrix}
            - \bm{\xi}^2 & \xi_0 \xi_1 \\
            \xi_0 \xi_1 & -4\bm{\xi}^2
        \end{bmatrix},    
    \end{align*}
so that $h_r$ is the $2\times 2$ identity matrix and 
\begin{align*}
A= 
         \begin{bmatrix}
            - \bm{\xi}^2 & \xi_0 \xi_1 \\
            \xi_0 \xi_1 & -4\bm{\xi}^2
        \end{bmatrix}   
\end{align*}    
has entries of degree at most $r-1=1$ in $\xi_0$, as required by Equation~\eqref{SH:2}.

The characteristic determinant is
\begin{align*}
    \det h = \xi_0^4 -(5\bm{\xi}^2+\xi_1^2)\xi_0^2 +4 \bm{\xi}^4.
\end{align*}
With $a:= 5\bm{\xi}^2+\xi_1^2$, this quartic polynomial has two roots
\begin{align*}
    z_\pm = \frac{a\pm\sqrt{a^2-16\bm{\xi}^4}}{2}
\end{align*}
in $z = \xi_0^2$, which are positive and distinct for $\bm{\xi}\neq 0$ since
\begin{align*}
    a \geq 5\bm{\xi}^2,\quad
    a^2 - 16\bm{\xi}^4 \geq 9 \bm{\xi}^4 > 0, \quad
    z_+ z_- = 4\bm{\xi}^4 > 0,
\end{align*}
so that Assumption \ref{SH:Assumption 1} is verified.

The matrix 
 $h_r^{-1} A$
is real and symmetric, thus diagonalizable\footnote{Thus, adding to Assumption \ref{SH:Assumption 2} the hypothesis that $h_r^{-1} A$ is diagonalizable does not rule out this counter-example when $r>1$. For $r=1$, adding such a hypothesis is equivalent to Assumption \ref{SH:Assumption_polynomials}.}. Its eigenvalues are
\begin{align*}
    \lambda_\pm = -\frac{5}{2}\bm{\xi}^2\pm \frac{1}{2}\sqrt{9\bm{\xi}^4+4\xi_0^2 \xi_1^2}
\end{align*}
which are distinct for $\bm{\xi}\neq 0$, thus $m=2$ and the characteristic polynomial has $4=mr$ distinct real roots, thus Assumption \ref{SH:Assumption 2} is satisfied. Moreover, Assumption \ref{SH:Assumption3} is also verified since $mr=4\geq 2-0+1=3$.

However, 
\begin{align*}
    p_\pm(\xi_0) = \xi_0^2 -\frac{5}{2}\bm{\xi}^2\pm \frac{1}{2}\sqrt{9\bm{\xi}^4+4\xi_0^2 \xi_1^2}
\end{align*}
is in general not a polynomial in $\xi_0$ if $\xi_1 \neq 0$. Indeed, taking $\bm{\xi}=(1,0,\dots,0)$, we have that the radicand is $9+4\xi_0^2$. For $p_\pm$ to be polynomials we would need  $9+4\xi_0^2=(B\xi_0+C)^2$ for some $B,C$, but this would give $B^2=4$, $C^2=9$, and $BC=0$, which is impossible.
\end{remark}

\begin{theorem}[Strict hyperbolicity]
\label{SH:Theorem1}
Under Assumptions \ref{SH:Assumption 1}, \ref{SH:Assumption 2},  \ref{SH:Assumption3}, and \ref{SH:Assumption_polynomials},  every sufficiently regular solution of Equation~\eqref{SH:1} satisfies a system of strictly hyperbolic partial differential equations of order $m\,r$.
\end{theorem}

Before we prove Theorem \ref{SH:Theorem1}, we need to establish some results. 
\begin{proposition}
\label{SH:Lemma2}
Under Assumptions \ref{SH:Assumption 1}, \ref{SH:Assumption 2}, and \ref{SH:Assumption_polynomials}, the number of distinct roots among the $m$ distinct functions $p_k(\xi_0)$ is exactly the $m\,r$ distinct characteristic roots $r_1(\bm{\xi}),\dots,r_{m\,r}(\bm{\xi})$. Moreover, $\det[h_r]=F(\partial^s\Psi)$.
\end{proposition}
\begin{proof}
From Assumption \ref{SH:Assumption_polynomials}, there exists a matrix $P$ such that $h_r^{-1}A=PJP^{-1}$, where $J=\lambda_1(\xi)\oplus\cdots\oplus \lambda_N(\xi)$ is the diagonal matrix containing the eigenvalues as entries. The repeated eigenvalues are labeled as follows: for any $i>m$ there exists a $k\le m$ such that $\lambda_i=\lambda_k$. Therefore, we may write
\begin{align}
\label{SH:Matrix_h}
h(\partial\Psi,\xi)=h_rP DP^{-1},
\end{align}
where $D=p_1(\xi_0)\oplus\cdots\oplus p_N(\xi_0)$. 

We then obtain that
\begin{align*}
\mathcal{P}(\xi_0)=\det[h_r]\Pi_{k=1}^m[p_k(\xi_0)]^{\alpha_k}=F(\partial^s\Psi)\Pi_{i=1}^{Nr}(\xi_0-r_i(\bm{\xi})).
\end{align*}
Consequently, $F(\partial^s\Psi)=\det[h_r]$ and, because $\mathcal{P}(\xi_0)$ has exactly $m\,r$ distinct roots, then the distinct $m$ functions $p_1(\xi_0),\cdots,p_m(\xi_0)$ must all have distinct roots. (Note that, although $h_r$ may depend on $\bm{\xi}$, its determinant cannot.) Furthermore, because the above factorization is unique, we obtain that
\begin{align*}
p_k(\xi_0)=\Pi_{i=1}^r(\xi_0-r_{k_i}(\bm{\xi})).
\end{align*} 
\end{proof}

\begin{remark}
The ordering of the polynomials $p_i(\xi_0)$ follows that of the eigenvalues $\lambda_i$ in the sense that $p_1(\xi_0),\,p_2(\xi_0),\,\cdots,\,p_m(\xi_0)$ are the distinct polynomials and, for $i>m$ there exists a $k\le m$ such that $p_i(\xi_0)=p_k(\xi_0)$. 
\end{remark}
Next, consider the inverse of $h(\partial\Psi,\xi)$, which now may be written as
\begin{align*}
h(\partial\Psi,\xi)^{-1}= P D^{-1} P^{-1} h_r^{-1}.
\end{align*}
Let $H(\partial\Psi,\xi)$ be the adjugate matrix of $h(\partial\Psi,\xi)$. Therefore,
\begin{align*}
H(\partial\Psi,\xi)&=\det[h(\partial\Psi,\xi)][h(\partial\Psi,\xi)]^{-1}\\
&= 
\det[h_r(\partial\Psi,\xi)]
P \begin{bmatrix}
\Pi_{i=2}^N p_i(\xi_0) & 0 & \cdots & 0\\
0 & \Pi_{i=1, i\ne 2}^N p_i(\xi_0) & \cdots & 0\\
\vdots & \vdots & \ddots & \vdots\\
0 & 0 & \cdots & \Pi_{i=1,i\neq N}^N p_i(\xi_0)
\end{bmatrix} P^{-1} h_r^{-1}.
\end{align*}
Define 
\begin{align}
\label{SH:q}
q(\xi_0)=\Pi_{k=1}^m[p_k(\xi_0)]^{\alpha_k-1}
\end{align}
which 
is a polynomial of order $(N-m)r$ that contains all repeated polynomials $p_i$.
\begin{lemma}
\label{SH:Lemma5}
Under Assumptions \ref{SH:Assumption 1}, \ref{SH:Assumption 2}, and \ref{SH:Assumption_polynomials}, the polynomial $q(\xi)$ is homogeneous in $\xi_\mu$.
\end{lemma}
\begin{proof}
Define
\begin{align*}
\Delta(\xi)=\frac{\det[h(\partial\Psi,\xi)]}{\det[h_r]}
=\Pi_{k=1}^m[p_k(\xi_0)]^{\alpha_k}.
\end{align*}
By Proposition \ref{SH:Lemma2}, each $p_k$ has $r$ distinct roots and, if $k\neq\ell$, then $p_k$ and $p_\ell$ have no common root. Therefore, considering $\bm{\xi}$ as a parameter and taking the greatest common divisor to be monic in $\xi_0$, we have
\begin{align*}
\operatorname{gcd}\left(\Delta,\frac{\partial\Delta}{\partial\xi_0}\right)
=\Pi_{k=1}^m[p_k(\xi_0)]^{\alpha_k-1}
=q(\xi),
\end{align*}
where $\operatorname{gcd}$ means greatest common divisor.
It remains to show that the greatest common divisor in this identity is a homogeneous polynomial in all the components of $\xi$. By Assumption \ref{SH:Assumption 1} and Proposition \ref{SH:Lemma2}, $\Delta$ is a homogeneous polynomial of degree $Nr$ and is monic as a polynomial in $\xi_0$. Factor $\Delta$ into irreducible polynomials in $\mathbb{R}[\xi_0,\xi_1,\ldots,\xi_d]$:
\begin{align*}
\Delta(\xi)=\Pi_{a=1}^{M}[f_a(\xi)]^{\beta_a},
\end{align*}
where the $f_a$ are distinct. Because $\Delta$ is homogeneous, the factors $f_a$ may be chosen homogeneous. Moreover, no $f_a$ is independent of $\xi_0$, since otherwise $f_a$ would divide the coefficient $1$ of $\xi_0^{Nr}$ in $\Delta$. Hence $\partial f_a/\partial\xi_0\neq0$. Since $f_a$ is irreducible, it does not divide $\partial f_a/\partial\xi_0$, whose degree in $\xi_0$ is strictly smaller than that of $f_a$. Consequently,
\begin{align*}
\operatorname{gcd}\left(\Delta,\frac{\partial\Delta}{\partial\xi_0}\right)
=\Pi_{a=1}^{M}[f_a(\xi)]^{\beta_a-1}.
\end{align*}
The polynomial on the right-hand side is homogeneous. The monic normalization of this greatest common divisor is $q(\xi)$ by the first identity, and the result follows.
\end{proof}

\begin{lemma}
\label{SH:Adjugate_divisibility}
Under Assumptions \ref{SH:Assumption 1}, \ref{SH:Assumption 2}, and \ref{SH:Assumption_polynomials}, every entry of the adjugate matrix $H(\partial\Psi,\xi)$ is divisible by the polynomial $q(\xi)$ defined in Equation~\eqref{SH:q}.
\end{lemma}

\begin{proof}
If $m=N$, then $\alpha_k=1$ for every $k=1,\dots,N$, and hence $q=1$. The result is immediate in this case. Assume, therefore, that $m<N$. For every $a=1,\dots,N$, let $k(a)$ be the unique integer in $\{1,\dots,m\}$ such that
\begin{align*}
p_a(\xi_0)=p_{k(a)}(\xi_0).
\end{align*}
The existence and uniqueness of $k(a)$ follow from the fact that $p_1,\dots,p_m$ are the distinct polynomials among $p_1,\dots,p_N$. Moreover,
\begin{align*}
\alpha_k=\#\{a\in\{1,\dots,N\}\,|\,k(a)=k\}.
\end{align*}
It follows from Equation~\eqref{SH:q} that, for every $a=1,\dots,N$,
\begin{align}
\label{SH:Product_q_factor}
\begin{split}
\Pi_{\substack{b=1\\b\neq a}}^N p_b(\xi_0)
&=[p_{k(a)}(\xi_0)]^{\alpha_{k(a)}-1}
\Pi_{\substack{\ell=1\\\ell\neq k(a)}}^m[p_\ell(\xi_0)]^{\alpha_\ell}
\\
&=q(\xi_0)\Pi_{\substack{\ell=1\\\ell\neq k(a)}}^m p_\ell(\xi_0).
\end{split}
\end{align}
For each fixed $\bm{\xi}\neq0$, Equation~\eqref{SH:Matrix_h}, the identity $H=\det[h]h^{-1}$, and Equation~\eqref{SH:Product_q_factor} give, whenever $\det[h(\partial\Psi,\xi)]\neq0$,
\begin{align}
\label{SH:Adjugate_q_factorization}
H(\partial\Psi,\xi)
={}&q(\xi_0)\det[h_r(\partial\Psi,\xi)]P
\operatorname{diag}\left(
\Pi_{\substack{\ell=1\\\ell\neq k(1)}}^m p_\ell(\xi_0),
\dots,
\Pi_{\substack{\ell=1\\\ell\neq k(N)}}^m p_\ell(\xi_0)
\right)P^{-1}h_r^{-1}.
\end{align}
Both sides of Equation~\eqref{SH:Adjugate_q_factorization} are polynomials in $\xi_0$, and hence the equality holds for every $\xi_0$.

It remains to verify that the quotient of each entry of $H$ by $q$ is a polynomial in all the components of $\xi$. By Lemma \ref{SH:Lemma5}, $q(\xi)$ is a polynomial in $\xi$. It is monic as a polynomial in $\xi_0$ by Equations~\eqref{SH:q} and \eqref{SH:Polynomials}. Each entry $H^I_J(\partial\Psi,\xi)$ is also a polynomial in $\xi$, since it is a cofactor of the polynomial matrix $h(\partial\Psi,\xi)$. Therefore, division by the monic polynomial $q$, with respect to $\xi_0$, gives polynomials $S^I_J(\partial\Psi,\xi)$ and $R^I_J(\partial\Psi,\xi)$ such that
\begin{align}
\label{SH:Adjugate_polynomial_division}
H^I_J(\partial\Psi,\xi)
=q(\xi)S^I_J(\partial\Psi,\xi)+R^I_J(\partial\Psi,\xi),
\end{align}
where
\begin{align*}
\deg_{\xi_0}R^I_J<(N-m)r=\deg_{\xi_0}q.
\end{align*}
For every fixed $\bm{\xi}\neq0$, Equation~\eqref{SH:Adjugate_q_factorization} and the uniqueness of the remainder in Equation~\eqref{SH:Adjugate_polynomial_division} imply that $R^I_J(\partial\Psi,\xi_0,\bm{\xi})=0$ for every $\xi_0$. Hence, every coefficient of $R^I_J$, considered as a polynomial in $\xi_0$, vanishes for every $\bm{\xi}\neq0$. These coefficients are polynomials in $\bm{\xi}$ and must therefore vanish identically. Thus $R^I_J=0$ for every $I,J=1,\dots,N$, which proves the result.
\end{proof}

\begin{remark}
A direct consequence is that the degree $rm$ polynomial $\Pi_{k=1}^m p_k(\xi_0)$ is also a homogeneous polynomial in $\xi_\mu$ and that 
\begin{align}
\label{SH:Fact_Det}
\det[h(\partial\Psi,\xi)]=\det[h_r(\partial\Psi,\xi)]q(\xi_0)\Pi_{k=1}^m p_k(\xi_0).
\end{align}
\end{remark}
\begin{lemma}
\label{SH:Lemma4}
Under Assumption \ref{SH:Assumption 1}, the adjugate matrix $H(\partial\Psi,\xi)$ satisfies
\begin{align*}
H(\partial\Psi,\xi)^I_J h(\partial\Psi,\xi)^J_K=\delta^I_K\det[h(\partial\Psi,\xi)].
\end{align*}
Then, 
\begin{align*}
H(\partial\Psi,\partial)^I_J=H(\partial^{m_K-\min_{L\ne J}n_L-1}\Psi^K,\partial^{Nr-m_I+n_J})^I_J,
\end{align*}
where, in the passage $H(\partial\Psi,\xi)\to H(\partial\Psi,\partial)$, we take the differentiation ordering 
\begin{align*}
H(\partial\Psi,\partial^{Nr-m_I+n_J})^I_J=
[H(\partial\Psi)^I_J]^{\mu_1\cdots\mu_{Nr-m_I+n_J}}\partial_{\mu_1}\cdots\partial_{\mu_{Nr-m_I+n_J}}.
\end{align*}
\end{lemma}
\begin{proof}
$H$ is, in fact, the adjugate of $h$ and, thus, the transpose of the cofactor matrix $C$ of $h$. Therefore,
\begin{align*}
H^I_J=C^J_I=\frac{1}{(N-1)!}\epsilon^{I\,\mu_1\mu_2\cdots\mu_{N-1}}\epsilon_{J\,\nu_1\nu_2\cdots\nu_{N-1}}h^{\nu_1}_{\mu_1}h^{\nu_2}_{\mu_2}\cdots h^{\nu_{N-1}}_{\mu_{N-1}},
\end{align*}
where there is an implicit summation over the indices $\mu_i$ and $\nu_j$ from 1 to $N$ and $\epsilon^{12\cdots N}=\epsilon_{12\cdots N}=1$ is the skew-symmetric Levi-Civita pseudo-tensor.
Hence, because  $h^{\nu_j}_{\mu_i}=h^{\nu_j}_{\mu_i}(\partial^{m_K-n_{\nu_{j}}-1}\Psi^K,\xi^{m_{\mu_i}-n_{\nu_j}})$, then $H^I_J(\partial^{\mathfrak{a}_{IK}}\Psi^K,\xi^{\mathfrak{b}_{IJ}})$ is such that 
\begin{align*}
\mathfrak{a}_{IK}&=m_K-\min_{L\ne J}n_L-1,\\
\mathfrak{b}_{IJ}&=\sum_{K\ne I}m_K-\sum_{K\ne J}n_K=Nr-m_I+n_J.
\end{align*}   
\end{proof}
Since $H(\partial\Psi,\xi)$ is divisible by the homogeneous polynomial $q(\xi_0)$ of degree $(N-m)r$ given in Equation~\eqref{SH:q}, we may define the matrix $\hat{H}(\partial\Psi,\xi)$ by the equation
\begin{align*}
H(\partial\Psi,\xi)= q(\xi)\hat{H}(\partial\Psi,\xi).
\end{align*}
Because of the factorization in Equation~\eqref{SH:Fact_Det}, 
\begin{align}
\label{SH:Identity1}
\hat{H}(\partial\Psi,\xi)^I_J h(\partial\Psi,\xi)^J_K=F(\partial^s\Psi)\Pi_{k=1}^m p_k(\xi_0)\delta^I_K,
\end{align}
where, by Proposition \ref{SH:Lemma2} and Lemma \ref{SH:Lemma5}, $\Pi_{k=1}^m p_k(\xi_0)$ is a homogeneous polynomial in $\xi_\mu$ of degree $mr$ with no repeated roots, $F(\partial^s\Psi)\ne 0$, and $s=\max_Km_K-\min_Ln_L-1$.  
\begin{remark}
Because $(N-m)r$ is the degree of the polynomial $q(\xi)$,
\begin{align*}
\hat{H}(\partial\Psi,\xi)^I_J&=\hat{H}(\partial^{m_K-\min_Ln_L-1}\Psi^K,\xi^{m\,r-m_I+n_J})^I_J.
\end{align*}
\end{remark}
\begin{definition}
In the passage $\xi\to\partial$, we use the differentiation ordering
\begin{align*}
\hat{H}(\partial\Psi,\xi^{mr-m_I+n_J})^I_J
=[\hat{H}(\partial\Psi)^I_J]^{\mu_1\cdots\mu_{mr-m_I+n_J}}\partial_{\mu_1}\cdots\partial_{\mu_{mr-m_I+n_J}}
\end{align*}
and, from the fact that $\Pi_{k=1}^m p_k(\xi_0)$ is homogeneous in $\xi_\mu$, 
\begin{align*}
\Pi_{k=1}^m p_k(\xi_0)=\mathcal{G}(\partial^s\Psi)^{\mu_1\cdots\mu_{m r}}\xi_{\mu_1}\cdots \xi_{\mu_{m r}}\to \mathcal{G}(\partial^s\Psi)^{\mu_1\cdots\mu_{m r}}\partial_{\mu_1}\cdots \partial_{\mu_{m r}},
\end{align*} 
where the latter is a strictly hyperbolic operator (in particular $\mathcal{G}\neq 0$).
\end{definition}
\begin{lemma}
\label{SH:Lemma6}
Under Assumptions \ref{SH:Assumption 1}, \ref{SH:Assumption 2},  \ref{SH:Assumption3}, and \ref{SH:Assumption_polynomials},  every sufficiently regular solution of Equation~\eqref{SH:1} satisfies
\begin{align}
\label{SH:4}
\nonumber
\\
F(\partial^s\Psi)\mathcal{G}(\partial^s\Psi)^{\mu_1\cdots\mu_{m r}}\partial_{\mu_1}\cdots \partial_{\mu_{m r}} \Psi^I
=
\tilde{B}^I\left(
\partial^{mr-m_I+m_K-1}\Psi^K;
\big(\partial^{mr-m_I+n_J+\chi_J}\mathcal{W}\big)_{J=1}^N
\right).
\end{align}
\end{lemma}
\begin{proof}
By applying the operator $\hat{H}(\partial\Psi,\partial)$ to Equation~\eqref{SH:1} using Equation~\eqref{SH:Identity1} we obtain that
\begin{align*}
\hat{H}( \partial^{m_K-\min_Ln_L-1}\Psi^K,\partial^{mr-m_I+n_J})^I_J
h(\partial^{m_K-n_J-1}\Psi^K,\partial^{m_M-n_J})^J_M\Psi^M\\
=F(\partial^s\Psi) \mathcal{G}^{\mu_1\cdots\mu_{mr}}\partial_{\mu_1}\cdots \partial_{\mu_{mr}} \Psi^I+\mathcal{O}^I(\partial^{m_K-\min_J n_J-1}\Psi^K,
\partial^{mr-m_I+m_K-1}\Psi^K,\partial^{m_K-\min_J n_J}\Psi^K)
\end{align*}
and
\begin{align*}
\nonumber
\\
\nonumber
\\
\hat{H}( \partial^{m_K-\min_Ln_L-1}\Psi^K,\partial^{mr-m_I+n_J})^I_J
B^J(\partial^{m_K-n_J-1}\Psi^K;\partial^{\chi_J}\mathcal{W})
\nonumber
\\
=\mathcal{O}^I\left(
\partial^{m_K-\min_{J}n_J-1}\Psi^K,
\partial^{mr-m_I+m_K-1}\Psi^K;
\big(\partial^{mr-m_I+n_J+\chi_J}\mathcal{W}\big)_{J=1}^N
\right).
\end{align*}
Therefore, the order of the derivative of the unknown $\Psi^K$ is the maximum between $mr-m_I+m_K-1$ and $m_K-\min_J n_J$ in each $I$. However, from Assumption \ref{SH:Assumption3} we have that
\begin{align*}
mr-m_I+m_K-1&=(mr-m_I+\min_Jn_J-1)+(m_K-\min_Jn_J)\\
&\ge m_K-\min_Jn_J.
\end{align*}
Therefore,
\begin{align*}
\max(mr-m_I+m_K-1, m_K-\min_J n_J)=mr-m_I+m_K-1.
\end{align*}  
\end{proof}
\begin{proof}[Proof of Theorem \ref{SH:Theorem1}]
By Lemma \ref{SH:Lemma6}, every sufficiently regular solution of Equation~\eqref{SH:1} satisfies Equation~\eqref{SH:4}. Next, we must show that there exist Leray indices $\tilde{m}_I$ and $\tilde{n}_I$ such that Equation~\eqref{SH:4} may be rewritten in the form
\begin{align}
\label{SH:6}
\nonumber
\\
\tilde{h}^I(\partial^{\tilde{m}_K-\tilde{n}_I-1}\Psi^K,\partial^{\tilde{m}_I-\tilde{n}_I})\Psi^I
=
\tilde{B}^I\left(
\partial^{\tilde{m}_K-\tilde{n}_I-1}\Psi^K;
\big(\partial^{mr-m_I+n_J+\chi_J}\mathcal{W}\big)_{J=1}^N
\right).
\quad(\text{No sum over }I)
\end{align}

Set
\begin{align*}
m_*&=\min_{1\leq L\leq N}m_L,\\
\tilde{n}_I&=m_I-m_*,\\
\tilde{m}_I&=mr+m_I-m_*,\qquad I=1,\dots,N.
\end{align*}
Then $\tilde{n}_I\geq0$ for every $I$ and $\min_I\tilde{n}_I=0$, in accordance with the convention adopted after Equation~\eqref{SH:1}. Moreover, for every $I,K=1,\dots,N$,
\begin{subequations}
\begin{align}
\label{E:New_Leray_indices_principal_order}
\tilde{m}_I-\tilde{n}_I&=mr,\\
\label{E:New_Leray_indices_right_hand_side_order}
\tilde{m}_K-\tilde{n}_I-1&=mr-m_I+m_K-1.
\end{align}
\end{subequations}
Equations~\eqref{E:New_Leray_indices_right_hand_side_order} and \eqref{SH:4} show that the right-hand side of the $I$th equation depends on at most $\tilde{m}_K-\tilde{n}_I-1$ derivatives of $\Psi^K$ for every $K$. The derivatives of the prescribed function $\mathcal{W}$ displayed in Equation~\eqref{SH:4} do not require Leray indices because $\mathcal{W}$ is not an unknown of the system.

Although the dependence of $F$ and $\mathcal{G}$ is abbreviated by $\partial^s\Psi$, their coefficients are obtained algebraically from the coefficients of $h$ in Equation~\eqref{SH:1}, without differentiating these coefficients. Their dependence on each $\Psi^K$ therefore involves at most $m_K-\min_L n_L-1$ derivatives of $\Psi^K$. Assumption \ref{SH:Assumption3} gives, for every $I,K=1,\dots,N$,
\begin{align*}
m_K-\min_L n_L-1
&\leq mr-m_I+m_K-2\\
&<\tilde{m}_K-\tilde{n}_I-1,
\end{align*}
where Equation~\eqref{E:New_Leray_indices_right_hand_side_order} was used in the second line. Hence the coefficients on the left-hand side of Equation~\eqref{SH:4} also have the dependence on $\Psi^K$ required in Equation~\eqref{SH:6}.

It follows from Equation~\eqref{E:New_Leray_indices_principal_order} that the principal differential operator in the $I$th equation in Equation~\eqref{SH:4} has order $\tilde{m}_I-\tilde{n}_I=mr$. Its symbol, up to the nonzero factor $F(\partial^s\Psi)$, is $\Pi_{k=1}^m p_k(\xi_0)$. By Proposition \ref{SH:Lemma2}, this polynomial has exactly the $mr$ distinct real roots $r_1(\bm{\xi}),\dots,r_{mr}(\bm{\xi})$ for every $\bm{\xi}\neq0$. Therefore, each equation in Equation~\eqref{SH:4} is a strictly hyperbolic equation of order $mr$, with the Leray indices defined above.
\end{proof}

\begin{theorem}
\label{Theorem_SH_LWP}
Let $\Sigma=\mathbb{R}^d$ or $\mathbb{T}^d$. Define
\begin{align*}
\mathcal{Z}[\Psi]
=
\big(\partial^\alpha\Psi^K\big)_{
1\leq K\leq N,\,
0\leq|\alpha|\leq m_K-\min_Ln_L-1
}
\end{align*}
and denote by $\mathcal{J}$ the finite-dimensional real vector space in which $\mathcal{Z}[\Psi]$ takes values. Components for which $m_K-\min_Ln_L-1<0$ are omitted.

When $N_{\mathcal{W}}\geq1$, set
\begin{align*}
\chi_{\max}
&=
\max_{1\leq I\leq N}\chi_I,
\\
\mathcal{Y}[\mathcal{W}]
&=
\big(\partial^\alpha\mathcal{W}^L\big)_{
1\leq L\leq N_{\mathcal{W}},\,
0\leq|\alpha|\leq\chi_{\max}
},
\end{align*}
and denote by $\mathcal{J}_{\mathcal{W}}$ the finite-dimensional real vector space in which $\mathcal{Y}[\mathcal{W}]$ takes values.

Suppose that there exist an open set $\mathcal{U}\subset\mathcal{J}$ and an integer $m$, $1\leq m\leq N$, such that the following conditions hold. Every coefficient of the differential operators
\begin{align*}
h(\partial^{m_K-n_I-1}\Psi^K,\partial^{m_J-n_I})^I_J
\end{align*}
is a real-analytic function of $\mathcal{Z}[\Psi]\in\mathcal{U}$.  When $N_{\mathcal{W}}\geq1$, suppose that there exists an open set $\mathcal{V}\subset\mathcal{J}_{\mathcal{W}}$ such that every function
\begin{align*}
B^I(\partial^{m_K-n_I-1}\Psi^K;\partial^{\chi_I}\mathcal{W})
\end{align*}
is a real-analytic function of $(\mathcal{Z}[\Psi],\mathcal{Y}[\mathcal{W}])\in\mathcal{U}\times\mathcal{V}$. When $N_{\mathcal{W}}=0$, suppose instead that $B^I(\partial^{m_K-n_I-1}\Psi^K)$ is a real-analytic function of $\mathcal{Z}[\Psi]\in\mathcal{U}$. For every $z\in\mathcal{U}$ and every $\bm{\xi}\neq0$, the principal symbol obtained by substituting $z$ for $\mathcal{Z}[\Psi]$ satisfies Assumptions \ref{SH:Assumption 1}, \ref{SH:Assumption 2}, and \ref{SH:Assumption_polynomials}, with the same integer $m$. The Leray indices satisfy Assumption \ref{SH:Assumption3} for this value of $m$.

If $\Sigma=\mathbb{R}^d$, assume in addition that $0\in\mathcal{U}$ . When $N_{\mathcal{W}}\geq1$, assume that $0\in\mathcal{V}$ and
\begin{align*}
B^I(0;0)
&=0,
\qquad
I=1,\dots,N.
\end{align*}
When $N_{\mathcal{W}}=0$, assume that $B^I(0)=0$ for every $I=1,\dots,N$.

Define
\begin{align*}
p&=m\,r,\\
m_*&=\min_{1\leq I\leq N}m_I,\\
S_I&=\ell+p+m_I-m_*,
\end{align*}
where $\ell$ is an integer such that $\ell\geq d$.

When $N_{\mathcal{W}}\geq1$, define
\begin{align}
\label{E:SH_Forcing_indices}
\chi_*
&=
\max_{1\leq I\leq N}(n_I+\chi_I),
\nonumber
\\
d_{\mathcal{W}}
&=
p-m_*+\chi_*,
\nonumber
\\
S_{\mathcal{W}}
&=
\ell+p-m_*+\chi_*+1.
\end{align}
Let $T_0>0$ and prescribe $\mathcal{W}$ on $[0,T_0]\times\Sigma$ so that
\begin{align}
\label{E:SH_Forcing_regularity}
\partial_t^a\mathcal{W}
&\in
C\big([0,T_0];H^{S_{\mathcal{W}}-a}(\Sigma)\big),
\qquad
0\leq a\leq d_{\mathcal{W}}.
\end{align}
Assume that there exists a compact set $\mathcal{L}\Subset\mathcal{V}$ such that
\begin{align}
\label{E:SH_Forcing_containment}
\overline{
\mathcal{Y}[\mathcal{W}]
\big([0,T_0]\times\Sigma\big)
}
&\subset
\mathcal{L}.
\end{align}
Assume, in addition, that there exist functions $\mathcal{W}_\nu$, real analytic on an open neighborhood of $[0,T_0]\times\Sigma$, such that
\begin{align}
\label{E:SH_Analytic_forcing_approximation}
\sup_{0\leq t\leq T_0}
\sum_{a=0}^{d_{\mathcal{W}}}
\left\|
\partial_t^a(\mathcal{W}_\nu-\mathcal{W})(t)
\right\|_{H^{S_{\mathcal{W}}-a}}^2
&\longrightarrow0
\end{align}
as $\nu\rightarrow\infty$. After discarding finitely many terms, assume that, for some compact set $\mathcal{L}_1\Subset\mathcal{V}$ with $\mathcal{L}\subset\operatorname{int}\mathcal{L}_1$,
\begin{align}
\label{E:SH_Analytic_forcing_containment}
\overline{
\mathcal{Y}[\mathcal{W}_\nu]
\big([0,T_0]\times\Sigma\big)
}
&\subset
\mathcal{L}_1
\end{align}
for every $\nu$.

Let
\begin{align*}
\Psi^I_{(j)}
=
\left.\partial_t^j\Psi^I\right|_{t=0},
\qquad
0\leq j\leq m_I-1,
\end{align*}
be initial data satisfying
\begin{align}
\label{E:SH_Initial_data_regularity}
\Psi^I_{(j)}
\in H^{S_I-j}(\Sigma),
\qquad
0\leq j\leq m_I-1,
\qquad
I=1,\dots,N.
\end{align}
If
\begin{align*}
\\
\mathcal{E}^I[\Psi;\mathcal{W}]
=
h(\partial^{m_K-n_I-1}\Psi^K,\partial^{m_J-n_I})^I_J\Psi^J
-
B^I(\partial^{m_K-n_I-1}\Psi^K;\partial^{\chi_I}\mathcal{W}),
\end{align*}
then the data are assumed to satisfy
\begin{align}
\label{E:SH_Compatibility_conditions}
\left.\partial_t^a\mathcal{E}^I[\Psi;\mathcal{W}]\right|_{t=0}
=0,
\qquad
0\leq a\leq n_I-1,
\end{align}
for every $I$ such that $n_I>0$. Every initial trace occurring in Equation~\eqref{E:SH_Compatibility_conditions} is evaluated using the data in Equation~\eqref{E:SH_Initial_data_regularity} and, when $N_{\mathcal{W}}\geq1$, the traces of $\mathcal{W}$ supplied by Equation~\eqref{E:SH_Forcing_regularity}.

Let $\mathcal{Z}_0$ be the value of $\mathcal{Z}[\Psi]$ determined by these data on $\{t=0\}$. Assume that there exists a compact set $\mathcal{K}\Subset\mathcal{U}$ such that
\begin{align}
\label{E:SH_Initial_state_containment}
\overline{\mathcal{Z}_0(\Sigma)}
\subset\mathcal{K}.
\end{align}
Assume, in addition, that there exist real-analytic initial data $\Psi^I_{(j),\nu}$ for which Equation~\eqref{E:SH_Compatibility_conditions} holds with $\Psi^I_{(j)}$ replaced by $\Psi^I_{(j),\nu}$ and, when $N_{\mathcal{W}}\geq1$, with $\mathcal{W}$ replaced by $\mathcal{W}_\nu$, and such that
\begin{align}
\label{E:SH_Analytic_data_approximation}
\Psi^I_{(j),\nu}
\longrightarrow
\Psi^I_{(j)}
\quad\text{in }H^{S_I-j}(\Sigma)
\end{align}
as $\nu\rightarrow\infty$, for every $I=1,\dots,N$ and $0\leq j\leq m_I-1$.

Then there exist $T>0$, with $T\leq T_0$ when $N_{\mathcal{W}}\geq1$, and a unique solution $\Psi$ of Equation~\eqref{SH:1} on $[0,T]\times\Sigma$ with the prescribed initial data such that
\begin{align}
\label{E:SH_Solution_regularity}
\partial_t^j\Psi^I
\in
L^\infty\big([0,T];H^{S_I-j}(\Sigma)\big),
\qquad
0\leq j\leq p,
\qquad
I=1,\dots,N.
\end{align}
For every $I=1,\dots,N$, $0\leq j\leq p$, and every real number $s<S_I-j$, the solution also satisfies
\begin{align*}
\partial_t^j\Psi^I
\in
C\big([0,T];H^s(\Sigma)\big).
\end{align*}
Moreover, $T$ may be chosen so that, for some compact set $\mathcal{K}_1\Subset\mathcal{U}$,
\begin{align*}
\mathcal{K}\subset\operatorname{int}\mathcal{K}_1
\end{align*}
and
\begin{align*}
\overline{
\mathcal{Z}[\Psi]\big([0,T]\times\Sigma\big)
}
\subset\mathcal{K}_1.
\end{align*}

The solution is unique among solutions satisfying Equation~\eqref{E:SH_Solution_regularity} whose values of $\mathcal{Z}[\Psi]$ remain in $\mathcal{U}$ and which solve Equation~\eqref{SH:1} with the same prescribed function $\mathcal{W}$. The time $T$ can be chosen uniformly for initial data in a sufficiently small neighborhood of the given data in
\begin{align*}
\prod_{I=1}^N\prod_{j=0}^{m_I-1}H^{S_I-j}(\Sigma)
\end{align*}
that satisfy Equation~\eqref{E:SH_Compatibility_conditions}, admit approximations satisfying Equation~\eqref{E:SH_Analytic_data_approximation}, and have their initial jets in $\mathcal{K}_1$. When $N_{\mathcal{W}}\geq1$, this uniform choice is for the fixed prescribed function $\mathcal{W}$, and its constants may depend on the norm in Equation~\eqref{E:SH_Forcing_regularity} and on $\mathcal{L}_1$.

The solution and its time derivatives through order $p-1$ depend continuously on the data in every Sobolev topology strictly below the corresponding top orders in Equation~\eqref{E:SH_Solution_regularity}. More precisely, let $\Psi^I_{(j),q}$, $q\in\mathbb{N}$, be a sequence of initial data satisfying Equation~\eqref{E:SH_Compatibility_conditions}, admitting approximations satisfying Equation~\eqref{E:SH_Analytic_data_approximation}, and having their initial jets in $\mathcal{K}_1$, such that
\begin{align*}
\Psi^I_{(j),q}
&\rightarrow
\Psi^I_{(j)}
\quad\text{in }H^{S_I-j}(\Sigma),
&&
0\leq j\leq m_I-1.
\end{align*}
When $N_{\mathcal{W}}\geq1$, let $\mathcal{W}_q$ satisfy Equations~\eqref{E:SH_Forcing_regularity}--\eqref{E:SH_Analytic_forcing_containment}, with the same compact set $\mathcal{L}_1$, and assume that
\begin{align*}
\sup_{0\leq t\leq T_0}
\sum_{a=0}^{d_{\mathcal{W}}}
\left\|
\partial_t^a(\mathcal{W}_q-\mathcal{W})(t)
\right\|_{H^{S_{\mathcal{W}}-a}}^2
&\rightarrow
0.
\end{align*}
After decreasing $T$ if necessary, the corresponding solutions $\Psi_q$ are defined on $[0,T]\times\Sigma$ for every sufficiently large $q$ and satisfy
\begin{align*}
\partial_t^j\Psi_q^I
&\rightarrow
\partial_t^j\Psi^I
\quad\text{in }C\big([0,T];H^s(\Sigma)\big)
\end{align*}
for every $I=1,\dots,N$, $0\leq j\leq p-1$, and $s<S_I-j$. When $N_{\mathcal{W}}=0$, the conditions involving $\mathcal{W}_q$ are omitted.
\end{theorem}

\begin{proof}
We construct exact nonlinear analytic solutions locally and use finite-order energy estimates to obtain a common real interval. Steps 6--7 supply the analytic extension needed for this construction, after the estimates of Steps 4--5. The analytic neighborhoods may depend on the approximation index. We then pass to the limit and prove uniqueness and continuous dependence. For the classical analytic approximation method, see \cite[Chapter VI, Sections 8--10]{Courant-Hilbert-Book-1989-2}, \cite[Chapter 5, section ``Symmetric Hyperbolic Systems'']{John-Book-1982}, \cite[Appendix A.2]{Disconzi-2024}, and \cite[Appendix C]{Cordeiro:2026jle}.

\medskip
\noindent \underline{Step 1} (\textbf{The analytic Cauchy problems and the compatibility conditions}).\nopagebreak[4]
We first verify that the analytic Cauchy problems used in the approximation argument are non-characteristic. Let
\begin{align*}
\tau=(1,0,\dots,0)
\end{align*}
and, for $z\in\mathcal{U}$, let $a(z)=h(z,\tau)$ be the matrix of coefficients of the pure time derivatives $\partial_t^{m_J-n_I}\Psi^J$ in the principal part of the $I$th equation. By Proposition \ref{SH:Lemma2},
\begin{align*}
\det a(z)=\mathcal{P}_z(\tau)=F(z)\neq0.
\end{align*}
Thus, $\{t=0\}$ is non-characteristic at every state in $\mathcal{U}$.

The role of the compatibility conditions can be seen directly from the Leray indices. Differentiating the $I$th equation in Equation~\eqref{SH:1} exactly $n_I$ times with respect to $t$ gives a system that can be written as
\begin{align}
\label{E:SH_CK_normal_form}
\nonumber
\\
a(\mathcal{Z}[\Psi])^I_J\partial_t^{m_J}\Psi^J
=
R^I\left(
\partial_x^\alpha\partial_t^b\Psi^K:
b\leq m_K-1;
\partial^\beta\mathcal{W}:
|\beta|\leq n_I+\chi_I
\right).
\end{align}
where $\partial_x^\alpha$ denotes derivatives with respect to the spatial coordinates on $\Sigma$; the arguments on the right-hand side are subject to the differential orders allowed by Equation~\eqref{SH:1}. Indeed, when the $n_I$ time derivatives fall on a coefficient, that coefficient initially depends on at most $m_K-n_I-1$ derivatives of $\Psi^K$, and hence the resulting term contains at most $m_K-1$ derivatives of $\Psi^K$. The same differentiation applies at most $n_I+\chi_I$ derivatives to $\mathcal{W}$. The functions $R^I$ are analytic because the coefficient functions in Equation~\eqref{SH:1} are analytic. For each analytic problem, the prescribed function in this expression is $\mathcal{W}_\nu$, which is analytic in the spacetime variables by hypothesis. The entries of $a^{-1}$ are analytic on $\mathcal{U}$ because the entries of $a$ are analytic there and $\det a$ does not vanish. Multiplying Equation~\eqref{E:SH_CK_normal_form} by $a(\mathcal{Z}[\Psi])^{-1}$ gives
\begin{align*}
\\
\partial_t^{m_J}\Psi^J
=
\big(a(\mathcal{Z}[\Psi])^{-1}\big)^J_I
R^I\left(
\partial_x^\alpha\partial_t^b\Psi^K:
b\leq m_K-1;
\partial^\beta\mathcal{W}:
|\beta|\leq n_I+\chi_I
\right).
\end{align*}
Thus, the derivatives $\partial_t^{m_J}\Psi^J$ are analytic functions of $\mathcal{Z}[\Psi]$ and of the derivatives displayed on the right-hand side, as required to apply the Cauchy--Kovalevskaya theorem.

For $0\leq a\leq n_I-1$, the quantity
\begin{align*}
\left.\partial_t^a\mathcal{E}^I[\Psi;\mathcal{W}]\right|_{t=0}
\end{align*}
contains only the prescribed traces $\Psi^J_{(j)}$, $0\leq j\leq m_J-1$, and the prescribed traces of $\mathcal{W}$ through order $n_I+\chi_I-1$. Thus, Equation~\eqref{E:SH_Compatibility_conditions} is precisely the condition that the solution of the differentiated system in Equation~\eqref{E:SH_CK_normal_form} also solve the original system. In fact, Equation~\eqref{E:SH_CK_normal_form} gives $\partial_t^{n_I}\mathcal{E}^I[\Psi;\mathcal{W}]=0$, while Equation~\eqref{E:SH_Compatibility_conditions} prescribes the first $n_I$ time traces of $\mathcal{E}^I[\Psi;\mathcal{W}]$ to vanish. Hence $\mathcal{E}^I[\Psi;\mathcal{W}]=0$. This is the usual Cauchy problem for a Leray system; see \cite[Appendix~IV]{Choquet-Bruhat-Book-2009} and \cite[Chapter 1, Section~6]{Lichnerowicz-Book-1967} for standard existence formulations in this setting.

\medskip
\noindent \underline{Step 2} (\textbf{The additional initial traces}).\nopagebreak[4]
Successive time differentiation of \eqref{E:SH_CK_normal_form} determines all the additional traces needed for the equations of order $p$. More precisely, for $0\leq q\leq p-1-m_*$, one obtains
\begin{align}
\label{E:SH_Higher_trace_recursion}
a(\mathcal{Z}_0)^I_J\Psi^J_{(m_J+q)}
=
R_q^I\left(
\partial_x^\alpha\Psi^K_{(b)}:
b\leq m_K+q-1;
\partial_x^\beta\mathcal{W}_{(c)}:
c+|\beta|\leq n_I+\chi_I+q
\right),
\end{align}
where
\begin{align*}
\mathcal{W}_{(c)}
&=
\left.\partial_t^c\mathcal{W}\right|_{t=0}.
\end{align*}
The right-hand side contains only traces determined at earlier stages and prescribed traces of $\mathcal{W}$. Moreover,
\begin{align}
\label{E:SH_Higher_trace_regularity}
S_J-(m_J+q)=\ell+p-m_*-q,
\end{align}
which is independent of $J$. If $c+|\beta|\leq n_I+\chi_I+q$, then Equations~\eqref{E:SH_Forcing_indices} and \eqref{E:SH_Forcing_regularity} give
\begin{align}
\label{E:SH_Forcing_initial_trace_count}
S_{\mathcal{W}}-c-|\beta|
&\geq
\ell+p-m_*+\chi_*+1-(n_I+\chi_I+q)
\nonumber
\\
&\geq
\ell+p-m_*+1-q.
\end{align}
Thus every prescribed trace of $\mathcal{W}$ occurring in Equation~\eqref{E:SH_Higher_trace_recursion} has at least one more spatial Sobolev derivative than the regularity in Equation~\eqref{E:SH_Higher_trace_regularity}.

When $\Sigma=\mathbb{R}^d$ and $n_I=0$, the expression $R_0^I$ in \eqref{E:SH_Higher_trace_recursion} contains the composition $B^I(\partial^{m_K-1}\Psi^K;\partial^{\chi_I}\mathcal{W})$ without a time derivative. The condition $B^I(0;0)=0$, together with \eqref{E:SH_Forcing_regularity}, is used here: the analytic-composition estimate then places this term in $H^{\ell+p-m_*}(\mathbb{R}^d)$, whereas, if $B^I(0;0)\neq0$, the composition has a nonzero limit as $|x|\rightarrow\infty$ and does not belong to $L^2(\mathbb{R}^d)$. The Sobolev product and analytic-composition estimates, together with inversion of $a(\mathcal{Z}_0)$, therefore show inductively that the traces obtained from \eqref{E:SH_Higher_trace_recursion} have the regularity in \eqref{E:SH_Higher_trace_regularity}. The same estimates, \eqref{E:SH_Analytic_data_approximation}, and \eqref{E:SH_Analytic_forcing_approximation}, show that the traces obtained from the analytic data and $\mathcal{W}_\nu$ converge to those obtained from the limiting data and $\mathcal{W}$ in these spaces. Taking $0\leq q\leq p-1-m_*$ produces the traces of $\Psi^J$ through order
\begin{align*}
m_J+p-1-m_*=\widetilde{m}_J-1,
\end{align*}
and, in particular, every trace $\Psi^I_{(j)}$ with $0\leq j\leq p-1$ that is required for the order-$p$ principal operators.

\medskip
\noindent \underline{Step 3} (\textbf{The analytic approximating solutions and the coefficients in Equation~\eqref{SH:4}}).\nopagebreak[4]
Sobolev embedding and Equation~\eqref{E:SH_Analytic_data_approximation} give uniform convergence of the initial physical jets to $\mathcal{Z}_0$. We discard finitely many terms so that these jets lie in a compact subset of $\mathcal{U}$. For each retained $\nu$, the Cauchy--Kovalevskaya theorem applied to Equation~\eqref{E:SH_CK_normal_form} with $\mathcal{W}_\nu$ gives an analytic solution near every point of the initial surface. The compatibility conditions imply Equation~\eqref{SH:1}, by Step 1. Uniqueness for the analytic Cauchy problem identifies these solutions near intersecting parts of the initial surface. A finite cover therefore gives a solution on a neighborhood of $\{0\}\times K$ for every compact spatial set $K$, with a positive time width depending on $\nu$ and $K$. On $\mathbb{T}^d$ take $K=\mathbb{T}^d$. On $\mathbb{R}^d$ we initially use only compact unions of closed balls. Steps 5--7 will give the common real interval and the solution on all of space. The local analytic theorem used here is stated in \cite[Chapter 3, section ``Real Analytic Functions and the Cauchy--Kowalevski Theorem'']{John-Book-1982}.

It is important that the analytic problem is solved for Equation~\eqref{SH:1}: Theorem \ref{SH:Theorem1} then implies that every $\Psi_\nu$ satisfies Equation~\eqref{SH:4}. We do not solve Equation~\eqref{SH:4} as an independent evolution system.

We next record why the coefficients in the strictly hyperbolic reduction have the regularity needed below. The coefficients of $F$ are analytic functions of $z$. The only less immediate point is the dependence of the polynomial $q$ in Equation~\eqref{SH:q}. With
\begin{align*}
\Delta_z(\xi)=\frac{\det h(z,\xi)}{\det h_r(z)},
\end{align*}
the polynomial $q_z$ is the monic greatest common divisor of $\Delta_z$ and $\partial_{\xi_0}\Delta_z$. Its degree is the constant $(N-m)r$ throughout $\mathcal{U}$. On a neighborhood on which this degree is constant, the coefficients of $q_z$ are obtained from a constant-rank Sylvester system whose coefficients depend analytically on $z$. They are therefore analytic in $z$. Uniqueness of the monic greatest common divisor makes these local expressions agree on overlaps. Monic polynomial division then shows that the coefficients of $\widehat{H}$ and $\mathcal{G}$, and consequently all the coefficient functions entering $\widetilde{B}^I$, are analytic on $\mathcal{U}$ with respect to the physical jet. The functions arising from $B^I$ are analytic on $\mathcal{U}\times\mathcal{V}$ with respect to the physical and prescribed forcing jets. The prescribed function does not enter $F$, $\widehat{H}$, or $\mathcal{G}$.

\medskip
\noindent \underline{Step 4} (\textbf{The reduced equations and the linear energy estimate}).\nopagebreak[4]
Choose compact sets $\mathcal{K}\subset\operatorname{int}\mathcal{K}_0$, $\mathcal{K}_0\subset\operatorname{int}\mathcal{K}_1$, and $\mathcal{K}_1\Subset\mathcal{U}$. The initial jets of $\Psi_\nu$ lie in $\mathcal{K}_0$ after finitely many terms have been discarded. All constants used in this step may depend on these compact sets, the coefficient functions, the dimension, and the indices. When a prescribed function is present, they may also depend on $\mathcal{L}_1$ and the finite bound
\begin{align}
\label{E:SH_Uniform_forcing_bound}
H=\sup_\nu\sup_{0\leq t\leq T_0}
\sum_{a=0}^{d_{\mathcal{W}}}
\norm{\partial_t^a\mathcal{W}_\nu(t)}_{H^{S_{\mathcal{W}}-a}(\Sigma)}^2<\infty.
\end{align}
Terms involving the prescribed function are omitted when $N_{\mathcal{W}}=0$, and then $H=0$.

We use the integers
\begin{align*}
s_0&=\ell+p-m_*+1,\\
r_I&=S_I-p=\ell+m_I-m_*,\\
s_I&=r_I+1,\\
d_{IK}&=p-m_I+m_K-1,\\
e_I&=p-m_I+\chi_*.
\end{align*}
In particular,
\begin{subequations}
\label{E:SH_Reduction_counts}
\begin{align}
S_K-(m_K-1)&=s_0,\\
S_K-d_{IK}&=s_I,\\
S_K-1-d_{IK}&=r_I,\\
S_{\mathcal{W}}-e_I&=s_I,\\
s_0-r_I&=p-m_I+1\geq2.
\end{align}
\end{subequations}
For a term with $a$ time derivatives of the prescribed function in row $J$, the required spatial order is
\begin{align}
\label{E:SH_Forcing_reduced_derivative_count}
\begin{split}
S_I-p+1+p-m_I+n_J+\chi_J-a
&=\ell+p-m_*+n_J+\chi_J+1-a\\
&\leq S_{\mathcal{W}}-a.
\end{split}
\end{align}
Here $r_I\geq\ell>\frac{d}{2}$. Also $m_K\geq1$: Assumption \ref{SH:Assumption3} excludes $r=0$, and a nonzero entry in the $K$th column of the invertible matrix $h_r$ requires $m_K-n_I\geq r$ for some $I$.

For clarity, we write the normal form obtained by inverting $a$ in \eqref{E:SH_CK_normal_form} as
\begin{align}
\label{E:SH_Inverted_normal_form}
\partial_t^{m_K}\Psi^K
=\mathcal{N}_K\left(
\big(\partial_x^\beta\partial_t^b\Psi^L\big)_{b<m_L,\,b+|\beta|\leq m_L};
\big(\partial^\gamma\mathcal{W}\big)_{|\gamma|\leq\chi_*}
\right).
\end{align}
The functions $\mathcal{N}_K$ are analytic in the physical and forcing jets and polynomial in the additional derivative arguments. In the case $\Sigma=\mathbb{R}^d$, they vanish when all their arguments vanish.

Write $q_{IJ}=p-m_I+n_J$, and expand the homogeneous operators with their coefficients to the left of the derivatives:
\begin{align*}
h^J_K(z,\partial)&=\sum_{|\gamma|=m_K-n_J}h^J_{K,\gamma}(z)\partial^\gamma,\\
\widehat{H}^I_J(z,\partial)&=\sum_{|\alpha|=q_{IJ}}\widehat{H}^I_{J,\alpha}(z)\partial^\alpha.
\end{align*}
An entry with negative order is absent. For a smooth solution, the product rule and \eqref{SH:Identity1} give the exact identity
\begin{align}
\label{E:SH_Expanded_reduction}
\begin{split}
F(\mathcal{Z}[\Psi])\mathcal{G}(\mathcal{Z}[\Psi],\partial)\Psi^I
={}&\sum_{J,\,|\alpha|=q_{IJ}}\widehat{H}^I_{J,\alpha}(\mathcal{Z}[\Psi])\partial^\alpha B^J\\
&-\sum_{J,K}\sum_{\substack{|\alpha|=q_{IJ}\\|\gamma|=m_K-n_J}}
\sum_{0<\eta\leq\alpha}\binom{\alpha}{\eta}
\widehat{H}^I_{J,\alpha}(\mathcal{Z}[\Psi])
\partial^\eta\big(h^J_{K,\gamma}(\mathcal{Z}[\Psi])\big)
\partial^{\alpha-\eta+\gamma}\Psi^K.
\end{split}
\end{align}
There are no derivatives of $\widehat{H}$ in \eqref{E:SH_Expanded_reduction}. The terms with $\eta=0$ combine by \eqref{SH:Identity1}. For $\eta\neq0$, the displayed derivative of $\Psi^K$ has order at most $d_{IK}$; every derivative of $\Psi^L$ arising from the differentiated coefficient has order at most
\begin{align*}
m_L-n_J-1+|\eta|\leq p-m_I+m_L-1=d_{IL}.
\end{align*}
The derivatives of $B^J$ contain derivatives of each $\Psi^L$ of total order at most $d_{IL}$ and contain at most $q_{IJ}+\chi_J\leq e_I$ derivatives of $\mathcal{W}$.

Every source factor $\partial_x^\beta\partial_t^q\Psi^K$ with $q\geq m_K$ can be replaced by $\partial_x^\beta\partial_t^{q-m_K}\mathcal{N}_K$ using \eqref{E:SH_Inverted_normal_form}. The product and chain rules show that each resulting derivative of $\Psi^L$ has time order at most $m_L-1+q-m_K$ and total order at most $m_L+|\beta|+q-m_K$. Thus, along each succession of replacements, the time order minus the index of its unknown decreases by at least one, whereas the total order minus that index does not increase. Induction on the nonnegative integer $q-m_K$ proves that the finite expansion terminates with time orders less than $m_L$. Starting with total order at most $d_{IK}$, all resulting derivatives of $\Psi^L$ have total order at most $d_{IL}$. The newly introduced forcing derivatives have order at most $\chi_*+|\beta|+q-m_K\leq e_I-1$; the forcing terms already in \eqref{E:SH_Expanded_reduction} have order at most $e_I$.

Since $\mathcal{G}$ is monic in the time variable, division by $F$ and these substitutions give
\begin{subequations}
\label{E:SH_Normalized_reduction}
\begin{align}
L_\Psi\Psi^I&=f_I(\Psi;\mathcal{W}),\\
L_\Psi&=\partial_t^p+
\sum_{\substack{j+|\beta|=p\\j<p}}
A_{j,\beta}(\mathcal{Z}[\Psi])\partial_x^\beta\partial_t^j.
\end{align}
\end{subequations}
Here $f_I$ is a fixed real-analytic function of the finite list
\begin{align}
\label{E:SH_Reduced_source_arguments}
\big(\partial_x^\beta\partial_t^b\Psi^K\big)_{b<m_K,\,b+|\beta|\leq d_{IK}},
\qquad
\big(\partial_x^\beta\partial_t^a\mathcal{W}\big)_{a+|\beta|\leq e_I}.
\end{align}
On $\mathbb{R}^d$, $f_I(0;0)=0$. The coefficient functions $A_{j,\beta}$ depend only on $\mathcal{Z}[\Psi]$. The reduced equation is an identity for original solutions; a converse will not be used. Indeed, the coefficient functions identified in Step 3 are analytic, and Equation~\eqref{E:SH_Expanded_reduction} and the finite substitutions from Equation~\eqref{E:SH_Inverted_normal_form} use only products, derivatives and compositions of these functions, and division by the nonvanishing $F$. These operations preserve analyticity on the set of admissible physical and forcing jets, with the additional derivative arguments unrestricted.

We record the quantitative linear estimate needed both at $s_I$ and at $r_I$. For a monic strictly hyperbolic operator $L$ of the form in \eqref{E:SH_Normalized_reduction}, suppose its coefficient vector $A$ has a fixed compact range with a positive minimum separation of its characteristic roots on $|\bm{\xi}|=1$. Suppose also, on a time interval of length at most one, that
\begin{align}
\label{E:SH_Linear_coefficient_bound}
\sup_t\left(
\norm{A(t)-A_*}_{H^{s_0}(\Sigma)}
+\norm{\partial_tA(t)}_{H^{s_0-1}(\Sigma)}\right)\leq C_0,
\end{align}
where $A_*$ is constant. Its size is included in $C_0$; on the torus it may be taken to be zero. For $k=r_I$ or $s_I$, put
\begin{align*}
E_k[v](t)=\sum_{j=0}^{p-1}\norm{\partial_t^jv(t)}_{H^{k+p-1-j}(\Sigma)}^2.
\end{align*}
Then, for a smooth solution with finite energies,
\begin{align}
\label{E:SH_Homogeneous_linear_estimate}
E_k[v](t)\leq C\left(E_k[v](0)+\int_0^t\norm{Lv(s)}_{H^k(\Sigma)}^2\,ds\right).
\end{align}
The constant depends only on $C_0$, the compact coefficient range and root separation, $d$, $p$, and $k$. In particular, there is no additive constant independent of $v$ and $Lv$.

We give the operator estimates establishing this dependence. Let $\Lambda=(1-\Delta_x)^{\frac{1}{2}}$ and $Y_j=\Lambda^{p-1-j}\partial_t^jv$, $0\leq j<p$. Then
\begin{align*}
\partial_tY_j&=\Lambda Y_{j+1},\qquad 0\leq j<p-1,\\
\partial_tY_{p-1}&=-\sum_{\substack{j+|\beta|=p\\j<p}}
A_{j,\beta}\partial_x^\beta\Lambda^{j+1-p}Y_j+Lv.
\end{align*}
Denote this system by $\partial_tY=\mathcal{A}Y+(0,\dots,0,Lv)$. Its degree-one symbol $\mathcal{A}_1$ has eigenvalues $\mathrm{i}c_a$, where $c_1,\dots,c_p$ are the distinct real characteristic roots. For $\bm{\xi}\neq0$, define
\begin{align*}
P_a&=\prod_{b\neq a}\frac{\mathcal{A}_1-\mathrm{i}c_bI}{\mathrm{i}(c_a-c_b)},\\
q&=\sum_{a=1}^pP_a^*P_a.
\end{align*}
The identities $\sum_aP_a=I$ and $P_a\mathcal{A}_1=\mathrm{i}c_aP_a$ imply
\begin{subequations}
\label{E:SH_Symbol_symmetrizer}
\begin{align}
\frac{1}{p}|v|^2&\leq v^*qv\leq C|v|^2,\\
(q\mathcal{A}_1)^*&=-q\mathcal{A}_1.
\end{align}
\end{subequations}
Replace $q$ by $q_0=(1-f_0(|\bm{\xi}|))I+f_0(|\bm{\xi}|)q$, where $0\leq f_0\leq1$ is smooth, zero on $[0,1]$, and one on $[2,\infty)$. The changes in the degree-one identities have bounded frequency support. Since $s_0-1\geq\ell+1>\frac{d}{2}+1$, \eqref{E:SH_Linear_coefficient_bound} bounds the coefficients and their time derivatives in $C^{1+e}(\Sigma)$ for a fixed $0<e<\min(1,s_0-2-\frac{d}{2})$. Root separation gives the same bounds for every fixed finite collection of angular derivatives of $q_0$ and $\partial_tq_0$.

We use the Sobolev boundedness, composition, and adjoint results in \cite[Proposition 2.1.D and Propositions 4.2.A--C]{Taylor-Book-1991}; the order-$p$ companion reduction is described there in Section 5.3, equations (5.3.3)--(5.3.8). The symbol regularity $C^{1+e}$ permits the Sobolev orders $-1,0,1$ in Proposition 2.1.D. Set $B=\operatorname{Op}(q_0^{\frac{1}{2}})$ and $Q=B^*B+\Lambda^{-2}$. The composition formula gives
\begin{align*}
\operatorname{Op}(q_0^{-\frac{1}{2}})B&=I+R_{-1},\\
\norm{R_{-1}v}_{L^2(\Sigma)}&\leq C\norm{v}_{H^{-1}(\Sigma)}.
\end{align*}
Consequently $\norm{v}_{L^2(\Sigma)}\leq C(\norm{Bv}_{L^2(\Sigma)}+\norm{v}_{H^{-1}(\Sigma)})$, which proves coercivity of $Q$. The composition and adjoint formulas, together with \eqref{E:SH_Symbol_symmetrizer}, give
\begin{subequations}
\label{E:SH_Operator_energy_bounds}
\begin{align}
c\norm{v}_{L^2(\Sigma)}^2&\leq (Qv,v)_{L^2(\Sigma)}\leq C\norm{v}_{L^2(\Sigma)}^2,\\
\norm{\partial_tQ}_{L^2(\Sigma)\rightarrow L^2(\Sigma)}&\leq C,\\
\label{E:SH_Operator_symmetric_part}
|2\operatorname{Re}(Q\mathcal{A}v,v)_{L^2(\Sigma)}|&\leq C\norm{v}_{L^2(\Sigma)}^2.
\end{align}
\end{subequations}
Indeed, $Q-\operatorname{Op}(q_0)$ maps $H^{-1}(\Sigma)$ to $L^2(\Sigma)$, $\mathcal{A}$ maps $L^2(\Sigma)$ to $H^{-1}(\Sigma)$, and the symmetric part of the order-one symbol $q_0\mathcal{A}_1$ has order zero by \eqref{E:SH_Symbol_symmetrizer}. These facts prove \eqref{E:SH_Operator_symmetric_part}. The same calculus holds for periodic symbols on $\mathbb{T}^d$.

For an integer spatial multi-index $\alpha$, $|\alpha|\leq k$, every term in $[\partial_x^\alpha,\mathcal{A}]Y$ is a constant multiple of
\begin{align*}
(\partial_x^\beta a)M(D)\partial_x^{\alpha-\beta}Y,
\qquad 1\leq|\beta|\leq|\alpha|,
\end{align*}
where $M(D)$ is a fixed multiplier of order one and $a$ is a coefficient. The two factors have Sobolev exponents $s_0-|\beta|\geq1$ and $k-|\alpha|+|\beta|-1\geq0$, whose sum is at least $s_0-1>\frac{d}{2}$. Sobolev embedding and H\"older's inequality therefore give
\begin{align}
\label{E:SH_Integer_commutator}
\norm{[\partial_x^\alpha,\mathcal{A}]Y}_{L^2(\Sigma)}\leq C\norm{Y}_{H^k(\Sigma)}.
\end{align}
Constant coefficient parts commute with $\partial_x^\alpha$. Thus no norm of a nonzero constant in $H^{s_0}(\mathbb{R}^d)$ is used. Differentiating
$\sum_{|\alpha|\leq k}(Q\partial_x^\alpha Y,\partial_x^\alpha Y)_{L^2(\Sigma)}$, \eqref{E:SH_Operator_energy_bounds}, \eqref{E:SH_Integer_commutator}, and $2ab\leq a^2+b^2$ give
\begin{align*}
\frac{d}{dt}\sum_{|\alpha|\leq k}(Q\partial_x^\alpha Y,\partial_x^\alpha Y)_{L^2(\Sigma)}
\leq C\left(\norm{Y}_{H^k(\Sigma)}^2+\norm{Lv}_{H^k(\Sigma)}^2\right).
\end{align*}
Gronwall's inequality proves \eqref{E:SH_Homogeneous_linear_estimate}. This calculation also applies when $Y\in L^\infty([0,T];H^{k+1}(\Sigma))$ and $\partial_tY\in L^\infty([0,T];H^k(\Sigma))$: every $\partial_x^\alpha Y$ used in the energy is Lipschitz into $L^2(\Sigma)$, the product rule holds almost everywhere, and the operator bounds extend from smooth functions by density. In particular, the calculation does not require $r_I>\frac{d}{2}+1$.

We will also need the norm form of the linear estimate. Put
\begin{align*}
\mathfrak{E}_k[v]=\sum_{|\alpha|\leq k}(Q\partial_x^\alpha Y,\partial_x^\alpha Y)_{L^2(\Sigma)}.
\end{align*}
Before the use of $2ab\leq a^2+b^2$, Equations~\eqref{E:SH_Operator_energy_bounds} and \eqref{E:SH_Integer_commutator} give
\begin{align*}
cE_k[v]&\leq\mathfrak{E}_k[v]\leq CE_k[v],\\
\frac{d}{dt}\mathfrak{E}_k[v]&\leq C\mathfrak{E}_k[v]
+C\mathfrak{E}_k[v]^{\frac{1}{2}}\norm{Lv}_{H^k(\Sigma)}.
\end{align*}
Division by $2(\mathfrak{E}_k[v]+\updelta)^{\frac{1}{2}}$, integration with the factor $e^{-Ct}$, and the limit $\updelta\downarrow0$ prove
\begin{align}
\label{E:SH_Linear_norm_estimate}
E_k[v](t)^{\frac{1}{2}}\leq C E_k[v](0)^{\frac{1}{2}}
+C\int_0^t\norm{Lv(s)}_{H^k(\Sigma)}\,ds.
\end{align}
The constant has exactly the dependence stated for Equation~\eqref{E:SH_Homogeneous_linear_estimate}.

\medskip
\noindent \underline{Step 5} (\textbf{Localized estimates and uniform Sobolev bounds}).\nopagebreak[4]
The estimates in this step are initially asserted only on domains where the analytic solutions already exist. We first treat $\Sigma=\mathbb{R}^d$. For an open spatial set $D$ and a nonnegative integer $q$, use the derivative norm
\begin{align*}
\norm{v}_{H^q(D)}^2=\sum_{|\alpha|\leq q}\int_D|\partial_x^\alpha v|^2\,dx.
\end{align*}
Define
\begin{align*}
E_D[v](t)&=\sum_{I=1}^N\sum_{j=0}^{p-1}\norm{\partial_t^jv^I(t)}_{H^{S_I-j}(D)}^2,\\
F_{D,\nu}(t)&=\sum_{a=0}^{d_{\mathcal{W}}}\norm{\partial_t^a\mathcal{W}_\nu(t)}_{H^{S_{\mathcal{W}}-a}(D)}^2.
\end{align*}
We set $F_{D,\nu}=0$ when $N_{\mathcal{W}}=0$. The bound $H$ in \eqref{E:SH_Uniform_forcing_bound} is enlarged, if necessary, for the equivalent derivative norms.

We need an extension from balls whose constants do not depend on their centers. Let $S_{\max}=\max_I S_I$ and fix an integer $K>4(S_{\max}+p+d+10)$. The numbers $c_1,\dots,c_K$ are the unique solution of
\begin{align}
\label{E:SH_Extension_moments}
\sum_{k=1}^Kc_k(-k)^j=1,\qquad 0\leq j<K.
\end{align}
The matrix in \eqref{E:SH_Extension_moments} is a Vandermonde matrix at the distinct points $-1,\dots,-K$. Let $f_1$ be a smooth function equal to one on $[0,1+\frac{1}{4K}]$, equal to zero on $[1+\frac{1}{2K},\infty)$, and taking values in $[0,1]$. The extension from $B_b(0)$ is
\begin{align}
\label{E:SH_Ball_extension}
\mathcal{E}_bf(y)=
\begin{cases}
f(y),&|y|\leq b,\\
\displaystyle f_1(\frac{|y|}{b})\sum_{k=1}^Kc_k
f\left(((k+1)b-k|y|)\frac{y}{|y|}\right),
&b<|y|<b(1+\frac{1}{2K}),\\
0,&|y|\geq b(1+\frac{1}{2K}).
\end{cases}
\end{align}
Translation defines the extension from $B_b(x_a)$. Every evaluation in the annulus has radius in $[\frac{b}{2},b]$. The identities \eqref{E:SH_Extension_moments} match the normal derivatives through order $K-1$ at $|y|=b$; tangential differentiation matches the mixed derivatives. On the annulus, the radial derivative of the $k$th evaluation map is $-k$, and its tangential eigenvalues have absolute value between $\frac{1}{3}$ and $1$. Differentiating \eqref{E:SH_Ball_extension} and changing variables in each resulting integral therefore give, for fixed $b_0>0$ and $\frac{b_0}{2}\leq b\leq b_0$,
\begin{subequations}
\label{E:SH_Extension_bounds}
\begin{align}
\norm{\mathcal{E}_bf}_{H^q(\mathbb{R}^d)}&\leq C_{b_0,K,d,q}\norm{f}_{H^q(B_b)},\qquad 0\leq q<K,\\
\norm{\mathcal{E}_bf}_{L^\infty(\mathbb{R}^d)}&\leq C_E\norm{f}_{L^\infty(B_b)},\qquad C_E=\sum_{k=1}^K|c_k|,\\
\norm{\partial_t(\mathcal{E}_{b(t)}f)}_{H^{q-1}(\mathbb{R}^d)}
&\leq C_{b_0,K,d,q,c}\left(\norm{\partial_tf}_{H^{q-1}(B_{b(t)})}+\norm{f}_{H^q(B_{b(t)})}\right),\qquad 1\leq q<K,
\end{align}
\end{subequations}
where $b(t)=b_0-ct$. In the time derivative, differentiation of an evaluation map adds $-c(k+1)(\frac{y}{|y|})\cdot\nabla f$, and differentiation of the cutoff adds a bounded multiple of $f$. These are precisely the terms estimated in \eqref{E:SH_Extension_bounds}.

Let $A$ denote the finite vector of coefficients in \eqref{E:SH_Normalized_reduction}. There is a number $a_0>0$ such that the closed $a_0$-neighborhood of $A(\mathcal{K}_0)$ consists of monic strictly hyperbolic polynomials with a common positive separation of the roots on the unit spatial sphere. Fix $c>0$ larger than the absolute values of all roots in this neighborhood on that sphere. The cone $\{\xi_0>c|\bm{\xi}|\}$ lies in their future hyperbolicity cones, so their forward propagation cones lie in $\{(v_0,v_x)\,|\,v_0\geq0,\ |v_x|\leq cv_0\}$. The initial trace bounds give a number $L_0$ such that
\begin{align*}
\sup_\nu\norm{\nabla_xA(\mathcal{Z}[\Psi_\nu](0))}_{L^\infty(\mathbb{R}^d)}\leq L_0.
\end{align*}
Fix
\begin{align}
\label{E:SH_Ball_radius}
0<b_0\leq\min\left(1,\frac{a_0}{8C_E(1+L_0)}\right).
\end{align}
This choice uses only the initial bounds, before any estimate on a ball is used.

Suppose on a closed shrinking ball $B_{b_0-ct}(x_a)$ that $E_{B_{b_0-ct}(x_a)}[\Psi_\nu]\leq R$ and $\mathcal{Z}[\Psi_\nu]\in\mathcal{K}_1$. Since $m_K\leq p-1$, \eqref{E:SH_Reduction_counts} gives
\begin{align*}
\norm{\mathcal{Z}[\Psi_\nu]}_{H^{s_0}(B_{b_0-ct}(x_a))}
+\norm{\partial_t\mathcal{Z}[\Psi_\nu]}_{H^{s_0-1}(B_{b_0-ct}(x_a))}
\leq C\sqrt{R}.
\end{align*}
The extension bounds, Sobolev embedding, and composition give constants $J_{R,b_0}$ and $L_{R,b_0}$ bounding respectively $\partial_t\mathcal{Z}[\Psi_\nu]$ and $\partial_tA(\mathcal{Z}[\Psi_\nu])$ in $L^\infty(B_{b_0-ct}(x_a))$. They are independent of $a$ and $\nu$. For $x$ in this ball the vertical segment back to time zero stays in the shrinking balls. Thus
\begin{align*}
|A(\mathcal{Z}[\Psi_\nu](t,x))-A(\mathcal{Z}[\Psi_\nu](0,x_a))|
\leq b_0L_0+tL_{R,b_0}.
\end{align*}
For $t\leq T_h=\frac{a_0}{8C_E(1+L_{R,b_0})}$, define
\begin{subequations}
\label{E:SH_Extended_coefficients}
\begin{align}
\widetilde{A}_{\nu,a}(t)
&=A^0_{\nu,a}+\mathcal{E}_{b_0-ct}\big(A(\mathcal{Z}[\Psi_\nu](t))-A^0_{\nu,a}\big),\\
A^0_{\nu,a}&=A(\mathcal{Z}[\Psi_\nu](0,x_a)).
\end{align}
\end{subequations}
Equations \eqref{E:SH_Extension_bounds} and \eqref{E:SH_Ball_radius} imply $|\widetilde{A}_{\nu,a}-A^0_{\nu,a}|\leq \frac{a_0}{4}$. This coefficient vector agrees with $A(\mathcal{Z}[\Psi_\nu])$ on the shrinking ball, equals $A^0_{\nu,a}$ outside $B_{2b_0}(x_a)$, and satisfies \eqref{E:SH_Linear_coefficient_bound} with a constant depending on $R,b_0,c$ and the fixed compact sets, independently of $a,\nu$. Its principal polynomial remains strictly hyperbolic and monic.

Let $w$ be smooth on a neighborhood of a closed shrinking domain
$\{(t,x)\,|\,0\leq t\leq t_1,\ |x-x_a|\leq b_0-ct\}$, where $t_1\leq\min(1,T_h,\frac{b_0}{2c})$, and suppose $L_{\Psi_\nu}w^I=h^I$ there. Extend its $p$ initial traces by $\mathcal{E}_{b_0}$ and its source at time $t$ by $\mathcal{E}_{b_0-ct}$. The linear Cauchy theory in \cite[Theorems 89, 93, 108--109 and Proposition 121]{Leray-Book-1953} supplies a solution $v_a$ for the operator with coefficients \eqref{E:SH_Extended_coefficients} and these extended data and sources. Here the slices are uniformly noncharacteristic, the coefficients are uniformly strictly hyperbolic and constant outside a fixed ball, and the data and source have compact support. Their finite differentiability is sufficient for all the Sobolev orders being used, by the choice of $K$. For fixed $\nu,a$ all their local derivatives before extension are finite by analyticity.

For precision, the high differentiability required for this auxiliary solution can be obtained from Proposition 121 of \cite{Leray-Book-1953} as follows. Put $q=S_{\max}+d+3$, let $g_0,\dots,g_{p-1}$ be the extended traces, and let $h_a$ be the extended source for one scalar row. Recursively define, for $0\leq i<q$,
\begin{align*}
g_{p+i}=\partial_t^ih_a(0)
-\sum_{\substack{j+|\alpha|=p\\j<p}}\sum_{b=0}^i\binom{i}{b}
\partial_t^b\widetilde{A}_{j,\alpha}(0)\partial_x^\alpha g_{j+i-b}.
\end{align*}
Every trace on the right has index less than $p+i$. Set $P_a=\sum_{j=0}^{p+q-1}\frac{t^j}{j!}g_j$. The source $h_a-\widetilde{L}P_a$ has its time derivatives through order $q-1$ zero at $t=0$, belongs to $H^q([0,t_1]\times\mathbb{R}^d)$, and has compact spatial support. The choice $K>2p+2q+2$ ensures all these derivatives exist after extension. Proposition 121, with its parameters $M=q+1$, $t=q$, and $s=q+p-1$, gives a zero-data solution in $H^{q+p-1}([0,t_1]\times\mathbb{R}^d)$. Adding $P_a$ gives $v_a$. Since $q+p-1>S_{\max}+1+\frac{d+1}{2}$, spacetime Sobolev embedding gives continuous derivatives through order $S_{\max}+1$. If the global-time formulation is used, the coefficients can be extended across the endpoints by the finite reflection formula \eqref{E:SH_Extension_moments}, on short intervals, and then cut off to $A^0_{\nu,a}$. The strict $a_0$ margin preserves hyperbolicity. No quantitative bound on these auxiliary exterior intervals or on higher analytic norms is used in \eqref{E:SH_Homogeneous_linear_estimate}.

Proposition 92 of \cite{Leray-Book-1953} gives $\operatorname{supp}v_a(t)\subset B_{2b_0+ct}(x_a)$, so its slice energies are finite and continuous. If $(t,x)$ lies in the shrinking domain, any backward causal path satisfies
\begin{align*}
|y(s)-x_a|\leq|x-x_a|+c(t-s)<b_0-cs,\qquad 0\leq s\leq t.
\end{align*}
Along that domain the two linear operators, sources, and initial traces agree. Local uniqueness and dependence on the past domain, as in Proposition 92(2), Theorem 93 and Note 93.2, and Theorems 108--109 of \cite{Leray-Book-1953}, imply $v_a=w$ there. This comparison requires only local regularity of $w$. Applying \eqref{E:SH_Homogeneous_linear_estimate} to $v_a$ and then \eqref{E:SH_Extension_bounds} gives
\begin{align}
\label{E:SH_Local_linear_estimate}
E_{B_{b_0-ct}(x_a)}[w](t)
\leq C_R\left(E_{B_{b_0}(x_a)}[w](0)
+\int_0^t\sum_I\norm{h^I(s)}_{H^{s_I}(B_{b_0-cs}(x_a))}^2\,ds\right).
\end{align}
The constant is independent of $a$, $\nu$, and $t_1$. In particular, no global integrability of $w$ has been assumed in obtaining \eqref{E:SH_Local_linear_estimate}.

Using Equation~\eqref{E:SH_Linear_norm_estimate} in this linear comparison gives
\begin{align}
\label{E:SH_Local_linear_norm_estimate}
\begin{split}
E_{B_{b_0-ct}(x_a)}[w](t)^{\frac{1}{2}}
\leq{}& C_L E_{B_{b_0}(x_a)}[w](0)^{\frac{1}{2}}\\
&+C_L\int_0^t\left(\sum_I\norm{L_{\Psi_\nu}w^I(s)}_{H^{s_I}(B_{b_0-cs}(x_a))}^2\right)^{\frac{1}{2}}\,ds.
\end{split}
\end{align}
For $w=\partial_x^\alpha\Psi_\nu$, the Sobolev orders and the extension integer $K$ are unchanged. For each fixed $\alpha$ and $t_1$ the additional local derivatives needed for the auxiliary linear solution exist by analyticity; their magnitudes do not enter $C_L$. Thus $C_L$ is independent of $|\alpha|$, as well as $\nu,a,t_1$.

We record local uniqueness for two already existing smooth original solutions $\Psi,\Phi$ with the same prescribed function. This fact will also be used at an analytic endpoint. On a sufficiently small shrinking ball $D_t$ where both solutions are defined, put
\begin{align*}
D_{D_t}[w](t)&=\sum_{I,j<p}\norm{\partial_t^jw^I(t)}_{H^{S_I-j-1}(D_t)}^2,\\
w&=\Psi-\Phi.
\end{align*}
The identity in Equation~\eqref{E:SH_Normalized_reduction} gives
\begin{align*}
L_\Psi w^I=f_I(\Psi;\mathcal{W})-f_I(\Phi;\mathcal{W})+(L_\Phi-L_\Psi)\Phi^I.
\end{align*}
Every difference of unknown source arguments has $H^{r_I}(D_t)$ norm at most $CD_{D_t}[w]^{\frac{1}{2}}$, since $S_K-1-d_{IK}=r_I$. For the coefficient difference use $S_K-1-(m_K-1)=s_0-1\geq r_I$; the accompanying derivative $\partial_x^\beta\partial_t^j\Phi^I$, $j+|\beta|=p$, $j<p$, belongs to $H^{r_I}(D_t)$ by the top energy of $\Phi$. Extend the finite-dimensional source and coefficient maps smoothly to the whole argument space by cutoffs equal to one near both compact argument ranges. Writing $F$ for any one of these extensions, the identity
\begin{align*}
F(u)-F(v)=\int_0^1DF(v+\uptheta(u-v))[u-v]\,d\uptheta
\end{align*}
for these extensions, the composition estimates cited after Equation~\eqref{E:SH_Local_source_estimate}, and the algebra property $r_I>\frac{d}{2}$ show that
\begin{align*}
\sum_I\norm{L_\Psi w^I}_{H^{r_I}(D_t)}^2\leq C D_{D_t}[w].
\end{align*}
No convexity of the admissible sets is required, since only the smooth extensions are evaluated along line segments. The linear comparison proving Equation~\eqref{E:SH_Local_linear_estimate}, now at $k=r_I$, and Gronwall's inequality yield
\begin{align}
\label{E:SH_Local_difference_uniqueness}
D_{D_t}[w](t)\leq C D_{D_0}[w](0).
\end{align}
Here the radius, time length, and constant may depend on the finite local top norms of this pair and on compact subsets of the admissible sets containing their jets. Continuity of the coefficients and strict hyperbolicity allow the ball and time length to be decreased so that the extensions in Equation~\eqref{E:SH_Extended_coefficients} remain hyperbolic. The proof uses only linear existence and estimates for two given solutions. It applies also when their derivatives have continuous one-sided limits at the base slice, by the finite reflection extension used in Step 5. It is invariant under reversing time and multiplying the equation by $(-1)^p$, which reverses the characteristic roots without changing their separation or absolute speed bound.

The derivative counts \eqref{E:SH_Reduction_counts} put every argument of $f_I$ in $H^{s_I}(B_b(x_a))$. Composition and multiplication in that algebra give, for $\frac{b_0}{2}\leq b\leq b_0$ under the same bound $R$,
\begin{align}
\label{E:SH_Local_source_estimate}
\sum_I\norm{f_I(\Psi_\nu;\mathcal{W}_\nu)}_{H^{s_I}(B_b(x_a))}^2
\leq C_R\big(E_{B_b(x_a)}[\Psi_\nu]+F_{B_b(x_a),\nu}\big).
\end{align}
The composition estimates used here are \cite[Section 3.1, equations (3.1.17)--(3.1.20), (3.1.53)--(3.1.55), and (3.1.59)]{Taylor-Book-1991}. To check the absence of an additive constant in \eqref{E:SH_Local_source_estimate}, extend the smooth source map by a cutoff equal to one on a neighborhood of the compact range of its arguments and of the zero list. It still vanishes at zero. Extend its spatial arguments by \eqref{E:SH_Ball_extension} and apply the global composition estimate for a function vanishing at zero. Restriction and \eqref{E:SH_Extension_bounds} give \eqref{E:SH_Local_source_estimate}. The bounds on the additional derivative arguments follow by Sobolev embedding from \eqref{E:SH_Reduction_counts}; thus the cutoff and its derivative bounds depend only on $R,H$ and the fixed compact sets. This cutoff is used only for a norm estimate, not to define a principal operator outside its hyperbolic set.

Let $h=\frac{b_0}{4\sqrt{d}}$, $x_a=ha$ for $a\in\mathbb{Z}^d$, and $N_0=(\lceil8\sqrt{d}\rceil+3)^d$. The balls $B_{\frac{b_0}{2}}(x_a)$ cover $\mathbb{R}^d$ and every point belongs to at most $N_0$ balls $B_{b_0}(x_a)$. Indeed, a nearest lattice point is at distance at most $\frac{b_0}{8}$, and each coordinate of a center whose ball contains a given point lies in an interval containing at most $\lceil8\sqrt{d}\rceil+3$ lattice coordinates. For a finite set $\mathcal{A}\subset\mathbb{Z}^d$, put
\begin{align}
\label{E:SH_Finite_energy_sum}
E_{\mathcal{A}}[w](t)=\sum_{a\in\mathcal{A}}E_{B_{b_0-ct}(x_a)}[w](t).
\end{align}
This quantity is finite and continuous for each analytic $\Psi_\nu$, before global integrability has been established. For globally Sobolev $w$, bounded overlap gives $E_{\mathcal{A}}[w](t)\leq N_0E_{\mathbb{R}^d}[w](t)$, and it gives the same bound for the forcing energy.

Set
\begin{align*}
M&=1+N_0\sup_\nu E_{\mathbb{R}^d}[\Psi_\nu](0),\\
R&=16M,\\
d_0&=\operatorname{dist}(\mathcal{K}_0,\mathcal{J}\backslash\operatorname{int}\mathcal{K}_1)>0.
\end{align*}
The initial energy is finite by Step 2. For each finite $\mathcal{A}$ and each $\nu$, impose the bootstrap conditions $E_{\mathcal{A}}[\Psi_\nu]\leq R$ and $\mathcal{Z}[\Psi_\nu]\in\mathcal{K}_1$ on the union of its closed shrinking balls. Continuity and the strict initial inequalities give a positive starting interval for each such pair; no uniform starting interval is needed. Summing \eqref{E:SH_Local_linear_estimate} and \eqref{E:SH_Local_source_estimate} gives
\begin{subequations}
\label{E:SH_Summed_local_estimates}
\begin{align}
E_{\mathcal{A}}[w](t)&\leq C_R\left(E_{\mathcal{A}}[w](0)+\int_0^t\sum_{a\in\mathcal{A},I}
\norm{L_{\Psi_\nu}w^I(s)}_{H^{s_I}(B_{b_0-cs}(x_a))}^2\,ds\right),\\
\sum_{a\in\mathcal{A},I}\norm{f_I(\Psi_\nu;\mathcal{W}_\nu)}_{H^{s_I}(B_{b_0-ct}(x_a))}^2
&\leq C_R(E_{\mathcal{A}}[\Psi_\nu]+N_0H).
\end{align}
\end{subequations}
These constants are independent of the number of balls.

Fix $C_R\geq1$ for \eqref{E:SH_Summed_local_estimates} and the coefficient multiplication bounds. Put $d_1=\frac{M}{64C_R}$. Denote the limiting initial traces through order $p-1$ by $a_j^I$. For $L\geq1$, define $J_L$ by
\begin{align}
\label{E:SH_Fourier_cutoff}
\widehat{J_Lv}(\xi)=f_1\left(\frac{|\xi|}{L}\right)\widehat{v}(\xi).
\end{align}
Here $\xi\in\mathbb{R}^d$ for the Fourier transform, and the same formula on the Fourier lattice defines $J_L$ on $\mathbb{T}^d$. Since $0\leq f_1\leq1$, $J_L$ is a contraction in every derivative norm, maps each Sobolev space into every higher Sobolev space, and converges strongly to the identity in each fixed Sobolev space as $L\rightarrow\infty$. Fix one $L$ large enough that $b_j^I=J_La_j^I$ satisfy
\begin{align*}
\sum_{I,j}\norm{a_j^I-b_j^I}_{H^{S_I-j}(\mathbb{R}^d)}^2\leq\frac{d_1}{4N_0}.
\end{align*}
By Step 2, finitely many more terms can be discarded so that the same bound holds for $\Psi^I_{(j),\nu}-a_j^I$. Define one fixed function
\begin{align}
\label{E:SH_Fixed_comparison_polynomial}
P^I(t,x)=\sum_{j=0}^{p-1}\frac{t^j}{j!}b_j^I(x).
\end{align}
The squared triangle inequality and overlap bound imply
\begin{align*}
E_{\mathcal{A}}[\Psi_\nu-P](0)\leq d_1.
\end{align*}
There is $t_P>0$ such that $N_0E_{\mathbb{R}^d}[P](t)\leq2M$ for $0\leq t\leq t_P$. Moreover the finite number
\begin{align*}
P_{\mathrm{high}}=\sup_{0\leq t\leq\min(1,t_P)}
\sum_I\sum_{|\gamma|=p}\norm{\partial^\gamma P^I(t)}_{H^{s_I}(\mathbb{R}^d)}^2
\end{align*}
and local multiplication give
\begin{align*}
\sum_{a\in\mathcal{A},I}\norm{L_{\Psi_\nu}P^I(t)}_{H^{s_I}(B_{b_0-ct}(x_a))}^2\leq C_RN_0P_{\mathrm{high}}.
\end{align*}
Indeed, each differentiated $P$ is multiplied by a coefficient with a fixed local $H^{s_I}$ and $L^\infty$ bound; $s_I>\frac{d}{2}$, and summation of its squared local norms uses the overlap bound. Together with \eqref{E:SH_Summed_local_estimates}, this gives a finite $K_{R,P}$, independent of $\nu,\mathcal{A}$, bounding the summed squared $H^{s_I}$ norms of $f_I-L_{\Psi_\nu}P^I$. Hence
\begin{align}
\label{E:SH_Uniform_energy_estimate}
E_{\mathcal{A}}[\Psi_\nu-P](t)\leq C_R(d_1+tK_{R,P}).
\end{align}
Define the positive number
\begin{align}
\label{E:SH_Bootstrap_time}
\overline{T}=\min\left(1,t_P,\frac{b_0}{2c},T_h,
\frac{d_0}{2(1+J_{R,b_0})},\frac{M}{64C_R(1+K_{R,P})}\right).
\end{align}
When a prescribed function is present, replace $\overline{T}$ by $\min(\overline{T},\frac{T_0}{2})$. These choices use no analytic radius and no prior common existence interval.
On every existing bootstrap interval in $[0,\overline{T})$, \eqref{E:SH_Uniform_energy_estimate} is at most $\frac{M}{32}$, and therefore
\begin{align}
\label{E:SH_Bootstrap_improvement}
E_{\mathcal{A}}[\Psi_\nu](t)
\leq2E_{\mathcal{A}}[\Psi_\nu-P](t)+2E_{\mathcal{A}}[P](t)
\leq\frac{65M}{16}<\frac{R}{2}.
\end{align}
For every point in a shrinking ball,
\begin{align}
\label{E:SH_Bootstrap_jet_improvement}
|\mathcal{Z}[\Psi_\nu](t,x)-\mathcal{Z}[\Psi_\nu](0,x)|
\leq tJ_{R,b_0}\leq\frac{d_0}{2}.
\end{align}
Both bootstrap conditions, including the finite sum in Equation~\eqref{E:SH_Finite_energy_sum}, are strictly improved on the existing domain. Hence a first exit through either inequality before $\overline{T}$ is impossible while the solution exists on a neighborhood of the relevant closed truncated cones. Extension through a possible endpoint of that domain is proved in Steps 6--7. The order of choices was $b_0$, $R$, $C_R$, $d_1$, $P$, $K_{R,P}$, and $\overline{T}$; none requires $R$ to dominate a function of itself.

On $\mathbb{T}^d$, the analytic solutions already have finite continuous energies on closed subintervals of their existing domains. Use the single global estimate \eqref{E:SH_Homogeneous_linear_estimate}, set $N_0=1$, and define $M,R,d_1,P$ by the same formulas with domain $\mathbb{T}^d$. The source estimate has the form $\sum_I\norm{f_I}_{H^{s_I}(\mathbb{T}^d)}^2\leq C_R(1+E_{\mathbb{T}^d}[\Psi_\nu]+H)$; the fixed volume term is absorbed into $K_{R,P}$. Equations \eqref{E:SH_Uniform_energy_estimate} and \eqref{E:SH_Bootstrap_improvement} hold on existing bootstrap intervals with $E_{\mathcal{A}}=E_{\mathbb{T}^d}$, and the bound on $\partial_t\mathcal{Z}$ gives Equation~\eqref{E:SH_Bootstrap_jet_improvement} on these intervals. The ball and propagation restrictions in \eqref{E:SH_Bootstrap_time} are unnecessary in this case.

\medskip
\noindent \underline{Step 6} (\textbf{Spatial analyticity on the existing interval}).\nopagebreak[4]
We prove the derivative bounds needed to extend each analytic solution. The factorial estimates follow the method of \cite[Sections 3.2--3.5, pp.~196--201]{Alinhac-Metivier-1984}; we give the argument for the indices in Equation~\eqref{E:SH_Reduction_counts}. Fix $\nu$ and one center $x_a$, and write
\begin{align*}
D_t&=B_{b_0-ct}(x_a),\qquad 0\leq t<\uptau\leq\overline{T}.
\end{align*}
Suppose that $\Psi=\Psi_\nu$ is analytic on a neighborhood of every closed truncated cone with upper time $t_1<\uptau$, that $E_{D_t}[\Psi](t)\leq R$, and that $\mathcal{Z}[\Psi]\in\mathcal{K}_1$. The initial coefficient bounds and the choices $b_0,c,T_h$ are those of Step 5. On the torus take $D_t=\mathbb{T}^d$. Constants denoted by $C$ in this step may depend on $R,H,b_0,c$, the fixed indices and coefficient functions, the admissible compact sets, and the root separation. They are independent of $\nu,a,t_1,\uptau$, and the number of spatial differentiations. Constants depending on the analytic data will be specified separately.

Besides Equation~\eqref{E:SH_Reduction_counts}, we use
\begin{align}
\label{E:SH_Analytic_derivative_margin}
S_K-m_K-s_I=p-m_I-1\geq0.
\end{align}
If $V_I$ denotes the list in Equation~\eqref{E:SH_Reduced_source_arguments}, its entries belong to $H^{s_I}(D_t)$ with norms at most $C(\sqrt{R}+\sqrt{H})$. Equation~\eqref{E:SH_Analytic_derivative_margin} gives this bound also for $\mathcal{Z}[\Psi]$ and $\partial_{x_l}\mathcal{Z}[\Psi]$. More generally, the spatial derivatives of order $q$ of their unknown entries have $H^{s_I}(D_t)$ norm at most $Ca_q(t)$, where
\begin{align}
\label{E:SH_Analytic_energy_sequence}
a_q(t)=\max_{|\alpha|=q}E_{D_t}[\partial_x^\alpha\Psi](t)^{\frac{1}{2}},
\qquad q\geq1,
\end{align}
and $C$ accounts for the fixed finite lists. Indeed, an unknown entry of total order $v$ and time order $b<p$ uses at most $s_I+v-b\leq S_K-b$ spatial derivatives after the $q$ specified differentiations. The inequalities required here are $v\leq d_{IK}$ for $V_I$ and $v\leq m_K$ for $\mathcal{Z}$ and $\partial_x\mathcal{Z}$.

Let $b_q(t)$ be the sum, over rows $I$ and prescribed entries $v$ of $V_I$, of $\max_{|\alpha|=q}\norm{\partial_x^\alpha v(t)}_{H^{s_I}(D_t)}$; put $b_q=0$ when there is no prescribed function. Analyticity of the initial traces and of $\mathcal{W}_\nu$ on a neighborhood of $[0,\overline{T}]\times\overline{D_0}$ gives numbers $M_{\mathrm{an}},B_{\mathrm{an}}\geq1$ such that
\begin{align}
\label{E:SH_Individual_analytic_data}
a_q(0)+\sup_{0\leq t\leq\overline{T}}b_q(t)
\leq M_{\mathrm{an}}B_{\mathrm{an}}^q q!,\qquad q\geq1.
\end{align}
To obtain these numbers, cover the compact initial ball and forcing cylinder by finitely many analytic neighborhoods and use the Cauchy estimates on smaller neighborhoods. The additional, fixed number of derivatives in the Sobolev norms is absorbed by increasing $B_{\mathrm{an}}$. The trace recursion in Step 2 supplies the analytic traces through order $p-1$. Both numbers in Equation~\eqref{E:SH_Individual_analytic_data} may depend on $\nu$ and $a$.

Set $\upmu_q=\frac{q!}{(q+1)^2}$ for $q\geq1$. For $n\geq2$,
\begin{subequations}
\label{E:SH_Factorial_convolutions}
\begin{align}
\label{E:SH_Factorial_product}
\sum_{k=1}^{n-1}\binom{n}{k}\upmu_k\upmu_{n-k}&\leq8\upmu_n,\\
\label{E:SH_Factorial_shifted_product}
\sum_{k=2}^{n}\binom{n}{k}\upmu_{k-1}\upmu_{n-k+1}&\leq16n\upmu_n.
\end{align}
\end{subequations}
For Equation~\eqref{E:SH_Factorial_product}, division by $n!$ and splitting at $\frac{n}{2}$ give an upper bound $\frac{8}{(n+1)^2}$, since $\sum_{k\geq1}(k+1)^{-2}<1$. For Equation~\eqref{E:SH_Factorial_shifted_product}, the summand after division by $n!$ is
\begin{align*}
\frac{n-k+1}{k^3(n-k+2)^2}.
\end{align*}
The sum over $k\leq\frac{n}{2}$ is at most $\frac{1}{n}$, and the sum over $k>\frac{n}{2}$ is at most $\frac{8}{n^2}$. Multiplication by $\frac{(n+1)^2}{n}$ bounds their sum by $12$ for $n\geq2$, which proves the asserted bound with $16$. For every multi-index $\alpha$ of length $n$ and positive integers $k_1+\cdots+k_q=n$, the multinomial theorem also gives
\begin{align}
\label{E:SH_Multiindex_convolution}
\sum_{\substack{\gamma_1+\cdots+\gamma_q=\alpha\\|\gamma_i|=k_i}}
\frac{\alpha!}{\gamma_1!\cdots\gamma_q!}
=\frac{n!}{k_1!\cdots k_q!}.
\end{align}
Thus Equation~\eqref{E:SH_Factorial_convolutions} applies to spatial multi-indices without an extra factor exponential in $n$.

We next give the composition estimate. Let $F$ be one of the analytic maps $f_I$ or
\begin{align*}
G_{j,\beta,l}(z,z_1,\dots,z_d)=DA_{j,\beta}(z)[z_l].
\end{align*}
The actual argument lists $v$ of these maps have uniformly bounded $H^{s_I}(D_t)$ norms and compact ranges, by Equations~\eqref{E:SH_Reduction_counts} and \eqref{E:SH_Analytic_derivative_margin}. Analyticity on a neighborhood of those compact ranges gives $|D^{q+r}F|\leq C A_0^{q+r}(q+r)!$ for $0\leq r\leq s_I$. The spatial chain rule through the fixed order $s_I$, H\"older's inequality, and the fixed-order composition estimates cited after Equation~\eqref{E:SH_Local_source_estimate} therefore give
\begin{align}
\label{E:SH_Analytic_map_derivatives}
\norm{D^qF(v)}_{H^{s_I}(D_t)}\leq C A_1^q q!,\qquad q\geq1.
\end{align}
Here and in products of derivative lists the norms include their finitely many components. More explicitly, the derivatives of $v$ in each chain-rule product have positive orders $r_i$ with $\sum_i r_i\leq s_I$. The Gagliardo--Nirenberg estimate on balls with radii in $[\frac{b_0}{2},b_0]$ bounds their $L^{\frac{2s_I}{r_i}}(D_t)$ norms by
\begin{align*}
C\norm{v}_{L^\infty(D_t)}^{1-\frac{r_i}{s_I}}\norm{v}_{H^{s_I}(D_t)}^{\frac{r_i}{s_I}}
+C\norm{v}_{L^\infty(D_t)}.
\end{align*}
H\"older's inequality and the bounded volume then bound the product in $L^2(D_t)$. There are finitely many terms at these fixed orders; the factor $\frac{(q+s_I)!}{q!}$ is absorbed into $A_1^q$. The extension bounds in Equation~\eqref{E:SH_Extension_bounds} make the constants uniform in the radius and center. On the torus the fixed volume and periodic Sobolev estimates give the same bound. No cutoff of $F$ is used in Equation~\eqref{E:SH_Analytic_map_derivatives}.

For $|\alpha|=n$ the exact chain rule is
\begin{align}
\label{E:SH_Analytic_chain_rule}
\begin{split}
\partial_x^\alpha F(v)
={}&DF(v)[\partial_x^\alpha v]\\
&+\sum_{q=2}^n\ \sum_{\substack{\gamma_1+\cdots+\gamma_q=\alpha\\|\gamma_i|\geq1}}
\frac{\alpha!}{q!\gamma_1!\cdots\gamma_q!}
D^qF(v)[\partial_x^{\gamma_1}v,\dots,\partial_x^{\gamma_q}v].
\end{split}
\end{align}
Suppose, for $1\leq k<n$, that the derivatives of $v$ of order $k$ have $H^{s_I}(D_t)$ norm at most $D H_*\upepsilon^{1-k}\upmu_k e^{\uplambda kt}$, where $D$ is fixed. By the algebra property, Equations~\eqref{E:SH_Factorial_convolutions}--\eqref{E:SH_Analytic_map_derivatives} bound the sum with $q\geq2$ in Equation~\eqref{E:SH_Analytic_chain_rule} by
\begin{align}
\label{E:SH_Analytic_composition_remainder}
C H_*^2\upepsilon^{2-n}\upmu_n e^{\uplambda nt},
\qquad \upepsilon H_*\leq\updelta_0.
\end{align}
Indeed, its terms for a fixed $q$ have total norm at most
$C A_2^q H_*^q\upepsilon^{q-n}8^{q-1}\upmu_n e^{\uplambda nt}$, where $A_2$ includes $D$ and the algebra constant. Fix $\updelta_0>0$ with $8A_2\updelta_0\leq\frac{1}{2}$ and sum this geometric series. If the derivative bound is also known at $k=n$, Equations~\eqref{E:SH_Analytic_chain_rule} and \eqref{E:SH_Analytic_composition_remainder} bound the full derivative by $C H_*\upepsilon^{1-n}\upmu_n e^{\uplambda nt}$. For $n=1$ the sum with $q\geq2$ is empty.

We claim that $H_*,\upepsilon,\uplambda$ can be fixed so that
\begin{align}
\label{E:SH_Factorial_solution_bound}
a_n(t)\leq B_n(t):=H_*\upepsilon^{1-n}\upmu_n e^{\uplambda nt},
\qquad n\geq1,\quad 0\leq t<\uptau.
\end{align}
Spatial differentiation of Equation~\eqref{E:SH_Normalized_reduction} gives
\begin{align}
\label{E:SH_Analytic_differentiated_equation}
\begin{split}
L_\Psi\partial_x^\alpha\Psi^I
={}&\partial_x^\alpha f_I(V_I)\\
&-\sum_{\substack{j+|\beta|=p\\j<p}}
\sum_{0<\gamma\leq\alpha}\binom{\alpha}{\gamma}
\partial_x^\gamma A_{j,\beta}(\mathcal{Z}[\Psi])
\partial_x^{\alpha-\gamma+\beta}\partial_t^j\Psi^I.
\end{split}
\end{align}
Write $n=|\alpha|$ and $k=|\gamma|$. Since $S_I-j=s_I+p-j-1$, the differentiated field in the commutator satisfies
\begin{align}
\label{E:SH_Analytic_commutator_field}
\norm{\partial_x^{\alpha-\gamma+\beta}\partial_t^j\Psi^I}_{H^{s_I}(D_t)}
\leq a_{n-k+1}(t).
\end{align}
In fact, $|\alpha-\gamma+\beta|=n-k+p-j$; removing any sub-multi-index of length $n-k+1$ leaves exactly $p-j-1$ derivatives, which are included in Equation~\eqref{E:SH_Analytic_energy_sequence}. For $k=1$, the $H^{s_I}(D_t)$ bound on $\partial_x A$ gives a total contribution at most $Cn a_n(t)$.

For $k\geq2$, take $l$ with $\gamma_l>0$ and write
\begin{align*}
\partial_x^\gamma A_{j,\beta}(\mathcal{Z})
=\partial_x^{\gamma-e_l}G_{j,\beta,l}(\mathcal{Z},\partial_{x_1}\mathcal{Z},\dots,\partial_{x_d}\mathcal{Z}).
\end{align*}
Assume Equation~\eqref{E:SH_Factorial_solution_bound} at every order less than $n$. For $1\leq q\leq k-1$, every derivative of an argument of $G$ of order $q$ is bounded by $Ca_q$ in $H^{s_I}(D_t)$, by Equation~\eqref{E:SH_Analytic_derivative_margin}. The arguments themselves have $H^{s_I}(D_t)$ norms at most $C\sqrt{R}$ by Equation~\eqref{E:SH_Analytic_derivative_margin} and the bootstrap bound. Equations~\eqref{E:SH_Analytic_chain_rule} and \eqref{E:SH_Analytic_composition_remainder} consequently give
\begin{align*}
\norm{\partial_x^\gamma A_{j,\beta}(\mathcal{Z})}_{H^{s_I}(D_t)}
\leq C H_*\upepsilon^{2-k}\upmu_{k-1}e^{\uplambda(k-1)t}.
\end{align*}
Equations~\eqref{E:SH_Analytic_commutator_field}, \eqref{E:SH_Multiindex_convolution}, and \eqref{E:SH_Factorial_convolutions} bound the sum over $k\geq2$ in Equation~\eqref{E:SH_Analytic_differentiated_equation} by
\begin{align*}
C nH_*^2\upepsilon^{2-n}\upmu_n e^{\uplambda nt}.
\end{align*}
For the source, Equations~\eqref{E:SH_Reduction_counts}, \eqref{E:SH_Analytic_chain_rule}, and \eqref{E:SH_Analytic_composition_remainder} give
\begin{align*}
\norm{\partial_x^\alpha f_I(V_I)}_{H^{s_I}(D_t)}
\leq C(a_n(t)+b_n(t))+CH_*^2\upepsilon^{2-n}\upmu_n e^{\uplambda nt}.
\end{align*}
Thus, if $b_k\leq B_k$ for $1\leq k\leq n$ and $\upepsilon H_*\leq\min(\updelta_0,1)$, Equation~\eqref{E:SH_Local_linear_norm_estimate} implies
\begin{align}
\label{E:SH_Factorial_integral_estimate}
a_n(t)\leq C_La_n(0)+C_0\int_0^t(n+1)(a_n(s)+B_n(s))\,ds,
\qquad n\geq2.
\end{align}
The constant $C_0$ is independent of $n,H_*,\upepsilon$, and $t_1$. The maximum over $\alpha$ is taken after the integral. The linear estimate is always used at the fixed Sobolev orders $s_I$ and $S_I-j$; its constant is therefore independent of $n$.

For $n=1$, the chain rule and the $k=1$ commutator instead give $C(a_1+b_1)$. Equation~\eqref{E:SH_Local_linear_norm_estimate}, Equation~\eqref{E:SH_Individual_analytic_data}, and Gronwall's inequality give a finite $A_1^*$ such that $a_1(t)\leq A_1^*$ for $t<\uptau$, independently of $t_1$. For example, after increasing the fixed constant $C$, one can take
\begin{align*}
A_1^*=C_L\big(a_1(0)+CM_{\mathrm{an}}B_{\mathrm{an}}\overline{T}\big)e^{C\overline{T}}.
\end{align*}
Fix
\begin{align*}
H_*&\geq\max\big(1,4A_1^*,16\max(C_L,1)M_{\mathrm{an}}B_{\mathrm{an}}\big),\\
0<\upepsilon&\leq\min\big(1,(4B_{\mathrm{an}})^{-1},\frac{\updelta_0}{H_*},\frac{1}{H_*}\big),\\
\uplambda&\geq\max(1,12C_0).
\end{align*}
Since $(n+1)^2(B_{\mathrm{an}}\upepsilon)^{n-1}\leq4$, Equation~\eqref{E:SH_Individual_analytic_data} gives $C_La_n(0)\leq\frac{B_n(0)}{4}$ and $b_n(t)\leq B_n(t)$. The bound on $A_1^*$ proves Equation~\eqref{E:SH_Factorial_solution_bound} for $n=1$. For $n\geq2$, Gronwall's inequality in Equation~\eqref{E:SH_Factorial_integral_estimate} gives
\begin{align*}
a_n(t)&\leq\frac{B_n(0)}{4}e^{C_0(n+1)t}
+C_0(n+1)\int_0^t e^{C_0(n+1)(t-s)}B_n(s)\,ds\\
&\leq\left(\frac{1}{4}+\frac{C_0(n+1)}{\uplambda n-C_0(n+1)}\right)B_n(t)
\leq\left(\frac{1}{4}+\frac{1}{7}\right)B_n(t)<B_n(t).
\end{align*}
This proves the induction. All computations were made on $[0,t_1]$ with $t_1<\uptau$, and their constants are independent of $t_1$. In particular,
\begin{align}
\label{E:SH_Positive_individual_radius}
\uprho_*:=\upepsilon e^{-\uplambda\overline{T}}>0
\end{align}
is available throughout the existing interval for this approximant on this ball. No lower bound for $\uprho_*$ uniform in $\nu$ or $a$ is needed.

If the coefficient maps also depend real-analytically on $(t,x)$, append these prescribed coordinates to the argument lists in Equation~\eqref{E:SH_Analytic_chain_rule}. Their spatial derivatives vanish at orders $q\geq2$, and their derivatives at order one are constant; their contribution can therefore be included in $b_1$ in Equation~\eqref{E:SH_Individual_analytic_data}. In the commutator use
\begin{align*}
G_{j,\beta,l}(t,x,z,z_1,\dots,z_d)
=\partial_{x_l}A_{j,\beta}(t,x,z)+D_zA_{j,\beta}(t,x,z)[z_l].
\end{align*}
On each compact cylinder the derivatives of these analytic maps satisfy Equation~\eqref{E:SH_Analytic_map_derivatives}. For this coordinate-dependent variant, all constants in the analytic estimates of this step may depend on the chosen compact cylinder; $C_L$ and the constants in Steps 4--5 are still fixed by the assumed uniform finite-order bounds. Equations~\eqref{E:SH_Factorial_integral_estimate}--\eqref{E:SH_Positive_individual_radius} still give a positive individual radius up to $\overline{T}$. Uniformity of the finite-order estimates in Steps 4--5 is the only spatial uniformity used in the choice of $\overline{T}$.

\medskip
\noindent \underline{Step 7} (\textbf{Extension and the common real interval}).\nopagebreak[4]
We first extend a solution satisfying the hypotheses of Step 6 through a possible endpoint $\uptau<\overline{T}$. Put $D_\uptau=B_{b_0-c\uptau}(x_a)$. Since $\overline{D_\uptau}\subset D_t$ for $t<\uptau$ and $S_I-j\geq s_I>\frac{d}{2}$ for $j<p$, Sobolev embedding and Equations~\eqref{E:SH_Factorial_solution_bound}--\eqref{E:SH_Positive_individual_radius} give
\begin{align}
\label{E:SH_Terminal_factorial_bound}
\sup_{t<\uptau}\norm{\partial_x^\alpha\partial_t^j\Psi^I(t)}_{C^0(\overline{D_\uptau})}
\leq C_*\uprho_*^{-|\alpha|}|\alpha|!,\qquad j<p.
\end{align}
Here $C_*$ may depend on $\nu,a$, and has been enlarged to cover $|\alpha|=0$. For $j\leq p-2$, Equation~\eqref{E:SH_Terminal_factorial_bound} also bounds the time derivative by taking $j+1$. For $j=p-1$, Equation~\eqref{E:SH_Normalized_reduction}, differentiated a fixed number of times in space, bounds $\partial_t^p\partial_x^\alpha\Psi^I$ in $C^0(\overline{D_\uptau})$: its mixed principal terms use only $j<p$ and finitely many additional spatial derivatives, and its source is the analytic function of the list in Equation~\eqref{E:SH_Reduced_source_arguments}. Hence, for every fixed $\alpha$ and $j<p$, $\partial_x^\alpha\partial_t^j\Psi^I$ is uniformly Lipschitz in time into $C^0(\overline{D_\uptau})$. There are traces $g_j^I$ such that
\begin{align*}
\partial_t^j\Psi^I(t)&\longrightarrow g_j^I
\quad\text{in }C^k(\overline{D_\uptau})\text{ for every }k,\\
\norm{\partial_x^\alpha g_j^I}_{C^0(\overline{D_\uptau})}
&\leq C_*\uprho_*^{-|\alpha|}|\alpha|!.
\end{align*}

These traces are analytic on a neighborhood of the closed ball. At each $x\in\overline{D_\uptau}$ their Taylor series have homogeneous sums bounded by
$C_*\big(\frac{|y-x|_1}{\uprho_*}\big)^n$. For $y\in\overline{D_\uptau}$, Taylor's formula along the segment from $x$ to $y$ has remainder at most $C_*\big(\frac{|y-x|_1}{\uprho_*}\big)^{n+1}$; boundary segments are obtained as limits of interior segments. The series therefore equal $g_j^I$ wherever $|y-x|_1<\uprho_*$. Use Euclidean balls of one radius $0<r<\frac{\uprho_*}{4\sqrt{d}}$ centered at points of $\overline{D_\uptau}$. Two such balls that intersect have a connected intersection containing a point of $D_\uptau$: their midpoint lies in the closed convex ball, and a small displacement into its interior remains in the intersection. The analytic identity theorem identifies their Taylor series on that intersection. They define one analytic extension to the union of these balls.

Every fixed higher time derivative has a finite expression in spatial derivatives of traces with time order less than $m_K$ and derivatives of $\mathcal{W}_\nu$. This follows by the substitution in Equation~\eqref{E:SH_Inverted_normal_form}: at each replacement the time order minus the index of the unknown decreases by at least one, as proved before Equation~\eqref{E:SH_Normalized_reduction}. These expressions have terminal limits by Equation~\eqref{E:SH_Terminal_factorial_bound}. Passing to the limit in the original equations and their finitely many time derivatives gives
\begin{align}
\label{E:SH_Terminal_compatibility}
\left.\partial_t^a\mathcal{E}^I[\Psi;\mathcal{W}_\nu]\right|_{t=\uptau}=0,
\qquad 0\leq a<n_I,
\end{align}
with the time derivatives assigned by that finite recursion. The identities hold in $D_\uptau$ and extend to the analytic neighborhoods by the identity theorem. The physical terminal jet belongs to $\mathcal{K}_1\Subset\mathcal{U}$; continuity permits restriction of these neighborhoods so that the extended jets remain admissible. The prescribed forcing is analytic there and has its jet in $\mathcal{L}_1\Subset\mathcal{V}$.

At $t=\uptau$, use the original traces $g_j^I$, $0\leq j<m_I$, in Equation~\eqref{E:SH_CK_normal_form}. Its time coefficient matrix remains invertible on $\mathcal{K}_1$. Cauchy--Kovalevskaya gives analytic solutions near the terminal slice, and compactness of $\overline{D_\uptau}$ gives a positive time width $\updelta$ after taking a finite cover. Uniqueness for the analytic Cauchy problem identifies these solutions on smaller intersecting cylinders. Equation~\eqref{E:SH_Terminal_compatibility} and $\partial_t^{n_I}\mathcal{E}^I=0$ show that they solve the original equations. Their traces through $p-1$ agree with $g_j^I$ by the recursion in Step 2.

We verify agreement with the old solution on the past side of the terminal slice. The finite-order local difference estimate in Equation~\eqref{E:SH_Local_difference_uniqueness}, with time reversed, applies on small backward cones based strictly inside $D_\uptau$. The terminal traces of the difference vanish through $p-1$, so the solutions agree on these cones. The terminal limits just established give the regularity required at their bases. Under time reversal and multiplication by $(-1)^p$, the characteristic roots change sign; their separation and absolute speed bound are preserved. Each terminal Cauchy--Kovalevskaya cylinder can be taken to have a ball as its spatial base. Its intersection with the old open shrinking cone is convex, and contains one of the backward cones just used, including when its center lies on $\partial D_\uptau$. The analytic identity theorem therefore gives equality on the entire past intersection. This proves that the new solution extends the old one. For $t>\uptau$, $D_t\subset D_\uptau$, so it contains a further segment of the required shrinking cone. The width $\updelta$ may depend on $\nu$ and $a$.

We now apply this extension to the finite unions in Step 5. Fix $\nu$ and a finite $\mathcal{A}\subset\mathbb{Z}^d$. The local Cauchy--Kovalevskaya solutions give a positive initial interval on the union of its closed base balls. Let $\uptau$ be the supremum of the heights less than $\overline{T}$ to which they extend analytically while satisfying the bootstrap bounds. Uniqueness identifies extensions wherever their domains overlap, so their union defines the solution for every $t<\uptau$. Equations~\eqref{E:SH_Bootstrap_improvement} and \eqref{E:SH_Bootstrap_jet_improvement} hold on that interval. If $\uptau<\overline{T}$, the argument of Steps 6--7 extends the solution across the terminal closed ball for every $a\in\mathcal{A}$. The minimum of the finitely many positive widths is positive. The strict energy and admissibility margins persist for a shorter positive width by continuity on this finite union, contradicting the definition of $\uptau$. Thus every such finite union has an analytic solution to height $\overline{T}$ with the stated bounds.

For each $a\in\mathbb{Z}^d$ denote its solution by $\Psi_{\nu,a}$ on
\begin{align*}
Q_a=\{(t,x)\,|\,0\leq t<\overline{T},\ |x-x_a|<b_0-ct\}.
\end{align*}
If $Q_a\cap Q_b$ is nonempty, its spatial base at $t=0$ is nonempty, the solutions agree near that base by analytic Cauchy uniqueness, and the intersection for $t>0$ is convex and connected. Analytic identity gives $\Psi_{\nu,a}=\Psi_{\nu,b}$ throughout the intersection. The covering in Step 5 therefore defines
\begin{align}
\label{E:SH_Analytic_global_gluing}
\Psi_\nu(t,x)=\Psi_{\nu,a}(t,x)
\quad\text{whenever }(t,x)\in Q_a.
\end{align}
This definition is independent of $a$, locally real analytic, and solves Equation~\eqref{SH:1} with the prescribed data. For every finite $\mathcal{A}$ it agrees with the solution on that union by the same uniqueness on connected intersections. Consequently the finite-sum bounds apply to Equation~\eqref{E:SH_Analytic_global_gluing}.

On $\mathbb{T}^d$, compactness gives the initial analytic interval. Step 6 uses the global norm estimate and the fixed volume. Equation~\eqref{E:SH_Terminal_factorial_bound} gives terminal traces in periodic coordinate balls; the Taylor extensions agree on their intersections and under lattice translations, so the terminal data are periodic. The normal form, terminal compatibility, and backward difference estimate then extend the periodic solution. The strict torus bootstrap bounds of Step 5 exclude an endpoint before $\overline{T}$. In either spatial domain we set
\begin{align}
\label{E:SH_Common_real_time}
T=\frac{\overline{T}}{2}>0.
\end{align}
All the exact nonlinear analytic approximants are now defined on the common closed real interval $[0,T]$. Equation~\eqref{E:SH_Bootstrap_time} involves only the finite-order bounds; the individual analytic constants used for extension do not enter this choice.

For $L\geq1$, take $\mathcal{A}_L=\{a\in\mathbb{Z}^d\,|\,|x_a|\leq L+b_0\}$. Its shrinking balls cover $B_L(0)$ because $b_0-ct\geq \frac{b_0}{2}$. Equation \eqref{E:SH_Bootstrap_improvement} bounds $E_{B_L(0)}[\Psi_\nu](t)$ by $\frac{65M}{16}$. Monotone convergence as $L\rightarrow\infty$ now proves
\begin{subequations}
\label{E:SH_Uniform_solution_bound}
\begin{align}
\sup_\nu\sup_{0\leq t\leq T}E_{\mathbb{R}^d}[\Psi_\nu](t)&\leq\frac{65M}{16},\\
\overline{\mathcal{Z}[\Psi_\nu]([0,T]\times\mathbb{R}^d)}&\subset\mathcal{K}_1.
\end{align}
\end{subequations}
This proves global integrability, rather than assuming it in a global energy inequality.

We also record strong continuity of each analytic approximant in its full energy norm on $\mathbb{R}^d$. Set $X=\prod_{I,j<p}H^{S_I-j}(\mathbb{R}^d)$ and $V_\nu=(\partial_t^j\Psi_\nu^I)_{I,j<p}$. The bound \eqref{E:SH_Uniform_solution_bound} allows \eqref{E:SH_Local_linear_estimate} to be restarted at any $t_0\in[0,T]$, with a common smaller radius and time length, using $A(\mathcal{K}_1)$ in place of $A(\mathcal{K}_0)$. Explicitly, take a positive hyperbolicity margin $a_1$ for this compact coefficient range and a speed bound $c_1$ on its closed $a_1$-neighborhood. The global bound gives finite constants $L_1,L_2$ for $\nabla_xA$ and $\partial_tA$ in $L^\infty(\mathbb{R}^d)$, independent of $\nu,t_0$. The choices
\begin{align*}
b_1&=\min\left(1,\frac{a_1}{8C_E(1+L_1)}\right),\\
t_1&=\min\left(1,\frac{b_1}{2c_1},\frac{a_1}{8C_E(1+L_2)}\right)
\end{align*}
give the required radius and time length by \eqref{E:SH_Extended_coefficients} and \eqref{E:SH_Extension_bounds}. For $t<t_0$, reverse time and multiply the equation by $(-1)^p$; the roots change sign and the same absolute speed and separation bounds hold. Finite summation, the overlap bound, and exhaustion give, for a smooth comparison polynomial $P$ based at $t_0$,
\begin{align}
\label{E:SH_Analytic_energy_continuity}
\norm{V_\nu(t)-V_P(t)}_X^2
\leq C\left(\norm{V_\nu(t_0)-V_P(t_0)}_X^2
+\int_{\min(t,t_0)}^{\max(t,t_0)}\sum_I\norm{L_{\Psi_\nu}(\Psi_\nu-P)^I(s)}_{H^{s_I}(\mathbb{R}^d)}^2\,ds\right)
\end{align}
for $|t-t_0|$ in that fixed time length. The constant $C$ is independent of $P$. Given $e>0$, smooth the traces at $t_0$ so that $\norm{V_\nu(t_0)-V_P(t_0)}_X\leq e$. The global source bound, now justified by exhaustion of \eqref{E:SH_Local_source_estimate}, and the coefficient multiplication bound give a finite $K_e$ for the integrand in \eqref{E:SH_Analytic_energy_continuity}. Hence the triangle inequality and continuity of $V_P$ give
\begin{align*}
\limsup_{t\rightarrow t_0}\norm{V_\nu(t)-V_\nu(t_0)}_X\leq(\sqrt{C}+1)e.
\end{align*}
Letting $e\rightarrow0$ proves $V_\nu\in C([0,T];X)$. Finally, solving \eqref{E:SH_Normalized_reduction} for $\partial_t^p\Psi_\nu^I$ gives its uniform $H^{r_I}(\Sigma)$ bound: the mixed order-$p$ terms use only time derivatives through $p-1$, and the source has $H^{s_I}(\Sigma)$ regularity.

\medskip
\noindent \underline{Step 8} (\textbf{The reduced identities at finite Sobolev regularity}).\nopagebreak[4]
We verify that \eqref{E:SH_Normalized_reduction} also holds for every distributional solution of \eqref{SH:1} in the class \eqref{E:SH_Solution_regularity}. This will justify the uniqueness argument without imposing analyticity on competing solutions.

First, for such a solution and $0\leq j<p$,
\begin{align}
\label{E:SH_Weak_class_time_regularity}
\partial_t^j\Psi^K\in W^{1,\infty}\big([0,T];H^{S_K-j-1}(\Sigma)\big).
\end{align}
Consequently $\mathcal{Z}[\Psi]\in L^\infty([0,T];H^{s_0}(\Sigma))$ and $\partial_t\mathcal{Z}[\Psi]\in L^\infty([0,T];H^{s_0-1}(\Sigma))$, so the jet is continuous in time with values in $C^0(\Sigma)$. On the torus its range on the compact time-space cylinder is a compact subset of $\mathcal{U}$. On $\mathbb{R}^d$, it is continuous into $C_0(\mathbb{R}^d)$, since $s_0-1>\frac{d}{2}$. Compactness of its image in $C_0(\mathbb{R}^d)$ gives uniform decay at spatial infinity: a finite cover by balls of radius $\frac{e}{2}$ in $C_0$ and the decay of their finitely many centers give
\begin{align*}
\sup_{0\leq t\leq T,\,|x|\geq R_e}|\mathcal{Z}[\Psi](t,x)|<e
\end{align*}
for some $R_e$. Take $e>0$ with $\overline{B_e(0)}\subset\mathcal{U}$. The jet image on $[0,T]\times\overline{B_{R_e}(0)}$, together with $\overline{B_e(0)}$, is a compact subset of $\mathcal{U}$ containing the entire range. Thus a compact coefficient range is a consequence of the stated uniqueness class and admissibility, not an additional hypothesis.

For any fixed pair of solutions, let $\mathcal{K}_c\Subset\mathcal{U}$ contain both ranges, including zero on $\mathbb{R}^d$. Constants in Steps 8--10 may depend on their finite top Sobolev bounds, $\mathcal{K}_c$, the compact forcing ranges, the coefficient functions on slightly larger compact sets, and the indices. They are uniform whenever these bounds and compact sets are uniform. No convexity of $\mathcal{U}$ or $\mathcal{V}$ is needed. Each smooth finite-jet function used in a norm estimate is extended by a compactly supported smooth cutoff equal to one near the actual compact argument ranges. Additional derivative variables range in a fixed finite ball by Sobolev embedding. The extension is a smooth function on the full finite-dimensional space. On $\mathbb{R}^d$ include zero in the cutoff region and separate the constant value of every coefficient at zero from its Sobolev part.

Differentiating row $I$ of \eqref{SH:1} $n_I$ times is valid in distributions and yields \eqref{E:SH_Inverted_normal_form}. To verify the weak product and chain rules here, put $k_0=\ell+p-m_*$. In the differentiated equation every derivative of an unknown has remaining Sobolev order at least $k_0$, while differentiated coefficient arguments and forcing arguments have order at least $k_0+1$. Since $H^{k_0}(\Sigma)$ is an algebra, spatial and temporal mollification, after the finite-dimensional extensions just described, proves these rules. For $0<a<b<T$, the factors converge in $L^q([a,b];H^{k_0}(\Sigma))$ for every finite $q$, with uniform $L^\infty([a,b];H^{k_0}(\Sigma))$ bounds. The finite product formulas therefore pass to the limit. This argument regularizes identities and their factors; it does not assert that the regularized functions solve the equations.

The original class supplies enough time derivatives for this normal form, but the direct application of $\widehat{H}$ may require more. They follow from \eqref{E:SH_Inverted_normal_form}. For $-1\leq E\leq p-m_*$, set $k_E=\ell+p-m_*-E$. We claim
\begin{align}
\label{E:SH_Additional_time_regularities}
\partial_t^q\Psi^K\in L^\infty\big([0,T];H^{S_K-q}(\Sigma)\big),
\qquad 0\leq q\leq m_K+E.
\end{align}
The case $E=-1$ is included in \eqref{E:SH_Solution_regularity}. Suppose the claim holds at $E-1$, where $E\geq0$. For an argument $g=\partial_x^\beta\partial_t^b\Psi^L$ of \eqref{E:SH_Inverted_normal_form} and $0\leq a\leq E$, one has
\begin{align*}
b+a&\leq m_L+E-1,\\
S_L-b-a-|\beta|&\geq\ell+p-m_*-a\geq k_E.
\end{align*}
Thus $g\in W^{E,\infty}([0,T];H^{k_E}(\Sigma))$. Every forcing argument has $E$ time derivatives because $\chi_*+E\leq d_{\mathcal{W}}$, with remaining spatial regularity at least $S_{\mathcal{W}}-\chi_*-E=k_E+1$. The algebra property $k_E\geq\ell>\frac{d}{2}$ and the finite chain rule give $\partial_t^E\mathcal{N}_K\in L^\infty([0,T];H^{k_E}(\Sigma))$. The chain rule follows by temporal mollification in this algebra and passage in $L^q([a,b];H^{k_E}(\Sigma))$ for $0<a<b<T$ and finite $q$, as in the normal-form argument. Differentiating the already established normal form therefore gives
\begin{align}
\label{E:SH_Weak_time_recursion}
\partial_t^{m_K+E}\Psi^K=\partial_t^E\mathcal{N}_K(\Psi;\mathcal{W})
\in L^\infty\big([0,T];H^{k_E}(\Sigma)\big).
\end{align}
Since $S_K-(m_K+E)=k_E$, the induction proves \eqref{E:SH_Additional_time_regularities}. In particular all mixed derivatives of total order at most $M_K=p+m_K-m_*$ are defined with remaining regularity at least $S_K-M_K=\ell$.

The largest derivative of $\Psi^K$ before cancellation in \eqref{E:SH_Expanded_reduction} has order
\begin{align*}
q_{IJ}+m_K-n_J=p-m_I+m_K\leq M_K.
\end{align*}
Hence every differentiated original equation is represented by Sobolev functions before multiplication by $\widehat{H}$. The product rule, justified in these algebras, proves \eqref{E:SH_Expanded_reduction} for the finite-regularity class. There is no multiplication of an arbitrary high-order distribution by a rough coefficient. The substitutions following \eqref{E:SH_Expanded_reduction} are also justified by \eqref{E:SH_Additional_time_regularities}; their integer time-order induction is unchanged. This proves \eqref{E:SH_Normalized_reduction}, with exactly the source function and arguments in \eqref{E:SH_Reduced_source_arguments}, for every solution under consideration.

We must also identify its additional initial traces. Equation \eqref{E:SH_Additional_time_regularities} gives
\begin{align}
\label{E:SH_Extended_trace_continuity}
\partial_t^q\Psi^K\in W^{1,\infty}\big([0,T];H^{S_K-q-1}(\Sigma)\big),
\qquad q\leq M_K-1.
\end{align}
For $0\leq E\leq p-1-m_*$, both sides of \eqref{E:SH_Weak_time_recursion} are continuous into $H^{k_E-1}(\Sigma)$. On its right, each unknown argument has time order at most $m_L+E-1$ and total order at most $m_L+E$; \eqref{E:SH_Extended_trace_continuity} gives continuity in at least $H^{k_E-1}(\Sigma)$. This is an algebra because $k_E-1\geq\ell$. The forcing arguments are continuous by hypothesis. Thus \eqref{E:SH_Weak_time_recursion} can be evaluated at $t=0$. Induction in $E$ determines the traces $\Psi^K_{(m_K+E)}$ from earlier traces and the prescribed function, giving precisely the recursion in Step 2. Evaluating that finite algebraic formula on the top-order initial data also puts the resulting trace in $H^{k_E}(\Sigma)$: unknown arguments retain at least $k_E$ spatial derivatives and forcing arguments retain at least $k_E+1$. This top-space trace agrees with the trace just identified in $H^{k_E-1}(\Sigma)$. Product, composition, and inversion are continuous in these algebras, so the recursion is continuous in the original top data spaces. In particular, every solution with the same original data and prescribed function has the same traces through order $p-1$.

\medskip
\noindent \underline{Step 9} (\textbf{Differences and convergence to a solution of \eqref{SH:1}}).\nopagebreak[4]
Let $\Psi,\Phi$ be two original solutions in the class of Step 8, with prescribed functions $\mathcal{W},\mathcal{W}'$, and set $w=\Psi-\Phi$. Define
\begin{align*}
D(t)&=\sum_{I=1}^N\sum_{j=0}^{p-1}\norm{\partial_t^jw^I(t)}_{H^{S_I-j-1}(\Sigma)}^2,\\
G_0(t)&=\sum_{a=0}^{d_{\mathcal{W}}}\norm{\partial_t^a(\mathcal{W}-\mathcal{W}')(t)}_{H^{S_{\mathcal{W}}-a-1}(\Sigma)}^2.
\end{align*}
Set $G_0=0$ in the absence of prescribed functions. The compactly supported extensions of Step 8 satisfy, for any two actual argument lists $u,v$,
\begin{align*}
F_{\mathrm{ext}}(u)-F_{\mathrm{ext}}(v)
=\int_0^1DF_{\mathrm{ext}}(v+a(u-v))(u-v)\,da.
\end{align*}
Thus the line segment is evaluated in the extension, even when it leaves $\mathcal{U}\times\mathcal{V}$. The composition estimates cited after \eqref{E:SH_Local_source_estimate} and the algebra property at $r_I>\frac{d}{2}$ apply. For each unknown argument of $f_I$, \eqref{E:SH_Reduction_counts} gives
\begin{align*}
\norm{\partial_x^\beta\partial_t^bw^K}_{H^{r_I}(\Sigma)}
\leq\norm{\partial_t^bw^K}_{H^{S_K-b-1}(\Sigma)}\leq D^{\frac{1}{2}}.
\end{align*}
The analogous bound for forcing arguments is $G_0^{\frac{1}{2}}$, since $S_{\mathcal{W}}-1-e_I=r_I$. Undifferenced arguments have at least $H^{r_I+1}(\Sigma)$ regularity. Consequently
\begin{align}
\label{E:SH_Source_difference_bound}
\norm{f_I(\Psi;\mathcal{W})-f_I(\Phi;\mathcal{W}')}_{H^{r_I}(\Sigma)}
\leq C(D^{\frac{1}{2}}+G_0^{\frac{1}{2}}).
\end{align}
The exact difference equation is
\begin{align}
\label{E:SH_Normalized_difference_equation}
L_\Psi w^I=f_I(\Psi;\mathcal{W})-f_I(\Phi;\mathcal{W}')+(L_\Phi-L_\Psi)\Phi^I.
\end{align}
Its operator difference contains no pure time derivative of order $p$. Each of its terms is
\begin{align*}
\big(A_{j,\beta}(\mathcal{Z}[\Phi])-A_{j,\beta}(\mathcal{Z}[\Psi])\big)
\partial_x^\beta\partial_t^j\Phi^I,
\qquad j+|\beta|=p,\quad j<p.
\end{align*}
The differentiated $\Phi^I$ is bounded in $H^{r_I}(\Sigma)$ by its top norm. The coefficient difference has that norm at most $CD^{\frac{1}{2}}$, since the jet difference is bounded by $CD^{\frac{1}{2}}$ in $H^{s_0-1}(\Sigma)$ and $s_0-1\geq r_I$. Multiplication in $H^{r_I}(\Sigma)$, \eqref{E:SH_Source_difference_bound}, and \eqref{E:SH_Normalized_difference_equation} imply
\begin{align}
\label{E:SH_Difference_source_energy}
\sum_I\norm{L_\Psi w^I}_{H^{r_I}(\Sigma)}^2\leq C(D+G_0).
\end{align}
The coefficients satisfy \eqref{E:SH_Linear_coefficient_bound} by Step 8. For the companion vector in Step 4 with $v=w^I$, the stated uniqueness class gives $Y\in L^\infty([0,T];H^{r_I+1}(\Sigma))$ and $\partial_tY\in L^\infty([0,T];H^{r_I}(\Sigma))$, including $j=p-1$ because $\partial_t^pw^I$ is in $H^{r_I}(\Sigma)$. Thus the nonsmooth energy calculation following \eqref{E:SH_Integer_commutator} applies directly. Summing over $I$, using \eqref{E:SH_Difference_source_energy}, and applying Gronwall gives
\begin{align}
\label{E:SH_Difference_estimate}
\sup_{0\leq t\leq T}D(t)\leq C_T\left(D(0)+\int_0^TG_0(s)\,ds\right).
\end{align}
The constants have the dependence specified in Step 8. This estimate is one derivative below the top energy.

Apply \eqref{E:SH_Difference_estimate} to $\Psi_\nu,\Psi_\mu$ and $\mathcal{W}_\nu,\mathcal{W}_\mu$. Steps 5--7 supply uniform top bounds, and Step 2 supplies convergence of all initial traces through $p-1$ in their top spaces. The forcing convergence implies $\sup_tG_0(t)\rightarrow0$. Thus completeness gives a global limit with
\begin{align}
\label{E:SH_Strong_convergence_below_top_order}
\partial_t^j\Psi_\nu^I\longrightarrow\partial_t^j\Psi^I
\quad\text{in }C\big([0,T];H^{S_I-j-1}(\Sigma)\big),
\qquad 0\leq j<p.
\end{align}
The notation for the derivatives of the limit follows by testing the convergence against compactly supported smooth functions in time and space. No compactness of the global Sobolev embedding on $\mathbb{R}^d$ is used. Every derivative in row $J$ of the original equation has total order at most $m_K-n_J\leq m_K\leq p-1$, and
\begin{align}
\label{E:SH_Passage_to_limit_count}
S_K-1-(m_K-n_J)=\ell+p-m_*-1+n_J>\frac{d}{2}.
\end{align}
Equation \eqref{E:SH_Passage_to_limit_count} places this convergence in a Sobolev algebra. The forcing hypothesis gives the same convergence for the forcing arguments. Products and compositions pass to the limit in \eqref{SH:1}, proving that $\Psi$ solves the original system with the original initial data and the prescribed function.

\medskip
\noindent \underline{Step 10} (\textbf{Regularity, time continuity, and uniqueness}).\nopagebreak[4]
For each fixed $t,I,j<p$, the sequence in \eqref{E:SH_Strong_convergence_below_top_order} is bounded in $H^{S_I-j}(\Sigma)$. A weakly convergent subsequence in that Hilbert space has limit $\partial_t^j\Psi^I(t)$ by the strong convergence one order below. Weak lower semicontinuity gives its top norm bounded by the lower limit of the approximating norms. These bounds are uniform in time. For $s=S_I-j$, the operator $J_L$ in \eqref{E:SH_Fourier_cutoff} satisfies $\norm{J_Lv}_{H^s(\Sigma)}\leq C(1+L)\norm{v}_{H^{s-1}(\Sigma)}$. Thus $J_L\partial_t^j\Psi^I$ is continuous into $H^s(\Sigma)$, and it converges there at every fixed time to $\partial_t^j\Psi^I$ as $L\rightarrow\infty$. This proves strong measurability in the top space. Hence the limit has the asserted $L^\infty$ bounds for $j<p$.

For $j=p$ we first pass to the limit in the normalized analytic equations \eqref{E:SH_Normalized_reduction} in distributions. Their mixed order-$p$ terms use time orders below $p$ and converge at least in $H^{r_I-1}(\Sigma)$; their coefficients converge uniformly with the Sobolev bounds of Step 5. The source arguments in \eqref{E:SH_Reduced_source_arguments} converge in $H^{r_I}(\Sigma)$ by \eqref{E:SH_Reduction_counts}. Thus the limit satisfies the normalized equation before invoking the reduction result for the uniqueness class. In that equation its mixed order-$p$ terms belong to $L^\infty([0,T];H^{r_I}(\Sigma))$ by the top bounds for $j<p$, and its source belongs to $L^\infty([0,T];H^{s_I}(\Sigma))$. Solving for $\partial_t^p\Psi^I$ proves the missing bound in \eqref{E:SH_Solution_regularity}.

The bounds for two successive time derivatives give \eqref{E:SH_Weak_class_time_regularity}. Interpolation with the top bound proves continuity of $\partial_t^j\Psi^I$ in every $H^s(\Sigma)$ with $s<S_I-j$, for $j<p$. For $j=p$, take $0<e<1$ with $r_I-e>\frac{d}{2}$ in \eqref{E:SH_Normalized_reduction}. All its mixed order-$p$ terms and source terms are continuous in $H^{r_I-e}(\Sigma)$ by the continuity just established, \eqref{E:SH_Reduction_counts}, and composition in that algebra. Solving for the pure time derivative proves its continuity there; embedding gives all smaller exponents. Uniform convergence of the physical jets and the distance bound in Step 5 give compact containment in $\mathcal{K}_1$.

Finally, if $\Psi$ and $\Phi$ belong to the uniqueness class with the same original data and prescribed function, Step 8 proves their normalized identities and equality of all reduced initial traces. In \eqref{E:SH_Difference_estimate} one has $D(0)=0$ and $G_0=0$. Therefore $D(t)=0$ and $\Psi=\Phi$. The compact containment needed for this comparison was derived for each competitor in Step 8. If the common lifespan is open at its upper endpoint, the argument applies on each closed subinterval.

\medskip
\noindent \underline{Step 11} (\textbf{The common time for nearby data and continuous dependence}).\nopagebreak[4]
We make explicit the uniformity of the choices in Step 5. The finite initial-trace recursion is continuous in the top data spaces. For data in a sufficiently small neighborhood of the given data, the initial energies and coefficient gradients therefore have common bounds, and their initial jets lie in a fixed compact subset of $\operatorname{int}\mathcal{K}_1$. Use these bounds to fix $b_0,R,C_R,d_1$ once. Smooth the limiting traces of the reference datum once to obtain the single polynomial $P$ in \eqref{E:SH_Fixed_comparison_polynomial}. Continuity of the recursion permits the neighborhood to be decreased so that its traces, and sufficiently accurate analytic approximations to each of them, satisfy $E_{\mathcal{A}}[\Psi_\nu-P](0)\leq d_1$ with this same $P$. Thus $t_P$ and $K_{R,P}$ are fixed for the family. Together with the common-interval construction in Step 7, Equations~\eqref{E:SH_Bootstrap_time} and \eqref{E:SH_Common_real_time} give a uniform positive $T$. For a convergent sequence of prescribed functions, their uniformly bounded norms and common compact forcing range give the same choices after finitely many terms have been discarded.

For the solutions $\Psi_q$ in the continuous-dependence assertion, Step 8 gives convergence of the additional initial traces in their top spaces. Their initial contribution to \eqref{E:SH_Difference_estimate} tends to zero, and the forcing convergence gives $\int_0^TG_0\,dt\rightarrow0$. It follows that $\partial_t^j(\Psi_q^I-\Psi^I)\rightarrow0$ in $C([0,T];H^{S_I-j-1}(\Sigma))$ for $j<p$. For $0<e<1$, interpolation gives
\begin{align*}
\norm{v}_{H^{S_I-j-e}(\Sigma)}
\leq C\norm{v}_{H^{S_I-j-1}(\Sigma)}^e
\norm{v}_{H^{S_I-j}(\Sigma)}^{1-e}.
\end{align*}
The uniform top bounds therefore give convergence in every strictly lower Sobolev topology; smaller exponents also follow by embedding. This proves the stated continuous dependence.

\end{proof}

\begin{remark}
\label{R:Analyticity_hypothesis}
The real-analytic dependence in Theorem \ref{Theorem_SH_LWP} concerns the dependence of the principal coefficient functions on $\mathcal{Z}[\Psi]$ and of the right-hand sides on $(\mathcal{Z}[\Psi],\mathcal{Y}[\mathcal{W}])$; it does not require the solution in Equation~\eqref{E:SH_Solution_regularity} to be real analytic in the spacetime variables. When a prescribed function is present, its Sobolev regularity and its analytic approximations are instead stated separately in Equations~\eqref{E:SH_Forcing_regularity}--\eqref{E:SH_Analytic_forcing_containment}. Thus, this hypothesis does not by itself prevent applications to systems coupled to Einstein's equations. If the metric is included among the unknowns and Einstein's equations are written in wave coordinates, then
\begin{align*}
g^{-1}
=
\frac{\operatorname{adj}(g)}{\det g}
\end{align*}
is a real-analytic function of the components of $g$ on the open set of nondegenerate Lorentzian metrics. In these coordinates, the principal part of the reduced Einstein equations is $g^{\alpha\beta}\partial_\alpha\partial_\beta g_{\mu\nu}$, while the remaining terms are finite sums of products of components of $g^{-1}$, first derivatives of $g$, and the matter variables and their derivatives. Consequently, the coefficient functions of the coupled Einstein--matter system are real analytic in the relevant finite collection of unknowns and their derivatives whenever the equation of state and the transport coefficients entering the matter equations are real analytic on the set under consideration. Subject to the remaining assumptions of Theorem \ref{Theorem_SH_LWP}, this gives the real-analytic dependence required there. It does not require the metric obtained from the Sobolev initial data to be an analytic function of the spacetime coordinates.

The same argument permits a fixed background metric whose coordinate components depend real-analytically on the spacetime coordinates: one allows the coefficient functions in Theorem \ref{Theorem_SH_LWP} to depend real-analytically on the spacetime point as well as on $\mathcal{Z}[\Psi]$. This explicit coordinate dependence does not change the proof on coordinate domains on which the bounds used in the proof are uniform. 

\end{remark}

\newpage
\draftpart{Numerical algorithms, implementation, and physical results}
\label{Part:Numberics}

\section{Setup}
\label{S:Numerical_setup}

The EOMs of the BDNK tensor in Equation~\eqref{BDNKTmunu} with an ideal current are second-order in the hydrodynamic variables $U^A$ or $\{\varepsilon,u^i\}$~(see Definition~\eqref{def:U^A}), and first-order in $n$. We recall the BDNK tensor with the ideal current,
\begin{align}
    \tag{\ref{BDNKTmunu}}
    T_{\mu\nu} &= \left(\varepsilon +\cA \right) u_\mu u_\nu + (P+\Pi)\Delta_{\mu\nu}-2\eta \sigma_{\mu\nu}+u_\mu \mathcal{Q}_\nu + u_\nu \mathcal{Q}_\mu\,,\\
    \tag{\ref{idealcurrent}}
    J^\mu&=nu^\mu\,.
\end{align}

Setting up a BDNK fluid system for numerical evolution requires specifying 
(i) an equation of state and associated thermodynamic relations, 
(ii) transport coefficients,
(iii) hydrodynamic frame parameters, 
(iv) consistent initial data for the \ac{FO} and \ac{MO} formulations, and 
(v) appropriate numerical methods for evolving the system.
In the following subsections, we describe the general procedures adopted in this work for each of these ingredients, highlighting any challenges involved in their specification.
Some of the choices made in this work may not generalize immediately to all systems of physical interest; for instance, incorporating arbitrary equations of state and generic transport coefficients, or dynamical spacetimes may require further development. 
Nevertheless, the framework presented here allows us to test a broad range of initial conditions across various background spacetimes, tabulated and analytical equations of state, and fluid configurations of varying symmetry and complexity. 
We will also briefly discuss ongoing and planned future work aimed at generalizing the treatment presented here.

The pressure $P(\varepsilon,n)$ and temperature $T(\varepsilon,n)$ appearing in the BDNK tensor in Equation~\eqref{BDNKTmunu} are specified by the EOS, a closure relation derived from the thermodynamic properties of the fluid of interest.
Furthermore, computing the thermal transport coefficients $\beta_\varepsilon$~[Equation~\eqref{def:beta_epsilon}] and $\beta_n$~[Equation~\eqref{def:beta_n}], which appear in the viscous correction $\mathcal{Q}_\nu$~[Equation~\eqref{defQ}], requires the corresponding thermodynamic derivatives of the pressure and chemical potential, evaluated consistently with the chosen EOS.
We recall the expressions for 
the viscous corrections,
\begin{align}
\tag{\ref{defA}}
\cA &= \tau_\varepsilon\left[ u^\mu \nabla_\mu \varepsilon + (\varepsilon+P)\nabla_\mu u^\mu\right] = -\tau_\varepsilon \left(u_\alpha \nabla_\beta T^{\alpha\beta}_0\right)\,,\\
\tag{\ref{definePi}}
\Pi &=  \frac{\tau_P}{\tau_\varepsilon} \mathcal{A} - \zeta \nabla_\mu u^\mu = - \zeta \nabla_\mu u^\mu - \tau_P \left(u_\alpha \nabla_\beta T^{\alpha\beta}_0\right)\,,\\
\tag{\ref{defQ}}
\mathcal{Q}_\nu &=\tau_Q (\varepsilon+P)u^\lambda \nabla_\lambda u_\nu + \beta_\varepsilon\Delta_{\nu}^\lambda \nabla_\lambda \varepsilon+ \beta_n\Delta_{\nu}^\lambda \nabla_\lambda n\,,
\end{align}
and the thermal transport coefficients,
\begin{align}
    \tag{\ref{def:beta_epsilon}}
    \beta_{\varepsilon}&=\tau_Q P_{\varepsilon}+\frac{\rho\sigma T}{n}\left (\frac{\partial(\mu/T)}{\partial\varepsilon}\right )_n\\
    \tag{\ref{def:beta_n}}
    \beta_{n}&=\tau_Q P_{n}+\frac{\rho\sigma T}{n}\left (\frac{\partial(\mu/T)}{\partial n}\right )_{\varepsilon},
\end{align}
In this work, we study fluids with three different EOSs: conformal, \ac{QCD} at zero baryon chemical potential, and ideal gas. 
The relevant thermodynamic relations corresponding to each EOS are stated in the following Section~\ref{SS:EOS}.

Additionally, the transport coefficients, namely the shear and bulk viscosities $\{\eta,\zeta\}$, must be specified or parameterized. 
Likewise, the BDNK frame parameters~$\{\tau_\varepsilon, \tau_P, \tau_Q, \sigma\}$ must be specified, subject to the constraints imposed by causality and linear stability.
Since the coefficients $\beta_\varepsilon$ and $\beta_n$, as well as the causality and stability conditions, require thermodynamic relations, 
the choice of functional forms for the frame parameters depends on the choice of the EOS.
For this reason, in Section~\ref{SS:EOS} we specify the transport coefficients $\{\eta,\zeta\}$ and frame parameters $\{\tau_\varepsilon,\tau_P,\tau_Q,\sigma\}$, alongside the respective causality and stability bounds, for each EOS in this study.

Given that the EOMs in Equations~\eqref{BDNK_EoM},
\begin{subequations}
    \begin{align}
        \nabla_\mu T^{\mu\nu} &= 0, 
        \tag{\ref{div_T_mumnu_DNK}}
        \\
        \nabla_\mu J^\mu & = 0,
        \tag{\ref{div_J_mu_BDNK}}
        \\
        g_{\mu\nu} u^\mu u^\nu & =-1.
        \tag{\ref{normalization_u_BDNK}}
    \end{align}
\end{subequations}
are second-order in $U^A$, 
solving the system requires specifying both the initial values of these variables and their initial time derivatives.
The \ac{FO} and \ac{MO} reductions adopted in this work both introduce auxiliary variables in lieu of these time derivatives, each requiring consistent initial data. 
We discuss the procedures adopted for constructing initial data for both formulations in Section~\ref{Sec:Initial Conditions}. Finally, the two formulations require distinct numerical considerations, which are described in Section~\ref{Sec:Num-methods}.

For numerical implementation, we employ a mathematically equivalent rescaled version of the \ac{FO} formulation introduced in Section~\ref{Sec:First_Order}. 
This rescaling improves numerical stability, particularly in low-energy-density regions, without changing the underlying BDNK dynamics. 
The corresponding expressions and their numerical advantages are discussed in Appendix~\ref{Appendix:Rescaled EOMs}. 
Unless stated otherwise, all \ac{FO} numerical results presented in this work are obtained using the rescaled formulation. 
Due to constraints, the corresponding rescaled \ac{MO} formulation was not implemented numerically, and all \ac{MO} results are therefore obtained using the original, unrescaled formulation of Section~\ref{Mixed_order_section}.

\subsection{Equations of State and Transport Coefficients}\label{SS:EOS}

\subsubsection{Transport Coefficients}\label{Transport coefficients}

Ideally, for a given EOS, the physical viscosities and frame parameters should be independently specifiable subject to their respective criteria: ${\eta, \zeta}$ constrained by theoretical calculations or experimental constraints, and ${\tau_\varepsilon, \tau_P, \tau_Q, \sigma}$ satisfying causality and stability conditions. 
In practice, this separation is complicated.
The causality and stability conditions are coupled inequalities involving ${\eta, \zeta}$, ${\tau_\varepsilon, \tau_P, \tau_Q, \sigma}$, and thermodynamic quantities simultaneously. 
Once an EOS and viscosities ${\eta, \zeta}$ are specified, the allowed values of the frame parameters are non-trivially constrained. 
If all frame parameters are left as arbitrary functions, the only feasible approach is to determine them locally throughout the computational domain by numerically solving the causality and stability constraints. 
This complicates even the specification of initial conditions, since the reconstruction of the equilibrium hydrodynamic fields depends on the frame parameters, which may themselves depend on the equilibrium fields.\footnote{If the frame parameters are not prescribed a priori, they must be determined simultaneously with the reconstruction of the equilibrium hydrodynamic fields from the stress--energy tensor by enforcing the causality and stability constraints while solving the differential relations between the BDNK tensor components and the hydrodynamic fields in Equations~\eqref{BDNKTmunu}, \eqref{defA}, \eqref{definePi}, and \eqref{defQ}.
While feasible in principle, this substantially increases the complexity of both the initialization procedure and the interpretation of the resulting equilibrium fields.}

Previous works have addressed this issue by parameterizing the frame parameters such that the constraints reduce to simple global conditions, guaranteeing causality and stability, independent of local thermodynamic quantities. 
While such parameterizations have been demonstrated for the conformal~\cite{Bemfica:2017wps} and ideal gas~\cite{Pandya_2022_IG} EOSs, extending this approach to general fluids presents additional challenges. 
The conformal EOS $P = \varepsilon/3$ greatly simplifies the causality and stability conditions not only because all thermodynamic identities take simple closed-form expressions, but also because the additional choice $\zeta = \sigma = 0$ eliminates several terms. 
For a general EOS, closed-form thermodynamic identities may not be available, and neither $\zeta$ nor $\sigma$ vanishes {\it{a priori}}.

We present our general solution to these difficulties for zero thermal conductivity, $\sigma=0$ 
through the following parameterizations
\begin{align}\label{eq:general-choice-frames-0sigma} \tau_\varepsilon &= \chi V/\rho\,,
\quad \tau_Q = \lambda V/\rho\,,
\quad \tau_P = \xi\tau_\varepsilon , 
\end{align}
where factors
\begin{equation}
V \equiv \frac{4}{3}\eta+\zeta\,, 
\end{equation}
and $1/\rho$ are introduced to simplify the resulting causality and stability conditions, recalling $\rho=\varepsilon+P$. 
This parameterization reduces the constraints to conditions on the three dimensionless quantities 
$\chi$, $\lambda$ and $\xi$. 
While the derivation is involved, the simplified conditions 
\begin{align} 
\label{cond:lambda-general-0sigma} 
\lambda &\geq 
\frac{\chi }{\chi (1-c_s^2)(1-\xi )-P'_\varepsilon}\\\label{cond:xi-general-0sigma}
\chi &\geq P'_\varepsilon 
\textrm{max} 
\left[ 
\frac{1}{(1-c_s^2)(1-\xi )}\,, \,\,\, \frac{1}{c_s^2\xi } 
\right] 
\end{align}
guarantee that causality and stability are satisfied for any valid EOS ($0 < c_s^2 < 1$ and $P'_\varepsilon > 0$) and arbitrary $\eta \geq 0$, $\zeta \geq 0$. 
Additionally, choosing \begin{equation}
0 < \xi < 1
\end{equation}
ensures positive lower bounds on both $\lambda$ and $\chi$, though this is not strictly required for causality and stability.
The key feature of the parameterization in Equation~\eqref{eq:general-choice-frames-0sigma} is that the resulting bounds in Equations~\eqref{cond:lambda-general-0sigma} and~\eqref{cond:xi-general-0sigma} are completely independent of the physical viscosities $\eta$ and $\zeta$, and valid for any EOS.
This results in a desirable separation between the physical and frame-parameters.

Thus, for a given EOS and $\sigma=0$, the parameters $\lambda$, $\chi$ and $\xi$ can be arbitrary functions provided they satisfy the inequalities, and are typically parameterized in terms of some dimensionless coefficients.
From the conditions in Equations~\eqref{cond:lambda-general-0sigma} and \eqref{cond:xi-general-0sigma}, a convenient choice of the functional forms of $\lambda$, $\chi$ and $\xi$ depends on the thermodynamic quantities $c_s^2$ and $P'_\varepsilon$.
Since $c_s^2$ and $P'_\varepsilon$ are local thermodynamic quantities, these conditions can be satisfied globally through a simple choice if their global extrema are {\it{a priori}} known
.
Indeed, such extrema are known for many EOSs and systems considered in relativistic hydrodynamics studies, including the \ac{QCD} equation of state considered in this study and discussed in Section~\ref{SSS:QCD-EOS}.
Regardless, the causality and stability bounds in Equations~\eqref{cond:lambda-general-0sigma} and \eqref{cond:xi-general-0sigma} resulting from the scaling in Equation~\eqref{eq:general-choice-frames-0sigma} are convenient for numerical implementation.
We note that we were unable to identify a similarly convenient parameterization for all frame parameters $\tau_\varepsilon, \tau_P, \tau_Q$, and $\sigma$ with non-zero thermal conductivity $\sigma\neq0$ that yields a comparable simplification. This problem is being addressed in ongoing work.
In the following subsections, we demonstrate how the parameterization in Equation~\eqref{eq:general-choice-frames-0sigma} specializes to conformal and \ac{QCD} fluids.

\subsubsection{Conformal}\label{SSS:conformal}
While the general BDNK theory admits arbitrary EOSs, conformal fluids with no baryon current have been the most widely studied in previous numerical work~\cite{Pandya-21,Pandya_2022,Bantilan:2022ech,Bea:2023rru,Clarisse:2025lli}. 
Conformal symmetry requires $g_{\mu\nu}T^{\mu\nu}=0$ and $P = \varepsilon/3$, with the energy density related to the temperature by $\varepsilon = \alpha T^4$, where $\alpha$ depends on the microphysical properties of the fluid; here we set $\alpha=1$ for simplicity.

Conformal symmetry further requires a constant $\eta/\varepsilon^{\frac{3}{4}}$, zero bulk viscosity $\zeta=0$, and we set the chemical potential and thermal conductivity to zero, $\sigma=\mu=0$. 
Since the conformal EOS is independent of $n$, the thermal coefficient $\beta_n$ defined in Equation~\eqref{def:beta_n} vanishes,
leading to a decoupling between the EOMs for the fluid variables and the current $J^\mu$.
These choices are summarized as
\begin{align}
    \label{Conformal EOS and transport choices}
    P&=\frac{\varepsilon}{3},\,
    \quad\mu=0,\,
    \quad c_s^2=\frac{1}{3},\,\nonumber\\
    \quad \frac{\eta}{\varepsilon^{\frac{3}{4}}}&= \eta_0,\,\quad\zeta=0\,,\quad \sigma=0\\
    \quad\beta_\varepsilon&=\frac{\tau_Q}{3},\,\quad\beta_n=0,\,\nonumber
\end{align}
where $\eta_0$ is a dimensionless parameter used to parameterize shear viscosity.
We adopt the frame parameters proposed in~\cite{Bemfica:2017wps} for conformal fluids that guarantee 
causality and stability, 
\begin{align}\label{Conformal relaxation time expressions}
    \tau_P &= \frac{\tau_\varepsilon}{3},\quad
    \tau_\varepsilon 
    =\frac{1}{\rho}\chi_0 \varepsilon^\frac{3}{4},
    \tau_Q 
    =\frac{1}{\rho}\lambda_0 \varepsilon^\frac{3}{4}\,.
\end{align}
where $(\chi_0,\lambda_0)$ are dimensionless constants, 
related to the set of frame parameters $(a_1,a_2)$ used in \cite{Clarisse:2025lli} by $(\chi_0,\lambda_0)=(a_1\eta_0,a_2\eta_0)$.
In terms of the coefficients of our general parameterization  in Equation~\eqref{eq:general-choice-frames-0sigma}, this corresponds to
\begin{align}\label{frame-coefficients-Conformal}
    \xi&=P'_\varepsilon=c_s^2=\frac{1}{3}\,,
    \qquad
    \chi=\frac{3}{4}\frac{\chi_0}{\eta_0}\,,
    \qquad
    \lambda=\frac{3}{4}\frac{\lambda_0}{\eta_0}\,,
\end{align}
which, upon substitution into the general conditions in Equations~\eqref{cond:lambda-general-0sigma} and \eqref{cond:xi-general-0sigma}, leads to    
\begin{equation}
    \label{conformal frame parameter caus and stab conditions}
    \frac{\chi_0}{\eta_0}\geq4,\,\qquad\frac{\lambda_0}{\eta_0}\geq\frac{3\chi_0}{\chi_0-\eta_0}\,.
\end{equation}
These conditions are consistent with those derived in~\cite{Bemfica:2017wps}, confirming that our general parameterization in Equation~\eqref{eq:general-choice-frames-0sigma} and the corresponding simplified conditions in Equations~\eqref{cond:lambda-general-0sigma} and \eqref{cond:xi-general-0sigma} reduce correctly in the conformal limit.

\subsubsection{{QCD equation of state}}\label{SSS:QCD-EOS}
As an application to non-conformal fluids, we consider strongly interacting \ac{QCD} matter at zero baryon chemical potential, as encountered in ultrarelativistic heavy-ion collisions.
The EOS for this type of matter can be computed from first principles using lattice \ac{QCD}~\cite{Borsanyi:2013bia, HotQCD:2014kol}. 
Here we use the lattice results for the EOS from the HotQCD collaboration~\cite{HotQCD:2014kol} and smoothly connect the lattice results to a hadron resonance gas EOS.

While the causality conditions in Equation~\eqref{conformal frame parameter caus and stab conditions} are strictly valid only for a conformal fluid, nothing prevents using the scaling laws for the transport coefficients $\tau_\varepsilon, \tau_Q$, and $\tau_P$ from Equation~\eqref{Conformal relaxation time expressions} for non-conformal fluids in the absence of conserved currents. 
For simplicity, we adopt the same scaling for $\eta, \tau_\varepsilon, \tau_P$, and $\tau_Q$ as in the conformal case, Equations~\eqref{Conformal EOS and transport choices} and \eqref{Conformal relaxation time expressions}, with
\begin{align}\label{frame-coefficients-QCD}
    \xi&=P'_\varepsilon=c_s^2\,,
    \qquad
    \chi=\frac{1}{4c_s^2}\frac{\chi_0}{\eta_0}\,,
    \qquad
    \lambda=\frac{1}{4c_s^2}\frac{\lambda_0}{\eta_0}\,,
\end{align}
where the only difference from the conformal case is that $c_s^2$ is no longer a global constant but varies with temperature.
Substituting into the general conditions in Equations~\eqref{cond:lambda-general-0sigma} and \eqref{cond:xi-general-0sigma}, we obtain \begin{equation}
    \frac{\chi_0}{\eta_0} \geq  4, \qquad \frac{\lambda_0}{\eta_0}\geq\frac{4c_s^2\chi_0}{\left(1-c_s^2\right)^2\chi_0-4(c_s^2)^2\eta_0}\,,
\end{equation}
where we simplified the first expression by using the fact that $c_s^2<1/2$ for the QCD EOS.
Since $c_s^2$ varies with the local thermodynamic state, these conditions must be verified locally. 
However, since the bounds on $c_s^2$ are known from the tabulated EOS, $c_{s,\min}^2 \leq c_s^2 \leq \frac{1}{3}$, the chosen values of $\chi_0/\eta_0$ and $\lambda_0/\eta_0$ are large enough to satisfy the conditions globally.

\subsubsection{Ideal Gas}\label{SSS:ideal-gas}
We consider the gamma-law EOS for a relativistic ideal gas, given by 
\begin{equation}
    P=(\Gamma-1)ne\,,
\end{equation}
where $\Gamma$ is the adiabatic index, $e$ is the specific internal energy, and we set the particle mass to unity, $m=1$.
The total energy density is thus related to the mass density by $\varepsilon=n(1+e)$. 

Fluid configurations for the ideal relativistic gas have been studied in \cite{Pandya_2022} for non-zero $\sigma$, with proposed scalings for BDNK frame parameters.
Here, we employ the general scalings from Section~\ref{Transport coefficients}, with relevant thermodynamic quantities given by
\begin{align}\label{eqs:thermal-coefficients-ideal-gas}
    P'_\varepsilon&=\Gamma-1\,,
    \quad
    P'_n=1-\Gamma\,,
    \quad
    c_s^2=\frac{\Gamma P}{\varepsilon+P}\,,
    \\
    \beta_\varepsilon&=({\Gamma-1})\tau_Q\,,
    \quad
    \beta_n=({1-\Gamma})\tau_Q\,.
\end{align}
We investigate ideal gas systems with vanishing thermal conductivity and bulk viscosity for simplicity, summarized by
\begin{equation}\label{eqs:visc-transport-coefficients-ideal-gas}
    \eta=\eta_0n^{1/\Gamma}\,,
    \qquad
    \zeta=0\,,
    \qquad
    \sigma=0\,.
\end{equation}
For the frame parameter coefficients, we adopt the same scalings as in the conformal case, Equations~\eqref{Conformal EOS and transport choices} and \eqref{Conformal relaxation time expressions}, given by
\begin{align}\label{eqs:frame-coefficients-ideal-gas}
    \xi&=(\Gamma-1)\,,
    \qquad
    \chi=\frac{3}{4}\frac{\chi_0}{\eta_0}\,,
    \qquad
    \lambda=\frac{3}{4}\frac{\lambda_0}{\eta_0}\,,
\end{align}
where the coefficients $\chi$ and $\lambda$ match those in the conformal and QCD cases.
Using the thermal coefficients in Equation~\eqref{eqs:thermal-coefficients-ideal-gas} with the transport coefficients in Equation~\eqref{eqs:visc-transport-coefficients-ideal-gas} and frame parameterizations in Equation~\eqref{eqs:frame-coefficients-ideal-gas}, the general parameterizations in Equation~\eqref{eq:general-choice-frames-0sigma} yield
\begin{align}\label{eqs:tau-expressions-ideal-gas}
    \tau_P = {(\Gamma-1)}{\tau_\varepsilon}\,,\quad
    \tau_\varepsilon 
    =\frac1{\rho}\chi_0 n^{1/\Gamma}\,,\quad
    \tau_Q
    =\frac1{\rho}\lambda_0 n^{1/\Gamma}\,.
\end{align}
The $n^{1/\Gamma}$ scaling is adopted for convenience for numerical tests.
For an ultra-relativistic electron gas with $\Gamma=4/3$, Equation~\eqref{eqs:tau-expressions-ideal-gas} recovers the $3/4$ factor analogous to the conformal and QCD cases.
We note that physically motivated scalings exist, such as the Shakura-Sunyaev $\alpha$-viscosity model for turbulence \cite{Shakura:1972te} or the constant kinematic viscosity model \cite{Shibata:2017jyf}, both studied widely in the context of accretion around black holes in astrophysics.
These can be substituted for $\eta$ in Equation~\eqref{eqs:visc-transport-coefficients-ideal-gas}, with the general expressions in Equation~\eqref{eq:general-choice-frames-0sigma} remaining valid without modification.

Using Equations~\eqref{eqs:thermal-coefficients-ideal-gas}--\eqref{eqs:frame-coefficients-ideal-gas}, the conditions in Equations~\eqref{cond:lambda-general-0sigma} and \eqref{cond:xi-general-0sigma} that ensure causality and stability become
\begin{align}
\frac{\lambda_0}{\eta_0}&>
\frac{
{4\chi_0}
}{
3{\chi_0}
(1-c_s^2)(2-\Gamma)
-4(\Gamma-1)\eta_0
}\,,
\label{cond:caus-ideal-gas-local-lambda0}
\\[2ex]
\frac{3\chi_0}{4\eta_0}&>
\max\!\left[
\frac{\Gamma-1}
{(1-c_s^2)(2-\Gamma)},
\,
\frac1{c_s^2}
\right]\,.
\label{cond:caus-ideal-gas-local-xi0}
\end{align}

As in the QCD case, these depend on local thermodynamic quantities, making it inconvenient to choose global constant values for $\chi_0$ and $\lambda_0$ {\it{a priori}}.
While local values can be parameterized to satisfy the conditions in Equations~\eqref{cond:caus-ideal-gas-local-lambda0}--\eqref{cond:caus-ideal-gas-local-xi0}, doing so provides little control over initial conditions or subsequent evolution.
This motivates establishing global bounds on the coefficients as discussed in the introduction~\ref{Transport coefficients}.

For an isentropic fluid obeying the polytropic EOS, $P=Kn^\Gamma$, with constant $K>0$, the sound speed increases monotonically with density $n$.
Thus $c_s^2<\Gamma-1$ provides an upper bound, and the first branch of Equation~\eqref{cond:caus-ideal-gas-local-xi0} has a global maximum, independent of any local thermodynamic quantity, given by
\begin{equation*}
    \frac{1}{1-c_s^2}<
    \left(\frac{1}{2-\Gamma}\right)\,,
    \quad\implies\quad
    \frac{\Gamma-1}
{(1-c_s^2)(2-\Gamma)}<
\frac{(\Gamma-1)}
{(2-\Gamma)^2}\,.
\end{equation*}
Subsequently, Equation~\eqref{cond:caus-ideal-gas-local-xi0} can be expressed as
\begin{equation}\label{cond:caus-ideal-gas-global-xi0}
    \frac{3\chi_0}{4\eta_0}>
\max\!\left[
\frac{\Gamma-1}
{(2-\Gamma)^2},
\,
\frac1{c_{s,\mathrm{min}}^2}
\right]\,,
\end{equation}
where a sufficiently large global $\chi_0$ is feasible only if a non-zero global minimum $c_{s,\mathrm{min}}^2$ exists.
We achieve this by introducing a number density floor $n_{\mathrm{floor}}$,\footnote{This is standard in hydrodynamic studies, where the floor typically represents vacuum or a dilute atmosphere surrounding the fluid of interest. 
While there is no rigorous \textit{a priori} guarantee that $n$ never drops below $n_{\mathrm{floor}}$, regulation procedures during primitive inversion routinely enforce this. 
Since a primitive inversion step is not involved in our study, we verify \textit{a posteriori} that $n\geq n_{\mathrm{floor}}$ throughout evolution.} 
which simplifies making a global choice for $\chi_0$ via Equation~\eqref{cond:caus-ideal-gas-global-xi0} given $\eta_0$, $\Gamma$, and polytropic constant $K$ are already chosen.
A corresponding sufficient global value for $\lambda_0$ is then given by
\begin{equation}
    \label{cond:caus-ideal-gas-global-lambda0}
    \frac{\lambda_0}{\eta_0}>
    \frac{
    {4\chi_0}
    }{
    3{\chi_0}
    (2-\Gamma)^2
    -4(\Gamma-1)\eta_0
    }\,.
\end{equation}

\subsection{Initial Conditions}\label{Sec:Initial Conditions}

Being a first-order gradient expansion theory of viscous fluids, the viscous corrections $\mathcal{A}$~[Equation~\eqref{defA}], ${\Pi}$~[Equation~\eqref{definePi}] and $\mathcal{Q}_\nu$~[Equation~\eqref{defQ}] in the BDNK fluid tensor involve
gradients of the five hydrodynamic variables $\{U^A,n\}$.
In 3+1 dimensions, the BDNK system in Equations~\eqref{BDNK_EoM} consists of four
second-order PDEs for $U^A$, coupled to a single first-order PDE for the current.
While a general first-order gradient expansion would
yield four independent components for $J^\mu$, we consider only an ideal current in this work.
Accordingly, the current sector has a single \ac{DOF}, $J^0$, or equivalently $n$, and its EOM is first-order in time.
Two 
first-order
reductions appropriate for numerical methods have been presented in this work: \ac{FO}~(Section~\ref{Sec:First_Order}) and the \ac{MO}~(Section~\ref{Mixed_order_section}) formulations. %

For the purposes of setting up initial conditions, one needs to specify 
initial data for the 4 fields $U^A$, and their initial time derivatives, as well as the current~($J^0$ or $n$).
We recall the definition (\ref{def:U^A}) of $U^A$ 
\begin{equation}
\label{New_variables-initial-conditions}
U^A=\begin{cases}\theta\equiv\ln\frac{\varepsilon}{\varepsilon_0},\; A=\theta\\ 
u^i,\; A=i \end{cases}\,,
\end{equation}
and the variables introduced for the time derivatives in Equations~\eqref{Y} and \eqref{Y_Mixed}, 
\begin{equation}\label{Y-initial-conditions}
    Y^\theta\equiv\partial_t\theta = \partial_tU^\theta = \partial_t\ln\varepsilon/\varepsilon_0\,,\qquad 
    Y^i \equiv \partial_tu^i=\partial_tU^i\,.
\end{equation}
In both formulations, the four time derivatives $\{Y^\theta, Y^i\}$ are traded for the four stress-energy components $T^{0\mu}$ (often referred to as conservative variables) as independent variables.

\subsubsection{Initial Time Derivatives}
\label{Sec:Initial Conditions, time derivatives}
As in typical relativistic hydrodynamic studies, the hydrodynamic variables $\{U^A(t_0,\vec{\mathbf{x}}), n(t_0,\vec{\mathbf{x}})\}$ on the initial hypersurface are specified by appropriate mathematical functions or numerical data describing the physical system of interest.
Additionally, one must specify the initial $T^{0\mu}(t_0,\vec{\mathbf{x}})$ or, equivalently, the time derivatives $Y^A(t_0,\vec{\mathbf{x}})$.
We present the two methods used in this work for initializing these time derivatives.
Note that the \ac{FO} reduction (Section~\ref{Sec:First_Order}) promotes the originally \ac{FO} current equation to a second-order one, which requires specifying $\partial_t J^0$ (or, equivalently, $\partial_t n$) as initial data. 
However, since the original current EOM is \ac{FO}, the corresponding time derivative cannot be arbitrary and is determined by the initial values of $\partial_t u^i$ (see Section~V.A in~\cite{Bemfica:2020zjp}).
The two initialization methods differ in how the initial $T^{0\mu}(t_0,\vec{\mathbf{x}})$~(equivalently~$Y^A(t_0,\vec{\mathbf{x}})$) are determined.
\paragraph{Method 1:}\label{IC-method-1}
In this approach, we compute the four independent components
$T_0^{0\mu}(t_0,\vec{\mathbf{x}})$ of the ideal fluid tensor from the initial values of $U^A$. We recall the definition for the ideal fluid tensor:
\begin{equation}
    \label{Def:Ideal fluid tensor}
    T_0^{\mu \nu} = \varepsilon u^\mu u^\nu + P\Delta^{\mu\nu}\,.
\end{equation}
We then set the BDNK conservative variables $T^{0\mu}(t_0,\vec{\mathbf{x}})$ equal to their ideal counterparts,
\begin{equation*}
    T^{0\mu}(t_0,\vec{\mathbf{x}})=
    T_0^{0\mu}(t_0,\vec{\mathbf{x}})\, 
    \implies \,
    Y^A(t_0,\vec{\mathbf{x}})\,,
\end{equation*}
which determines the initial time derivatives of the hydrodynamic variables $U^A$ via Equation~\eqref{Relations},
\begin{equation}\tag{\ref{Relations}}
Y^A  =\tilde{P}^{A\lambda}M^{-1}_{\lambda\mu}\left [T^{0\mu}- T_0^{0\mu} - H^{\mu0 k \beta}P_{B\beta}X_k^B-2\beta_n u^{(\mu} \Omega^{k0)} Z_k - H^{\mu 0 \alpha\beta} C_{k\beta} \Gamma^k_{\alpha \lambda}u^\lambda\right ]    
\end{equation}

\paragraph{Method 2:}\label{IC-method-2}
In this approach, we directly prescribe the initial time derivatives $Y^A$:
\begin{equation*}
    Y^A(t_0,\vec{\mathbf{x}})\implies\,T^{0\mu}(t_0,\vec{\mathbf{x}})\,,
\end{equation*}
which determine the initial values of the variables $T^{0\mu}$ via Equation~\eqref{defineBDNKnovo2} for the \ac{FO} formulation,
\begin{align}
\tag{\ref{defineBDNKnovo2}}
T^{\mu\nu} = T_0^{\mu\nu} + H^{\mu\nu\alpha\beta}P_{B\beta}\nabla_\alpha U^B+2\beta_n u^{(\mu} \Omega^{j\nu)}\partial_j n,
\end{align}
and Equation~\eqref{Tmunu_Mixed} for the \ac{MO} formulation,
\begin{align}
\tag{\ref{Tmunu_Mixed}}
T^{\mu\nu} &= T_0^{\mu\nu} + H^{\mu\nu 0\beta}P_{A\beta} Y^A
+H^{\mu\nu j \beta}P_{A\beta} \partial_j U^A +2\beta_n u^{(\mu} \Omega^{j\nu)} \partial_j n + H^{\mu\nu\alpha\beta} C_{j\beta} \Gamma^j_{\alpha \lambda}u^\lambda.
\end{align}
In most of the examples discussed in this work, we set the initial time derivatives to zero, $Y^A(t_0,\vec{\mathbf{x}})=0$, unless appropriate initial configurations for the time derivatives are available~(see for example, the details of Bjorken~(Section~\ref{Subsubsec:Bjorken}) and Gubser~(Section~\ref{Subsubsec:Gubser}) flows).

\subsubsection{Summaries for the \ac{FO} and \ac{MO} formulations}
\label{S:Initial_data_summaries}
We now summarize the initial data setup for each formulation:
\paragraph{\ac{FO}:}\label{sec:FO-init-procedure}
The \ac{FO} formulation has 24 dynamical variables~$\{Y^A,X_{i}^A,Z_i,U^A,n\}$, where spatial derivatives are promoted to independent variables, defined in Equations~\eqref{X} and \eqref{Z}, 
\begin{align}\tag{\ref{X}}
X_{i}^A&\equiv \partial_i U^A\,,
\\
\tag{\ref{Z}}
Z_i&\equiv\partial_in\,,
\end{align}
and 20 auxiliary variables~$\{\varphi_{i}^A,\phi^A,\Psi_i,\psi\}$ for the cleaning fields.
\begin{enumerate}
    \item Set $\{U^A(t_0,\vec{\mathbf{x}}),n(t_0,\vec{\mathbf{x}})\}$ using analytical expressions or numerical data describing the physical system of interest.
    \item Set the spatial derivative variables $X_i^A(t_0,\vec{\mathbf{x}})$ and $Z_i(t_0,\vec{\mathbf{x}})$ using analytical derivatives, or the CWENO numerical differentiation scheme if initial $\{U^A(t_0,\vec{\mathbf{x}}),n(t_0,\vec{\mathbf{x}})\}$ data is set numerically.
    \item Set the auxiliary variables associated with the cleaning fields to zero: 
    $$\varphi_{i}^A(t_0,\vec{\mathbf{x}})=
    \phi^A(t_0,\vec{\mathbf{x}})=
    \Psi_i(t_0,\vec{\mathbf{x}})=
    \psi(t_0,\vec{\mathbf{x}})=0\,.$$ 
    See Section~\ref{Sec:Cleaning_formulation} for a discussion on the cleaning tools.
    \item Set initial values of $T^{0\mu}(t_0,\vec{\mathbf{x}})$~(equivalently $Y^A(t_0,\vec{\mathbf{x}})$) using Method-1~(Section~\ref{IC-method-1}) or Method-2~(Section~\ref{IC-method-2}). 
\end{enumerate}

\paragraph{\ac{MO}:}\label{sec:MO-init-procedure}
The EOMs in the \ac{MO} system in Equation~\eqref{MO:eoms-Matrix} are \ac{FO} in time and second-order in space, with 9 dynamical variables~$\{T^{0\mu}(t_0,\vec{\mathbf{x}}),U^A(t_0,\vec{\mathbf{x}}),J^0(t_0,\vec{\mathbf{x}})\}$.
The initial data is constructed as follows:
\begin{enumerate}
    \item Set $\{U^A(t_0,\vec{\mathbf{x}}),n(t_0,\vec{\mathbf{x}})\}$ as in the \ac{FO} case.
    \item Compute spatial derivatives of $U^A(t_0,\vec{\mathbf{x}})$ required for
      $T^{0\mu}(t_0,\vec{\mathbf{x}})$ in Equation~\eqref{Tmunu_Mixed} analytically where the initial profiles are given by closed-form expressions, or via the CWENO numerical differentiation scheme otherwise.
    \item Set initial values of $T^{0\mu}(t_0,\vec{\mathbf{x}})$ as in the \ac{FO} case.
\end{enumerate}

\subsubsection{Discussion}
\label{Sec:Initial Conditions, discussion}
There are various challenges involved in formulating well-posed initial data for BDNK; see, in particular, the discussion in Section~VII.C of~\cite{Bemfica:2017wps}.
Method-1~(\ref{IC-method-1}) follows a common approach from \cite{Bemfica:2020zjp,Clarisse:2025lli} in which initial configurations are chosen such that the viscous corrections to the components $T^{0\mu}(t_0,\vec{\mathbf{x}})$ vanish\footnote{Note that, if the initial flow is non-zero, each individual viscous correction $\cA, \Pi, Q^\mu$ or $-2\eta\sigma^{\mu\nu}$ can still be non-zero even though the net viscous contribution to the BDNK tensor vanishes. },
and thus these components are equal to their ideal counterparts.
The corresponding time derivatives $Y^A$ can then be determined as discussed in Section~(\ref{IC-method-1}). 

Method-2~(\ref{IC-method-2}) instead treats the initial time derivatives as free data, as is natural for any second-order differential system, and one specifies both $U^A(t_0,\vec{\mathbf{x}})$ and $Y^A(t_0,\vec{\mathbf{x}})$ on the initial hypersurface. 
In general, this implies that the viscous corrections computed from the initial data do not vanish, and thus the initial stress-energy tensor strongly depends on the choice of $Y^A$. 
This sensitivity is tied to the fact that the definitions of $\{\varepsilon, u^i,n\}$ in BDNK depend on the hydrodynamic frame. 
A practical consequence is that it becomes a non-trivial task to construct initial data $\{U^A(t_0,\vec{\mathbf{x}}), Y^A(t_0,\vec{\mathbf{x}})\}$ which precisely~(or even approximately) reproduces a given target $T^{\mu\nu}(t_0,\vec{\mathbf{x}})$, such as IP-Glasma initial conditions commonly used in heavy-ion-collision simulations.

These difficulties reflect a more general problem in relativistic hydrodynamics: determining initial data for the dynamical variables of a given hydrodynamic theory from a physical description of matter given in the form of some stress-energy tensor.
A general $T^{\mu\nu}$ has 10 and a current $J^\mu$ has 4 independent components.
In ideal relativistic hydrodynamics, there exist 4 hydrodynamical variables $\{\varepsilon, u^i\}$, and one current variable $n$ or, equivalently $J^0$.
Starting from a general $T^{\mu\nu}$ and $J^\mu$ that may not correspond to an ideal fluid, one must choose which information to discard to determine the 5 required variables.
In IS-type theories, there are 14 hydrodynamical variables that can be reconstructed from the 14 independent components of $T^{\mu\nu}$ and $J^{\mu}$.
In both cases, convenient prescriptions exist for determining the respective dynamical variables from a given stress-energy tensor, owing to the algebraic relationship between the dynamical variables and $T^{\mu\nu}$. 
However, even though the mapping exists, this does not guarantee that one is describing a physically correct system~(see discussion in Section~V.A in \cite{Bemfica:2020zjp} for an example).

On the other hand, the BDNK theory with an ideal current has 9 dynamical \ac{DOF}s ($U^A$, $Y^A$, and $n$). 
Moreover, unlike in ideal hydrodynamics or Israel-Stewart–type theories, where the constitutive relations are algebraic, the relations between the BDNK stress-energy tensor and the hydrodynamic variables are differential. 
Thus, the system is overdetermined, and the differential nature of the constitutive relations prevents straightforward local inversion. 
Consequently, there is no general one-to-one mapping from an arbitrary physical $T^{\mu\nu}$ (10 independent components) and a current $J^\mu$ (4 components) to the 9 BDNK variables.

Method-1~(Section~\ref{IC-method-1}) is overly restrictive, since it requires specifying the 5 variables $\{U^A, n\}$ which then fix the time derivatives, exploring only a small subspace of the full BDNK phase space. 
Method-2~(Section~\ref{IC-method-2}) allows one to specify all dynamical variables freely, but the challenge lies in choosing a global configuration of $\{U^A, Y^A, n\}$ whose gradients produce the target $T^{\mu\nu}$ everywhere correctly.
This is related to the choice of the hydrodynamic frame, 
which defines how the total energy and momentum densities are distributed among the equilibrium fields and viscous corrections.

While both methods, Method-1~(Section~\ref{IC-method-1}) and Method-2~(Section~\ref{IC-method-2}), adopted in this work are adequate for demonstrating that the BDNK reductions proposed in this study can be reliably evolved numerically, a general algorithm for determining consistent initial data ${U^A, Y^A, n}$ from a given target $T^{\mu\nu}$ remains an open problem. 
We plan to explore a general approach in future work.

\subsection{Numerical Methods}\label{Sec:Num-methods}

Finite-volume high-resolution shock-capturing (HRSC) schemes are routinely employed to simulate both ideal and dissipative (MIS-type) relativistic hydrodynamics in astrophysical and heavy-ion contexts. However, the presence of dissipative corrections involving gradients of the hydrodynamic variables, as is the case for the BDNK theory, requires additional considerations, as discussed in Ref.~\cite{Pandya_2022}. 
In particular, standard finite-volume methods require the system of PDEs to be cast in \ac{FO} conservative form, which is not manifest in the original formulation of the BDNK equations.

Various \ac{FO} reductions of the BDNK equations have been proposed and explored in numerical studies~\cite{Pandya-21, Shum:2025jnl}, often under simplifying assumptions, e.g., fixed metric backgrounds, analytic equations of state, or lower-dimensional spacetimes.
In this work, we instead employ two fully general reductions of the BDNK system: the \ac{FO} and \ac{MO} formulations introduced in Sections~\ref{Sec:First_Order} and \ref{Mixed_order_section}, respectively. 
The governing PDEs for both formulations, given in matrix form by Equations~\eqref{FO:Matrices} and \eqref{MO:eoms-Matrix}, take the general balance law form 
\begin{equation} \label{eq:numerics-general-balance-law} \partial_t\mathcal{Q}+\partial_a F^a=\mathcal{S}\,, \end{equation} where $\mathcal{Q}$ denotes the state vector of evolved variables, $F^a$ is the spatial flux vector in the $a$-direction with $a \in {x, y, z}$, and $\mathcal{S}$ is the source vector.

In the finite-volume approach, the spatial domain is partitioned into non-overlapping cells. For simplicity, we consider a rectangular domain of size $L_x\times L_y \times L_z$, uniformly discretized into $N_x\times N_y\times N_z$ cells, each of volume $\delta V = \delta x\,\delta y\,\delta z$. The cell-averaged variables at a cell $\mathcal{C}_{i,j,k}$ centered at $(x_i, y_j, z_k)$ are denoted by a subscript $(i,j,k)$, while cell interfaces are labeled by half-integer indices; for example, the surfaces orthogonal to the $x$-direction bounding cell $\mathcal{C}_{i,j,k}$ are denoted by $\Sigma_{i\pm\frac{1}{2},j,k}$.

The semi-discrete evolution equations are obtained by integrating Equation~\eqref{eq:numerics-general-balance-law} over the volume of each cell and applying the divergence theorem to the flux terms:
\begin{equation}
\label{discrete-FV-eom-general}
\partial_t\bar{\mathcal{Q}}_{i,j,k}+
\left(
{\langle F^x_{i+\frac{1}{2},j,k}\rangle - \langle F^x_{i-\frac{1}{2},j,k}\rangle}
\right)
+
\left(
{\langle F^y_{i,j+\frac{1}{2},k}\rangle - \langle F^y_{i,j-\frac{1}{2},k}\rangle}
\right)
+
\left(
{\langle F^z_{i,j,k+\frac{1}{2}}\rangle - \langle F^z_{i,j,k-\frac{1}{2}}\rangle}
\right)
=
\bar{\mathcal{S}}_{i,j,k}\,, 
\end{equation} 
where overbars denote cell-averaged quantities and angle brackets denote surface-integrated fluxes, defined respectively as 
\begin{align} \bar{\mathcal{Q}}_{i,j,k} &= \frac{1}
{\delta V}
\int
\mathcal{Q}\, dx\, dy\, dz\,,
\label{discrete-FV-cell-average}\\ 
\langle F^x_{i\pm\frac{1}{2},j,k}\rangle &= \frac{1}{\delta V}\int_{\Sigma_{i\pm\frac{1}{2},j,k}}F^x\, dy\, dz\,.
\label{discrete-FV-surface-integral}
\end{align} 
Thus $\langle F^x_{i\pm\frac{1}{2},j,k}\rangle$ represents the integrated flux through the cell interface $\Sigma_{i\pm\frac{1}{2},j,k}$.
For simplicity, we only discuss, as an example, $x$-components of the flux at the interfaces $\Sigma_{i\pm\frac{1}{2},j,k}$.
The corresponding expressions for the $y$ and $z$ components at the interfaces $\Sigma_{i,j\pm\frac{1}{2},k}$ and $\Sigma_{i,j,k\pm\frac{1}{2}}$ can be obtained in a straightforward manner.

The semi-discrete system in Equation~\eqref{discrete-FV-eom-general} is evolved in time to advance the cell-averaged variables $\bar{\mathcal{Q}}_{i,j,k}$ from one discrete time level $t^n$ to the next, $t^{n+1}$. 
Evaluating the surface flux terms in Equation~\eqref{discrete-FV-surface-integral} requires two ingredients: (i) 
a numerical flux function that approximates the surface integral %
, and 
(ii) a reconstruction procedure that interpolates the cell-averaged data to the cell interfaces, $\bar{\mathcal{Q}}_{i,j,k} \to \mathcal{Q}_{i\pm\frac{1}{2},j,k}$, since computing $\langle F^a_{i\pm\frac{1}{2},j,k}\rangle$ requires evaluating the flux function $F^a$ in Equation~\eqref{discrete-FV-surface-integral} at the cell interfaces.
In this work, we employ the central-upwind Kurganov-Tadmor (KT) scheme~\cite{Tadmor2000} for the numerical flux evaluation, together with two reconstruction algorithms: the minmod slope limiter~\cite{Sweby1984} and the fifth-order weighted essentially non-oscillatory scheme with Z-type smoothness indicators (WENO-Z)~\cite{WENOZ-Borges}. 
For completeness, we summarize the KT scheme in Appendix~\ref{Appendix:KT}, and the reconstruction algorithms for the minmod slope limiter and WENO-Z in Appendices~\ref{Appendix:MINMOD} and \ref{Appendix:WENO-Z}, respectively.
For time integration, both first-order and \ac{MO} formulations 
use Heun's method, a second-order total-variation-diminishing (TVD) Runge-Kutta scheme~\cite{SHU1988439} 
commonly used with finite-volume discretizations for hyperbolic conservation laws.
We now describe the specific numerical treatment for the first-order and \ac{MO}
formulations introduced in Sections~\ref{Sec:First_Order} and \ref{Mixed_order_section}.

\paragraph{\ac{FO} formulation.} 
The full \ac{FO} reduction described in Section~\ref{Sec:First_Order} promotes all gradients of the hydrodynamic variables to independent evolved fields. 
In this formulation, the flux functions $F^a_{(\textrm{FO})}(\mathcal{Q})$ and source terms $\mathcal{S}_{(\textrm{FO})}(\mathcal{Q})$ given by Equation~\eqref{FO:Matrices} are purely algebraic in $\mathcal{Q}$, i.e., they contain no spatial derivatives. 
The resulting system consists of \ac{FO} hyperbolic conservation laws for the hydrodynamic variables and their promoted gradients. The system is supplemented with additional auxiliary fields and corresponding equations that dynamically enforce divergence and curl constraints among the gradient fields.
A general discussion of the mathematical formulation and numerical considerations for such constraints was provided in Section~\ref{Sec:CC}. 
In Section~\ref{Sec:Cleaning_formulation}, we described the curl- and divergence-cleaning implementation for the \ac{FO} BDNK formulation.

The numerical implementation proceeds as follows. 
\begin{enumerate}
    \item
    The discrete source terms are computed by evaluating the source functions at the cell-averaged variables: $\bar{\mathcal{S}}_{i,j,k} = \mathcal{S}_{(\textrm{FO})}(\bar{\mathcal{Q}}_{i,j,k})$.
    \item 
    The surface-averaged fluxes $\langle F^a_{i\pm\frac{1}{2},j,k}\rangle_{(\textrm{FO})}$ are evaluated using the KT numerical flux formulae. A brief description is provided in Appendix~\ref{Appendix:KT}, with the specific numerical flux functions given by Equation~\eqref{KT-flux-terms}.
\end{enumerate}

This procedure closely mirrors standard HRSC finite volume methods used in numerical relativistic hydrodynamics, with one key distinction: no conservative-to-primitive inversion is required. 
This is a deliberate consequence of our \ac{FO} reduction formulated in Section~\ref{Sec:First_Order}, wherein the fluxes $F^a_{(\textrm{FO})}$ and sources $\mathcal{S}_{(\textrm{FO})}$ are given as explicit analytical functions of the evolved variables $\mathcal{Q}$. 

\paragraph{\ac{MO} formulation.} 
In contrast, the \ac{MO} formulation introduced in Section~\ref{Mixed_order_section} is not strictly \ac{FO}: both the flux and source terms in Equation~\eqref{MO:eoms-Matrix} retain spatial derivatives of the evolved variables, i.e., $F^a_{(\textrm{MO})}(\mathcal{Q},\partial_b\mathcal{Q})$ and $\mathcal{S}_{(\textrm{MO})}(\mathcal{Q},\partial_b\mathcal{Q})$. 
As a result, standard HRSC methods designed for conservative systems of PDEs are not directly applicable, and have to be supplemented with numerical methods to handle the derivatives. 
In particular, approximating these spatial derivatives with standard centered finite differences can introduce spurious oscillations near discontinuities, undermining the shock-capturing properties of the overall scheme.

We address this by using the KT scheme formulated for diffusive systems in~\cite{Tadmor2000} to handle the flux terms.
For the flux-terms, we split the flux functions $F^a_{(\textrm{MO})}$ in Equation~\eqref{MO:eoms-Matrix} into 
\begin{equation*}
F^a_{(\textrm{MO})} = 
\tilde{F}^a_{(\textrm{MO})}(\mathcal{Q})+
\tilde{\mathcal{K}}^a_{(\textrm{MO})}(\mathcal{Q},\partial_b\mathcal{Q})\,.
\end{equation*}
The corresponding KT scheme for diffusion-type problems is outlined in Appendix~\ref{Appendix:KT}.
The numerical flux functions $\langle \tilde{F}^a_{i\pm\frac{1}{2},j,k}\rangle_{(\textrm{MO})}$ and $\langle \tilde{\mathcal{K}}^a_{i\pm\frac{1}{2},j,k}\rangle_{(\textrm{MO})}$, corresponding to the flux terms $\tilde{F}^a_{(\textrm{MO})}$ and $\tilde{\mathcal{K}}^a_{(\textrm{MO})}$, are computed using Equations~\eqref{KT-flux-terms} and \eqref{KT-diff-EOM-P-flux-terms}, respectively. 
As in the \ac{FO} approach, the non-derivative variables at the cell interfaces, $\mathcal{Q}_{i\pm\frac{1}{2},j,k}$, are reconstructed via minmod or WENO-Z. 
As highlighted in Section~\ref{Mixed_order_section}, one major advantage of employing the KT scheme for diffusion-type problems is that the method implicitly handles the derivatives, and no separate reconstruction method to approximate derivatives $\partial_a\mathcal{Q}$ at the interfaces is required.

However, unlike the \ac{FO} method, the source terms $\mathcal{S}_{(\textrm{MO})}(\mathcal{Q},\partial_b\mathcal{Q})$ in the \ac{MO} formulation also contain derivatives, since not all the derivative terms in the BDNK EOMs could be conveniently collected inside flux-like terms. 
Following the approach proposed in~\cite{Pandya_2022} for the BDNK PDE system, we use reconstruction-based derivative evaluation to handle the derivatives in the source terms. 
Accordingly, we employ the Central-WENO (CWENO) reconstruction scheme~\cite{CWENO}, which provides local interpolation polynomials centered about each cell interface. 
The required derivatives at cell centers are then obtained by analytically differentiating these polynomials and evaluating the result at the cell center. 
Once computed, these derivative quantities are treated on the same footing as the other cell-centered variables.
This approach ensures that the derivative approximations inherit the non-oscillatory characteristics of the CWENO reconstruction. 
Additional details of the CWENO-based algorithm used to approximate derivatives, as well as a BDNK-specific discussion of errors arising from numerical curl-type constraint violations associated with this approach, can be found in Section~III.D of ~\cite{Pandya_2022}.

The complete algorithm for a single time step of the \ac{MO} formulation can be summarized as follows: 
\begin{enumerate} 
\item \textbf{Derivative evaluation:} 
Given the cell-averaged variables $\bar{\mathcal{Q}}_{i,j,k}$ at time level $t^n$, compute spatial derivatives $\overline{\partial_a{\mathcal{Q}}}_{i,j,k}$ at cell centers using CWENO reconstruction. 
\item \textbf{Source computation:} 
Evaluate the numerical source terms from the cell-centered variables and their derivatives: $\bar{\mathcal{S}}_{i,j,k} = \mathcal{S}_{(\textrm{MO})}(\bar{\mathcal{Q}}_{i,j,k},, \overline{\partial_a{\mathcal{Q}}}_{i,j,k})$. 
\item \textbf{Interface reconstruction:} 
Using WENO-Z (or minmod), reconstruct the evolved variables at the cell interfaces: 
$\bar{\mathcal{Q}}_{i,j,k}, 
\to \mathcal{Q}_{i\pm\frac{1}{2},j,k}\,.$
\item \textbf{Flux evaluation:} 
Using the reconstructed quantities $\mathcal{Q}_{i\pm\frac{1}{2},j,k}$, evaluate 
$\tilde{F}^a_{(\textrm{MO})}$ and $\tilde{\mathcal{K}}_{(\textrm{MO})}^a$ at cell-interfaces to 
compute $\langle \tilde{F}^a_{i\pm\frac{1}{2},j,k}\rangle_{(\textrm{MO})}$ and $\langle \tilde{\mathcal{K}}^a_{i\pm\frac{1}{2},j,k}\rangle_{(\textrm{MO})}$
with the numerical flux functions in Equations~\eqref{KT-flux-terms} and \eqref{KT-diff-EOM-P-flux-terms}, respectively.
\item \textbf{Time integration:} 
With the numerical fluxes and source terms now determined, advance the solution from $t^n$ to $t^{n+1}$ using Heun's method (second-order TVD Runge-Kutta). 
\end{enumerate}

\section{Numerical tests in flat spacetime}
\label{S:Flat_spacetime_tests}

\subsection{Cartesian}
\label{S:Cartesian_tests}
\subsubsection{Smooth shock}
\label{S:Smooth_shock}
Numerical setups of relativistic hydrodynamics commonly involve shock tests.
We consider a 1D smooth shocktube setup in the energy density for a conformal fluid, as discussed in \cite{Clarisse:2025lli}, to test our numerical setup. We consider the initial condition
\begin{align}
    \label{smooth-energy-density-shocktube-initial-condition}
    \varepsilon=\varepsilon_R+\frac{\varepsilon_L-\varepsilon_R}{1+e^{x/\Delta}}\,,
\end{align}
where $\varepsilon_L$ and $\varepsilon_R$ are the left and right states of the energy density, respectively, and $\Delta$ is the length scale over which the left and right states are smoothly connected. The fluid is initially at rest with 
$u^x=0$ 
at $t=0\,\GeV^{-1}$ everywhere on the grid, and we adopt the profile parameters $\varepsilon_L=1.3\,\GeV^4,\,\varepsilon_R=0.3\,\GeV^4$ and $\Delta=1.0\,\GeV^{-1}$. 
The initial conditions are set up on a 1D Cartesian grid with $x\in (-64,64)\,{\GeV^{-1}}$, cell width $\delta x=(1/8)\,\GeV^{-1}$ and $\delta t = \delta x/100$. 

\begin{figure}
    \centering
    \includegraphics[width=\linewidth]{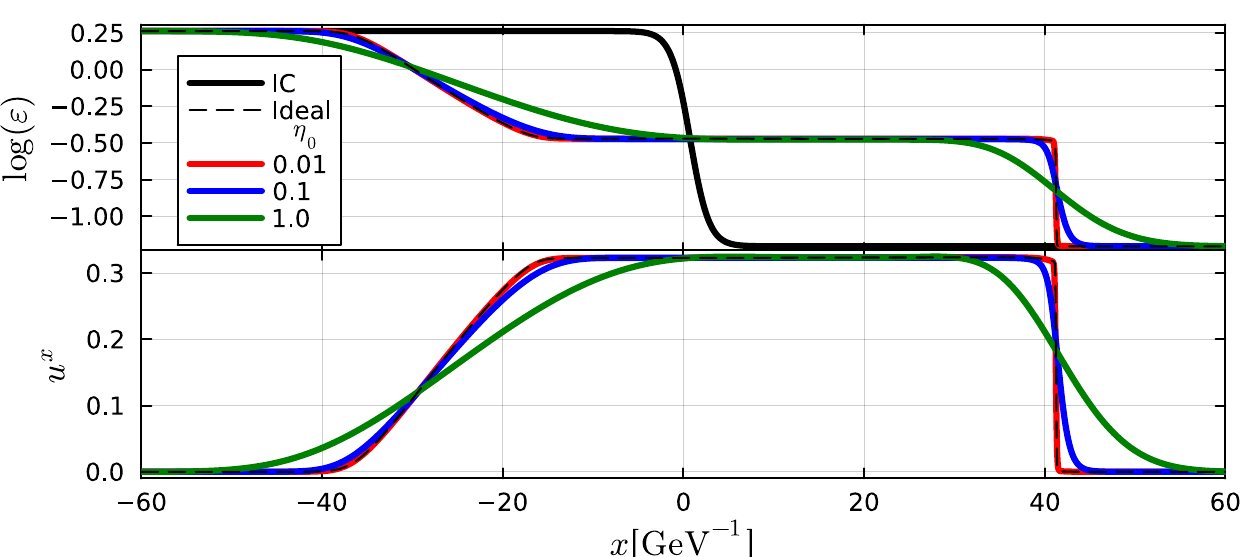}
    \caption{Snapshots from the evolution of the smooth shocktube system for the conformal EOS. The upper panel shows $\log \varepsilon$, while the lower panel shows the velocity $u^x$. The solid black curve corresponds to the initial condition at $t_0=0\,\GeV^{-1}$ in $\log\varepsilon$, the solid colored curves correspond to the numerical solution at $t_f=60\,\GeV^{-1}$, with the corresponding values of $\eta_0$ indicated in the legend, and the dashed curves represent the inviscid solution.
    The same frame parameters here are $\chi_0/\eta_0=25/4$ and $\lambda_0/\eta_0=25/7$.
    }
    \label{fig:smooth-energy-shoctube-compare-eta0}
\end{figure}

In Figure~\ref{fig:smooth-energy-shoctube-compare-eta0}, we compare the evolution of the smooth shock for several values of the viscosity parameter $\eta_0$ with the ideal inviscid case. Starting from the shock initial conditions in Equation~\eqref{smooth-energy-density-shocktube-initial-condition}, the system evolves into the expected structure, consisting of a left-moving rarefaction region and a right-moving compression front. In the ideal fluid limit, these features remain relatively sharp, whereas the inclusion of viscosity progressively smooths the gradients and broadens the transition regions.

This behavior is evident in both the energy-density and velocity profiles. 
For the smallest viscosity parameter, $(\eta_0=0.01)$, the solution remains very close to the ideal inviscid solution.
For the largest value $\eta_0=1.0$, viscous effects are more pronounced, with the rarefaction wave spread over a wider spatial interval, and the right-moving front also significantly smoothed out. 
The same trend is visible in the velocity profile, where increasing viscosity lowers the peak velocity.

The results of our setup exhibit the same viscous regularization of shocklike structures discussed in \cite{Clarisse:2025lli}. We also note the discussion in previous works \cite{Pandya_2022,Friestuhler-BDNK-shock,Clarisse:2025lli} on an important caveat on the subject of formation of discontinuous shock like structures in a gradient-expansion theory like BDNK.

\begin{figure}
     \centering
     \begin{subfigure}[b]{0.49\textwidth}
         \centering
         \includegraphics[width=\textwidth]{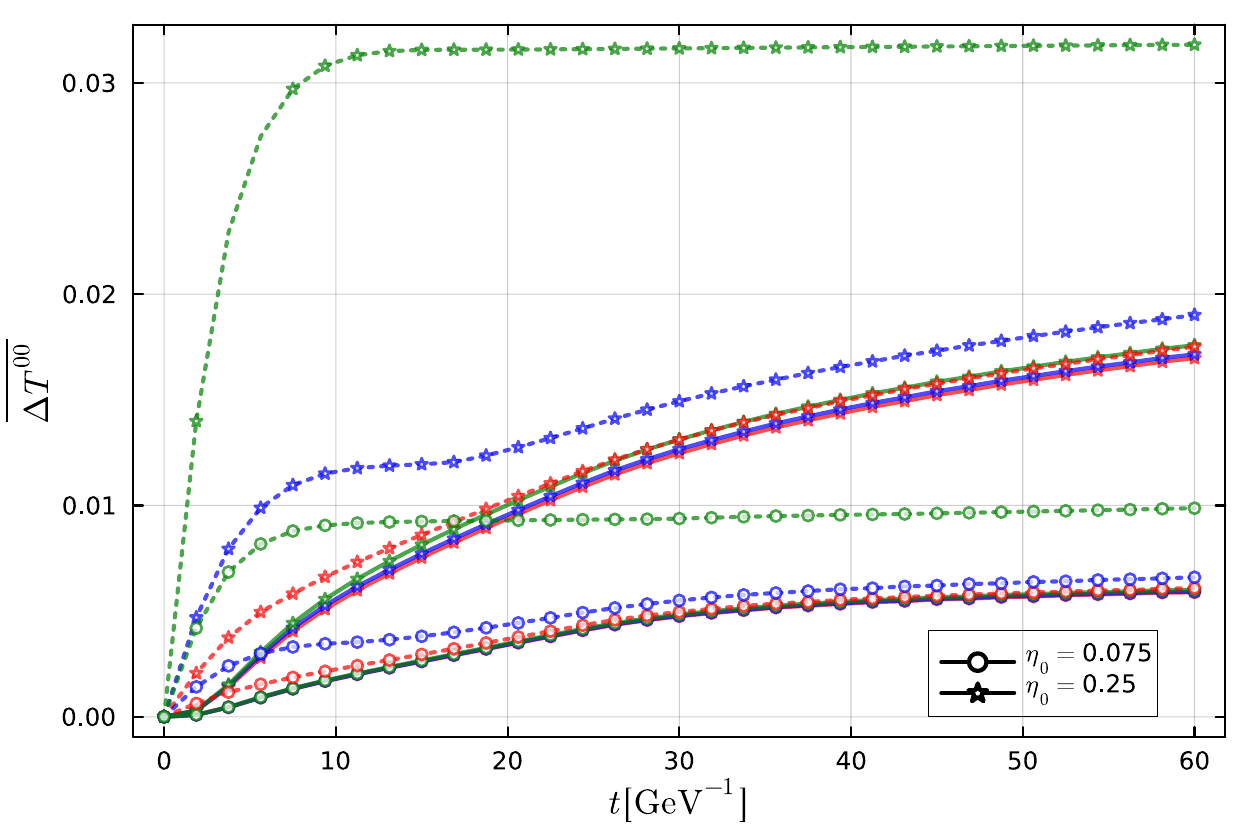}
     \end{subfigure}
     \hfill
     \begin{subfigure}[b]{0.49\textwidth}
         \centering
         \includegraphics[width=\textwidth]{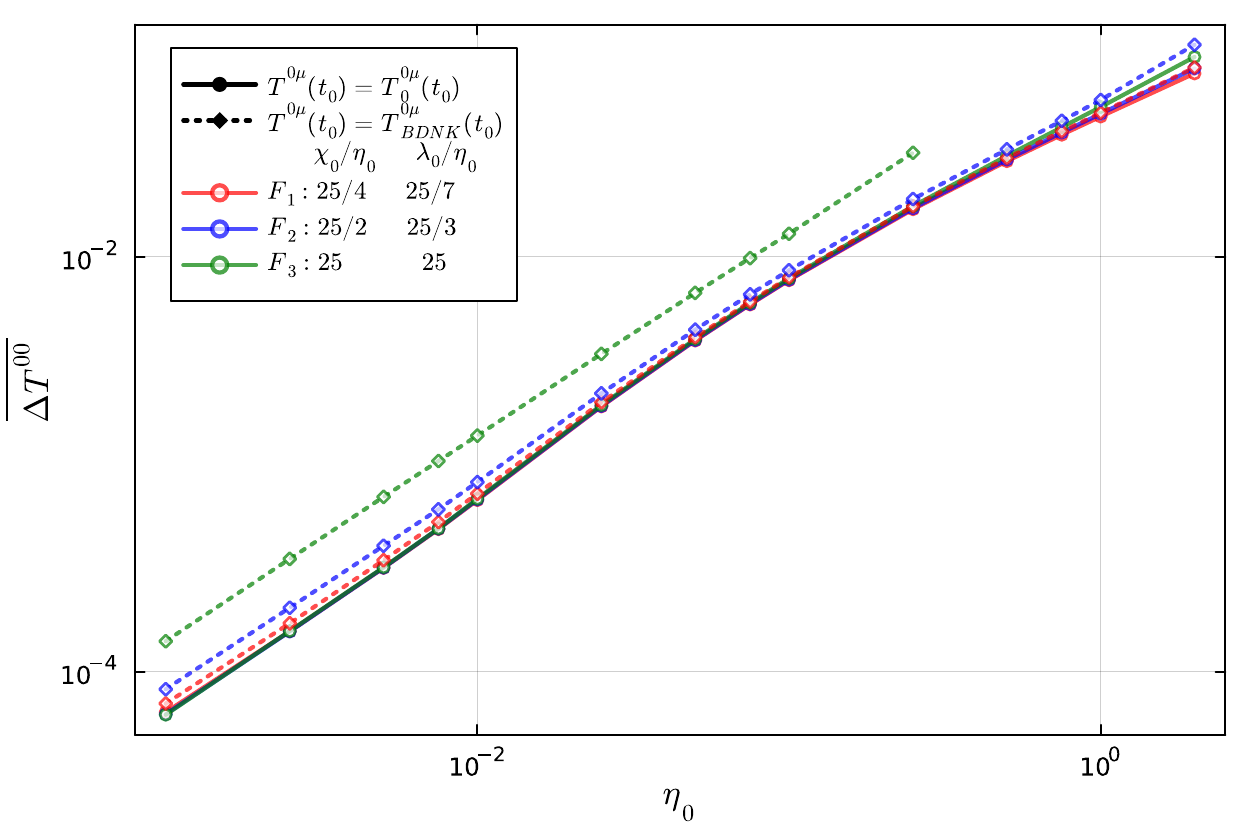}
     \end{subfigure}
     \hfill
        \caption{
        Normalized mean absolute differences, $\overline{\Delta T^{00}}$ between the viscid BDNK variable $T^{00}$ and the corresponding ideal inviscid counterpart for the smooth shock setup. 
        In the left panel, $\overline{\Delta T^{00}}$ is displayed as a function of time for a selected $\eta_0$ value.
        In the right, the value of $\overline{\Delta T^{00}}$ at the end of the simulations, $t_f$, is shown for a range of $\eta_0$.
        The colors represent different frame parameters $(\chi_0,\lambda_0)$ as indicated in legends.
        Solid and dashed curves correspond to initialization Methods~1 (Section~\ref{IC-method-1}) and~2 (Section~\ref{IC-method-2}), respectively.
        }
        \label{fig:shock-bdnk-ideal-init-diff-compare}
\end{figure}

In previous numerical studies of BDNK hydrodynamics \cite{Pandya_2022, Clarisse:2025lli}, the evolution of the fluid variables has been compared with that of the ideal inviscid theory in the regime of small but nonzero viscous parameters. This provides a useful consistency check, since the dynamics of the viscous BDNK system should approach those of the ideal fluid when viscous effects are sufficiently weak. While the snapshots in Figure~\ref{fig:smooth-energy-shoctube-compare-eta0} already illustrate this behavior qualitatively, here we examine it more systematically by quantifying the difference between the BDNK solution and the corresponding ideal solution.

To quantify this difference, we use the normalized mean absolute difference $\overline{\Delta T^{00}}$, defined in Equation~\eqref{def:normalized-mean-absolute-difference}.
The results are shown in Figure~\ref{fig:shock-bdnk-ideal-init-diff-compare} for a range of values of $\eta_0$, together with several choices of the frame parameters $\chi_0$ and $\lambda_0$. 
For ease of comparison, we adopt the same three frame-parameter choices $(\chi_0/\eta_0,\lambda_0/\eta_0)$ used in the recent analysis of Ref.~\cite{Clarisse:2025lli}. In the left panel, we show the differences at several times for the smallest and largest values of $\eta_0$, for all three frame choices, and for both initialization schemes. In the right panel, we show the dependence of the late-time difference on the viscous parameter $\eta_0$ after evolving the system to $t_f=60\,\GeV^{-1}$.

As discussed in Ref.~\cite{Clarisse:2025lli}, choosing initial data such that $T^{0\mu}(t_0)=T^{0\mu}_0$, with nonzero initial time derivatives $Y^A(t_0)$ determined through Equation~\eqref{Relations}, leads to frame-independent evolution. This corresponds to initialization Method~1 (Section~\ref{IC-method-1}), represented by the solid curves in Figure~\ref{fig:shock-bdnk-ideal-init-diff-compare}. In the left panel, the curves corresponding to different frame choices show close agreement throughout the evolution. This behavior is also reflected in the right panel, where the late-time differences for the three frame choices overlap.

By contrast, initialization Method~2 (Section~\ref{IC-method-2}), represented by the dashed curves, imposes vanishing initial time derivatives, $Y^A(t_0)=0$, with $T^{0\mu}(t_0)=T^{0\mu}_{\rm BDNK}(t_0;\eta_0,\chi_0,\lambda_0)$. This procedure exhibits a more pronounced dependence on the frame parameters, consistent with the discussion in Section~\ref{Sec:Initial Conditions}.

Nevertheless, for both initialization procedures and all frame choices considered, the differences decrease as $\eta_0$ is reduced, consistent with the expected recovery of ideal inviscid hydrodynamics in the vanishing-viscosity limit.

\subsubsection{Kelvin--Helmholtz}
\label{S:Kelvin_Helmholtz}

\begin{figure}
    \centering
    \includegraphics[width=\linewidth]{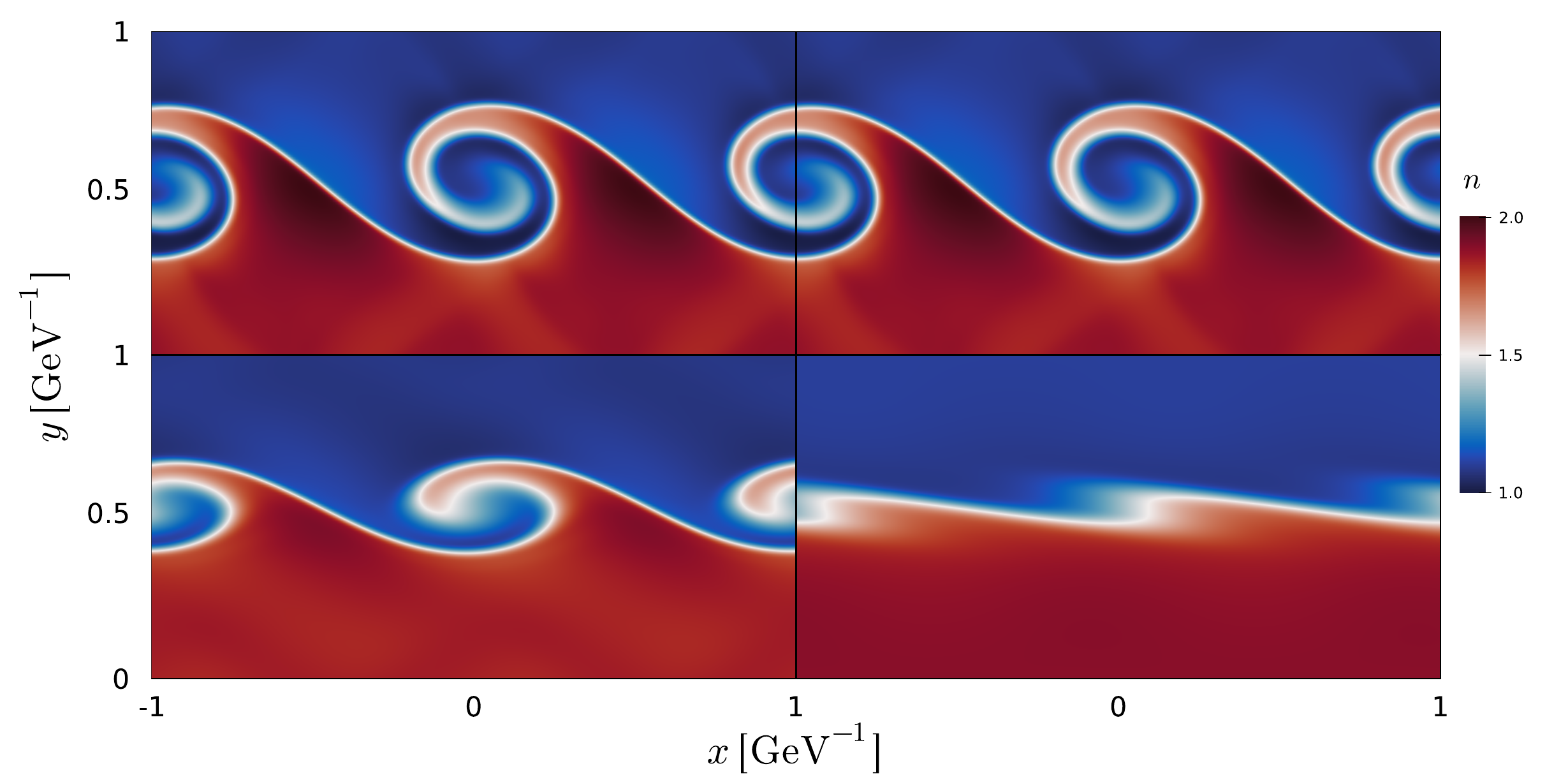}
    \caption{Comparison of the density $n$ for the Kelvin--Helmholtz instability between ideal inviscid hydrodynamics (top left) and viscous BDNK hydrodynamics with $\eta_0=10^{-5}$ (top right), $10^{-3}$ (bottom left), and $10^{-2}$ (bottom right).
    The panels show two-dimensional heatmaps of $n$ in the $xy$-plane at $t=5,\GeV^{-1}$, with the corresponding color scales indicated by the colorbars.
    All results here use the same frame parameters, $\chi_0/\eta_0=25/4$ and $\lambda_0/\eta_0=25/7$ and the Method-1~\ref{IC-method-1} for initial conditions.
    }
    \label{fig:Kelvin Helmholtz snapshots}
\end{figure}

The Kelvin--Helmholtz instability is a standard test of numerical stability in relativistic hydrodynamics codes.
The issue of numerical instability in resolving BDNK dynamics for small viscous transport coefficients was discussed in \cite{Pandya_2022} wherein the authors employed an adaptive method that switched between BDNK and ideal primitive recovery to effectively resolve BDNK dynamics for small viscous coefficients.
We investigate our \ac{FO} and \ac{MO} numerical implementations using the same conformal fluid configuration considered in~\cite{Pandya_2022}, with initial conditions given by
\begin{align}
    \label{eq:kelvin-helmholtz-ic}
    \varepsilon&=1\,,\\ \nonumber
    n &= 1+\frac{1}{2}\left[\tanh\left(\frac{y-y_1}{a}\right)-\tanh\left(\frac{y-y_2}{a}\right)\right]\,,\\ \nonumber
    u^x&=u_{\rm{flow}}\left[\tanh\left(\frac{y-y_1}{a}\right)-\tanh\left(\frac{y-y_2}{a}\right)-1\right]\,,\\ \nonumber
    u^y&=A\sin(2\pi x)\left[\exp\left(-\left[\frac{y-y_1}{\sigma}\right]^2\right)+\exp\left(-\left[\frac{y-y_2}{\sigma}\right]^2\right)\right]\,,
\end{align}
where $A=0.01$, $a=0.05$, $\sigma=0.2$, $y_1=-y_2=-0.5$ and $u_{\rm{flow}}=c_s/4$. We also employ a rectangular domain with extents $x\in[-L,L]$, $y\in[-2L,2L]$ and identical cell widths along the two directions, $\delta x=\delta y$.
The configuration consists of a dense fluid layer moving relative to its surroundings, with small perturbations introduced at the interfaces to trigger the Kelvin--Helmholtz instability.

\begin{figure}
     \centering
     \begin{subfigure}[b]{0.49\textwidth}
         \centering
         \includegraphics[width=\textwidth]{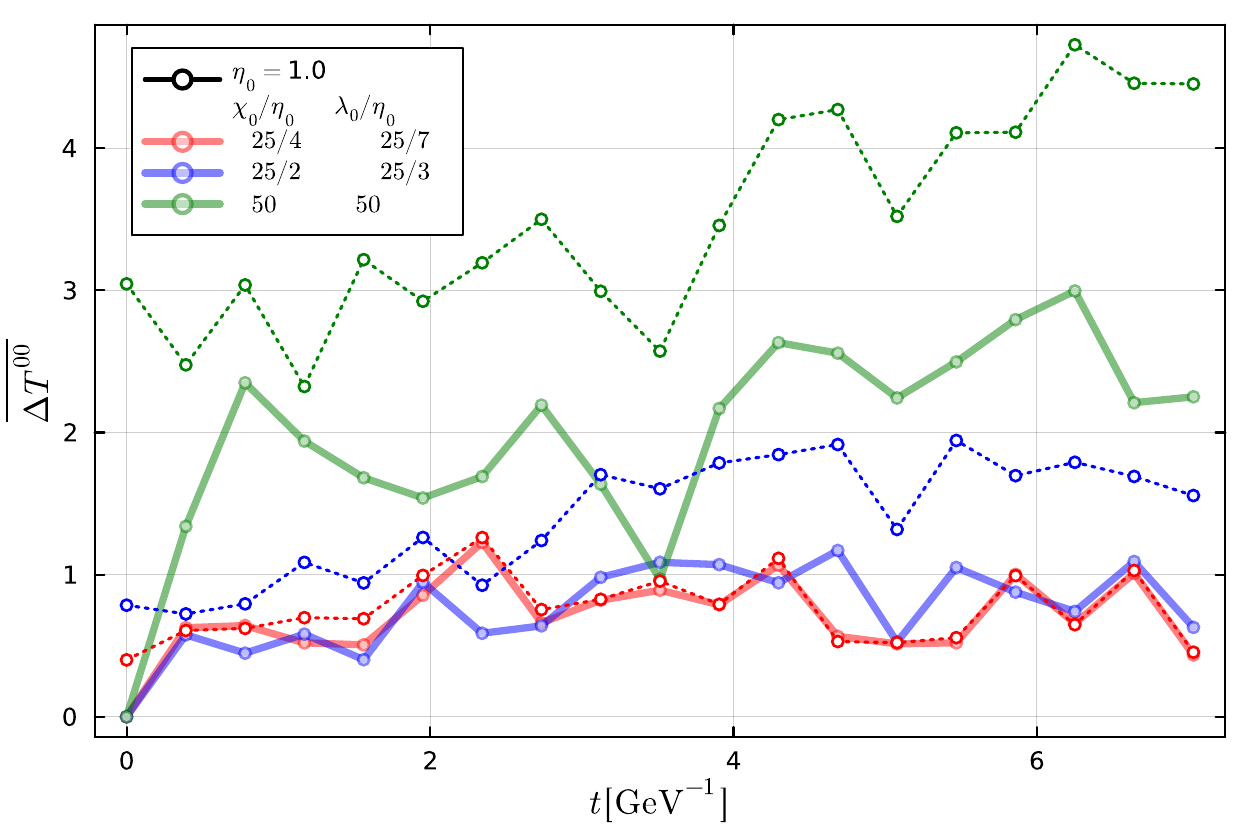}
     \end{subfigure}
     \hfill
     \begin{subfigure}[b]{0.49\textwidth}
         \centering
         \includegraphics[width=\textwidth]{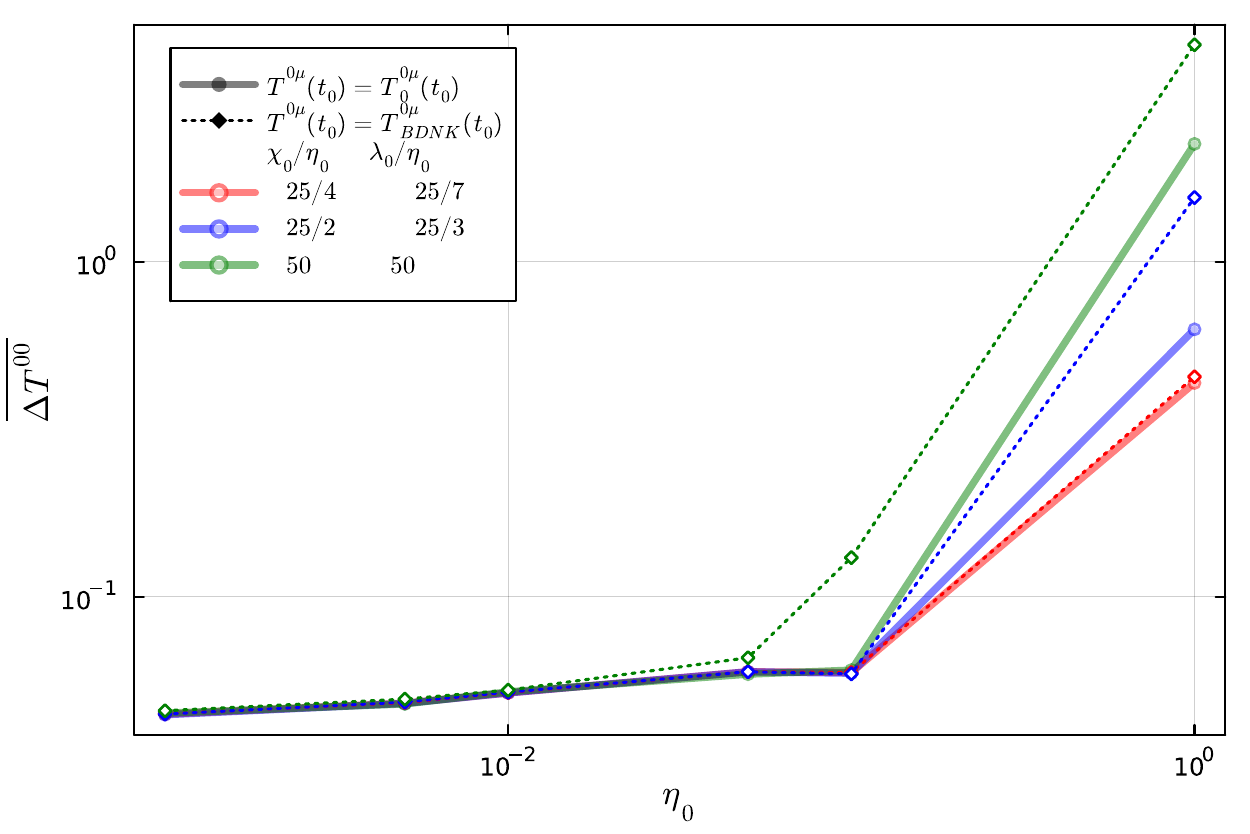}
     \end{subfigure}
     \hfill
        \caption{
        Kelvin--Helmholtz: Mean absolute differences, $\overline{\Delta T^{00}}$ between the viscid BDNK variable $T^{00}$ and the corresponding ideal inviscid counterpart. 
        In the left panel, $\overline{\Delta T^{00}}$ is displayed as a function of time for a selected $\eta_0$ value.
        In the right, the value of $\overline{\Delta T^{00}}$ at the end of the simulations, $t_f=7\,\GeV^{-1}$, is shown for a range of $\eta_0$.
        The colors represent different frame parameters $(\chi_0,\lambda_0)$, as indicated in the legend, while the solid (dashed) lines correspond to initialization Method-1 (Section~\ref{IC-method-1}) and Method-2 (Section~\ref{IC-method-2}), respectively.
        }
        \label{fig:kelvin-helmholtz-bdnk-ideal-init-diff-compare}
\end{figure}

Ideal hydrodynamic simulations of such configurations are known to exhibit non-convergent behavior as the spatial resolution is increased.
In contrast, viscous dissipation in BDNK hydrodynamics suppresses the formation of vortices on arbitrarily small spatial scales.
We present snapshots of the evolution in %
Figure~\ref{fig:Kelvin Helmholtz snapshots},
comparing the density $n$ for several values of the shear viscosity parameter $\eta_0$ with the corresponding ideal hydrodynamic solution.
For the smallest value tested, $\eta_0=10^{-5}$, the density profile is visually indistinguishable from the ideal solution, consistent with the expected recovery of ideal hydrodynamics in the limit of vanishing viscosity.
While this qualitatively agrees with the results in \cite{Pandya_2022}, we note that we strictly employ our \ac{FO} and \ac{MO} BDNK PDE systems, without the adaptive switching between ideal and BDNK primitive recovery, at a resolution of $N_x=2^8$. 
We evolve the system using both the \ac{FO} and \ac{MO} formulations, employing both initialization procedures, Method-1~(\ref{IC-method-1}) and Method-2~\ref{IC-method-2}, with periodic boundary conditions. 

As in the smooth shock test of Section~\ref{S:Smooth_shock}, we quantify the effects of viscosity by comparing the BDNK evolution with the corresponding ideal hydrodynamic reference solution. 
We examine the normalized mean absolute difference $\overline{\Delta T^{00}}$, defined in Equation~\eqref{def:normalized-mean-absolute-difference}, for a range of shear viscosity parameters $\eta_0$, frame parameters $(\chi_0,\lambda_0)$, and both initialization procedures.
The results are presented in Figure~\ref{fig:kelvin-helmholtz-bdnk-ideal-init-diff-compare}. 
The left panel shows $\overline{\Delta T^{00}}$ as a function of time for the largest shear viscosity parameter considered, $\eta_0=1.0$, while the right panel shows its dependence on $\eta_0$ at the final simulation time. 
Both panels compare the same three choices of frame parameters $(\chi_0,\lambda_0)$.
We choose the largest $\eta_0$ for the time-evolution (left) panel to make the differences between the frame choices and initialization procedures more apparent.

Unlike the smooth-shock configuration discussed in Section~\ref{S:Smooth_shock}, 
the Kelvin--Helmholtz system exhibits noticeable frame dependence at the largest values of $\eta_0$ considered, even with initialization Method-1~(Section~\ref{IC-method-1}), which ensures that the initial conservative variables coincide with their ideal counterparts, $T^{0\mu}(t_0)=T^{0\mu}_0(t_0)$ and have no viscous contributions by construction.
However, the condition from this method does not require the remaining components of the BDNK energy-momentum tensor, $T^{ij}$, to equal their ideal counterparts, and these can have nonvanishing, frame-dependent viscous corrections.
In the smooth-shock setup, the fluid is initially at rest, whereas the Kelvin--Helmholtz configuration has a non-vanishing and a non-uniform initial velocity profile in Equation~\eqref{eq:kelvin-helmholtz-ic}, which can produce nonzero viscous corrections to $T^{ij}$ even when the initial conserved quantities agree with their ideal values. 
These additional contributions affect the subsequent evolution through the conservation equations (see, e.g., Equation~\eqref{Relations-Tij-FO}).
Thus, the observed frame dependence is sensitive not only to the initialization procedure but also to the physical configuration being considered, consistent with the discussion in Section~\ref{Sec:Initial Conditions}.
Nevertheless, the observed frame dependence 
decreases rapidly for smaller $\eta_0$ and becomes negligible at sufficiently small $\eta_0$, consistent with the expected recovery of ideal hydrodynamics in the vanishing-viscosity limit.

\subsection{Milne coordinates}
\label{S:Milne_coordinates}

\begin{table}
\centering
\renewcommand{\arraystretch}{1.3}
\setlength{\extrarowheight}{2pt}
\begin{tabular}{||c|c||c|c||c|c||}
\hline
\multirow{3}{*}{$\delta t$} &
\multirow{3}{*}{$\delta x$}
& \multicolumn{2}{c||}{{First Order}}
& \multicolumn{2}{c||}{{Mixed Order}} \\
\cline{3-6}
&
& \multicolumn{2}{c||}{WENO-Z}
& \multicolumn{2}{c||}{WENO-Z}
\\
\cline{3-6}
&
&
$\Delta_{\rm rel}(T)$ &
$\Delta_{\rm rel}^{\max}(T)$ &
$\Delta_{\rm rel}(T)$ &
$\Delta_{\rm rel}^{\max}(T)$ 
\\
\hline\hline

$9.4\times10^{-3}$
&
\multirow{2}{*}{
0.1
}
& $3.30\times10^{-6}$%
&
$3.30\times10^{-6}$%
&
$3.30\times10^{-6}$
& $3.30\times10^{-6}$
\\
\cline{1-1}\cline{3-6}
$9.4\times10^{-4}$
&
&
$2.88\times10^{-8}$%
&
$2.88\times10^{-8}$
&
$2.88\times10^{-8}$
&
$2.88\times10^{-8}$
\\
\hline

\multirow{2}{*}{
$0.01$
}
&
$0.2$
& $3.57\times10^{-6}$%
& $3.57\times10^{-6}$%
& $3.57\times10^{-6}$   %
& $6.94\times10^{-6}$   %
\\
\cline{2-6}
&
$0.02$
&
$3.57\times10^{-6}$
&
$3.57\times10^{-6}$%
&
$3.57\times10^{-6}$   %
& $6.94\times10^{-6}$   %
\\
\hline

\end{tabular}
\caption{Absolute relative errors at the final time $t_f$ for temporal and spatial refinement studies of the Bjorken flow problem. 
}
\label{tab:Bjorken_convergence}
\end{table}

One of the most widely studied applications of relativistic viscous hydrodynamics is for \ac{QGP} produced in heavy-ion collisions.
The evolution of such systems is often approximated as being boost-invariant along the collision (``longitudinal'') axis, with spatial variations in the transverse plane.
We adopt Milne coordinates, which are commonly used in heavy-ion literature and well suited to capture the dynamics of interest under exact or approximate longitudinal boost invariance. 
These coordinates are defined as $x^{\mu}=(\tau,x,y,\eta_s)$, where $\tau=\sqrt{t^2-z^2}$ is the longitudinal proper time, $\eta_s=\mathrm{arctanh}(z/t)$ is the spacetime rapidity, and $z$ is the Cartesian coordinate along the collision axis. 
The $x$ and $y$ coordinates form the transverse plane, and the line element is given by $ds^2 = -d\tau^2 + dx^2 + dy^2 + \tau^2 d\eta_s^2$.

In heavy-ion collisions, a \ac{QGP} is formed at some initial time $\tau_0$ and its rapid cooldown is initially driven by expansion in the longitudinal direction, and subsequently by additional expansion in the transverse plane. 
The cooling of the plasma due to longitudinal expansion is well described by ``Bjorken'' flow~\cite{Bjorken:1982qr}.
Another exact solution for IS-type theories is the ``Gubser'' solution~\cite{Gubser:2010ze,Gubser:2010ui}, first obtained in \cite{Marrochio:2013wla}, which exhibits radial symmetry in the transverse plane and accounts for expansion in both longitudinal and transverse directions.

We study these two systems with their respective symmetries, a conformal equation of state~(Section~\ref{SSS:conformal}), and known semi-analytical BDNK solutions \cite{Bemfica:2017wps} as convergence tests of our numerical simulations. 

\subsubsection{Bjorken flow}\label{Subsubsec:Bjorken}

\begin{figure}
     \centering
     \begin{subfigure}[b]{0.49\textwidth}
         \centering
         \includegraphics[width=\textwidth]{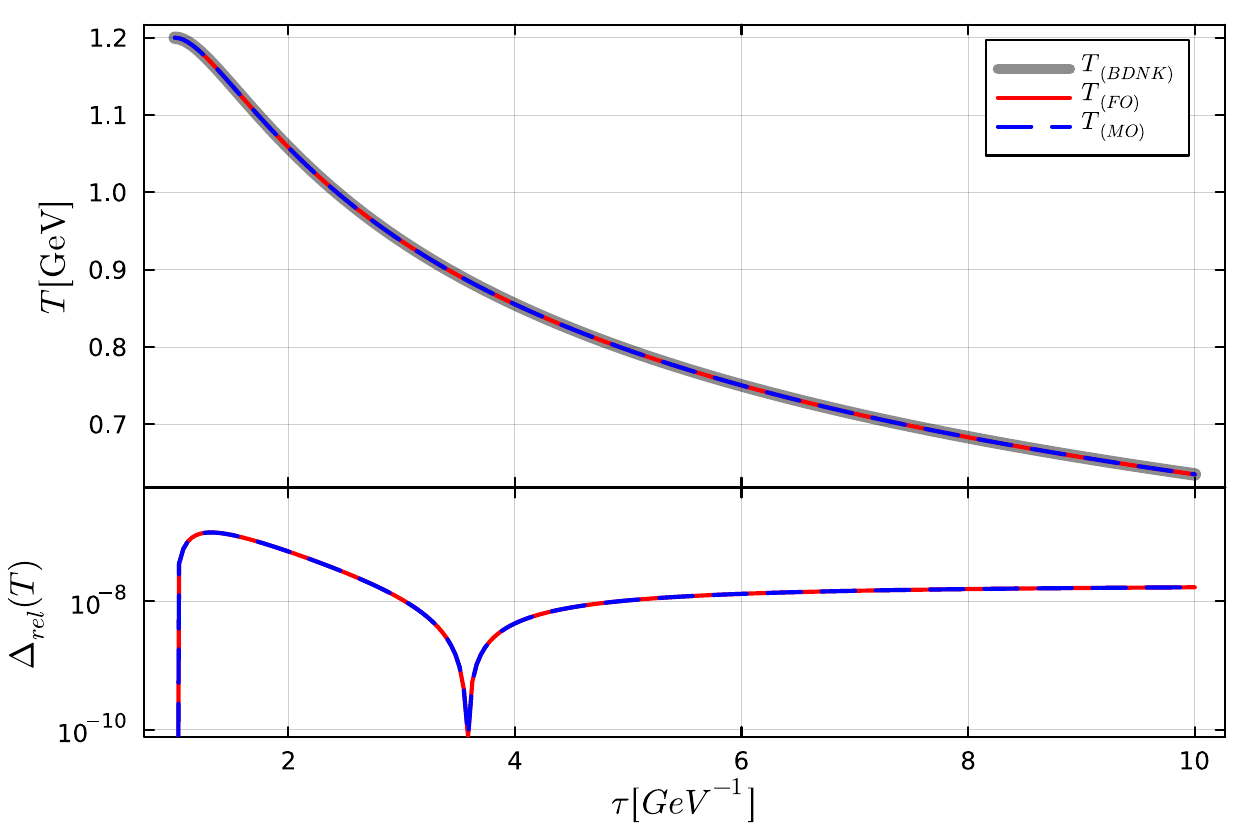}
     \end{subfigure}
     \hfill
     \begin{subfigure}[b]{0.49\textwidth}
         \centering
         \includegraphics[width=\textwidth]{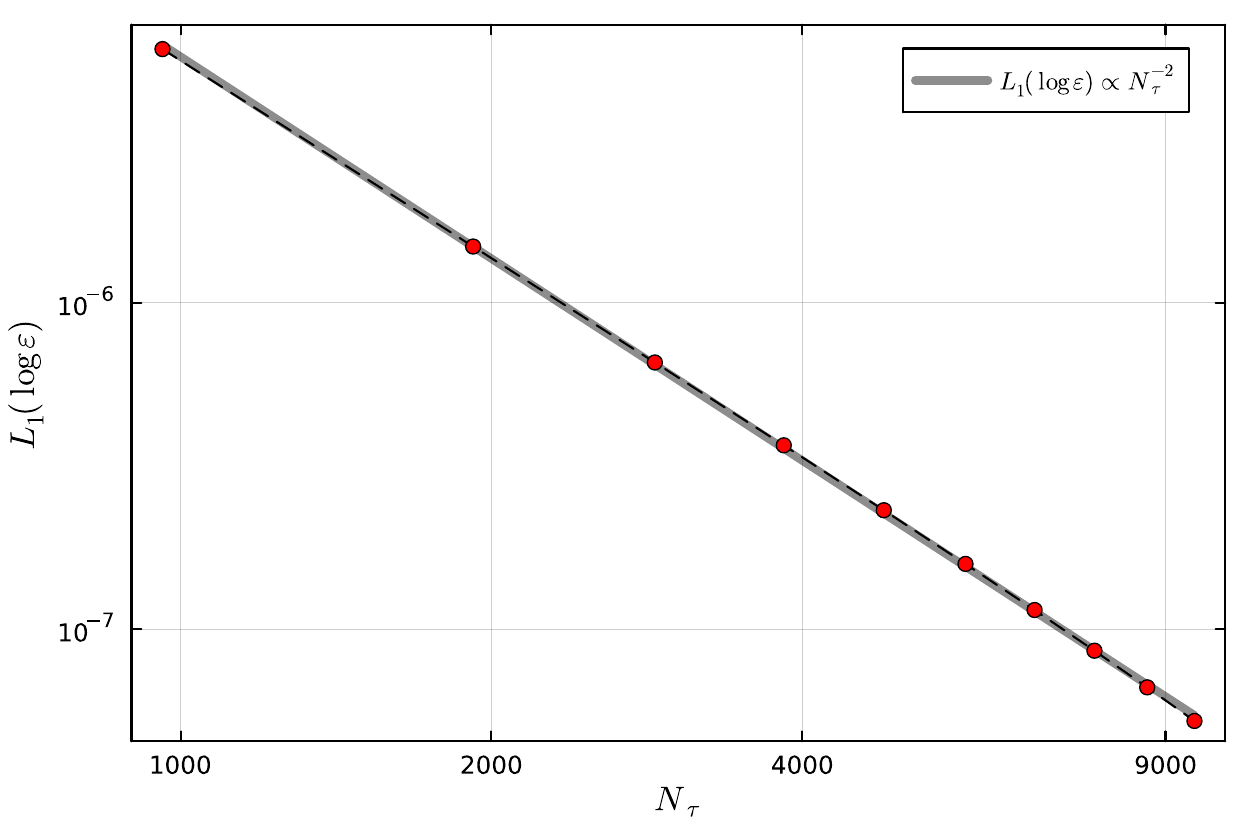}
     \end{subfigure}
     \hfill
        \caption{
        Results for the Bjorken flow test in Milne coordinates. The system is initialized at $\tau_0 = 1\,\GeV^{-1}$ and evolved up to $\tau_f=10\,\GeV^{-1}$ on a homogeneous spatial grid with fixed spatial extent and resolution.
        In the left panel, the temperature evolution from the simulations, with $N_\tau = 9600$ timesteps ($\delta \tau = (\tau_f-\tau_0)/N_\tau$), is shown for the \ac{FO} (solid red) and \ac{MO} (dashed blue) systems, and the semi-analytical solution is displayed as the grey curve.
        The curves in the bottom panel represent the relative absolute difference between simulation and semi-analytical temperatures.
        In the right panel, the $L_1$ norm of the absolute error in $\log\varepsilon$ for simulations with time resolution $N_\tau$ is shown as red circles connected by dashed lines.
        The grey line corresponds to a power law fit, and the fitted exponent is denoted in the legend.
        The results here were obtained for 
        $\eta_0=0.1$, $\chi_0/\eta_0=25/4$ and $\lambda_0/\eta_0=25/7$.
        }
        \label{fig:Bjorken-ev-convergence}
\end{figure}

We first consider the Bjorken flow configuration, which corresponds to a fluid that is homogeneous in the $xy$ plane and rapidly expanding in the longitudinal $z$ direction with a velocity $v^z=z/t$. 
In Milne coordinates, this reduces to a homogeneous fluid with $u^\mu=(1,0,0,0)$ and a single unknown variable $\varepsilon(\tau)$ that depends only on the longitudinal proper time $\tau$.
This configuration and its semi-analytical BDNK solution were considered in detail in \cite{Bemfica:2017wps}.
While the Milne metric corresponds to flat spacetime, the nontrivial component $g_{\eta\eta}=\tau^2$ captures the geometric expansion of the QGP in the longitudinal direction, providing a test case for our implementation of BDNK expressions with arbitrary $g^{\mu\nu}$.

Following \cite{Bemfica:2017wps}, we adopt the conformal EOS, transport coefficients, and frame parameters stated in Section~\ref{SSS:conformal}.
The temperature evolution $T(\tau)$ is obtained from the energy-momentum conservation $u_\nu\nabla_\mu T^{\mu\nu}=0$:
\begin{equation}\label{eq:bjorken-bdnk-ode}
    \tau{\chi_0}\frac{\partial^2_\tau T}{T}+2\tau{\chi_0}\left(\frac{\partial_\tau T}{T}\right)^2 
    +\frac{7}{3}{\chi_0}\frac{\partial_\tau T}{T}
    +\frac{{\chi_0}}{9\tau}
    +\frac{4}{3}\tau \partial_\tau T 
    - \frac{4}{9}\frac{{\eta_0}}{\tau}
    +\frac{4}{9}{T}=0\,.
\end{equation}
Since spatial dynamics are absent for the homogeneous plasma, this reduces to a purely temporal evolution problem. 
We verify that our numerical implementation of both the \ac{FO} and \ac{MO} produces consistent results and maintains spatial homogeneity throughout the evolution.

We set up a cubic computational domain with spatial extents $x,y\in[-1,1]\,\mathrm{GeV}^{-1}$  and $\eta_s\in[-1,1]$, and uniform resolution $N_x=N_y=N_{\eta_s}$ cells in each direction.
The system is initialized at $\tau_0 = 1\,\mathrm{GeV}^{-1}$ with uniform plasma temperature $T_0=1.2\,\mathrm{GeV}$ and zero initial time derivative $\partial_\tau T|_{\tau_0}=0$; all spatial gradients vanish due to homogeneity.
For comparison with the semi-analytical BDNK solution of Equation~\eqref{eq:bjorken-bdnk-ode}, we employ the initialization procedure in Section~\ref{IC-method-2}, setting the initial values of the variables $T^{0\mu}(\vec{x},\tau_0)$ to those given by the BDNK tensor.
We evaluate the evolution up to $\tau_f=10\,\mathrm{GeV}^{-1}$ with transport coefficients, $\eta_0=0.1$ and $\zeta=0$, and frame parameters, $\chi_0=(25/4)\eta_0$ and $\lambda_0=(25/7)\eta_0$.

We perform a spatial convergence test by varying the spatial resolution at fixed time step $\delta \tau = (1/100)\,\mathrm{GeV}^{-1}$. 
The results can be found in Table~\ref{tab:Bjorken_convergence}.
As can be inferred from the results therein, the errors evaluated at the final time $\tau_f$ remain uniformly small across all numerical approaches considered. 
The typical relative error, computed from the $L_1$ norm of the absolute error, is of order $10^{-4}\%$, regardless of whether the FO or MO formalism is used.
These results demonstrate that the numerical implementation robustly preserves the expected spatial homogeneity across all methods considered.

The homogeneous configuration enables us to test our second-order RK time integration scheme.
Figure~\ref{fig:Bjorken-ev-convergence} displays the convergence behavior for both \ac{FO} and \ac{MO} formulations for a fixed number of cells, $N_x=10$. 
The left panel shows the temperature evolution at the domain center for both schemes alongside the semi-analytical solution from Equation~\eqref{eq:bjorken-bdnk-ode}, using $N_\tau = 9600$ time steps ($\delta \tau = (\tau_f-\tau_0)/N_\tau\simeq9.4\times10^{-4}\,\GeV^{-1}$).
The detailed error results are presented in Table~\ref{tab:Bjorken_convergence}. 
The relative error in the temperature at the lattice center remains $\lesssim10^{-5}\%$ up to the final time $\tau_f$.
The right panel shows the $L_1$ norm of the absolute error in $\log\varepsilon$ at final time $\tau_f$ as a function of the number of timesteps $N_\tau$.
The power-law fit confirms the expected second-order convergence of the RK2 time integrator.

We have verified these convergence rates for both minmod and WENO-Z reconstruction schemes, across \ac{FO} and \ac{MO} formulations, and for a range of BDNK parameters $\eta_0$, $\chi_0$, and $\lambda_0$ (results not shown).
The Bjorken flow tests confirm that implementations of both \ac{FO} and \ac{MO} formulations maintain homogeneity and are fully consistent with each other.

\subsubsection{Gubser flow}\label{Subsubsec:Gubser}

A complementary system to Bjorken flow is Gubser flow, which is boost-invariant along the collision axis but exhibits a radial temperature and velocity profile in the transverse plane.
Gubser flow solutions have been found for the conformal equation of state in both relativistic Navier Stokes \cite{gubser2010symmetry} and IS theories \cite{Marrochio:2013wla}, with semi-analytical BDNK solutions discussed in \cite{Bemfica:2017wps}.
The Gubser flow symmetry pattern requires a purely radial 4-velocity \cite{gubser2010symmetry}, given by
\begin{align}
\label{Gubser-radial-U}
u_r = \sinh\left[\tanh^{-1}\left(\frac{2\tau rq^2}{1+q^2\tau^2+q^2r^2}\right)\right]\, ,
\end{align}
where $r=\sqrt{x^2+y^2}$ is the radial coordinate and $q$ is a free parameter describing the spatial extent in the transverse plane.\footnote{We note a typo in section VII.B of~\cite{Bemfica:2017wps}: the expressions for $u_\tau$ and $u_r$ have incorrect factors of $q$ in the numerators, and the corrected expressions with factors $q^2$ are stated in Equation~\eqref{Gubser-radial-U}.}
We set $q=1\,\mathrm{GeV}$ without loss of generality.
With the velocity determined by symmetry, the coupled \ac{FO} differential equations for temperature in BDNK \cite{Bemfica:2017wps}, using the conformal EOS and parameters from Section~\ref{SSS:conformal}, are
\begin{align}
    \label{Gubser-BDNK-EOMs}
    \frac{1}{\hat{T}}\frac{d\hat{T}}{d\rho}+
    \frac{2}{3}\tanh \rho = 
    \hat{\mathcal{F}}(\rho)\,, \\ \nonumber
    {\chi_0}\frac{d\hat{\mathcal{F}}}{d\rho}
    +3\chi_0\hat{\mathcal{F}}^2
    +\frac{2}{3}{\chi_0}\hat{\mathcal{F}}\tanh\rho+
    \frac{4}{3}\hat{T}\hat{\mathcal{F}}-
    \frac{4}{9}{\eta_0}(\tanh\rho)^2=0\,,
\end{align}
where
\begin{align*}
    \sinh\rho=-\frac{1-\tau^2+r^2}{2\tau}\,,\quad\text{and} \quad
    \hat{T}(\rho(\tau,r))=\tau T(\tau,r)\,.
\end{align*}

\begin{figure}
     \centering
     \begin{subfigure}[b]{0.49\textwidth}
         \centering
         \includegraphics[width=\textwidth]{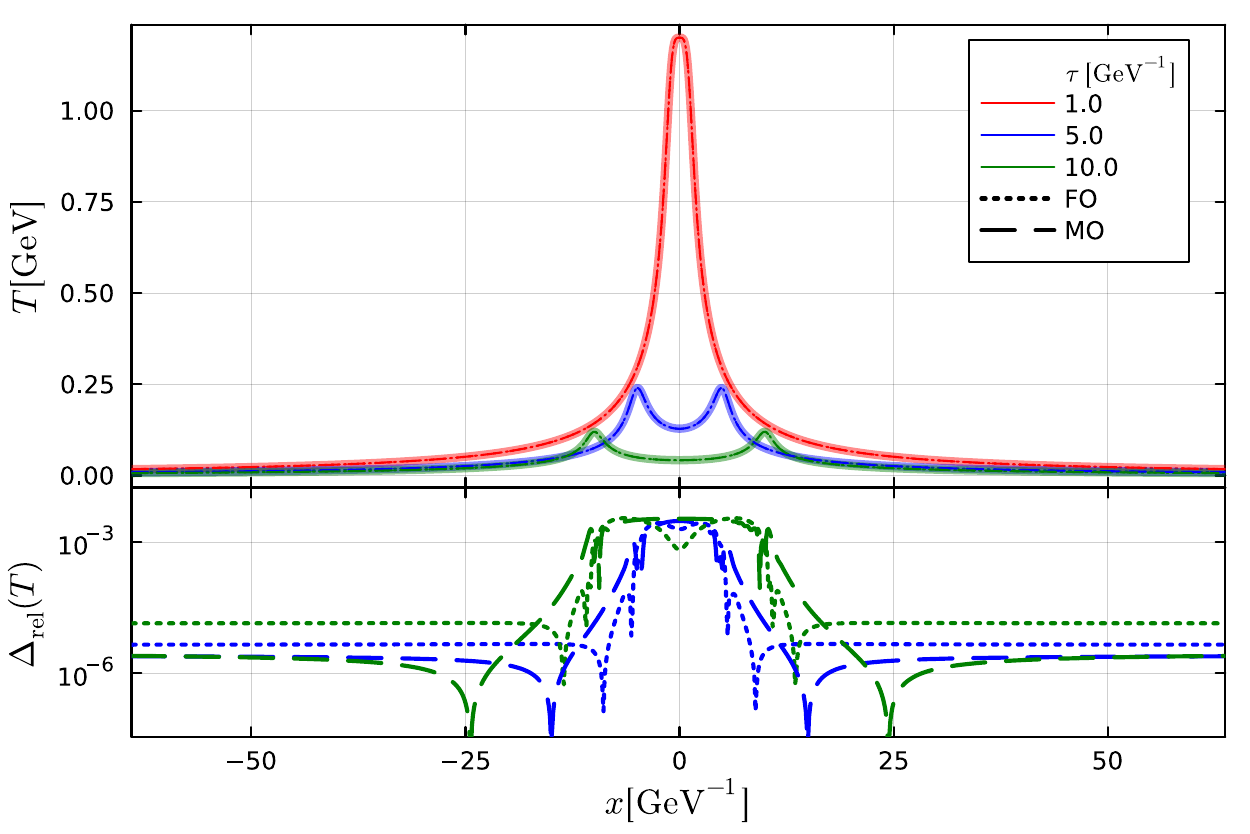}
     \end{subfigure}
     \hfill
     \begin{subfigure}[b]{0.49\textwidth}
         \centering
         \includegraphics[width=\textwidth]{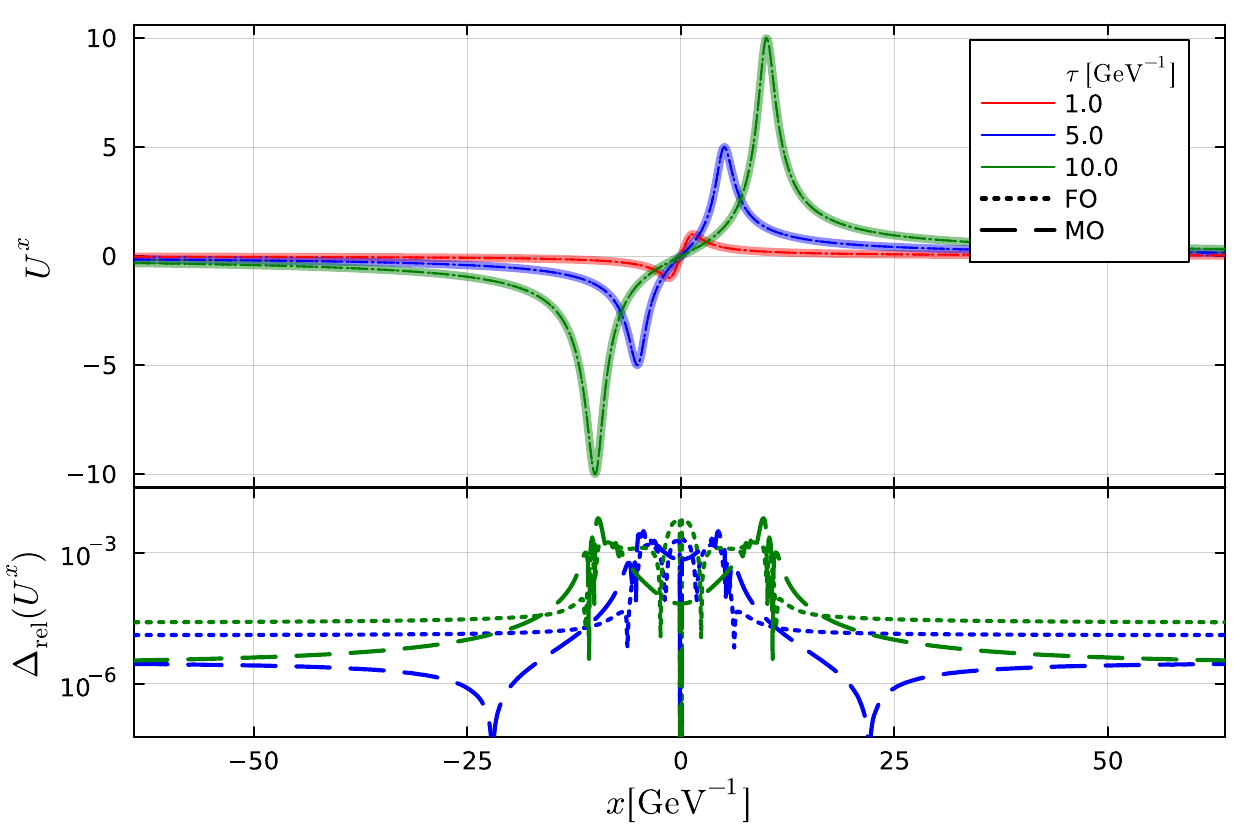}
     \end{subfigure}
     \hfill
        \caption{The curves in the top panels represent slices of the temperature (left) and 4-velocity component $u^x$ (right) along the x-axis for Gubser flow.
        The curves in the bottom panels represent the absolute relative error between simulation and semi-analytical values for the temperature (left) and 4-velocity component $u^x$ (right).
        The colors correspond to Milne coordinate times $\tau=1,5,\,\rm{ and }\,10\,\GeV^{-1}$ as indicated in the legend.
        The thick lines represent the semi-analytical solutions described in Section~\ref{Subsubsec:Gubser} and the dashed lines correspond to the simulation results.
        The results here were obtained for $\delta x=(1/8)\,\GeV^{-1}$, $\delta \tau = 0.01\,\GeV^{-1}$, $\eta_0=0.1$, $\chi_0/\eta_0=25/4$, $\lambda_0/\eta_0=25/7$.}
        \label{fig:gubser_slices_w_err}
\end{figure}

\begin{figure}
     \centering
     \includegraphics[width=0.5\textwidth]{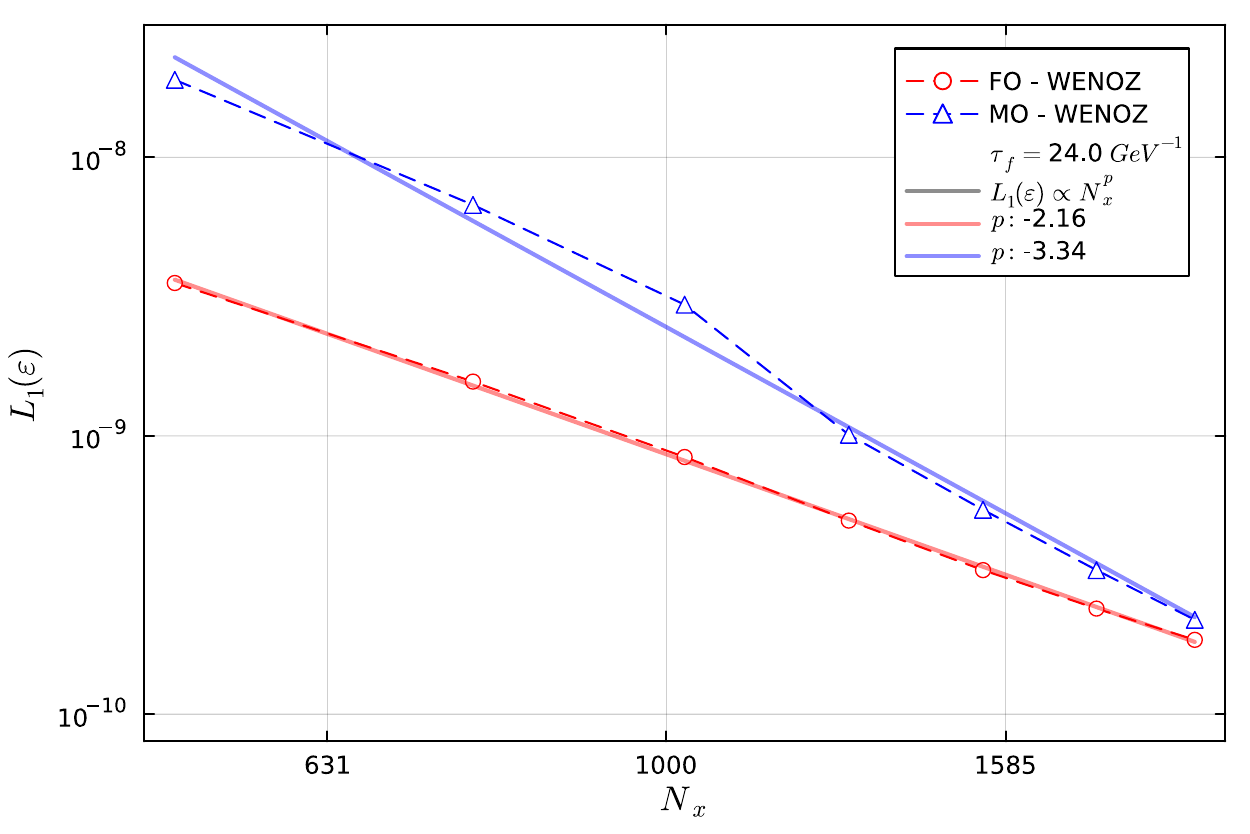}
     \centering
     
    \caption{
    $L_1$ norm of the absolute error for $\varepsilon$ 
    with respect to the number of cells $N_x = L_x/\delta x$ for Gubser flow with the \ac{FO} (red) and the \ac{MO} (blue) formalisms.
    The markers correspond to the simulation results and the connected dashed lines are for visual guidance.
    The thick solid lines correspond to a power law fit, with the corresponding exponents displayed in the legends.
    The simulation parameters used here are $\eta_0=0.1$, $\chi_0/\eta_0=25/4$, $\lambda_0/\eta_0=25/7$.
    The initial and final times used here are $\tau_0=1\,\GeV^{-1}$ and $\tau_f=24\,\GeV^{-1}$, respectively, with the time interval fixed at $\delta \tau = 0.01\,\GeV^{-1}$ for all $N_x$. 
    }
    \label{fig:gubser_err_NX_scaling}
\end{figure}

The semi-analytical solution is obtained by solving the coupled ODE system in Equation~\eqref{Gubser-BDNK-EOMs} with initial conditions $\hat{T}(\rho(\tau_0,r_0))=1.2$ and $\hat{\mathcal{F}}(\rho(\tau_0,r_0))=0$ at initial time $\tau_0=1,\mathrm{GeV}^{-1}$ and origin $r_0=0$~\cite{Bemfica:2017wps}.
We construct the initial conditions for our numerical simulations by transforming the semi-analytical solution from Gubser to Milne coordinates, and subsequently use the same solution as a reference for validating the numerical evolution.
Throughout this section, we use transport coefficients $\eta_0=0.1$ and $\zeta=0$.

Gubser flow features a sharp central peak in the $xy$-plane and a non-vanishing initial velocity profile that rapidly becomes ultra-relativistic, providing a strong test of our discretized EOMs, finite volume methods, and boundary handling.
We initialize at $\tau_0$ with velocity from Equation~\eqref{Gubser-radial-U} and temperature from Equation~\eqref{Gubser-BDNK-EOMs} solved numerically on the 2D $xy$-plane with $\delta x=(1/8)\,\mathrm{GeV}^{-1}$ and $\delta \tau = (1/100)\,\mathrm{GeV}^{-1}$.
While cylindrical coordinates would reduce this to a 1+1D problem, we intentionally use Cartesian coordinates and set up a 2+1D problem to test whether our implementation preserves cylindrical symmetry under evolution.
Furthermore, the non-trivial pressure gradients in the $x$ and $y$ directions test our curl-cleaning procedure in the \ac{FO} formalism and numerical derivative schemes implemented for the \ac{MO} formalism.

An important numerical consideration arises from the nonvanishing spatial gradients of the Gubser solution near the computational boundaries. We found the \ac{MO} formulation, where spatial derivatives are evaluated numerically, to be numerically unstable.
To address this issue, we smoothly transition the initial Gubser profiles to a spatially homogeneous atmosphere near the boundaries, allowing periodic boundary conditions to be imposed (see Appendix~\ref{appendix:gubser-boundary-handling} for details). 
This modification is applied sufficiently far from the central region of interest to minimize its influence on the interior evolution over the simulated time interval.
We note that the \ac{FO} formulation does not require this smoothing procedure and is evolved using the unmodified Gubser flow initial conditions.

Figure~\ref{fig:gubser_slices_w_err} compares snapshots of temperature and the $x$-component of four-velocity alongside the semi-analytical solution.
The \ac{MO} and \ac{FO} solutions (respectively wide-dashed and narrow-dashed lines) are essentially indistinguishable from each other or from the semi-analytical solution (transparent thick solid line) for the temperature and the flow velocity (top panels). 
Indeed, as shown in the lower panels, the absolute relative errors of \ac{MO} and \ac{FO} with the semi-analytical solution do not exceed $1\%$ even after a long evolution time.

We perform comprehensive convergence tests for both \ac{FO} and \ac{MO} formulations at various spatial resolutions $\delta x$, evolving from $\tau_0=1\,\mathrm{GeV}^{-1}$ to $\tau_f=24\,\GeV^{-1}$ with fixed time step $\delta \tau = (1/100)\,\mathrm{GeV}^{-1}$.
Figure~\ref{fig:gubser_err_NX_scaling} shows the $L_1$ norm of the absolute error in $\varepsilon$ as a function of the number of grid points $N_x$, comparing the two formulations. 
Both exhibit errors scaling as $~N_x^{-2}$, consistent with the expected accuracy of the second-order KT scheme.
Additional comparisons of the numerical solutions with the semi-analytical Gubser solution are presented in Appendix~\ref{app:more-gubser-plots}, including the spatial derivatives $\partial_x T$ and $\partial_x u^x$, analogous to the temperature and velocity profiles 
and the corresponding error scalings with spatial resolution 
, providing further tests of numerical convergence for both formulations.
Despite the similar convergence rates, the \ac{FO} formulation yields smaller errors than the \ac{MO} formulation at comparable resolutions. 

The availability of analytical expressions for the spatial derivatives in the Gubser solution also provides an additional test of the \ac{FO} formulation, in which these derivatives are evolved as independent dynamical variables. 
In Appendix~\ref{app:more-gubser-plots}, we examine the effectiveness and accuracy of the curl-cleaning formulation and its numerical implementation by comparing the evolved spatial derivatives against their analytical counterparts and examining the dependence of the numerical errors on the curl-cleaning parameters.
The results therein demonstrate that the derivative variables in the \ac{FO} formalism exhibit convergence similar to the hydrodynamic variables, with only weak dependence on the curl-cleaning parameters.
These tests thus simultaneously validate our numerical implementation, demonstrate the practical applicability of the mathematical \ac{FO} formulation developed in Section~\ref{Sec:First_Order}, and provide numerical evidence supporting its equivalence to the original second-order BDNK PDE system, as established analytically in the mathematical proof Section~\ref{FO:Equivalence}.

The \ac{FO} implementation exhibits greater numerical stability, allowing evolution up to $\tau_f=50\,\GeV^{-1}$, whereas the \ac{MO} implementation becomes unstable beyond approximately $\tau_f=24\,\GeV^{-1}$ with WENO-Z reconstruction.
The differences between the two formulations are particularly significant when using minmod reconstruction. 
While the \ac{FO} formulation can be evolved using either WENO-Z or minmod reconstruction, the \ac{MO} implementation with minmod becomes unstable at significantly earlier times, with the maximum evolution time limited to approximately $\tau_f=3\,\GeV^{-1}$. 
Moreover, the onset of instability occurs earlier as the spatial resolution is increased, preventing numerical convergence. 
Consequently, we restrict the \ac{MO} convergence results presented here to WENO-Z reconstruction.

The smoothing procedure introduced for the \ac{MO} formulation allows stable evolution of the \ac{MO} formulation over significantly longer timescales, but also introduces a limitation. 
Since the modified initial profiles are no longer exact solutions of the Gubser system in the transition region, the resulting deviations can eventually propagate into the central region of interest, affecting comparisons with the semi-analytical solution. 
The errors for the different formulations and reconstruction schemes are summarized in Table~\ref{tab:Gubser_comparison}, including results for the \ac{FO} formulation with smoothed initial conditions to provide a reference for the \ac{MO} results.

\section{Numerical tests in curved spacetime}
\label{S:Curved_spacetime_tests}
The two tests presented in this section are stationary solutions of the ideal relativistic hydrodynamic equations on a fixed black hole  background, and are widely used as benchmarks for relativistic hydrodynamics codes \cite{Rezzolla-Zanotti-Book-2013}.
In both cases, standard tests of ideal hydrodynamics numerical implementations consist of initializing the simulation with the analytic stationary solution and verifying that the code preserves stationarity over a sufficiently long evolution time~\cite{1984ApJ...277..296H,DelZanna:2007pk}.
Such tests are typically used to check the finite-volume discretization, the boundary handling and numerical dissipation of numerical implementations.

While viscous accretion onto black holes has been investigated using other relativistic dissipative hydrodynamic theories~\cite{Chabanov:2021dee,Shibata:2025ycs}, exact solutions for the corresponding BDNK systems considered here are not known.
We therefore initialize our simulations using stationary solutions of ideal relativistic hydrodynamics and investigate their departures from stationarity under small but finite viscous corrections.
Although arbitrarily small values of the transport coefficients~$(\eta_0\rightarrow0)$ are not numerically accessible in our implementation, we evolve the ideal initial data with the smallest numerically stable values.
Any residual, resolution-independent departure from the ideal solution in regions sufficiently away from the boundaries\footnote{In inviscid ideal hydrodynamics simulations, the systems are exact stationary solutions. Thus the dynamics near the boundaries would be trivial, and discretization and numerical noise dominate the error. The same system would no longer be stationary in BDNK and the interaction of the dynamics with the boundary would contribute to numerical artifacts at the boundary that lead to numerical instability in time.
}
can thus be interpreted as a genuine viscous effect.

Unless stated otherwise, both tests adopt the following conventions.
We use the ideal-gas EOS of Section~\ref{SSS:ideal-gas}, together with the corresponding viscous transport coefficients and BDNK frame parameters.
We choose  $\chi_0/\eta_0=\lambda_0/\eta_0=10^4\,, 10^5$ and $ 10^6$, which satisfy the global causality and stability bounds in Equations~\eqref{cond:caus-ideal-gas-global-xi0} and \eqref{cond:caus-ideal-gas-global-lambda0} throughout the evolution of both systems considered here.
We set the thermal conductivity and bulk viscosity to zero, $\sigma=0$ and $\zeta=0$, and vary the shear viscosity parameter $\eta_0$, with the smallest value considered, $\eta_0\sim10^{-5}$, being limited by numerical stability.
We test both the \ac{FO}~(Section~\ref{Sec:First_Order}) and \ac{MO}~(Section~\ref{Mixed_order_section}) formulations using WENO-Z reconstruction. Since we are interested in comparing the viscous evolution against the stationary ideal hydrodynamic solutions, we employ initialization Method-1~(Section~\ref{IC-method-1}), which sets the initial conservative variables equal to their ideal counterparts, $T^{0\mu}(t_0)=T^{0\mu}_0(t_0)$.
The initial spatial derivative variables required by the \ac{FO} formulation are computed from the analytic stationary profiles using the CWENO-based numerical differentiation procedure described in Section~\ref{Sec:Num-methods}.

\subsection{Michel accretion}\label{Sec:Michel-Accretion}

The Bondi accretion solutions~\cite{Bondi..1952MNRAS.112..195B} describe steady-state, spherically symmetric radial accretion onto a  point mass in the Newtonian limit of ideal hydrodynamics.
Their relativistic generalization to accretion onto a Schwarzschild black hole, the Michel solution~\cite{Michel...1972}, provides a simple stationary background on which to test our formulation and implementation.

\begin{figure}
     \centering
     \begin{subfigure}[b]{0.49\textwidth}
         \centering
         \includegraphics[width=\textwidth]{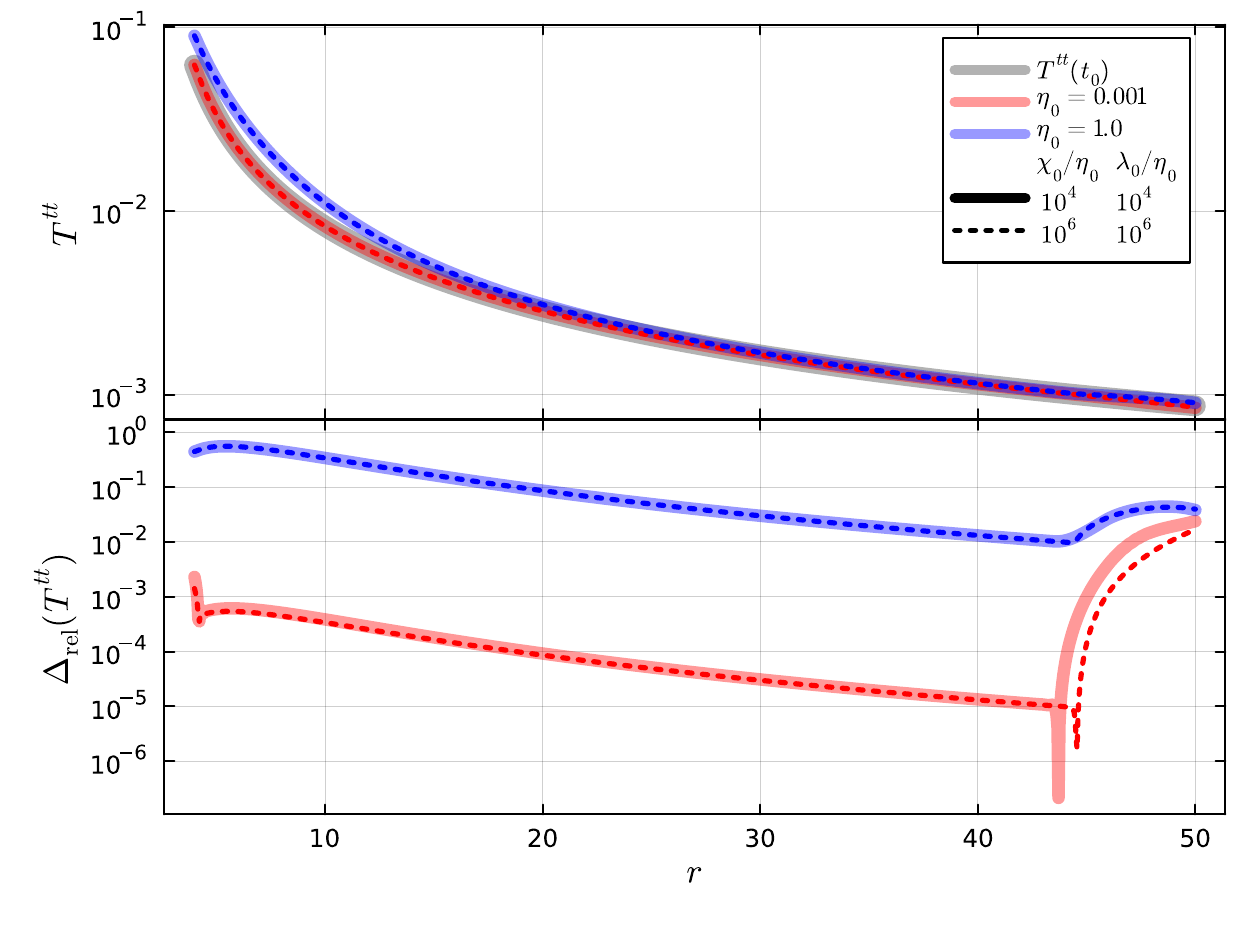}
     \end{subfigure}
     \hfill
     \begin{subfigure}[b]{0.49\textwidth}
         \centering
         \includegraphics[width=\textwidth]{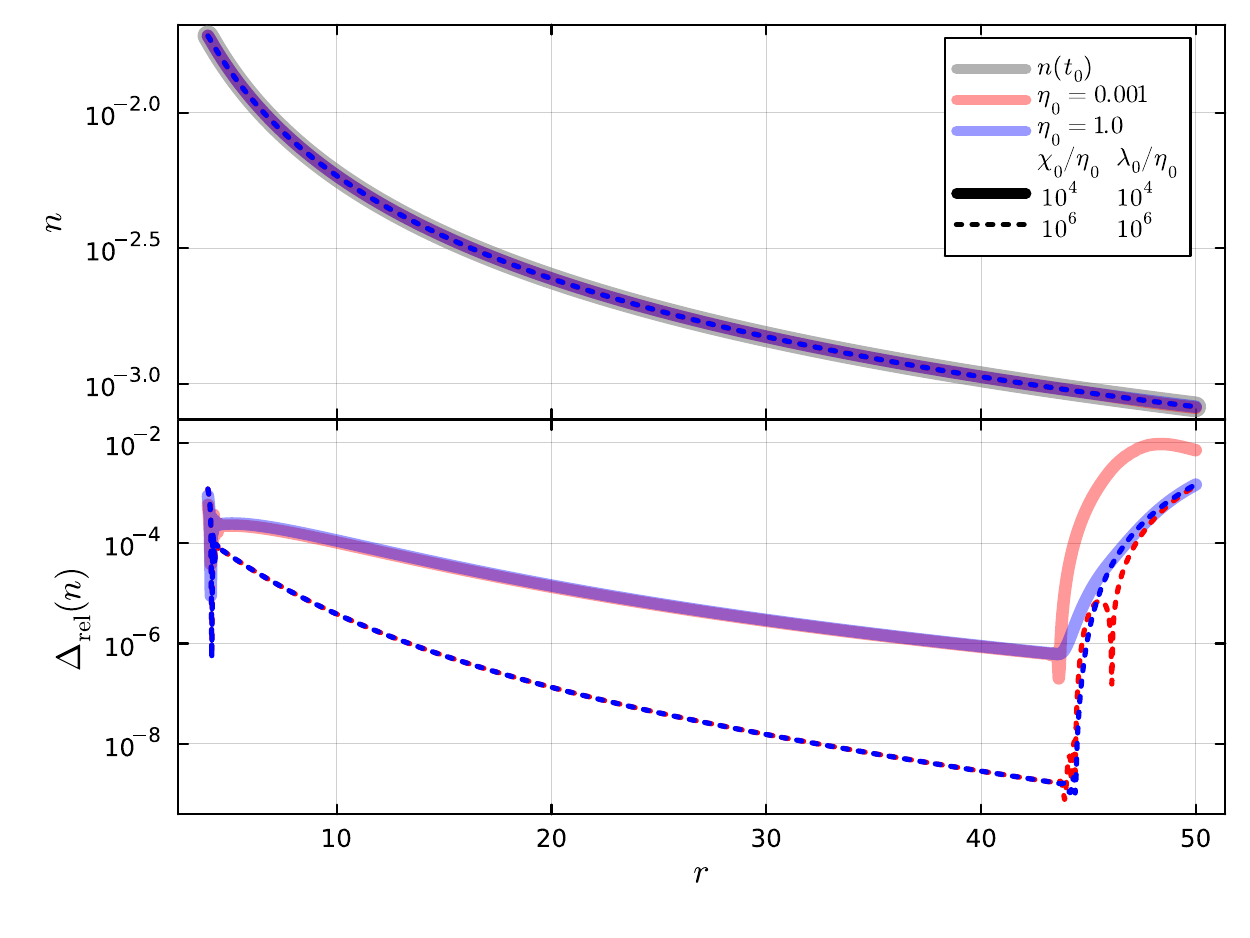}
     \end{subfigure}
     \hfill
        \caption{Comparison of the initial and final radial profiles of $T^{tt}$ (left) and the density $n$ (right) for radial accretion onto a Schwarzschild black hole. The upper panels show the corresponding profiles, while the lower panels show the absolute relative differences between the final and initial states, $\Delta_{\rm rel}(T^{tt})$ and $\Delta_{\rm rel}(n)$ for the left and right panels, respectively. The colors correspond to different values of the shear viscosity parameter $\eta_0$, while the line styles distinguish the choices of frame parameters $\chi_0/\eta_0=\lambda_0/\eta_0$, as indicated in the legend. The initial profiles, identical for all parameter choices, are shown as thick gray curves in the upper panels.}
        \label{fig:michel-acc-snapshots}
\end{figure}

We follow the setup of \cite{1984ApJ...277..296H} for an ideal gas and adopt the Schwarzschild metric in Schwarzschild coordinates,
\begin{equation}
  ds^2 = -f(r)\,dt^2 + f(r)^{-1} dr^2 + r^2 d\Omega^2\,,
  \qquad f(r) \equiv 1 - \frac{2M}{r}\,,
\end{equation}
where $M$ is the black hole mass.
For steady spherically symmetric flow, the four-velocity reduces to
$u^\mu=(u^t,u^r,0,0)$, with $u^t$ fixed by the normalization $u^\mu u_\mu=-1$.
Conservation of the particle current then gives
\begin{align}
    \label{eq:ideal-eos-current-conservation}
    T^{n_\Gamma} u^r r^2 = C_1\,,
\end{align}
while the $t$-component of the ideal energy-momentum conservation,
$\nabla_\mu T_0^{t\mu}=0$, yields the relativistic Bernoulli relation
\begin{align}
    \label{eq:ideal-eos-energy-convservation}
    \big[1+(1+n_\Gamma)T\big]^2\left[1-\frac{2M}{r}+(u^r)^2\right]=C_2\,,
\end{align}
where $C_1$ and $C_2$ are integration constants and $n_\Gamma=1/(\Gamma-1)$ is the polytropic index.
As in \cite{1984ApJ...277..296H}, the polytropic EOS $P=Kn^\Gamma$ is used alongside the ideal gas law with the polytropic constant set to unity, $K=1$.
The flow becomes transonic at the critical radius $r_c$, where
\begin{align}
    \label{eq:Michel-critical-values}
    u_c^2 = \frac{M}{2r_c}\,,\qquad c^2_{s,c}=\frac{u_c^2}{1-3u_c^2}\,,
\end{align}
with $u_c \equiv u^r(r_c)$ and $c_{s,c}$ the sound speed at the critical
radius~\cite{Michel...1972}.
Specifying $r_c$ therefore fixes $C_1$ and $C_2$, after which $T(r)$ can be computed by numerically solving for the roots of the non-linear Equation~\eqref{eq:ideal-eos-energy-convservation} and $u^r(r)$ from Equation~\eqref{eq:ideal-eos-current-conservation}.
\begin{figure}
     \centering
     \begin{subfigure}[b]{0.49\textwidth}
         \centering
         \includegraphics[width=\textwidth]{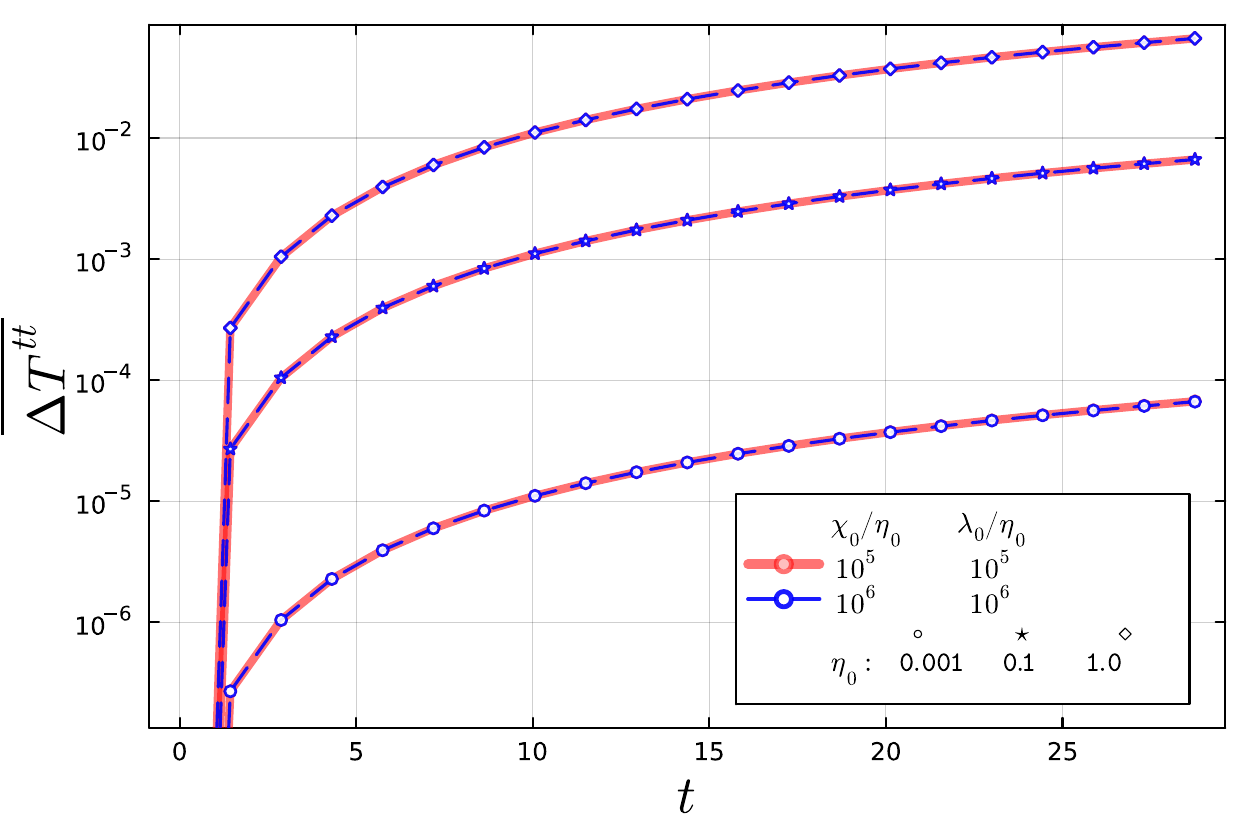}
     \end{subfigure}
     \hfill
     \begin{subfigure}[b]{0.49\textwidth}
         \centering
         \includegraphics[width=\textwidth]{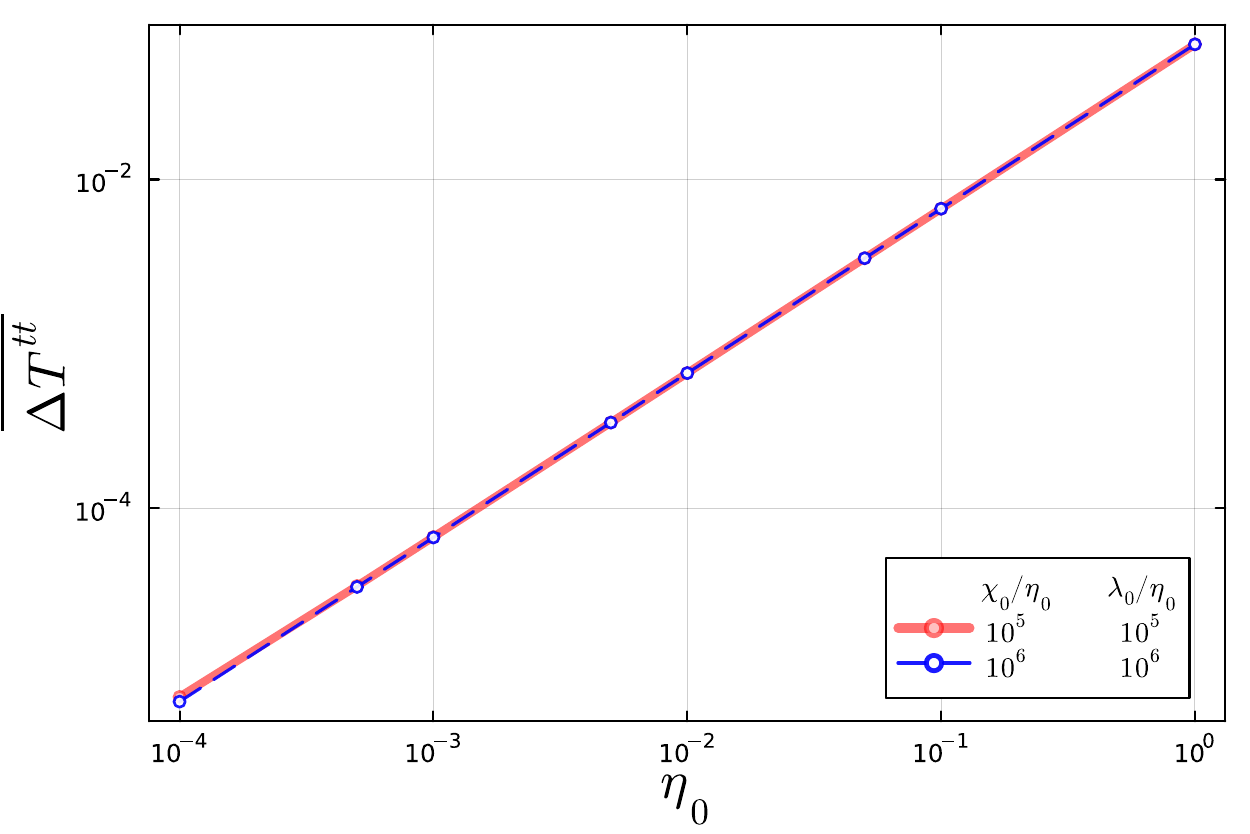}
     \end{subfigure}
        \caption{
        Michel accretion: mean absolute difference $\overline{\Delta T^{00}}$ between
        the viscous BDNK variable $T^{00}$ and its ideal, inviscid counterpart.
        \emph{Left}: $\overline{\Delta T^{00}}$ as a function of time for  different values of $\eta_0$, as indicated in the legend.
        \emph{Right}: $\overline{\Delta T^{00}}$ at the final time $t_f$ as a function of $\eta_0$.
        Colors denote different frame parameters $(\chi_0,\lambda_0)$ as indicated in the legend.
        }
        \label{fig:michel-acc-bdnk-ideal-init-diff-compare}
\end{figure}
We employ a uniform radial grid $2R_{\rm sch}<r<50$ with $N_r=1025$ cells ($\delta r=0.045$), where $R_{\rm sch}=2$, $r_c=8\,R_{\mathrm{sch}}$, and $\Gamma=5/3$. 
We evolve the system up to $t_f=28.75$ with a fixed time step $\delta t=\delta r/10$. For the boundary conditions, we linearly extrapolate the variables from the computational interior into the exterior ghost zones.

Figure~\ref{fig:michel-acc-snapshots} compares the initial and final radial profiles of $T^{tt}$ and the density $n$, obtained using the \ac{FO} formulation with initialization Method-1 (Section~\ref{IC-method-1}). 
For $T^{tt}$ (left panels), the departure from the initial stationary profile increases with the shear viscosity parameter $\eta_0$, while it is only weakly dependent on the frame parameters $(\chi_0,\lambda_0)$. 
In contrast, the density $n$ (right panels) exhibits a clear dependence on the frame parameters, while remaining relatively insensitive to variations in $\eta_0$. 
Nevertheless, the deviations in $n$ remain small, indicating that the density profile remains approximately stationary over the simulated time interval.

We note that the numerical evolution is limited by the extrapolating boundary conditions, as evident from the radial profiles. 
These effects become increasingly important at later times, limiting our ability to simulate the long-term evolution of the accretion flow. 
While the density profile may be expected to undergo more significant changes over longer evolution times, or for other transport coefficient choices, this has not been established in the present simulations.

To further quantify the departure from the stationary Michel solution, we examine the normalized mean absolute difference $\overline{\Delta T^{tt}}$ in Equation~\eqref{def:normalized-mean-absolute-difference}, using the initial stationary profile as the reference.
The results are presented in Figure~\ref{fig:michel-acc-bdnk-ideal-init-diff-compare}, where the left panel shows the time evolution of $\overline{\Delta T^{tt}}$ for selected values of $\eta_0$, while the right panel shows its dependence on $\eta_0$ at the final simulation time $t_f=28.75$ for several choices of the frame parameters $(\chi_0,\lambda_0)$.
The differences increase with $\eta_0$ while exhibiting only weak dependence on the frame parameters across the parameter choices considered, consistent with the behavior observed in the radial profiles of Figure~\ref{fig:michel-acc-snapshots}.

\subsection{Kerr tori}\label{Sec:Kerr-Fat-Disk}

\begin{figure}
    \centering
    \includegraphics[width=\linewidth]{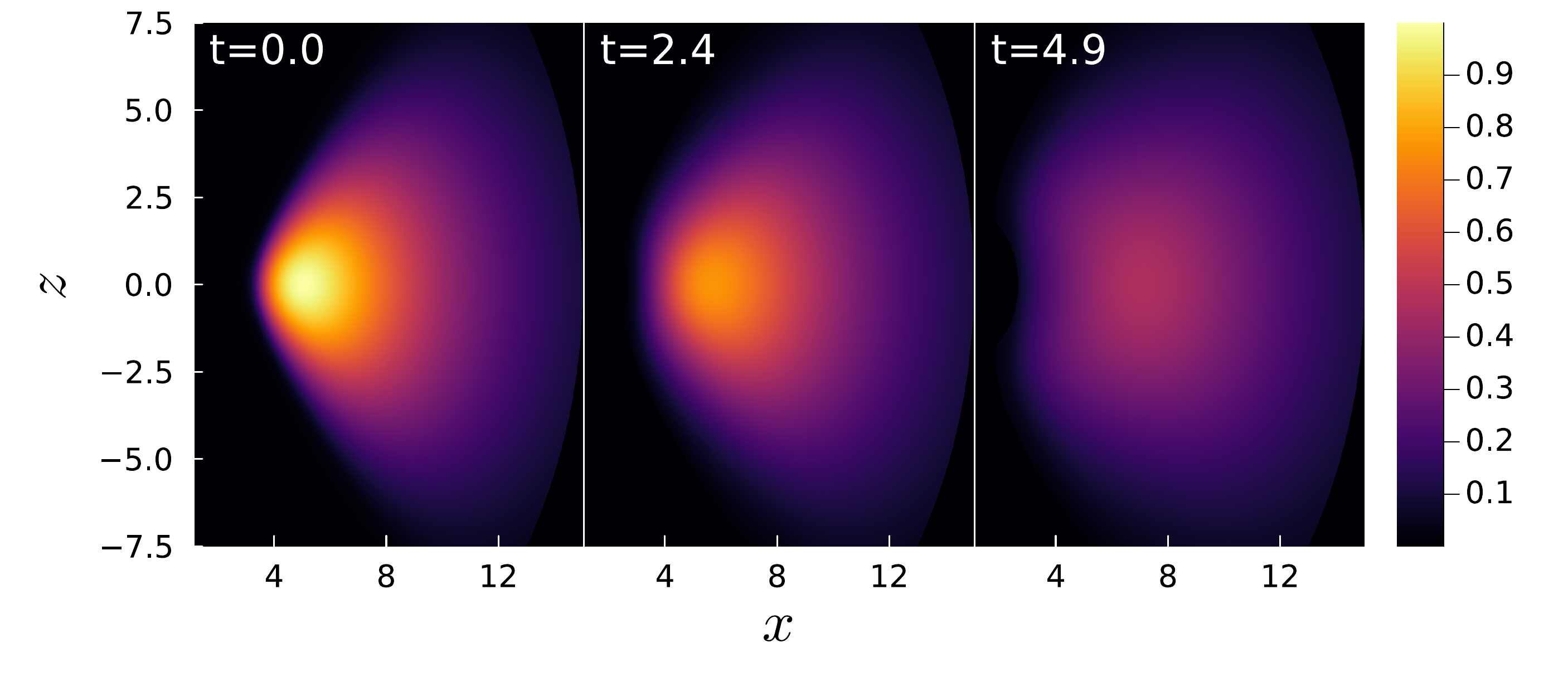}
    \caption{Snapshots of the mass density $n$ for the Kerr torus evolution at three different times, as indicated in the panels. The results are displayed in Cartesian coordinates $(x,z)$. The color scale represents the mass density $n$. The results are obtained using the viscosity parameter $\eta_0=10^{-3}$ and frame parameters $\chi_0=\lambda_0=10^3$.}
    \label{fig:kerr-slices}
\end{figure}

The second curved-spacetime configuration we consider is stationary axisymmetric flow of a geometrically thick relativistic torus around a Kerr black hole~\cite{Fishbone-Moncrief,1978A&A....63..209K}.
In comparison to all the systems considered in this work, this setup is distinctive in two respects.
First, the background spacetime is that of a 
spinning Kerr black hole, whose non-diagonal metric makes this the most non-trivial spacetime considered in this work.
Second, although the ideal hydrodynamical solution is analytic and the setup is simple, thick tori around spinning black holes are routinely used as initial data in astrophysical models of accreting systems, 
and characterizing their behavior under small but finite viscous corrections in the BDNK framework may provide qualitative understanding of what role, if any, BDNK viscosity may play in such systems.

We adopt Boyer-Lindquist coordinates, in which the line element is given by
\begin{align} \label{eq:Kerr-BL-metric}
ds^2 = -\left(1-\frac{2Mr}{\Sigma}\right) dt^2
- \frac{4Mar\sin^2\theta}{\Sigma}\, dt\, d\phi
+ \frac{\Sigma}{\Delta} dr^2 + \Sigma\, d\theta^2
+ \left(r^2 + a^2 + \frac{2Ma^2 r \sin^2\theta}{\Sigma}\right) \sin^2\theta\, d\phi^2\,,
\end{align}
where $M$ and $a$ are the black hole mass and spin parameter, respectively, and
$\Sigma = r^2 + a^2 \cos^2\theta$, $\Delta = r^2 - 2Mr + a^2$.
The analytic ideal-hydrodynamics solution for a stationary axisymmetric torus was derived in \cite{Fishbone-Moncrief}~(see also the modern textbook discussion in \cite{Rezzolla-Zanotti-Book-2013}), and we use the setup described for testing the ECHO ideal MHD code in \cite{DelZanna:2007pk}.
The flow is purely azimuthal, so that 
\begin{align} \label{eq:kerr-thick-torus-4-velocity}
u_\mu=u_t\,(1,0,0,-l)\,,
\end{align}
where $l\equiv-u_\phi/u_t$ is the specific angular momentum.
The normalization $u_\mu u^\mu=-1$ then gives
\begin{align}
\label{eq:expression-for-Ut-kerr-thick-torus}
u_t(r,\theta)=-\left[\frac{g_{t\phi}^2-g_{tt}g_{\phi\phi}}
{g_{\phi\phi}+2lg_{t\phi}+l^2g_{tt}}\right]^{1/2}\, .
\end{align}

The geometrically thick torus solution follows from assuming a constant specific angular momentum distribution, $l=l_0$.
For barotropic fluids the surfaces of constant $l$ coincide with the surfaces of constant angular velocity $\Omega=u^\phi/u^t$ (see the discussion on von Zeipel cylinders in \cite{Rezzolla-Zanotti-Book-2013}), and integrating the relativistic Euler equation gives the solution most compactly in terms of the specific enthalpy $h$,
\begin{equation}
\mathcal{W}_\textrm{in}-\mathcal{W}(r,\theta)= \ln h(r,\theta)=
\ln|(u_t)_\textrm{in}|-\ln|u_t(r,\theta)|\,,
\end{equation}
where $\mathcal{W}\equiv\ln|u_t|$ is the effective potential and the subscript ``$\textrm{in}$'' denotes evaluation at the inner edge of the torus in the equatorial plane, $(u_t)_\textrm{in}=u_t(r_\textrm{in},\pi/2)$.
Here we have imposed $h_\textrm{in}=1$, corresponding to vanishing density and pressure at the torus surface.
Combined with the polytropic EOS $P=Kn^\Gamma$, this yields the analytic density profile
\begin{equation}
\label{eq:exp-rho-kerr-thick-torii}
\rho(r,\theta)
= \left\{
\left(\frac{\Gamma-1}{K\Gamma}\right)
\left[
\exp\!\big(\mathcal{W}_\textrm{in}-\mathcal{W}(r,\theta)\big)-1
\right]
\right\}^{1/(\Gamma-1)}\,.
\end{equation}

Following \cite{DelZanna:2007pk}, we fix the torus geometry by choosing the radius $r_c$ of its center in the equatorial plane, which coincides with the pressure and density maximum, and setting the specific angular momentum there to the Keplerian value for equatorial circular orbits in Kerr~\cite{Daigne:2003tf},
\begin{equation}
l_K^\pm(r_c,\pi/2)=\pm\frac{r_c^2\mp2a\sqrt{r_c}+a^2}{(r_c-2)\sqrt{r_c}\pm a}\,,
\end{equation}
where the upper (lower) signs correspond to prograde (retrograde) orbits.
Taking the prograde branch, $l_0=l_K^+(r_c,\pi/2)$, the four-velocity follows from Equations~\eqref{eq:kerr-thick-torus-4-velocity} and \eqref{eq:expression-for-Ut-kerr-thick-torus} and the density from Equation~\eqref{eq:exp-rho-kerr-thick-torii}.
The two radii $r_c$ and $r_\textrm{in}$, together with the EOS parameters, therefore determine all hydrodynamic fields in the torus interior, where $\mathcal{W}(r,\theta)\leq\mathcal{W}_\textrm{in}$.
Since numerical evolution schemes cannot handle true vacuum, the exterior region $\mathcal{W}(r,\theta)>\mathcal{W}_\textrm{in}$ is filled with a low-density static atmosphere, constructed from a one-dimensional radial hydrostatic solution for an isentropic fluid as in \cite{DelZanna:2007pk}, and smoothly matched to the torus across its surface. The construction of the atmosphere is detailed in Appendix~\ref{Appendix:kerr-atm-tor-smoothing}.

For the numerical setup we take $K=1$ and normalize the density at the center of the
torus, $n(r_c,\pi/2)=1$, a free choice that can be made  because the spacetime is fixed and the fluid self-gravity is neglected~\cite{DelZanna:2007pk}.
We use a uniform two-dimensional grid in the $r$-$\theta$ plane at $\phi=0$, with $2.5<r<15$ and $0.45<\theta<\pi-0.45$, with $dr\approx 6.1\times10^{-3}$ and $dt=dx/10$.
We set the black hole spin to $a=0.99$ and $\Gamma=5/3$, and the torus geometry parameter values to $r_c=7$, $r_\textrm{in}=5$.

\begin{figure}
     \centering
     \begin{subfigure}[b]{0.49\textwidth}
         \centering
         \includegraphics[width=\textwidth]{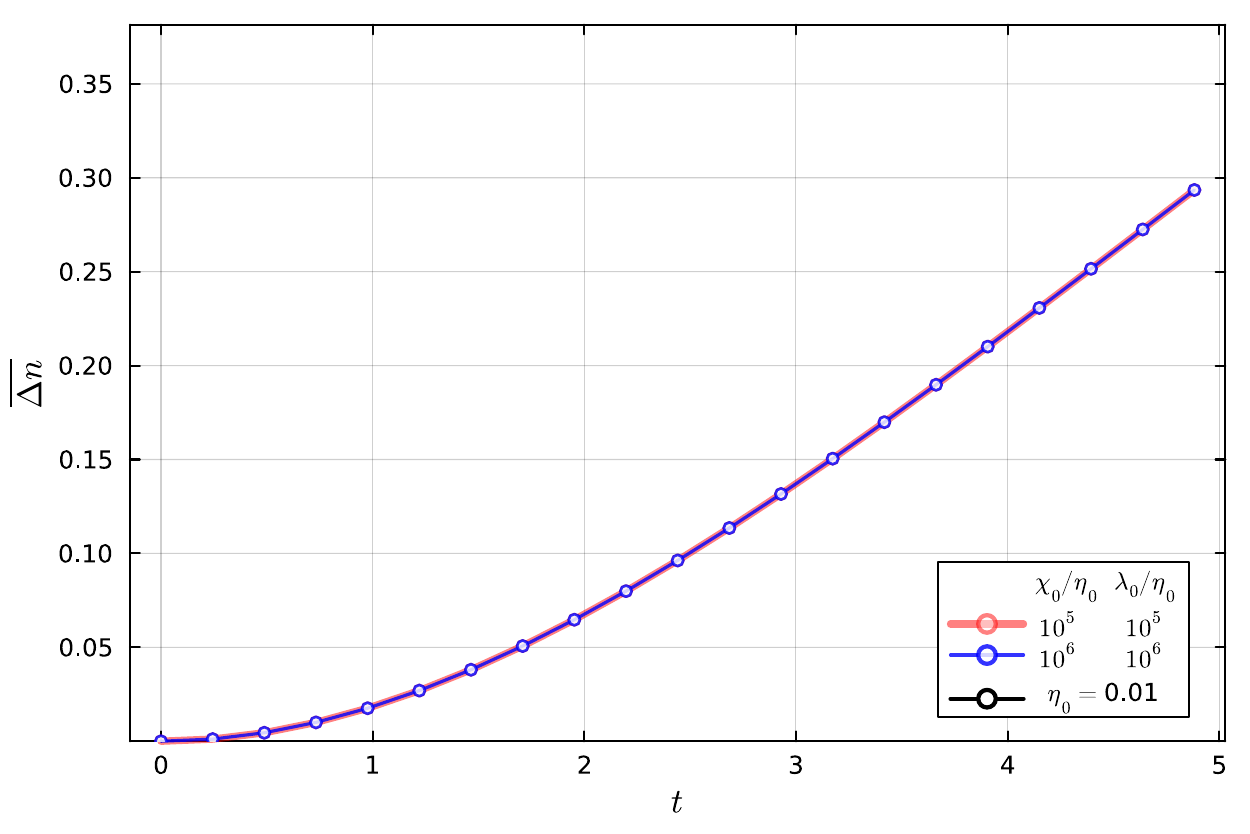}
     \end{subfigure}
        \caption{
        Normalized mean absolute difference, $\overline{\Delta n}$, between the BDNK  density $n$ 
        and the corresponding initial condition constructed with the stationary ideal hydrodynamics Kerr torus solution. It shows $\overline{\Delta n}$ over time for a selected value of $\eta_0$~(indicated in the legends).
        The colors distinguish different choices of frame parameters $(\chi_0,\lambda_0)$, as indicated in the legends.}
        \label{fig:kerr-bdnk-ideal-init-diff-compare}
\end{figure}

Figure~\ref{fig:kerr-slices} shows snapshots of the density $n$ at selected times, obtained using the \ac{FO} formulation with initialization Method~1 (Section~\ref{IC-method-1}). 
The shear viscosity parameter is set to $\eta_0=10^{-3}$, and the system is evolved up to $t_f=4.9$.

To quantify the departure from the initial stationary torus solution, we examine the normalized mean absolute difference $\overline{\Delta n}$ in Equation~\eqref{def:normalized-mean-absolute-difference}, using the initial stationary density profile as the reference. 
The results are presented in Figure~\ref{fig:kerr-bdnk-ideal-init-diff-compare}, which shows the differences $\overline{\Delta n}$ at various times for $\eta_0=0.01$, and for two choices of the frame parameters $(\chi_0,\lambda_0)$.
Unlike the radial accretion problem discussed in Section~\ref{Sec:Michel-Accretion}, the differences exhibit negligible frame dependence, even at the largest values of $\eta_0$ considered.
Nevertheless, the departure from the initial stationary solution is substantially more pronounced in the Kerr torus system, as evident from density snapshots in Figure~\ref{fig:kerr-slices} and the magnitudes of the differences in Figure~\ref{fig:kerr-bdnk-ideal-init-diff-compare}.

\section{Comparisons with Israel-Stewart in Milne coordinates}
\label{S:Israel_Stewart_comparisons}
In this section, we compare the BDNK hydrodynamics against the Israel-Stewart (IS) theory~\cite{Stewart-1977,Israel-Stewart-1979,Stewart-Stewart-1976,Israel:1976tn,Israel:1979wp} in the Milne coordinates without any conserved currents. As discussed in the introduction, there exist many variations of IS theories, but all of them are characterized by the fact that they introduce additional dynamical fields $\Pi$, $\pi^{\mu \nu}$ and $q^\mu$ that describe the deviations from the equilibrium. Here, we compare with the minimal version of the IS theory in the Landau hydrodynamic frame, in which the decomposition of the energy-momentum tensor is given by 
\begin{equation}
    \label{eq:IS_Tmunu}
    T^{\mu\nu}_{\rm IS} = \varepsilon u^\mu u^\nu + (P+\Pi)\Delta^{\mu\nu}+\pi^{\mu \nu}.
\end{equation}
The evolution equations for the bulk pressure $\Pi$ and shear viscous tensor $\pi^{\mu \nu}$ are given by relaxation type equations: 
\begin{align}
    \label{eq:IS_corrections_bulk}
    \tau_\Pi u^\mu \nabla_\mu \Pi +\Pi &= -\zeta \nabla_\mu u^\mu, \\
    \label{eq:IS_corrections_shear}
    \tau_\pi \Delta_{\alpha \beta }^{\mu \nu} u^\sigma \nabla_\sigma  \pi^{\alpha \beta} +\pi^{\mu \nu} &=-2\eta\sigma^{\mu \nu},   
\end{align}
where $\tau_\Pi$ and $\tau_\pi$ are relaxation times of the bulk and shear viscosities, respectively. The relaxation times in the minimal IS formalism describe how quickly the viscous corrections relax towards their Navier-Stokes values. 

The system in Equations~\eqref{eq:IS_Tmunu}-\eqref{eq:IS_corrections_shear} differs from the BDNK formulation in three major ways. First, the decomposition of the energy-momentum tensor in the Landau frame does not include either heat flow or any out-of-equilibrium correction to the energy density, both of which are included in the BDNK (see Equation~\eqref{BDNKTmunu}). Therefore, the meaning of the energy density or 4-velocity generally differs between the two theories if the viscous corrections are large. Consequently, it is more meaningful to compare the components $T^{0\mu}$ rather than the equilibrium fields. 

Second, the BDNK theory only contains four dynamical fields $u^i$ and $\varepsilon$, and the out-of-equilibrium corrections to $T^{\mu\nu}$ are directly proportional to the gradients of these variables. On the other hand, in the IS theory, there are six additional independent fields from $\Pi$ and $\pi^{\mu \nu}$, which respond to the gradients in $u^i$ on the relaxation time scales. Thus, it takes approximately one relaxation time before the gradients show an effect in the energy-momentum tensor. In principle, one could recover instantaneous response if $\tau_\pi \to 0$, but this will lead to an acausal and unstable system of equations. 

Finally, the initial data needed for both theories are somewhat different. Both theories require knowledge of $u^i({\bf x})$ and $\varepsilon({\bf x})$ at the initial time. However, BDNK also requires knowledge of the time gradients of $u^i(\bf x)$ and $\varepsilon(\bf x)$ while the IS requires the data for $\Pi ({\bf x})$, and $\pi^{\mu\nu}({\bf x})$. Therefore, finding a mapping between BDNK and IS initial conditions can be challenging. The most natural way to do the mapping would be to choose some initial $T^{\mu \nu}$ and then initialize both of the theories based on that. For the IS theory in the Landau frame, this is relatively straightforward, since the $T^{\mu \nu}$ and hydrodynamic fields ($u^i(\bf x), \varepsilon(\bf x), \Pi,\pi^{\mu\nu}({\bf x})$) both have 10 independent components. In addition, $u^i(\bf x)$ and $ \varepsilon(\bf x)$ can be solved conveniently as a timelike eigenvector and eigenvalue of the $T^{\mu \nu}$ respectively. Therefore, the physically relevant solutions are always guaranteed to exist as long as $T^{\mu \nu}$ has a timelike eigenvector with a negative eigenvalue.
However, as discussed in Section~\ref{Sec:Initial Conditions, discussion}, in the BDNK theory, the situation is significantly more complicated. Since the initial data consist only of eight \ac{DOF}s, namely $u^i({\bf x})$, $\varepsilon({\bf x})$ and their time derivatives, one cannot map arbitrary $T^{\mu \nu}$ into corresponding BDNK initial conditions. In this paper, we do not solve this problem in full generality, but instead focus on converting BDNK initial conditions into valid IS initial conditions. We accomplish this by first choosing the initial values for $u^i({\bf x})$ and $\varepsilon({\bf x})$ and their time derivatives, computing the corresponding BDNK $T^{\mu\nu}$, and finally computing the hydrodynamic fields for IS theory from this as described earlier. Even in this approach, it is possible that the resulting IS initial conditions are not physical, e.g., the weak energy condition is violated.\footnote{The weak energy condition states that any observer moving along a timelike path will measure positive total energy density, i.e., $T^{\mu \nu} a_\mu a_\nu\geq0$, for any future-pointing timelike vector $a^\mu$.} This only happens in regions where the gradient terms in $T^{\mu \nu}$ are dominating over equilibrium terms, i.e., in the regions where the first-order hydrodynamic gradient expansion is no longer valid. For all the tests done here, we have checked that the weak energy condition is satisfied for all different initial conditions. 

Even though there are quite significant differences between BDNK and IS theories, they still share important common features. Both of the theories will approach ideal relativistic hydrodynamics when $\eta \to 0$ and $\zeta \to 0$. They both also share the same linearized non-hydro modes that also match with the relativistic extension of the Navier-Stokes theory~\cite{Bemfica:2020zjp, Hiscock:1987zz}. Therefore, one would expect similar behavior from both theories when the gradients are small enough and the system is close to equilibrium. 

All IS simulations performed in this section are performed using a modified version of the MUSIC hydrodynamics code~\cite{Schenke:2010nt,Schenke:2010rr,Paquet:2015lta} with the simplified relaxation Equations~\eqref{eq:IS_corrections_bulk} and \eqref{eq:IS_corrections_shear}. We assume that the system is boost-invariant, meaning that in the Milne coordinates all of the fields are independent of the $\eta_s$ coordinate and that $u^{\eta}=0$. Throughout this section, we also set the bulk viscosity to zero and use the shear relaxation time $\tau_\pi=5 \eta/(\varepsilon + P)$~\cite{Denicol:2012cn} 
unless mentioned otherwise. 
As an initialization method, we use Method-1 presented in Section~\ref{Sec:Initial Conditions, time derivatives}, i.e., the initial time gradients are initialized in such a way that $T^{0\mu}(\tau_0)=T^{0\mu}_0(\tau_0)$, and match IS variables to match the full BDNK $T^{\mu \nu}$ as discussed above. 
Note that this type of initialization does not mean that $T^{\mu \nu}(\tau_0)=T^{\mu \nu}_0(\tau_0)$. This is very apparent for a boost-invariant system in the Milne coordinates if the fluid velocity is zero and $\zeta=0$. In this case, the condition $T^{0\mu}(\tau_0)=T^{0\mu}_0(\tau_0)$ is equivalent to saying that $\mathcal{A} = 0$, and $Q^{\mu} = 0$. However, due to the longitudinal flow introduced by the geometry of the coordinate system $\sigma^{\eta \eta} = 2/(3\tau_0^3) \neq 0$, i.e., there is anisotropy between the longitudinal and transverse pressures arising from the geometry of the system, which scales as $\sim\eta/\tau_0$. Therefore, $T^{\eta \eta} = T^{\eta \eta}_{0} - 2\eta\sigma^{\eta \eta} \neq T^{\eta \eta}_{0}$ if $\eta$ is non-zero.

\subsection{Transverse Gaussian}
\label{S:Transverse_Gaussian}

\begin{figure}
    \centering
    \includegraphics[width=0.48\linewidth]{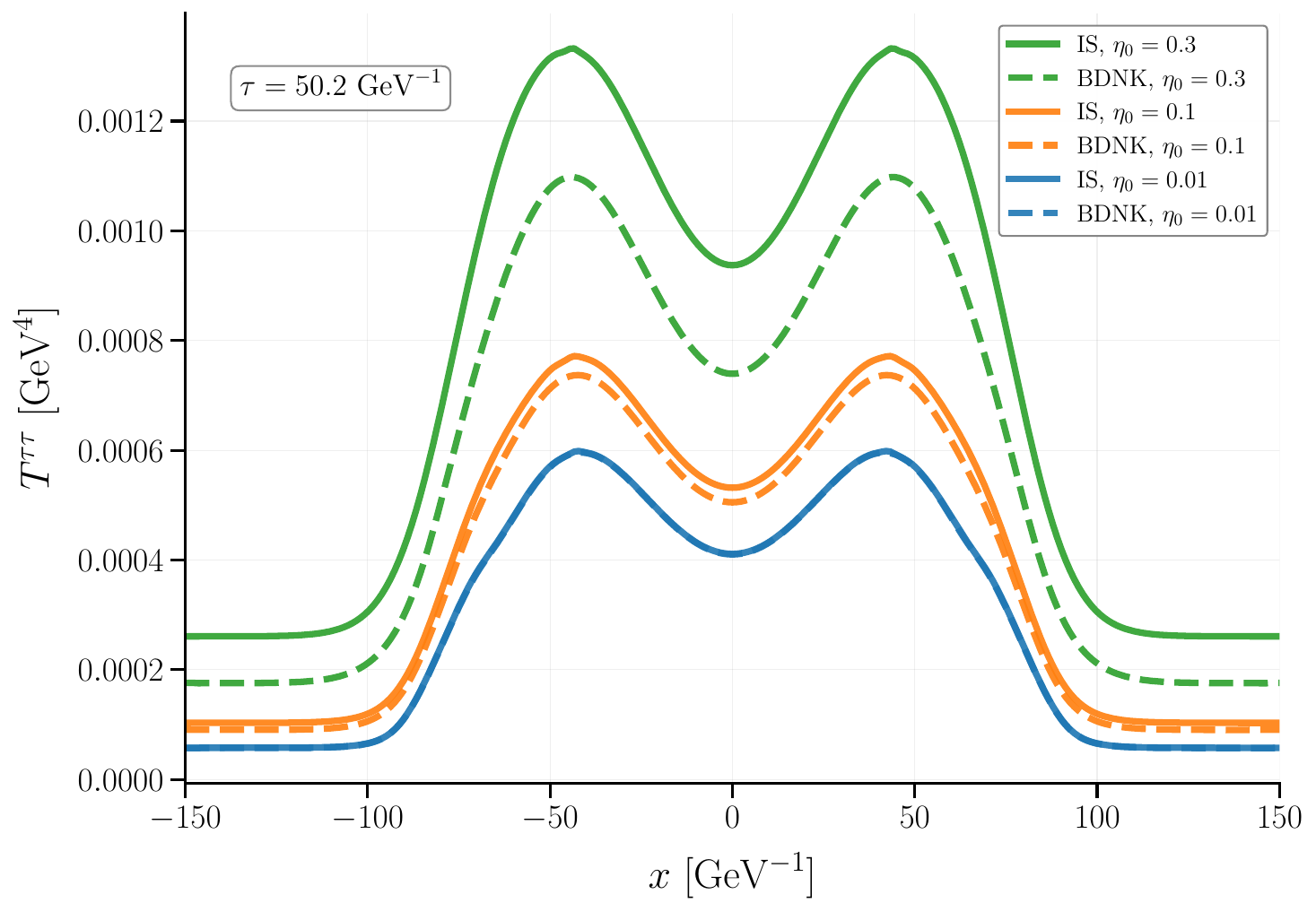}
    \includegraphics[width=0.48\linewidth]{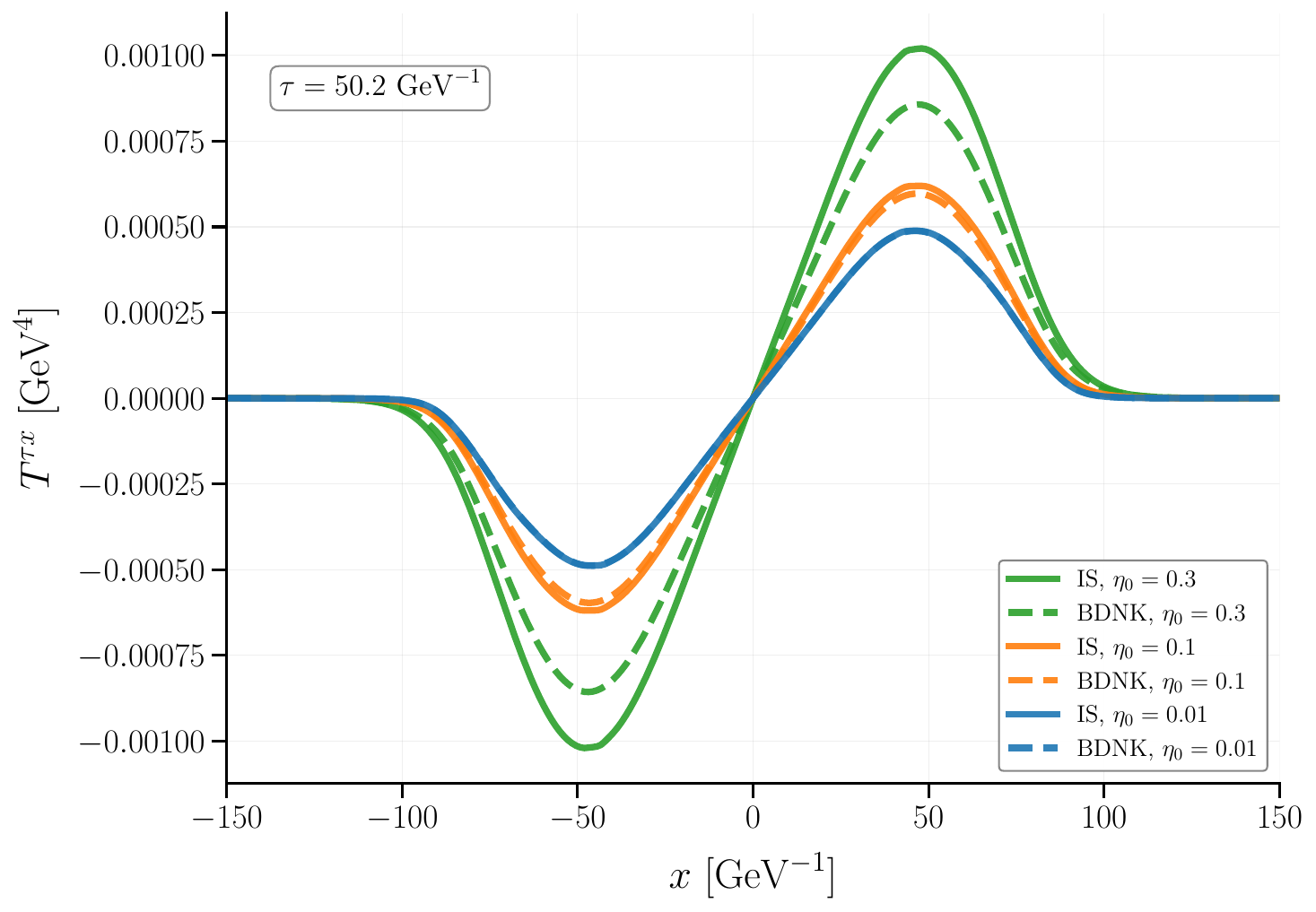}
    \caption{Comparison of the BDNK solutions against the IS solutions for a transverse Gaussian initial condition in Milne coordinates with conformal EoS and various shear viscosities at $\tau = 50.2~\rm GeV^{-1}$. All the BDNK results use $\chi_0 = (25/4) \eta_0$ and $\lambda_0 =  (25/7)\eta_0$.}
    \label{fig:IS_comp_Gaussian_conformal}
\end{figure}

We begin by comparing BDNK and IS in the Milne coordinates with a simple Gaussian transverse profile for energy density:
\begin{equation}
    \varepsilon({\bf r}, \tau_0 ) = A \exp \left( -\frac{{\bf r}^2}{2\sigma^2}\right) + \varepsilon_0,
\end{equation}
where $\bf r$ is the transverse position vector. We use values $A=0.6~\rm GeV^4\approx 78.1~GeV/fm^3$ for the amplitude of the Gaussian, $2\sigma^2=900~\rm GeV^{-2}\approx 35.0~fm^2$ for the Gaussian width, and $\varepsilon_0= 10^{-2}~\rm GeV^4\approx 1.3~GeV/fm^3$ for the floor. The width and the amplitude are chosen to approximately match the size and maximum temperature of \ac{QGP} formed in a central heavy-ion collision. In addition, we set the initial transverse velocity to zero and the initial proper time to $\tau_0 = 1~ \GeV^{-1}$.

\begin{figure}
    \centering
    \includegraphics[width=0.48\linewidth]{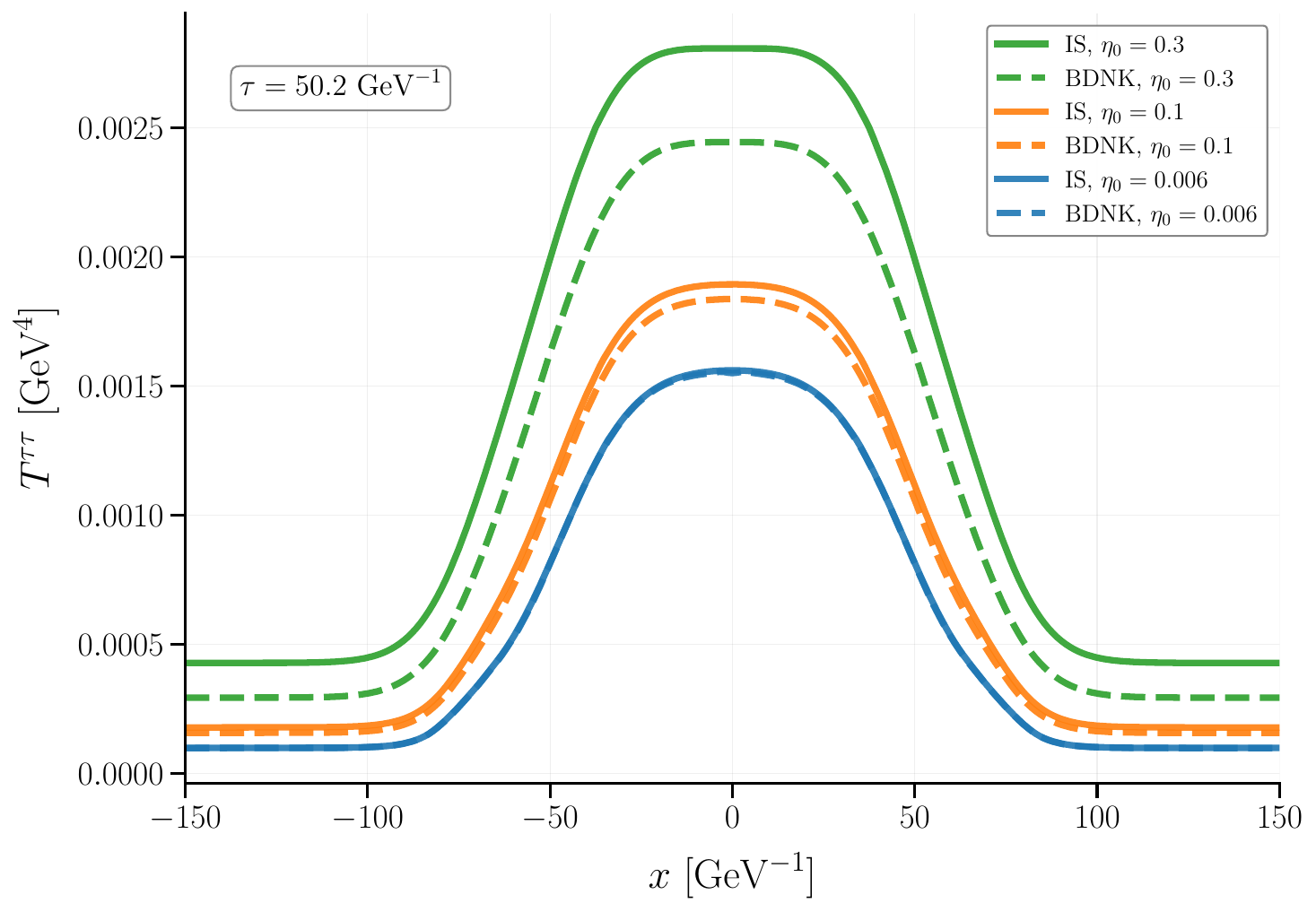}
    \includegraphics[width=0.48\linewidth]{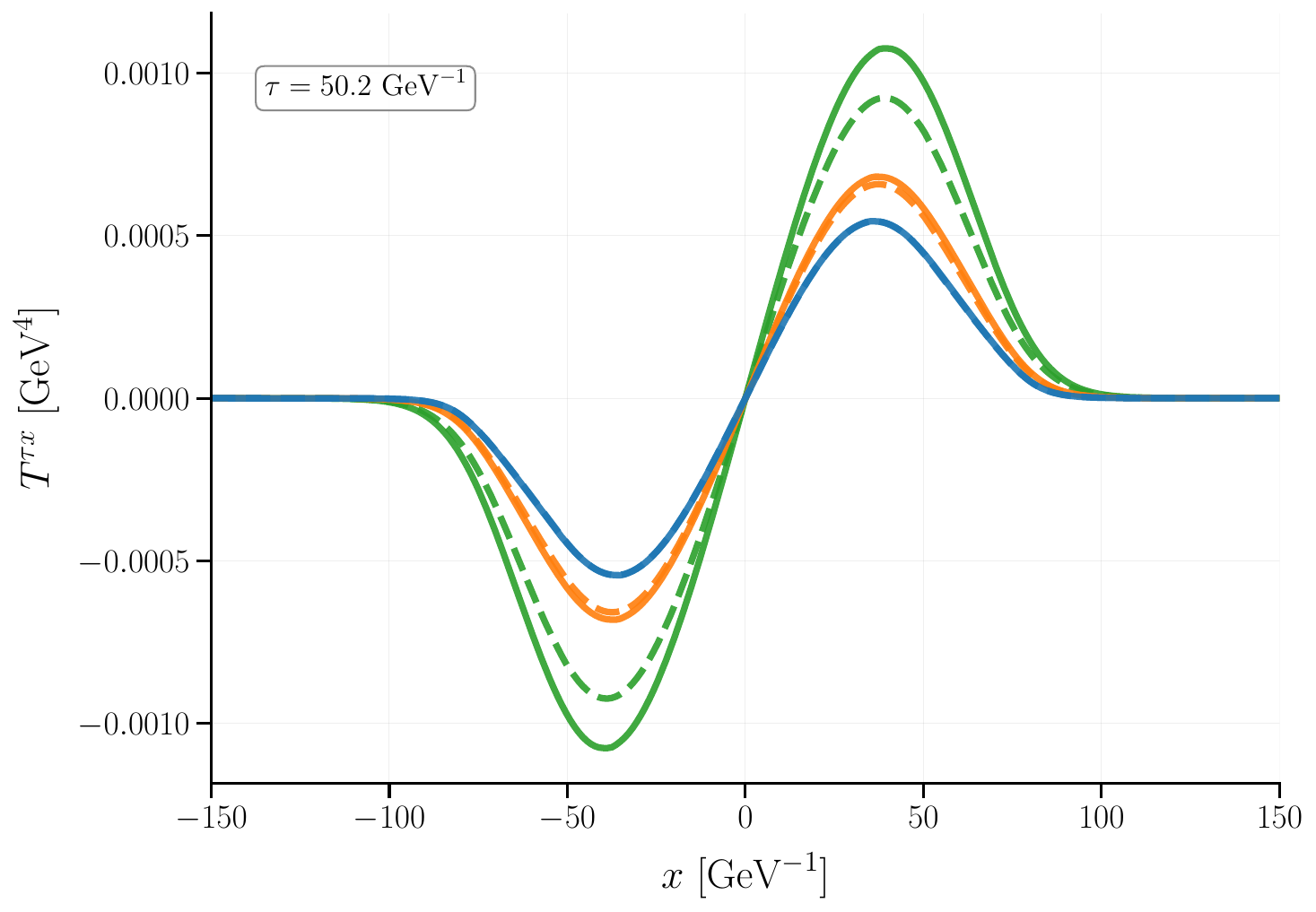}
    \caption{Comparison of the BDNK solutions against the IS solutions for a transverse Gaussian initial condition in Milne coordinates with \ac{QCD} EoS and various shear viscosities at $\tau = 50.2~\rm GeV^{-1}$. All the BDNK results use $\chi_0 = (25/4) \eta_0$ and $\lambda_0 =  (25/7)\eta_0$. }
    \label{fig:IS_comp_Gaussian_QCD}
\end{figure}

We begin by showing results with the conformal equation of state.
The comparison of $T^{\tau \tau}$ and $T^{\tau x}$ between BDNK and IS for various values of shear viscosity using Gaussian initial conditions is shown in Figure~\ref{fig:IS_comp_Gaussian_conformal}. When $\eta_0$ is small, we can see that the BDNK agrees almost exactly with the IS results. 
This is in line with the expectation that the two theories should be similar at low viscosity, as discussed above. Increasing the shear viscosity significantly amplifies the differences. 
However, we note that the difference is mostly in normalization rather than the transverse shape, which only changes slightly with increasing viscosity. This indicates that most of the differences are driven by the dynamics in the rapid longitudinal expansion at early times.

\begin{figure}
    \centering
    \includegraphics[width=0.48\linewidth]{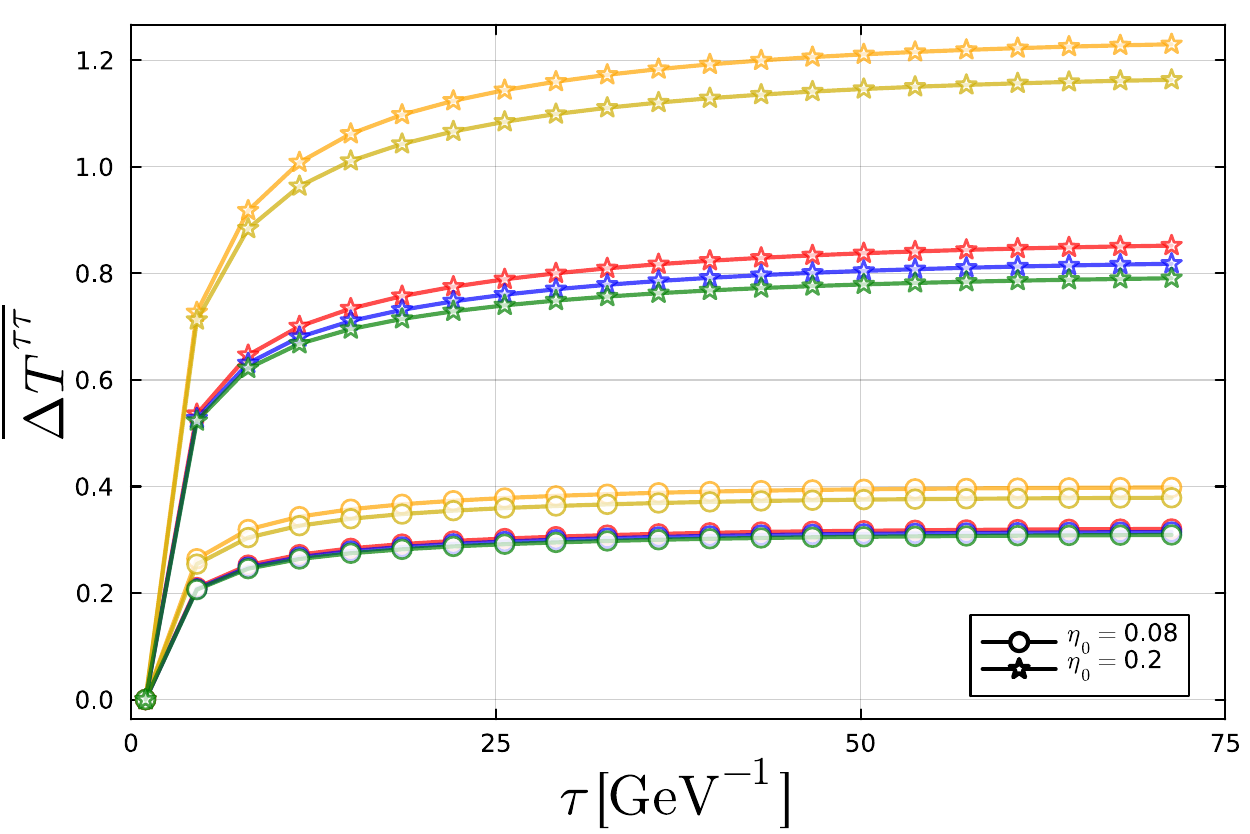}
    \includegraphics[width=0.48\linewidth]{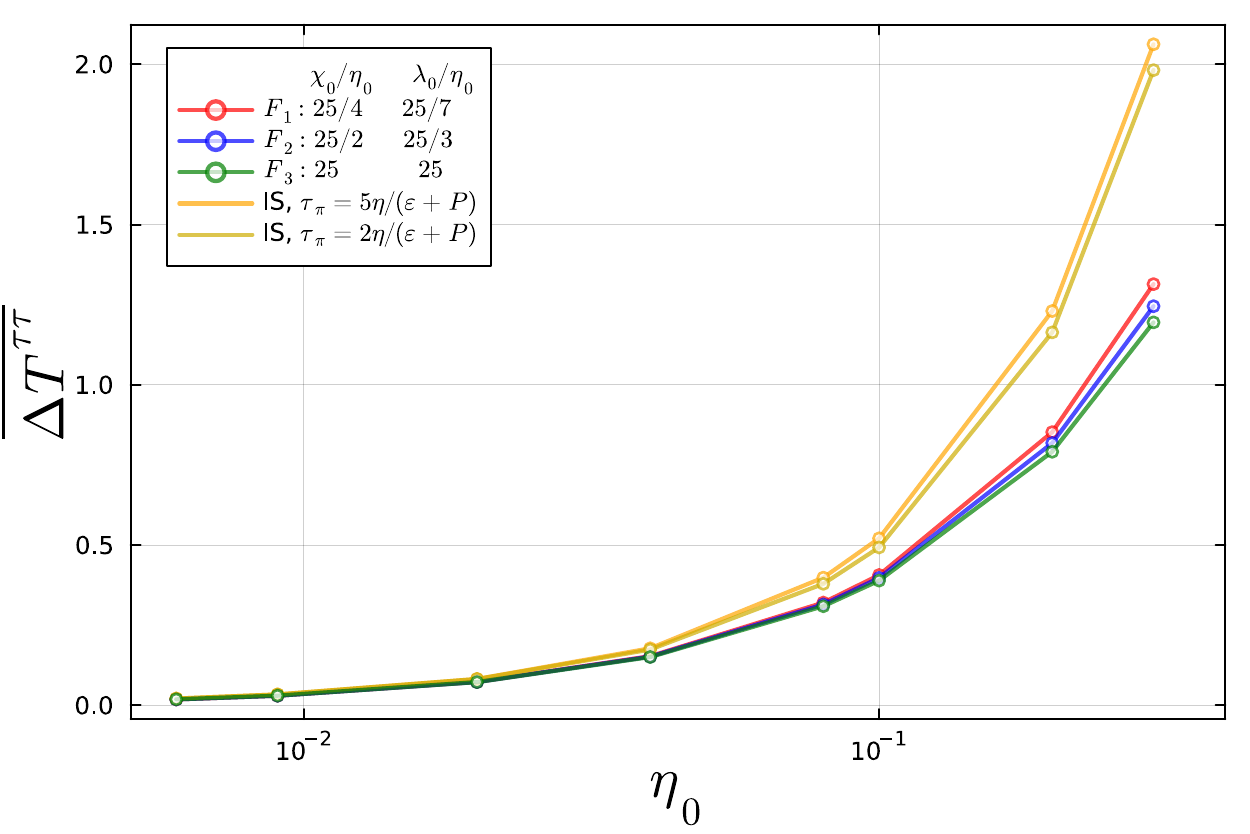}
    \caption{The normalized averaged differences $\overline{\Delta T^{\tau \tau}}$ from the ideal fluid reference using Gaussian initial conditions and \ac{QCD} EoS with $\tau_0 = \rm 1~GeV^{-1}$. The left panel shows the longitudinal proper time evolution of $\overline{\Delta T^{\tau \tau}}$, while the right panel shows $\overline{\Delta T^{\tau \tau}}$ evaluated at the last time step as a function of $\eta_0$. The red, blue, and green curves correspond to different values of BDNK frame parameters, and the orange and yellow curves are the IS results with different relaxation times (see legend). On the left panel, the different markers indicate different values of $ \eta_0$. }
    \label{fig:differences_from_ideal_Gaussian_QCD}
\end{figure}

The same comparison, but this time with the \ac{QCD} equation of state, is shown in Figure~\ref{fig:IS_comp_Gaussian_QCD}. The \ac{QCD} EoS has a speed of sound close to the conformal one at high temperatures ($T \gtrsim 0.5$~GeV), but has a much smaller speed of sound at low temperatures, which makes the system expand significantly more slowly. The relative behavior of BDNK and IS remains very similar to the conformal case. To more systematically quantify how the differences between the two theories arise, in Figure~\ref{fig:differences_from_ideal_Gaussian_QCD} we show normalized average differences $\overline{\Delta T^{\tau \tau}}$ (defined in Equation~\eqref{def:normalized-mean-absolute-difference}) from the ideal fluid reference. 
For the BDNK results, we show three different curves corresponding to different choices of frame parameters. The left panel shows how the differences behave as a function of proper time $\tau$ for $\eta_0$ values 0.08 and 0.2. We see that at the early proper times, the viscous results start to deviate very quickly from the ideal hydrodynamics solution. This effect is mostly driven by fast geometric expansion in the longitudinal direction that induces large longitudinal pressure anisotropy through shear-viscous corrections. Note that these corrections are also present in the initial conditions as discussed earlier.
While the quantitative behavior is similar for both BDNK and IS theories, the IS theory will get pushed farther from the ideal solution due to the relaxation time formalism, which keeps the values of viscous corrections lagging behind the quickly decaying gradients. On the right panel of Figure~\ref{fig:differences_from_ideal_Gaussian_QCD}, we show how the differences on the final timestep behave as a function of $\eta_0$. We see that both BDNK and IS solutions are very close to each other and the ideal fluid solution when the viscosity is small. We also see that making the shear relaxation time smaller drives the IS result closer towards the BDNK results, as expected. At higher viscosities, the differences between BDNK and IS become more relevant to the point where at $\eta_0=0.3$, the $\overline{\Delta T^{\tau \tau}}$ of the IS solution is approximately 1.5 times farther away from the ideal solution than the BDNK solution. Notably, the differences between different BDNK frame choices are significantly smaller than this. 
  
\subsection{TRENTo initial conditions}
\label{S:Trento_initial_conditions}
\begin{figure}
    \centering
    \includegraphics[width=\linewidth]{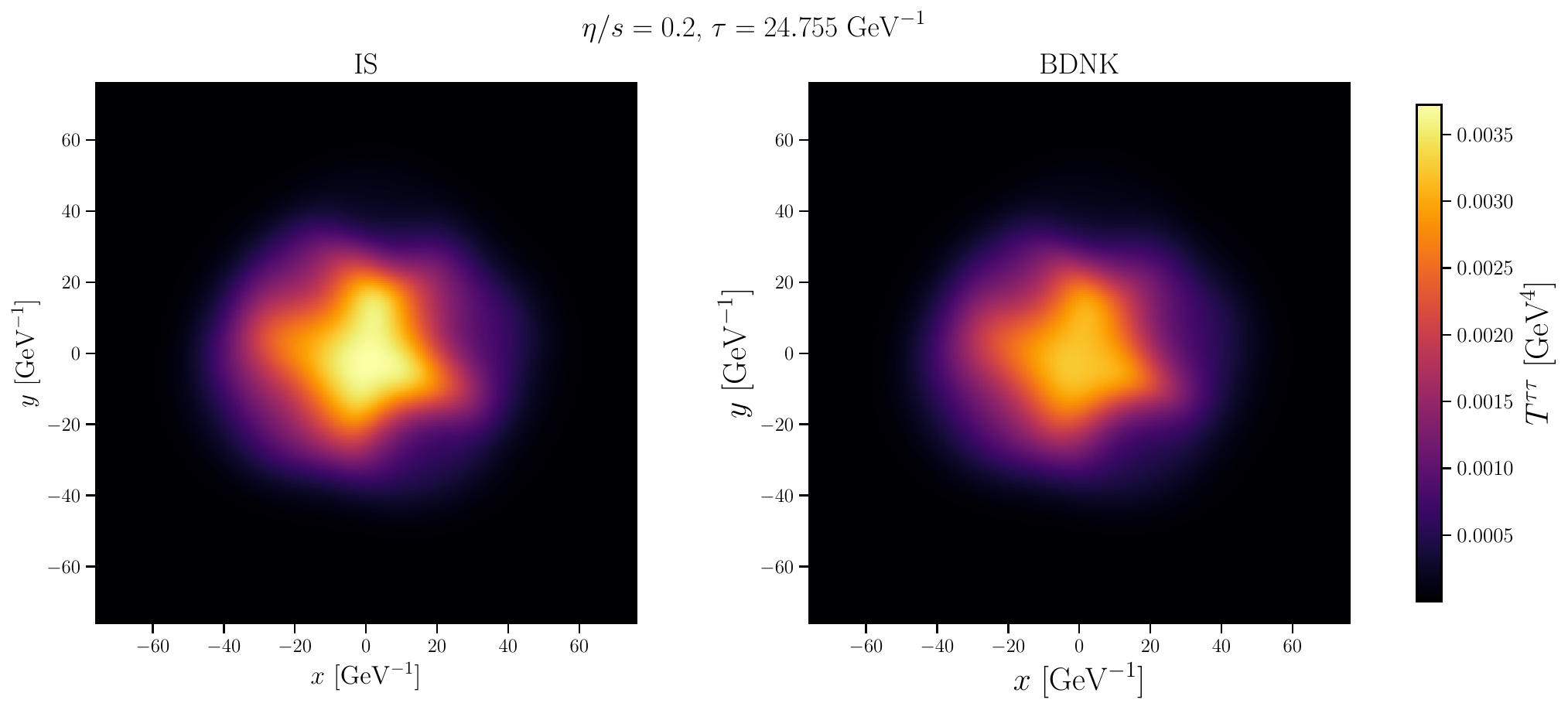}
    \includegraphics[width=\linewidth]{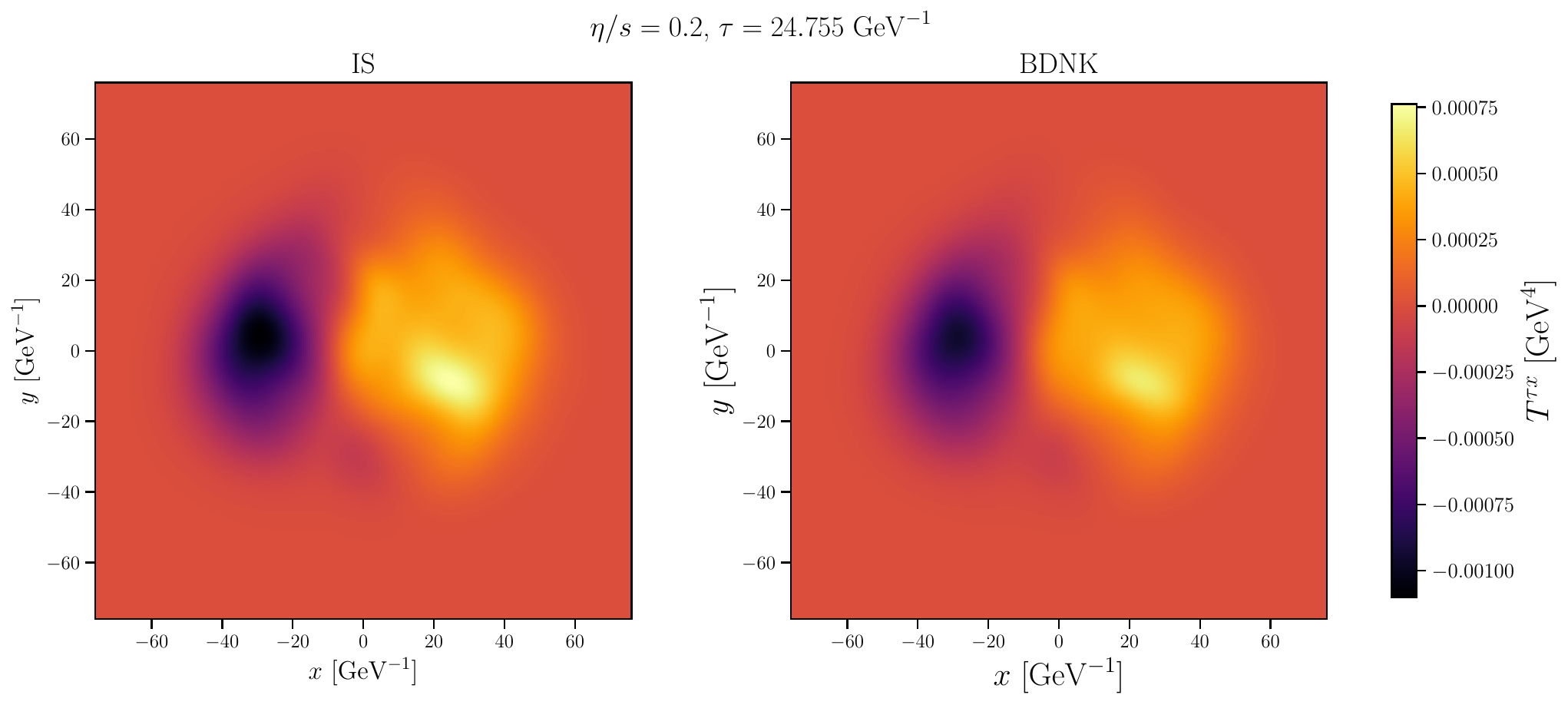}
    \caption{Comparison of the transverse profiles of the BDNK solution against the IS solution for the TRENTo initial conditions with \ac{QCD} EoS at $\tau=24.755~\rm GeV^{-1}$. The BDNK result uses $\chi_0 = (25/4) \eta_0$ and $\lambda_0 =  (25/7)\eta_0$.}
    \label{fig:IS_comp_Trento}
\end{figure}
In a realistic heavy-ion collision simulation, the initial conditions do not have exact symmetries in the transverse plane due to the fluctuating positions of nucleons inside the nuclei. To test the behavior of the BDNK theory under these circumstances, we perform simulations using TRENTo initial conditions that are widely used in modern heavy-ion collision simulations~\cite{JETSCAPE:2020mzn, Nijs:2020roc, Parkkila:2021yha}. 

The TRENTo model~\cite{Moreland:2014oya} is a parametric initial condition model, where the energy deposition for a collision of two nuclei with mass numbers $A$ and $B$ is based on the locations of the colliding nucleons. These locations are sampled from the expected spatial density of nucleons in each nucleus. The degree of overlap of the nuclei is controlled by impact parameter $b$, which is the distance between the centers of the colliding nuclei.

Formally, the energy deposition in the TRENTo model can be written as 
\begin{equation}
\label{eq:trento}
        \varepsilon({\bf r}, \tau_0) = N \left( \frac{T_A^p({\bf r}) +T_B^p({\bf r})}{2} \right)^{1/p}, 
\end{equation}
where $N$ and $p$ are free model parameters, which we set to values $N=0.1~\rm GeV^{4}fm^{2}$ and $p=0$ to approximately match realistic heavy-ion initial conditions. The functions $T_A({\bf r})$ and $T_B({\bf r})$ are often referred to as nuclear thickness functions, which describe the participant nucleon areal densities for nuclei $A$ and $B$, respectively. They can be further written as a sum of the contributions from each nucleon as
\begin{equation}
      T_{A/B}({\bf r})= \sum_i T_n ({\bf r}_i -{\bf r}),
\end{equation}
where ${\bf r}_i$ is the position vector of nucleon $i$ inside the nucleus, and the sum goes through all the participant nucleons. The nucleon thickness function $T_n$ is assumed to have a Gaussian form 
\begin{equation}
    T_n({\bf r}) = \frac{1}{2\pi \sigma^2} \exp\left(-\frac{{\bf r}^2}{2\sigma^2}\right),
\end{equation}
where we choose nucleon width $\sigma=1~{\rm fm}\approx 5.067~\rm GeV^{-1}$. 

There are many variations and extensions of the Trento model, e.g., including multiplicity fluctuations or nucleon substructure~\cite{Moreland:2018jos}. In addition, before hydrodynamic evolution, one often evolves the TRENTo initial conditions via free streaming~\cite{Liu:2015nwa}, which leads to a nonzero initial transverse velocity. However, initializing BDNK variables after free streaming is complicated since this would require matching full $T^{\mu \nu}$ with the BDNK variables~(see discussion in Section~\ref{Sec:Initial Conditions}). 
Therefore, here we only focus on the simpler setup described above, and assume that the initial transverse velocity is zero. To make numerics stable in the vacuum region, we add a floor of $10^{-6}~\rm GeV^4$ to the energy density.

Figure~\ref{fig:IS_comp_Trento} shows the comparison between BDNK and IS for the transverse profiles of $T^{\tau \tau}$ and $T^{\tau x}$ using a Trento initial condition for a Pb+Pb collision with an impact parameter $b=7~\rm fm\approx35.5~\rm GeV^{-1}$ and initial proper time of $\tau_0 = 1~\rm GeV^{-1}$. Despite a more non-trivial initial geometry, the differences between BDNK and IS behave in a similar way as with the Gaussian initial conditions, i.e., the largest differences come from the normalization, while there are only minor differences in the shapes of the transverse profiles. This is not totally surprising since, for both initial conditions, at the initial time, all the other corrections except those connected with the anisotropy between the longitudinal and transverse pressure are zero. Therefore, one would expect the largest differences between theories to arise from the different types of relaxation of these corrections towards equilibrium.   
\begin{figure}
    \centering
    \includegraphics[width=0.48\linewidth]{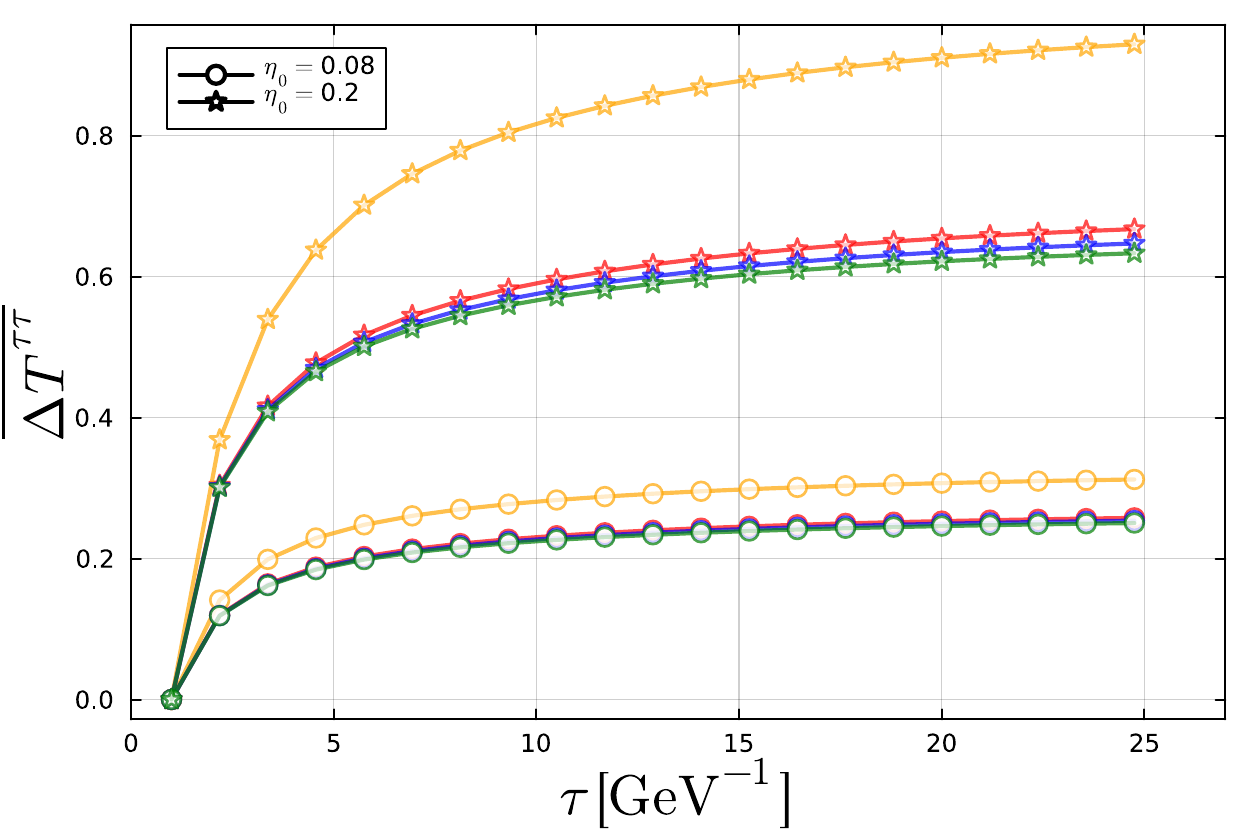}
    \includegraphics[width=0.48\linewidth]{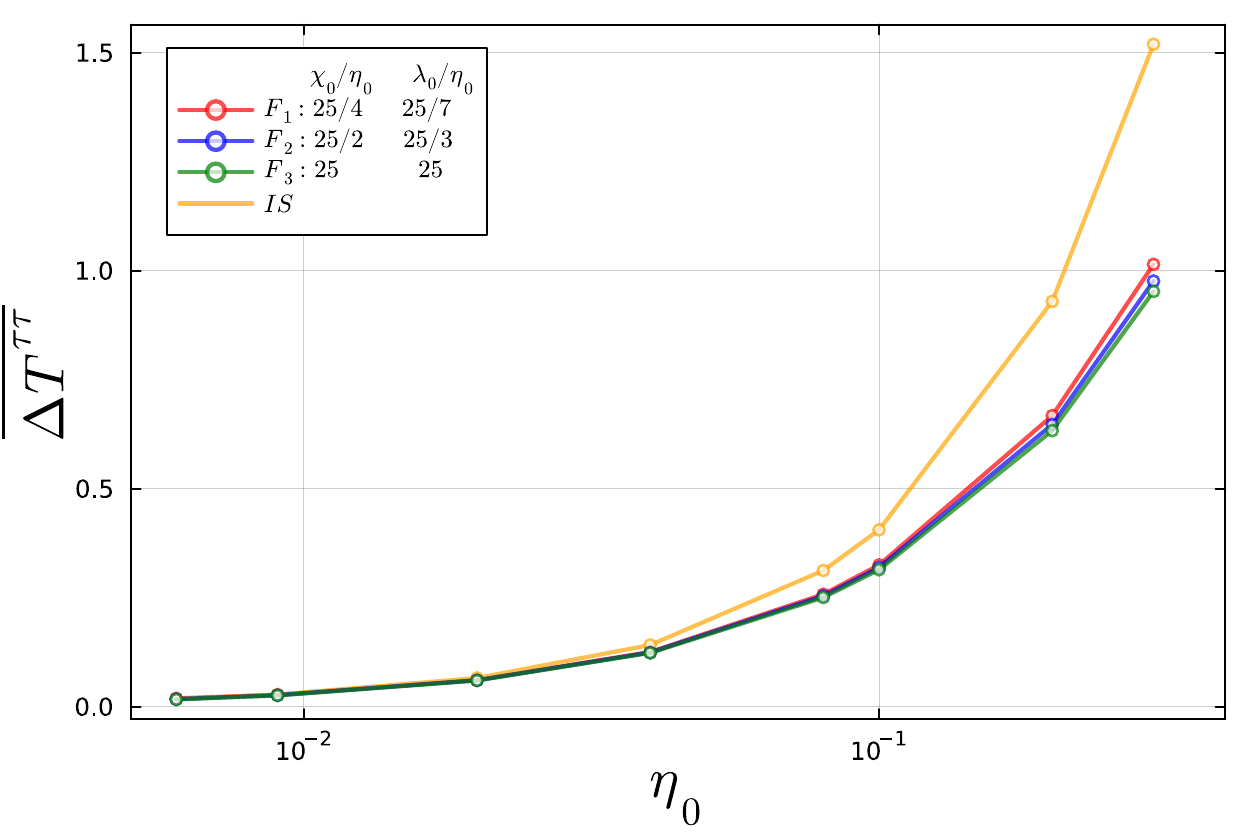}
    \caption{Same as Figure~\ref{fig:differences_from_ideal_Gaussian_QCD}, but using TRENTo initial conditions and \ac{QCD} EoS. }
    \label{fig:differences_from_ideal_Trento}
\end{figure}

\begin{figure}
    \centering
   \includegraphics[width=0.32\linewidth]{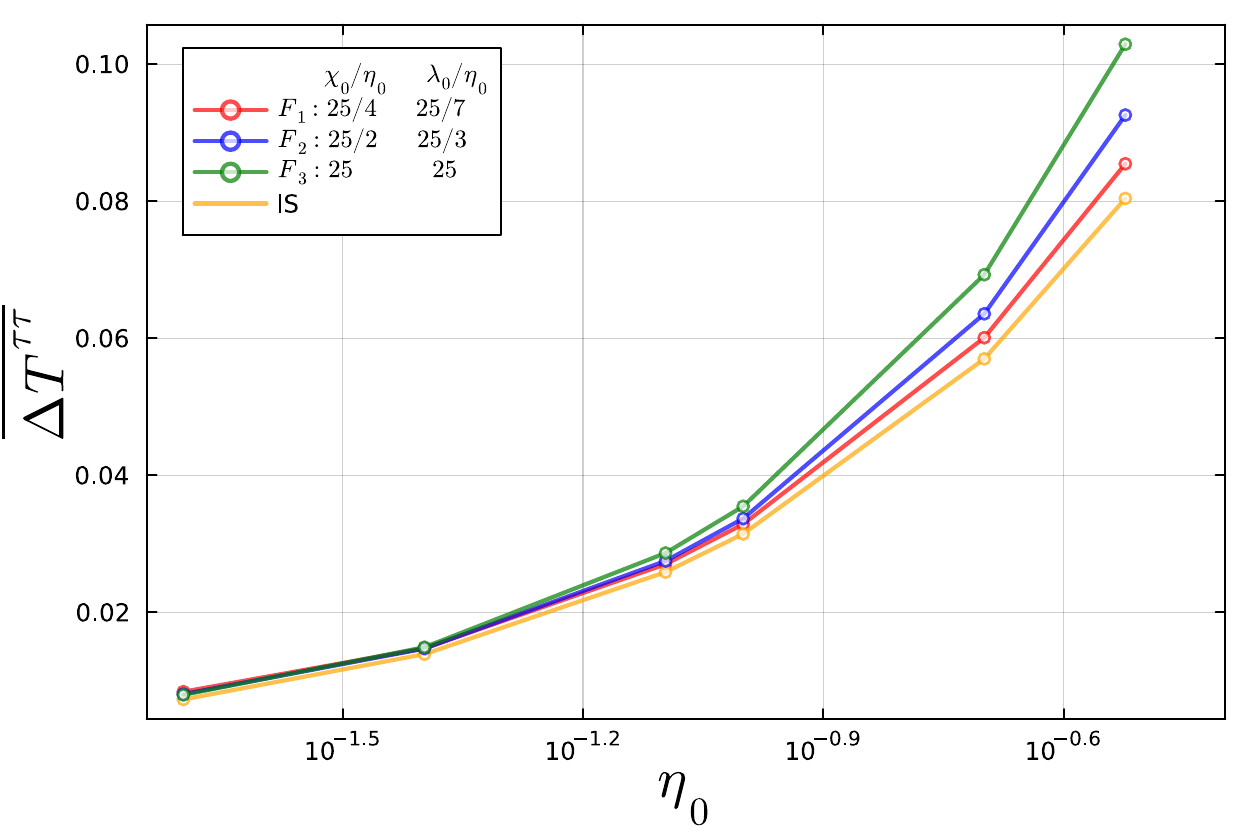}
    \includegraphics[width=0.32\linewidth]{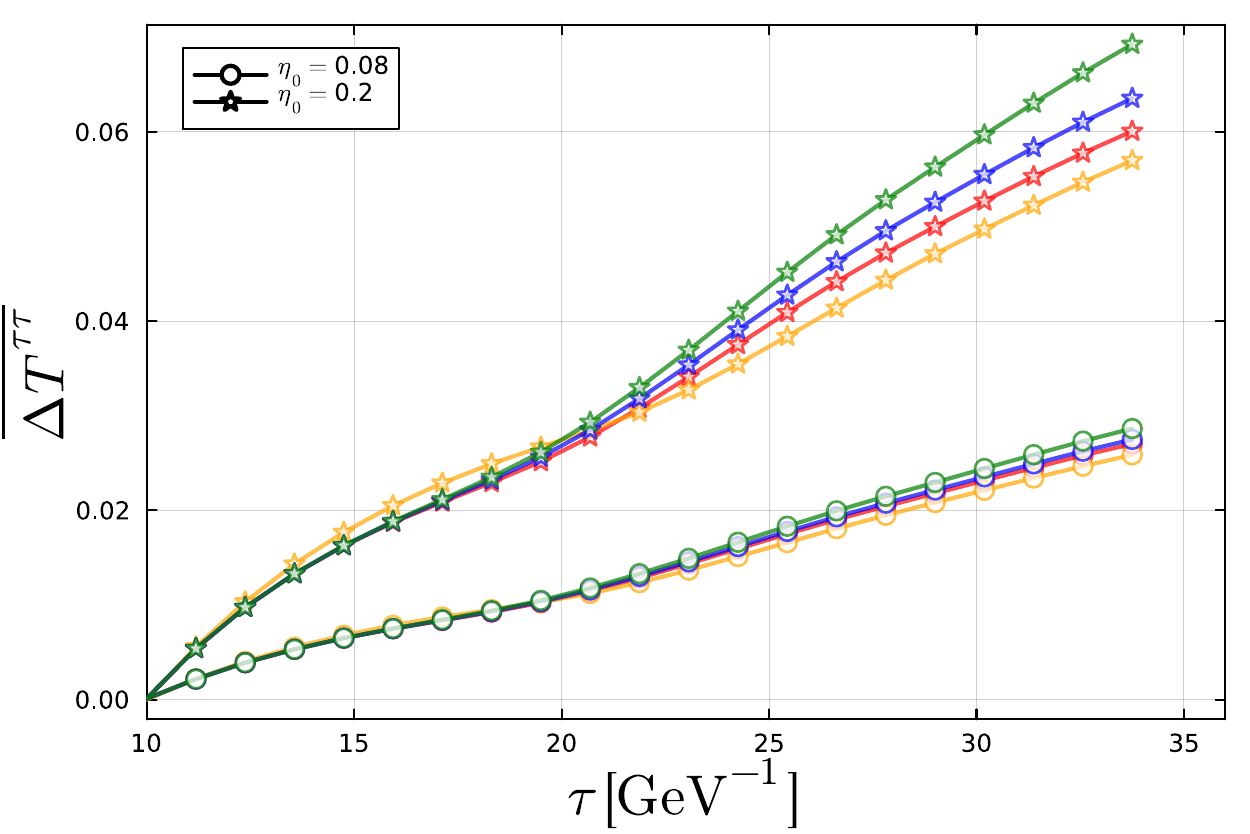}
    \includegraphics[width=0.32\linewidth]{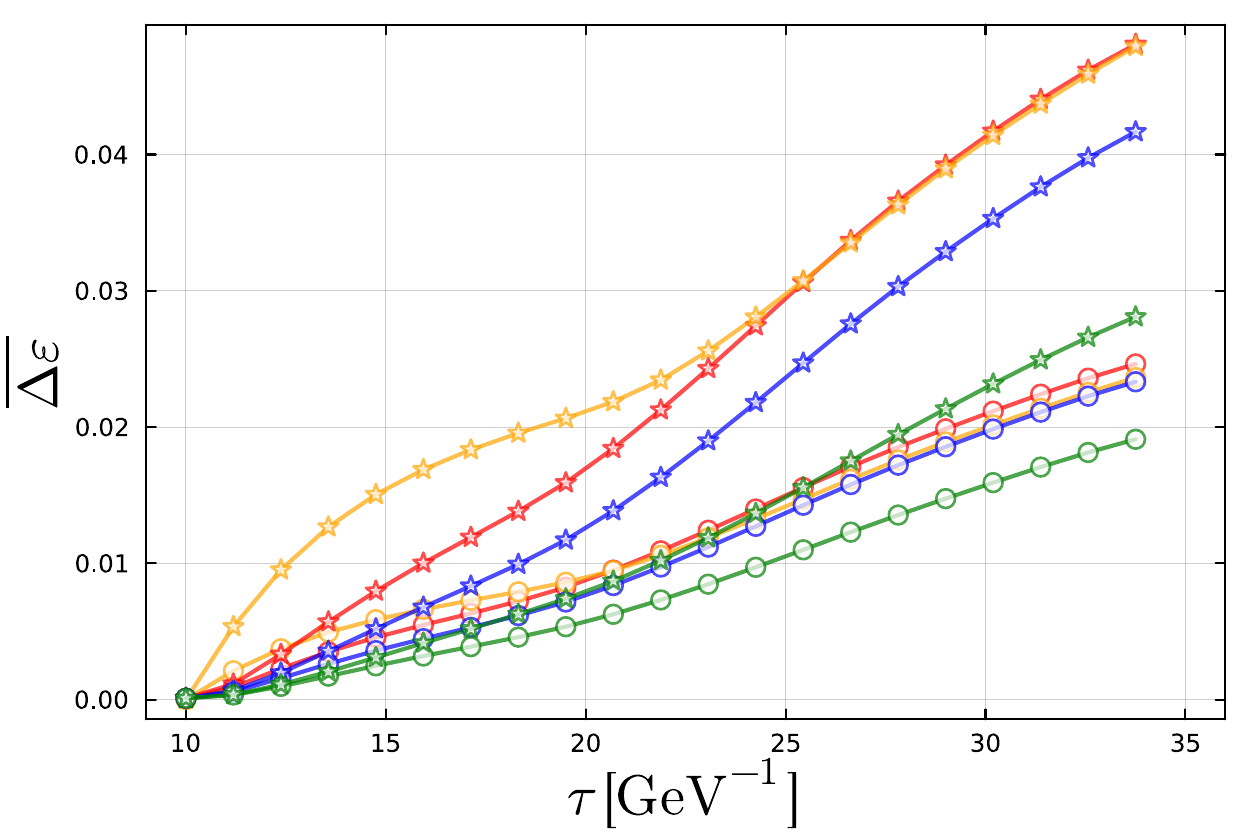}
    
    \caption{The normalized averaged differences from the ideal fluid reference using TRENTo initial conditions and \ac{QCD} EoS with $\tau_0 = \rm 10~GeV^{-1}$. The left panel shows $\overline{\Delta T^{\tau \tau}}$ evaluated at the last time step as a function of $\eta_0$, while the middle panel shows the longitudinal proper time evolution of $\overline{\Delta T^{\tau \tau}}$. The right panel shows the same as the middle one but for $\overline{\Delta \varepsilon}$. The red, blue, and green curves correspond to different values of BDNK frame parameters, and the orange curve is the IS result. On the middle and right panels, the different markers indicate different values of $ \eta_0$.}
    \label{fig:differences_from_ideal_Trento_tau0_10}
\end{figure}

The behavior of the normalized average differences from the ideal fluid reference is shown in Figure~\ref{fig:differences_from_ideal_Trento}. Both the difference as a function of time (left panel) and the difference as a function of $\eta_0$ (right panel) show a very similar picture as in the case of Gaussian initial conditions. While both theories approach ideal fluid when the viscosity is small, at larger viscosities, the IS theory solution deviates significantly more from the ideal solution compared to the BDNK. 

To test the behavior of the theories in a setting where the system is initially very close to equilibrium, we also perform the simulations with an initial proper time of $\tau_0 = 10~\rm GeV^{-1}$, in which case the transverse flow mostly dominates the dynamics.
The results are shown in Figure~\ref{fig:differences_from_ideal_Trento_tau0_10}. The left panel demonstrates how $\overline{\Delta T^{\tau \tau}}$ as a function of $\eta_0$ behaves very similarly between the BDNK and IS theories, and that the frame dependence in the BDNK theory is very modest. More detailed information is contained in the middle panel, which shows $\overline{\Delta T^{\tau \tau}}$ as a function of longitudinal proper time for $\eta_0 = 0.08$ and $\eta_0 = 0.2$. Even though the difference between BDNK and IS theories is small, we again notice how the relaxation time-based IS formalism responds to a change in gradients more slowly. At the early times, the IS deviates more from the ideal solution as before, but this time with significantly more modest effects due to weak longitudinal flow. At the later times, when transverse velocity gradients start to build up, it takes longer before viscous corrections in IS theory catch up with these changes, and therefore the difference from the ideal solution increases more slowly compared to the BDNK solution. Lastly, on the right panel, we show the normalized mean average difference from the ideal fluid reference for $\varepsilon$ as a function of longitudinal proper time. We see that the results vary greatly between the different BDNK frames. This clearly demonstrates how changes in frame parameters significantly alter the meaning of the equilibrium quantities, even though $T^{\tau \mu}$ remains nearly unchanged. We want to emphasize that even though the differences between the BDNK and IS seem to vanish when the longitudinal flow is small, this does not mean that there cannot be large differences coming from transverse dynamics. In fact, one would expect the differences to always be notable when the system is initially far from equilibrium. Therefore, the differences arising from transverse flow are expected to be more significant in a situation where the initial $\sigma^{\mu \nu}$ or $\partial_\mu u^\mu$ have large contributions from the transverse components of the velocity.

\begin{figure}
    \centering
    \includegraphics[width=0.49\linewidth]{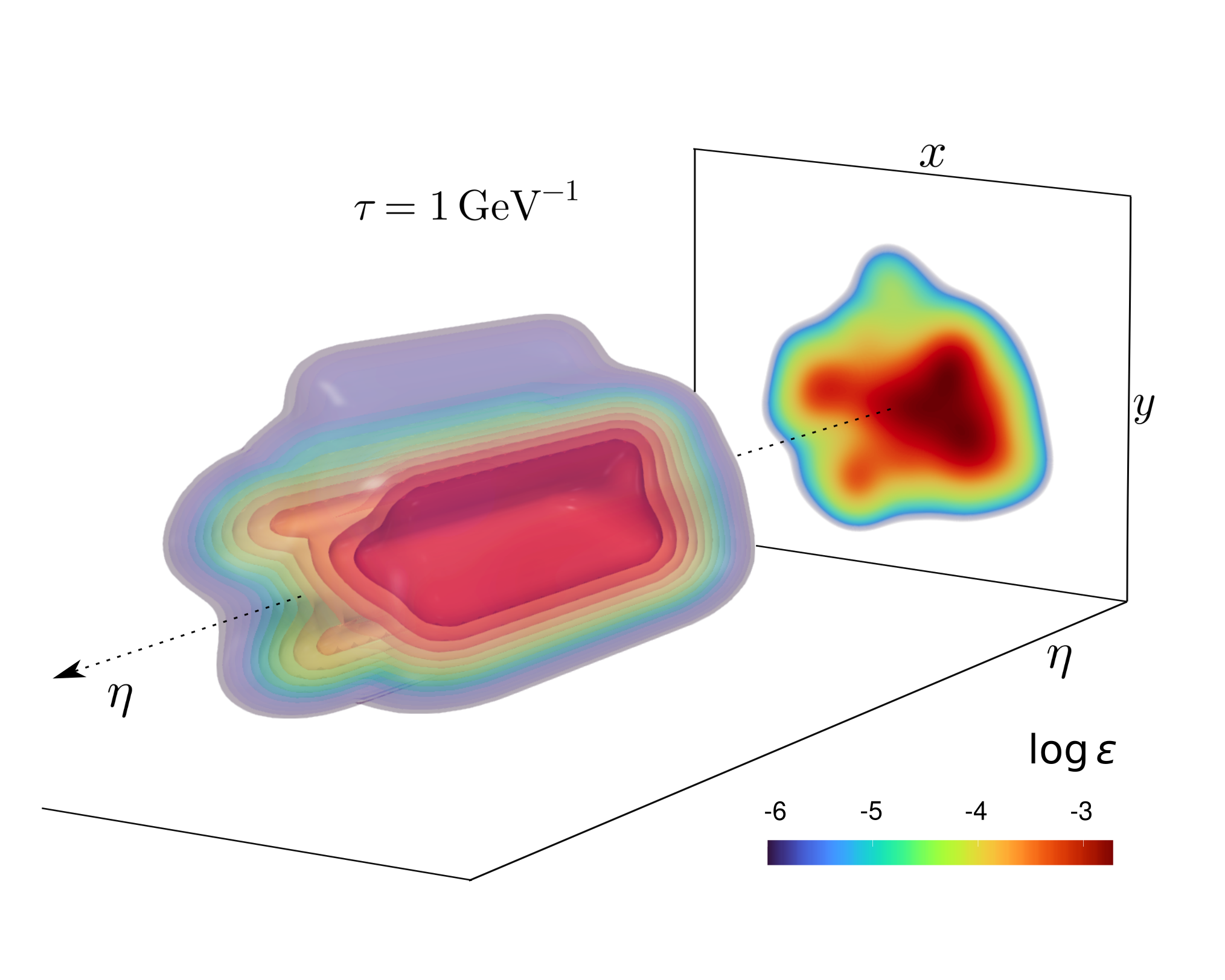}
    \includegraphics[width=0.49\linewidth]{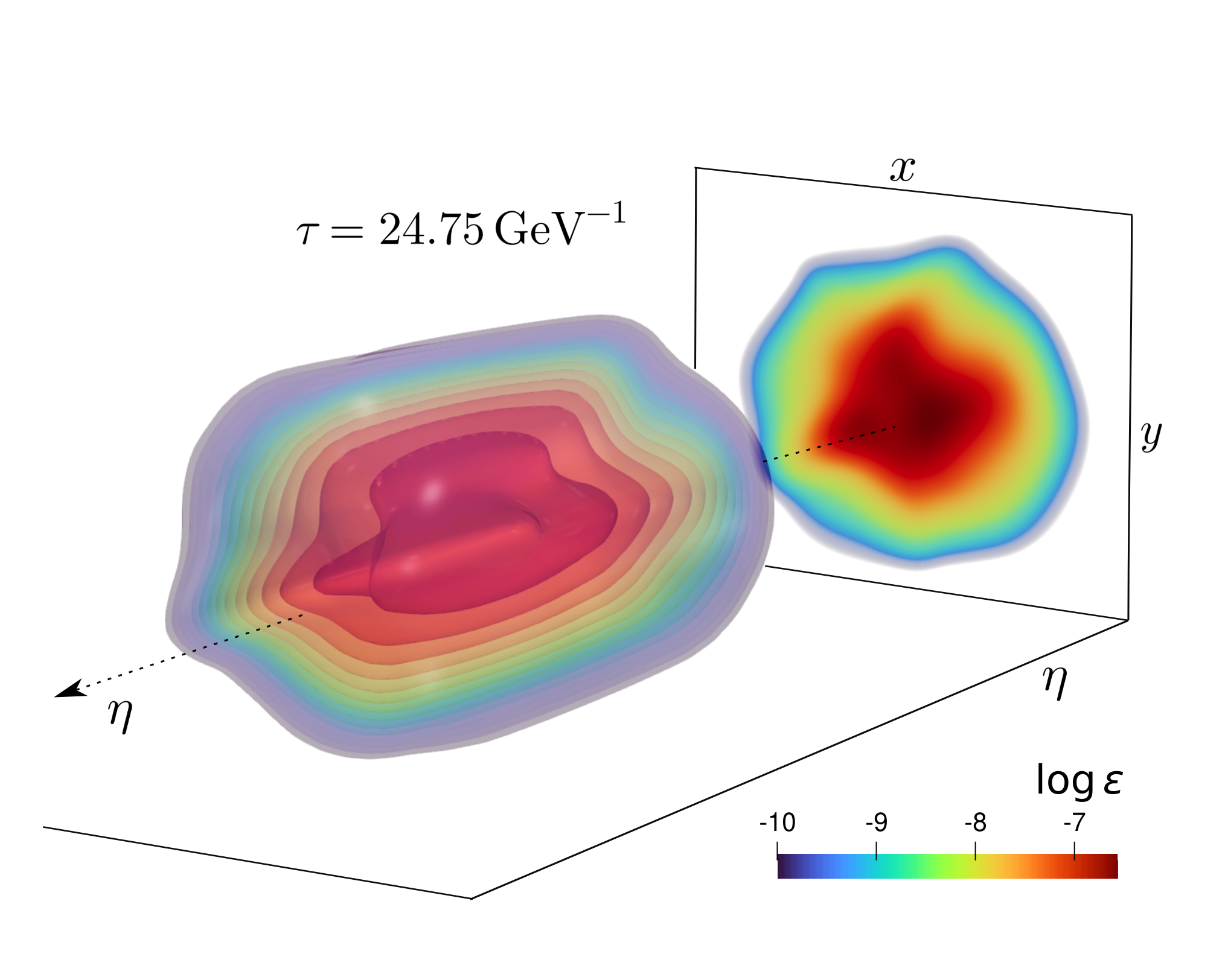}
    \caption{3D contours of $\log\varepsilon$ for the TRENTo initial conditions and the HotQCD-based EOS~(Sec.~\ref{SSS:QCD-EOS}) at the initial and final simulation times, as indicated in the legends.
    The space-time rapidity coordinate $\eta$ is indicated by the arrow originating at the center of the $xy$-plane.
    A 2D heatmap at midrapidity, $\log\varepsilon(\eta=0)$, is projected onto the back face of the simulation box for visualization.
    The simulations are performed on the $(x,y)\in[-15,15]\,\mathrm{fm}$ and $\eta\in[-10,10]$ domain, with spatial resolutions $\delta x=\delta y=(15/128)\,\mathrm{fm}$ and $\delta\eta=(20/63)$, and $\delta\tau=\delta x/10$.
    Here, $\eta_0=0.2$, $\chi_0=(25/4)\eta_0$ and $\lambda_0=(25/7)\eta_0$.}
    
    \label{fig:Trento_3D_viz}
\end{figure}

For illustrative purposes, we also performed an initial test for a simple 3D initial condition in Milne, where the transverse energy density profile $\varepsilon(\bf r, \tau_0)$ is given by the TRENTo profile from Eq.~\ref{eq:trento}, and the $\eta_s$-dependence is given by:
\begin{equation}
    \varepsilon({\bf r}, \eta_s, \tau_0) = \varepsilon({\bf r}, \tau_0) \exp \left[ -\frac{\left(|\eta_s|-\eta_{\rm flat}/2\right)^2}{2\sigma_\eta^2} \theta\left(|\eta_s|-\eta_{\rm flat}/2\right)\right],
\end{equation}
where $\theta$ is the Heaviside step function. For the parameters describing the shape of the $\eta_s$ profile, we used values $\eta_{\rm flat} = 4.0$, and $\sigma_\eta=1.0$. In order to make the initial data to obey the weak energy condition at large rapidities and small densities, we had to regulate the behavior of the shear viscosity coefficient $\eta$ at low densities as:
\begin{equation}
    \eta_{\rm regulated} = \eta\left[1-\exp\left(-\frac{(\varepsilon/\varepsilon_0)^{1/4}}{2}-\frac{\varepsilon}{\varepsilon_{\eta,0}}\right)\right],
\end{equation}
where $\varepsilon_{\eta,0}=(0.06 \rm~GeV)^4$. Note that $\eta_{\rm regulated} \approx \eta$, when $\varepsilon\gg \varepsilon_{\eta,0}$.  The Fig.~\ref{fig:Trento_3D_viz} shows the 3D profile of $\log \varepsilon$ at the initial and final simulation times. This demonstrates that our BDNK implementation can handle realistic initial conditions for heavy-ion collisions also in 3D. However, one has to be cautious of initializing BDNK hydrodynamics at very early times or large value of $\eta$ , since this can very easily lead to the violation of the weak energy condition at large rapidities and small densities, due to the large viscous corrections originating from the geometric expansion.

\newpage

\draftpart{Conclusions and Outlook}
\label{Part:Conclusion}

We have presented two mathematically equivalent formulations of the BDNK relativistic viscous hydrodynamics framework: the first-order (\ac{FO}, Section~\ref{Sec:First_Order}) and mixed-order (\ac{MO}, Section~\ref{Mixed_order_section}) formulations.
The causality and local well-posedness of these formulations are established on prescribed spacetime backgrounds under the stated hypotheses. The formulations can, in principle, be coupled to Einstein's equations, but extending the local well-posedness results to the coupled system requires separate verification of the remaining hypotheses (see Remark~\ref{R:Analyticity_hypothesis}). Only stationary spacetimes were studied numerically in the present work.

This work presented a number of new mathematical results that establish a rigorous basis for the system of equations implemented in our simulations. In fact, under the characteristic, regularity, and analyticity assumptions stated in Sections~\ref{Sec:LWP} and~\ref{Sec:Proof_MO}, in this work we proved local existence and uniqueness for Sobolev initial data, together with continuous dependence on those data in Sobolev topologies strictly below the highest controlled regularity. It is important to stress that these results apply to the nonlinear systems in $3+1$ dimensions without imposing spatial symmetries, on prescribed spacetime backgrounds satisfying the stated hypotheses. As such, they establish the existence of local, well-defined solutions for the continuum formulation of the theory, which is crucial for the numerical implementations pursued in this work.

Concerning the first-order formulation, we mathematically proved that the constraints introduced by promoting spatial derivatives to independent variables are preserved, which in turn establishes equivalence with the original BDNK equations for compatible initial data. Furthermore, we also established local well-posedness of the system augmented by the needed cleaning fields. Thus, our mathematical analysis includes the full set of variables that may be used to monitor and control (inevitable) numerical violations of the constraints. The corresponding result for the rescaled formulation is established in Appendix~\ref{S:Rescaled_local_well_posedness}. Moreover, for the mixed-order formulation, we also proved equivalence with the original BDNK equations, though in this case there was no need to introduce independent spatial-derivative variables or additional constraints that require cleaning.

From a mathematical perspective, a central new ingredient provided in this work is the general strictly hyperbolic reduction theorem presented in Section~\ref{Sec:SH} where, for a class of quasilinear mixed-order systems involving equations of arbitrary differential order, we give sufficient structural conditions under which their solutions satisfy a system with a diagonal, strictly hyperbolic principal part. It is important to remark that this construction removes repeated characteristic factors while retaining the distinct characteristic roots, which, within the class covered by our structural assumptions, strengthens the famous weakly hyperbolic reduction result of Choquet-Bruhat~\cite{Choquet-Bruhat-1966}. We further combine this reduction with analytic approximation and Sobolev energy estimates to establish local existence, uniqueness, and continuous dependence for the original mixed-order system. Therefore, we believe that the resulting new theorems are of independent general interest, as they can be applied in other systems beyond our original fluid mechanics motivation.

Since our developed formalisms are flux conservative, we can directly leverage standard high-resolution shock-capturing techniques commonly used for computational relativistic fluid dynamics. 
In fact, 
using
(i) WENO-Z or minmod reconstruction,
(ii) Kurganov-Tadmor~(KT) flux solver, and
(iii) Heun's Runge-Kutta scheme time-stepping,
we have numerically implemented these two formulations.
The \ac{MO} implementation separates the flux into terms that includes spatial derivatives and those that do not, using the KT scheme adapted for diffusion-type problems to treat the former. Spatial derivatives appearing in the source terms are evaluated separately using CWENO.
Our results highlight the contrasting advantages and limitations in terms of numerical implementations. 
While the \ac{FO} system involves considerably more dynamical variables, it avoids the additional numerical algorithms and considerations associated with higher-order spatial derivatives. Furthermore, although both implementations exhibit numerical convergence, the \ac{MO} formulation generally yields larger errors than the \ac{FO} formulation at same resolutions in the tests considered here.

We performed an array of tests to validate our numerical implementations of \ac{FO} and \ac{MO}. With the smooth shock and the Kelvin-Helmholtz instability tests (Section~\ref{S:Cartesian_tests}), we verified that our results converge to the ideal limit at small shear viscosity --- a non-trivial limiting case for viscous hydrodynamic implementations. Additionally, we remark that the smooth-shock calculations found in this work reproduce the viscous broadening of shock profiles discussed in \cite{Clarisse:2025lli}, while the Kelvin--Helmholtz simulations show the expected suppression of vortical structures as the viscosity increases. These tests therefore probe both the approach to ideal hydrodynamics and the physical effects of finite viscosity.
For both examples, we also verified that our results did not depend meaningfully on choice of hydrodynamic frame parameters at sufficiently small viscosity, although the dependence increased with viscosity. 
We  introduced a practical prescription for selecting admissible hydrodynamic frames beyond conformal fluids. For vanishing thermal conductivity and $V=4\eta/3+\zeta>0$, the parametrization $\tau_\varepsilon=\chi V/(\varepsilon+P)$, $\tau_Q=\lambda V/(\varepsilon+P)$, and $\tau_P=\xi\tau_\varepsilon$ nicely reduces the nonlinear causality and linear stability conditions to inequalities for dimensionless coefficients, with no explicit dependence on the shear or bulk viscosity. This is very useful because, given bounds on the relevant equation-of-state derivatives over the thermodynamic domain of interest, one can have an \textit{a priori} choice of frame parameters satisfying these conditions throughout that domain. It would be interesting to extend this procedure to also consider the case of a nonzero thermal conductivity in future simulations.

We further compared our \ac{FO} and \ac{MO} formulations with the exact BDNK Bjorken and Gubser solutions, which can be calculated from reduced ODEs for given choices of frame parameters (Section~\ref{S:Milne_coordinates}). The solutions enabled us to study the accuracy of the \ac{FO} and \ac{MO} formalisms, as well as the performance of the WENO-Z and minmod reconstruction methods. The \ac{FO} formulation with WENO-Z was found to perform particularly well in terms of stability and numerical accuracy. Our resolution studies also test numerical convergence under refinement, which is distinct from the approach to ideal hydrodynamics as the viscosity decreases. 
For curved spacetimes, we studied radial accretion onto a Schwarzschild black hole (Michel accretion), finding negligible departures from stationarity at small shear viscosity and only weak dependence on the frame parameters. 
We also investigated the Kerr torus system, which similarly exhibits negligible frame dependence but substantially larger departures from the initial stationary solution, even at the smallest shear viscosity parameters that we studied. 
This may be associated with limitations of our numerical setup, and addressing these limitations to enable reliable simulations at smaller transport coefficients and over longer evolution times remains an important direction for future work.

We also used our numerical framework to investigate heavy-ion collision applications by considering boost-invariant simulations in $2+1$ dimensions with a realistic, nonconformal equation of state obtained by matching lattice-QCD results to a hadron resonance gas. Together with fluctuating TRENTo initial conditions, these calculations extend our tests to transverse configurations without imposed spatial symmetries and with a temperature-dependent speed of sound, which provides a new starting point for future applications of the BDNK equations in realistic hydrodynamic simulations of the QGP.

For the sake of completeness, in this paper we also compared BDNK and Israel-Stewart for initial conditions relevant for heavy-ion collisions (this was done using the minimal Israel--Stewart theory specified in Section~\ref{S:Israel_Stewart_comparisons}). As expected, BDNK and Israel-Stewart were found to converge at small viscosity. The frame parameter dependence of the BDNK solution was small in our tests and, in fact, it was of similar size to the effect of varying the relaxation time in Israel-Stewart solutions. At large viscosity, the frame parameter dependence of BDNK remained modest, much smaller than the difference between BDNK and Israel-Stewart. 
In this context, we remind the reader that in Israel-Stewart theory, the shear stresses are additional dynamical variables whose initial values can be specified independently of the initial velocity gradients. However, in BDNK, the dissipative stresses are determined by derivatives of the hydrodynamic fields and, thus, they cannot be prescribed independently of those fields and their corresponding derivatives. Our comparisons thus account for this distinction by initializing the two theories with the same full energy-momentum tensor.

We note that the choice of initial data also affects the observed frame dependence within BDNK. The initialization Method-1, which sets the initial $T^{0\mu}$ components equal to their ideal-fluid values, shows particularly weak frame dependence in the smooth-shock tests. However, this prescription restricts the initial data by fixing the initial time derivatives once the hydrodynamic fields are specified. It also does not generally set the spatial stresses $T^{ij}$ equal to their ideal-fluid values. Therefore, it would be important to develop more general initialization procedures and check their corresponding hydrodynamic frame dependence. Of course, in this case, one must determine which target energy-momentum tensors can actually be represented by admissible BDNK initial data so that one can then construct the corresponding hydrodynamic fields and time derivatives in a consistent fashion.

Unavoidably, statements about frame parameter independence are system dependent, both at the level of the system's configuration, and its equation of state and transport coefficients, and general statements cannot be made.
In this regard, our conclusions regarding frame independence are specifically for components of the energy-momentum tensor since, as we have demonstrated here, in various examples the hydrodynamic fields themselves, such as the energy density for example, can display large dependence on frame parameters. This is, of course, expected by construction since the frame parameters are intrinsically related to the out-of-equilibrium definition of the hydrodynamic variables used in the simulation. In fact, the same statement applies to temperature, whose value is inferred from the equation of state using the hydrodynamic fields defined in the given chosen frame. Thus, away from equilibrium, different prescriptions can give different temperature fields even when the evolved energy-momentum tensor components are frame robust.

Our work connects the deep and rich underlying mathematical structure of BDNK hydrodynamics to robust formulations suitable for numerical evolution. It provides a starting point for future systematic  investigations of BDNK theory in more realistic heavy-ion collision applications which involve comparison to data. Finally, the series of new developments reported in this work paves the way for future simulations of the coupled Einstein--BDNK system of equations, which could allow one to investigate how causal dissipation shapes black-hole formation triggered by collapsing viscous fluids, providing a new connection between the microscopic transport properties of relativistic matter and the dynamics of spacetime.

\section*{Acknowledgments}

We thank N.~Clarisse, L.~Gavassino, M.~Hippert, E.~Most, E.~O.~Pinho, G.~Soares Rocha and M.~Singh for useful discussions. We also acknowledge the open source code accompanying~\cite{Pandya_2022} as a crucial starting point/reference.
TP gratefully acknowledges support from the Vanderbilt Initiative for Gravity, Waves, and Fluids, supported by Vanderbilt's College of Arts \& Sciences.
HH gratefully acknowledges support from DOE grant DE-SC0024711.
FSB gratefully acknowledges support from the Vanderbilt Initiative for Gravity, Waves, and Fluids, supported by Vanderbilt's College of Arts \& Sciences. Part of this work was completed as FSB was a Research Visiting Professor at Vanderbilt University.
MMD gratefully acknowledges support from NSF grants DMS-2406870 and NRT-2125764, and from DOE grant DE-SC0024711.
JN was partly supported by the US Department of Energy, Office of Science, Office for Nuclear Physics under award no.~DE-SC0023861.
JFP\ gratefully acknowledges support by Vanderbilt University and by
the U.S. Department of Energy, Office of Science under Award Numbers DE-SC0024711 and DE-SC-0024347.
This research used resources of the National Energy Research Scientific Computing Center (NERSC), a Department of Energy User Facility using NERSC award NP-ERCAP 35437.

\bigskip
\section*{Disclosure on the use of Artificial Intelligence (AI)} AI tools (ChatGPT Pro 6 Sol and Astra and Claude Opus 5.5) were used to check the text for grammatical errors, spelling mistakes, and typos, and to improve the wording in a few places. 
AI has also been used to generate the table of Section~\ref{S:How_to_read} and to convert some handwritten calculations and proofs into LaTeX code. 
Certain plotting scripts, particularly towards the end for minor adjustments to the figures, used AI generated code.
Finally, after the manuscript was largely ready, AI was used to review the theoretical and mathematical arguments. Minor mistakes were then identified and corrected by the authors. All the intellectual content of this work is the authors' own.

\appendix

\section{Rescaled Expressions}
\label{Appendix:Rescaled EOMs}
We introduce a rescaling of the \ac{FO} (Equation~\eqref{FO:Matrices}) and \ac{MO} (Equation~\eqref{MO:eoms-Matrix}) equations of motion (EOMs) used in our numerical implementation. 
The rescaling was developed to improve numerical stability. 
Local well-posedness of the rescaled formulation is proved in Section~\ref{S:Rescaled_local_well_posedness}.
We define the rescaled BDNK energy-momentum tensor by
\begin{equation}\label{tilde_T0mu_def}
\tilde{T}^{\mu \nu }=T^{\mu \nu }\varepsilon^{-3/4}\,.
\end{equation}
Applying this definition to the expressions of Section~\ref{S:Auxiliary_definitions}, we obtain
\begin{align}
\label{defineBDNKnovo_rescaled}
\tilde{T}^{\mu \nu } &= \tilde{T}_0^{\mu\nu} + \frac{1}{\varepsilon}\tilde{H}^{\mu\nu\alpha\beta}\nabla_\alpha (\varepsilon u_\beta)+2\varepsilon^{-3/4}\beta_n u^{(\mu} \Omega^{j\nu)}\partial_j n\nonumber\\
&= \tilde{T}_0^{\mu\nu} + \tilde{H}^{\mu\nu\alpha\beta}\left (\nabla_\alpha  u_\beta+u_\beta\nabla_\alpha\theta\right )+2\varepsilon^{-3/4}\beta_n u^{(\mu} \Omega^{j\nu)}\partial_j n,
\end{align}
where
\begin{align*}
\tilde{T}_0^{\mu\nu} = \varepsilon^{1/4}\left[u^\mu u^\nu + \frac{P}{\varepsilon} \Delta^{\mu\nu}\right],
\end{align*}
and
\begin{align}
\tilde{H}^{\mu\nu\alpha\beta} &= \tilde{\tau}_\varepsilon u^\mu u^\nu \left[(1+\frac{P}{\varepsilon}) \Delta^{\alpha\beta}- u^\alpha u^\beta\right]+\Delta^{\mu\nu}\left[(\tilde{\tau}_P(1+\frac{P}{\varepsilon})-\tilde{\zeta}) \Delta^{\alpha\beta}- \tilde{\tau}_P u^\alpha u^\beta\right]- 2\tilde{\eta} \Delta^{\mu\nu\alpha\beta} \nonumber \\ 
&+ \tilde{\tau}_Q(1+\frac{P}{\varepsilon})u^\mu u^\alpha \Delta^{\beta\nu}+\tilde{\tau}_Q(1+\frac{P}{\varepsilon}) u^\nu u^\alpha \Delta^{\mu\beta}-\tilde{\beta}_\varepsilon u^\mu u^\beta \Delta^{\alpha\nu}-\tilde{\beta}_\varepsilon u^\nu u^\beta \Delta^{\mu\alpha}\nonumber\\
&-\frac{ n \tilde \beta_n}{u^0}\left (u^\mu \Delta^{\nu0}+u^\nu \Delta^{\mu0}\right )\Delta^{\alpha\beta}.
\label{defineBDNK_H_rescaled}
\end{align} 
The rescaled transport coefficients are defined as
\begin{align}
\label{Defs:rescaled_transport_coefficients}
    \tilde{\tau}_a = \tau_a \varepsilon^{1/4}, \tilde{\eta}=\eta\varepsilon^{-3/4}, \tilde{\zeta}=\zeta\varepsilon^{-3/4}, \tilde{\beta}_\varepsilon = \beta_\varepsilon \varepsilon^{1/4}, \tilde \beta_n = \beta_n \varepsilon^{-3/4}
\end{align}
where $a\in\{\varepsilon,P,Q\}$.
Using the \ac{FO} variables defined in Equation~\eqref{New_variables}, the rescaled energy-momentum tensor becomes
\begin{align}
\label{Rescaled_Tmunu_}
\tilde T^{\mu\nu} &= \tilde T_0^{\mu\nu} + \tilde H^{\mu\nu 0\beta} P_{A\beta}Y^A+\tilde H^{\mu\nu j \beta} P_{A\beta}X_j^A
+2\tilde \beta_n u^{(\mu} \Omega^{j\nu)} Z_j 
+ \tilde H^{\mu\nu\alpha\beta} C_{j\beta} \Gamma^j_{\alpha \lambda}u^\lambda
\end{align}
The rescaled counterpart of the matrix $M$ introduced in Section~\ref{Sec:First_Order} is
\begin{align}
\label{M rescaled def}
\tilde M^{\mu\beta}\equiv \tilde H^{\mu 00 \beta}.
\end{align}
Following the inversion procedure of Section~\ref{Inverse_M_section}, we write
\be
\label{Inverse_M_rescaled}
\tilde M^{-1}_{\lambda\mu}=\tilde{A} u_\lambda u_\mu+\tilde{B} u_\lambda \Delta^0_\mu+\tilde{C}\Delta^0_\lambda u_\mu+\tilde{D}\Delta^0_\lambda \Delta^0_\mu+\tilde{E}\Delta_{\lambda\mu},
\ee
where
\bml\label{Inverse_M_rescaled_coefficients}
\bea
\tilde{A}&=&\frac{\Delta^{00} \left[3(1+\frac{P}{\varepsilon}) \tilde\tau_P-3 \tilde\zeta -4\tilde\eta-3n\tilde\beta_n \right]+3(u^0)^2 (1+\frac{P}{\varepsilon})\tilde\tau_Q}{ \tilde{F}},\\
\tilde{B}&=&\frac{3\left [(u^0)^2 (1+\frac{P}{\varepsilon})\left(\tilde\tau_Q+\tilde\tau_{\varepsilon }\right)-\Delta^{00}n \tilde\beta _n\right ]}{ u^0 \tilde{F}},\\
\tilde{C}&=& \frac{- 3 u^0 (\tilde\beta_\varepsilon+\tilde\tau_P)}{\tilde{F}},\\
\tilde{D}&=&\frac{ \left\{(u^0)^2  \left[3 (1+\frac{P}{\varepsilon})\tilde\tau_Q\left (\tilde\tau_P + \tilde\beta _{\varepsilon }\right )+\tilde\tau_{\varepsilon } \left(3 \tilde\zeta+3 (1+\frac{P}{\varepsilon}) \tilde\beta_s +\tilde\eta\right ) \right]+\Delta^{00}\left [\tilde\beta_{\varepsilon } \left(3\tilde\zeta +\tilde\eta \right)- 3(1+\frac{P}{\varepsilon})\tilde\tau_P \tilde\beta_s\right ]\right\}}{(\Delta^{00} \tilde\eta -(u^0)^2 (1+\frac{P}{\varepsilon})\tilde\tau_Q)\tilde{F}}, \nonumber \\ \\
\tilde{E}&=& \frac{1}{(1+\frac{P}{\varepsilon})\tilde\tau_Q(u^0)^2-\tilde\eta \Delta^{00}}\\
\tilde{F}&=&(\Delta^{00})^2\left [\tilde\beta_{\varepsilon } \left(3 \tilde\zeta +4 \tilde\eta \right)-3 (1+\frac{P}{\varepsilon})\tilde\tau_P\tilde\beta_s\right ]+\Delta^{00}(u^0)^2 \left[\tilde\tau_{\varepsilon } \left(3 \tilde\zeta+3 (1+\frac{P}{\varepsilon})\tilde\beta _{s} +4 \tilde\eta \right)+3 (1+\frac{P}{\varepsilon})\tilde\tau _P \tilde\tau_Q\right ] \nonumber \\
&&-3 (u^0)^4 (1+\frac{P}{\varepsilon})\tilde\tau_Q \tilde\tau_{\varepsilon },
\eea
\eml
where $\tilde\beta_s=\tilde\beta_\varepsilon+n \tilde\beta_n/(1+\frac{P}{\varepsilon})$.
The numerical motivation for the rescaling in Equations~\eqref{tilde_T0mu_def} and \eqref{Defs:rescaled_transport_coefficients} is apparent upon comparing the inverse-matrix coefficients in Equations~\eqref{Inverse_M} and \eqref{Inverse_M_coefficients} with their rescaled counterparts in Equations~\eqref{Inverse_M_rescaled} and \eqref{Inverse_M_rescaled_coefficients}. 
We observed that in one of the iterations of our numerical implementation of the original unscaled system in Equation~\eqref{FO:Matrices}, explicit inverse powers of $\varepsilon$ in the original coefficients contributed to numerical instabilities, particularly near sharp transitions to low energy density ($\varepsilon\rightarrow0$) in simulations using TRENTo initial conditions (Section~\ref{S:Trento_initial_conditions}). 
The rescaled coefficients eliminate these explicit factors from the denominators, improving numerical stability.

The relations in Equations~\eqref{Relations} and \eqref{Relations-Tij-FO} that close the system of equations become
\begin{align}
\label{Relations-rescaled}
Y^A  &=\tilde{P}^{A\lambda}\tilde M^{-1}_{\lambda\mu}\left [\tilde T^{0\mu}- \tilde T_0^{0\mu} - \tilde H^{\mu0 k \beta}P_{B\beta}X_k^B-2\tilde\beta_n u^{(\mu} \Omega^{k0)} Z_k - \tilde H^{\mu 0 \alpha\beta} C_{k\beta} \Gamma^k_{\alpha \lambda}u^\lambda\right ],\\
\label{Relations-Tij-FO-rescaled}
\tilde T^{ij} &= \tilde T_0^{ij} + \tilde H^{ij 0\beta} P_{A\beta} Y^A
+\tilde H^{ij k \beta}P_{A\beta}X_k^A+2\tilde\beta_n u^{(i} \Omega^{kj)} Z_k + \tilde H^{ij\alpha\beta} C_{k\beta} \Gamma^k_{\alpha \lambda}u^\lambda.
\end{align}
The relations for $Y^A$ remain unchanged w.r.t. Equation~\eqref{Relations} and the expressions for $T^{ij}$ retain the functional form in Equation~\eqref{Relations-Tij-FO}, with the constituents in the R.H.S. simply replaced by appropriate rescaled quantities.

The rescaling also modifies the conservation equations through the identity $\partial_\mu T^{\mu \nu}= \partial_\mu (\varepsilon^{3/4}\tilde T^{\mu \nu})=\varepsilon^{3/4}  \partial_\mu \tilde T^{\mu \nu}+\frac{3}{4}\varepsilon^{-1/4}\tilde T^{\mu \nu}\partial_\mu \varepsilon$. The resulting flux-conservative system is

\begin{align}
\label{Matrix EOM general form rescaled}
\partial_t \mathcal{\tilde{Q}}+\partial_i \tilde{F}_i=\mathcal{\tilde{S}},
\end{align}
where 
\begin{align}
\label{FO:Matrices-rescaled}
\mathcal{\tilde{Q}}=\bbm \tilde T^{0\mu}\\ X_{1}^A\\ X_{2}^A \\ X_{3}^A \\ Z_1 \\ Z_2 \\ Z_3 \\ U^A \\ n \\  \varphi_{1}^A \\ \varphi_{2}^A \\ \varphi_{3}^A \\ \phi^A \\  \Psi_1\\ \Psi_2\\ \Psi_3\\ \psi 
\ebm, \quad 
\tilde{F}_i=\bbm \tilde T^{i\mu} \\ 
\epsilon_{1ik}\varphi_{k}^A-\delta_{1i} Y^A \\ 
\epsilon_{2ik}\varphi_{k}^A-\delta_{2i}Y^A \\ 
\epsilon_{3ik}\varphi_{k}^A-\delta_{3i}Y^A \\ 
\epsilon_{1ik}\Psi_k-\delta_{1i}S_n\\
\epsilon_{2ik}\Psi_k-\delta_{2i}S_n\\
\epsilon_{3ik}\Psi_k-\delta_{3i}S_n\\
0_{4\times 1}\\
0\\
-\kappa_{1}^A \epsilon_{1ik}X_{k}^A+\delta_{1i}\phi^A\\ 
-\kappa_{1}^A \epsilon_{2ik}X_{k}^A+\delta_{2i}\phi^A\\
-\kappa_{1}^A \epsilon_{3ik}X_{k}^A+\delta_{3i}\phi^A\\
\kappa_{2}^A \varphi_{i}^A \\ 
-\lambda_{1} \epsilon_{1ik}  Z_k +\delta_{1i}\psi\\
-\lambda_{1} \epsilon_{2ik}  Z_k +\delta_{2i}\psi\\
-\lambda_{1} \epsilon_{3ik}  Z_k +\delta_{3i}\psi\\
\lambda_{2}\Psi_i
 \ebm,\quad\text{and}\quad 
\tilde{\mathcal{S}}=\bbm \tilde S^\mu \\ 0_{4\times 1} \\ 0_{4\times 1} \\ 0_{4\times 1} \\  0 \\ 0 \\ 0  \\  Y^A \\ S_n \\ -\beta_{1}^A \varphi_{1}^A \\ -\beta_{1}^A \varphi_{2}^A \\ -\beta_{1}^A \varphi_{3}^A \\ 
-\beta_{2}^A \phi^A \\-\sigma_{1} \Psi_1\\-\sigma_{1} \Psi_2\\-\sigma_{1} \Psi_3\\-\sigma_{2} \psi \ebm
\end{align} 
are real $44\times 1$ column vectors. 
The quantity $S_n$ is defined in Equation~\eqref{Sn} and remains unchanged.
The rescaled source term is 
\begin{equation}\label{souce-FO-rescaled}
\tilde  S^\mu\equiv -\Gamma^\nu_{\nu 0}\tilde T^{0\mu}-\Gamma^\mu_{0\nu}\tilde T^{0\nu}
-\Gamma^\nu_{\nu i}\tilde T^{i\mu}-\Gamma^\mu_{i\nu}\tilde T^{i\nu}-\frac{3}{4}(
\tilde T^{0\mu} Y^\theta
+\tilde T^{i\mu}X_i^\theta).     
\end{equation}

In our numerical implementation, we evolve the transformed variables defined by Equations~\eqref{tilde_T0mu_def} and \eqref{Defs:rescaled_transport_coefficients}. 
The resulting flux-conservative formulation, Equation~\eqref{Matrix EOM general form rescaled}, retains the structure of the original system, with $Y^A$ unchanged,  the rescaled quantities $\tilde{T}^{\mu\nu}$, and $\tilde{S}^\mu$ and the additional source contributions in Equation~\eqref{souce-FO-rescaled}.

In summary,
\begin{itemize}
    \item In our numerical implementation, we adopt the redefinitions in Equations~\eqref{tilde_T0mu_def} and \eqref{Defs:rescaled_transport_coefficients} and solve for the corresponding rescaled variables.
    Under these redefinitions, all terms appearing in the resulting equations of motion remain finite in the low-energy-density limit.
    Although the original expressions do not contain genuine physical singularities, their explicit dependence on inverse powers of $\varepsilon$ can introduce numerical instabilities when evaluated at very small energy densities.
    The rescaled formulation avoids these difficulties by eliminating divisions by vanishing quantities in our numerical implementation. 

    \item The equations of motion for the rescaled variables, given in Equation~\eqref{Matrix EOM general form rescaled}, retain the same flux-conservative structure as the original \ac{FO} formulation~\eqref{FO:Matrices}. 
    The expressions in equations~\eqref{FO:Matrices-rescaled} and \eqref{souce-FO-rescaled} use $Y^A$ unchanged and the rescaled quantities $\tilde{T}^{\mu\nu}$ and $\tilde{S}^\mu$.

    \item The rescaled formulation also allows the equations of motion to be expressed in terms of the ratio $P/\varepsilon$ and the rescaled transport coefficients defined in Equation~\eqref{Defs:rescaled_transport_coefficients}.
    Besides yielding more compact expressions, this formulation offers numerical advantages for both analytical and tabulated EoS.
    In particular, the ratio $P/\varepsilon$ is well defined in the low-energy-density limit for a broad class of fluids. 
    This is particularly advantageous when using tabulated EoS, such as the HotQCD-based EoS employed in this work, where interpolation errors in the low-density regime can further amplify numerical inaccuracies.

\end{itemize}

\subsection{Local well-posedness of the rescaled formulation}
\label{S:Rescaled_local_well_posedness}

We prove local well-posedness of Equation~\eqref{Matrix EOM general form rescaled} with the source \eqref{souce-FO-rescaled}, allowing nonzero constraint errors and cleaning fields. We use the exact rescalings
\begin{subequations}
\label{E:Rescaled_exact_coefficients}
\begin{align}
\tilde T_0^{\mu\nu}&=\varepsilon^{-\frac{3}{4}}T_0^{\mu\nu},
\\
\tilde H^{\mu\nu\alpha\beta}&=\varepsilon^{-\frac{3}{4}}H^{\mu\nu\alpha\beta},
\\
\tilde\beta_n&=\varepsilon^{-\frac{3}{4}}\beta_n,
\\
\tilde M^{\mu\beta}&=\varepsilon^{-\frac{3}{4}}M^{\mu\beta},
\\
\tilde M^{-1}_{\lambda\mu}&=\varepsilon^{\frac{3}{4}}M^{-1}_{\lambda\mu}.
\end{align}
\end{subequations}
In Equation~\eqref{souce-FO-rescaled}, $Y^\theta$ is the $\theta$ component of $Y^A$ defined by Equation~\eqref{Relations-rescaled}. Indeed, substitution of Equations~\eqref{E:Rescaled_exact_coefficients} and $T^{0\mu}=\varepsilon^{\frac{3}{4}}\tilde T^{0\mu}$ into Equation~\eqref{Relations-rescaled} gives exactly Equation~\eqref{Relations}. Thus, $Y^A$ is not rescaled. Equation~\eqref{Relations-Tij-FO-rescaled} also gives $T^{ij}=\varepsilon^{\frac{3}{4}}\tilde T^{ij}$.

\begin{proposition}
Let $\Sigma=\mathbb{T}^3$ and let $\ell\geq4$ be an integer. Suppose that the rescaled initial data, after the substitution
\begin{align*}
T^{0\mu}_{(0)}
&=\big(\varepsilon_0e^{U^\theta_{(0)}}\big)^{\frac{3}{4}}
\tilde T^{0\mu}_{(0)},
\end{align*}
with every other initial field unchanged, satisfy the hypotheses of Theorem \ref{Theorem_1_precise}. Here $\varepsilon_0>0$ is the constant in $\varepsilon=\varepsilon_0e^\theta$.

Then Equation~\eqref{Matrix EOM general form rescaled}, with Equations~\eqref{E:Rescaled_exact_coefficients} and \eqref{souce-FO-rescaled}, has a local solution satisfying Equations~\eqref{E:FO_CD_theorem_solution_regularity} and \eqref{E:FO_CD_theorem_time_continuity}, with $T^{0\mu}$ replaced by $\tilde T^{0\mu}$. The solution is unique in this class, with the admissibility condition on the physical jet imposed after the substitution $T^{0\mu}=\varepsilon^{\frac{3}{4}}\tilde T^{0\mu}$. The continuous-dependence conclusion of Theorem \ref{Theorem_1_precise} holds with this replacement as well. In particular, continuous dependence is asserted strictly below the top solution orders, and no additional derivative loss is introduced by the rescaling.
\end{proposition}

\begin{proof}
We retain the constraint errors
\begin{align*}
C_i^A&=X_i^A-\partial_iU^A,
\\
D_i&=Z_i-\partial_i n,
\end{align*}
defined in Equations~\eqref{E:FO_CD_definitions}. The equations for $X,Z,U,n$ and the cleaning fields in Equation~\eqref{FO:Matrices-rescaled} are unchanged after setting $T^{0\mu}=\varepsilon^{\frac{3}{4}}\tilde T^{0\mu}$. Consequently, $C,D$ and the cleaning fields satisfy Equations~\eqref{E:FO_CD_cleaning_system}.

For the energy-momentum equation, the product rule gives
\begin{align*}
\nabla_\nu T^{\mu\nu}
&=\varepsilon^{\frac{3}{4}}
\left(\nabla_\nu\tilde T^{\mu\nu}
+\frac{3}{4}\tilde T^{\mu\nu}\partial_\nu\theta\right).
\end{align*}
Since $\partial_t\theta=Y^\theta$, equation~\eqref{souce-FO-rescaled} therefore yields
\begin{align}
\label{E:Rescaled_unscaled_conservation}
\nabla_\nu T^{\mu\nu}
&=-\frac{3}{4}T^{i\mu}C_i^\theta.
\end{align}
Thus the rescaled system is equivalent to Equations~\eqref{Complete_set_Eq} with Equation~\eqref{Complete_set_Eqa} replaced by Equation~\eqref{E:Rescaled_unscaled_conservation}. This equivalence does not require $C$ or $D$ to vanish.

We now use the mixed-order variables $V=(T^{0\mu},U^A,J^0)$ and the prescribed function $W=(C,D)$ from the proof of Theorem \ref{Theorem_1_precise}. By Equations~\eqref{E:FO_CD_constitutive_identities} and \eqref{E:FO_CD_R_time_definition}, equation~\eqref{E:Rescaled_unscaled_conservation} adds the term
\begin{align}
\label{E:Rescaled_additional_source}
-\frac{3}{4}
\left(T_{\mathrm{MO}}^{i\mu}+\mathcal{R}^{i\mu}[C,D]\right)C_i^\theta
\end{align}
to the right-hand side of Equation~\eqref{E:FO_CD_modified_T_equation}, where $T_{\mathrm{MO}}^{i0}=T^{0i}$. Equations~\eqref{E:FO_CD_modified_U_equation} and \eqref{E:FO_CD_modified_J_equation} remain unchanged.

The formulas \eqref{MO:Transformations} show that $T_{\mathrm{MO}}^{i\mu}$ depends on $T^{0\nu}$ without derivatives and on at most first derivatives of $U^A,J^0$. The expression $\mathcal{R}^{i\mu}[C,D]$ contains no derivative of $C,D$, by Equations~\eqref{E:FO_CD_corrections}. Hence Equation~\eqref{E:Rescaled_additional_source} is a real-analytic function of the physical jet and $W$, involving no derivative of $W$. For the Leray indices
\begin{align*}
m_T&=1,
\\
m_U=m_J&=2,
\\
n_T&=0,
\\
n_U=n_J&=1,
\end{align*}
the allowed source orders in the conservation equations are
\begin{align*}
m_T-n_T-1&=0,
\\
m_U-n_T-1=m_J-n_T-1&=1.
\end{align*}
Thus Equation~\eqref{E:Rescaled_additional_source} satisfies the source-order requirements of Theorem \ref{Theorem_SH_LWP} and leaves the principal matrix \eqref{E:FO_CD_principal_matrix} unchanged. The principal-symbol hypotheses verified in the proof of Theorem \ref{Theorem_1_precise} therefore continue to hold.

The solution of Equations~\eqref{E:FO_CD_cleaning_system} with the prescribed initial data supplies $W$ and satisfies Equation~\eqref{E:FO_CD_shifted_cleaning_regularity}. Its analytic approximations satisfy Equation~\eqref{E:FO_CD_shifted_cleaning_approximation}. The forcing indices remain $\chi_T=1$ and $\chi_U=\chi_J=0$, since Equation~\eqref{E:Rescaled_additional_source} contains no derivative of $W$. With $p=7$, $m_*=1$, and the integer $\ell-1$ in Theorem \ref{Theorem_SH_LWP}, this gives
\begin{align*}
\chi_*&=1,
\\
d_W&=7,
\\
S_W&=\ell+7,
\\
S_T&=\ell+6,
\\
S_U=S_J&=\ell+7.
\end{align*}
In particular, the derivative requirement for the prescribed function remains exactly Equation~\eqref{E:FO_CD_shifted_forcing_count}. The term \eqref{E:Rescaled_additional_source} is polynomial of degree at most two in $W$ and real analytic in the physical jet, so it is defined on the physical open set in Theorem \ref{Theorem_1_precise} and on the entire finite-dimensional space of forcing jets. The compact-containment conclusions for the prescribed functions are unchanged.

There are no new initial compatibility conditions. Indeed, Equation~\eqref{E:Rescaled_additional_source} occurs only in the rows with $n_T=0$, whereas the conditions in the rows with $n_U=n_J=1$ are still precisely Equations~\eqref{E:FO_CD_theorem_initial_traces}. The analytic initial approximations in Equations~\eqref{E:FO_CD_full_data_convergence} and \eqref{E:FO_CD_first_trace_convergence} therefore satisfy the hypotheses of Theorem \ref{Theorem_SH_LWP} for this modified source. Theorem \ref{Theorem_SH_LWP} gives a solution $V$ with the regularity \eqref{E:FO_CD_shifted_physical_output} and the stated continuity strictly below these orders.

Define
\begin{align*}
n&=\frac{J^0}{u^0(U)},
\\
X_i^A&=\partial_iU^A+C_i^A,
\\
Z_i&=\partial_i n+D_i,
\\
\tilde T^{0\mu}&=\varepsilon^{-\frac{3}{4}}T^{0\mu}.
\end{align*}
The constitutive identities \eqref{E:FO_CD_constitutive_identities}, the equations~\eqref{E:FO_CD_cleaning_system}, and \eqref{E:Rescaled_unscaled_conservation} show that these fields solve Equation~\eqref{Matrix EOM general form rescaled} with the prescribed data.

We verify that the rescaling preserves the asserted Sobolev regularity. For every $s>\frac{3}{2}$ and $R>0$, Sobolev multiplication and composition give
\begin{align}
\label{E:Rescaled_composition_estimate}
\left\|
e^{a\theta_1}v_1-e^{a\theta_2}v_2
\right\|_{H^s(\Sigma)}
&\leq K_{s,R}
\left(
\|v_1-v_2\|_{H^s(\Sigma)}
+\|\theta_1-\theta_2\|_{H^s(\Sigma)}
\right),
\end{align}
for $a\in\{-\frac{3}{4},\frac{3}{4}\}$ whenever $\|\theta_b\|_{H^s(\Sigma)}+\|v_b\|_{H^s(\Sigma)}\leq R$, $b=1,2$. The constant $K_{s,R}$ is independent of the two pairs of functions. Fixed factors $\varepsilon_0^a$ do not affect the regularity. For time derivatives,
\begin{align}
\label{E:Rescaled_time_derivatives}
\partial_t^j\big(e^{a\theta}T^{0\mu}\big)
&=\sum_{b=0}^j\binom{j}{b}
\partial_t^b(e^{a\theta})\,
\partial_t^{j-b}T^{0\mu}.
\end{align}
The chain rule expresses $\partial_t^b(e^{a\theta})$ as a finite sum of products of time derivatives of $\theta$ of total order $b$, multiplied by $e^{a\theta}$. By Equations~\eqref{E:FO_CD_shifted_physical_output}, every factor on the right-hand side of Equation~\eqref{E:Rescaled_time_derivatives} belongs to $H^{\ell+6-j}(\Sigma)$ for $0\leq j\leq7$. Since $\ell+6-j\geq3$, this space is a Sobolev algebra. This proves the required regularity of $\tilde T^{0\mu}$ and of the inverse rescaling. Equations~\eqref{E:Rescaled_composition_estimate} and \eqref{E:Rescaled_time_derivatives}, together with Sobolev embedding, also preserve time continuity in $H^s(\Sigma)$ for every $s<\ell+6-j$. The reconstruction of $n,X,Z$ has the regularity already established in the proof of Theorem \ref{Theorem_1_precise}.

For uniqueness, two rescaled solutions with identical initial data give identical solutions of Equations~\eqref{E:FO_CD_cleaning_system}, by Equation~\eqref{E:FO_CD_cleaning_energy_identity} applied to their difference. They therefore give the same prescribed $W$. Their unscaled mixed-order variables solve the system with the additional source \eqref{E:Rescaled_additional_source}, with identical data and identical $W$. Uniqueness in Theorem \ref{Theorem_SH_LWP} gives equality of these variables, and reconstruction gives equality of the rescaled solutions.

Finally, convergence of initial data in the spaces of Theorem \ref{Theorem_1_precise} gives convergence of the unscaled initial data by Equation~\eqref{E:Rescaled_composition_estimate}. The estimate \eqref{E:FO_CD_shifted_cleaning_regularity}, applied to differences, gives convergence of $W$ in the forcing spaces required by Theorem \ref{Theorem_SH_LWP}. Its continuous-dependence conclusion then gives convergence of $V$ strictly below the orders in Equations~\eqref{E:FO_CD_shifted_physical_output}. Reconstruction and Equation~\eqref{E:Rescaled_composition_estimate} give the asserted continuous dependence of the rescaled fields on a common local interval.
\end{proof}

\begin{remark}
If the initial constraint errors and cleaning fields vanish, Equation~\eqref{E:FO_CD_cleaning_energy_identity} implies that they vanish throughout the existence interval. Equation~\eqref{E:Rescaled_unscaled_conservation} then becomes $\nabla_\nu T^{\mu\nu}=0$, and the equivalence argument in Section~\ref{FO:Equivalence} identifies the solution with a solution of the original BDNK equations. For general initial constraint errors, local well-posedness follows from the proof without imposing their vanishing.
\end{remark}

\section{Gubser flow: Additional plots}\label{app:more-gubser-plots}
We present additional numerical results for the Gubser flow configuration discussed in Section~\ref{Subsubsec:Gubser}. 
In particular, we compare the spatial derivatives of the hydrodynamic variables obtained using the \ac{FO} and \ac{MO} formulations against the corresponding semi-analytical solutions.
These comparisons help us investigate the effectiveness of the curl-cleaning procedure for the evolution of the independent spatial derivative variables in the \ac{FO} formulation, which are subject to the dynamical constraint-preserving equations introduced in Section~\ref{Sec:First_Order}.
All results are obtained from the same simulations discussed in Section~\ref{Subsubsec:Gubser}.

\begin{figure}
     \centering
     \begin{subfigure}[b]{0.49\textwidth}
         \centering
         \includegraphics[width=\textwidth]{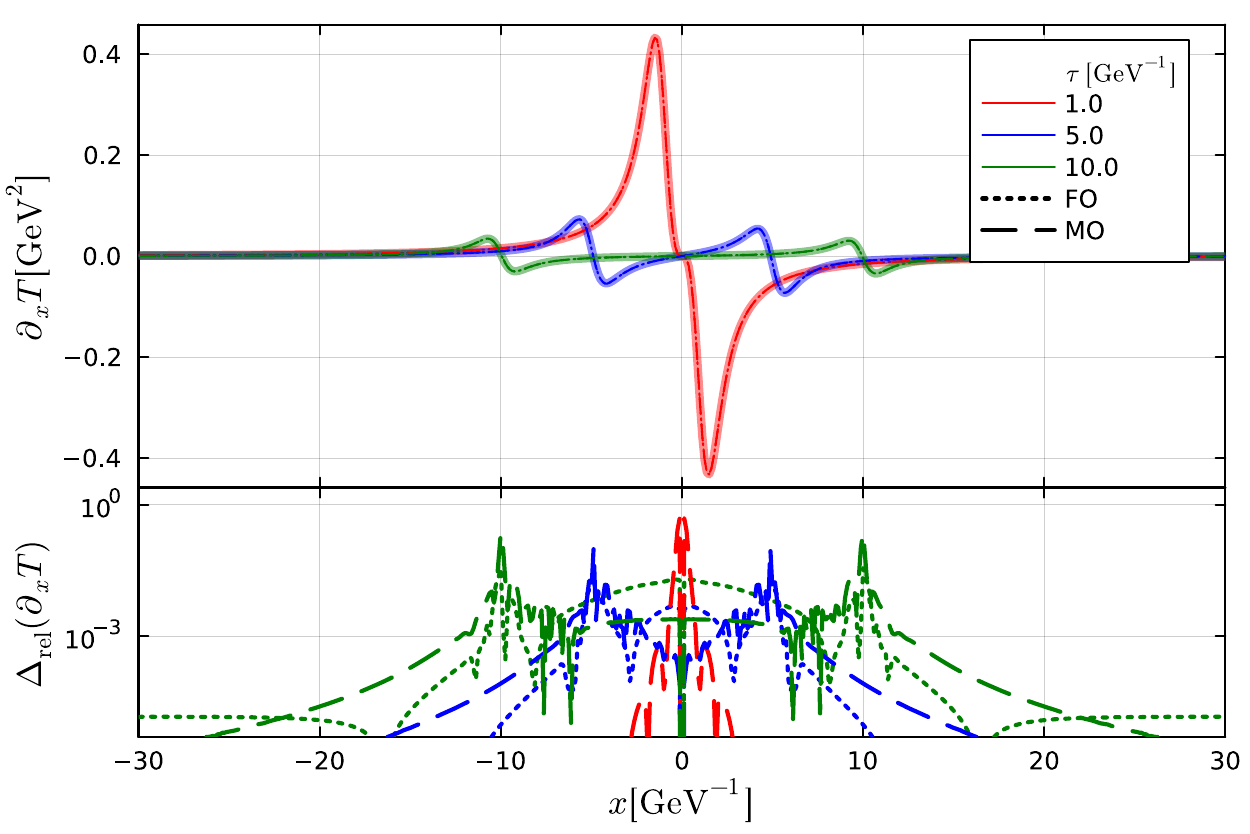}
     \end{subfigure}
     \hfill
     \begin{subfigure}[b]{0.49\textwidth}
         \centering
         \includegraphics[width=\textwidth]{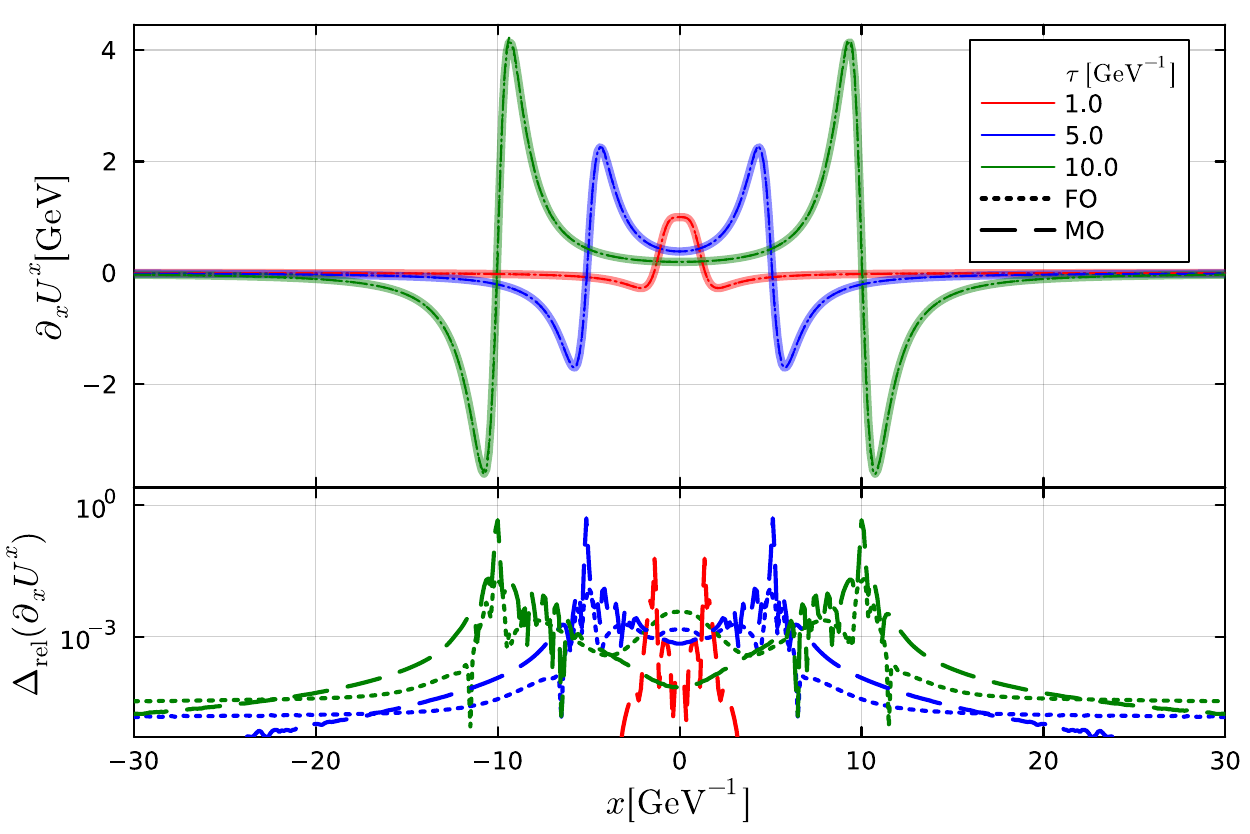}
     \end{subfigure}
     \hfill
        \caption{
        The curves in the top panels represent slices of the derivatives of the temperature (left) $\partial_xT$ and 4-velocity component $\partial_xu^x$ (right), plotted along the x-axis for Gubser flow.
        The curves in the bottom panels represent the relative 
        error between simulation and semi-analytical values for the derivatives of temperature (left) and 4-velocity component $u^x$ (right).
        The colors correspond to Milne coordinate times $\tau=1,5,\,\rm{ and }\,10\,\GeV^{-1}$ as indicated in the legend.
        The thick lines represent the semi-analytical solutions described in Section~\ref{Subsubsec:Gubser} and the dashed lines correspond to the simulation results.
        The results here were obtained for $\delta x=1/8\,\GeV^{-1}$, $\delta t = 0.01\,\GeV^{-1}$, $\eta_0=0.1$, $\chi_0=(25/4)\eta_0$, $\lambda_0=(25/7)\eta_0$.}
        \label{fig:gubser_slices_derivatives_w_err}
\end{figure}

\begin{figure}
     \centering
     \includegraphics[width=0.5\textwidth]{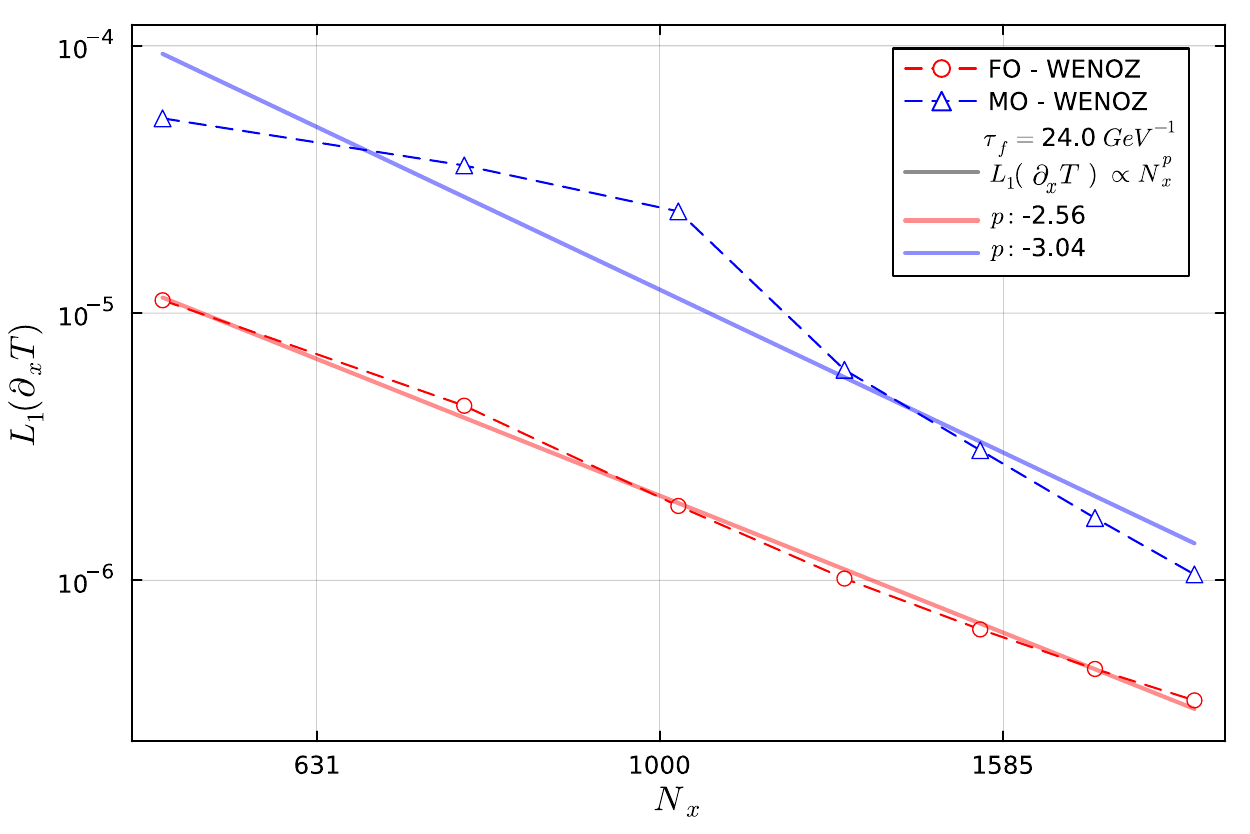}   
    \caption{
    $L_1$ norm of the absolute error for the spatial derivative $\partial_xT$ (left) with respect to the number of grid points $N_x = L_x/\delta x$ for Gubser flow.
    The red and blue data points correspond to simulations 
    with the \ac{FO} (red) and the \ac{MO} (blue) formulations.
    The markers correspond to the simulation results and the connected dashed lines are for visual guidance.
    The thick solid lines correspond to a power law fit, with the corresponding exponents displayed in the legends.
    The simulation parameters used here are $\eta_0=0.1$, $\chi_0/\eta_0=25/4$, $\lambda_0/\eta_0=25/7$.
    The initial and final times used here are $\tau_0=1\,\GeV^{-1}$ and $\tau_f=24\,\GeV^{-1}$, respectively, with the time interval fixed at $\delta t = 0.01\,\GeV^{-1}$ for all $N_x$.
    }
    \label{fig:gubser_derivatives_err_NX_scaling}
\end{figure}

Figure~\ref{fig:gubser_slices_derivatives_w_err} compares the simulation derivative fields with their semi-analytical counterparts and shows the corresponding local errors.
Figure~\ref{fig:gubser_derivatives_err_NX_scaling} presents the dependence of these errors on spatial resolution for both the \ac{FO} and \ac{MO} formulations.
We recall that, unlike the \ac{FO} formulation, the spatial derivatives in the \ac{MO} formulation are evaluated using CWENO rather than evolved as independent variables, providing a useful reference for comparison.

\begin{figure}
     \centering
     \begin{subfigure}[b]{0.49\textwidth}
         \centering
         \includegraphics[width=\textwidth]{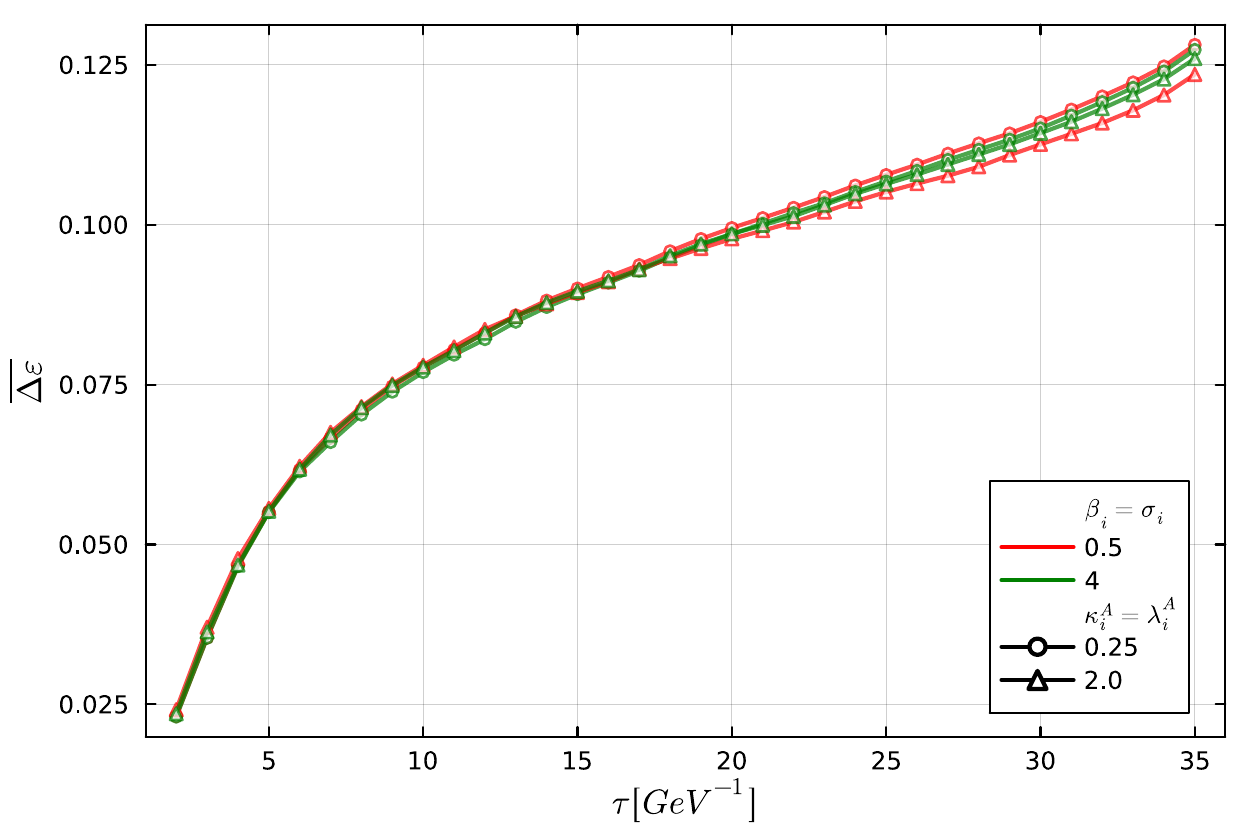}
     \end{subfigure}
     \hfill
     \begin{subfigure}[b]{0.49\textwidth}
         \centering
         \includegraphics[width=\textwidth]{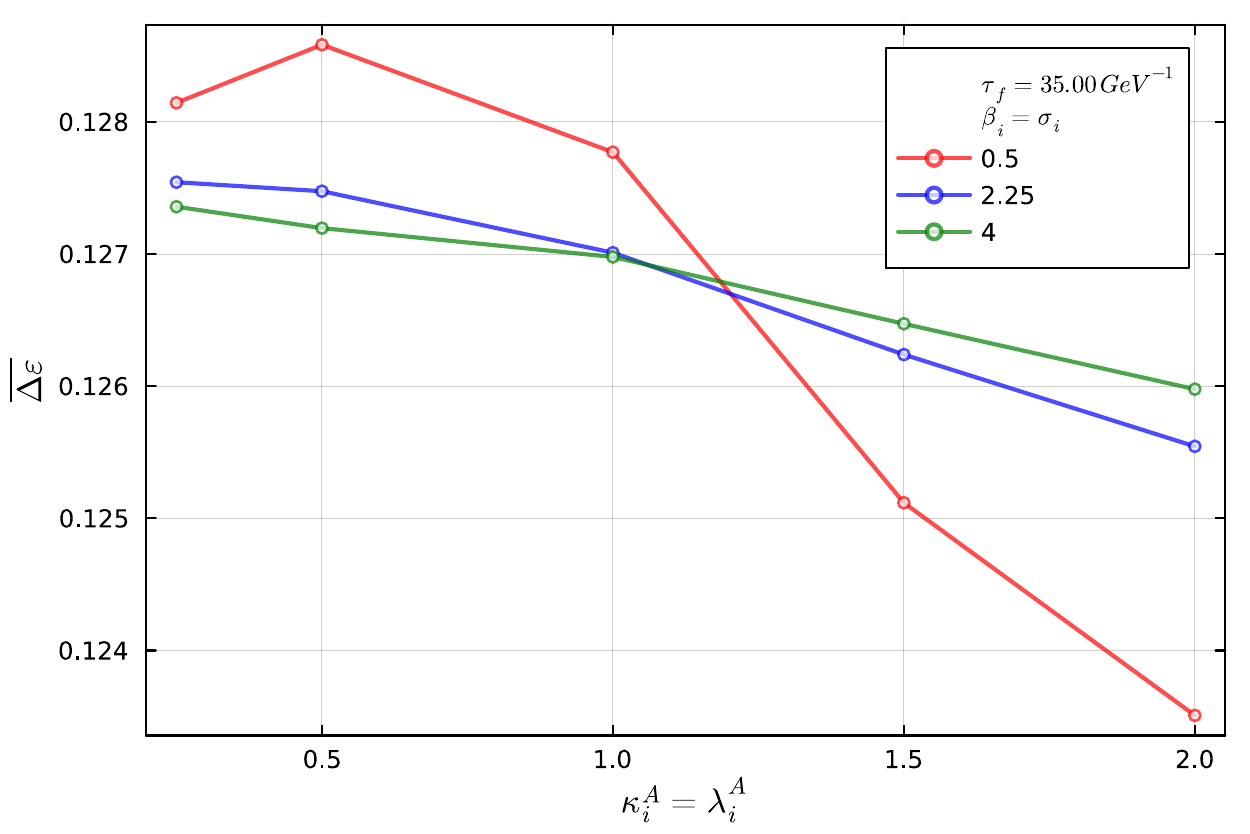}
     \end{subfigure}
     \hfill
        \caption{
        Normalized mean absolute differences $\overline{\Delta\varepsilon}$ between the numerical Gubser flow solutions and the corresponding semi-analytical solution for different choices of the curl-cleaning parameters. The left panel shows the time evolution of $\overline{\Delta\varepsilon}$, while the right panel shows its dependence on the curl-cleaning parameters at the final simulation time $\tau_f$. The different curves correspond to the parameter choices indicated in the legend, with $\beta_i^A=\sigma_i$ and $\kappa_i^A=\lambda_i$, for $i\in{1,2}$.
        Here, $\eta_0=0.1\,,\chi_0/\eta_0=25/4\,,\lambda_0/\eta_0=25/7$.}
        \label{fig:gubser_errs_w.r.t_CC_paras}
\end{figure}

Finally, Figure~\ref{fig:gubser_errs_w.r.t_CC_paras} examines the sensitivity of the numerical errors to the curl-cleaning parameters, 
$$\{\kappa_1^A,\kappa_2^A,\lambda_1,\lambda_2,\beta_1^A,\beta_2^A,\sigma_1,\sigma_2\}\,,$$ introduced in Section~\ref{Sec:First_Order}.
These parameters control the propagation and damping of constraint violations and can, in principle, be chosen independently.
For simplicity, we restrict discussion of results to equal amplitudes and propagation speeds, $\beta_i^A=\sigma_i$ and $\kappa_i^A=\lambda_i$, respectively, for $i\in\{1,2\}$.
While the results presented in the main text employ parameter choices satisfying the criteria discussed in Section~\ref{S:Bounds_cleaning_speeds}, here we explore a broader range to assess their influence on the numerical evolution.
The resulting errors show relatively weak dependence on the curl-cleaning parameters over the range considered.
Nevertheless, as discussed in Section~\ref{S:Bounds_cleaning_speeds}, appropriate choices of these parameters can be important for numerical stability, particularly in strongly curved spacetimes.

\section{Convergence Tables}
\label{S:Convergence_tables}
We define the discretized value of the variable $u(t,{x})$ at time level $n$ and the $i$-th cell as $U^n_i:=u(x=x_i,t=t_n)$. Assuming that the error due to the time discretization is negligible and dropping the time index $n$, the local error is given by
\begin{align*}
    E^{(h)}_i=U_i^{(h)}-u_i=Ch^{p_i}+\mathcal{O}(h^{p_i+1})\, ,
\end{align*}
where $h:=\delta x$ is the grid spacing and $u_i=u(x_i)$. We define the error ratio at two different grid spacings $k$ and $h$, with $k<h$ as
\begin{align*}
    R_i(h,k):=\frac{E^{(h)}}{E^{(k)}}=\frac{h^{p_i}}{k^{p_i}}+\mathcal{O}(h)\, ,
\end{align*}
and the corresponding measured local convergence rate as
\begin{align*}
    p_i:=\frac{\log|R_i(h,k)|}{\log(h/k)}\,.
\end{align*}
A global measure of the convergence rate can be computed using the norm of the error $E^{(h)}:=||U_i^{(h)}-u_i||$ and the corresponding error ratio $R(h,k):={E^{(h)}}/{E^{(k)}}$, given by
\begin{align*}
    p=:\frac{\log R(h,k)}{\log (h/k)}\, .
\end{align*}
We further denote the $L_1$-norm of the error for a variable $u$ on a one-dimensional computational grid with $N$ lattice points as
\begin{align*}
    L_1(u)=\frac{1}{N}\sum_i^N|U_i-u_i|\, ,
\end{align*}
and the corresponding convergence rate as $p^{(L_1)}$.
In the absence of a continuum solution, the self-convergence rates are computed using three resolutions $l<k<h$ and $h,k=h/\alpha,l=h/\alpha^2$, given by
\begin{align*}
    q:=\frac{\log R(h,k,l)}{\log \alpha}\,, 
\end{align*}
where the error ratios are given by
\begin{align*}
    R(h,k,l):=\frac{||U^{(h)}_i-U^{(k)}_i||}{||U^{(k)}_i-U^{(l)}_i||}\,,
\end{align*}
with the corresponding convergence rate using the $L_1$ norm of the error as $q^{(L_1)}$.

For the purposes of comparison against an ideal hydro reference, we define the normalized mean of the absolute difference of a variable $u$ between the discrete counterparts of the viscid~$U^n_i$ and the inviscid variables~$U^n_{0,i}$, at time level $n$, as
\begin{equation}\label{def:normalized-mean-absolute-difference}
    \overline{\Delta u }(t_n) := \frac{\sum^N_i|U^n_i-U^n_{0,i}|}{\sum^N_i|U^n_{0,i}|}\,,
\end{equation}
where the absolute difference is averaged over the computational domain and normalized by the average of the absolute value of the reference ideal variable $U_{0,i}$.

For simplicity, all the definitions stated in this appendix are for one-dimensional scalar fields~$u(t,x)$, but can be trivially generalized to arbitrary dimensions, and to vector and tensor fields.

\begin{table}[t]
\centering
\small
\setlength{\tabcolsep}{4pt}
\renewcommand{\arraystretch}{1.2}
\begin{tabular}{|l|c|c|r|c|c|c|c|}
\hline
Scheme & Rescaled & Smoothing & $\tau_f$ & $\overline{\Delta T}$ & $\overline{\Delta u^x}$ & $\overline{\Delta \partial_x T}$ & $\overline{\Delta \partial_x u^x}$ \\
\hline
\multicolumn{8}{|l|}{\textbf{\ac{FO} system}} \\
\hline
\multirow{3}{*}{WENO-Z}
 & Yes & No  & 50 & $1.7\times10^{-3}$ & $4.4\times10^{-3}$ & $5.5\times10^{-3}$ & $3.0\times10^{-3}$ \\
\cline{2-8}
 & No  & No  & 50 & $6.5\times10^{-4}$ & $4.6\times10^{-3}$ & $4.1\times10^{-3}$ & $2.9\times10^{-3}$ \\
\cline{2-8}
 & No  & Yes & 38 & $1.6\times10^{-3}$ & $1.6\times10^{-3}$ & $6.7\times10^{-3}$ & $5.9\times10^{-3}$ \\
\hline
\multirow{3}{*}{minmod}
 & Yes & No  & 29 & $1.7\times10^{-2}$ & $6.5\times10^{-2}$ & $2.4\times10^{-1}$ & $4.1\times10^{-1}$ \\
\cline{2-8}
 & No  & No  & 22 & $1.4\times10^{-2}$ & $3.5\times10^{-2}$ & $2.2\times10^{-1}$ & $3.5\times10^{-1}$ \\
\cline{2-8}
 & No  & Yes & 22 & $1.4\times10^{-2}$ & $3.5\times10^{-2}$ & $2.2\times10^{-1}$ & $3.5\times10^{-1}$ \\
\hline
\multicolumn{8}{|l|}{\textbf{\ac{MO} system}} \\
\hline
WENO-Z & No & Yes & 36 & $2.5\times10^{-3}$ & $3.7\times10^{-3}$ & $1.8\times10^{-2}$ & $3.5\times10^{-2}$ \\
\hline
minmod & No & Yes & $4^\dagger$ & $1.9\times10^{-3}$ & $2.0\times10^{-3}$ & $1.2\times10^{-1}$ & $6.3\times10^{-2}$ \\
\hline
\end{tabular}
\caption{Errors for the Gubser flow test with $\delta x=1/8\,\GeV^{-1}$, $\delta t=1/100\,\GeV^{-1}$. The simulation parameters used here are $\eta_0=0.1$, $\chi_0=(25/4)\eta_0$, $\lambda_0=(25/7)\eta_0$. $\overline{\Delta f}$ is the mean absolute difference from the analytic solution normalized by the mean absolute analytic value, $\sum|f-f_{\rm sol}|/\sum|f_{\rm sol}|$, over $r<0.5\,R$, evaluated at the final time~$\tau_f$ of each run. All MO runs require smoothing for numerical stability, as simulations without smoothing become unstable at early times. The superscript $\dagger$ indicates an unstable run that terminates prematurely.}
\label{tab:Gubser_comparison}
\end{table}

\section{Gubser flow boundary handling}\label{appendix:gubser-boundary-handling}

As discussed in Section~\ref{Subsubsec:Gubser}, gradients in the hydrodynamical fields, $X^A_i$, near the boundaries of the computational domain are small but finite. These are particularly problematic for the \ac{MO} formulation, where numerically computed spatial derivatives, combined with spurious dynamics near the boundaries, were found to lead to numerical instabilities.
To address this, we smoothly transition the initial Gubser profile to a constant-temperature atmosphere with vanishing transverse velocity and spatial gradients.

Taking advantage of the cylindrical symmetry and localized temperature and energy-density profiles of Gubser flow, we introduce a radial transition region $R<r<R+\delta R$, where $r=\sqrt{x^2+y^2}$ and $\delta R>0$, to smoothly connect the original initial conditions to a stationary, homogeneous atmosphere.
The transition is constructed using the quintic smoothstep polynomial
\begin{equation}
S_5(s)=6s^5-15s^4+10s^3, \qquad s=\frac{r-R}{\delta R},
\end{equation}
which defines the radial smoothing function
\begin{equation}
w(r)=
\begin{cases}
1, & r\leq R\,,\\
1-S_5(s), & R<r<R+\delta R\,,\\
0, & r\geq R+\delta R\,.
\end{cases}
\label{eq:gubser-taper}
\end{equation}
The quintic polynomial is chosen as it allows for a choice of $w(r)$ and its first and second derivatives which are continuous across both endpoints of the transition region, with the derivatives vanishing at $r=R$ and $r=R+\delta R$.

Denoting the original Gubser temperature and transverse four-velocity components by $T$ and $u^i$, respectively, the smoothed profiles are defined as
\begin{align}
T_{\mathrm{sm}} &= T_{\mathrm{floor}}+w(r)\left(T-T_{\mathrm{floor}}\right), \label{eq:gubser-temperature-blend}\\
u^i_{\mathrm{sm}} &= w(r)u^i,
\label{eq:gubser-velocity-blend}
\end{align}
where $T_{\mathrm{floor}}=T(R+\delta R)$ is the temperature of the Gubser solution at the outer edge of the transition region.

Smoothing is introduced to address instabilities in the \ac{MO} formulation, and is unnecessary for the \ac{FO} formulation.
However, we also apply it to selected \ac{FO} simulations to enable direct comparisons of numerical errors under identical initial conditions, and the results are presented in Table~\ref{tab:Gubser_comparison}.
For the \ac{FO} formulation, the spatial derivatives of the smoothed fields must be initialized consistently.
Since the radial derivative of the smoothing function is known analytically,
\begin{equation}
\frac{dw}{dr}=-\frac{30}{\delta R}s^2(1-s)^2\,,
\end{equation}
within the transition region, the required spatial derivatives can be obtained directly by differentiating the smoothed profiles in Equations~\eqref{eq:gubser-temperature-blend} and \eqref{eq:gubser-velocity-blend}.

By construction, the original Gubser profile is preserved for $r\leq R$, while for $r\geq R+\delta R$ the transverse velocity and its gradients vanish, and the temperature takes the constant value $T_{\mathrm{floor}}$.
Furthermore, the resulting homogeneous, stationary atmosphere allows us to impose periodic boundary conditions without introducing discontinuities across the domain boundaries.

However, the smoothed profiles are no longer exact solutions of the Gubser system in the modified exterior region~$r>R$, and the resulting deviations can propagate into the interior as the system evolves.
The transition radius $R$ is therefore chosen sufficiently large that the resulting deviations do not significantly influence the central region of interest during the simulated time interval, while remaining sufficiently far from the domain boundary $L$ to minimize interactions with the boundaries and their subsequent influence on the interior.
Similarly, the transition width $\delta R$ is chosen to be sufficiently large to avoid introducing sharp gradients into the smoothed profiles.
For our simulations, we adopt $R=0.6L$ and $\delta R=0.3L$.

\section{Kerr Torus: Atmosphere and Interface Treatment}\label{Appendix:kerr-atm-tor-smoothing}

In this appendix, we describe the construction of the static atmosphere exterior to the torus and the smoothing procedure applied at the torus--atmosphere interface for the Kerr thick torus test presented in Section~\ref{Sec:Kerr-Fat-Disk}.

\subsection{Atmosphere Construction}
\label{app:atmosphere}

Numerical evolution schemes require finite density and pressure throughout the computational domain.
Following~\cite{DelZanna:2007pk}, we fill the region exterior to the torus, defined by $\mathcal{W}(r,\theta) > \mathcal{W}_\textrm{in}$, with a low-density static atmosphere constructed from a one-dimensional radial hydrostatic equilibrium solution.

Following~\cite{DelZanna:2007pk}, the atmosphere is modeled as a static polytropic fluid 
with 
$l_\textrm{atm} = 0$.
Thus the four-velocity has only a temporal component, and Equation~\eqref{eq:expression-for-Ut-kerr-thick-torus} with $l=0$ simplifies to
\begin{equation}\label{eq:atm-ut-kerr}
-(u_\textrm{atm})_t = \sqrt{\frac{(g_{t\phi})^2 - g_{tt}g_{\phi\phi}}{g_{\phi\phi}}}\,.
\end{equation}

The density profile for this static ideal polytropic fluid in hydrostatic equilibrium can be obtained analytically by integrating the ideal hydrodynamics relativistic Euler equation in the Kerr spacetime~(see~\cite{DelZanna:2007pk}), and is most conveniently expressed in terms of the conserved quantity
\begin{equation}\label{eq:atm-enthalpy-kerr}
h_\textrm{atm}\,(-(u_\textrm{atm})_t) = \textrm{const}\,.
\end{equation}
This relation, together with the polytropic EOS, determines the atmosphere density everywhere once a reference spatial position and a corresponding density for the atmosphere are specified.

We fix the atmosphere density at a reference radius $r_\textrm{out}$, chosen to be the outer edge of the radial grid on the equatorial plane~$(\theta=\pi/2)$.
The atmosphere density at $r_\textrm{out}$ is set to a small fraction of the torus density as
\begin{equation}
    \rho_\textrm{atm}(r_\textrm{out}, \pi/2) = f\, \rho_\textrm{torus}(r_\textrm{out}, \pi/2)\,,
\end{equation}
where $f$ is a free parameter that determines the atmosphere-to-torus density ratio at the reference point, and $\rho_\textrm{torus}$ can be computed using the thick disk solution in Equation~\eqref{eq:exp-rho-kerr-thick-torii}.
Using this reference value, the radial density profile is determined everywhere through Equation~\eqref{eq:atm-enthalpy-kerr}.
For the simulations presented in Section~\ref{Sec:Kerr-Fat-Disk}, we set $f = 10^{-2}$, which ensures that the atmosphere density remains subdominant to the torus density throughout the computational domain.

\subsection{Smoothing at the Torus--Atmosphere Interface}
\label{app:smoothing}

At the boundary between the torus and the atmosphere ($\mathcal{W}(r,\theta) = \mathcal{W}_\textrm{in}$), the density and velocity fields are discontinuous.
The BDNK formulation involves spatial gradients of the hydrodynamic fields, and experimentation with our setup revealed numerical instabilities.
To mitigate this, we apply an \textit{ad hoc} localized iterative smoothing procedure in a narrow band around the torus edge, which removes sharp kinks at the interface while preserving the profiles in the torus interior and the atmosphere far from the torus edge.

A discrete signed distance $d(x_i,\theta_j)$ from the mask boundary is computed on the discrete $(r,\theta)$ grid using nearest-neighbor graph distances, with the convention
\begin{equation}
    d > 0 \;\text{(inside mask)}\,,\qquad d < 0 \;\text{(outside mask)}\,,\qquad d = 0 \;\text{(interface cells)}\,.
\end{equation}
Smoothing is applied only within a narrow band of half-width
\begin{equation}
    d_\textrm{band} = f_\textrm{band} \times \min(N_r, N_\theta)\,,
\end{equation}
where $f_\textrm{band}$ is a tunable parameter.
Points with $|d(i,j)| > d_\textrm{band}$ are left unchanged throughout the smoothing procedure.
For the simulations in this work, we use $f_\textrm{band} = 0.1$.

Within the interface band, the density is iteratively smoothed using a weighted average with the four nearest neighbors:
\begin{align}
    \bar{\rho}_{i,j}^{(n)} &= \frac{1}{4}\left(\rho_{i+1,j}^{(n)} + \rho_{i-1,j}^{(n)} + \rho_{i,j+1}^{(n)} + \rho_{i,j-1}^{(n)}\right)\,,\\[4pt]
    \rho_{i,j}^{(n+1)} &= (1-\alpha_s)\,\rho_{i,j}^{(n)} + \alpha_s\,\bar{\rho}_{i,j}^{(n)}\,,
\end{align}
where $\alpha_s \in (0,1]$ is a relaxation parameter controlling the smoothing strength per iteration.
We construct the initial conditions on a high-resolution $2049^2$ grid, applying the smoothing procedure with $n_\textrm{iter}=5000$ iterations and $\alpha_s=0.95$, before downsampling the resulting profiles to the simulation grid. 
The same procedure is independently applied to the $(U_t)$ field, producing a smoothed velocity profile consistent with the smoothed density interface.

The result is a continuous initial profile with no sharp kinks, suitable for the numerical differentiation required to initialize the BDNK derivative variables and viscous tensor components.
We note that the modification will cause the resulting density and velocity profiles to no longer be stationary ideal hydrodynamical solutions.
However, this modification is small and located near the vacuum region whose dynamics are not the focus of our interest.
While any dynamics resulting from this artificial modification may propagate inwards towards the central torus region of interest,
we observe that the dynamics in the central torus region are far more dominant and the dynamics at the torus edge mostly dictate the numerical stability.

\section{Kurganov-Tadmor}\label{Appendix:KT}

We briefly illustrate the central-upwind Kurganov--Tadmor (KT) scheme used in our numerical implementation by considering a one-dimensional, \ac{FO} evolution equation for a scalar variable $u$, with flux $F(u)$ and source $S(u)$,
\begin{equation}
    \partial_t u+\partial_x F(u)=S(u)\,.
\end{equation}
The corresponding semi-discrete evolution equation for a cell
$i$ of width $\delta x$ is
\begin{equation}
    \partial_t u_i
    +\frac{1}{\delta x}
    \left[
    \langle F_{i+\frac{1}{2}}\rangle
    -\langle F_{i-\frac{1}{2}}\rangle
    \right]
    =S(u_i)\,,
\end{equation}
where $u_i$ denotes the cell-averaged value of $u$, and $\langle F_{i\pm\frac{1}{2}}\rangle$ are the numerical fluxes at the cell interfaces. The left and right reconstructed states, $u^L_{i\pm\frac{1}{2}}$ and $u^R_{i\pm\frac{1}{2}}$, are obtained using a reconstruction scheme.

The KT numerical fluxes are given by
\begin{align}
    \label{KT-flux-terms}
    \langle F_{i-\frac{1}{2}}\rangle
    &=
    \frac{1}{2}
    \left\{
    F(u^R_{i-\frac{1}{2}})
    +F(u^L_{i-\frac{1}{2}})
    -a_{i-\frac{1}{2}}
    \left[
    u^R_{i-\frac{1}{2}}
    -u^L_{i-\frac{1}{2}}
    \right]
    \right\}\,,
    \nonumber\\
    \langle F_{i+\frac{1}{2}}\rangle
    &=
    \frac{1}{2}
    \left\{
    F(u^R_{i+\frac{1}{2}})
    +F(u^L_{i+\frac{1}{2}})
    -a_{i+\frac{1}{2}}
    \left[
    u^R_{i+\frac{1}{2}}
    -u^L_{i+\frac{1}{2}}
    \right]
    \right\}\,,
\end{align}
where $a_{i\pm\frac{1}{2}}$ denote the local propagation speeds at the cell interfaces. These expressions are also commonly referred to as the local Lax--Friedrichs (or Rusanov) fluxes.

The generalization to systems with diffusion terms presented in the work for the KT scheme of~\cite{Tadmor2000} can also accommodate equations containing diffusion terms, of the form
\begin{equation}
    \label{KT-diff-general-eom-scalar}
    \partial_t u+\partial_x F(u)
    =
    \partial_x\mathcal{K}(u,\partial_xu)\,.
\end{equation}
For such equations, the semi-discrete evolution equation takes the form
\begin{equation}
    \label{KT-diff-EOM-central-diff-approx}
    \partial_t u_i
    +\frac{1}{\delta x}
    \left[
    \langle F_{i+\frac{1}{2}}\rangle
    -\langle F_{i-\frac{1}{2}}\rangle
    \right]
    =
    \frac{1}{\delta x}
    \left[
    \langle\mathcal{K}_{i+\frac{1}{2}}\rangle
    -\langle\mathcal{K}_{i-\frac{1}{2}}\rangle
    \right]\,,
\end{equation}
where the derivative-free numerical fluxes $\langle F_{i\pm\frac{1}{2}}\rangle$ are given by Equation~\eqref{KT-flux-terms}, while the diffusive
fluxes are approximated by
\begin{align}
    \label{KT-diff-EOM-P-flux-terms}
    \langle\mathcal{K}_{i+\frac{1}{2}}\rangle
    &=
    \frac{1}{2}
    \left[
    \mathcal{K}\left(
    u_i,\frac{u_{i+1}-u_i}{\delta x}
    \right)
    +
    \mathcal{K}\left(
    u_{i+1},\frac{u_{i+1}-u_i}{\delta x}
    \right)
    \right]\,,
    \nonumber\\
    \langle\mathcal{K}_{i-\frac{1}{2}}\rangle
    &=
    \frac{1}{2}
    \left[
    \mathcal{K}\left(
    u_{i-1},\frac{u_i-u_{i-1}}{\delta x}
    \right)
    +
    \mathcal{K}\left(
    u_i,\frac{u_i-u_{i-1}}{\delta x}
    \right)
    \right]\,.
\end{align}

\subsection{Characteristics}

In the KT scheme, the local propagation speeds $a_{i\pm\frac{1}{2}}$ are determined by the maximum absolute eigenvalues of the flux Jacobian $\frac{\partial F(u)}{\partial u}$~\cite{KurganovLevy2000},
\begin{equation}
\label{KT-local-propagation-speeds}
a_{i\pm\frac{1}{2}}
=
\max\left\{
\rho\left(
\left.\frac{\partial F}{\partial u}\right|_{u^L_{i\pm\frac{1}{2}}}
\right),
\rho\left(
\left.\frac{\partial F}{\partial u}\right|_{u^R_{i\pm\frac{1}{2}}}
\right)
\right\}\,,
\end{equation}
where $\rho$ denotes the spectral radius of the flux Jacobian, corresponding to the maximum absolute value of its eigenvalues.

The propagation speeds of the characteristic modes of BDNK are instrumental in determining the causality conditions and are given by Equations~\eqref{Characteristics}. For a given characteristic mode $b$, with squared characteristic velocity $c_b^2$ in the local fluid rest frame, the corresponding characteristic velocities in an arbitrary coordinate frame are given by
\begin{equation}
\label{eq:KT_characteristic_velocities}
\Lambda_{b,l}^\pm
=
\Lambda_b^\pm(\xi_l)
=
\frac{
-c_b^2 \Delta^{0l}\xi_l
+u^0 u^l\xi_l
\pm \sqrt{\mathcal{Z}_b}
}{
(u^0)^2-c_b^2\Delta^{00}
}\,,
\end{equation}
where
\begin{align}
\label{eq:KT_discriminant}
\mathcal{Z}_b
= c_b^2 \Big[
&(u^0)^2\Delta^{lm}\xi_l\xi_m
+\Delta^{00}u^l u^m\xi_l\xi_m
-2u^0\Delta^{0l}u^m\xi_l\xi_m
\nonumber\\
&-c_b^2\left(
\Delta^{00}\Delta^{lm}\xi_l\xi_m
-\Delta^{0l}\Delta^{0m}\xi_l\xi_m
\right)
\Big]\,.
\end{align}
Here, $u^\mu$ is the fluid four-velocity, $\Delta^{\mu\nu}=g^{\mu\nu}+u^\mu u^\nu$ is the projector orthogonal to the fluid four-velocity, and $\xi_l$ specifies the spatial direction along which the characteristic velocities are evaluated. The index $b\in\{1,\pm\}$ denotes one of the three BDNK characteristic speeds in Equations~\eqref{Characteristics}\footnote{We use $b$ here to denote the characteristic instead of $a$ as in Equations~\eqref{Characteristics} to distinguish it from the KT local propagation speeds $a_{i\pm\frac{1}{2}}$.}.

For the one-dimensional KT scheme described above, we choose $\xi_l=(1,0,0)$, corresponding to the $x$-direction, and denote the resulting characteristic velocities as $\Lambda_{b,x}^{\pm}$.

Thus, the local KT propagation speed $a_{i+\frac{1}{2}}$ is computed as follows:
\begin{itemize}
    \item Compute all fluid rest-frame BDNK characteristic speeds $c_b\big(u^{L/R}_{i+\frac{1}{2}}\big)$, with $b\in\{1,\pm\}$, using Equations~\eqref{Characteristics} for the left and right reconstructed states $u^{L/R}_{i+\frac{1}{2}}$ at the cell interface $i+\frac{1}{2}$.

    \item Using the values of $c_b$ computed above, evaluate the corresponding characteristic velocities $\Lambda_{b,x}^\pm\big(u^{L/R}_{i+\frac{1}{2}}\big)$ using Equations~\eqref{eq:KT_characteristic_velocities} and \eqref{eq:KT_discriminant} for each characteristic mode and reconstructed state.

    \item The KT propagation speed $a_{i+\frac{1}{2}}$, used to compute the numerical flux $\langle F_{i+\frac{1}{2}}\rangle$ in Equation~\eqref{KT-flux-terms}, is then given by the maximum absolute value of all characteristic velocities evaluated at the left and right reconstructed states,
    \begin{equation}
        \label{eq:KT_BDNK_local_speed}
        a_{i+\frac{1}{2}}
        =
        \max_{\substack{b\in\{1,\pm\}\\s\in\{L,R\}}}
        \left[
        \left|\Lambda_{b,x}^+\left(u^s_{i+\frac{1}{2}}\right)\right|,
        \left|\Lambda_{b,x}^-\left(u^s_{i+\frac{1}{2}}\right)\right|
        \right]\,.
    \end{equation}
\end{itemize}

The generalization of the KT scheme to three spatial dimensions in our $3+1$ implementation is straightforward, with the characteristic velocities evaluated along $\xi_l=(1,0,0)$, $(0,1,0)$, and $(0,0,1)$ corresponding to the three respective spatial coordinates~$x^l$.

\section{Reconstruction}
\label{S:Reconstruction}
We provide an overview of the two reconstruction schemes used to reconstruct values of the variables at the cell interfaces.

\subsection{MINMOD}\label{Appendix:MINMOD}
A flux limiter scheme is the simplest method used for reconstructing the values at the interfaces, given by 
\begin{align}
    u^R_{i+\frac{1}{2}} = \bar{u}_{i+1} - \frac{\delta x}{2}(\partial_x u)_{i+1},\\\nonumber
    u^L_{i+\frac{1}{2}} = \bar{u}_{i} + \frac{\delta x}{2}(\partial_x u)_{i},\nonumber    
\end{align}
where the spatial derivatives are computed using the minmod flux limiter
\begin{equation}
    (\partial_x u )_{i} = \rm{minmod}\left(\theta \frac{\bar{u}_{i+1}-\bar{u}_i}{\delta x},
    \frac{\bar{u}_{i+1}-\bar{u}_{i-1}}{2\delta x},
    \theta \frac{\bar{u}_{i}-\bar{u}_{i-1}}{\delta x}
    \right)\,,
\end{equation}
where $1\leq\theta\leq2$ is the parameter that controls the amount of numerical diffusion and oscillatory behavior around sharp gradients.
Thus, the minmod limiter effectively switches between a second-order accurate method for higher accuracy in smooth regions and a first-order accurate method around sharp gradients necessary for capturing shock fronts.

\subsection{WENO-Z}\label{Appendix:WENO-Z}
WENO-Z, an adaptation of the WENO scheme, is a fifth-order scheme that uses a 5-point stencil.
The general WENO upwind reconstruction for the left state $u^L_{i+\frac{1}{2}}$ is given by 
\begin{equation}
    u^L_{i+\frac{1}{2}} = \sum^3_{k=1}w_k u^{(k)}_{i+\frac{1}{2}}\,,  
\end{equation}
where $w_k$ are the weights, and $u^{(k)}_{i+\frac{1}{2}}$ are the third-order ENO polynomials given by
\begin{align}
    u^{(1)}_{i+\frac{1}{2}} = \frac{1}{3} \bar{u}_{i-2}
                    -\frac{7}{6} \bar{u}_{i-1}
                    +\frac{11}{6} \bar{u}_{i}
    \,,\\ \nonumber
        u^{(2)}_{i+\frac{1}{2}} = -\frac{1}{6} \bar{u}_{i-1}
                    +\frac{5}{6} \bar{u}_{i}
                    +\frac{1}{3} \bar{u}_{i+1}
    \,,\\ \nonumber
    u^{(3)}_{i+\frac{1}{2}} = \frac{1}{3} \bar{u}_{i}
                    +\frac{5}{6} \bar{u}_{i+1}
                    -\frac{1}{6} \bar{u}_{i+2}
    \,,\\ \nonumber
\end{align}
where $\bar{u}_i$ is the cell averaged variable in cell $i$.

The non-linear WENO-Z weights $w_k$ are given by 
\begin{equation}
w_k=\frac{\alpha_k}{\sum^3_{l=1}\alpha_l}\,,\quad \alpha_k = d_k\left[1+\left(\frac{\tau_5}{\epsilon_w+\beta_k}\right)^2\right]\,,\quad \tau_5=|\beta_1-\beta_3|\,,
\end{equation}
where $d_k=\left(\frac{1}{10},\frac{3}{5},\frac{3}{10}\right)$ are the linear weights, $\epsilon_w$ is a small value that prevents division by zero, and $\beta_k$ are the smoothness indicators, given by
\begin{align}
    \beta_1 &= 
    \frac{13}{12}(\bar{u}_{i-2}-2\bar{u}_{i-1}+\bar{u}_i)^2
    +\frac{1}{4}(\bar{u}_{i-2}-4\bar{u}_{i-1}+3\bar{u}_i)^2
    \,,\\\nonumber    
    \beta_2 &= 
    \frac{13}{12}(\bar{u}_{i-1}-2\bar{u}_{i}+\bar{u}_{i+1})^2
    +\frac{1}{4}(\bar{u}_{i-1}-\bar{u}_{i+1})^2
    \,,\\\nonumber    
    \beta_3 &= 
    \frac{13}{12}(\bar{u}_{i}-2\bar{u}_{i+1}+\bar{u}_{i+2})^2
    +\frac{1}{4}(3\bar{u}_{i}-4\bar{u}_{i+1}+\bar{u}_{i+2})^2
    \,.\\\nonumber    
\end{align}

The downwind reconstruction for $u^R_{i-\frac{1}{2}}$,
\begin{equation}
    u^R_{i-\frac{1}{2}} = \sum^3_{k=1} w'_k u^{(k)}_{i-\frac{1}{2}}\,, \quad 
    w'_k=\frac{\alpha'_k}{\sum^3_{l=1}\alpha'_l}\,,\quad \alpha'_k = d_k\left[1+\left(\frac{\tau'_5}{\epsilon_w+\beta'_k}\right)^2\right]\,,\quad \tau'_5=|\beta'_1-\beta'_3|\,,
\end{equation} 
is obtained by mirroring the expressions for $u^L_{i+\frac{1}{2}}$ and the smoothness indicators $\beta_k$ about $i$. 
The smoothness indicators are given by $\beta'_1=\beta_3$, $\beta'_2=\beta_2$ and $\beta'_3=\beta_1$ and the corresponding ENO polynomials are given by 
\begin{align}
    u^{(1)}_{i-\frac{1}{2}} = \frac{1}{3} \bar{u}_{i+2}
                    -\frac{7}{6} \bar{u}_{i+1}
                    +\frac{11}{6} \bar{u}_{i}
    \,,\\ \nonumber
        u^{(2)}_{i-\frac{1}{2}} = -\frac{1}{6} \bar{u}_{i+1}
                    +\frac{5}{6} \bar{u}_{i}
                    +\frac{1}{3} \bar{u}_{i-1}
    \,,\\ \nonumber
    u^{(3)}_{i-\frac{1}{2}} = \frac{1}{3} \bar{u}_{i}
                    +\frac{5}{6} \bar{u}_{i-1}
                    -\frac{1}{6} \bar{u}_{i-2}
    \,.\\ \nonumber
\end{align}

\bibliography{References}

@article{Sroczinski-2024,
    AUTHOR = {Sroczinski, Matthias},
     TITLE = {Global existence and decay of small solutions for quasi-linear
              second-order uniformly dissipative hyperbolic-hyperbolic
              systems},
   JOURNAL = {J. Diff. Eq.},
  FJOURNAL = {Journal of Differential Equations},
    VOLUME = {383},
      YEAR = {2024},
     PAGES = {130--162},
      ISSN = {0022-0396,1090-2732},
   MRCLASS = {35A01 (35B35 35L52 35L72 35Q35 35Q75 35S50)},
  MRNUMBER = {4669576},
       DOI = {10.1016/j.jde.2023.10.056}
}

@article{Freistuhler-Sroczinski-2025,
    AUTHOR = {Freist\"{u}hler, Heinrich and Sroczinski, Matthias},
     TITLE = {Nonlinear stability of first-order relativistic viscous
              hydrodynamics},
   JOURNAL = {J. Math. Phys.},
  FJOURNAL = {Journal of Mathematical Physics},
    VOLUME = {66},
      YEAR = {2025},
    NUMBER = {9},
     PAGES = {Paper No. 093101},
      ISSN = {0022-2488},
   MRCLASS = {Prelim},
  MRNUMBER = {4961002},
       DOI = {10.1063/5.0243169}
}

@article{Alinhac-Metivier-1984,
  author = {Alinhac, Serge and M{\'e}tivier, Guy},
  title = {Propagation de l'analyticit{\'e} des solutions de syst{\`e}mes hyperboliques non-lin{\'e}aires},
  journal = {Inventiones Mathematicae},
  volume = {75},
  pages = {189--204},
  year = {1984},
  doi = {10.1007/BF01388563}
}

@article{Rendall-2002,
    AUTHOR = {Rendall, Alan D.},
     TITLE = {Theorems on existence and global dynamics for the {E}instein
              equations},
   JOURNAL = {Living Rev. Relativ.},
  FJOURNAL = {Living Reviews in Relativity},
    VOLUME = {5},
      YEAR = {2002},
     PAGES = {6},
       DOI = {10.12942/lrr-2002-6},
       URL = {https://doi.org/10.12942/lrr-2002-6}
}

@article{Sarbach-Tiglio-2012,
    AUTHOR = {Sarbach, Olivier and Tiglio, Manuel},
     TITLE = {Continuum and discrete initial-boundary value problems and
              {E}instein's field equations},
   JOURNAL = {Living Rev. Relativ.},
  FJOURNAL = {Living Reviews in Relativity},
    VOLUME = {15},
      YEAR = {2012},
     PAGES = {9},
       DOI = {10.12942/lrr-2012-9},
       URL = {https://doi.org/10.12942/lrr-2012-9}
}

@article {Ifrim-Tataru-2023,
    AUTHOR = {Ifrim, Mihaela and Tataru, Daniel},
     TITLE = {Local well-posedness for quasi-linear problems: a primer},
   JOURNAL = {Bull. Amer. Math. Soc. (N.S.)},
  FJOURNAL = {American Mathematical Society. Bulletin. New Series},
    VOLUME = {60},
      YEAR = {2023},
    NUMBER = {2},
     PAGES = {167--194},
      ISSN = {0273-0979,1088-9485},
   MRCLASS = {35L45 (35L50 35L60)},
  MRNUMBER = {4557379},
       DOI = {10.1090/bull/1775},
       URL = {https://doi.org/10.1090/bull/1775}
}

@article{Einstein-1939,
    AUTHOR = {Einstein, Albert},
     TITLE = {On a stationary system with spherical symmetry consisting of many gravitating masses},
   JOURNAL = {Ann. of Math. (2)},
  FJOURNAL = {Annals of Mathematics. Second Series},
    VOLUME = {40},
      YEAR = {1939},
    NUMBER = {4},
     PAGES = {922--936},
       DOI = {10.2307/1968902},
       URL = {https://doi.org/10.2307/1968902}
}

@article{Tychonoff-1935,
    AUTHOR = {A. Tychonoff},
     TITLE = {Th\'eor\`emes d'unicit\'e pour l'\'equation de la chaleur},
   JOURNAL = {Mat. Sb.},
  FJOURNAL = {Matematicheskii Sbornik},
    VOLUME = {42},
    NUMBER = {2},
    PAGES = {199--216},
      YEAR = {1935}
}

@book {Engel-Nagel-Book-2000,
    AUTHOR = {Engel, Klaus-Jochen and Nagel, Rainer},
     TITLE = {One-parameter semigroups for linear evolution equations},
    SERIES = {Graduate Texts in Mathematics},
    VOLUME = {194},
      NOTE = {With contributions by S. Brendle, M. Campiti, T. Hahn, G.
              Metafune, G. Nickel, D. Pallara, C. Perazzoli, A. Rhandi, S.
              Romanelli and R. Schnaubelt},
 PUBLISHER = {Springer-Verlag, New York},
      YEAR = {2000},
     PAGES = {xxii+586},
      ISBN = {0-387-98463-1},
   MRCLASS = {47D06 (34G10 35K90 47N20)},
  MRNUMBER = {1721989},
MRREVIEWER = {Charles\ Batty}
}

@book{Teschl-Book-2014,
    AUTHOR = {Teschl, Gerald},
     TITLE = {Mathematical methods in quantum mechanics},
    SERIES = {Graduate Studies in Mathematics},
    VOLUME = {157},
   EDITION = {Second},
      NOTE = {With applications to Schr\"{o}dinger operators},
 PUBLISHER = {American Mathematical Society, Providence, RI},
      YEAR = {2014},
     PAGES = {xiv+358},
      ISBN = {978-1-4704-1704-8},
   MRCLASS = {81-02 (47N50 81Q10 81Q12 81U10)},
  MRNUMBER = {3243083},
MRREVIEWER = {Rupert\ L.\ Frank},
       DOI = {10.1090/gsm/157}
}

@incollection{Secchi-2016,
    AUTHOR = {Secchi, Paolo},
     TITLE = {On the {N}ash-{M}oser iteration technique},
 BOOKTITLE = {Recent developments of mathematical fluid mechanics},
    SERIES = {Adv. Math. Fluid Mech.},
     PAGES = {443--457},
 PUBLISHER = {Birkh{\"a}user/Springer, Basel},
      YEAR = {2016},
   MRCLASS = {35A35 (35F60 35L60 47J25)},
  MRNUMBER = {3643501},
MRREVIEWER = {Mariarosaria Padula},
       DOI = {10.1007/978-3-0348-0939-9_23},
       URL = {https://doi.org}
}

@book{Rodino-Book-1993,
    AUTHOR = {Rodino, Luigi},
     TITLE = {Linear partial differential operators in {G}evrey spaces},
 PUBLISHER = {World Scientific Publishing Co., Inc., River Edge, NJ},
      YEAR = {1993},
     PAGES = {x+251},
      ISBN = {981-02-0845-6},
   MRCLASS = {35-02 (35B27 35D05 35H05 35S05 46Fxx)},
  MRNUMBER = {1249275},
MRREVIEWER = {Todor V. Gramchev},
       DOI = {10.1142/9789814360036}
}

@book{Halmos-Book-1974-1,
    AUTHOR = {Halmos, Paul R.},
     TITLE = {Measure theory},
    SERIES = {Graduate Texts in Mathematics},
    VOLUME = {18},
      NOTE = {Reprint of the 1950 edition},
 PUBLISHER = {Springer-Verlag, New York-Heidelberg},
      YEAR = {1974},
     PAGES = {xi+304},
      ISBN = {0-387-90088-8},
   MRCLASS = {28-02},
  MRNUMBER = {0354999},
MRREVIEWER = {A. H. Stone}
}

@book{Hartman-Book-2002,
    AUTHOR = {Hartman, Philip},
     TITLE = {Ordinary differential equations},
    SERIES = {Classics in Applied Mathematics},
    VOLUME = {38},
      NOTE = {Corrected reprint of the second (1982) edition [Birkh\"{a}user,
              Boston, MA;  MR0658490 (83e:34002)],
              With a foreword by Peter Bates},
 PUBLISHER = {Society for Industrial and Applied Mathematics (SIAM),
              Philadelphia, PA},
      YEAR = {2002},
     PAGES = {xx+612},
      ISBN = {0-89871-510-5},
   MRCLASS = {34-01 (37-01)},
  MRNUMBER = {1929104},
       DOI = {10.1137/1.9780898719222}
}

@article{Disconzi-Ifrim-Tataru-2022,
    AUTHOR = {Disconzi, Marcelo M. and Ifrim, Mihaela and Tataru, Daniel},
     TITLE = {The relativistic {E}uler equations with a physical vacuum
              boundary: {H}adamard local well-posedness, rough solutions,
              and continuation criterion},
   JOURNAL = {Arch. Ration. Mech. Anal.},
  FJOURNAL = {Archive for Rational Mechanics and Analysis},
    VOLUME = {245},
      YEAR = {2022},
    NUMBER = {1},
     PAGES = {127--182},
      ISSN = {0003-9527},
   MRCLASS = {35Q31 (76 83C05)},
  MRNUMBER = {4444071},
       DOI = {10.1007/s00205-022-01783-3}
}

@book{Ringstrom-Book-2009,
    AUTHOR = {Ringstr\"{o}m, Hans},
     TITLE = {The {C}auchy problem in general relativity},
    SERIES = {ESI Lectures in Mathematics and Physics},
 PUBLISHER = {European Mathematical Society (EMS), Z\"{u}rich},
      YEAR = {2009},
     PAGES = {xiv+294},
      ISBN = {978-3-03719-053-1},
   MRCLASS = {83-02 (35Q75 53C80 83C05)},
  MRNUMBER = {2527641},
MRREVIEWER = {Paul T. Allen},
       DOI = {10.4171/053}
}

@book{Christodoulou-Klainerman-Book-1993,
    AUTHOR = {Christodoulou, Demetrios and Klainerman, Sergiu},
     TITLE = {The global nonlinear stability of the {M}inkowski space},
    SERIES = {Princeton Mathematical Series},
    VOLUME = {41},
 PUBLISHER = {Princeton University Press, Princeton, NJ},
      YEAR = {1993},
     PAGES = {x+514},
      ISBN = {0-691-08777-6},
   MRCLASS = {83C05 (35Q75 58G16 83C35)},
  MRNUMBER = {1316662},
MRREVIEWER = {Alan D. Rendall}
}

@BOOK{Alcubierre-Book-2008,
  author = {Alcubierre, Miguel},
  title = {Introduction to 3+1 Numerical Relativity},
  series = {International Series of Monographs on Physics},
  volume = {140},
  publisher = {Oxford University Press},
  address = {Oxford},
  year = {2008},
  isbn = {978-0-19-920567-7},
  doi = {10.1093/acprof:oso/9780199205677.001.0001}
}

@article{Roache-1997,
    AUTHOR = {Roache, Patrick J.},
     TITLE = {Quantification of uncertainty in computational fluid dynamics},
   JOURNAL = {Annu. Rev. Fluid Mech.},
  FJOURNAL = {Annual Review of Fluid Mechanics},
    VOLUME = {29},
      YEAR = {1997},
     PAGES = {123--160},
       DOI = {10.1146/annurev.fluid.29.1.123}
}

@BOOK{Baumgarte-Shapiro-Book-2021,
  author = {Baumgarte, Thomas W. and Shapiro, Stuart L.},
  title = {Numerical Relativity: Starting from Scratch},
  publisher = {Cambridge University Press},
  address = {Cambridge},
  year = {2021},
  doi = {10.1017/9781108933445}
}

@book{Hawking-Ellis-Book-1973,
    AUTHOR = {Hawking, S. W. and Ellis, G. F. R.},
     TITLE = {The large scale structure of space-time},
    SERIES = {Cambridge Monographs on Mathematical Physics, No. 1},
 PUBLISHER = {Cambridge University Press, London-New York},
      YEAR = {1973},
     PAGES = {xi+391},
   MRCLASS = {83.58},
  MRNUMBER = {0424186},
MRREVIEWER = {Michael P. Ryan, Jr.}
}

@book{Baumgarte:2010ndz,
    author = "Baumgarte, Thomas W. and Shapiro, Stuart L.",
    title = "{Numerical Relativity: Solving Einstein's Equations on the Computer}",
    doi = "10.1017/CBO9781139193344",
    publisher = "Cambridge University Press",
    year = "2010"
}

@book{Courant-Hilbert-Book-1989-2,
    AUTHOR = {Courant, R. and Hilbert, D.},
     TITLE = {Methods of mathematical physics. {V}ol. {II}: Partial differential equations},
    SERIES = {Wiley Classics Library},
      NOTE = {Reprint of the 1962 original},
 PUBLISHER = {Wiley},
 address = {New York},
      YEAR = {1989},
      doi = {10.1002/9783527617234},
     PAGES = {xxii+830},
      ISBN = {0-471-50439-4},
   MRCLASS = {35-00 (00A05 01A75)},
  MRNUMBER = {1013360}
}

@article{Cordeiro:2026jle,
    author = "Cordeiro, Ian and Speranza, Enrico and Bemfica, F{\'a}bio S. and Disconzi, Marcelo M. and Noronha, Jorge",
    title = "{Nonlinear Causality and Strong Hyperbolicity of Einstein-Israel-Stewart Theories of Transient Relativistic Fluid Dynamics}",
    journal = "arXiv:2607.05639 [nucl-th]",
    eprint = "2607.05639",
    archivePrefix = "arXiv",
    primaryClass = "nucl-th",
    month = "7",
    year = "2026"
}

@article{alic2012z4,
  title = {Conformal and covariant formulation of the Z4 system with constraint-violation damping},
  author = {Alic, Daniela and Bona-Casas, Carles and Bona, Carles and Rezzolla, Luciano and Palenzuela, Carlos},
  journal = {Phys. Rev. D},
  volume = {85},
  issue = {6},
  pages = {064040},
  numpages = {12},
  year = {2012},
  month = {Mar},
  publisher = {American Physical Society},
  doi = {10.1103/PhysRevD.85.064040},
  url = {https://link.aps.org/doi/10.1103/PhysRevD.85.064040}
}

@article{Bemfica:2017wps,
    author = "Bemfica, F\'abio S. and Disconzi, Marcelo M. and Noronha, Jorge",
    title = "{Causality and existence of solutions of relativistic viscous fluid dynamics with gravity}",
    eprint = "1708.06255",
    archivePrefix = "arXiv",
    primaryClass = "gr-qc",
    doi = "10.1103/PhysRevD.98.104064",
    journal = "Phys. Rev. D",
    volume = "98",
    number = "10",
    pages = "104064",
    year = "2018"
}

@article{Bemfica:2019knx,
    author = "Bemfica, F\'abio S. and and Disconzi, Marcelo M.  and Noronha, Jorge",
    title = "{Nonlinear Causality of General First-Order Relativistic Viscous Hydrodynamics}",
    eprint = "1907.12695",
    archivePrefix = "arXiv",
    primaryClass = "gr-qc",
    doi = "10.1103/PhysRevD.100.104020",
    journal = "Phys. Rev. D",
    volume = "100",
    number = "10",
    pages = "104020",
    year = "2019",
    note = "[Erratum: Phys.Rev.D 105, 069902 (2022)]"
}

@article{Bemfica:2020zjp,
    author = "Bemfica, Fabio S. and Disconzi, Marcelo M. and Noronha, Jorge",
    title = "{First-Order General-Relativistic Viscous Fluid Dynamics}",
    eprint = "2009.11388",
    archivePrefix = "arXiv",
    primaryClass = "gr-qc",
    doi = "10.1103/PhysRevX.12.021044",
    journal = "Phys. Rev. X",
    volume = "12",
    number = "2",
    pages = "021044",
    year = "2022"
}

@article{courant1928partial,
  author    = {Courant, Richard and Friedrichs, Kurt and Lewy, Hans},
  title     = {{\"U}ber die partiellen Differenzengleichungen der mathematischen Physik},
  journal   = {Mathematische Annalen},
  volume    = {100},
  number    = {1},
  pages     = {32--74},
  year      = {1928},
  doi       = {10.1007/BF01448839},
  url       = {https://doi.org/10.1007/BF01448839},
  note      = {English translation: IBM Journal (1967), 215--234.}
}

@article{courant1967partial,
  author    = {Courant, Richard and Friedrichs, Kurt and Lewy, Hans},
  title     = {On the Partial Difference Equations of Mathematical Physics},
  journal   = {IBM Journal of Research and Development},
  volume    = {11},
  number    = {2},
  pages     = {215--234},
  year      = {1967},
  month     = {mar},
  publisher = {IBM},
  doi       = {10.1147/rd.112.0215},
  url       = {https://doi.org/10.1147/rd.112.0215},
  note      = {English translation of the original 1928 German paper.}
}

@article{DEDNER2002645,
title = {Hyperbolic Divergence Cleaning for the MHD Equations},
journal = {Journal of Computational Physics},
volume = {175},
number = {2},
pages = {645-673},
year = {2002},
issn = {0021-9991},
doi = {https://doi.org/10.1006/jcph.2001.6961},
url = {https://www.sciencedirect.com/science/article/pii/S002199910196961X},
author = {A. Dedner and F. Kemm and D. Kröner and C.-D. Munz and T. Schnitzer and M. Wesenberg}
}

@article{DUMBSER2020109088,
title = {On GLM curl cleaning for a first order reduction of the CCZ4 formulation of the Einstein field equations},
journal = {Journal of Computational Physics},
volume = {404},
pages = {109088},
year = {2020},
issn = {0021-9991},
doi = {https://doi.org/10.1016/j.jcp.2019.109088},
url = {https://www.sciencedirect.com/science/article/pii/S0021999119307934},
author = {Michael Dumbser and Francesco Fambri and Elena Gaburro and Anne Reinarz}
}

@book{evans2010pde,
  author    = {Evans, Lawrence C.},
  title     = {Partial Differential Equations},
  publisher = {American Mathematical Society},
  year      = {2010},
  edition   = {2nd},
  isbn      = {978-0-8218-4974-3},
  series    = {Graduate Studies in Mathematics},
  volume    = {19}
}

@article{godel1949example,
  title={An example of a new type of cosmological solutions of Einstein's field equations of gravitation},
  author={G{\"o}del, Kurt},
  journal={Reviews of Modern Physics},
  volume={21},
  number={3},
  pages={447},
  year={1949},
  publisher={APS},
  doi={10.1103/RevModPhys.21.447}
}

@article{gubser2010symmetry,
  title={Symmetry constraints on generalizations of Bjorken flow},
  author={Gubser, Steven S.},
  journal={Physical Review D},
  volume={82},
  number={8},
  pages={085027},
  year={2010},
  publisher={APS},
  doi={10.1103/PhysRevD.82.085027}
}

@article{kerr1963gravitational,
  title={Gravitational field of a spinning mass as an example of algebraically special metrics},
  author={Kerr, Roy P.},
  journal={Physical Review Letters},
  volume={11},
  number={5},
  pages={237},
  year={1963},
  publisher={APS},
  doi={10.1103/PhysRevLett.11.237}
}

@article{Kovtun:2019hdm,
    author = "Kovtun, Pavel",
    title = "{First-order relativistic hydrodynamics is stable}",
    eprint = "1907.08191",
    archivePrefix = "arXiv",
    primaryClass = "hep-th",
    doi = "10.1007/JHEP10(2019)034",
    journal = "JHEP",
    volume = "10",
    pages = "034",
    year = "2019"
}

@article{Tadmor2000,
    author = {Kurganov, Alexander and Tadmor, Eitan},
    title = {New High-Resolution Central Schemes for Nonlinear Conservation Laws and Convection-Diffusion Equations},
    journal = {Journal of Computational Physics},
    volume = {160},
    number = {1},
    pages = {241--282},
    year = {2000},
    doi = {10.1006/jcph.2000.6459},
    url = {https://doi.org/10.1006/jcph.2000.6459}
}

@article{KurganovLevy2000,
    author = {Kurganov, Alexander and Levy, Doron},
    title = {A Third-Order Semidiscrete Central Scheme for Conservation Laws and Convection-Diffusion Equations},
    journal = {SIAM Journal on Scientific Computing},
    volume = {22},
    number = {4},
    pages = {1461--1488},
    year = {2000},
    doi = {10.1137/S1064827599360236},
    url = {https://doi.org/10.1137/S1064827599360236}
}

@article{taub1951empty,
  title={Empty space-times admitting a three parameter group of motions},
  author={Taub, A. H.},
  journal={Annals of Mathematics},
  pages={472--490},
  year={1951},
  publisher={JSTOR},
  doi={10.2307/1969787}
}

@book{taylor2011pde3,
  author    = {Taylor, Michael E.},
  title     = {Partial Differential Equations III: Nonlinear Equations},
  series    = {Applied Mathematical Sciences},
  volume    = {117},
  publisher = {Springer},
  year      = {2011},
  edition   = {2nd},
  isbn      = {978-1-4419-7048-0},
  doi       = {10.1007/978-1-4419-7049-7},
  url       = {https://doi.org/10.1007/978-1-4419-7049-7}
}

@book{Taylor-Book-1991,
    AUTHOR = {Taylor, Michael E.},
     TITLE = {Pseudodifferential operators and nonlinear {PDE}},
    SERIES = {Progress in Mathematics},
    VOLUME = {100},
 PUBLISHER = {Birkh\"{a}user},
 address = {Boston},
      YEAR = {1991},
     PAGES = {213},
      ISBN = {0-8176-3595-5},
   MRCLASS = {35Sxx (35G20 46E35 46F99 47G30 58G15)},
  MRNUMBER = {1121019},
MRREVIEWER = {William Margulies},
       DOI = {10.1007/978-1-4612-0431-2}
}

@book{wald1984general,
  title={General Relativity},
  author={Wald, Robert M.},
  publisher={University of Chicago Press},
  year={1984},
  chapter={6.4}
}

@book{weinberg1972gravitation,
  title={Gravitation and Cosmology: Principles and Applications of the General Theory of Relativity},
  author={Weinberg, Steven},
  publisher={Wiley},
  year={1972},
  chapter={14}
}

@article{Disconzi-2024,
    AUTHOR = {Disconzi, Marcelo Mendes},
     TITLE = {Recent developments in mathematical aspects of relativistic fluids},
   JOURNAL = {Living Rev. Relativ.},
  FJOURNAL = {Living Reviews in Relativity},
     YEAR = {2024},
     VOLUME = {27},
    NUMBER = {6},
     PAGES = {218},
       DOI = {10.1007/s41114-024-00052-x},     
    eprint = {2308.09844},
    archivePrefix = {arXiv},
    primaryClass = {math.AP}
}

@book{Choquet-Bruhat-Book-2009,
    AUTHOR = {Choquet-Bruhat, Yvonne},
     TITLE = {General relativity and the {E}instein equations},
    SERIES = {Oxford Mathematical Monographs},
 PUBLISHER = {Oxford University Press, Oxford},
      YEAR = {2009},
     PAGES = {xxvi+785},
      ISBN = {978-0-19-923072-3},
   MRCLASS = {83-02 (35Q75 83C05)},
  MRNUMBER = {2473363},
MRREVIEWER = {Hans-Peter K\"{u}nzle}
}

@book{John-Book-1982,
    AUTHOR = {John, Fritz},
     TITLE = {Partial differential equations},
    SERIES = {Applied Mathematical Sciences},
    VOLUME = {1},
   EDITION = {Fourth},
 PUBLISHER = {Springer-Verlag, New York},
      YEAR = {1982},
     PAGES = {x+249},
      ISBN = {0-387-90609-6},
   MRCLASS = {35-01},
  MRNUMBER = {831655},
       DOI = {10.1007/978-1-4684-9333-7}
}

@BOOK{Rezzolla-Zanotti-Book-2013,
  author = {Rezzolla, Luciano and Zanotti, Olindo},
  title = {Relativistic Hydrodynamics},
  publisher = {Oxford University Press},
  address = {New York},
  year = {2013},
  doi = {10.1093/acprof:oso/9780198528906.001.0001}
}

@book{Romatschke:2017ejr,
    author = "Romatschke, Paul and Romatschke, Ulrike",
    title = "{Relativistic Fluid Dynamics In and Out of Equilibrium}",
    eprint = "1712.05815",
    archivePrefix = "arXiv",
    primaryClass = "nucl-th",
    doi = "10.1017/9781108651998",
    isbn = "978-1-108-48368-1, 978-1-108-75002-8",
    publisher = "Cambridge University Press",
    series = "Cambridge Monographs on Mathematical Physics",
    month = "5",
    year = "2019"
}

@book{Weinberg:1972kfs,
    author = "Weinberg, Steven",
    title = "{Gravitation and Cosmology}: {Principles and Applications of the General Theory of Relativity}",
    isbn = "978-0-471-92567-5, 978-0-471-92567-5",
    publisher = "John Wiley and Sons",
    address = "New York",
    year = "1972"
}

@BOOK{Landau-Lifshitz-Book-Fluids-1987,
    author = {Landau, Lev Davidovich and Lifshitz, Evgeny},
	edition = {Second},
	isbn = {0750627670},
	location = {Oxford, UK},
	pages = {552},
	publisher = {Butterworth-Heinemann},
	title = {Fluid Mechanics - Volume 6 (Course of Theoretical Physics)},
	year = {1987}
}

@BOOK{Lichnerowicz-Book-1955,
  author = {A. Lichnerowicz},
  title = {Th\'eories Relativistes de la Gravitation et de l'\'Electromagn\'etisme},
  publisher = {Masson et Cie},
  address = {Paris},
  year = {1955}
}

@article{Eckart:1940te,
    author = "Eckart, Carl",
    title = "{The Thermodynamics of irreversible processes. 3. Relativistic theory of the simple fluid}",
    doi = "10.1103/PhysRev.58.919",
    journal = "Phys. Rev.",
    volume = "58",
    pages = "919--924",
    year = "1940"
}

@article{Adcox:2004mh,
    author = "Adcox, K. and others",
    collaboration = "PHENIX",
    title = "{Formation of dense partonic matter in relativistic nucleus-nucleus collisions at RHIC: Experimental evaluation by the PHENIX collaboration}",
    eprint = "nucl-ex/0410003",
    archivePrefix = "arXiv",
    doi = "10.1016/j.nuclphysa.2005.03.086",
    journal = "Nucl. Phys. A",
    volume = "757",
    pages = "184--283",
    year = "2005"
}

@article{Back:2004je,
    author = "Back, B.B. and others",
    collaboration = "PHOBOS",
    title = "{The PHOBOS perspective on discoveries at RHIC}",
    eprint = "nucl-ex/0410022",
    archivePrefix = "arXiv",
    doi = "10.1016/j.nuclphysa.2005.03.084",
    journal = "Nucl. Phys. A",
    volume = "757",
    pages = "28--101",
    year = "2005"
}

@article{Arsene:2004fa,
    author = "Arsene, I. and others",
    collaboration = "BRAHMS",
    title = "{Quark gluon plasma and color glass condensate at RHIC? The Perspective from the BRAHMS experiment}",
    eprint = "nucl-ex/0410020",
    archivePrefix = "arXiv",
    doi = "10.1016/j.nuclphysa.2005.02.130",
    journal = "Nucl. Phys. A",
    volume = "757",
    pages = "1--27",
    year = "2005"
}

@article{Gyulassy:2004zy,
    author = "Gyulassy, Miklos and McLerran, Larry",
    editor = "Rischke, D. and Levin, G.",
    title = "{New forms of QCD matter discovered at RHIC}",
    eprint = "nucl-th/0405013",
    archivePrefix = "arXiv",
    doi = "10.1016/j.nuclphysa.2004.10.034",
    journal = "Nucl. Phys. A",
    volume = "750",
    pages = "30--63",
    year = "2005"
}

@article{Alford:2017rxf,
    author = "Alford, Mark G. and Bovard, Luke and Hanauske, Matthias and Rezzolla, Luciano and Schwenzer, Kai",
    title = "{Viscous Dissipation and Heat Conduction in Binary Neutron-Star Mergers}",
    eprint = "1707.09475",
    archivePrefix = "arXiv",
    primaryClass = "gr-qc",
    doi = "10.1103/PhysRevLett.120.041101",
    journal = "Phys. Rev. Lett.",
    volume = "120",
    number = "4",
    pages = "041101",
    year = "2018"
}

@article{Chabanov:2023blf,
    author = "Chabanov, Michail and Rezzolla, Luciano",
    title = "{Impact of bulk viscosity on the post-merger gravitational-wave signal from merging neutron stars}",
    eprint = "2307.10464",
    archivePrefix = "arXiv",
    primaryClass = "gr-qc",
    month = "7",
    year = "2023",
    journal = "arXiv e-prints"
}

@article{Shibata:2017xht,
    author = "Shibata, Masaru and Kiuchi, Kenta",
    title = "{Gravitational waves from remnant massive neutron stars of binary neutron star merger: Viscous hydrodynamics effects}",
    eprint = "1705.06142",
    archivePrefix = "arXiv",
    primaryClass = "astro-ph.HE",
    doi = "10.1103/PhysRevD.95.123003",
    journal = "Phys. Rev. D",
    volume = "95",
    number = "12",
    pages = "123003",
    year = "2017"
}

@article{Shibata:2017jyf,
    author = "Shibata, Masaru and Kiuchi, Kenta and Sekiguchi, Yu-ichiro",
    title = "{General relativistic viscous hydrodynamics of differentially rotating neutron stars}",
    eprint = "1703.10303",
    archivePrefix = "arXiv",
    primaryClass = "astro-ph.HE",
    doi = "10.1103/PhysRevD.95.083005",
    journal = "Phys. Rev. D",
    volume = "95",
    number = "8",
    pages = "083005",
    year = "2017"
}

@article{Alford:2020lla,
    author = "Alford, Mark and Harutyunyan, Arus and Sedrakian, Armen",
    title = "{Bulk Viscous Damping of Density Oscillations in Neutron Star Mergers}",
    eprint = "2006.07975",
    archivePrefix = "arXiv",
    primaryClass = "nucl-th",
    doi = "10.3390/particles3020034",
    journal = "Particles",
    volume = "3",
    number = "2",
    pages = "500--517",
    year = "2020"
}

@article{Alford:2019kdw,
    author = "Alford, Mark and Harutyunyan, Arus and Sedrakian, Armen",
    title = "{Bulk viscosity of baryonic matter with trapped neutrinos}",
    eprint = "1907.04192",
    archivePrefix = "arXiv",
    primaryClass = "astro-ph.HE",
    doi = "10.1103/PhysRevD.100.103021",
    journal = "Phys. Rev. D",
    volume = "100",
    number = "10",
    pages = "103021",
    year = "2019"
}

@article{Alford:2019qtm,
    author = "Alford, Mark G. and Harris, Steven P.",
    title = "{Damping of density oscillations in neutrino-transparent nuclear matter}",
    eprint = "1907.03795",
    archivePrefix = "arXiv",
    primaryClass = "nucl-th",
    doi = "10.1103/PhysRevC.100.035803",
    journal = "Phys. Rev. C",
    volume = "100",
    number = "3",
    pages = "035803",
    year = "2019"
}

@article{Alford:2020pld,
    author = "Alford, Mark G. and Haber, Alexander",
    title = "{Strangeness-changing Rates and Hyperonic Bulk Viscosity in Neutron Star Mergers}",
    eprint = "2009.05181",
    archivePrefix = "arXiv",
    primaryClass = "nucl-th",
    doi = "10.1103/PhysRevC.103.045810",
    journal = "Phys. Rev. C",
    volume = "103",
    number = "4",
    pages = "045810",
    year = "2021"
}

@article{Most:2022yhe,
    author = "Most, Elias R. and Haber, Alexander and Harris, Steven P. and Zhang, Ziyuan and Alford, Mark G. and Noronha, Jorge",
    title = "{Emergence of Microphysical Bulk Viscosity in Binary Neutron Star Postmerger Dynamics}",
    eprint = "2207.00442",
    archivePrefix = "arXiv",
    primaryClass = "astro-ph.HE",
    reportNumber = "INT-PUB-22-017",
    doi = "10.3847/2041-8213/ad454f",
    journal = "Astrophys. J. Lett.",
    volume = "967",
    number = "1",
    pages = "L14",
    year = "2024"
}

@article{Most:2021zvc,
    author = "Most, Elias R. and Harris, Steven P. and Plumberg, Christopher and Alford, Mark G. and Noronha, Jorge and Noronha-Hostler, Jacquelyn and Pretorius, Frans and Witek, Helvi and Yunes, Nicol\'as",
    title = "{Projecting the likely importance of weak-interaction-driven bulk viscosity in neutron star mergers}",
    eprint = "2107.05094",
    archivePrefix = "arXiv",
    primaryClass = "astro-ph.HE",
    reportNumber = "INT-PUB-21-015",
    doi = "10.1093/mnras/stab2793",
    journal = "Mon. Not. R. Astron. Soc.",
    volume = "509",
    number = "1",
    pages = "1096--1108",
    year = "2021"
}

@article{Ripley:2023qxo,
    author = "Ripley, Justin L. and Hegade K. R., Abhishek and Yunes, Nicolas",
    title = "{Probing internal dissipative processes of neutron stars with gravitational waves during the inspiral of neutron star binaries}",
    eprint = "2306.15633",
    archivePrefix = "arXiv",
    primaryClass = "gr-qc",
    doi = "10.1103/PhysRevD.108.103037",
    journal = "Phys. Rev. D",
    volume = "108",
    number = "10",
    pages = "103037",
    year = "2023"
}

@article{Camelio:2022ljs,
    author = "Camelio, Giovanni and Gavassino, Lorenzo and Antonelli, Marco and Bernuzzi, Sebastiano and Haskell, Brynmor",
    title = "{Simulating bulk viscosity in neutron stars. I. Formalism}",
    eprint = "2204.11809",
    archivePrefix = "arXiv",
    primaryClass = "gr-qc",
    doi = "10.1103/PhysRevD.107.103031",
    journal = "Phys. Rev. D",
    volume = "107",
    number = "10",
    pages = "103031",
    year = "2023"
}

@article{Camelio:2022fds,
    author = "Camelio, Giovanni and Gavassino, Lorenzo and Antonelli, Marco and Bernuzzi, Sebastiano and Haskell, Brynmor",
    title = "{Simulating bulk viscosity in neutron stars. II. Evolution in spherical symmetry}",
    eprint = "2204.11810",
    archivePrefix = "arXiv",
    primaryClass = "gr-qc",
    doi = "10.1103/PhysRevD.107.103032",
    journal = "Phys. Rev. D",
    volume = "107",
    number = "10",
    pages = "103032",
    year = "2023"
}

@article{Huth:2021bsp,
    author = "Huth, S. and others",
    title = "{Constraining Neutron-Star Matter with Microscopic and Macroscopic Collisions}",
    eprint = "2107.06229",
    archivePrefix = "arXiv",
    primaryClass = "nucl-th",
    reportNumber = "LA-UR-21-22072",
    doi = "10.1038/s41586-022-04750-w",
    journal = "Nature",
    volume = "606",
    pages = "276--280",
    year = "2022"
}

@article{Dexheimer:2020zzs,
    author = "Dexheimer, Veronica and Noronha, Jorge and Noronha-Hostler, Jacquelyn and Ratti, Claudia and Yunes, Nicol\'as",
    title = "{Future physics perspectives on the equation of state from heavy ion collisions to neutron stars}",
    eprint = "2010.08834",
    archivePrefix = "arXiv",
    primaryClass = "nucl-th",
    doi = "10.1088/1361-6471/abe104",
    journal = "J. Phys. G",
    volume = "48",
    number = "7",
    pages = "073001",
    year = "2021"
}

@article{MUSES:2023hyz,
    author = "Kumar, Rajesh and others",
    collaboration = "MUSES",
    title = "{Theoretical and experimental constraints for the equation of state of dense and hot matter}",
    eprint = "2303.17021",
    archivePrefix = "arXiv",
    primaryClass = "nucl-th",
    doi = "10.1007/s41114-024-00049-6",
    journal = "Living Rev. Rel.",
    volume = "27",
    number = "1",
    pages = "3",
    year = "2024"
}

@article{Huang:2022mqp,
    author = "Huang, Yong-Jia and Baiotti, Luca and Kojo, Toru and Takami, Kentaro and Sotani, Hajime and Togashi, Hajime and Hatsuda, Tetsuo and Nagataki, Shigehiro and Fan, Yi-Zhong",
    title = "{Merger and Postmerger of Binary Neutron Stars with a Quark-Hadron Crossover Equation of State}",
    eprint = "2203.04528",
    archivePrefix = "arXiv",
    primaryClass = "astro-ph.HE",
    reportNumber = "RIKEN-iTHEMS-Report-22",
    doi = "10.1103/PhysRevLett.129.181101",
    journal = "Phys. Rev. Lett.",
    volume = "129",
    number = "18",
    pages = "181101",
    year = "2022"
}

@article{Hammond:2022uua,
    author = "Hammond, Peter and Hawke, Ian and Andersson, Nils",
    title = "{Impact of nuclear reactions on gravitational waves from neutron star mergers}",
    eprint = "2205.11377",
    archivePrefix = "arXiv",
    primaryClass = "astro-ph.HE",
    doi = "10.1103/PhysRevD.107.043023",
    journal = "Phys. Rev. D",
    volume = "107",
    number = "4",
    pages = "043023",
    year = "2023"
}

@article{Hiscock:1985zz,
    author = "Hiscock, William A. and Lindblom, Lee",
    title = "{Generic instabilities in first-order dissipative relativistic fluid theories}",
    doi = "10.1103/PhysRevD.31.725",
    journal = "Phys. Rev. D",
    volume = "31",
    pages = "725--733",
    year = "1985"
}

@article{Pichon-1965,
    AUTHOR = {Pichon, Guy},
     TITLE = {\'{E}tude relativiste de fluides visqueux et charg\'{e}s},
   JOURNAL = {Ann. Inst. H. Poincar\'{e} Sect. A (N.S.)},
  FJOURNAL = {Annales de l'Institut Henri Poincar\'{e}. Section A. Physique
              Th\'{e}orique. Nouvelle S\'{e}rie},
    VOLUME = {2},
      YEAR = {1965},
     PAGES = {21--85},
      ISSN = {0246-0211},
   MRCLASS = {76.00},
  MRNUMBER = {204029},
MRREVIEWER = {N. Coburn}
}

@article{Stewart-1977,
    AUTHOR = {Stewart, J. M.},
     TITLE = {On transient relativistic thermodynamics and kinetic theory},
   JOURNAL = {Proc. R. Soc. A},
  FJOURNAL = {Proceedings of the Royal Society. London. Series A.
              Mathematical, Physical and Engineering Sciences},
    VOLUME = {357},
      YEAR = {1977},
    NUMBER = {1688},
     PAGES = {59--75},
      ISSN = {0962-8444},
   MRCLASS = {80.60},
  MRNUMBER = {468685},
MRREVIEWER = {Ole J. Heilmann},
       DOI = {10.1098/rspa.1977.0155}
}

@article{Israel-Stewart-1979,
    AUTHOR = {Israel, W. and Stewart, J. M.},
     TITLE = {On transient relativistic thermodynamics and kinetic theory.
              {II}},
   JOURNAL = {Proc. R. Soc. A},
  FJOURNAL = {Proceedings of the Royal Society. London. Series A.
              Mathematical, Physical and Engineering Sciences},
    VOLUME = {365},
      YEAR = {1979},
    NUMBER = {1720},
     PAGES = {43--52},
      ISSN = {0962-8444},
   MRCLASS = {83C99 (35Q20 76Y05 80A99)},
  MRNUMBER = {527592},
       DOI = {10.1098/rspa.1979.0005}
}

@article{Stewart-Stewart-1976,
    AUTHOR = {Israel, W. and Stewart, J. M.},
     TITLE = {Thermodynamics of nonstationary and transient effects in a relativistic gas},
   JOURNAL = {Phys. Lett. A},
  FJOURNAL = {Physics Letters A},
    VOLUME = {58},
      YEAR = {1976},
    NUMBER = {4},
     PAGES = {213--215}
}

@article{Mueller-1967,
  author = {M\"uller, I.},
  title = {Zum {P}aradox der {W}{\"a}rmeleitungstheorie},
  journal = {Z. Phys},
  volume = {198},
  pages = {329-344},
  year = {1967},
  doi = {10.1007/BF01326412}
}

@article{Israel:1976tn,
    author = "Israel, W.",
    title = "{Nonstationary irreversible thermodynamics: A Causal relativistic theory}",
    doi = "10.1016/0003-4916(76)90064-6",
    journal = "Ann. Phys.",
    volume = "100",
    pages = "310--331",
    year = "1976"
}

@article{Israel:1979wp,
    author = "Israel, W. and Stewart, J.M.",
    title = "{Transient relativistic thermodynamics and kinetic theory}",
    doi = "10.1016/0003-4916(79)90130-1",
    journal = "Ann. Phys.",
    volume = "118",
    pages = "341--372",
    year = "1979"
}

@article{Grad-1949,
    AUTHOR = {Grad, Harold},
     TITLE = {On the kinetic theory of rarefied gases},
   JOURNAL = {Commun. Pure Appl. Math.},
  FJOURNAL = {Communications on Pure and Applied Mathematics},
    VOLUME = {2},
      YEAR = {1949},
     PAGES = {331--407},
      ISSN = {0010-3640},
   MRCLASS = {76.1X},
  MRNUMBER = {33674},
MRREVIEWER = {H. S. Tsien},
       DOI = {10.1002/cpa.3160020403}
}

@article{Denicol:2012cn,
    author = "Denicol, G.S. and Niemi, H. and Molnar, E. and Rischke, D.H.",
    title = "{Derivation of transient relativistic fluid dynamics from the Boltzmann equation}",
    eprint = "1202.4551",
    archivePrefix = "arXiv",
    primaryClass = "nucl-th",
    doi = "10.1103/PhysRevD.85.114047",
    journal = "Phys. Rev. D",
    volume = "85",
    pages = "114047",
    year = "2012",
    note = "[Erratum: Phys.Rev.D 91, 039902 (2015)]"
}

@article{Baier:2007ix,
    author = "Baier, Rudolf and Romatschke, Paul and Son, Dam Thanh and Starinets, Andrei O. and Stephanov, Mikhail A.",
    title = "{Relativistic viscous hydrodynamics, conformal invariance, and holography}",
    eprint = "0712.2451",
    archivePrefix = "arXiv",
    primaryClass = "hep-th",
    reportNumber = "BI-TP-2007-29, INT-PUB-07-45, SHEP-07-47",
    doi = "10.1088/1126-6708/2008/04/100",
    journal = "JHEP",
    volume = "04",
    pages = "100",
    year = "2008"
}

@incollection{Leray-Ohya-1964,
    AUTHOR = {Leray, Jean and Ohya, Yujiro},
     TITLE = {Syst\`emes lin\'{e}aires, hyperboliques non stricts},
 BOOKTITLE = {Deuxi\`eme {C}olloq. l'{A}nal. {F}onct},
     PAGES = {105--144},
 PUBLISHER = {Centre Belge Recherches Math., Librairie Universitaire,
              Louvain},
      YEAR = {1964},
   MRCLASS = {35.53},
  MRNUMBER = {0190544},
MRREVIEWER = {D. Ludwig}
}

@article{Leray-Ohya-1967,
    AUTHOR = {Leray, Jean and Ohya, Yujiro},
     TITLE = {\'{E}quations et syst\`emes non-lin\'{e}aires, hyperboliques nonstricts},
   JOURNAL = {Math. Ann.},
  FJOURNAL = {Mathematische Annalen},
    VOLUME = {170},
      YEAR = {1967},
     PAGES = {167--205},
      ISSN = {0025-5831},
   MRCLASS = {35.57},
  MRNUMBER = {208136},
MRREVIEWER = {C. Harvey},
       DOI = {10.1007/BF01350150}
}

@book{Leray-Book-1953,
    AUTHOR = {Leray, Jean},
     TITLE = {Hyperbolic differential equations},
 PUBLISHER = {Institute for Advanced Study (IAS), Princeton, N.J.},
      YEAR = {1953},
     PAGES = {240},
   MRCLASS = {35.0X},
  MRNUMBER = {0063548},
MRREVIEWER = {H. G. Garnir}
}

@BOOK{Lichnerowicz-Book-1967,
  author = {A. Lichnerowicz},
  title = {Relativistic Hydrodynamics and Magnetohydrodynamics: Lectures on the Existence of Solutions},
  publisher = {W. A. Benjamin},
  address = {New York},
  year = {1967}
}

@article{Choquet-Bruhat-1966,
    AUTHOR = {Choquet-Bruhat, Yvonne},
     TITLE = {Diagonalisation des syst\`emes quasi-lin\'{e}aires et hyperbolicit\'{e} non stricte},
   JOURNAL = {J. Math. Pures Appl. (9)},
  FJOURNAL = {Journal de Math\'{e}matiques Pures et Appliqu\'{e}es. Neuvi\`eme S\'{e}rie},
    VOLUME = {45},
      YEAR = {1966},
     PAGES = {371--386},
      ISSN = {0021-7824},
   MRCLASS = {35.09},
  MRNUMBER = {216131},
MRREVIEWER = {A. K. Azia}
}

@book{Mizohata-Book-1985,
    AUTHOR = {Mizohata, Sigeru},
     TITLE = {On the {C}auchy problem},
    SERIES = {Notes and Reports in Mathematics in Science and Engineering},
    VOLUME = {3},
 PUBLISHER = {Academic Press, Inc., Orlando, FL; Science Press Beijing,
              Beijing},
      YEAR = {1985},
     PAGES = {vi+177},
      ISBN = {0-12-501660-3},
   MRCLASS = {35Axx (35Gxx)},
  MRNUMBER = {860041},
MRREVIEWER = {G. Roach}
}

@book{Mizohata-1965,
    AUTHOR = {Mizohata, Sigeru},
     TITLE = {Lectures on {C}auchy problem},
    SERIES = {Tata Institute of Fundamental Research Lectures on Mathematics
              and Physics, No. 35},
      NOTE = {Notes by M. K. Venkatesha Murthy and B. V. Singbal},
 PUBLISHER = {Tata Institute of Fundamental Research, Bombay},
      YEAR = {1965},
     PAGES = {ii+226},
   MRCLASS = {35.37},
  MRNUMBER = {0219881},
MRREVIEWER = {J. Kop\'{a}\v{c}ek},
  url = {http://www.math.tifr.res.in/~publ/ln/tifr35.pdf}
}

@book{Hormander-Book-1976,
    AUTHOR = {H\"{o}rmander, Lars},
     TITLE = {Linear partial differential operators},
 PUBLISHER = {Springer-Verlag, Berlin-New York},
      YEAR = {1976},
     PAGES = {vii+285},
   MRCLASS = {35-XX (46FXX)},
  MRNUMBER = {0404822}
}

@ARTICLE{1984ApJ...277..296H,
       author = {{Hawley}, J.~F. and {Smarr}, L.~L. and {Wilson}, J.~R.},
        title = "{A numerical study of nonspherical black hole accretion. I Equations and test problems}",
      journal = {\apj},
         year = 1984,
        month = feb,
       volume = {277},
        pages = {296-311},
          doi = {10.1086/161696},
       adsurl = {https://ui.adsabs.harvard.edu/abs/1984ApJ...277..296H}
}

@ARTICLE{Michel...1972,
       author = {{Michel}, F. Curtis},
        title = "{Accretion of Matter by Condensed Objects}",
      journal = {Astrophysics and Space Science},
         year = 1972,
        month = jan,
       volume = {15},
       number = {1},
        pages = {153-160},
          doi = {10.1007/BF00649949},
       adsurl = {https://ui.adsabs.harvard.edu/abs/1972Ap&SS..15..153M}
}

@ARTICLE{Bondi..1952MNRAS.112..195B,
       author = {{Bondi}, H.},
        title = "{On spherically symmetrical accretion}",
      journal = {Mon. Not. R. Astron. Soc.},
         year = 1952,
        month = jan,
       volume = {112},
        pages = {195},
          doi = {10.1093/mnras/112.2.195},
       adsurl = {https://ui.adsabs.harvard.edu/abs/1952MNRAS.112..195B}
}

@article{Clarisse:2025lli,
    author = "Clarisse, Nicolas and Pinho, Eduardo O. and Patel, Teerthal and Bemfica, Fabio S. and Hippert, Mauricio and Noronha, Jorge",
    title = "{Flux-conservative Bemfica-Disconzi-Noronha-Kovtun hydrodynamics and shock regularization}",
    eprint = "2510.16603",
    archivePrefix = "arXiv",
    primaryClass = "gr-qc",
    doi = "10.1103/f8y1-3yck",
    journal = "Phys. Rev. D",
    volume = "113",
    number = "2",
    pages = "024051",
    year = "2026"
}

@article{Friestuhler-BDNK-shock,
  title = {Nonexistence and existence of shock profiles in the Bemfica-Disconzi-Noronha model},
  author = {Freist\"uhler, Heinrich},
  journal = {Phys. Rev. D},
  volume = {103},
  issue = {12},
  pages = {124045},
  numpages = {7},
  year = {2021},
  month = {Jun},
  publisher = {American Physical Society},
  doi = {10.1103/PhysRevD.103.124045},
  url = {https://link.aps.org/doi/10.1103/PhysRevD.103.124045}
}

@article{Pandya-21,
  title = {Numerical exploration of first-order relativistic hydrodynamics},
  author = {Pandya, Alex and Pretorius, Frans},
  journal = {Phys. Rev. D},
  volume = {104},
  issue = {2},
  pages = {023015},
  numpages = {23},
  year = {2021},
  month = {Jul},
  publisher = {American Physical Society},
  doi = {10.1103/PhysRevD.104.023015},
  url = {https://link.aps.org/doi/10.1103/PhysRevD.104.023015}
}

@article{Shum:2025jnl,
    author = "Shum, Harry L. H. and Abalos, Fernando and Bea, Yago and Bezares, Miguel and Figueras, Pau and Palenzuela, Carlos",
    title = "{Neutron star evolution with the Bemfica-Disconzi-Noronha-Kovtun viscous hydrodynamics framework}",
    eprint = "2509.15303",
    archivePrefix = "arXiv",
    primaryClass = "gr-qc",
    doi = "10.1103/mthy-4119",
    journal = "Phys. Rev. D",
    volume = "113",
    number = "8",
    pages = "084029",
    year = "2026"
}

@article{WENOZ-Borges,
title = {An improved weighted essentially non-oscillatory scheme for hyperbolic conservation laws},
journal = {Journal of Computational Physics},
volume = {227},
number = {6},
pages = {3191-3211},
year = {2008},
issn = {0021-9991},
doi = {https://doi.org/10.1016/j.jcp.2007.11.038},
url = {https://www.sciencedirect.com/science/article/pii/S0021999107005232},
author = {Rafael Borges and Monique Carmona and Bruno Costa and Wai Sun Don}
}

@misc{bea2025relativisticnavierstokesdescriptionquarkgluon,
      title={Relativistic Navier-Stokes description of the quark-gluon plasma radial flow}, 
      author={Yago Bea},
      year={2025},
      eprint={2503.02931},
      archivePrefix={arXiv},
      primaryClass={hep-ph},
      url={https://arxiv.org/abs/2503.02931}, 
}

@ARTICLE{1978A&A....63..209K,
       author = {{Kozlowski}, M. and {Jaroszynski}, M. and {Abramowicz}, M.~A.},
        title = "{The analytic theory of fluid disks orbiting the Kerr black hole.}",
      journal = {Astronomy and Astrophysics},
         year = {1978},
       volume = {63},
       number = {1-2},
        pages = {209-220},
       adsurl = {https://ui.adsabs.harvard.edu/abs/1978A&A....63..209K}
}

@article{ CWENO,
	author = {{Levy, Doron} and {Puppo, Gabriella} and {Russo, Giovanni}},
	title = {Central WENO schemes for hyperbolic systems  of conservation laws},
	DOI= "10.1051/m2an:1999152",
	url= "https://doi.org/10.1051/m2an:1999152",
	journal = {ESAIM: M2AN},
	year = 1999,
	volume = 33,
	number = 3,
	pages = "547-571",
	month = "",
}

@article{Pandya_2022_IG,
   title={Causal, stable first-order viscous relativistic hydrodynamics with ideal gas microphysics},
   volume={106},
   ISSN={2470-0029},
   url={http://dx.doi.org/10.1103/PhysRevD.106.123036},
   DOI={10.1103/physrevd.106.123036},
   number={12},
   journal={Physical Review D},
   publisher={American Physical Society (APS)},
   author={Pandya, Alex and Most, Elias R. and Pretorius, Frans},
   year={2022},
   month=Dec }

@article{SHU1988439,
title = {Efficient implementation of essentially non-oscillatory shock-capturing schemes},
journal = {Journal of Computational Physics},
volume = {77},
number = {2},
pages = {439-471},
year = {1988},
issn = {0021-9991},
doi = {https://doi.org/10.1016/0021-9991(88)90177-5},
url = {https://www.sciencedirect.com/science/article/pii/0021999188901775},
author = {Chi-Wang Shu and Stanley Osher}
}

@article{DelZanna:2007pk,
    author = "Del Zanna, L. and Zanotti, O. and Bucciantini, N. and Londrillo, P.",
    title = "{ECHO: an Eulerian Conservative High Order scheme for general relativistic magnetohydrodynamics and magnetodynamics}",
    eprint = "0704.3206",
    archivePrefix = "arXiv",
    primaryClass = "astro-ph",
    doi = "10.1051/0004-6361:20077093",
    journal = "Astron. Astrophys.",
    volume = "473",
    pages = "11--30",
    year = "2007"
}

@article{Daigne:2003tf,
    author = "Daigne, Frederic and Font, Jose A.",
    title = "{The runaway instability of thick discs around black holes. II. non constant angular momentum discs}",
    eprint = "astro-ph/0311618",
    archivePrefix = "arXiv",
    doi = "10.1111/j.1365-2966.2004.07547.x",
    journal = "Mon. Not. Roy. Astron. Soc.",
    volume = "349",
    pages = "841",
    year = "2004"
}

@article{Shakura:1972te,
    author = "Shakura, N. I. and Sunyaev, R. A.",
    title = "{Black holes in binary systems. Observational appearance}",
    journal = "Astron. Astrophys.",
    volume = "24",
    pages = "337--355",
    year = "1973"
}

@article{Armas:2026kfp,
    author = "Armas, Jay",
    title = "{Stable and causal frames for BDNK hydrodynamics}",
    eprint = "2610.28984",
    archivePrefix = "arXiv",
    primaryClass = "hep-th",
    month = "10",
    year = "2026"
}

@article{Sweby1984,
    author  = {Sweby, P. K.},
    title   = {High Resolution Schemes Using Flux Limiters for Hyperbolic Conservation Laws},
    journal = {SIAM Journal on Numerical Analysis},
    volume  = {21},
    number  = {5},
    pages   = {995--1011},
    year    = {1984},
    doi     = {10.1137/0721062}
}

@article{Shibata:2025ycs,
    author = "Shibata, Masaru and Kawaguchi, Kyohei and Lam, Alan Tsz-Lok",
    title = "{Viscous accretion and ejection from tori around black holes in general relativity}",
    eprint = "2502.17566",
    archivePrefix = "arXiv",
    primaryClass = "astro-ph.HE",
    doi = "10.1103/PhysRevD.111.083042",
    journal = "Phys. Rev. D",
    volume = "111",
    number = "8",
    pages = "083042",
    year = "2025"
}

@article{Chabanov:2021dee,
    author = "Chabanov, Michail and Rezzolla, Luciano and Rischke, Dirk H.",
    title = "{General-relativistic hydrodynamics of non-perfect fluids: 3+1 conservative formulation and application to viscous black hole accretion}",
    eprint = "2102.10419",
    archivePrefix = "arXiv",
    primaryClass = "gr-qc",
    doi = "10.1093/mnras/stab1384",
    journal = "Mon. Not. Roy. Astron. Soc.",
    volume = "505",
    number = "4",
    pages = "5910--5940",
    year = "2021"
}

@article{Moreland:2014oya,
      author         = "Moreland, J. Scott and Bernhard, Jonah E. and Bass,
                        Steffen A.",
      title          = "{Alternative ansatz to wounded nucleon and binary
                        collision scaling in high-energy nuclear collisions}",
      journal        = "Phys.Rev.",
      number         = "1",
      volume         = "C92",
      pages          = "011901",
      doi            = "10.1103/PhysRevC.92.011901",
      year           = "2015",
      eprint         = "1412.4708",
      archivePrefix  = "arXiv",
      primaryClass   = "nucl-th",
      SLACcitation   = "%%CITATION = ARXIV:1412.4708;%%",
}

@article{JETSCAPE:2020mzn,
    author = "Everett, D. and others",
    collaboration = "JETSCAPE",
    title = "{Multisystem Bayesian constraints on the transport coefficients of QCD matter}",
    eprint = "2011.01430",
    archivePrefix = "arXiv",
    primaryClass = "hep-ph",
    doi = "10.1103/PhysRevC.103.054904",
    journal = "Phys. Rev. C",
    volume = "103",
    number = "5",
    pages = "054904",
    year = "2021"
}

@article{Nijs:2020roc,
    author = {Nijs, Govert and van der Schee, Wilke and G{\"u}rsoy, Umut and Snellings, Raimond},
    title = "{Bayesian analysis of heavy ion collisions with the heavy ion computational framework Trajectum}",
    eprint = "2010.15134",
    archivePrefix = "arXiv",
    primaryClass = "nucl-th",
    reportNumber = "CERN-TH-2020-175, MIT-CTP/5251",
    doi = "10.1103/PhysRevC.103.054909",
    journal = "Phys. Rev. C",
    volume = "103",
    number = "5",
    pages = "054909",
    year = "2021"
}

@article{Parkkila:2021yha,
    author = "Parkkila, J. E. and Onnerstad, A. and Taghavi, S. F. and Mordasini, C. and Bilandzic, A. and Virta, M. and Kim, D. J.",
    title = "{New constraints for QCD matter from improved Bayesian parameter estimation in heavy-ion collisions at LHC}",
    eprint = "2111.08145",
    archivePrefix = "arXiv",
    primaryClass = "hep-ph",
    doi = "10.1016/j.physletb.2022.137485",
    journal = "Phys. Lett. B",
    volume = "835",
    pages = "137485",
    year = "2022"
}

@article{Moreland:2018jos,
    author = "Moreland, J. Scott and Bernhard, Jonah E. and Bass, Steffen A.",
    editor = "Antinori, Federico and Dainese, Andrea and Giubellino, Paolo and Greco, Vincenzo and Lombardo, Maria Paola and Scomparin, Enrico",
    title = "{Estimating nucleon substructure properties in a unified model of p-Pb and Pb-Pb collisions}",
    eprint = "1806.04802",
    archivePrefix = "arXiv",
    primaryClass = "nucl-th",
    doi = "10.1016/j.nuclphysa.2018.07.011",
    journal = "Nucl. Phys. A",
    volume = "982",
    pages = "503--506",
    year = "2019"
}

@article{Liu:2015nwa,
    author = "Liu, Jia and Shen, Chun and Heinz, Ulrich",
    title = "{Pre-equilibrium evolution effects on heavy-ion collision observables}",
    eprint = "1504.02160",
    archivePrefix = "arXiv",
    primaryClass = "nucl-th",
    doi = "10.1103/PhysRevC.91.064906",
    journal = "Phys. Rev. C",
    volume = "91",
    number = "6",
    pages = "064906",
    year = "2015",
    note = "[Erratum: Phys.Rev.C 92, 049904 (2015)]"
}

@article{Hiscock:1987zz,
    author = "Hiscock, William A. and Lindblom, Lee",
    title = "{Linear plane waves in dissipative relativistic fluids}",
    doi = "10.1103/PhysRevD.35.3723",
    journal = "Phys. Rev. D",
    volume = "35",
    pages = "3723--3732",
    year = "1987"
}

@article{Borsanyi:2013bia,
    author = "Borsanyi, Szabocls and Fodor, Zoltan and Hoelbling, Christian and Katz, Sandor D. and Krieg, Stefan and Szabo, Kalman K.",
    title = "{Full result for the QCD equation of state with 2+1 flavors}",
    eprint = "1309.5258",
    archivePrefix = "arXiv",
    primaryClass = "hep-lat",
    doi = "10.1016/j.physletb.2014.01.007",
    journal = "Phys. Lett. B",
    volume = "730",
    pages = "99--104",
    year = "2014"
}

@article{HotQCD:2014kol,
    author = "Bazavov, A. and others",
    collaboration = "HotQCD",
    title = "{Equation of state in ( 2+1 )-flavor QCD}",
    eprint = "1407.6387",
    archivePrefix = "arXiv",
    primaryClass = "hep-lat",
    reportNumber = "BNL-105928-2014-JA",
    doi = "10.1103/PhysRevD.90.094503",
    journal = "Phys. Rev. D",
    volume = "90",
    pages = "094503",
    year = "2014"
}

@ARTICLE{Fishbone-Moncrief,
       author = {{Fishbone}, L.~G. and {Moncrief}, V.},
        title = "{Relativistic fluid disks in orbit around Kerr black holes.}",
      journal = {\apj},
         year = 1976,
        month = aug,
       volume = {207},
        pages = {962-976},
          doi = {10.1086/154565},
       adsurl = {https://ui.adsabs.harvard.edu/abs/1976ApJ...207..962F}
}

@article{STAR:2005gfr,
    author = "Adams, John and others",
    collaboration = "STAR",
    title = "{Experimental and theoretical challenges in the search for the quark gluon plasma: The STAR Collaboration's critical assessment of the evidence from RHIC collisions}",
    eprint = "nucl-ex/0501009",
    archivePrefix = "arXiv",
    doi = "10.1016/j.nuclphysa.2005.03.085",
    journal = "Nucl. Phys. A",
    volume = "757",
    pages = "102--183",
    year = "2005"
}

@article{Pandya_2022,
   title={Conservative finite volume scheme for first-order viscous relativistic hydrodynamics},
   volume={105},
   ISSN={2470-0029},
   url={http://dx.doi.org/10.1103/PhysRevD.105.123001},
   DOI={10.1103/physrevd.105.123001},
   number={12},
   journal={Physical Review D},
   publisher={American Physical Society (APS)},
   author={Pandya, Alex and Most, Elias R. and Pretorius, Frans},
   year={2022} }

@article{Molnar:2013lta,
    author = "Moln{\'a}r, E. and Niemi, H. and Denicol, G. S. and Rischke, D. H.",
    title = "{Relative importance of second-order terms in relativistic dissipative fluid dynamics}",
    eprint = "1308.0785",
    archivePrefix = "arXiv",
    primaryClass = "nucl-th",
    doi = "10.1103/PhysRevD.89.074010",
    journal = "Phys. Rev. D",
    volume = "89",
    number = "7",
    pages = "074010",
    year = "2014"
}

@article{Denicol:2014vaa,
    author = "Denicol, G. S. and Jeon, S. and Gale, C.",
    title = "{Transport Coefficients of Bulk Viscous Pressure in the 14-moment approximation}",
    eprint = "1403.0962",
    archivePrefix = "arXiv",
    primaryClass = "nucl-th",
    doi = "10.1103/PhysRevC.90.024912",
    journal = "Phys. Rev. C",
    volume = "90",
    number = "2",
    pages = "024912",
    year = "2014"
}

@article{Schianchi:2026ayx,
    author = "Schianchi, Federico and Abalos, Fernando",
    title = "{A strongly hyperbolic viscous relativistic hydrodynamics theory with first-order charge current}",
    eprint = "2602.20254",
    archivePrefix = "arXiv",
    primaryClass = "gr-qc",
    month = "2",
    year = "2026"
}

@misc{redondoyuste2025superradiantamplificationrotatingviscous,
      title={Superradiant amplification by rotating viscous compact objects}, 
      author={Jaime Redondo-Yuste and Vitor Cardoso},
      year={2025},
      eprint={2506.13850},
      archivePrefix={arXiv},
      primaryClass={gr-qc},
      url={https://arxiv.org/abs/2506.13850}, 
}

@misc{caballero2025neutronstarradialperturbations,
      title={Neutron Star Radial Perturbations for Causal, Viscous, Relativistic Fluids}, 
      author={Daniel A. Caballero and Nicolás Yunes},
      year={2025},
      eprint={2506.09149},
      archivePrefix={arXiv},
      primaryClass={gr-qc},
      url={https://arxiv.org/abs/2506.09149}, 
}

@article{Bhambure:2024axa,
    author = "Bhambure, Jay and Mazeliauskas, Aleksas and Paquet, Jean-Francois and Singh, Rajeev and Singh, Mayank and Teaney, Derek and Zhou, Fabian",
    title = "{Relativistic viscous hydrodynamics in the density frame: Numerical tests and comparisons}",
    eprint = "2412.10303",
    archivePrefix = "arXiv",
    primaryClass = "nucl-th",
    doi = "10.1103/PhysRevC.111.064910",
    journal = "Phys. Rev. C",
    volume = "111",
    number = "6",
    pages = "064910",
    year = "2025"
}

@article{Lier:2026owd,
    author = "Lier, Ruben and Armas, Jay and Porth, Oliver",
    title = "{Finite-volume scheme for first-order viscoresistive relativistic magnetohydrodynamics}",
    eprint = "2606.22691",
    archivePrefix = "arXiv",
    primaryClass = "astro-ph.HE",
    month = "6",
    year = "2026"
}

@article{Bhattacharyya:2007vjd,
    author = "Bhattacharyya, Sayantani and Hubeny, Veronika E and Minwalla, Shiraz and Rangamani, Mukund",
    title = "{Nonlinear Fluid Dynamics from Gravity}",
    eprint = "0712.2456",
    archivePrefix = "arXiv",
    primaryClass = "hep-th",
    reportNumber = "TIFR-TH-07-44, DCPT-07-73, NI07097",
    doi = "10.1088/1126-6708/2008/02/045",
    journal = "JHEP",
    volume = "02",
    pages = "045",
    year = "2008"
}

@article{Gavassino:2025bsn,
    author = "Gavassino, Lorenzo and Hirvonen, Henry and Paquet, Jean-Fran{\c{c}}ois and Singh, Mayank and Soares Rocha, Gabriel",
    title = "{Acausality-driven instabilities in relativistic viscous hydrodynamics}",
    eprint = "2508.04918",
    archivePrefix = "arXiv",
    primaryClass = "nucl-th",
    doi = "10.1103/np53-f6cl",
    journal = "Phys. Rev. D",
    volume = "112",
    number = "7",
    pages = "076018",
    year = "2025"
}

@article{Bemfica:2020xym,
    author = "Bemfica, F{\'a}bio S. and Disconzi, Marcelo M. and Hoang, Vu and Noronha, Jorge and Radosz, Maria",
    title = "{Nonlinear Constraints on Relativistic Fluids Far from Equilibrium}",
    eprint = "2005.11632",
    archivePrefix = "arXiv",
    primaryClass = "hep-th",
    doi = "10.1103/PhysRevLett.126.222301",
    journal = "Phys. Rev. Lett.",
    volume = "126",
    number = "22",
    pages = "222301",
    year = "2021"
}

@article{Bemfica:2019cop,
    author = "Bemfica, Fabio S. and Disconzi, Marcelo M. and Noronha, Jorge",
    title = "{Causality of the Einstein-Israel-Stewart Theory with Bulk Viscosity}",
    eprint = "1901.06701",
    archivePrefix = "arXiv",
    primaryClass = "gr-qc",
    doi = "10.1103/PhysRevLett.122.221602",
    journal = "Phys. Rev. Lett.",
    volume = "122",
    number = "22",
    pages = "221602",
    year = "2019"
}

@article{Floerchinger:2017cii,
    author = "Floerchinger, Stefan and Grossi, Eduardo",
    title = "{Causality of fluid dynamics for high-energy nuclear collisions}",
    eprint = "1711.06687",
    archivePrefix = "arXiv",
    primaryClass = "nucl-th",
    doi = "10.1007/JHEP08(2018)186",
    journal = "JHEP",
    volume = "08",
    pages = "186",
    year = "2018"
}

@article{Pu:2009fj,
    author = "Pu, Shi and Koide, Tomoi and Rischke, Dirk H.",
    title = "{Does stability of relativistic dissipative fluid dynamics imply causality?}",
    eprint = "0907.3906",
    archivePrefix = "arXiv",
    primaryClass = "hep-ph",
    doi = "10.1103/PhysRevD.81.114039",
    journal = "Phys. Rev. D",
    volume = "81",
    pages = "114039",
    year = "2010"
}

@article{Denicol:2008ha,
    author = "Denicol, G. S. and Kodama, T. and Koide, T. and Mota, Ph.",
    title = "{Stability and Causality in relativistic dissipative hydrodynamics}",
    eprint = "0807.3120",
    archivePrefix = "arXiv",
    primaryClass = "hep-ph",
    doi = "10.1088/0954-3899/35/11/115102",
    journal = "J. Phys. G",
    volume = "35",
    pages = "115102",
    year = "2008"
}

@article{Olson:1990rzl,
    author = "Olson, Timothy S.",
    title = "{STABILITY AND CAUSALITY IN THE ISRAEL-STEWART ENERGY FRAME THEORY}",
    reportNumber = "PRINT-90-0081",
    doi = "10.1016/0003-4916(90)90366-V",
    journal = "Annals Phys.",
    volume = "199",
    pages = "18",
    year = "1990"
}

@article{Hiscock:1983zz,
    author = "Hiscock, W. A. and Lindblom, L.",
    title = "{Stability and causality in dissipative relativistic fluids}",
    doi = "10.1016/0003-4916(83)90288-9",
    journal = "Annals Phys.",
    volume = "151",
    pages = "466--496",
    year = "1983"
}

@article{Plumberg:2021bme,
    author = "Plumberg, Christopher and Almaalol, Dekrayat and Dore, Travis and Noronha, Jorge and Noronha-Hostler, Jacquelyn",
    title = "{Causality violations in realistic simulations of heavy-ion collisions}",
    eprint = "2103.15889",
    archivePrefix = "arXiv",
    primaryClass = "nucl-th",
    doi = "10.1103/PhysRevC.105.L061901",
    journal = "Phys. Rev. C",
    volume = "105",
    number = "6",
    pages = "L061901",
    year = "2022"
}

@article{Chiu:2021muk,
    author = "Chiu, Cheng and Shen, Chun",
    title = "{Exploring theoretical uncertainties in the hydrodynamic description of relativistic heavy-ion collisions}",
    eprint = "2103.09848",
    archivePrefix = "arXiv",
    primaryClass = "nucl-th",
    doi = "10.1103/PhysRevC.103.064901",
    journal = "Phys. Rev. C",
    volume = "103",
    number = "6",
    pages = "064901",
    year = "2021"
}

@article{Gavassino:2021owo,
    author = "Gavassino, Lorenzo",
    title = "{Can We Make Sense of Dissipation without Causality?}",
    eprint = "2111.05254",
    archivePrefix = "arXiv",
    primaryClass = "gr-qc",
    doi = "10.1103/PhysRevX.12.041001",
    journal = "Phys. Rev. X",
    volume = "12",
    number = "4",
    pages = "041001",
    year = "2022"
}

@article{Das:2020fnr,
    author = "Das, Arpan and Florkowski, Wojciech and Noronha, Jorge and Ryblewski, Radoslaw",
    title = "{Equivalence between first-order causal and stable hydrodynamics and Israel-Stewart theory for boost-invariant systems with a constant relaxation time}",
    eprint = "2001.07983",
    archivePrefix = "arXiv",
    primaryClass = "nucl-th",
    doi = "10.1016/j.physletb.2020.135525",
    journal = "Phys. Lett. B",
    volume = "806",
    pages = "135525",
    year = "2020"
}

@article{Keeble:2026zms,
    author = "Keeble, Lennox S. and C{\'a}rdenas-Avenda{\~n}o, Alejandro",
    title = "{Breakdown of hydrodynamic universality in causal first-order viscous neutron star oscillations}",
    eprint = "2610.21966",
    archivePrefix = "arXiv",
    primaryClass = "gr-qc",
    month = "10",
    year = "2026"
}

@article{Hoult:2021gnb,
    author = "Hoult, Raphael E. and Kovtun, Pavel",
    title = "{Causal first-order hydrodynamics from kinetic theory and holography}",
    eprint = "2112.14042",
    archivePrefix = "arXiv",
    primaryClass = "hep-th",
    doi = "10.1103/PhysRevD.106.066023",
    journal = "Phys. Rev. D",
    volume = "106",
    number = "6",
    pages = "066023",
    year = "2022"
}

@article{Rocha:2022ind,
    author = "Rocha, Gabriel S. and Denicol, Gabriel S. and Noronha, Jorge",
    title = "{Perturbative approaches in relativistic kinetic theory and the emergence of first-order hydrodynamics}",
    eprint = "2205.00078",
    archivePrefix = "arXiv",
    primaryClass = "nucl-th",
    doi = "10.1103/PhysRevD.106.036010",
    journal = "Phys. Rev. D",
    volume = "106",
    number = "3",
    pages = "036010",
    year = "2022"
}

@article{Rocha:2023hts,
    author = "Rocha, Gabriel S. and de Brito, Caio V. P. and Denicol, Gabriel S.",
    title = "{Hydrodynamic theories for a system of weakly self-interacting classical ultrarelativistic scalar particles: Microscopic derivations and attractors}",
    eprint = "2306.07423",
    archivePrefix = "arXiv",
    primaryClass = "nucl-th",
    doi = "10.1103/PhysRevD.108.036017",
    journal = "Phys. Rev. D",
    volume = "108",
    number = "3",
    pages = "036017",
    year = "2023"
}

@article{HegadeKR:2023glb,
    author = "Hegade K. R., Abhishek and Ripley, Justin L. and Yunes, Nicol{\'a}s",
    title = "{Nonrelativistic limit of first-order relativistic viscous fluids}",
    eprint = "2305.09725",
    archivePrefix = "arXiv",
    primaryClass = "gr-qc",
    doi = "10.1103/PhysRevD.107.124029",
    journal = "Phys. Rev. D",
    volume = "107",
    number = "12",
    pages = "124029",
    year = "2023"
}

@article{Bhattacharyya:2023srn,
    author = "Bhattacharyya, Sayantani and Mitra, Sukanya and Roy, Shuvayu",
    title = "{Frame transformation and stable-causal hydrodynamic theory}",
    eprint = "2312.16407",
    archivePrefix = "arXiv",
    primaryClass = "nucl-th",
    month = "12",
    year = "2023"
}

@article{Bantilan:2022ech,
    author = "Bantilan, Hans and Bea, Yago and Figueras, Pau",
    title = "{Evolutions in first-order viscous hydrodynamics}",
    eprint = "2201.13359",
    archivePrefix = "arXiv",
    primaryClass = "hep-th",
    reportNumber = "HIP-2022-1/TH",
    doi = "10.1007/JHEP08(2022)298",
    journal = "JHEP",
    volume = "08",
    pages = "298",
    year = "2022"
}

@article{Redondo-Yuste:2024vdb,
    author = "Redondo-Yuste, Jaime",
    title = "{Perturbations of relativistic dissipative stars}",
    eprint = "2411.16841",
    archivePrefix = "arXiv",
    primaryClass = "gr-qc",
    doi = "10.1088/1361-6382/adbfef",
    journal = "Class. Quant. Grav.",
    volume = "42",
    number = "7",
    pages = "075012",
    year = "2025"
}

@article{Jain:2023obu,
    author = "Jain, Akash and Kovtun, Pavel",
    title = "{Schwinger-Keldysh effective field theory for stable and causal relativistic hydrodynamics}",
    eprint = "2309.00511",
    archivePrefix = "arXiv",
    primaryClass = "hep-th",
    doi = "10.1007/JHEP01(2024)162",
    journal = "JHEP",
    volume = "01",
    pages = "162",
    year = "2024"
}

@article{Gavassino:2024ufs,
    author = "Gavassino, Lorenzo and Abboud, Nick and Speranza, Enrico and Noronha, Jorge",
    title = "{First-order relativistic hydrodynamics with an information current}",
    eprint = "2401.13852",
    archivePrefix = "arXiv",
    primaryClass = "nucl-th",
    reportNumber = "CERN-TH-2024-012",
    doi = "10.1103/PhysRevD.109.085013",
    journal = "Phys. Rev. D",
    volume = "109",
    number = "8",
    pages = "085013",
    year = "2024"
}

@article{Gavassino:2024vyu,
    author = "Gavassino, Lorenzo and Mullins, Nicki and Hippert, Mauricio",
    title = "{Consistent inclusion of fluctuations in first-order causal and stable relativistic hydrodynamics}",
    eprint = "2402.06776",
    archivePrefix = "arXiv",
    primaryClass = "nucl-th",
    doi = "10.1103/PhysRevD.109.125002",
    journal = "Phys. Rev. D",
    volume = "109",
    number = "12",
    pages = "125002",
    year = "2024"
}

@article{Panda:2026lbp,
    author = "Panda, Ankit Kumar and Biswas, Rajesh",
    title = "{Relativistic Bemfica-Disconzi-Noronha-Kovtun MHD evolution in a boost-invariant medium and its impact on dilepton production}",
    eprint = "2605.01807",
    archivePrefix = "arXiv",
    primaryClass = "nucl-th",
    doi = "10.1103/3p8k-88xr",
    journal = "Phys. Rev. D",
    volume = "114",
    number = "1",
    pages = "014035",
    year = "2026"
}

@article{Hoult:2024qph,
    author = "Hoult, Raphael E. and Kovtun, Pavel",
    title = "{Causality in dissipative relativistic magnetohydrodynamics}",
    eprint = "2411.04966",
    archivePrefix = "arXiv",
    primaryClass = "hep-th",
    doi = "10.1007/JHEP04(2025)009",
    journal = "JHEP",
    volume = "04",
    pages = "009",
    year = "2025"
}

@article{Speranza:2021bxf,
    author = "Speranza, Enrico and Bemfica, Fabio S. and Disconzi, Marcelo M. and Noronha, Jorge",
    title = "{Challenges in solving chiral hydrodynamics}",
    eprint = "2104.02110",
    archivePrefix = "arXiv",
    primaryClass = "hep-th",
    doi = "10.1103/PhysRevD.107.054029",
    journal = "Phys. Rev. D",
    volume = "107",
    number = "5",
    pages = "054029",
    year = "2023"
}

@article{Abboud:2023hos,
    author = "Abboud, Nick and Speranza, Enrico and Noronha, Jorge",
    title = "{Causal and stable first-order chiral hydrodynamics}",
    eprint = "2308.02928",
    archivePrefix = "arXiv",
    primaryClass = "hep-th",
    doi = "10.1103/PhysRevD.109.094007",
    journal = "Phys. Rev. D",
    volume = "109",
    number = "9",
    pages = "094007",
    year = "2024"
}

@article{Armas:2022wvb,
    author = "Armas, Jay and Camilloni, Filippo",
    title = "{A stable and causal model of magnetohydrodynamics}",
    eprint = "2201.06847",
    archivePrefix = "arXiv",
    primaryClass = "hep-th",
    doi = "10.1088/1475-7516/2022/10/039",
    journal = "JCAP",
    volume = "10",
    pages = "039",
    year = "2022"
}

@article{Fantini:2025gnm,
    author = "Fantini, Delfina and Rubio, Marcelo E.",
    title = "{Constraint evolution in first-order viscous relativistic fluids}",
    eprint = "2506.06430",
    archivePrefix = "arXiv",
    primaryClass = "gr-qc",
    doi = "10.1103/31vh-1fjb",
    journal = "Phys. Rev. D",
    volume = "112",
    number = "6",
    pages = "063038",
    year = "2025"
}

@article{ExTrEMe:2023nhy,
    author = "Krupczak, Renata and others",
    collaboration = "ExTrEMe",
    title = "{Causality violations in simulations of large and small heavy-ion collisions}",
    eprint = "2311.02210",
    archivePrefix = "arXiv",
    primaryClass = "nucl-th",
    doi = "10.1103/PhysRevC.109.034908",
    journal = "Phys. Rev. C",
    volume = "109",
    number = "3",
    pages = "034908",
    year = "2024"
}

@article{Bea:2023rru,
    author = "Bea, Yago and Figueras, Pau",
    title = "{Field redefinitions and evolutions in relativistic Navier-Stokes}",
    eprint = "2312.16671",
    archivePrefix = "arXiv",
    primaryClass = "hep-th",
    doi = "10.1007/JHEP11(2024)110",
    journal = "JHEP",
    volume = "11",
    pages = "110",
    year = "2024"
}

@article{Gubser:2010ze,
    author = "Gubser, Steven S.",
    title = "{Symmetry constraints on generalizations of Bjorken flow}",
    eprint = "1006.0006",
    archivePrefix = "arXiv",
    primaryClass = "hep-th",
    reportNumber = "PUPT-2340",
    doi = "10.1103/PhysRevD.82.085027",
    journal = "Phys. Rev. D",
    volume = "82",
    pages = "085027",
    year = "2010"
}

@article{Gubser:2010ui,
    author = "Gubser, Steven S. and Yarom, Amos",
    title = "{Conformal hydrodynamics in Minkowski and de Sitter spacetimes}",
    eprint = "1012.1314",
    archivePrefix = "arXiv",
    primaryClass = "hep-th",
    reportNumber = "PUPT-2358",
    doi = "10.1016/j.nuclphysb.2011.01.012",
    journal = "Nucl. Phys. B",
    volume = "846",
    pages = "469--511",
    year = "2011"
}

@article{Bjorken:1982qr,
    author = "Bjorken, J. D.",
    title = "{Highly Relativistic Nucleus-Nucleus Collisions: The Central Rapidity Region}",
    reportNumber = "FERMILAB-PUB-82-044-THY, FERMILAB-PUB-82-044-T",
    doi = "10.1103/PhysRevD.27.140",
    journal = "Phys. Rev. D",
    volume = "27",
    pages = "140--151",
    year = "1983"
}

@article{Schenke:2010nt,
    author = "Schenke, Bjoern and Jeon, Sangyong and Gale, Charles",
    title = "{(3+1)D hydrodynamic simulation of relativistic heavy-ion collisions}",
    eprint = "1004.1408",
    archivePrefix = "arXiv",
    primaryClass = "hep-ph",
    doi = "10.1103/PhysRevC.82.014903",
    journal = "Phys. Rev. C",
    volume = "82",
    pages = "014903",
    year = "2010"
}

@article{Paquet:2015lta,
    author = {Paquet, Jean-Fran\c{c}ois and Shen, Chun and Denicol, Gabriel S. and Luzum, Matthew and Schenke, Bj\"orn and Jeon, Sangyong and Gale, Charles},
    title = "{Production of photons in relativistic heavy-ion collisions}",
    eprint = "1509.06738",
    archivePrefix = "arXiv",
    primaryClass = "hep-ph",
    doi = "10.1103/PhysRevC.93.044906",
    journal = "Phys. Rev. C",
    volume = "93",
    number = "4",
    pages = "044906",
    year = "2016"
}

@article{Schenke:2010rr,
    author = "Schenke, Bjorn and Jeon, Sangyong and Gale, Charles",
    title = "{Elliptic and triangular flow in event-by-event (3+1)D viscous hydrodynamics}",
    eprint = "1009.3244",
    archivePrefix = "arXiv",
    primaryClass = "hep-ph",
    doi = "10.1103/PhysRevLett.106.042301",
    journal = "Phys. Rev. Lett.",
    volume = "106",
    pages = "042301",
    year = "2011"
}

@article{Yang:2023ogo,
    author = "Yang, Yumu and Hippert, Mauricio and Speranza, Enrico and Noronha, Jorge",
    title = "{Far-from-equilibrium bulk-viscous transport coefficients in neutron star mergers}",
    eprint = "2309.01864",
    archivePrefix = "arXiv",
    primaryClass = "nucl-th",
    doi = "10.1103/PhysRevC.109.015805",
    journal = "Phys. Rev. C",
    volume = "109",
    number = "1",
    pages = "015805",
    year = "2024"
}

@article{Yang:2025yoo,
    author = "Yang, Yumu and Hippert, Mauricio and Speranza, Enrico and Noronha, Jorge",
    title = "{Symmetry-Energy Dependence of the Bulk Viscosity of Nuclear Matter}",
    eprint = "2504.07805",
    archivePrefix = "arXiv",
    primaryClass = "nucl-th",
    reportNumber = "CERN-TH-2025-078",
    doi = "10.1103/9nck-dm66",
    journal = "Phys. Rev. Lett.",
    volume = "135",
    number = "22",
    pages = "222702",
    year = "2025"
}

@article{HegadeKR:2026iou,
    author = "Hegade K. R., Abhishek and Yang, Yumu and Hippert, Mauricio and Noronha-Hostler, Jacquelyn and Noronha, Jorge and Yunes, Nicol{\'a}s",
    title = "{Dynamical tidal response of neutron stars as a probe of dense-matter properties}",
    eprint = "2603.26886",
    archivePrefix = "arXiv",
    primaryClass = "gr-qc",
    doi = "10.1103/92sp-krfc",
    journal = "Phys. Rev. D",
    volume = "114",
    number = "4",
    pages = "044014",
    year = "2026"
}

@article{Ripley:2023lsq,
    author = "Ripley, Justin L. and Hegade K. R., Abhishek and Chandramouli, Rohit S. and Yunes, Nicolas",
    title = "{A constraint on the dissipative tidal deformability of neutron stars}",
    eprint = "2312.11659",
    archivePrefix = "arXiv",
    primaryClass = "gr-qc",
    doi = "10.1038/s41550-024-02323-7",
    journal = "Nature Astron.",
    volume = "8",
    number = "10",
    pages = "1277--1283",
    year = "2024"
}

@article{Gavassino:2023xkt,
    author = "Gavassino, Lorenzo and Noronha, Jorge",
    title = "{Relativistic bulk rheology: From neutron star mergers to viscous cosmology}",
    eprint = "2305.04119",
    archivePrefix = "arXiv",
    primaryClass = "gr-qc",
    doi = "10.1103/PhysRevD.109.096040",
    journal = "Phys. Rev. D",
    volume = "109",
    number = "9",
    pages = "096040",
    year = "2024"
}

@article{Rocha:2023ilf,
    author = "Rocha, Gabriel S. and Wagner, David and Denicol, Gabriel S. and Noronha, Jorge and Rischke, Dirk H.",
    title = "{Theories of Relativistic Dissipative Fluid Dynamics}",
    eprint = "2311.15063",
    archivePrefix = "arXiv",
    primaryClass = "nucl-th",
    doi = "10.3390/e26030189",
    journal = "Entropy",
    volume = "26",
    number = "3",
    pages = "189",
    year = "2024"
}

@article{Gavassino:2023odx,
    author = "Gavassino, Lorenzo and Disconzi, Marcelo M. and Noronha, Jorge",
    title = "{Universality Classes of Relativistic Fluid Dynamics: Foundations}",
    eprint = "2302.03478",
    archivePrefix = "arXiv",
    primaryClass = "nucl-th",
    doi = "10.1103/PhysRevLett.132.222302",
    journal = "Phys. Rev. Lett.",
    volume = "132",
    number = "22",
    pages = "222302",
    year = "2024"
}

@article{Cordeiro:2023ljz,
    author = "Cordeiro, Ian and Speranza, Enrico and Ingles, Kevin and Bemfica, F{\'a}bio S. and Noronha, Jorge",
    title = "{Causality Bounds on Dissipative General-Relativistic Magnetohydrodynamics}",
    eprint = "2312.09970",
    archivePrefix = "arXiv",
    primaryClass = "astro-ph.HE",
    reportNumber = "CERN-TH-2023-215",
    doi = "10.1103/PhysRevLett.133.091401",
    journal = "Phys. Rev. Lett.",
    volume = "133",
    number = "9",
    pages = "091401",
    year = "2024"
}

@article{Mullins:2023ott,
    author = "Mullins, Nicki and Hippert, Mauricio and Gavassino, Lorenzo and Noronha, Jorge",
    title = "{Relativistic hydrodynamic fluctuations from an effective action: Causality, stability, and the information current}",
    eprint = "2309.00512",
    archivePrefix = "arXiv",
    primaryClass = "hep-th",
    doi = "10.1103/PhysRevD.108.116019",
    journal = "Phys. Rev. D",
    volume = "108",
    number = "11",
    pages = "116019",
    year = "2023"
}

@article{Bemfica:2022dnk,
    author = "Bemfica, F{\'a}bio S. and Disconzi, Marcelo M. and Noronha, Jorge and Scherrer, Robert J.",
    title = "{Cosmological consequences of first-order general-relativistic viscous fluid dynamics}",
    eprint = "2210.13372",
    archivePrefix = "arXiv",
    primaryClass = "gr-qc",
    doi = "10.1103/PhysRevD.107.023512",
    journal = "Phys. Rev. D",
    volume = "107",
    number = "2",
    pages = "023512",
    year = "2023"
}

@article{Noronha:2021syv,
    author = "Noronha, Jorge and Spali{\'n}ski, Micha{\l} and Speranza, Enrico",
    title = "{Transient Relativistic Fluid Dynamics in a General Hydrodynamic Frame}",
    eprint = "2105.01034",
    archivePrefix = "arXiv",
    primaryClass = "nucl-th",
    doi = "10.1103/PhysRevLett.128.252302",
    journal = "Phys. Rev. Lett.",
    volume = "128",
    number = "25",
    pages = "252302",
    year = "2022"
}

@article{Noronha-Hostler:2014dqa,
    author = "Noronha-Hostler, J. and Noronha, J. and Grassi, F.",
    title = "{Bulk viscosity-driven suppression of shear viscosity effects on the flow harmonics at energies available at the BNL Relativistic Heavy Ion Collider}",
    eprint = "1406.3333",
    archivePrefix = "arXiv",
    primaryClass = "nucl-th",
    doi = "10.1103/PhysRevC.90.034907",
    journal = "Phys. Rev. C",
    volume = "90",
    number = "3",
    pages = "034907",
    year = "2014"
}

@article{Noronha-Hostler:2013gga,
    author = "Noronha-Hostler, Jacquelyn and Denicol, Gabriel S. and Noronha, Jorge and Andrade, Rone P. G. and Grassi, Frederique",
    title = "{Bulk Viscosity Effects in Event-by-Event Relativistic Hydrodynamics}",
    eprint = "1305.1981",
    archivePrefix = "arXiv",
    primaryClass = "nucl-th",
    doi = "10.1103/PhysRevC.88.044916",
    journal = "Phys. Rev. C",
    volume = "88",
    number = "4",
    pages = "044916",
    year = "2013"
}

@article{Pala:2025qoa,
    author = "Pala, Kevin P. and others",
    title = "{NuclearConfectionery: Multi-stage Simulation Framework for Modeling Relativistic Heavy-ion Collisions}",
    eprint = "2511.22852",
    archivePrefix = "arXiv",
    primaryClass = "nucl-th",
    month = "11",
    year = "2025"
}

@article{Song:2007ux,
    author = "Song, Huichao and Heinz, Ulrich W.",
    title = "{Causal viscous hydrodynamics in 2+1 dimensions for relativistic heavy-ion collisions}",
    eprint = "0712.3715",
    archivePrefix = "arXiv",
    primaryClass = "nucl-th",
    reportNumber = "CERN-PH-TH-2007-269",
    doi = "10.1103/PhysRevC.77.064901",
    journal = "Phys. Rev. C",
    volume = "77",
    pages = "064901",
    year = "2008"
}

@article{Karpenko:2013wva,
    author = "Karpenko, Iu. and Huovinen, P. and Bleicher, M.",
    title = "{A 3+1 dimensional viscous hydrodynamic code for relativistic heavy ion collisions}",
    eprint = "1312.4160",
    archivePrefix = "arXiv",
    primaryClass = "nucl-th",
    doi = "10.1016/j.cpc.2014.07.010",
    journal = "Comput. Phys. Commun.",
    volume = "185",
    pages = "3016--3027",
    year = "2014"
}

@article{DelZanna:2013eua,
    author = "Del Zanna, L. and Chandra, V. and Inghirami, G. and Rolando, V. and Beraudo, A. and De Pace, A. and Pagliara, G. and Drago, A. and Becattini, F.",
    title = "{Relativistic viscous hydrodynamics for heavy-ion collisions with ECHO-QGP}",
    eprint = "1305.7052",
    archivePrefix = "arXiv",
    primaryClass = "nucl-th",
    doi = "10.1140/epjc/s10052-013-2524-5",
    journal = "Eur. Phys. J. C",
    volume = "73",
    pages = "2524",
    year = "2013"
}

@article{Song:2010aq,
    author = "Song, Huichao and Bass, Steffen A. and Heinz, Ulrich",
    title = "{Viscous QCD matter in a hybrid hydrodynamic+Boltzmann approach}",
    eprint = "1012.0555",
    archivePrefix = "arXiv",
    primaryClass = "nucl-th",
    doi = "10.1103/PhysRevC.83.024912",
    journal = "Phys. Rev. C",
    volume = "83",
    pages = "024912",
    year = "2011"
}

@article{Shen:2014vra,
    author = "Shen, Chun and Qiu, Zhi and Song, Huichao and Bernhard, Jonah and Bass, Steffen and Heinz, Ulrich",
    title = "{The iEBE-VISHNU code package for relativistic heavy-ion collisions}",
    eprint = "1409.8164",
    archivePrefix = "arXiv",
    primaryClass = "nucl-th",
    doi = "10.1016/j.cpc.2015.08.039",
    journal = "Comput. Phys. Commun.",
    volume = "199",
    pages = "61--85",
    year = "2016"
}

@article{Pang:2018zzo,
    author = "Pang, Long-Gang and Petersen, Hannah and Wang, Xin-Nian",
    title = "{Pseudorapidity distribution and decorrelation of anisotropic flow within the open-computing-language implementation CLVisc hydrodynamics}",
    eprint = "1802.04449",
    archivePrefix = "arXiv",
    primaryClass = "nucl-th",
    doi = "10.1103/PhysRevC.97.064918",
    journal = "Phys. Rev. C",
    volume = "97",
    number = "6",
    pages = "064918",
    year = "2018"
}

@article{Du:2019obx,
    author = "Du, Lipei and Heinz, Ulrich",
    title = "{(3+1)-dimensional dissipative relativistic fluid dynamics at non-zero net baryon density}",
    eprint = "1906.11181",
    archivePrefix = "arXiv",
    primaryClass = "nucl-th",
    doi = "10.1016/j.cpc.2019.107090",
    journal = "Comput. Phys. Commun.",
    volume = "251",
    pages = "107090",
    year = "2020"
}

@article{Marrochio:2013wla,
    author = "Marrochio, Hugo and Noronha, Jorge and Denicol, Gabriel S. and Luzum, Matthew and Jeon, Sangyong and Gale, Charles",
    title = "{Solutions of Conformal Israel-Stewart Relativistic Viscous Fluid Dynamics}",
    eprint = "1307.6130",
    archivePrefix = "arXiv",
    primaryClass = "nucl-th",
    doi = "10.1103/PhysRevC.91.014903",
    journal = "Phys. Rev. C",
    volume = "91",
    number = "1",
    pages = "014903",
    year = "2015"
}

@article{Armas:2020mpr,
    author = "Armas, Jay and Jain, Akash",
    title = "{Effective field theory for hydrodynamics without boosts}",
    eprint = "2010.15782",
    archivePrefix = "arXiv",
    primaryClass = "hep-th",
    doi = "10.21468/SciPostPhys.11.3.054",
    journal = "SciPost Phys.",
    volume = "11",
    number = "3",
    pages = "054",
    year = "2021"
}

\end{document}